%% file: arxiv-NG-VG-RR-SJ.tex
\documentclass{memo-l}

\PassOptionsToPackage{linktocpage}{hyperref}

\newtheorem{theorem}{Theorem}[chapter]
\newtheorem{lemma}[theorem]{Lemma}

\theoremstyle{definition}
\newtheorem{definition}[theorem]{Definition}

\usepackage{amsthm}
\usepackage{thmtools}

\theoremstyle{definition}
\declaretheorem[name=Remark,numberlike=theorem,qed=$\triangle$]{remark}

\theoremstyle{definition}
\declaretheorem[
name=Example,
numberlike=theorem,
qed=$\triangle$]{example}

\theoremstyle{theorem}
\newtheorem{corollary}[theorem]{Corollary}
\newtheorem{proposition}[theorem]{Proposition}

\theoremstyle{definition}
\newcounter{openproblemcounter}
\newtheorem{openproblem}[openproblemcounter]{Open problem}

\numberwithin{section}{chapter}
\numberwithin{equation}{chapter}

\usepackage[dvipsnames, table]{xcolor}
\usepackage{comment}
\usepackage{macros, StJohnMacros}

\usepackage{booktabs}
\usepackage{threeparttable}
\usepackage{amsmath}
\usepackage{amsthm}
\usepackage{thmtools}

\usepackage{lineno}
\usepackage{amsfonts}

\usepackage{bussproofs}
\usepackage{amsmath}
\usepackage{multirow}
\usepackage{makecell}
\usepackage{lscape}
\usepackage{stmaryrd}
\usepackage{graphicx}
\usepackage{comment}
\usepackage{mdframed}
\usepackage{enumerate}
\usepackage{pdflscape}
\usepackage{tikz}
\usepackage{float}

\usepackage{amssymb}

\usepackage[graphicx]{realboxes}

\usepackage{hyperref}

\usetikzlibrary{fit,shapes.geometric}

\numberwithin{section}{chapter}
\numberwithin{equation}{chapter}

\makeindex

\begin{document}

\frontmatter

\title{Complexities of Well-Quasi-Ordered\\Substructural Logics}

%    Remove any unused author tags.
%    author one information
% AUTHORS WERE REMOVED FOR THE SUBMISSION VERSION
\begin{comment}
\author{Nikolaos Galatos}
\address{Department of Mathematics, University of Denver, 2390 S. York St., Denver, 80208, CO, USA}
\curraddr{}
\email{ngalatos@du.edu}
\thanks{}

%    author two information
\author{Vitor Greati}
\address{Bernoulli Institute, University of Groningen, Nijenborgh 9, 9747 AG, Groningen, The Netherlands}
\curraddr{}
\email{v.rodrigues.greati@rug.nl}
\thanks{}

%    author two information
\author{Revantha Ramanayake}
\address{Bernoulli Institute \& CogniGron, University of Groningen, Nijenborgh 9, 9747 AG, Groningen, The Netherlands}
\curraddr{}
\thanks{}

%    author two information
\author{Gavin St.\,John}
\address{Department of Mathematics and Computer Science, University of Cagliari, Via Università 40, Cagliari, 09124, CA, Italy}
\curraddr{}
\email{gavinstjohn@gmail.com}
\thanks{}
\end{comment}

%    \date is required; it is the date received by the editor.
\date{10 April 2025}

%    Recognition of the 2010 edition of the Mathematics Subject
%    Classification requires a version of amsbook.cls from July 2009
%    or later.  If "2010" is not recognized, please upgrade.


%\dedicatory{Dedication text (use \\[2pt] for line break if necessary)}

\begin{titlepage}
    \centering
    \vspace*{1in}
    
    % Title
    {\Huge\bfseries Complexities of Well-Quasi-Ordered\\
    Substructural Logics \par}
    \vspace{1.5cm}
    
    % Author name
    \vspace{0.5cm}
    {\Large 
    Nikolaos Galatos\textsuperscript{1}\\
    Vitor Greati\textsuperscript{2}\\
    Revantha Ramanayake\textsuperscript{2,3}\\
    Gavin St.\,John\textsuperscript{4}\\
    \par}

    \vfill 

    {
    {\small 
    \textsuperscript{1}%
    Department of Mathematics, University of Denver, USA\\
    \texttt{ngalatos@du.edu}
    }\\[1em]
    {\small 
    \textsuperscript{2}%
    Bernoulli Institute, University of Groningen, The Netherlands\\
    \textsuperscript{3}%
    CogniGron, University of Groningen, The Netherlands\\
    \texttt{v.rodrigues.greati@rug.nl};
\texttt{d.r.s.ramanayake@rug.nl}
    }\\[1em]
    {\small
    \textsuperscript{4}%
    Department of Mathematics and Computer Science, University of Cagliari, Italy\\
    \texttt{gavinstjohn@gmail.com}
    }
    }

    \vfill 
    
    % Institution / Affiliation (optional)
    {\large Preprint submitted to arXiv\par}
    \vspace{0.3cm}
    
    % Date
    {\large \today\par}
    
\end{titlepage}

\let\cleardoublepage\clearpage
\begin{center}
{\Large \textbf{Abstract}}
\end{center}
\vspace{1em}
\input{tex/abstract}

\tableofcontents

%    Include unnumbered chapters (preface, acknowledgments, etc.) here.
\chapter*{Introduction}
\label{s: introduction}
\input{tex/introduction}

\mainmatter
%    Include main chapters here.
\chapter{Preliminaries}
\label{sec:preliminaries}
\input{tex/preliminaries}

\chapter{Finite embeddability property for $\mathcal{P}_3^\flat$ axioms}
\label{sec:fepP3}
\input{tex/fepP3}

\chapter{Knotted well-quasi-orders and fast-growing complexities}
\label{sec:knotted-wqos}
\input{tex/knotted-wqos}

\chapter{Upper bounds for commutative logics with knotted contraction}
\label{sec:ub-wc}
\input{tex/weak-contraction-ub}

\chapter{Upper bounds for commutative logics with knotted weakening}
\label{sec:ub-ww}
\input{tex/weak-weakening-ub}

\chapter{Upper bounds for weakly commutative logics with knotted axioms}
\label{sec:noncom-ub}
\input{tex/noncommutative-upper-bounds}

\chapter{Upper bounds for noncommutative integral logics}
\label{sec:noncom-int-ub}
\input{tex/noncommutative-integral-ub}

\chapter{Algebraic counter machines}
\label{sec:lowerbounds}
\input{tex/lb-algebraic-machines}

\chapter{Lower (and tight) bounds for joinand-increasing equations}
\label{s: joinand-incresing-lb}
\input{tex/lb-joinand-increasing}

\chapter{Lower bounds for joinand-decreasing equations}
\label{s: joinand-decreasing}
\input{tex/lb-joinand-decreasing}

\chapter{Conclusion}
\label{s: conclusion}
\input{tex/conclusion}

\begin{landscape}
\input{tex/contrib-table-new}
\end{landscape}
\clearpage  %This prevents the figure showing up in the references in case !tbh doesn't place it where we want

% VG: I am commenting out appendices
%\appendix
%    Include appendix "chapters" here.
%\include{}

\backmatter
%    Bibliography styles amsplain or author-year (using natbib) are
%    also acceptable.
%\bibliographystyle{amsplain}

\bibliographystyle{amsalpha}
\bibliography{cas-refs}
%    See note above about multiple indexes.
%\printindex

\end{document}

%% file: tex/abstract.tex
Substructural logics are formal logical systems that omit familiar structural rules of classical and intuitionistic logic such as contraction, weakening, exchange (commutativity), and associativity. 
This leads to a resource-sensitive logical framework that has proven influential beyond mathematical logic and its algebraic semantics, across theoretical computer science, linguistics, and philosophical logic.
The set of theorems---i.e., the set of formulas provable in the formal logical system---of a substructural logic are recursively enumerable and, in many cases, recursive. These logics also possess an intricate mathematical structure that has been the subject of research for over six decades.

We undertake a comprehensive study of substructural logics possessing an underlying well-quasi-order (wqo), using established ordinal-indexed fast-growing complexity classes %of \cite{schmitz2016hierar} 
to classify the complexity of their deducibility (quasiequational) and provability (equational) problems. This includes substructural logics with weak variants of contraction and weakening, 
and logics with weak or even no exchange. We further consider infinitely many axiomatic extensions over the base systems.

We establish a host of decidability and complexity bounds---many of them tight---by developing new techniques in proof theory, well-quasi-order theory (contributing new length theorems), the algebraic semantics of substructural logics via residuated lattices, algebraic proof theory, and novel encodings of counter machines. Classifying the computational complexity of substructural logics (and the complexity of the word problem and of the equational theory of their algebraic semantics, as in~\cite{Andrka1997}) reveals how subtle variations in their design influence their algorithmic behavior, with the decision problems often reaching Ackermannian or even hyper-Ackermannian complexity.

Our work echoes classical results relating undecidability with first-order logic, charting the realm of high computational complexity that lies just within the boundary of decidability. The low (primitive recursive) complexity classes on one hand, and the classifications of undecidability via the arithmetical hierarchy and relative computability on the other, are well studied and populated by concrete decision problems. Our results contribute natural logical problems that inhabit the gap between them.

%% file: tex/introduction.tex
Classical logic studies the notion of \emph{truth}, familiar  through its Boolean truth-table semantics for example, where statements are assigned values of either true or false and logical connectives are interpreted accordingly. A familiar feature of reasoning in classical logic is that assumptions in an implication may be freely reused or introduced.
 The ability to reuse assumptions is known as \emph{contraction} ($\mathbf{c}$), while the option to introduce additional assumptions is called \emph{weakening} ($\mathbf{w}$), following Gentzen's  terminology  introduced in his seminal work~\cite{Gentzen1935} from the 1930's (see \cite[Ch.~3]{Gen69} for an English translation). Additionally, the order in which assumptions are used is considered irrelevant, {giving rise to the property of} \emph{exchange} ($\mathbf e$). 
These three properties also hold in intuitionistic logic, where the central notion is that of \emph{constructive provability}, instead of truth, making the logic suitable for the foundation of constructive mathematics and theoretical aspects of  computer science. The standard proof-theoretic formulations of classical and intuitionistic logic---via the calculi $\mathsf{LK}$ and $\mathsf{LJ}$, respectively---typically incorporate these three properties implicitly. However, they can also be expressed explicitly as \emph{structural rules}, as in Gentzen’s original formulation~\cite{Gentzen1935}.
While the assumptions underlying these properties may seem intuitive and straightforward, they fail to hold in many contexts where truth or constructive provability are not the only objects of study. For instance, when reasoning about consumable resources such as time, energy, or computational steps, duplicating or ignoring resources is illogical, and the order in which they are used may be crucial. Also, when considering pieces of information, the information content of two confirming sources is often not the same as that of a single source. {As a result, in  resource-sensitive situations we are often required to abandon} some of the traditional structural rules. In such settings, we must distinguish between the usual, additive, extensional, idempotent, lattice-theoretic conjunction $\phi \mt \psi$ and the multiplicative, intentional, strong conjunction $\phi \cdot \psi$.

Logics that omit any of the three familiar structural rules are known as \emph{substructural logics}\footnote{One could also remove \emph{associativity}, but this goes beyond the scope of this work.} and they formalize the concept of reasoning in resource-sensitive contexts (interpreted in a broad sense). Examples of such logics (such as \L ukasiewicz many-valued logic) date back to the early 1900s, and they were studied individually ever since. Over the past decades, the diverse approaches to substructural logics, along with the theoretical challenges and the applications they present, have garnered significant attention from researchers in fields such as classical algebra, model theory, order algebra, proof theory, theoretical computer science, quantum physics, engineering, linguistics, and philosophy. The unifying designation \emph{substructural} first appears in Do\v{s}en~\cite{dosen1993} in 1990. 
Since then, a number of influential textbooks on substructural logics have been published, such as \cite{Pa, Restall, GalJipKowOno07, MPT, Ono2019}.

The algebraic semantics of substructural logics, known as \emph{residuated lattices}, have their independent history rooted in areas such as ring theory (as algebras of ideals) \cite{Di, Wa, WD}, order algebra (lattice-ordered groups and $f$-rings) \cite{AF, KM, Da}, point-free topology (as frames and locales) \cite{Jo, PP}, and in the study of relation and cylindric algebras (as introduced by Tarski) \cite{Ta, HT, HMT}.
Moreover, the same proof-theoretic formalisms that arise in the study of substructural logics also appear in the study of mathematical linguistics (both theoretical and applied) and are intimately linked to the expressive power of automata and context-free grammars, but also to the parsing of sentences in natural languages \cite{vBe, La1, La2, Bu1, Bu2}. 
In philosophy, various reasoning systems that attempt to capture the notion of relevance between assumptions and conclusion also end up being formalizable as substructural logics \cite{AB1, AB2, Be}. 
The study of logics with multiple truth values \cite{Mu, Ha} is inspired by combining information for many agents, as well as from applications of many-valued reasoning in engineering. 
Connections to quantum physics include the notions of a quantic nucleus and of a phase space, as well as the quantale of closed left ideals of a noncommutative C$^*$-algebra~\cite{Mu1, Mu2, Mu3}. 
Finally, the applications to computer science are abundant.
Via linear logic~\cite{Gi,llcs2004,Troelstra1992}---modalities over a substructural logic---we see the genesis and proliferation of the functional programming paradigm and the development of various programming languages \cite{darlington1993,brauner1997,Mackie1994,DellaRocca1997}; via relational methods they have been useful in the formalization of pre and postconditions for the sequential execution of programs \cite{dexter2003,Sedlr2024,Bochman2011} or in adding tests for specification and verification \cite{KoKA, KoAA, Pratt}, and in 
studies in mathematical morphology \cite{Morphology, Stell};
 via separation logic they have been instrumental in the study of dynamic memory allocation and pointer arithmetic for concurrent programming over a shared memory space \cite{reynolds2002,Pym, OPym}, leading to industry-level static-analysis tools, such as Infer~\cite{CalDisOHeYan2011,OHe2018}.

The central concern of this work is the computational complexity of the consequence relations associated with a wide range of substructural logics. A prominent decision problem in this direction is the provability problem: determining whether a given statement (formula) is a theorem of the logic. The provability problem is well-known to be \coNP-complete for classical propositional logic and \PSPACE-complete~\cite{statman1979} for intuitionistic propositional logic.
In contrast, many foundational substructural logics exhibit complexities that are complete for \emph{fast-growing complexity classes}~\cite{schmitz2016hierar}. Hence, although the problems are decidable, the associated decision procedures are non-primitive recursive. As we shall see, provability for intuitionistic propositional logic without weakening already serves as an example.
This means that their running time---as a function of the input length---grows faster than any polynomial, exponential, or even towers of exponentials of variable height, making them significantly more complex to compute.
Moreover, slight variations of these highly complex yet decidable substructural logics often lead to undecidability---for instance, provability becomes undecidable in intuitionistic propositional logic without weakening and exchange~\cite{horcik2016}. Many of the substructural logics in this work are situated just within the boundary of decidability, in the realm of non-primitive recursive complexities.

In proof-theoretic terms, starting from Gentzen's 1936 proof calculus $\mathsf{LJ}$ for intuitionistic propositional logic~\cite[Ch.~3]{Gen69}, we obtain the foundational substructural logic known as the Full Lambek calculus ($\mathbf{FL}$)  by removing the structural rules of exchange, contraction, and weakening. 
This ends up extending \emph{Lambek's calculus} of syntactic types~\cite{lambek1958} by incorporating lattice connectives and their corresponding rules. 
Through language expansion or axiomatic extension of $\mathbf{FL}$, one can generate many well-studied families of logics, such as linear logic, many-valued logics, and relevant logics.  
Here we will study infinitely many axiomatic extensions of $\mathbf{FL}$, of various sorts.
An alternative, and crucial, characterization of substructural logics for our investigation is as logics whose models are classes of \emph{pointed residuated lattices}, or \emph{$\m{FL}$-algebras}.
Such classes form \emph{(quasi-)varieties}, i.e., classes of algebras axiomatized by \emph{(quasi-)equations}; these correspond to the axiomatization of the logic via an algebraization result. 
From this algebraic perspective, our results determine the computational complexity of deciding the word problem, as well as the quasiequational and equational theories of these varieties.
Indeed, the aforementioned proof theory can be seen from this perspective as the combinatorial study of words (or terms) in the free algebra.

We establish complexity upper bounds by applying \emph{well-quasi-ordering theory} to proof-theoretic deductive systems for the logics, while lower bounds are derived using \emph{algebraic counter machines}.
The latter approach enables us to leverage advanced techniques involving \emph{residuated frames} and their dual algebras constructed from the associated Galois connections.
Additionally, we prove the \emph{finite embeddability property} for these classes of algebras, which in turn ensures the finite model property for the corresponding logics.

\section*{An overview of the logics under consideration}

In the context of this work, a logical system, or derivation system, or a \emph{logic} is defined via a \emph{consequence
relation} $\vdash$ over a propositional language. 
The decision problem for $\vdash$ asks whether there exists an algorithm that, given a finite set~$\Gamma$ of formulas and a formula $\varphi$, decides whether $\varphi$ is a consequence of (or deducible from) $\Gamma$ (i.e., whether $\Gamma \vdash \varphi$); this is referred to as the \emph{deducibility problem} for $\vdash$. 
A notable special case is the \emph{provability problem} for 
$\vdash$, which asks whether theoremhood---i.e., the set of formulas $\UniFmA$ such that $\varnothing \vdash \UniFmA$---is decidable.
In classical and intuitionistic logic, these problems are equivalent since the deduction theorem reduces an instance of deducibility to a (similar sized) instance of provability.
However, the deduction theorem fails in many substructural logics. For example, in the absence of contraction, one encounters a fundamental obstacle when attempting to encode a deduction witnessing $\Gamma \vdash \varphi$ as a single formula (i.e., a theorem) whose size is bounded by a fixed function. In such cases, deducibility can exhibit significantly higher computational complexity than provability, or even become undecidable.

We investigate axiomatic extensions of $\mathbf{FL}$ by considering weaker forms of exchange, contraction, and weakening, along with further extensions by axioms up to the level $\mathcal{P}_3^\flat$ in the \emph{substructural hierarchy}~\cite{CiaGalTer08}. 
This hierarchy organizes axioms according to the alternation of 
logical connectives in positive and negative position (at each position, a logical connective is either invertible---it decomposes into its immediate subformulas without any loss of logical information---or not, so the alternation of positive and negative connectives relates to the proof-theoretic structure that is needed to obtain a reasonable proof-system). In that sense, it parallels the arithmetical hierarchy, which is based on alternations of quantifiers.
The axioms within this level allow us to reach many logics of specific and independent interest, e.g., 
monoidal t-norm based logic $\m{MTL}$~\cite{EstGod01}, a fundamental logic in the study of mathematical fuzzy logics, as well as its extensions that have received attention (see e.g., \cite{HorNogPet07,NogEstGis08,EstGodNog10,CiaMetMon10}).
A notable axiom that lies at the subsequent level (and hence beyond our reach) is distributivity. 
A compelling external reason why our results do not extend to distributivity is its tendency to lead to undecidability in substructural settings:  %; (see e.g.,~\cite{Urq1984}).
\cite{Urq1984} shows that adding distributivity to $\m {FL}$ together with exchange (or even further also with contraction) leads to undecidability of deducibility, \cite{JT02} observe that adding distributivity to $\m {FL}$ by itself has the same effect, \cite{Ga02b} proves that the same holds for all logics between these two extremes, and \cite{GR04} shows that this holds even if we add involutivity to distributivity and exchange.

The modified versions of contraction and weakening we will consider are known as \emph{knotted axioms} and have received significant attention in the literature~\cite{vanalten2005,hori1994,horcik2011,gavin2019}; in particular, they strike a balance between full/strict control of the resources and no control at all, by providing different levels of {freedom} depending on the application. 
Contraction corresponds to the axiom $\UniFmA \to \UniFmA \cdot \UniFmA$ (or, equivalently, to 
$(\UniFmA \to (\UniFmA\to \UniFmB)) \to (\UniFmA \to \UniFmB)$), while weakening\footnote{This axiom is \emph{integrality} but we will call it \emph{weakening} here, to simplify the presentation.} corresponds to $\UniFmA \to 1$. 
A knotted axiom takes the form $\UniFmA^m \to \UniFmA^n$, for some fixed $m,n \in \mathbb{N}$; here the values of $m$ and $n$ regulate the level of control we want to delegate to strong conjunction. 
Specifically, when $m < n$, we have \emph{knotted contraction} axioms, and when $m > n$, we have \emph{knotted weakening} axioms.

Regarding weak exchange axioms, the simplest example is the axiom $\UniFmA \UniFmA \UniFmB \leftrightarrow \UniFmA  \UniFmB \UniFmA$, which modifies the standard exchange axiom $\UniFmA \UniFmB \leftrightarrow \UniFmB \UniFmA$ so that a leftmost occurrence of $\UniFmA$ is required to permit $\UniFmA$ and $\UniFmB$ to swap positions; in that sense, this axiom can be  viewed as a conditional form of exchange and it is just one instance of the infinitely many forms of weak exchange that we will consider. 
While these axioms have been extensively studied from an algebraic perspective~\cite{Cardona2015}, their computational complexity has remained unexplored.

Each choice of weak exchange and knotted axiom over $\mathbf{FL}$ generates a distinct base substructural logic, resulting in an infinite number of such logics. We systematically explore their computational properties---along with those of a broad class of their axiomatic extensions---using a uniform approach.

As we will show, most of these logics have non-primitive-recursive complexity, so we employ a classification framework beyond the familiar primitive recursive complexity classes. 
In particular, we utilize the ordinal-indexed family of complexity classes $\{ \mathbf{F}_\alpha \}_{\alpha < \varepsilon_0}$, introduced by Schmitz~\cite{schmitz2016hierar}, where $\EpsZero$ is the smallest epsilon number (see Section~\ref{s: lenght theorems} for the formal definition). 
Within this hierarchy, problems of elementary complexity fall under $\mathbf{F}_2$, while $\mathbf{F}_3$ corresponds to non-elementary problems (also known as the \TOWER class). 
The classes $\mathbf{F}_{k}$ with $k < \omega$ capture primitive recursive problems, and $\mathbf{F}_\omega$ marks the first class of non-primitive recursive problems, often referred to as \ACK because the complexity of its problems is controlled by a composition of a single application of an Ackermann function and any number of primitive recursive functions.
As one ascends the hierarchy further, the complexity of problems increases significantly. 
This classification allows us to rigorously situate the computational difficulty of the logics we study within a precise framework.

A milestone in the computational study of logical systems with fast-growing complexities was achieved by Urquhart~\cite{urquhart1999} in 1999, when he proved that $\mathbf{FL_{ec}}$ (i.e., $\mathbf{FL}$ with exchange and contraction) is \ACK-complete. 
Since then, some consequence relations within the family of substructural logics have been proven to be \ACK-hard (St.\,John~\cite{gavin2019}), though \ACK-completeness has been established only rarely; for example, Greati and Ramanayake~\cite{GreatiRamanayake2024} recently showed $\mathbf{F}_{\omega^\omega}$-completeness, or hyper-Ackermannian completeness, for deducibility in $\mathbf{FL}$ with weakening. Meanwhile, slight variations of these logics, such as $\mathbf{FL_{c}}$ (Chvalovsk\'y and Hor\v{c}\'{\i}k~\cite{horcik2016}) and infinitely many extensions of $\mathbf{FL_{e}}$ (Galatos and St.\,John~\cite{galatos2022}) are in fact undecidable. 
In this work, we show that \ACK-completeness actually arises in infinitely many consequence relations, specifically in the presence of weak exchange and knotted axioms. 
Furthermore, we reveal that the complexity can be significantly worse when no form of exchange is present.

A summary of the results presented in this work appears in Table~\ref{tab:table-contrib} (page~\pageref{tab:table-contrib}).

\section*{Methods for proving upper and lower bounds}

To establish upper bounds, we utilize \emph{Proof Theory}, an approach that investigates logical systems presented via \emph{proof calculi}, allowing for a detailed analysis of derivations/\emph{proofs}.
The decision procedure systematically explores the space of potential minimal proofs, a process known as \emph{proof search}. 
This method leverages the fact that the substructural logics under study possess well-behaved proof calculi, all of which enjoy the \emph{subformula property}.
To identify these suitable analytic calculi on the substructural hierarchy, we use the results of \cite{CiaGalTer08, CiaGalTer12, CiaGalTer17} on \emph{algebraic proof theory}; the objects manipulated by the calculi are \emph{sequents} or \emph{hypersequents}. 
Crucially, the subformula property restricts the proof-search space to objects involving only the finitely many subformulas of the input; the multiplicity of occurrences of formulas within {these objects in} the proof search is still unbounded.

A key component in achieving termination and obtaining run time estimates is the use of \emph{well-quasi-ordered} (wqo) sets; these are mathematical structures that have been successfully applied, among many other areas, to computer science in order to prove decidability and analyze fast-growing complexity in various settings, such as program verification~\cite{finkel2001}. 
Every sequence of elements from a wqo set that lacks increasing pairs---i.e., no element in larger than a prior element in the sequence---is finite.
A sequence with this property is called a \emph{bad sequence}.
We will define wqos of sequents and hypersequents, and show that it will be enough to consider proofs whose branches are bad sequences, thereby guaranteeing that proof search will terminate. 
Additionally, bounding the length of such bad sequences (\emph{controlled bad sequences}) allows us to upper bound the size of the proof-search space, leading to complexity upper bounds.

Urquhart implicitly employed controlled bad sequences in his proof of the Ackermannian complexity of $\mathbf{FL_{ec}}$, while Balasubramanian \textit{et al.}~\cite{BalLanRam21LICS} and Ramanayake~\cite{revantha2020} extended these methods to obtain upper bounds for many axiomatic extensions of $\mathbf{FL_{ec}}$ and $\mathbf{FL_{ew}}$---here $\mathbf{FL_{ew}}$ is the substructural version of intuitionistic logic that rejects contraction rather than weakening. 
However, these proofs rely on the presence of exchange ($\mathbf{e}$) and either contraction ($\mathbf{c}$) or weakening ($\mathbf{w}$).
Since the logics we investigate do not assume the full strength of these rules, we end up exploring alternative wqos and developing new proof strategies to establish decidability and upper bounds for the problems at hand.

A natural question to ask is how sensitive these techniques are to the presence of exchange. 
In the presence of exchange/commutativity,  the (left-hand sides of) sequents that can be constructed from a finite supply of formulas fits in finitely many types. 
For example, under commutativity the words over $\{a,b,c\}$ are in the form $a^rb^sc^t$, for $r,s,t \in \mathbb{N}$. 
This reduces the infinite nature of the set of such words to the exponents ($r,s,t$) of each letter. 
We will see that, in conjunction with a knotted equation, a suitable well-ordered set over tuples of natural numbers arises and this can be used to get finiteness (decidability via finite embeddability property) and also complexity results; the latter are obtained via a detailed proof-theoretic analysis of suitably modified proof systems and a study of the branches in proof searches.
It turns out that giving up commutativity fully can lead to undecidability, as for $\m{FL_c}$, but any generalized commutativity saves the day because it guarantees that words over a finite supply of letters fit in only finitely many types, as described in Cardona and Galatos~\cite{Cardona2015}. 
For example, in the context of the equation $xyx \UniEq xxy$, the words over $\{a,b,c\}$ are in one of the following (finitely many) forms: $a^{\UniIdxA}b^{\UniIdxB}c^{\UniIdxC}$, $a^{\UniIdxA}c^{\UniIdxC}b^{\UniIdxB}$, $b^{\UniIdxB}a^{\UniIdxA}c^{\UniIdxC}$,  $b^{\UniIdxB}c^{\UniIdxC}a^{\UniIdxA}$,$c^{\UniIdxC}a^{\UniIdxA}b^{\UniIdxB}$ and $c^{\UniIdxC}b^{\UniIdxB}a^{\UniIdxA}$. 
The well-ordered set now needs to code not only the total amount of copies of each letter, but also the type of the word. 
The proof-theoretic analysis of the situation is, admittedly, quite a bit more complex, but it can be carried out, as we show.
Finally, in the setting where contraction and commutativity are fully given up and only weakening is retained, i.e., $\m{FL_w}$, decidability of provability and deducibility hold for all the axiomatic extensions under consideration.
In these cases, the complexity analysis still fits in the framework of wqos, but now it requires the more complex Higman ordering over finite alphabets (and iterations thereof).
 
For the lower bounds, we consider suitable classes of \emph{and-branching counter machines} and we reduce their \emph{reachability problem} to the word problem of particular classes of residuated lattices, thus transferring the corresponding complexity lower bound. 
Counter machines are models of computation that consist of a finite set of \emph{states}, a finite set of \emph{counters}, and a finite set of \emph{instructions}. 
In the case of and-branching counter machines, these instructions fall into three categories: 
\begin{itemize}
\item \emph{Increment}: {increase the value of a counter by 1 and} possibly change the state. 
\item \emph{Decrement}:  
{decrease the value of a non-zero counter by 1 and possibly change the state}.  
\item \emph{Branch}: split the computation into two branches, with a possible state change in each branch.
\end{itemize}
Due to the fact that $\mathbf{FL_{ec}}$ has contraction in its axiomatization, the encoding of general and-branching counter machines in the logic, although sound, is not complete/faithful. 
For this reason Urquhart considers such machines that further allow  \emph{expansion} \emph{glitches} during their computations: any non-empty counter can be incremented at any time without affecting the state; he further gives a proof-theoretic argument that the encoding in the logic of suitable classes of such machines with expansion glitches is complete and establishes the \ACK-hardness of $\mathbf{FL_{ec}}$. 
In Galatos and St.\,John~\cite{galatos2022}, a corresponding suitable glitch is identified for every knotted rule/equation, or more generally for every \emph{simple} rule/equation; there the encoding establishes undecidability of the investigated logics.

Using these ideas, we consider and-branching counter machines with glitches corresponding to the rules of the logic, we ensure that the complexity of their reachability problem is non-primitive recursive, and we establish the completeness of the encoding with respect to the given logic, thus obtaining \ACK-hardness for the logic. 
In the encoding itself, an instance of the reachability problem is reduced to an instance of the word problem by representing configurations of the machines as terms in our language over a finite set of generators, and instructions as a finite set of equations over those terms.

Our proof of the completeness of the encoding is not proof-theoretic as the one of Urquhart. 
We first note that, due to the absence of \emph{zero-test} instructions in the definition of and-branching counter machines, the configurations of the machine can be equivalently presented as terms over the language of disjunction and fusion and, further, that the computation relation is compatible with these two connectives/operations. 
This way of viewing/presenting the machines together with the compatibility of the operations is the justification of why they are referred to as \emph{algebraic counter machines} in~\cite{gavin2019}, and their modularity is capitalized in \cite{galatos2022}.
In particular, this algebraic formulation allows us to easily define a \emph{residuated frame}. 
The latter, introduced in Galatos and Jipsen~\cite{GalJip13}, are shown to be versatile tools that serve as relational semantics for substructural logics and they fit well with both the proof theory and the algebraic semantics of the logics. 
For example, they allow for establishing representation theorems for the algebras, conservativity results for the logics, and  semantical, short proofs of cut elimination for the proof-theoretic calculi, among many other applications. 
In our case, they help us prove the completeness of the encoding (and we also use them to obtain finite models in other parts of the monograph). 
 
\section*{Overview of the chapters}

From a broad perspective, this paper provides a detailed exploration of how modern techniques from algebra, proof theory, relational semantics, computational complexity through ordinal-indexed fast-growing classes, and well-quasi-order theory can be effectively integrated to study the computational properties of logics and their algebraic semantics. 
This includes both the study of algebras and their associated consequence relations, reflecting the current state of the art in fast-growing/well-quasi-ordered substructural logics. 
In doing so, the paper also uncovers the limitations of existing methods, opening up new avenues for future research.

Aside from obtaining concrete results on complexity bounds and on the existence of finite models for our logics, one of our goals is to showcase the different tools that we use from all of the above areas as well as the ways they can be creatively combined, in hopes that our general methodology will be applicable to other, different, contexts by the readers. 
At the same time, we also want to show the intricate and implicit connections between all of these research areas, working towards making the point that only via a comprehensive and holistic approach to all of them  can we obtain a deep understanding of substructural logics. 
Finally, although our motivation comes from the study of knotted rules, our results actually cover much more general \emph{simple rules} and in many cases they also work in the absence of the full strength of exchange. 
We hope that this will draw interest from not only proof-theory, algebra and complexity theory, but also from parts of computer science, philosophy and linguistics.

Given the interdisciplinary nature of our work, we aim to make the paper self-contained. 
In Chapter~\ref{sec:preliminaries}, we introduce the fundamental concepts from logic, proof theory, and algebra that we will use. We provide the precise definitions of the basic systems of substructural logics with weak exchange and knotted axioms---weakly commutative knotted substructural logics---in terms of both their corresponding (hyper)sequent proof calculi and their algebraic semantics. 
In particular, we mention the algebraizability connection between substructural logics and residuated latices, and we give the basics of algebraic proof theory that are needed for {our} work. 
After presenting normal forms for axioms in the lowest two levels of the substructural hierarchy, we describe the  algorithmic translations between axioms and equivalent first-order clauses, as well as the corresponding translation to structural rules in the (hyper)sequent calculus.
%Moreover, we describe the different data types used in these proof-theoretic formalisms, according to the location  of their defining axioms within the substructural hierarchy. 

In Chapter~\ref{sec:fepP3}, we establish the finite embeddability property (FEP) for extensions of $\m{FL}$-algebras and residuated lattices with axioms in the $\mathcal{P}_3^\flat$-level of the hierarchy, leading to the decidability of deducibility for all related logics. 
To that goal, we recall relational semantics, known as \emph{residuated frames}, which are suitable for sequent calculi. 
We also discuss their generalizations---\emph{residuated hyperframes}---which are suitable for hypersequent calculi. 
In the proof of the FEP, we pivot from one to the other (since the former yields finiteness, while the latter preserves the $\mathcal{P}_3^\flat$ axioms of the hypersequent level), and we show that the two produce the same algebra in our context. 
Our results imply that the corresponding varieties are generated by their finite members and that the logics enjoy the \emph{strong finite model property}. 

In Chapter~\ref{sec:knotted-wqos}, we investigate the well-quasi-orders (wqos) required to derive upper bounds for these logics, formalizing the notion of controlled bad sequences within normed wqos (nwqos) and presenting theorems regarding their length. 
In particular, we investigate the preservation properties of the direct product and disjoint sum of (well) quasi-ordered sets and their normed versions, the constructions of the majoring and minoring wqos, and the particular orders coming from knotted rules.
Finally, we show that when the growth rate of the increase of the norms of the elements of bad sequences is controlled then there is a sequence of maximal length; this length, as a function of the initial norm, is then bounded by certain fast-growing complexity functions, yielding  length theorems that will be used in the following sections. 

In Chapter~\ref{sec:ub-wc} we focus on hypersequent extensions of knotted contraction logics with exchange. 
We start by defining an auxiliary calculus with the same provable sequents, but which supports a finite backward proof search. 
To obtain this finiteness, we establish the height-preserving admissibility of weak contraction and deduce Curry's lemma for these calculi. 
Then we connect the proof search to an nwqo, use this nwqo to define minimal proofs, and show that considering such proofs suffices and that minimal-proof search is finite. 
Finally, we apply the length theorems of Chapter~\ref{sec:knotted-wqos} to obtain bounds on branches of these, control the overall space required for such proof search, and conclude with showing the membership of the deducibility and provability problems in specific complexity classes: for sequent extensions we get an Ackermannian upper bound and for hypersequent extensions we get a hyper-Ackermannian bound.

Chapter~\ref{sec:ub-ww} contains the same results for extensions with knotted weakening, but they are obtained with different methods. 
Due to the lack of a deduction theorem, deducibility and provability do not coincide, so we first replace hypotheses by structural rules in the system. 
We define a similar nwqo, but instead of a backward proof search starting from the hypersequent at hand, we rather consider a forward search from the axioms and define increasingly larger derived sets. 
The question becomes whether the hypersequent is in one of these sets, but also how many of these sets are enough to consider. 
We realize branches in the forward proof search as controlled bad sequences and leverage suitable length theorems of Chapter~\ref{sec:knotted-wqos} to ultimately obtain Ackermannian and  hyper-Ackermannian upper bound as in Chapter~\ref{sec:ub-wc}.

Chapter~\ref{sec:noncom-ub} extends the results of Chapters~\ref{sec:ub-wc} and~\ref{sec:ub-ww} to the case where the exchange rule is generalized to a weak exchange rule. 
It is known that removing exchange altogether leads to undecidability, so we make tight use of the existing strength of the generalized rules. 
In particular, we identify useful derivable rules which allow for better control of the level and positions of commutativity in a sequent and show that their basic instances suffice (where only formulas are permuted each time). 
The associated nwqos keep track not only of the total amount of each formula in the hypersequent, but also of a \emph{type}, and we show that there are only finitely many such classifying types. 
The height-preserving admissibility argument is much more complicated, but we achieve the same bounds as in the previous sections for both types of knotted rules.

In Chapter~\ref{sec:noncom-int-ub} we consider the case where no exchange at all is present, but we assume the strongest of all knotted rules: weakening. 
As in the commutative case, we also employ forward proof search, but now the employed nwqo is based on the Higman ordering (also known as \emph{subword embedding}) on words over finite alphabets, which makes the argument a bit more involved.
We rely on much looser existing length theorems for them, yielding upper bounds which are two levels higher than in the previous cases.

In Chapter~\ref{sec:lowerbounds} we shift our goal to obtaining lower complexity bounds, which we achieve via encodings of various types of abstract machines. 
We start with and-branching counter machines,  presenting them and their computations in an algebraic way, and we offer a visual representation of them via computation forests. 
Using the latter, we obtain intuitive proofs of auxiliary lemmas.
Then we discuss the problems that knotted equations pose for the encodings, running the risk of over-computing when applying such a spontaneous knotted step (a glitch) in a computation. 
To guarantee that computations are impervious to such glitches, we provide a completeness argument based on residuated frames. 
This allows us to establish computational complexity lower bounds for any quasivariety of residuated lattices that contains the countermodel algebras of the residuated frames corresponding to a suitable class of machines. 
This general sufficient condition is phrased in terms of classes of equations much more general than that of knotted ones.

In Chapter~\ref{s: joinand-incresing-lb} we show that this condition is satisfied for joinand-increasing equations, a class that includes in particular knotted contraction equations, thus obtaining lower bounds for a vast collection of substructural logics. 
To do this, we leverage results of Urquhart proved for $\m{FL_{ec}}$ to the much broader classes of logics that we consider. 
In particular, coupled with the results of the previous chapters, we prove that the computational complexity of deducibility is exactly Ackermannian for every logic axiomatized by any knotted contraction and any weak commutativity equation. 
We sharpen this result even further, by identifying such logics (we need two related weak commutativity axioms this time) whose provability also has the same complexity; on the way we establish suitable deduction theorems for many substructural logics so as to relate deducibility with provability. 
We conclude the section with an interesting, alternative, proof-theoretic take on establishing the complexity for deducibility of our main logics.

In Chapter~\ref{s: joinand-decreasing}, we prove the same results for joinand-decreasing axioms, which include knotted weakening. 
As an Urquhart-type result is not available, we take a much longer route, via usual counter machines. 
The latter have zero-test as one of their instructions, which prohibits their algebraic treatment, so we first discuss their simulation via and-branching counter machines, even in the presence of glitches. 
We further discuss lossy counter machines and show that they end up working  for all joinand-decreasing axioms. 
In particular, we leverage existing lower complexity results about lossy machines and show how they can be safely imported to our glitchy machines, while being sensitive also to the run time of the computations despite having different types of instruction in the definitions. 
This takes us through a closer look at the functions used to define the fast-growing hierarchy and their direct connection of computations of their values to the various types of machines we consider. 
The resulting technical lemma that we establish actually also subsumes Urquhart's approach (thus can serve as an alternative basis for the work that was carried out in Chapter~\ref{s: joinand-incresing-lb}). 
The main result of this section is  that every logic axiomatized by any knotted weakening (other than weakening itself) and any weak commutativity equation has precisely Ackermannian complexity. 
We conclude by a discussion of the exact range of applicability of our results, by explaining why the case of weakening is special (and thus has lower complexity) and by identifying the smallest variety above weakening in the language of the equations we have been considering. 
We prove that there is a smallest variety, which thus serves as this important boundary (we call its members almost integral residuated lattices) and we establish three different simple finite axiomatizations for it. 

Finally, in the Conclusion (Chapter~\ref{s: conclusion}), we summarize the main results of the paper and delineate the limitations of our methods. 
In particular, we mention numerous concrete open problems for which new ideas are needed.

%% file: tex/preliminaries.tex
Given a set $\UniSetA$, let $\UniPowerSet{\UniSetA}$ denote its power set, $\UniPowerSetFin{\UniSetA}$ the collection of all its finite subsets, and $\UniSetCard{\UniSetA}$ its cardinality.
The first limit ordinal is denoted by $\omega$. 
The \emph{cartesian product} of the sets $\UniSetA_1,\ldots,\UniSetA_k$, for $k \geq 1$, is denoted by $\prod_{i=1}^k \UniSetA_i$ or $\UniSetA_1 \times \cdots \times \UniSetA_k$, and $\UniSetA^k$ is just $\prod_{i=1}^k \UniSetA$; its elements are called \emph{$k$-tuples (over $\UniSetA$)}.
Sometimes we write an element $(a_1,\ldots,a_k)$ of a cartesian product using vector notation, i.e., $\vec a$.
As usual, $\UniNaturalSet$ denotes the set~$\UniSet{0,1,2,\ldots}$ of \emph{natural numbers}, and we set $\UniNaturalSetNN \UniSymbDef \UniNaturalSet\setminus\UniSet{0}$  and $\UniNaturalInitSeg{\beta} \UniSymbDef \UniSet{\UniValA \in \UniNaturalSet \mid \UniValA < \beta}$, for $\beta \in \UniNaturalSet \cup \UniSet{\omega}$---the initial segments of $\UniNaturalSet$.
A \emph{multiset} over a set $\UniSetA$ is a mapping $M : \UniSetA \to \UniNaturalSet$ assigning finitely many elements of $\UniSetA$ to a non-zero \emph{multiplicity}  and can be seen as a collection that may contain multiple copies of some elements.

A \emph{signature} or \emph{language} is a finite collection of symbols, each of them having an associated \emph{arity} (a natural number).
In the context of propositional logic, these symbols are called \emph{connectives} and in the context of algebra they are called \emph{operation symbols}.
An \emph{algebra} is a first-order structure without relational symbols, i.e., a set with a collection of operations (of various arities) on it (refer to~\cite{BuSa00} for the relevant definitions).

\section{Full Lambek logic and its extensions}
\label{subsec:prelim-fl-logic-ext}

Let $\UniPropVars$ be a countably infinite set of \emph{propositional variables}. We define \emph{(propositional)} \emph{formulas} by the following grammar:
\[
\UniFmA:=\UniPropA\in \UniPropVars \VL 1 \VL 0 \VL \UniFmA\land \UniFmA \VL \UniFmA\lor \UniFmA \VL \UniFmA\fus \UniFmA \VL \UniFmA\ld \UniFmA \VL \UniFmA\rd \UniFmA
\]
and we denote their set by $\UniLangSet{\UniPropVars}$.
Note that we consider the signature $\{1, 0, \land,\lor,\fus,\allowbreak\ld,\rd\}$.
Elements of $\UniLangSet{\UniPropVars}$ also serve as the \emph{terms} generated by $P$ in the setting of the algebras over the above signature (in this case, we usually denote the variables by lowercase letters, e.g., $x,y,z$).
The \textit{fusion} connective~$\fus$ is referred to as a \emph{multiplicative}/resource conjunction to contrast it with the familiar \emph{additive}/truth conjunction~$\land$.
The role of the connectives is evident via the algebraic semantics of \emph{residuated lattices} (see Chapter~\ref{sec:fepP3}): $\land,\lor$ are the lattice connectives; $\fus$ is a monoidal operation with unit~$1$;~$0$ is an arbitrary element of the lattice called the \emph{negation constant}; and the \emph{left  division} $\ld$ and \emph{right division} $\rd$ connectives, which are the residuals of fusion.
In the presence of commutativity of fusion (provided by the exchange rule) the two divisions coincide ($x \ld y=y \rd x$) and we denote the common value by $x \imp y$, where $\imp$ is the \emph{implication} connective.

\begin{remark}
    Sometimes Full Lambek logic is presented with the constants $\top$ and~$\bot$, representing, respectively, the top and bottom element of the corresponding lattice.
    Here we opt to not include them to simplify the presentation. 
    It will be evident that all of our results hold when these constants are present.
\end{remark}

Recall that a \emph{consequence relation} {(over formulas)} is a relation $\vdash$ between sets of formulas and single formulas that is \emph{inclusive} ($\UniFmA \in \Gamma \Longrightarrow \Gamma \vdash \UniFmA$), \emph{monotone} ($\Gamma \vdash \UniFmA \text{ and } \Gamma \subseteq \Delta \Longrightarrow \Delta \vdash \UniFmA$), \emph{transitive} ($\Delta \vdash \gamma$ for all $\gamma \in \Gamma \text{ and } \Gamma \vdash \UniFmA \Longrightarrow \Delta \vdash \UniFmA$) and {\emph{substitution invariant} ($\Gamma \vdash \UniFmA \Longrightarrow \sigma[\Gamma] \vdash \sigma(\UniFmA)$, for every substitution $\sigma$)}.
In this context, sets of formulas appearing on the left of $\vdash$ are read as \emph{assumptions}.
Additionally, a consequence relation may be \emph{finitary} (if $\Gamma \vdash \UniFmA$, then $\Gamma_0 \vdash \UniFmA$ for some finite $\Gamma_0 \subseteq \Gamma$).
The set of \emph{theorems} of $\vdash$ is defined to be  $\UniThm{\vdash} \UniSymbDef \{ \UniFmA \in \UniLangSet{\UniPropVars}\mid \varnothing \vdash \UniFmA \}$.

Given a consequence relation $\vdash$, the problems of \emph{deducibility} and \emph{provability} refer, respectively, to membership in $\vdash$ (for finite sets of assumptions) and in $\UniThm{\vdash}$.

A \emph{Hilbert calculus} $\UniHilCalcA{}$ is defined in the usual way as a collection of \emph{rule schemas} of the form $\frac{\UniFmA_1,\ldots,\UniFmA_k}{\UniFmB}$, where $\UniFmA_1,\ldots,\UniFmA_k,\UniFmB$ are formulas over schematic variables.
Schematic variables can then be instantiated with formulas in $\UniLangSet{\UniPropVars}$, yielding \emph{rule instances}.
A \emph{derivation} of $\UniFmA$ from $\UniSetFmA$ in $\UniHilCalcA{}$ is a finite sequence of formulas $\UniFmC_1,\ldots,\UniFmC_d$ where $\UniFmC_d = \UniFmA$ and each formula in the sequence is either an instance of an axiom, or an element of $\UniSetFmA$ or the result of applying a rule instance of a rule schema of $\UniHilCalcA{}$ to previous formulas in the sequence.
When such derivation exists, we write $\UniMSetFmA \vdash_{\UniHilCalcA{}} \UniFmA$.
It is well-known that $\vdash_{\UniHilCalcA{}}$ is a finitary consequence relation (and we say that it is \emph{determined by $\UniHilCalcA{}$}).

We define the consequence relation
$\vdash_{\UniFLExtLogic{}}$ as the one determined by the  Hilbert calculus\footnote{There are various choices for such calculus. We chose the one from~\cite{GalJipKowOno07}; see \cite{GalOno10,AltenClintRaft2004} for variants.} presented in Figure~\ref{fig:hilbert-calculus-fl}. 

Given a consequence relation $\vdash$ and a finite set of axiom schemas $\UniAxiomSetA$ (called \emph{axioms} in this context), we define the \emph{axiomatic extension of $\vdash$ by $\UniAxiomSetA$} as the consequence relation $\vdash'$ such that $\Gamma \vdash' \UniFmA$ if and only if $\Gamma,\Sigma[\UniAxiomSetA] \vdash \UniFmA$, where $\Sigma[\UniAxiomSetA]$ is the set of all substitution instances of $\UniAxiomSetA$.
We will be denoting the axiomatic extension of $\vdash_{\UniFLExtLogic{}}$ by $\UniAxiomSetA$ using $\vdash_{\UniFLExtLogic{\UniAxiomSetA}}$ and  $\vdash_{\UniAxiomExt{\UniFLExtLogic{}}{\UniAxiomSetA}}$.

In this manuscript, \emph{full Lambek logic}, $\UniFLExtLogic{}$, will refer to both $\vdash_{\UniFLExtLogic{}}$ and $\UniThm{\vdash_{\UniFLExtLogic{}}}$, relying on the context to disambiguate.
\emph{Substructural logics (over $\UniFLExtLogic{}$)} are axiomatic extensions of $\UniFLExtLogic{}$; when the set of axioms is $\UniAxiomSetA$ we will denoted the extension by $\UniFLExtLogic{\UniAxiomSetA}$ and by $\UniAxiomExt{\UniFLExtLogic{}}{\UniAxiomSetA}$.

Note that, by using $\vdash_{\UniFLExtLogic{}}$ as a base consequence relation, $\vdash_{\UniAxiomExt{\UniFLExtLogic{}}{\UniAxiomSetA}}$ can be recovered from $\UniThm{\vdash_{\UniAxiomExt{\UniFLExtLogic{}}{\UniAxiomSetA}}}$ by taking the least consequence relation that contains $\UniThm{\vdash_{\UniAxiomExt{\UniFLExtLogic{}}{\UniAxiomSetA}}}$ and is closed under $\vdash_{\UniFLExtLogic{}}$.
Therefore, in our context the use of the term `logic' for both a consequence relation and set of theorems will not cause any confusion.
This  allows us to refer to the former (when talking about the decidability/complexity of deducibility in $\UniAxiomExt{\UniFLExtLogic{}}{\UniAxiomSetA}$) or to the latter (when talking about the decidability/complexity of provability in $\UniAxiomExt{\UniFLExtLogic{}}{\UniAxiomSetA}$) depending on the context.

The axiom corresponding to exchange ($\UniEProp$) is $(\UniFmA\fus \UniFmB)\ld (\UniFmB\fus \UniFmA)$; to contraction ($\UniCProp$) is $\UniFmA \ld (\UniFmA \fus \UniFmA)$; to integrality ($\UniIProp$) is $\UniFmA \ld 1$; to zero-boundedness ($\UniOProp$) is $0\ld \UniFmA$. 
Weakening ($\UniWProp$) corresponds to the set $\{ \UniIProp,\UniOProp \}$.
In this way, by $\UniFLExtLogic{\UniEProp}$ we mean  the axiomatic extension of $\UniFLExtLogic{}$ by $\{ \UniEProp \}$.
Axiomatic extensions of $\UniFLExtLogic{\UniEProp}$ are called \emph{commutative substructural logics}.
Similarly, axiomatic extensions of $\UniFLExtLogic{\UniCProp}$, $\UniFLExtLogic{\UniIProp}$ and $\UniFLExtLogic{\UniOProp}$ are called, respectively, \emph{contractive}, \emph{integral} and \emph{zero-bounded} substructural logics.
Axiomatic extensions of $\UniFLExtLogic{\UniWProp}$ are called substructural logics \emph{with weakening}.
In Section~\ref{s: P3/N2}, we present via the \emph{substructural hierarchy} some classes of axioms that will be of interest for us in this work.

\begin{figure}
    \centering
    \begin{tabular}{llll}
        (id) & $\UniFmA\ld\UniFmA$ & (pf$_\ell$) & $(\UniFmA\ld\UniFmB)\ld[(\UniFmC\ld\UniFmA)\ld(\UniFmC\ld\UniFmB)]$\\[.5em]
        %%%%%
        (as$_{\ell\ell}$) & $\UniFmA\ld[(\UniFmB\rd\UniFmA)\ld\UniFmB]$ & (a) & $[(\UniFmB\ld\UniFmC)\rd\UniFmA]\ld[\UniFmB\ld(\UniFmC\rd\UniFmA)]$\\[.5em]
        %%%%%%
        ($\fus\ld\rd$)&$[(\UniFmB\fus(\UniFmB\ld\UniFmA))\rd\UniFmB]\ld(\UniFmA\rd\UniFmB)$&($\fus\land$)&$[(\UniFmA\land 1)\fus(\UniFmB\land 1)]\ld(\UniFmA\land\UniFmB)$\\[.5em]
        %%%%%%
        ($\land\ld$)&$(\UniFmA\land\UniFmB)\ld\UniFmA$&($\land\ld$)&$(\UniFmA\land\UniFmB)\ld\UniFmB$\\[.5em]
        %%%%%%
        ($\ld\land$)&$[(\UniFmA\ld\UniFmB)\land(\UniFmA\ld\UniFmC)]\ld[\UniFmA\ld(\UniFmB\land\UniFmC)]$&($\ld\lor$)&$\UniFmA\ld(\UniFmA \lor \UniFmB)$\\[.5em]
        %%%%%%%
        ($\ld\lor$)&$\UniFmB\ld(\UniFmA \lor \UniFmB)$&($\lor\ld$)&$[(\UniFmA\ld\UniFmC) \land (\UniFmB\ld\UniFmC)]\ld[(\UniFmA\lor\UniFmB)\ld\UniFmC]$\\[.5em]
        %%%%%
        ($\ld\fus$)&$\UniFmB\ld(\UniFmA\ld(\UniFmA\fus\UniFmB))$&($\fus\ld$)&$[\UniFmB\ld(\UniFmA\ld\UniFmC)]\ld((\UniFmA\fus\UniFmB)\ld\UniFmC)$
        \\[.5em]
        %%%%%%
        (1)&1&($1\ld$)&$1\ld(\UniFmA\ld\UniFmA)$\\[.5em]
        ($\ld 1$)&$\UniFmA\ld(1\ld\UniFmA)$&\\[.5em]
    \end{tabular}
    
    \vspace{.5em}
    
    \begin{tabular}{cccc}
         %%%%% RULES
         \AxiomC{$\UniFmA \quad \UniFmA \ld \UniFmB$}
         \RightLabel{(mp$_\ell$)}
         \UnaryInfC{$\UniFmB$}
         \DisplayProof
         & 
         \AxiomC{$\UniFmA$}
         \RightLabel{(adj$_1$)}
         \UnaryInfC{$\UniFmA \land 1$}
         \DisplayProof
         &
         \AxiomC{$\UniFmA$}
         \RightLabel{(pn$_\ell$)}
         \UnaryInfC{$\UniFmB\ld(\UniFmA\fus\UniFmB)$}
         \DisplayProof
         & 
         \AxiomC{$\UniFmA$}
         \RightLabel{(pn$_r$)}
         \UnaryInfC{$(\UniFmB\cdot\UniFmA)\rd\UniFmB$}
         \DisplayProof
         \\
    \end{tabular}

    \caption{Hilbert calculus for the Full Lambek logic.}
    \label{fig:hilbert-calculus-fl}
\end{figure}

\section[{Proof theory of substructural logics: (hyper)sequent calculi}]{Proof theory of substructural logics: (hyper)sequent calculi}
\label{sec:prelims-proof-theory}

A \emph{sequent} is an expression of the form $\UniSequent{\UniMSetFmA}{\UniMSetSucA}$, where~$\UniMSetFmA$ (the \emph{antecedent}) is a finite list of formulas and~$\UniMSetSucA$ (the \emph{succedent}) is a list that is either empty or consists of a single formula (we call it a \emph{succedent-list} or \emph{stoup}).
Note that we employ the same symbols (namely $\UniMSetFmA, \UniMSetFmB, \UniMSetFmC, \UniMSetSucA$) to denote sets, lists and, later on, multisets of formulas; we count on the context of usage to eliminate any potential confusion.

Gentzen introduced sequents as a means of formalizing natural deduction proof systems; the idea was to make the assumptions in the deduction local and explicit. Sequents also correspond to algebraic inequalities (where the list formulas on the left-hand side is interpreted as their product/fusion). 
%The data type of a sequent supports the definition of sequent rules, representing derivations among sequents, and \emph{sequent calculi}, defined as sets of sequent rules. 
%The majority of proof theory studies sequents and their calculi (or variations thereof). 
%We assume that the reader is acquainted with sequent calculi and \emph{sequent-style derivations} (for a detailed account, see~\cite{negri2001}). 

The remainder of this subsection offers a brief introduction to the proof theory of substructural logics, focusing on hypersequent calculi---a generalization of the standard sequent calculus framework. For a gentle introduction to sequent calculi, see~\cite{negri2001}.

A \textit{hypersequent} is a (possibly empty) finite multiset of sequents. It is written explicitly as follows, for $k\geq 0$,
\begin{equation}
\label{hypersequent}
\UniSequent{\UniMSetFmA_{1}}{  \UniMSetSucA_{1}} \VL \ldots \VL \UniSequent{\UniMSetFmA_{k}}{  \UniMSetSucA_{k}}
\end{equation}
Each sequent~$\UniSequent{\UniMSetFmA_{i}}{\UniMSetSucA_{i}}$ is called a \textit{component} of the hypersequent.
For a hypersequent~$h$ and sequent~$s$, we write $s \in h$ to mean that $s$ is a component of $h$.

\begin{definition}
    \label{omega-hyper}
    Given a finite
    set  $\UniSubfmlaHyperseqSet$ of formulas, an \emph{$\UniSubfmlaHyperseqSet$-hypersequent}
    is a hypersequent whose  formulas are all in $\UniSubfmlaHyperseqSet$.
\end{definition}

\emph{Hypersequent calculi}~\cite{Min68,Pot83,Avr87} generalize sequent calculi and consist of finite sets of \textit{hypersequent rule schemas} (see Example~\ref{ex:rule-instances}). 
They provide a proof-theoretic setting for the analysis of many logics that do not admit a useful sequent calculus formulation.
Each rule schema has a \textit{conclusion} hypersequent and some number of \textit{premise} hypersequent(s); rule schemata, and rules, are usually presented in fraction notation with the premises listed above the fraction line and the conclusion below the line (see Figure~\ref{figure-HFLec}).
A rule schema with no premises is an \textit{initial hypersequent}.
To properly define rule schemata we need to consider variables of different types, depending on whether they will be instantiated by formulas, list of formulas, sequents, or hypersequents.
The conclusion and every premise of a rule schema has the form $\UniHyperSequentCompA_{1}|\ldots|\UniHyperSequentCompA_{k}$ where each~$\UniHyperSequentCompA_{i}$ is (i)~a hypersequent-variable (denoted by~$\UniHyperMSetA$), or (ii)~a sequent-variable (denoted by $s$), or (iii)~$\UniSequent{\UniMSetVariableListA}{\UniMSetSucA}$, or (iv)~$\UniSequent{\UniMSetVariableListA}{\phantom{\UniMSetSucA}}$.
Here $\UniMSetVariableListA$ is a finite list of list-variables (we use $\UniMSetFmA,\UniMSetFmB,\UniMSetFmC$ to denote list-variables) and formulas built from formula-variables $\UniFmA,\UniFmB,\UniFmC$ and proposition-variables $\UniSchPropA,\UniSchPropB,\UniSchPropC$; also, $\UniMSetSucA$ is a  succedent-list-variable.
The variables in the rule schema are collectively referred to as \textit{schematic variables}.
We refer to $\UniHyperSequentCompA_{1}|\ldots|\UniHyperSequentCompA_{k}$ as a \emph{hypersequent} and to each~$\UniHyperSequentCompA_{i}$ as a \textit{component} even though they are built from schematic variables, unlike in (\ref{hypersequent}) where they are built from formulas; this overloading of terminology is standard.

In Figure~\ref{figure-HFLec}, we list the rule schemas of the hypersequent calculus $\UniFLExtHCalc{}$; this is the most important calculus for us in this work, since its associated consequence relation is $\vdash_{\UniFLExtLogic{}}$, as we will see. 
Also  the axiomatic extensions of interest in our work will correspond to calculi obtained by adding rules to $\UniFLExtHCalc{}$ (see Section~\ref{s: P3/N2}). 
Specializing this calculus to the sequent setting yields the sequent calculus denoted by $\m {FL}$ (overloading the symbol we used for Full Lambek logic); in detail, we omit the rules (EC) and (EW) and we allow only instances where $H$ is instantiated to the empty hypersequent.  
Moreover, the two calculi are essentially equivalent: the hypersequents provable in $\m {HFL}$ are precisely the ones of the form $g \mid s$, where $s$ is a sequent provable in $\m {FL}$. 
In particular, the two calculi prove the same sequents and actually these are exactly the ones of the form $\UniSequent{\phi_1, \phi_2, \ldots, \phi_n }{\phi}$, where the formula $(\phi_1\fus \phi_2 \cdots \phi_n)\ld \phi$ is contained (as a theorem) in the logic $\m {FL}$; this is why we overload the symbol $\m {FL}$ to denote both the logic and the calculus, relying on the context to disambiguate. 
The benefit and payoff of considering the more complicated hypersequent formalism is that it supports many more logics as extensions by rule schemas.

A rule schema containing neither formula-variables nor proposition-variables is called a \textit{structural rule schema}.
Well-known examples of such rules are the following:

\begin{center}
\begin{tabular}{r@{\hspace{1cm}}r}
     \AxiomC{$\UniSequent{H \VL \UniMSetFmA,
    \UniFmA,\UniFmB,\UniMSetFmB}{\UniMSetSucA}$}
                \RightLabel{($\UniERule$)}
                \UnaryInfC{$ H \VL \UniSequent{\UniMSetFmA,\UniFmB,\UniFmA,\UniMSetFmB}{\UniMSetSucA}$}
                \DisplayProof &  \AxiomC{$H \VL \UniSequent{\UniMSetFmA,\UniMSetFmB}{\UniMSetSucA}$}
                \RightLabel{($\UniIRule$)}
                \UnaryInfC{$
                H \VL \UniSequent{\UniMSetFmA,\UniFmA,\UniMSetFmB}{\UniMSetSucA}$}
                \DisplayProof\\[1.5em]
    \AxiomC{$H \VL \UniSequent{\UniMSetFmA}{}$}
                \RightLabel{($\UniORule$)}
                \UnaryInfC{$
                H\VL\UniSequent{\UniMSetFmA}{0}$}
                \DisplayProof &  \AxiomC{$H \VL\UniSequent{\UniMSetFmA,
    \UniFmA,\UniFmA,\UniMSetFmB}{\UniMSetSucA}$}
                \RightLabel{($\UniCRule$)}
                \UnaryInfC{$H \VL \UniSequent{\UniMSetFmA,\UniFmA,\UniMSetFmB}{\UniMSetSucA}$}
                \DisplayProof
\end{tabular}
\end{center}

We write $\UniWRule$ to refer to the two rules $\UniIRule$ and $\UniORule$.
The \textit{extension of $\UniFLExtHCalc{}$} by a set~$\UniSetRuleSchA$ of rule schemas is the hypersequent calculus~$\UniFLExtHCalc{}\cup \UniSetRuleSchA$, which we write as~$\UniCalcExt{\UniFLExtHCalc{}}{\UniSetRuleSchA}$ following standard practice.
For some rules, we denote the corresponding extensions of $\UniFLExtHCalc{}$ by writing the names of the rules as subscripts.
The fact that they have the same names as the axioms presented in Section~\ref{subsec:prelim-fl-logic-ext} is no coincidence: adding them to $\UniFLExtHCalc{}$ yields calculi for the corresponding axiomatic extensions of $\UniFLExtLogic{}$.
For instance, the hypersequent calculi for $\UniFLeExtLogic{\UniCProp}$ and $\UniFLeExtLogic{\UniWProp}$ are denoted respectively by $\UniFLeExtHCalc{\UniCProp} \UniSymbDef \UniCalcExt{\UniFLExtHCalc{}}{\UniSet{\UniERule,\UniCRule}}$ and $\UniFLeExtHCalc{\UniWProp} \UniSymbDef \UniCalcExt{\UniFLExtHCalc{}}{ \UniSet{\UniERule,\UniIRule,\UniORule}}$.

A \textit{rule instance} is obtained from a rule schema by uniformly instantiating the schematic variables with concrete objects of the corresponding type: a hypersequent-variable with a hypersequent, a sequent-variable with a sequent, a list-variable with a list of formulas, and so on. A succedent-list-variable is instantiated by a \emph{stoup} (i.e., recall, a list that is either empty or consists of a single formula).
 
Any component in the premise or conclusion of a rule schema that is not a hypersequent-variable is called an \textit{active component}; we use the same terminology for the corresponding component in a rule instance.

\begin{example}
\label{ex:rule-instances}
We consider the rule schema~($\land$R) given by
\begin{center}
\AxiomC{$\UniHyperMSetA \VL  \UniActiveCompHighlight{\UniSequent{\UniMSetFmA}{ \UniFmA}}$}
\AxiomC{$\UniHyperMSetA \VL  \UniActiveCompHighlight{\UniSequent{\UniMSetFmA}{ \UniFmB}}$}
\RightLabel{($\land$R)}
\BinaryInfC{$\UniHyperMSetA \VL  \UniActiveCompHighlight{\UniSequent{\UniMSetFmA}{ \UniFmA\land\UniFmB}}$}
\DisplayProof
\end{center}
where the active components have been highlighted with a surrounding box. Its premises are $\UniHyperMSetA \VL  \UniSequent{\UniMSetFmA}{ \UniFmA}$ and $\UniHyperMSetA \VL \UniSequent{\UniMSetFmA}{ \UniFmB}$, and its conclusion is~$\UniHyperMSetA \VL  \UniSequent{\UniMSetFmA}{ \UniFmA\land\UniFmB}$.
Below we give some rule instances of ($\land$R), where their active components have been highlighted.
Following standard notation, the comma between list variables is interpreted as list concatenation.

\begin{center}
\begin{footnotesize}
\begin{tabular}{c@{\hspace{1em}}c}
\AxiomC{$\UniActiveCompHighlight{\UniSequent{}{\UniPropA}}$}
\AxiomC{$\UniActiveCompHighlight{\UniSequent{}{\UniPropB}}$}
\BinaryInfC{$\UniActiveCompHighlight{\UniSequent{}{\UniPropA\land\UniPropB}}$}
\DisplayProof
&
\AxiomC{$\UniSequent{\UniPropA}{\UniPropA} \VL \UniActiveCompHighlight{
\UniSequent{\UniPropC,\UniPropC}{\UniPropA\land \UniPropB}
}$}
\AxiomC{$\UniSequent{\UniPropA}{\UniPropA} \VL \UniActiveCompHighlight{\UniSequent{\UniPropC,\UniPropC}{\UniPropB}}$}
\BinaryInfC{$\UniSequent{\UniPropA}{\UniPropA} \VL \UniActiveCompHighlight{
    \UniSequent{\UniPropC,\UniPropC}{(\UniPropA\land \UniPropB)\land \UniPropB}
}$}
\DisplayProof
\\[3em]
\multicolumn{2}{c}{
\AxiomC{$\UniSequent{}{\UniPropA\fus\UniPropA} \VL \UniSequent{\UniPropB\imp \UniPropA}{\UniPropA} \VL \UniActiveCompHighlight{\UniSequent{\UniPropC}{ \UniPropA\land \UniPropB}}$}
\AxiomC{$\UniSequent{}{\UniPropA\fus\UniPropA} \VL \UniSequent{\UniPropB\imp \UniPropA}{\UniPropA} \VL \UniActiveCompHighlight{\UniSequent{\UniPropC}{\UniPropB}}$}
\BinaryInfC{$\UniSequent{}{\UniPropA\fus\UniPropA} \VL \UniSequent{\UniPropB\imp \UniPropA}{\UniPropA} \VL \UniActiveCompHighlight{\UniSequent{\UniPropC}{(\UniPropA\land \UniPropB)\land \UniPropB}}$}
\DisplayProof
}
\end{tabular}
\end{footnotesize}
\end{center}
\end{example}

Let $\UniHyperCalcA$ be an extension of $\UniFLExtHCalc{}$. A \emph{derivation} in $\UniHyperCalcA$ is a finite labelled rooted tree such that the labels of a non-leaf node and its child node(s) are respectively the conclusion and premise(s) of an instance of some rule schema in the calculus.
A \textit{deduction} in $\UniHyperCalcA$ of a hypersequent~$\UniHypersequentA$ from a set $\UniHyperseqSetA$ of hypersequents is a derivation in which the root is labelled with~$\UniHypersequentA$ and the leaves are labelled with instances of initial hypersequents or with hypersequents in $\UniHyperseqSetA$.
Deductions from $\varnothing$ are called \emph{proofs}.
We write $\UniHyperseqSetA \UniHyperDerivRel{\UniHyperCalcA} \UniHypersequentA$ when such a derivation exists, giving rise to a consequence relation over hypersequents.

The consequence relation $\UniConseqRel{\UniHyperCalcA}$ over formulas is then defined by: 
$$\UniMSetFmA \UniConseqRel{\UniHyperCalcA} \UniFmB \text{ iff }\UniSet{(\UniSequent{}{\UniFmA}) \mid \UniFmA \in \UniMSetFmA} \UniHyperDerivRel{\UniHyperCalcA} (\UniSequent{}{\UniFmB}),$$ 
for all $\UniMSetFmA \cup\UniSet{\UniFmB} \subseteq \UniLangSet{\UniPropVars}$.
We say that $\UniHyperCalcA$ is \emph{a hypersequent calculus for a substructural logic} $\UniLogicA$ when $\UniConseqRel{\UniLogicA} \;=\; \UniConseqRel{\UniHyperCalcA}$; note that in this case $\UniThm{\UniLogicA} = \UniSet{\UniFmA \in \UniLangSet{\UniPropVars} \mid \UniEmptySet\UniConseqRel{\UniHyperCalcA}\UniFmA}$. It is well-known that $\vdash_{\UniFLExtLogic{}}\;=\;\UniConseqRel{\UniFLExtHCalc{}}$~\cite{metcalfe2009,CiaGalTer17}.

We denote by $\UniDelCut{\UniHyperCalcA}$ the calculus $\UniHyperCalcA$ without the $\UniCutRule$ rule, and define $\UniHyperDerivRel{\UniDelCut{\UniHyperCalcA}}$ and $\UniConseqRel{\UniDelCut{\UniHyperCalcA}}$ just as above, noting that now they are not consequence relations.
However, when $\UniHyperCalcA$ satisfies \emph{cut elimination} (i.e., whenever a hypersequent is provable then it has a proof in which no applications of $\UniCutRule$ appear), we have $\UniHyperDerivRel{\UniDelCut{\UniHyperCalcA}} \UniHypersequentA$ iff $\UniHyperDerivRel{{\UniHyperCalcA}} \UniHypersequentA$. 
So, when discussing provability, we may use the calculus $\UniDelCut{\UniHyperCalcA}$ instead of ${\UniHyperCalcA}$. 
This is the case for $\UniFLExtHCalc{}$ and, as we will see, for all the extensions we will be considering in this work~\cite{CiaGalTer17}.

A \textit{branch} in a derivation is a path from the root to a leaf.
The \textit{height} of a derivation is the maximum number of nodes on a branch.
For an index set~$J \UniSymbDef \{j_{1},\ldots,j_{k}\}$, denote the hypersequent $\UniHyperMSetA \VL \UniSequent{\UniMSetFmA_{j_{1}}}{\UniMSetSucA_{j_{1}}} \VL \ldots \VL \UniSequent{\UniMSetFmA_{j_{k}}}{\UniMSetSucA_{j_{k}}}$ by $\UniHyperMSetA\VL \UniSequent{\UniMSetFmA_{j}}{ \UniMSetSucA_{j}} (j\in J)$.

\begin{figure}
    \textbf{Initial hypersequents}\\[1em]
    
    \begin{center}
    \begin{tabular}{ccc}
        \AxiomC{}
        \UnaryInfC{
        $\UniHyperMSetA 
        \VL 
        \UniSequent{\UniSchPropA}{\UniSchPropA}$
        }
        \DisplayProof
        &
        \AxiomC{}
        \UnaryInfC{
        $\UniHyperMSetA 
        \VL   
        \UniSequent{0}{}$
        }
        \DisplayProof
        &
        \AxiomC{}
        \UnaryInfC{
        $\UniHyperMSetA 
        \VL   
        \UniSequent{}{1}$
        }
        \DisplayProof\\[1em]
        \end{tabular}
        \end{center}
        
        \textbf{Structural rules}\\[1em]

        \begin{center}
            \begin{tabular}{rr}
                \AxiomC{$H \VL \UniSequent{\UniMSetFmC}{\UniFmA}$}
                \AxiomC{$H \VL \UniSequent{\UniMSetFmA,\UniFmA,\UniMSetFmB}{\UniMSetSucA}$}
                \RightLabel{(cut)}
                \BinaryInfC{$
                H \VL \UniSequent{\UniMSetFmA,\UniMSetFmC,\UniMSetFmB}{\UniMSetSucA}$}
                \DisplayProof
                &
                \\[1.5em]
                \AxiomC{$\UniHyperMSetA \VL   
                \UniSequent{\UniMSetFmA}{\UniFmA} \VL \UniSequent{\UniMSetFmA}{\UniFmA}$}
                \RightLabel{(EC)}
                \UnaryInfC{$\UniHyperMSetA \VL  \UniSequent{\UniMSetFmA}{\UniFmA}$}
                \DisplayProof
                \quad
                \AxiomC{$\UniHyperMSetA$}
                \RightLabel{(EW)}
                \UnaryInfC{$\UniHyperMSetA \VL   
                \UniSequent{\UniMSetFmA}{\UniFmA}$}
                \DisplayProof\\[1em]
            \end{tabular}
        \end{center}

        \textbf{Logical rules}

        \begin{center}
            \begin{tabular}{rr}
            \AxiomC{$
            \UniHyperMSetA \VL \UniSequent{\UniMSetFmA}{}
            $}
            \RightLabel{($0$R)}
            \UnaryInfC{$\UniHyperMSetA 
            \VL
            \UniSequent{\UniMSetFmA}{0}$}
            \DisplayProof
            &
             \AxiomC{$
            \UniHyperMSetA 
            \VL   
            \UniSequent{\UniMSetFmA,\UniMSetFmB}{\UniMSetSucA}$}
            \RightLabel{($1$L)}
            \UnaryInfC{$\UniHyperMSetA 
            \VL   
            \UniSequent{\UniMSetFmA,1,\UniMSetFmB}{\UniMSetSucA}$}
            \DisplayProof 
            \\[.7cm]
            \AxiomC{$\UniHyperMSetA 
            \VL   
            \UniSequent{\UniMSetFmA,\UniFmA,\UniFmB,\UniMSetFmB}{\UniMSetSucA}$}
            \RightLabel{($\fus$L)}
            \UnaryInfC{$\UniHyperMSetA 
            \VL   
            \UniSequent{\UniMSetFmA,\UniFmA\fus \UniFmB,
            \UniMSetFmB}{ \UniMSetSucA}$}
            \DisplayProof
            &
            \AxiomC{$\UniHyperMSetA \VL 
            \UniSequent{\UniMSetFmA}{\UniFmA}$}
            \AxiomC{$\UniHyperMSetA \VL 
            \UniSequent{\UniMSetFmB}{\UniFmB}$}
            \RightLabel{($\fus$R)}
            \BinaryInfC{$\UniHyperMSetA \VL   
            \UniSequent{\UniMSetFmA,\UniMSetFmB}{\UniFmA\fus \UniFmB}$}
            \DisplayProof
            \\[0.7cm]
            \AxiomC{$\UniHyperMSetA \VL  
            \UniSequent{\UniMSetFmA,\UniFmA,\UniMSetFmB}{\UniMSetSucA}$
            }
            \AxiomC{$\UniHyperMSetA \VL   
            \UniSequent{\UniMSetFmA,\UniFmB,\UniMSetFmB}{\UniMSetSucA}
            $
            }
            \RightLabel{($\lor$L)}
            \BinaryInfC{$\UniHyperMSetA \VL  
            \UniSequent{\UniMSetFmA,\UniFmA\lor\UniFmB,\UniMSetFmB}{\UniMSetSucA}$}
            \DisplayProof
            &
            \AxiomC{$\UniHyperMSetA \VL   
            \UniSequent{\UniMSetFmA}{\UniFmA_i}
            $}
            \RightLabel{($\lor$R)}
            \UnaryInfC{$\UniHyperMSetA 
            \VL   
            \UniSequent{\UniMSetFmA}{\UniFmA_1\lor\UniFmA_2}$}
            \DisplayProof
            \\[0.7cm]
            \AxiomC{$\UniHyperMSetA 
            \VL   
            \UniSequent{\UniMSetFmA,\UniFmA_i,\UniMSetFmB}{\UniMSetSucA}$}
            \RightLabel{($\land$L)}
            \UnaryInfC{$\UniHyperMSetA \VL   
            \UniSequent{\UniMSetFmA,\UniFmA_1\land \UniFmA_2,
            \UniMSetFmB}{\UniMSetSucA}$}
            \DisplayProof
            &
            \AxiomC{$\UniHyperMSetA \VL   
            \UniSequent{\UniMSetFmA}{\UniFmA}$}
            \AxiomC{$\UniHyperMSetA \VL   
            \UniSequent{\UniMSetFmA}{\UniFmB}$}
            \RightLabel{($\land$R)}
            \BinaryInfC{$\UniHyperMSetA \VL   
            \UniSequent{\UniMSetFmA}{\UniFmA\land \UniFmB}$}
            \DisplayProof
            \\[.7cm]
            \AxiomC{$\UniHyperMSetA \VL   
            \UniSequent{\UniMSetFmA}{\UniFmA}$}
            \AxiomC{$\UniHyperMSetA \VL   
            \UniSequent{\UniMSetFmB,\UniFmB,\UniMSetFmC}{\UniMSetSucA}$}
            \RightLabel{(${\rd}$L)}
            \BinaryInfC{$\UniHyperMSetA \VL   
            \UniSequent{\UniMSetFmB,\UniFmB\rd\UniFmA,\UniMSetFmA,\UniMSetFmC}{\UniMSetSucA}$}
            \DisplayProof
            &
            \AxiomC{$\UniHyperMSetA \VL   
            \UniSequent{\UniMSetFmA,\UniFmA}{\UniFmB}$}
            \RightLabel{(${\rd}$R)}
            \UnaryInfC{$\UniHyperMSetA \VL   
            \UniSequent{\UniMSetFmA}{\UniFmB\rd \UniFmA}$}
            \DisplayProof
             \\[.7cm]
            \AxiomC{$\UniHyperMSetA \VL   
            \UniSequent{\UniMSetFmA}{\UniFmA}$}
            \AxiomC{$\UniHyperMSetA \VL   
            \UniSequent{\UniMSetFmB,\UniFmB,\UniMSetFmC}{\UniMSetSucA}$}
            \RightLabel{(${\ld}$L)}
            \BinaryInfC{$\UniHyperMSetA \VL   
            \UniSequent{\UniMSetFmB,\UniMSetFmA,\UniFmA\ld\UniFmB,\UniMSetFmC}{\UniMSetSucA}$}
            \DisplayProof
            &
            \AxiomC{$\UniHyperMSetA \VL   
            \UniSequent{\UniFmA,\UniMSetFmA}{\UniFmB}$}
            \RightLabel{(${\ld}$R)}
            \UnaryInfC{$\UniHyperMSetA \VL   
            \UniSequent{\UniMSetFmA}{\UniFmA \ld \UniFmB}$}
            \DisplayProof
        \end{tabular}
    \end{center}
    \caption{The hypersequent calculus~$\UniFLExtHCalc{}$
    for~$\UniFLExtLogic{}$.}
    \label{figure-HFLec}
\end{figure} 

Ciabattoni \textit{et al.}~\cite{CiaGalTer08,CiaGalTer17} provide hypersequent calculi for infinitely many axiomatic extensions of the logic~$\UniFLExtLogic{}$ via \emph{analytic structural rule extensions} of~$\UniFLExtHCalc{}$ (whose key properties we discuss below). 
More precisely, by using terminology about levels of the substructural hierarchy given in those papers, we mention that these are extensions by axioms up to the level $\mathcal{P}_3^\flat$ of this hierarchy~\cite[Sec. 4.1]{CiaGalTer17}. 
When commutativity is present, this class coincides with the level $\mathcal{P}_3^\flat$ and when integrality is also present, it corresponds to the level $\mathcal{P}_3$~\cite[Sec. 3]{CiaGalTer08}.
We will soon detail in Section~\ref{s: FEPvarieties} these levels of the substructural hierarchy and their connection to analytic (sequent and hypersequent) structural rules.

Analytic structural rules satisfy the properties of \emph{linear conclusion}, \emph{separation}, \emph{coupling} and \emph{strong subformula property} (\hypertarget{def-analytic-main}{see}~\cite[Sec.~6]{CiaGalTer08} for details); see Figure~\ref{fig-str-rules-app} {on page~\pageref{fig-str-rules-app}} for examples of analytic structural rules in presence of the exchange rule. 
The first three properties ensure that an extension of $\UniFLExtHCalc{}$ by analytic structural rules has cut-elimination.
The strong subformula property ensures that only subformulas of the end formula appear in a backward proof search, resulting in a crucial restriction on the proof-search space.
Given a (hyper)sequent, the \emph{backward proof search} in the calculus describes the algorithm where at every step, starting with the given (hyper)sequent, we explore all of the rule instances that could produce the (hyper)sequent as conclusion and for each one of them (via branching in the algorithm) we shift our focus to each of the (hyper)sequents in the premise of the rule instance, and we repeat the same procedure. 
We will see that such strategies may be used over contractive substructural logics (see Chapters~\ref{sec:ub-wc} and~\ref{sec:noncom-ub}).

\begin{remark}
In this work, for simplicity we opted to present the proof search   and upper bound analysis in full detail for extensions of $\UniFLeExtLogic{}$, and then show how to extend the results to weakly commutative/noncommutative logics. 
This means that we will be  often working with extensions of the hypersequent calculus $\UniFLeExtHCalc{}$. The presence of commutativity allows for a simplified presentation of this calculus.
First of all, we may assume that the antecedents of sequents are  multisets instead of lists, as the order of formulas there does not matter. 
Also, the difference between the left and right implication (i.e., division) connectives vanishes, and we only need to consider the connective $\imp$. Finally, most of the rules are simplified, since it is not necessary to consider contexts on the left and on the right of a given formula in the antecedent of a sequent. 
For example, in $\UniFLeExtHCalc{}$ the rules $\text{(\rd R)}$, $\text{(\rd L)}$, $\text{($\ld$R)}$ and $\text{($\ld$L)}$ are replaced by the following, simpler rules:
\begin{center}
            \AxiomC{$\UniHyperMSetA \VL   
        \UniSequent{\UniMSetFmA}{\UniFmA}$}
        \AxiomC{$\UniHyperMSetA \VL   
        \UniSequent{\UniMSetFmB,\UniFmB}{\UniFmC}$}
        \RightLabel{(${\imp}$L)}
        \BinaryInfC{$\UniHyperMSetA \VL   
        \UniSequent{\UniMSetFmA,\UniMSetFmB,\UniFmA\imp \UniFmB}{\UniFmC}$}
        \DisplayProof
        \qquad
        \AxiomC{$\UniHyperMSetA \VL   
        \UniSequent{\UniMSetFmA,\UniFmA}{\UniFmB}$}
        \RightLabel{(${\imp}$R)}
        \UnaryInfC{$\UniHyperMSetA \VL   
        \UniSequent{\UniMSetFmA}{\UniFmA\imp \UniFmB}$}
        \DisplayProof
\end{center}
\end{remark}

\begin{remark}
    In order to include the bounds $\bot$ and $\top$, it is enough to add the following rules to the calculus presented in Figure~\ref{figure-HFLec}:
    \begin{center}
        \AxiomC{}
        \UnaryInfC{
        $\UniHyperMSetA 
        \VL   
        \UniSequent{\UniMSetFmA, \bot, \UniMSetFmB}{\UniMSetSucA}$
        }
        \DisplayProof
        \quad
        \AxiomC{}
        \UnaryInfC{
        $\UniHyperMSetA 
        \VL   
        \UniSequent{\UniMSetFmA}{\top}$
        }
        \DisplayProof 
    \end{center}
\end{remark}

Since we are interested in complexity bounds, we need to be precise about what is the size of the input for the algorithms we will construct. Below is the definition of the size of a hypersequent, as well as what we mean by the size of rules and of a calculus, notions that will be used directly in some of our algorithms (see, for example, Chapter~\ref{sec:ub-ww}) and will be essential for the complexity analyses throughout this work.

\begin{definition}
\label{def:size-hypersequent}
Let $\UniHypersequentA$ be a hypersequent, $\UniRuleSchemaA$ be a rule schema and $\UniHyperCalcA$ be a hypersequent calculus.
Then $\UniSizeHyper{\UniHypersequentA}$ and $\UniSizeHyper{\UniRuleSchemaA}$ are the number of symbols in the written representation of $\UniHypersequentA$ and $\UniRuleSchemaA$, respectively. 
Also, $\UniSizeHyper{\UniHyperCalcA} \UniSymbDef \sum_{\UniRuleSchemaA \in \UniHyperCalcA} \UniSizeHyper{\UniRuleSchemaA}$ and, for a finite set $D$ of hypersequents, we let $\UniSizeHyper{D} \UniSymbDef  \sum_{\UniHypersequentA \in D} \UniSizeHyper{\UniHypersequentA}$.
\end{definition}

For example, $\UniSizeHyper{\UniSequent{\UniPropA}{}\VL\UniSequent{}{}} = 4$, $\UniSizeHyper{\UniSequent{\UniPropA,\UniPropB}{} \VL \UniSequent{}{\UniPropB \land \UniPropA}} = 9$ and $\UniSizeHyper{\text{\normalfont(EC)}} = 14$.

\section{Algebraic semantics: residuated lattices and $\UniFLExtLogic{}$-algebras}\label{s: AlgSem}
A \emph{residuated lattice} is an algebra $\UniResLatticeA\UniSymbDef \UniStruct{\UniResLatSetA; \land, \lor, \fus,\ld, \rd, 1 }$, where $\UniStruct{\UniResLatSetA; \land,\lor}$ is a lattice, $\UniStruct{\UniResLatSetA; \fus, 1}$ is a monoid and $\UniResLatticeA$ satisfies the \emph{residuation law}: for all $\UniValAlgA,\UniValAlgB,\UniValAlgC\in \UniResLatSetA$  
$$
\UniValAlgA \cdot \UniValAlgB \leq \UniValAlgC 
\iff \UniValAlgB\leq \UniValAlgA\ld \UniValAlgC
\iff \UniValAlgA\leq \UniValAlgC\rd \UniValAlgB,
$$
where the lattice order is defined as usual by: $x\leq y$ iff $x\vee y=y$. 
It follows that in residuated lattices multiplication distributes over join; in particular, multiplication is order-preserving.
Examples of residuated lattices include lattice-ordered groups, relation algebras, and lattices of ideals of rings, among others; it is in this latter setting that residuated lattices were first defined in \cite{WD} in the 1930's. 
Therefore, residuated lattices have an independent importance and history in the context of ordered algebras; see~\cite{GalJipKowOno07}. 
Our interest in them comes from the fact that they provide algebraic semantics for various substructural logics, as we explain below.

An \emph{$\UniFLExtLogic{}$-algebra} (also known as a \emph{pointed residuated lattice}) is an expansion of a residuated lattice by an additional constant/designated element $0$, known as the \emph{negation constant}, as it can be used to define negation operations; examples include  Boolean, Heyting and MV-algebras. 

It is well-known that residuated lattices and $\UniFLExtLogic{}$-algebras form varieties of algebras, which we denote by $\UniRLV$ and $\UniFLV$, respectively.
Recall that a class of algebraic structures over the same signature is called a \emph{variety} if it is closed under the three fundamental algebraic processes: direct products, subalgebras and homomorphic images. 
Equivalently, a class of algebras is a variety if it is  an \emph{equational class}: a class defined by equations (refer to~\cite{BuSa00} for detailed definitions and proofs). 

It turns out that residuated lattices and $\UniFLExtLogic{}$-algebras each are the  equivalent algebraic semantics, in the sense of \cite{BlokPig1989}, of {$\m {RL}$} (defined as the $0$-free fragment of $\UniFLExtLogic{}$) and of $\UniFLExtLogic{}$, respectively. 
In particular, there are  \emph{transformers} $\tau$ and $\rho$ transforming formulas to sets of equations and equations to sets of formulas, given by  $\tau(\UniFmA)\UniSymbDef\{1\leq\UniFmA\}$ and $\rho(\UniFmA = \UniFmB)\UniSymbDef\{(\UniFmA\imp \UniFmB)\land(\UniFmB\imp \UniFmA)\}$, where $\UniFmA$ and $\UniFmB$ are formulas, such that for all $\UniMSetFmA\cup\UniSet{\UniFmA}\subseteq\UniLangSet{\UniPropVars}$:

\begin{itemize}
\item $\UniMSetFmA\vdash_{\UniFLExtLogic{}}\UniFmA$ if and only if $\UniSet{\tau(\UniFmB):\UniFmB\in \UniMSetFmA } \UniEqRel_{\UniFLV}\tau(\UniFmA)$;
\item $\UniFmA = \UniFmB\UniEqConvRel\UniEqRel_{\UniFLV}\tau[\rho(\UniFmA = \UniFmB)]$,
\end{itemize}
or, equivalently, for all sets  $\UniEqSetA\cup\{\UniEqA\}\subseteq \UniLangSet{\UniPropVars}^2$ of equations:

\begin{itemize}
\item $\UniEqSetA\UniEqRel_{\UniFLV}\UniEqA$ iff $\{\rho(\UniEqA): \UniEqA\in \UniEqSetA\} \vdash_{\UniFLExtLogic{}}\rho(\UniEqA)$;
\item $\UniFmA\UniConseqConvRel{}\UniConseqRel{\UniFLExtLogic{}}\rho[\tau(\UniFmA)]$. 
\end{itemize}
Here, the algebraic consequence relation $ \UniEqRel_{\UniFLV}$ on equations is defined as usual by: $\UniEqSetA\UniEqRel_{\UniFLV} \UniFmA = \UniFmB$ iff for every algebra $\UniResLatticeA \in \UniFLV$ and \emph{valuation} $f$ into $\UniResLatticeA$---i.e., a homomorphism from the algebra of formulas to $\UniResLatSetA$, which is fully specified by how it maps the set~$\UniPropVars$ of propositional variables to $\UniResLatSetA$---if $f(\UniFmC)=f(\UniFmD)$ for all $(\UniFmC = \UniFmD) \in \UniEqSetA$ then $f(\UniFmA)=f(\UniFmB)$. 
In effect the algebraization result above allows for logical derivability to be faithfully reduced to algebraic consequence and vice versa. 
This entails, among other things, that the lattice of axiomatic extensions of $\UniFLExtLogic{}$ is dually isomorphic to the lattice of subvarieties of $\UniFLV$ (i.e., subclasses that are also varieties); we say that the axiomatic extension and the subvariety \emph{correspond} under this dual isomorphism. 
For example, the equivalent algebraic semantics of classical propositional logic (as an extension of $\UniFLExtLogic{}$) is the variety of Boolean algebras (viewed as a subvariety of $\UniFLV$); the same holds between intuitionistic logic and Heyting algebras, and between \L ukasiewicz infinite-valued logic and the variety of MV-algebras, to name a few examples of axiomatic extensions and subvarieties that correspond.
Also, it entails that the algebraization connection also holds between the relations $\vdash_{\UniLogicA}$ and $ \UniEqRel_{\mathcal{V}}$, where $\UniLogicA$ is an axiomatic extension of ${\UniFLExtLogic{}}$ and $\mathcal{V}$ is the corresponding subvariety.
A detailed account of residuated lattices and the algebraization result above can be found in \cite{GalJipKowOno07}.

As mentioned above, the algebraization result shows that deductions in logic and deductions in algebra are mutually interpretable in one another, and this allows for the transfer of many properties from a logic to its algebraic semantics and back, and also allows for the use of tools from universal algebra in the study of substructural logics.

For a class~$\mathcal{K}$ of algebras in some formal language, the \emph{universal theory} of~$\mathcal{K}$ is the set of universal formulas (i.e., first-order sentences $\forall x_1\ldots\forall x_n\varphi$ with $\varphi$ quantifier-free) that hold on every algebra in~$\mathcal{K}$ and every valuation into it. The \emph{equational theory} and \emph{quasiequational theory} replace universal formulas with, respectively, universally closed equations of the form $t=u$, and universally closed quasiequations of the form $t_1=u_1 \text{ and }\ldots \text{ and } t_n=u_n \Longrightarrow t_0=u_0$. In our setting of residuated lattices, we may use inequations instead of equations, since $t\leq u$ was already defined as $t\lor u=u$.

\begin{lemma}\label{l: alg_tanslation}
A decision procedure for $ \UniEqRel_{\mathcal{V}}$ automatically yields a decision procedure for the relation $\vdash_{\UniLogicA}$, where $\UniLogicA$ is an axiomatic extension of ${\UniFLExtLogic{}}$ and $\mathcal{V}$ is the corresponding subvariety of residuated lattices, and vice versa. 
The translation, via $\tau$ and $\rho$, between the two consequence relations is done in {linear} time, so the complexity of the equational/quasiequational theory of the variety is the same as that of the provability/deducibility of the corresponding logic.
\end{lemma}
In Chapter~\ref{sec:fepP3} we establish this decidability result for the consequence relations associated to infinitely many varieties of residuated lattices, and as a corollary we obtain the decidability of the deducibility for the corresponding substructural logics. 
Actually, we obtain more than that: we show that if a formula is not provable from a set of formulas, then there is a finite countermodel that exhibits this failure. 

\section{Levels of the substructural hierarchy: $\mathcal{N}_2$ and $\mathcal{P}_3^\flat$}
\label{s: FEPvarieties}\label{s: P3/N2}

In this section we discuss the formulas that are contained in the first three levels of the substructural formula hierarchy, defined in \cite{CiaGalTer17}, by providing a very condensed account of the required definitions and facts for our purposes. 
In particular, we will discuss how the set $\mathcal{P}_3^\flat$ of the hierarchy connects to structural rules in hypersequent calculi and how the set $\mathcal{N}_2$ connects to structural rules in sequent calculi. 
Moreover, we will explain how any equation in the language $\{\jn, \cdot, 1\}$ is equivalent to an $\mathcal{N}_2$ formula and to a set of so-called \emph{simple equations}.

In the following we will conflate a formula $\varphi$ and the equation $1 \leq \varphi$. 

\subsection{The $\mathcal{P}_3^\flat$ level}
\label{s: P3}

The set of $\mathcal{P}_3^\flat$ formulas/equations is defined as part of the substructural hierarchy, which keeps track of alternations of the positive connectives $\{\jn, \cdot, 1\}$ and the negative connectives $\{\mt, \ld, \rd\}$; we will not need the full definition of the hierarchy, which is defined in detail in \cite{CiaGalTer08}.  
First, the normal forms of $\mathcal{N}_2$ formulas have the form $\bigwedge_{1\leq i \leq k} \ell_i \ld u_i \rd r_i$, where $u_i$ is either $0$ or a join of products of variables, and $\ell_i$ and $r_i$ each is a product of  meets of terms of the form $\ell'_i \ld u'_i \rd r'_i$, where $u'_i$ is either $0$ or a variable and $\ell'_i$ and $r'_i$  are products of variables. 
The formulas in  $\mathcal{N}_2$ are the ones that can be transformed into sequent rules that can be  added to $\m {FL}$. 
 
The \emph{left} and \emph{right conjugate} by $a$ are defined as the unary (parameterized) terms $\lambda_a(x):= a \ld xa \mt 1$ and $\rho_a(x):= ax \rd a \mt 1$. 
An \emph{iterated conjugate} is a composition of various left and right conjugates by potentially different conjugating elements, e.g., $\lambda_a(\rho_b(\rho_a(x)))$.

A \emph{$\mathcal{P}_3^\flat$ formula/equation} in normal form is an infinite set of equations of the form $1 \leq \bigvee_i \prod_j \gamma_{ij}(s_{ij})$, where the $s_{ij}$'s are fixed terms in $\mathcal{N}_2$ (in normal form) and the $\gamma_{ij}$'s range over all (infinitely many) iterated conjugates (over an increasing supply of variables as conjugating elements). 
For example, the set of all equations $1 \leq \gamma_1(xy \ld 0) \jn \gamma_2((x \mt y) \ld xy)$ collectively constitute a $\mathcal{P}_3^\flat$ equation, where $\gamma_1, \gamma_2$ range over iterated conjugates. 
By Theorem~3.3(2) of  \cite{Ga04a}, in \emph{subdirectly irreducible algebras} (see Section~\ref{sec:fep-def} for the formal definition) this equation is equivalent to the universal sentence ($xy \leq 0$ or $x \mt y \leq xy$).

We have that
$\mathcal{P}_3^\flat$ formulas give rise to structural hypersequent rules via structural clauses.
We define a \emph{clause} as a universal first-order formula
\begin{equation*}
  t_1 \leq u_1 \text{ and } \ldots \text{ and } t_{\UniIdxA} \leq u_{\UniIdxA} \Longrightarrow t_{\UniIdxA+1} \leq u_{\UniIdxA+1} \text{ or }  \ldots \text{ or } t_{\UniIdxB} \leq u_{\UniIdxB}  \tag{q}
\end{equation*}
where $t_i$ is a product of variables and $u_i$ is either $0$ or a variable. 
It is shown in \cite{CiaGalTer17} that for every $\mathcal{P}_3^\flat$ formula there is a corresponding clause such that the two are equivalent over subdirectly irreducible algebras.
  
The clause (q) is called \emph{analytic} if every variable occurs at most once in the conclusion (\emph{linearity}); and the variables in $t_1, \ldots, t_{\UniIdxA}$ are included in the variables of $t_{\UniIdxA+1}, \ldots, t_{\UniIdxB}$ and 
the variables in $u_1, \ldots, u_{\UniIdxA}$ are included in the variables of $u_{\UniIdxA+1}, \ldots, u_{\UniIdxB}$ (\emph{inclusion}).

The clause ($xy \leq 0$ or $x \mt y \leq xy$) that we saw above is not analytic, as it is not linear. 
However, it is acyclic: (q) is called \emph{acyclic} if there are no cycles in the graph with vertices the variables of the antecedent and {directed edges $(x, y)$} for $\ell x  r \leq y$ in the antecedent. 
It is shown in \cite{CiaGalTer17} that every acyclic clause, e.g., ($xy \leq 0$ or $x \mt y \leq xy$), is equivalent to an analytic one, and a transformation algorithm is provided. 
We call a $\mathcal{P}_3^\flat$ formula \emph{acyclic} or \emph{analytic} if the corresponding clause is acyclic or analytic, respectively. 

A hypersequent \emph{structural rule (schema)} is of the form
$$\infer[\mathrm{(r)}]{H \mathrel{|} \Gamma_{\UniIdxA+1} \Rightarrow \Pi_{\UniIdxA+1} \mathrel{|} \ldots  \mathrel{|} \Gamma_{\UniIdxB} \Rightarrow \Pi_{\UniIdxB}}{H \mathrel{|} \Gamma_1 \Rightarrow \Pi_1 \;\;  \cdots \;\;  H \mathrel{|} \Gamma_{\UniIdxA} \Rightarrow \Pi_{\UniIdxA}}$$
where, for each $i$, $\Gamma_i$ is a possibly empty sequence of formula-variables and sequence-variables, and $\Pi_i$ is empty  or a formula-variable, or a stoup-variable (recall that a \emph{stoup} is either a formula or empty).

Given an analytic clause, the corresponding structural rule is obtained by first associating to each inequality $t_i\leq u_i$ a sequent $S_i$ as follows: to $x_1\cdots x_k \leq y_i$  we associate  the sequent $\Gamma_i, \Sigma_1, \ldots, \Sigma_k, \Delta_i \Rightarrow \Pi_i$ and to  $x_1\cdots x_k \leq 0$ we associate $\Gamma_i, \Sigma_1, \ldots, \Sigma_k, \Delta_i \Rightarrow \;$. 
Then to (q) we associate the structural rule: 
$$\infer[\mathrm{(q^\circ)}]{H \mathrel{|} S_{\UniIdxA+1} \mathrel{|} \ldots  \mathrel{|} S_{\UniIdxB}}{H \mathrel{|} S_1 \;\;  \cdots \;\;  H \mathrel{|} S_{\UniIdxA}}$$ 
where $H$ is a hypersequent-variable.

Also, given a structural rule (r) as above, the corresponding  clause is obtained by first associating to each (formula, sequence, stoup)-variable in the $\Gamma$'s and $\Pi$'s a (distinct) variable to obtain $\Gamma^\bullet$ (by replacing comma with multiplication) and $\Pi^\bullet$; if $\Gamma$ is empty, then $\Gamma^\bullet \UniSymbDef 1$ and if $\Pi$ is empty then $\Pi^\bullet \UniSymbDef 0$. 
Then the clause corresponding to (r) is defined to be
\begin{equation}
  \Gamma_1^\bullet \leq \Pi^\bullet_1 \text{ and } \ldots \text{ and } \Gamma_{\UniIdxA}^\bullet \leq \Pi^\bullet_{\UniIdxA} \Longrightarrow \Gamma^\bullet_{\UniIdxA+1} \leq \Pi^\bullet_{\UniIdxA+1} \text{ or }  \ldots \text{ or } \Gamma^\bullet_{\UniIdxB} \leq \Pi^\bullet_{\UniIdxB}  \tag{q$^\bullet$}
\end{equation}

Recall that a set of hypersequents $\UniHyperseqSetA$ is called \emph{elementary} if each hypersequent in $\UniHyperseqSetA$ consists of atomic formulas and it is closed under cuts: if $\UniHyperseqSetA$ contain $H|\Sigma \Rightarrow p$ and $H | \Gamma, p, \Delta \Rightarrow \Pi$, then it also contains  $H | \Gamma, \Sigma, \Delta \Rightarrow \Pi$.
A hypersequent calculus is called \emph{strongly analytic} if for every elementary set of hypersequents $\UniHyperseqSetA$ and hypersequent $\UniHypersequentA$, if $\UniHypersequentA$ is derivable in the calculus from  $\UniHyperseqSetA$, then it actually has a cut-free derivation with the subformula propery. 
A hypersequent structural rule is called \emph{strongly analytic} if its addition to $\UniFLExtHCalc{}$ yields a strongly analytic calculus.

It is shown in \cite{CiaGalTer17} that if (q) is an analytic clause then the corresponding  structural rule (q$^\circ$) is strongly analytic and that if (r) is a strongly analytic structural rule then the corresponding clause (r$^\bullet$) is analytic. 
Also, it is shown that if $\UniSetRuleSchA$ is a set of structural rules such that $\UniCalcExt{\UniFLExtHCalc{}}{\UniSetRuleSchA}$ is strongly analytic, then $\UniSetRuleSchA$ is actually equivalent to a set of rules corresponding to analytic clauses. 
Therefore, to capture all strongly analytic extensions of $\UniFLExtHCalc{}$ we only need to consider structural rules coming from analytic clauses. 
Also, the rules obtained from acyclic $\mathcal{P}_3^\flat$ formulas/equations are always strongly analytic. 
Given this correspondence, in the following we will by the extension of a logic by clauses we will understand the extension by the corresponding rules.

\subsection{Simple equations and $\mathcal{N}_2^{-0}$}
\label{s: N2}

We mentioned that every acyclic $\mathcal{P}_3^\flat$-formula axiomatizes the same logic as (an equivalent) analytic structural clause and as an analytic structural hypersequent rule. 
Since $\mathcal{N}_2$ is a subclass of $\mathcal{P}_3^\flat$, the same applies to acyclic/analytic $\mathcal{N}_2$-axioms, but the special clauses we get are actually (analytic) \emph{quasiequations} (they do not contain disjunctions) and the corresponding rules are (analytic) \emph{sequent} structural rules; the study of $\mathcal{N}_2$ predates that of $\mathcal{P}_3^\flat$, so these facts are well known, but we present them for clarity and completeness. 
So, every acyclic/analytic $\mathcal{N}_2$-axiom is equivalent to a quasiequation 
\begin{equation*}
  t_1 \leq u_1 
  \text{ and } \ldots 
  \text{ and } t_k \leq u_k \Longrightarrow t_{0} \leq u_{0} \tag{q}
\end{equation*}
where $t_i$ is a product of variables and $u_i$ is either $0$ or a variable, that is further analytic: every variable occurs at most once in the conclusion (\emph{linearity}); and the variables in $t_1, \ldots, t_k$ are included in the variables of $t_0$ and 
the variables in $u_1, \ldots, u_k$ are included in the variables of $u_0$ (\emph{inclusion}); recall Section~\ref{s: P3}. 

We define $\mathcal{N}_2^{-0}$ to be the subclass of $\mathcal{N}_2$ containing formulas where the negation constant $0$ does not appear, hence the $u_i$'s above are all variables. 
Because of linearity, $t_0$ is a product $x_1 x_2 \cdots x_\ell$ of distinct variables and because of inclusion we get that all the $u_i$'s are all the same variable, say $x_0$, and that all the variables in the $t_i$'s are among the  variables $x_1, x_2, \ldots , x_\ell$. 
Such a quasiequation 
\begin{equation*}
  t_1 \leq x_0 \text{ and } \ldots \text{ and } t_k \leq x_0 \Longrightarrow x_1 x_2 \cdots x_\ell \leq x_{0} \tag{q}
\end{equation*}
is, in turn, equivalent to the equation $x_1 x_2 \cdots x_\ell \leq t_1 \jn  \cdots \jn t_n$.

A particular case is the \emph{integrality} equation $x \leq 1$.
Also, note that if there is a variable, say $x_i$, that does not appear on the right-hand side of this inequality, then by setting all the other variables equal to $1$ we get $x_i \leq 1$, while by setting $x_i$ equal to $1$, we get $x_1 \cdots x_{i-1} x_{i+1} x_\ell \leq t_1 \jn  \cdots \jn t_k$. 
Conversely, by order preservation of multiplication, these two equations imply $x_1 x_2 \cdots x_\ell \leq x_1 \cdots x_{i-1} 1 x_{i+1} x_\ell  \leq t_1 \jn  \cdots \jn t_k$, which is the original equation. 
So, by doing this process for all the variables that appear only on the left-hand side, we get that (q) is equivalent (up to renaming of variables) to the conjunction of integrality (if there are such variables at all) and an inequality of the form $x_1 x_2 \cdots x_\ell \leq t_1 \jn  \cdots \jn t_k$, where every variable appears on \emph{both} sides. 

For the purposes of this paper, a \emph{simple equation} has the form
\begin{equation}
    \label{eq:simple-equation}
  x_1 x_2 \cdots x_\ell \leq t_1 \jn  \cdots \jn t_k,  
  \tag{$\varepsilon$}
\end{equation}
where the variables $x_1,\ldots,x_\ell$ are pairwise distinct, each $t_i$ is a product of variables, and any variable in the equation occurs on both sides; we also consider integrality as a simple rule, by exception. 
Thus every $\mathcal{N}_2^{-0}$-equation is equivalent to a conjunction of simple equations.
The corresponding sequent \emph{structural rule (schema)} is of the form
$$\infer[(r)]{\Gamma, \Sigma_0, \Delta \Rightarrow \varphi}{ \Gamma, \Sigma_1, \Delta \Rightarrow \varphi \;\;  \ldots \;\;  \Gamma, \Sigma_k, \Delta \Rightarrow \varphi}$$
where, $\Gamma, \Delta$ are sequence-variables, $\Sigma_0, \ldots, \Sigma_k$  are possibly empty sequences of sequence-variables and $\varphi$ is a formula-variable, $\Sigma_0$ is a sequence of distinct sequence-variables, and each sequence-variable appears in both the conclusion and in (some of) the premises. 
Here, $\Sigma_0$ is obtained from $x_1 x_2 \cdots x_\ell$ and each other $\Sigma_i$ is obtained from $t_i$ by replacing multiplication with comma and variables by sequence-variables. 
We call such structural rules \emph{simple}; we also consider weakening a simple rule by exception.

We have already argued that every  extension of $\UniFLExtLogic{}$ by a set of acyclic/analytic $\mathcal{N}_2^{-0}$ axioms can be alternatively axiomatized by a set of simple equations and that it can be alternatively axiomatized by a set of simple sequent rules; also all these implications are reversible. 
The next result also shows that extensions of $\UniFLExtLogic{}$ by sets of non-trivializing $\{\jn, \cdot, 1\}$-equations also come down to the same thing. 
A set of equations is called \emph{trivializing} if it has only the trivial (one-element) model; for example $\{x=y\}$, $\{x \leq y, y \leq x\}$, $\{x=1\}$ and $\{1 \leq x\}$ are trivializing sets. 
One of the steps in the proof of the following lemma defines a process of \emph{linearization} that will often occasionally be invoked in the sections that follow.

\begin{lemma}\label{l: N_2^-0}
The following collections of logics are equal.  
\begin{enumerate}
    \item The extensions of  $\UniFLExtLogic{}/{\m{RL}}$ by sets of acyclic/analytic $\mathcal{N}_2^{-0}$ equations.
    \item The extensions of  $\UniFLExtLogic{}/{\m{RL}}$ by sets of simple equations.
        \item The extensions of   $\UniFLExtLogic{}/{\m{RL}}$ by sets of simple rules.
    \item The extensions of  $\UniFLExtLogic{}/{\m{RL}}$ by sets of non-trivializing $\{\jn, \cdot, 1\}$-equations.
        \item The extensions of  ${\UniFLExtLogic{}}/\m {RL}$ by sets of {strongly} analytic {structural} rules.
\end{enumerate}
In particular, these extensions can be axiomatized by acyclic/analytic $\mathcal{P}_3^\flat$-equations.
\end{lemma}
\begin{proof}
The proofs of different parts of this lemma can be found in \cite{GalJip13}, \cite{CiaGalTer17} and \cite{CiaGalTer08}, but we provide a combined proof here for completeness. 
The equivalence of the first three statements was established above. 
Clearly (2) implies (4). 
For the converse direction, note that the equation $u= v$ is equivalent to the conjunction of $u\leq v$ and $v\leq u$. By distributing multiplication over joins, each of these equations can be written in the form $s_1 \jn \cdots \jn s_r \leq t_1 \jn \cdots \jn t_k$, where $s_i, t_j$ are all monoid terms. 
Such an inequality is, in turn, equivalent to the conjunction of the equations: $s_1 \leq t_1 \jn \cdots \jn t_k, \ldots,$ and $ s_r \leq t_1 \jn \cdots \jn t_k$.
Given one of these equations, say $t_0 \leq t_1 \jn \cdots \jn t_k$, if $t_0$ is not linear, say with variable $x$ occurring $p$~times, then we substitute $x$ by $x_1 \jn \cdots \jn x_p$ (on both sides). 
After  multiplying out, one of the joinands, say $t$, on the left-hand side will be linear so, by transitivity, we obtain the linear equation $t \leq t'_1 \jn \cdots \jn t'_k$, where the $t'_i$'s are obtained from the $t_i$'s by the above substitution; we can then multiply out each of the terms $t'_i$ so that the right-hand side is a join of monoid terms. 
Now, conversely, this equation implies $t_0 \leq t_1 \jn \cdots \jn t_k$, by instantiating $x_1= \ldots = x_p:=x$. This process is known as \emph{linearization}.
 
If in one of these (linearized) equations, say $t_0 \leq t_1 \jn \cdots \jn t_k$, there is a variable, say $x$, that occurs only on the right-hand side, then (after renumbering) the equation is of the form $t_0 \leq t'_1xt''_1 \jn \cdots \jn t'_pxt''_p \jn t_{p+1} \jn \cdots \jn t_k$, where $x$ does not appear in $t_{p+1}, \ldots , t_k$ (but it may appear in  $t'_1,t''_1, \ldots, t'_p, t''_p$). 
Note that we cannot have $p=k$, as then we could substitute $x$  by $(t'_1 \ld x \rd t''_1) \mt \cdots \mt (t'_p \ld x \rd t''_p)$ to obtain $1 \leq x$, which is a trivializing equation (and we assume that $u=v$ is not trivializing); hence $p<k$. 
Now we substitute $x$ in the equation by $(t'_1 \ld t \rd t''_1) \mt \cdots \mt (t'_p \ld t \rd t''_p)$, where $t=t_{p+1} \jn \cdots \jn t_k$, and using the fact $t'_i(t'_1 \ld t \rd t''_1)t''_i\leq t$ we obtain $t_0 \leq t_{p+1} \jn \cdots \jn t_k$; this implies back the equation $t_0 \leq t_1 \jn \cdots \jn t_k$ as the latter involves more joinands, so the two equations are equivalent. 
By repeating this process we can eliminate all joinands involving variables that appear only in the right-hand side.

Finally, if in one of the resulting equations there is a variable that occurs only on the left-hand side, by the process described before the lemma it can be replaced by a conjunction of simple equations (one of which will be integrality, if there are such variables). 
Therefore, $u = v$ is equivalent to a set of simple equations, thus establishing that (4) implies (2). 
\end{proof}

\begin{remark}
We note that in the literature \cite{GalJip13} simple equations are defined without the demand that all variables on the left-hand side appear also on the right-hand side; such a definition naturally includes integrality, as well as equations such as  $xy\leq x^2$. 
With our definition, the latter is intentionally considered not simple, as  it is somewhat vexing to handle it directly in our work, while integrality is considered simple by exception. 
Nevertheless, we are able to handle $xy\leq x^2$ indirectly, as it is equivalent to the conjunction of the simple (in our sense) equations $x \leq x^2$ and $y \leq 1$---our arguments are more streamlined when we consider sets of simple equations (in our sense).

Our slightly modified definition of simple rules is one of the reasons why we opted to give a full proof of Lemma~\ref{l: N_2^-0}. 
As our definition of simple equation is narrower than in the literature, all the existing results about simple equations apply to our context, and we make free use of them.
\end{remark}

\section{Weakly commutative knotted substructural logics}
\label{sec:wck-substructural-logics}

We are ready to formally define the logics that are the focus of this work.

A \emph{knotted equation}\index{knotted equation} is an equation of the form $x^n \leq x^m$ and a \emph{knotted axiom}\index{knotted axiom} is one of the form $\UniFmA^n\ld\UniFmB^m$, {denoted by $\UniWeakKProp{m}{n}$}, where $m$ and $n$ are \emph{distinct} integers with $n > 0$.
(Note: $n=0$ results in trivial logics. Also, $m=0$ is the same as taking $m=0$, $n=1$~\cite{gavin2019}, so one could restrict to the case $m,n \geq 1$.)
Given the already discussed connections between equations and formulas, in this section we will focus  on equations for simplicity.
When $n<m$, the equation is called a \emph{knotted contraction equation} and denoted by $\UniWeakCProp{m}{n}$, and when $n>m$ it is called a \textit{knotted weakening equation} and denoted by $\UniWeakWProp{m}{n}$. (The corresponding axioms are denoted in the same way.)
For example, integrality ($x \leq 1$), contraction ($x \leq x^2$) and mingle ($x^2\leq x$) are knotted equations. 
Note that knotted equations are not simple equations except when $n=1$, but the linearization process given in the proof of Lemma~\ref{l: N_2^-0} yields the following simple equation equivalent to $x^n \leq x^m$:
$$x_1 x_2 \cdots x_n \leq \bigvee \{x_{i_1}\cdots x_{i_m}: i_1, \ldots, i_m \in \{1,\ldots,n\}\}.$$

\begin{example}\label{ex: lin}
As an example, consider the linearization of the equation $x^2\leq x^3$. Using the substitution $x\mapsto u\vee v$ and  distributing multiplication of joins, we obtain
$$u^2\vee uv\vee vu \vee v^2 \leq u^3\vee u^2v\vee uvu \vee uv^2\vee vu^2\vee vuv\vee v^2 u\vee v^3.$$
Since $\vee$ is a $\vee$-semilattice operation, we have $uv\leq u^2\vee uv\vee vu \vee v^2$, so the above entails the simple equation
$$uv\leq u^3\vee u^2v\vee uvu \vee uv^2\vee vu^2\vee vuv\vee v^2 u\vee v^3.$$
This equation, conversely, implies $x^2\leq x^3$ via the substitution $u,v\mapsto x$; so the two equations are equivalent relative to $\UniRLV$. 
Likewise, the linearization of $x^3\leq x^2$ (via $x\mapsto u\vee v\vee w$) is the simple equation:
\begin{equation*}
uvw \leq u^2 \vee uv \vee uw \vee vu \vee v^2 \vee vw \vee wu \vee wv \vee w^2.\tag*{\qedhere}
\end{equation*}
\end{example}

As we have already discussed, the hypersequent rules corresponding to $x^n \leq x^m$  and to the simple equation above are given  (in their hypersequent rendering) by

\begin{center}
\AxiomC{$\UniHyperMSetA \VL 
\UniSequent{\UniMSetFmA,\UniFmA^m,\UniMSetFmB}{\UniMSetSucA}$}
\RightLabel{$\UniWeakKRule{m}{n}$}
\UnaryInfC{$\UniHyperMSetA \VL
\UniSequent{\UniMSetFmA,\UniFmA^n,\UniMSetFmB}{\UniMSetSucA}$}
\DisplayProof
\; 
\AxiomC{$\{\UniHyperMSetA\VL
\UniSequent{\UniMSetFmA, \UniMSetFmB_{1}^{r_1}, \ldots, \UniMSetFmB_{n}^{r_n},\UniMSetFmC}{\UniMSetSucA}\}_{\sum r_i = m}$}
\RightLabel{$\UniWeakKAnaRule{m}{n}$}
\UnaryInfC{$\UniHyperMSetA \VL
\UniSequent{\UniMSetFmA, \UniMSetFmB_1, \ldots, \UniMSetFmB_n, \UniMSetFmC}{\UniMSetSucA}$}
\DisplayProof
\end{center}

We call \emph{knotted rules} the rules of the form $\UniWeakKAnaRule{m}{n}$. 
Instead of $\UniWeakKRule{m}{n}$ and $\UniWeakKAnaRule{m}{n}$, we may write $\UniWeakCRule{m}{n}$ and $\UniWeakCAnaRule{m}{n}$ when $n<m$, i.e., these rules correspond to knotted contraction equations (the latter being called \emph{knotted contraction rules}) and analogously when $m<n$ for $\UniWeakWRule{m}{n}$ and $\UniWeakWAnaRule{m}{n}$ (the latter called \emph{knotted weakening rules}).
The reason is that the proof argument sometimes depends on whether the knotted rule is has contraction type or weakening type.
For the purely sequent form, we omit the hypersequent variable $H$. 
In general, if, for a set $\UniAxiomSetA$  of $\mathcal{N}_2$ formulas, $\UniAxiomSetA^{\mathsf{seq}}$ denotes the set of corresponding sequent rules and $\UniAxiomSetA^{\mathsf{hseq}}$ denotes the set of corresponding hypersequent rules (by adding a hypersequent context variable to the rules in $\UniAxiomSetA^{\mathsf{seq}}$), then: a  sequent $s$ is provable in $\UniCalcExt{\m{FL}}{\UniAxiomSetA^{\mathsf{seq}}}$ iff the hypersequent $h\mid s$ is provable in $\UniCalcExt{\m{HFL}}{\UniAxiomSetA^{\mathsf{hseq}}}$, for some hypersequent $h$.

Another type of $\{\jn, \cdot, 1\}$-equations that will play a prominent role in our study (see Chapters~\ref{sec:fepP3}, Chapter~\ref{sec:noncom-ub} and Chapter~\ref{s: joinand-incresing-lb}) is the following.
A \emph{weak commutativity equation} is an equation of the form:
$$xy_1xy_2\cdots y_k x \UniEq x^{a_0}y_1 x^{a_1}y_2 \cdots y_k x^{a_k},$$
where $a_0+a_1+\cdots +a_k=k+1$ and not all $a_i$ are equal to $1$ (if all $a_i$ are equal to $1$, the two sides of the equation are identical and the equation is trivial).
For example, $xyx \UniEq xxy$, $xyx \UniEq yxx$ and $xyxzx \UniEq yx^2zx$ are generalized commutativity equations. 
We also allow commutativity itself to be considered a generalized commutativity equation. 
We call the corresponding axiom \emph{weak exchange axiom} or \emph{weak commutativity axiom}.

In the form of a hypersequent rule, these equations translate {(recall Section~\ref{s: P3})} to (the double line indicates that we have two rules, the one below and its converse):

    \begin{center}
    \AxiomC{$\UniHyperMSetA \VL
    \UniSequent{\UniMSetFmC_1,\UniMSetFmB, \UniMSetFmA_1, \UniMSetFmB, \UniMSetFmA_2, \UniMSetFmB, \ldots, 
        \UniMSetFmB, \UniMSetFmA_k, \UniMSetFmB, \UniMSetFmC_2}{\UniMSetSucA}
        $
    }
    \RightLabel{}
    \doubleLine
    \UnaryInfC{$\UniHyperMSetA \VL
        \UniSequent{\UniMSetFmC_1,\UniMSetFmB^{a_0}, \UniMSetFmA_1, \UniMSetFmB^{a_1}, \UniMSetFmA_2, \UniMSetFmB^{a_2}, \ldots, 
        \UniMSetFmB^{a_{k-1}}, \UniMSetFmA_k, \UniMSetFmB^{a_k}, \UniMSetFmC_2}{\UniMSetSucA}
        $
    }
    \DisplayProof
    \end{center}

By Lemma~\ref{l: N_2^-0}, we can always obtain the linearized form of weak commutative equations.
For example, in the case of $xyx=xxy$, we obtain equations $x_1yx_2 \leq x_1^2y \lor x_1x_2y \lor x_2x_1y \lor x_2^2y$ (equivalent to $xyx\leq xxy$) and $x_1x_2y \leq x_1yx_1 \lor x_1 y x_2 \lor x_2 y x_1 \lor x_2 y x_2$ (equivalent to $xxy \leq xyx$).
The corresponding hypersequent rules are:
\begin{center}
\AxiomC{ 
$\{\UniHyperMSetA\VL
\UniSequent{\UniMSetFmA,C_1,C_2,\UniMSetFmD,\UniMSetFmC}{\UniMSetSucA}\}_{(C_1,C_2) \in \{ \UniMSetFmB_1,\UniMSetFmB_2\}^2}$}
\RightLabel{}
\UnaryInfC{$\UniHyperMSetA \VL
\UniSequent{\UniMSetFmA, \UniMSetFmB_1, \UniMSetFmD, \UniMSetFmB_2, \UniMSetFmC}{\UniMSetSucA}$}
\DisplayProof
\end{center}
\begin{center}
\AxiomC{ 
$\{\UniHyperMSetA\VL
\UniSequent{\UniMSetFmA,C_1,\UniMSetFmD,C_2,\UniMSetFmC}{\UniMSetSucA}\}_{(C_1,C_2) \in \{ \UniMSetFmB_1,\UniMSetFmB_2\}^2}$}
\RightLabel{}
\UnaryInfC{$\UniHyperMSetA \VL
\UniSequent{\UniMSetFmA, \UniMSetFmB_1,  \UniMSetFmB_2, 
\UniMSetFmD,\UniMSetFmC}{\UniMSetSucA}$}
\DisplayProof
\end{center}
We call \emph{weak exchange rules} or \emph{generalized exchange rules} the rules that correspond to weak commutativity equations (like the ones above).

A \emph{weakly commutative knotted substructural logic} is then an axiomatic extension of a substructural logic of the form $\UniFLExtLogic{\UniWEProp{\vec a}\UniWeakKProp{m}{n}}$, that is, of $\UniFLExtLogic{}$, axiomatically extended with one weak exchange axiom and one knotted axiom.

Regarding the computational properties of these logics, Hori \emph{et al.}~\cite{hori1994} (1994)  established decidability of provability  via proof theory only for the special cases $\UniFLeExtLogic{\UniWeakCProp{m}{1}}$ and $\UniFLeExtLogic{\UniWeakWProp{1}{n}}$ and mentioned ``great difficulties in the remaining cases''. Van Alten~\cite{vanalten2005} (2005) extended the decidability results to $\UniFLeExtLogic{\UniWeakCProp{m}{n}}$ and $\UniFLeExtLogic{\UniWeakWProp{m}{n}}$ by  showing that their algebraic models satisfy the finite embeddability property (FEP), discussed in detail in the next section. 
Decidability of provability and deducibility of the logics $\UniFLExtLogic{\UniWEProp{\vec a}\UniWeakKProp{m}{n}}$ has been established also via the FEP in~\cite{Cardona2015}.
In the next sections, we will fill many gaps regarding the decidability (FEP and proof theory) and complexity of axiomatic extensions of $\UniFLExtLogic{\UniWEProp{\vec a}\UniWeakKProp{m}{n}}$ by $\mathcal{P}_3^\flat$ axioms (see Table~\ref{tab:table-contrib} on page \pageref{tab:table-contrib} for an overview).

%% file: tex/fepP3.tex
As a corollary of the results in this chapter we obtain the decidability of the  deducibility relation of (infinitely) many extensions of $\UniFLExtLogic{}$, including extensions by a knotted axiom and a weak commutativity equation.
For knotted extensions over $\m {FL_e}$ the finite embeddability property was proved in \cite{vanalten2005} and its generalization to knotted extensions over $\m {FL}$ plus a weak commutativity equation was shown in \cite{Cardona2015}. 
Here we extend these results to include arbitrary $\mathcal{P}_3^\flat$ equations.

As discussed in Section~\ref{s: AlgSem}, the deducibility relation $\vdash_{\UniFLExtLogic{}}$ of the substructural logic $\UniFLExtLogic{}$ (as well as all of its extensions) is algebraizable and its algebraic semantics is the variety of residuated lattices. 
In this section we prove that the finite embeddability property holds for all varieties of residuated lattices axiomatized by a knotted axiom plus a generalized commutativity equation plus any set (possibly empty) of analytic $\mathcal{P}_3^\flat$ equations. 
This implies the decidability of the universal theory of these (infinitely many) varieties and, via the algebraization result, it entails the decidability of the deducibility relations of the corresponding extensions of $\UniFLExtLogic{}$. 
Actually, every time a deduction in these logics fails, we are able to produce a finite countermodel (\emph{strong finite model property}). 
In particular, by specializing the result to theorems only, we get that all of these logics enjoy the finite model property. 

Our main tools in this section will be residuated frames and hyperframes. 
These are relational semantics for logics defined over $\mathcal{N}_2$ and  $\mathcal{P}_3^\flat$ axioms, respectively. 
These structures have the power and generality to encompass situations from both algebra and from proof theory and yield new and interesting results on both of these areas. 
Each such structure gives rise to a residuated lattice, called its \emph{Galois algebra}. 
The Galois algebra of the residuated frame we will consider will be finite and the  Galois algebra of the residuated hyperframe we will consider will be back in our variety (i.e., a model of the logic). 
We will then further prove that the two algebras are the same and this will be our countermodel enjoying all of the desired properties.

\section{Finite embeddability property and decidability}
\label{sec:fep-def}

We say that a class $\mathcal{K}$ of algebras over a signature $\mathcal{L}$ has the \emph{finite embeddability property} (FEP, for short) if for every algebra $\m A \in \mathcal{K}$ and finite subset $B$ of $A$, there is a finite algebra $\m D \in \mathcal{K}$ such that $\m B$ embeds in $\m D$ (as a partial algebra), where $\m B$ is the partial subalgebra  of $\m A$ with underlying set $B$. 
This means that, for example, if $a,b \in B$ and $a \bullet^{\m A} b \in B$, where $\bullet$ is a binary operation in $\mathcal{L}$, then $f(a \bullet^{\m A} b) = f(a) \bullet^{\m D} f(b)$; if $a \bullet^{\m A} b \not \in B$ then no condition is imposed.
In other words, the graph of the partial algebra $\m B$ embeds in $\m D$ via the injective map $f$.

The FEP for a class $\mathcal{K}$ implies that if a universal first-order formula fails in $\mathcal{K}$, then it actually already fails in a finite algebra of $\mathcal{K}$. 
Indeed, if the formula is not valid, then it will fail in some algebra $\m A \in \mathcal{K}$ under some valuation; let $B$ be the subset of $A$ consisting of all of the values of the subterms of the formula under the valuation (hence the finite set $B$ witnesses the failure). 
By the FEP, there exists a finite algebra $\m D \in \mathcal{K}$ such that $\m B$ embeds in $\m D$ as a partial algebra. 
Therefore, the same failure  is exhibited in $\m D$: all the calculations performed in $\m A$ yielding elements of $B$ can be faithfully performed in $\m D$ due to the partial algebra embedding. 

If a universal class {(a class defined by universal formulas)} $\mathcal{K}$ is finitely axiomatized and has the FEP, then its universal theory (the set of the universal formulas that hold in $\mathcal{K}$) is decidable. 
Indeed, given a universal first-order formula, if it is valid in $\mathcal{K}$ then it can be deduced in finite time by running a theorem prover (enumerating all of the first-order consequences of the finitely many axioms of $\mathcal{K}$). 
On the other hand, if it fails in $\mathcal{K}$  then by the preceding argument it fails in a finite algebra of $\mathcal{K}$. 
Now we run a model-checker: for every natural number {$n$}, we sequentially consider all possible algebras of
size $n$ in the signature of $\mathcal{K}$, and for each one we check if the (finitely many) axioms of $\mathcal{K}$ are satisfied. 
(In case that the signature is infinite, there are infinitely many algebras of each finite size, but we only need the interpretations of the operations in the language that are involved in the axiomatization, thus yielding finitely many pertinent reducts.) 
For each such finite algebra we check if the universal formula fails, so eventually we will find a finite countermodel for it. 

Therefore, the FEP (combined with finite axiomatizability) yields a very strong decidability result. 
We will describe infinitely many  varieties of residuated lattices that have the FEP (and are finitely-axiomatized), so all of them have decidable consequence relations. 
By algebraizability, this can then be transferred to the corresponding substructural logics.

In the following, we will employ one important notion from universal algebra. 
A \emph{subdirect product} is a subalgebra $\m A$ of a product $\prod_{i \in I} \m A_i$ of a family of algebras $(\m A_i)_{i \in I}$ such that the restriction $\pi_j: \m A \to \m A_j$  to $\m A$ of each projection is onto. 
An algebra $\m B$ is called \emph{subdirectly irreducible} if, whenever it is isomorphic to a subdirect product $\m A$ of algebras $(\m A_i)_{i \in I}$, say by an isomorphism $f: \m B \to \m A$, there is an $i \in I$ such that $\pi_j \circ f: \m B \to \m A_i$ is an isomorphism; in other words, an algebra is subdirectly irreducible if it cannot be written as a subdirect product in a non-trivial way. 
Subdirectly irreducible algebras admit an internal characterization in terms of their lattice of congruences and they play an important role in studying varieties of algebras: every variety is fully determined by the subdirectly irreducible algebras that it contains. 
In the following we will first prove the FEP for subdirectly irreducible algebras and then extend the property to the whole variety. 

\section{Relational semantics: residuated frames and hyperframes}
\label{s: RelSem}

It is well known that intuitionistic propositional logic, aside from its algebraic semantics (Heyting algebras), also has relational semantics, known as Kripke frames. 
Kripke frames are essentially dual structures to Heyting algebras (to be precise this works for the subclass of perfect Heyting algebras) for which they serve as their posets of join irreducibles (for the finite case; prime filters are needed in the infinite case). 
The join irreducibles are enough to obtain the algebra (in the finite case), because Heyting algebras have distributive lattice reducts.

For arbitrary residuated lattices (with no stipulation of lattice distributivity) two-sorted relational semantics are needed; in the finite case the two sorts correspond to the sets of join irreducible and meet irreducible elements of the lattice. 
In proof-theoretic terms, the two sorts correspond to the left-hand sides and the right-hand sides of sequents in a calculus. 
Therefore, the resulting two-sorted relational structures, which are known as residuated frames, can capture aspects of both the algebraic semantics and of the proof-theory of a logic. 
We will make crucial use of them in the proof of the FEP (and also later in Chapter~\ref{sec:lowerbounds} where they will also connect to computations of counter machines).

For our purposes, a \textit{residuated frame} $\mathbf{W}\UniSymbDef(W,W', N, \circ, \ldd, \rdd, e)$ is a two-sorted structure, where $(W, \circ, e)$ is a monoid, $W'$ is a set, $\ldd: W' \times W \rightarrow W'$ and  $\rdd: W \times W' \rightarrow W'$ are functions and $N$ is a subset of $W \times W'$, such that the relation $N$ is \emph{nuclear}: for all $x,y \in W$ and $z \in W'$,
$$x \circ y \mathrel{N} z \iff  x \mathrel{N} z \rdd y \iff y \mathrel{N} x \ldd z.$$
For $X\cup\{x\}\subseteq W$ and $Y\cup\{y\}\subseteq W'$ we define $X \mathrel{N} y$ to mean $x \mathrel{N} y$ for all $x \in X$, and 
$x \mathrel{N} Y$ to mean $x \mathrel{N} y$ for all $y \in Y$; also, we define $X^\triangleright:=\{y\in W':X \mathrel{N}y\}$ and $Y^\triangleleft:=\{x\in W: x \mathrel{N} Y\}$. The pair $({}^\triangleright,{}^\triangleleft)$ forms a {\em Galois connection} and it gives rise to the closure operator $\gamma(X):= X^{\triangleright\triangleleft}$. We write $\gamma(x)=\gamma(\{x\})$ for $x\in W$, and set $\UniPowerSet{W}_\gamma:=\gamma[\UniPowerSet{W}]$. It follows from \cite{GalJip13} that the algebra 
$$\mathbf{W}^+:=(\UniPowerSet{W}_{\gamma},\cap,\cup_{\gamma} , \circ_{\gamma},\ld, \rd, \gamma(e))$$ is a residuated lattice, where $X\cup_\gamma Y:=\gamma(X\cup Y)$, $X\circ_\gamma Y:=\gamma(X\cdot Y)$,  $X\ld Y:=\{z\in W: X\circ\{z\}\subseteq Y\}$ and $Y\rd X:=\{z\in W: \{z\} \circ X \subseteq Y\}$. 
We call  $\mathbf{W}^+$ the \emph{Galois algebra} of $\m W$ and also write $W^+:=\UniPowerSet{W}_\gamma=\gamma[\UniPowerSet{W}]$ for its underlying set. 
Every $X \in W^+$ is the intersection of \emph{basic closed} subsets, i.e., subsets of $W$ of the form $z^\triangleleft:=\{z\}^\triangleleft$, where $z \in W'$.

A (pointed) residuated frame is called \emph{commutative} if the monoid $(W, \circ, e)$  is commutative; in this case we keep only $\ldd$ in the signature, as $y \rdd x$ can be taken to be $x \ldd y$. 

\begin{example}
\label{e: frames}
Given a residuated lattice $\m A=(A,\mt, \jn, \cdot, \ld, \rd, 1)$, we  obtain a residuated frame $\m W_{\m A}$ by taking  $W=W'=A$, $\circ = \cdot$, $\varepsilon =1$, $\ldd=\ld$, $\rdd=\rd$, and $N=\,\leq$. 
It is shown in \cite{GalJip13} that this is indeed a residuated frame. 
Its Galois algebra $\m W_{\m A}^+$ is actually the Dedekind-McNeille completion of $\m A$~\cite[Sec.~3.4.12]{GalJipKowOno07} and the map $a \mapsto \{a\}^\triangleleft$ is an embedding of $\m A$ in $\m W_{\m A}^+$. 

Starting from the calculus $\m {FL}$ we obtain a  residuated frame $\m W_{\m {FL}}$, where $W$ is the set of finite sequences of formulas (i.e., the free monoid over the set of formulas $\UniLangSet{P}$), $\circ$ is concatenation of sequences, $\varepsilon$ is the empty sequence, $W'=W \times \UniLangSet{P} \times W$, and $N$ is defined by $x \mathrel{N} (y,a, z)$ iff the sequent $y \circ x \circ z \Rightarrow a$ is provable in $\m {FL}$. 
It is shown \cite{GalJip13} that $\m W_{\m {FL}}^+$ generates the variety of residuated lattices, thus yielding a completeness theorem; actually $\m W_{\m {FL}}^+$ is the Dedekind-McNeille completion of the free residuated lattice. 

Similarly, the residuated frame $\m W_{\UniDelCut{\m {FL}}}$ is defined in the same way, but $N$ is taken with respect to provability in the cut-free version $\UniDelCut{\m {FL}}$ of $\m {FL}$. 
Using this frame \cite{GalJip13} obtains a  short proof of cut elimination for $\m {FL}$.

If $\m A$ is a residuated lattice and $B$ is a subset of $A$, it is shown in \cite{GalJip13} that $\m W_{\m A, \m B}$ is a residuated frame, where $(W, \cdot, 1)$ is the submonoid  of $\m A$ generated by $B$, $W’=W \times B \times W$, the relation $N \subseteq W \times W’$ is defined by $x \mathrel{N} (y,b,z)$ iff $yxz \leq^{\m A} b$, and  $x \ldd (y,b,z):=(yx, b, z)$ and $(y, b, z) \rdd x:=(y, b, xz)$. 
It is further shown that if $B$ is finite then the Galois algebra $\m W_{\m A, \m B}^+$ is finite. 
We will use this frame in the proof of the FEP below.
\end{example}

Residuated frames have many more applications in the study of extensions by $\mathcal{N}_2$ axioms and we will also see them used in Chapter~\ref{sec:lowerbounds} to obtain lower complexity bounds. 
In our case, we consider \emph{pointed} residuated frames: they are expansions with one more constant $d \in W'$, and the Galois algebra is an $\m {FL}$-algebra, where the negation constant is the closed set $d^\triangleleft$.

For extensions by $\mathcal{P}_3^\flat$ axioms, a more complex type of relational semantics is needed. 
The elements of the set $H$ below correspond to hypersequents in a calculus. 

A \emph{residuated hyperframe} is a structure $\m H=(W, W', {\Vdash}, \circ, \ldd, \rdd, e)$, where $(W, \circ, e)$ is a monoid, $W'$ is a set,   $\ldd: W' \times W \rightarrow W'$ and  $\rdd: W \times W' \rightarrow W'$ are functions, and $\Vdash$ is a subset of $H$, such that $\Vdash$ is \emph{nuclear} and satisfies \emph{external weakening} and \emph{external contraction}: for all $x,y \in W$, $z \in W'$ and $h, g \in H$,
\begin{itemize}
 \item $\Vdash h \mathrel{|} x \circ y \hra z \iff \Vdash h \mathrel{|}  y \hra x \ldd z \iff \Vdash h \mathrel{|} x \hra z \rdd x\; $  (nuclear).
 \item $\Vdash h$ implies $\Vdash h \mathrel{|} g$ (EW).
 \item  $\Vdash h \mathrel{|} g\mathrel{|}g \;$ implies $\Vdash h \mathrel{|} g \;$ (EC).
\end{itemize}
Here we write $\Vdash h$ for $h \in {\Vdash}$ and $x \hra y$ for $(x,y)$, when $x \in W$ and $y \in W'$; also $|$ denotes the concatenation operation of $H$ and we will denote its neutral element (the empty word) by $e_H$. 
Again, a pointed version of such frames contains a constant $d \in W'$.

Given such a residuated hyperframe $\m H$, we can obtain a residuated frame $r(\m H):=(H\times W, H\times W',N,\circ, \ldd, \rdd, (e_H , e_W))$, where 
\begin{center}
$(h_x, x) \circ (h_y, y)=(h_x|h_y, x \circ y)$\\
$(h_x, x) \ldd (h_z, z)=(h_x|h_z, x \ldd z)$\\
$(h_z, z) \rdd (h_y, y)=(h_z|h_y, z \rdd y)$\\
$(h_x, x) \mathrel{N} (h_z, z) \iff \;\; \Vdash h_x \mathrel{|} h_z \mathrel{|} x \hra z$.
\end{center}
The \emph{Galois algebra of $\m H$} is defined to be that of $r(\m H)$, i.e., $\m H^+:=(r(\m H))^+$; we write $H^+$ for the underlying set of $\m H^+$. 
In the pointed version, the new constant is $(e_H, d)$. 
In the rest of the section, we will work with the pointed version of these notions and obtain results about $\m{FL}$-algebras, but the same results can be obtained for residuated lattices by ignoring the additional constant of the pointed frames.

In \cite{CiaGalTer17} it is shown that residuated hyperframes are suitable for studying $\mathcal{P}_3^\flat$ axioms. 
Also an analogue of the Dedekind-McNeille completion $\m W_{\m A}^+$ of a residuated lattice $\m A$ is given, called \emph{hyper DM-completion} and is used to show the intimate connection between cut elimination in hypersequent calculi and closure of the corresponding variety under completions. 
Some of this work is behind the results we presented in Section~\ref{s: P3/N2}.

\section{Proof of the FEP}
We are now ready to state the result and carry out the proof of the FEP for extensions by $\mathcal{P}_3^\flat$ axioms. 
Recall the definitions of knotted equations and weak commutativity from Section~\ref{s: P3/N2}. 

\begin{theorem}\ \label{t: FEP} 
\begin{enumerate}
\item Any variety of residuated lattices or $\m {FL}$-algebras axiomatized by a knotted equation and a weak commutativity equation and any (possibly empty) set of analytic $\mathcal{P}_3^\flat$ equations has the FEP. 
\item Any variety of integral residuated lattices or $\m {FL_i}$-algebras or $\m {FL_w}$-algebras  axiomatized by any (possibly empty) set of analytic $\mathcal{P}_3^\flat$ equations has the FEP. 
\end{enumerate}    
\end{theorem}

\begin{proof}
We first note that it suffices to show that the FEP holds for subdirectly irreducible algebras. 
Indeed, by a basic result in universal algebra, every algebra $\m A$ can be decomposed as a subdirect product of subdirectly irreducibles:  $\m A$ is isomorphic to a subalgebra of $\prod_{i \in I} \m A_i$, for some family  $(\m A_i)_{i \in I}$ of subdirectly irreducible algebras such that the restriction $\pi_j: \m A \to \m A_j$ of each projection  to $\m A$ is onto. 
So each $\m A_j$ is a homomorphic image of $\m A$, which is in the variety, so all of the $\m A_j$'s are in the variety.  
If $B$ is a finite subset of $\m A$, then  finitely many coordinates in $I$ suffice to separate the points of $B$; let $J\subseteq I$ be such a subset  of coordinates.   
Since $B$ is finite,  the subset $B_i:=\pi_i[B]$ of $\m A_i$ is also finite for every $i \in I$. 
By applying the FEP for subdirectly irreducibles,  for every $i \in J$ there exists a finite algebra $\m D_i$ in the variety such that the partial algebra $\m B_i$ embeds in $\m D_i$, say via $f_i: B_i \to D_i$. 
Therefore, the direct product $\m D:=\prod_{i\in J}\m D_i$ is finite and it is in the variety. 
Also, it is easy to see that the function $f:B \to D$, defined by $f(b) := (f_i(\pi_i(b)))_{i\in J}$, is an embedding of the partial algebra $\m B$ into $\m D$. 

So, we consider a subdirectly irreducible residuated lattice $\m A$ in the variety and $\m B$ a partial subalgebra of $\m A$.  
As we mentioned in Example~\ref{e: frames}, it is shown in \cite{GalJip13} that the following structure $\m W:=\m W_{\m A, \m B}$ is a residuated frame: $\m W_{\m A, \m B}$ consists of the submonoid $(W, \cdot, 1)$ of $\m A$ generated by $B$, the set $W’=W \times B \times W$, the nuclear relation $N \subseteq W \times W’$ such that $x \mathrel{N} (y,b,z)$ iff $yxz \leq^{\m A} b$, and the functions given by $x \ldd (y,b,z):=(yx, b, z)$ and $(y, b, z) \rdd x:=(y, b, xz)$.

By results of \cite{GalJip13} and \cite{Cardona2015} we get that, in both cases (1) and (2) of the theorem, the Galois algebra $\m W^+$ of the residuated frame $\m W$ is a finite residuated lattice that satisfies the knotted equation and the weak commutativity equation, and also that $\m B$ embeds in $\m W^+$. 
It remains to show that $\m W^+$ satisfies the analytic $\mathcal{P}_3^\flat$ equations. 
Since analytic $\mathcal P_3^\flat$ equations are preserved by residuated hyperframes, rather than residuated frames, we pivot our analysis to a suitable residuated hyperframe whose Galois algebra ends up being isomorphic to $\m W^+$.

We consider the residuated hyperframe\footnote{This can be considered as the FEP analogue of the residuated hyperframe $\m H_{\m A}$ of \cite{CiaGalTer17}.} $\m H$ with the same $(W, \circ, e)$, $W’$, $\ldd$ and $\rdd$ as in $\m W_{\m A, \m B}$ above and where 
\begin{center}
$\Vdash a_1 \hra (y_1, b_1, z_1) \mathrel{|} \cdots \mathrel{|}   a_n \hra (y_n, b_n, z_n)$\\ iff \\ 
$y_1a_1z_1 \leq^{\m A} b_1 \; $ or $\ldots$ or $\; y_na_nz_n \leq^{\m A} b_n$.
\end{center} 
In particular, $\Vdash e_H$ does not hold (recall that $e_H$ is the empty word). 
It is straightforward to verify that $\m H$ is indeed a residuated hyperframe, using the residuation of $\m A$.

We will show that $\m W$ and $\m H$ have isomorphic Galois algebras\footnote{The proof is vaguely inspired by Lemma~6.5 and Proposition~6.6 of \cite{CiaGalTer17}.}. 
Recall that $\m H^+$ is defined to be $(r(\m H))^+$; we denote by $N_H$ the nuclear relation of $r(\m H)$.
Note that by the definition of $\Vdash$,  for all $h,g \in H$, $a \in W$ and $z=(x,b,y)\in W'$ we have $(h,a) \mathrel{N_H} (g,z)$ iff 
 \begin{center}
    $\Vdash h \mathrel{|} g \mathrel{|} a \hra (x,b,y)$ iff ($\Vdash h$ or $\Vdash g$ or $xay \leq^{\m A} b$) iff ($h \in T$ or $g\in T$ or $a \mathrel{N} z$),  
 \end{center}
where $T:=\{h \in H: \; \Vdash h\}$. 
This calculation of $N_H$ shows that if $g \in T$ and $z \in W'$, then $(g,z)^{\triangleleft}=H \times W$. 
Also, if  $g \not \in T$ and $z \in W'$, then $(g,z)^{\triangleleft}=(e_H,z)^{\triangleleft}$. 
So, the only basic closed sets of $\m H^+$ are $H \times W$ and  $(e_H,z)^{\triangleleft}$, for $z \in W'$. 
The calculation also shows that for all $z \in W'$, if $h \in T$ and $a \in W$, then $(h,a) \in (e_H,z)^{\triangleleft}$, so $T \times W \subseteq  (e_H,z)^{\triangleleft}$. 
Finally, if $h \not \in T$, then $(h,a) \in (e_H,z)^{\triangleleft}$ iff $(e_H,a) \in (e_H,z)^{\triangleleft}$ iff $a \mathrel{N} z$ iff $a \in z^\triangleleft$, where the latter is computed in $\m W$. In summary, 
$$(g,z)^{\triangleleft}=
\begin{cases}
H \times W & \text{if } g \in T\\
(g,z)^{\triangleleft}=(T \times W) \cup (H \times z^\triangleleft) & \text{if } g \not \in T
\end{cases}
$$

Recall that every closed subset is an intersection of basic closed {subsets} and note that $\bigcap_{z \in Z} (T \times W) \cup (H \times z^\triangleleft)=(T \times W) \cup (H \times Z^\triangleleft)$, for every $Z \subseteq W'$. 
Also, recall that $W^+=\{Z^\triangleleft: Z \subseteq W'\}$, and note that $H \times W$ is subsumed in the more general form of elements, since $H \times W=(T \times W) \cup (H \times W)$ and $W \in W^+$. 
So, we can describe all the closed subsets: 
$$H^+= \{(T \times W) \cup (H \times X): X \in W^+\}.$$ 
As a result, the map $X \mapsto (T \times W) \cup (H \times X)$ is an order isomorphism from $\m W^+$ to $\m H^+$. 
To see that it is a monoid isomorphism, note that for $X_1, X_2 \in W^+$, 
$$((T \times W) \cup (H \times X_1)) \circ ((T \times W) \cup (H \times X_2))=(T \times W) \cup (H \times (X_1 \circ X_2)),$$ since 
$h | g \in T$ if  $h\in T$ or $g \in T$.
Moreover, for $X \subseteq W$ and $ Y \in W^+$, we have: 
$$\gamma((T \times W) \cup (H \times X)) \subseteq (T \times W) \cup (H \times Y)$$ iff
$$(T \times W) \cup (H \times X) \subseteq (T \times W) \cup (H \times Y)$$ iff
$X \subseteq Y$; hence 
$$\gamma((T \times W) \cup (H \times X))=(T \times W) \cup (H \times \gamma(X)).$$ 
Therefore, for $X_1, X_2 \in W^+$, 
$$\gamma(((T \times W) \cup (H \times X_1)) \circ ((T \times W) \cup (H \times X_2)))=(T \times W) \cup (H \times \gamma(X_1 \circ X_2)),$$ 
i.e., the map is a multiplicative homomorphism. 
Since the residuated lattice structure is definable by the order and the multiplication, we obtain that $\m H^+\cong \m W^+$.
 
Let $R$ be a set of analytic $\mathcal P_3^\flat$ equations/formulas and $\mathcal{R}$ the corresponding set of analytic structural rules. 
In particular, the formulas in $R$ are provable in the system $\UniCalcExt{\UniFLExtHCalc{}}{\mathcal{R}}$, so Theorem~5.16 of \cite{CiaGalTer17} is applicable:  to show that $\m H^+(=\m W^+)$ satisfies the $\mathcal P_3^\flat$ equations in $R$, it suffices to show that $\m H$ satisfies the corresponding rules $\mathcal{R}$ \emph{pointwise}, i.e., if it satisfies the assumption evaluated to elements of the corresponding sets, then it also satisfies the conclusion; see below. 

The proof that the clauses corresponding to the $\mathcal P_3^\flat$ equations hold pointwise in $\m H$ relies on the subdirect irreducibility of $\m A$ and is essentially the same as that of Lemma~6.7 of \cite{CiaGalTer17}. 
For completeness, we give the proof for a particular example, as is done in \cite{CiaGalTer17}: $xy\leq z \Longrightarrow x\leq 0$ or $y \leq z$ $(em)$. 
So, we need to verify that 
$$\infer[(em^\circ)]{\Vdash g \mathrel{|} x  \hra d \mathrel{|} y \hra z}{\Vdash g \mathrel{|} x \circ y \hra z}$$
for all $x,y \in W$, $z=(z_1,b,z_2)\in W'$ and $g \in H$.
Indeed, from $\Vdash g \mathrel{|} x \circ y \hra z$ we get ($\Vdash g$ or $z_1 \circ x \circ y \circ z_2 \leq b$), so ($\Vdash g$ or $x \leq 0$ or $z_1 \circ y \circ z_2 \leq b$), by $(em)$; hence $\Vdash g \mathrel{|} x  \hra d \mathrel{|} y \hra z$ holds.
\end{proof}

Since $\mathcal{P}_3^\flat$ equations include all $\mathcal{N}_2$ equations, which in turn include all equations of the form $t_0 \leq t_1 \jn \cdots \jn t_n$, where $t_i$ is a product of variables, the addition of further knotted rules and of further weak commutativity equations in the axiomatization of the variety is allowed by the statement of the theorem.

In particular, all the above varieties have the \emph{finite model property} (FMP): if an equation fails in the variety, then it fails in some finite algebra in the variety (equivalently, the variety is generated by its finite algebras). 
Also, if the variety is finitely axiomatizable then its equational theory (i.e., theoremhood in the corresponding logic), and even its quasiequational theory, is decidable.

Recall that an algebraizable consequence relation $\vdash$ has the \emph{strong finite model property}, sFMP (or the finite model property for deductions), if whenever a deductions in $\vdash$  fails, then there is a finite counter model in the algebraic semantics of $\vdash$ exhibiting the failure.

\begin{corollary}\label{c: algebraicFEP}\
\begin{enumerate}
\item Any logic corresponding to an extension of $\m {HFL}$ by a knotted rule and a generalized exchange rule and any (possibly empty) set of rules corresponding to analytic $\mathcal P_3^\flat$ formulas has decidable deducibility relation and the strong finite model property.
\item Any logic corresponding to an extension of $\m {HFL_i}$ or $\m {HFL_w}$ by any (possibly empty) set of rules corresponding to analytic $\mathcal P_3^\flat$ formulas has decidable deducibility relation and the strong finite model property.
\end{enumerate} 
\end{corollary}

In summary, not only do the above substructural logics enjoy very strong decidability properties, but also we have a way of obtaining finite (and concrete) countermodels for derivations that fail. 
On the other hand, the proof of this result does not provide any sense of the complexity of the decision procedure. 
In the following sections, we will provide upper and also lower complexity bounds for many of these logics, and in some cases these bounds are shown to be tight.  

We note that the results from \cite{GalJip13} and \cite{Cardona2015} that we used in the proof of Theorem~\ref{t: FEP} make crucial use of the notion of wqos and employ techniques involving them; we have not introduced wqo's up to now, as we used the results from these papers without reviewing their quite complicated proofs. 
Our  proofs of the upper complexity bounds in the next sections will also rely on wqos, this time in a proof-theoretic setting. As we will need to use wqos in sophisticated ways for our arguments, we introduce them in detail, review some basic facts about them and also prove results about the wqos that we will be using.

%% file: tex/knotted-wqos.tex
In this section, we recall the basics on wqos, we define the wqos that will be pertinent to our study (called \emph{knotted wqos}) and we establish useful (order reflection) results about them. 
We then present just enough of the theory of subrecursive and fast-growing hierarchies in order to prove length theorems for these wqos via the obtained order reflections, and to use them in complexity analysis of algorithms. 
The results of this chapter will be used in each of the following three chapters in order to obtain complexity bounds for various knotted substructural logics.

\section{Basics of wqo theory}

We begin by reviewing the notions of
quasi orders, bad sequences 
and well-quasi-orders; see~\cite{schuster2020} for a more detailed and accessible treatment.

A \emph{quasi order (qo)} $\UniWqoRel{}$ on a non-empty set $\UniWqoSet$ is a  reflexive and transitive  relation on $\UniWqoSet$. 
A \emph{quasi-ordered set (qoset)} is a structure $\UniWqoA\UniSymbDef\UniStruct{\UniWqoSet, \UniWqoRel{}}$,	where $\UniWqoSet$ is non-empty set and $\UniWqoRel{}$ is quasi order on $\UniWqoSet$. 
At times we abuse terminology and refer to $\UniWqoA$ as a qo (instead of a qoset).

A \emph{bad sequence over} a qo $\UniWqoA$ is a sequence $\{\UniValA_i\}_{i \in \UniNaturalInitSeg{\UniLengthSequence}}$ of elements of $\UniWqoA$ such that, for all $i < j$, $a_i \not\UniWqoRel{\UniWqoA} a_j$.
As usual, $\UniLengthSequence \in \UniNaturalSet \cup \UniSet{\omega}$ is called the \emph{length} of the sequence. 
Note that when $\UniLengthSequence=\omega$ we have an infinite sequence.

A qo $\UniWqoA$ is called a \emph{well-quasi-ordered set} (\emph{wqoset} or just \emph{wqo}) when every bad sequence over it is finite. 

For example, bad sequences in $\UniStruct{\mathbb{N}, \leq}$ are precisely the strictly decreasing sequences, and every such sequence is finite. 
So, $\UniStruct{\mathbb{N}, \leq}$ is a well-quasi-ordered set (actually a well-ordered set).
Also, in $\UniStruct{\mathbb{N}^2, \leq}$ with the usual direct product order, strictly decreasing sequences are examples of bad sequences and also \emph{antichains} (collections of pairwise incomparable elements) give rise to bad sequences. 

Because the lengths of such sequences cannot be upper bounded in arbitrary wqos
(as a simple example, it is easy to see that decreasing sequences over $\UniStruct{\mathbb{N}, \leq}$ may be arbitrarily long), 
we restrict our attention to
\emph{normed (w)qos} (or \emph{n(w)qos}) here, and to \emph{controlled bad sequences} in a following section.

The idea of nwqos is to be able to control how the \emph{size} of the elements grows according to their positions in a bad sequence.
The notion of size of an element in a (w)qo is realized via a \emph{proper norm}.

\begin{definition}
	A \emph{normed quasi order (nqo)} is a structure $\UniWqoA \UniSymbDef \UniStruct{\UniWqoSet, \UniWqoRel{}, \UniNorm{\cdot}{}}$, where $\UniStruct{\UniWqoSet, \UniWqoRel{}}$ is a qo and $\UniNorm{\cdot}{} : \UniWqoSet \to \UniNaturalSet$ is a \emph{proper norm}, i.e., for every $n \in \UniNaturalSet$, $\UniSet{\UniValA \in \UniWqoSet : \UniNorm{\UniValA}{} \leq n}$ is finite. 
    If $\UniStruct{\UniWqoSet, \UniWqoRel{}}$ happens to be a wqo, then we refer to $\UniWqoA$ as an nwqo.
\end{definition}

Note that proper norms are bijective to  partitions of $\UniWqoSet$ where every part is finite and where each part is labelled by a distinct natural number. 
(Given a proper norm, the parts of the partition are simply $\UniSet{\UniValA \in \UniWqoSet : \UniNorm{\UniValA}{} = n}$ for $n\in \mathbb{N}$.)

An example of a nwqo is the one over the natural numbers. 
Below we use lambda notation; for example, the function $x \mapsto x^2$ is denoted by $\lambda x.x^2$.

\begin{definition}
\label{def:wqo-natural}
$\UniWqoModNatural \UniSymbDef \UniStruct{\UniNaturalSet, \UniWqoNatural, \lambda \UniValA. \UniValA}$, where $\UniWqoNatural$ is the usual total order on the natural numbers.
\end{definition}

There are some standard ways of building (n)qos from other (n)qos, which will be important in many of our results.

\begin{definition}
\label{def:wqo-constructions}
    Let $n \in \UniNaturalSetNN$, $\{\UniWqoA_i \UniSymbDef \UniStruct{\UniWqoSet_i, \UniWqoRel{i}, \UniNorm{\cdot}{i}}\}_{i=1}^{n}$ be a family of nqos and
    $
	\UniWqoA \UniSymbDef \UniStruct{\UniWqoSet, \UniWqoRel{}, \UniNorm{\cdot}{}}
	$
	be a nqo. 
    Then we define the following structures:
	\begin{enumerate}
		\item $\prod_{i=1}^n \UniWqoA_i 
			\UniSymbDef 
			\UniStruct{\prod_{i=1}^{n} \UniWqoSet_i,
				\UniWqoRel{\Pi},
				\UniInfNorm{\cdot}}
		      $, the \emph{direct product of $\UniWqoA_1,\ldots,\UniWqoA_n$},
			where
			\begin{itemize}
				\item $\UniTuple{\UniValA_1,\ldots,\UniValA_n} \UniWqoRel{\prod} \UniTuple{\UniValB_1,\ldots,\UniValB_n}$ iff $\UniValA_i \UniWqoRel{i} \UniValB_i$, for all $1 \leq i \leq n$; and
				\item $\UniInfNorm{\UniTuple{\UniValA_1,\ldots,\UniValA_n}} \UniSymbDef \max_i \UniNorm{\UniValA_i}{i}$ (the \emph{infinity norm}).
			\end{itemize}
		\item $\UniWqoA^k \UniSymbDef 
   \prod_{i=1}^{k} \UniWqoA$, the \emph{$k$-power} of $\UniWqoA$.
		\item $\UniDisjUnion_{i=1}^{n} \UniWqoA_i \UniSymbDef 
			\UniStruct{
				\UniDisjUnion_{i=1}^{n} \UniWqoSet_i,
				\UniWqoRel{\UniDisjUnion},
				\UniNorm{\cdot}{\UniDisjUnion},
			}$, the \emph{disjoint union} of $\UniWqoA_1,\ldots,\UniWqoA_n$,
			where
			\begin{itemize}
                \item $\UniDisjUnion_{i=1}^{n} \UniWqoSet_i \UniSymbDef \bigcup_{i=1}^n \UniSet{\UniPair{\UniValA}{i} \mid \UniValA \in \UniWqoSet_i}$ is the \emph{disjoint union 
                of $\UniWqoSet_1,\ldots,\UniWqoSet_n$};
				\item $\UniPair{i}{\UniValA} \UniWqoRel{\UniDisjUnion} \UniPair{j}{\UniValB}$ iff $i = j$ and $\UniValA \UniWqoRel{{i}} \UniValB$; %and
				\item $\UniNorm{\UniPair{i}{\UniValA}}{\UniDisjUnion} = \UniNorm{\UniValA}{i}$.
			\end{itemize}
        \item $k \cdot \UniWqoA \UniSymbDef \UniDisjUnion_{i=1}^k \UniWqoA$,
    the \emph{$k$-sum of $\UniWqoA$}.
	\end{enumerate}
	We employ these operations also for qos 
   (by omitting the norm) without explicit mention.
\end{definition}

\begin{proposition}[\cite{schmitz2012notes}]
	\label{fact:operations_preserve_wqo}
	The property of being a (n)wqo is preserved under taking disjoint unions and direct products (and hence also $k$-powers and $k$-sums).
\end{proposition}

This immediately yields what is best known as \emph{Dickson's Lemma}~\cite{dickson1913}, namely that, for each $k \in \UniNaturalSetNN$, the $k$-power $\UniWqoModNatural^k$ (see Definition~\ref{def:wqo-natural}) is a (n)wqo.

We say that the nwqos $\UniWqoA_1$ and $\UniWqoA_2$ are \emph{isomorphic via $f$}, denoted by $\UniWqoA_1 \UniNwqoIsomorph_{f} \UniWqoA_2$, if there is a bijection $f : \UniSetA_1 \to \UniSetA_2$ such that $a \UniWqoRel{1} b$ iff $f(a) \UniWqoRel{2} f(b)$ for all $a,b \in A_1$, and $\UniNorm{a}{1} = \UniNorm{f(a)}{2}$ for all $a \in A_1$; such an $f$ is called an \emph{nwqo isomorphism} and we write $\UniWqoA_1 \UniNwqoIsomorph \UniWqoA_2$ if such an $f$ exists.
If we do not insist that $f$ satisfies the second requirement, $f$ is called an \emph{order isomorphism} and we write $\UniWqoA_1 \UniOrderIsomorph \UniWqoA_2$ when the two (n)wqos are order isomorphic.

The following nwqo isomorphisms involving direct products and disjoint unions will be useful in the next section.

\begin{proposition}
\label{fact:order-isomorph-powers}
If $n, d,s,k \in \UniNaturalSetNN$ and $\UniWqoA, \UniWqoA_1, \ldots, \UniWqoA_n$ are nwqos, then
    \begin{enumerate}
        \item $(\prod_{i=1}^n \UniWqoA_i)^d 
        \UniNwqoIsomorph
        \prod_{i=1}^n \UniWqoA_i^d$;
        \item $(\UniWqoA^k)^d \UniNwqoIsomorph \UniWqoA^{k \cdot d}$; and
        \item $(s \cdot \UniWqoA)^k \UniNwqoIsomorph s^k \cdot \UniWqoA^{k}$.
    \end{enumerate}
\end{proposition}
\begin{proof}
\sloppy
For (1), we use $f(\UniTuple{\vec\UniValA_1,\ldots,\vec\UniValA_d}) \UniSymbDef \UniTuple{
\UniTuple{\UniValA_{11},\ldots,\UniValA_{d1}},
\ldots,
\UniTuple{\UniValA_{1n},\ldots,\UniValA_{dn}}
}$.
(2) is just a particular case of (1).
For (3), we use
$f(\UniTuple{\UniTuple{i_1, a_1},\ldots,\UniTuple{i_k,a_k}})
\UniSymbDef
\UniTuple{e(i_1,\ldots,i_k),\UniTuple{a_1,\ldots,a_k}}$,
where $e(i_1,\ldots,i_k)$ is a number in $\UniSet{0,\ldots,s^k-1}$ uniquely identifying the tuple $(i_1,\ldots,i_k)$---in fact, there are $s^k$-many of them.
Both functions are bijections and clearly preserve the corresponding norms.
\end{proof}

We will also be interested in the following standard constructions, which we call \emph{power set extensions} of a given nwqo. 
Recall that for a subset $B$ of a qoset $\UniWqoA$ we set  ${\downarrow} B \UniSymbDef \{a \in  \UniWqoSet : a \UniWqoRel{} b,$ for some $b \in B\}$ and ${\uparrow} B \UniSymbDef \{a \in  \UniWqoSet : b \UniWqoRel{} a,$ for some $b \in B\}$.

\begin{definition}
For an nqo $\UniWqoA \UniSymbDef \UniStruct{\UniWqoSet, \UniWqoRel{}, \UniNorm{\cdot}{}}$ the \emph{power norm} is given by $\UniPowerFntWqoNorm{\UniSetB}{\UniWqoA} \UniSymbDef \max(\UniSet{\UniSetCard{\UniSetB}} \cup \UniSet{\UniNorm{\UniValB}{} : \UniValB \in \UniSetB})$, for each finite $\UniSetB \subseteq \UniWqoSet$. 
We then define:
	\begin{itemize}
		\item $\UniMajoringFntWqo{\UniWqoA} \UniSymbDef \UniStruct{\UniPowerSetFin{\UniWqoSet}, \UniMajoringWqoRel{}{}, \UniPowerFntWqoNorm{\cdot}{\UniWqoA}}$, where $\UniSetB \UniMajoringWqoRel{}{} \UniSetC$ if, and only if, $B \subseteq {\downarrow} C$, i.e., for all $\UniValB \in \UniSetB$ there is $\UniValC \in \UniSetC$ such that $\UniValB \UniWqoRel{} \UniValC$. 
        This is the \emph{majoring order (over $\UniWqoA$)}.
		\item $\UniMinoringFntWqo{\UniWqoA} \UniSymbDef \UniStruct{\UniPowerSetFin{\UniWqoSet}, \UniMinoringWqoRel{}{}, \UniPowerFntWqoNorm{\cdot}{\UniWqoA}}$, where $\UniSetB \UniMinoringWqoRel{}{} \UniSetC$ if, and only if,  ${\uparrow} B \supseteq C$, i.e., for all $\UniValC \in \UniSetC$ there is $\UniValB \in \UniSetB$ such that $\UniValB \UniWqoRel{} \UniValC$. 
        This is the \emph{minoring order (over $\UniWqoA$)}.
	\end{itemize}
  For a qo $\UniWqoA$, we set $\UniMajoringWqo{\UniWqoA}:= \UniStruct{\UniPowerSet{\UniWqoSet}, \UniMajoringWqoRel{}{}}$ and $\UniMinoringWqo{\UniWqoA}:=\UniStruct{\UniPowerSet{\UniWqoSet}, \UniMinoringWqoRel{}{}}$.
\end{definition}

It is easy to see that the above are nqos and qos respectively. 
Also, (see~\cite[Section 2]{abriola2015} for details) we have the following.

\begin{proposition}
    \label{fact:maj-is-nwqo}
    If $\UniWqoA$ is a (n)wqo then so is $\UniMajoringFntWqo{\UniWqoA}$.
\end{proposition}

The analogous result for $\UniMinoringFntWqo{\UniWqoA}$ fails (a famous counter-example is \emph{Rado's structure}~\cite{rado1954,schmitz2012notes}).
However, it follows from results in~\cite{marcone2001} that $\UniMinoringFntWqo{\UniWqoModNatural^k}$ is a (n)wqo, for all $k \in \UniNaturalSetNN$.
The following result can also be obtained using~\cite{marcone2001} and it will be instrumental to our work.

\begin{proposition}
    \label{fact:min-preserved-products-sums}
	If $\UniWqoA_1,\ldots,\UniWqoA_n$, with $n \geq 2$, are (n)wqos such that $\UniMinoringFntWqo{\UniWqoA_i}$ is a (n)wqo for each $1 \leq i \leq n$, then $\UniMinoringFntWqo{\prod_{i=1}^{n} \UniWqoA_i}$ and $\UniMinoringFntWqo{\UniDisjUnion_{i=1}^{n} \UniWqoA_i}$ are also (n)wqos.
\end{proposition}

Given nwqos $\UniWqoA_1$ and $\UniWqoA_2$, we say that $f : \UniWqoSet_1 \to \UniWqoSet_2$ is \emph{norm-decreasing} when $\UniNorm{f(a)}{\UniWqoA_2} \leq \UniNorm{a}{\UniWqoA_1}$ for all $a \in \UniWqoSet_1$.

We present below some useful order isomorphisms involving products and disjoint unions of majoring and minoring extensions of wqos, in which the isomorphisms are norm-decreasing. 
This will be essential for us in the next subsections, when we discuss reflections for knotted nwqos.

\begin{proposition}
    \label{fact:order-isomorph-powerset-extensions}
If $n, s_1, \ldots, s_n, s, d \in \UniNaturalSetNN$ and $\UniWqoA_1, \ldots,  \UniWqoA_n, \UniWqoA$ are wqos, then
    \begin{enumerate}
        \item $\UniPowerMaj{\UniMajoringFntWqo{\UniDisjUnion_{i=1}^{n} s_i \cdot \UniWqoA_i}}{d} \UniOrderIsomorph \prod_{i=1}^{n}\UniPowerMaj{\UniMajoringFntWqo{\UniWqoA_i}}{d \cdot s_i}$;
        \item $\UniPowerMaj{\UniMinoringFntWqo{\UniDisjUnion_{i=1}^{n} s_i \cdot \UniWqoA_i}}{d} \UniOrderIsomorph \prod_{i=1}^{n }
        \UniPowerMaj{\UniMinoringFntWqo{\UniWqoA_i}}{d \cdot s_i}$;
        \item $\UniPowerMaj{\UniMajoringFntWqo{s \cdot \UniWqoA}}{d} \UniOrderIsomorph \UniPowerMaj{\UniMajoringFntWqo{\UniWqoA}}{d\cdot s}$; and
        \item $\UniPowerMaj{\UniMinoringFntWqo{s \cdot \UniWqoA}}{d} \UniOrderIsomorph \UniPowerMaj{\UniMinoringFntWqo{\UniWqoA}}{d\cdot s}$.  
    \end{enumerate}
Moreover, the isomorphisms are norm-decreasing.
\end{proposition}
\begin{proof}
(3) and (4) are particular cases of (1) and (2) respectively, so we focus in proving (1) and (2).
We show first that $\UniMajoringFntWqo{\UniDisjUnion_{i=1}^{n} s_i \cdot \UniWqoA_i} \UniIsomorph \prod_{i=1}^{n}\UniPowerMaj{\UniMajoringFntWqo{\UniWqoA_i}}{s_i}$ and $\UniMinoringFntWqo{\UniDisjUnion_{i=1}^{n} s_i \cdot \UniWqoA_i} \UniIsomorph \prod_{i=1}^{n } \UniPowerMaj{\UniMinoringFntWqo{\UniWqoA_i}}{s_i}$.
We consider the mapping $f$ such that 
$f(X) \UniSymbDef (\vec X_1,\ldots,\vec X_{n}),$ 
where, for each 
$1 \leq i \leq n$,
$\vec X_{i} = (X_{i 1},\ldots,X_{i s_i})$ 
with 
$X_{ij} \UniSymbDef \UniSet{\UniValA \in \UniSetA_i \mid (i, (j, \UniValA)) \in X},$
for each 
$0 \leq j < s_i$.
We will see that $f$ provides the desired order isomorphism
for both items.

It is easy to see that $f$ is bijective, so we proceed to show that $f$ is order preserving with respect to the minoring ordering; the case for the majoring ordering can be proved analogously.
To simplify notation, let 
$\UniWqoA_{\mathsf{Min}} \UniSymbDef\UniMinoringFntWqo{\UniDisjUnion_{i=1}^{n} s_i \cdot \UniWqoA_i}$ 
and
$\UniWqoA_{\Pi} \UniSymbDef \prod_{i=1}^{n}\UniPowerMaj{\UniMinoringFntWqo{\UniWqoA_i}}{s_i}$.
Suppose that 
$X \UniWqoRel{\UniWqoA_{\mathsf{Min}}} Y$ 
and let 
$f(X) \UniSymbDef {(\vec X_1,\ldots, \vec X_n)}$ 
and
$f(Y) \UniSymbDef {(\vec Y_1,\ldots, \vec Y_n)}$.
For $1 \leq i \leq n$, $1 \leq j < s_i$ and  $\UniValB \in Y_{ij}$ we have $(i, (j, \UniValB)) \in Y$, so there is $(i, (j, \UniValA)) \in X$, i.e., $\UniValA \in X_{ij}$, such that $\UniValA \UniWqoRel{i} \UniValB$.
Therefore $X_{ij} \UniWqoRel{\UniMinoringFntWqo{\UniWqoA_i}} Y_{ij}$ as desired. 
Conversely, suppose that
$f(X) \UniWqoRel{\UniWqoA_{\Pi}} f(Y)$ and that $(i, (j, \UniValB)) \in Y$.
Then $\UniValB \in Y_{ij}$, and thus there is $\UniValA \in X_{ij}$ with $\UniValA \UniWqoRel{i} \UniValB$.
So, $(i, (j, \UniValA)) \in X$, thus $X \UniWqoRel{\UniWqoA_{\mathsf{Min}}} Y$. 

Regarding the norm requirement, we have:
\begin{align*}
\UniNorm{f(X)}{\UniWqoA_{\Pi}}
=&
\UniNorm{{(\vec X_1,\ldots, \vec X_n)}}{\UniWqoA_{\Pi}}\\
=&
\max\left\{ \max_{1 \leq i \le n, 1 \leq j < s_i}\UniSetCard{X_{ij}},
\max_{1 \leq i \le n, 1 \leq j < s_i} \max_{a \in X_{ij}} \UniNorm{a}{\UniWqoA_i}
\right\}\\
\leq&
\max\left\{ \sum_{1 \leq i \le n, 1 \leq j < s_i}\UniSetCard{X_{ij}},
\max_{1 \leq i \le n, 1 \leq j < s_i} \max_{a \in X_{ij}} \UniNorm{a}{\UniWqoA_i}
\right\}\\
=&
\max\left\{\UniSetCard{X},
\max_{1 \leq i \le n, 1 \leq j < s_i} \max_{a \in X_{ij}} \UniNorm{a}{\UniWqoA_i}
\right\}\\
=&
\max\left\{\UniSetCard{X},
\max_{(i, (j, a)) \in X} \UniNorm{a}{\UniWqoA_i}
\right\}\\
=&
\UniNorm{X}{\UniWqoA_{\mathsf{Min}}}.
\end{align*}
The proof is the same for the majoring order, as the norm is the same.
We combine the result proved above with Proposition~\ref{fact:order-isomorph-powers} to conclude the proof for the $d$-powers as stated in (1) and (2).
\end{proof}

\section{Knotted orders on $\mathbb{N}$}
In this section, we define a nwqo on $\UniNaturalSet$ that expresses the effect of successive backwards applications of $\UniWeakCtr$ or forward applications of $\UniWeakWkn$ on the multiplicity of a fixed formula on the left of a sequent (see Section~\ref{sec:wck-substructural-logics} for the definition of these rules).
The corresponding relation associated with $\UniWeakKRule{m}{n}$ is denoted by $\UniExtModRel{m}{n}$ ($m > n$).

The algebraic rendering of $\UniWeakKRule{m}{n}$, as given in Section~\ref{s: FEPvarieties}, is $x^n \leq x^m$, and $\UniExtModRel{m}{n}$ is defined to be the compatible (with respect to multiplication) order generated by $x^n \leq x^m$ and is given in Definition~\ref{d: knottedorder}; this order was first considered in \cite{vanalten2005}.

The intuitive proof-theoretic meaning of $\UniValA \UniExtModRel{m}{n} \UniValB$ for $\UniValA,\UniValB \in \UniNaturalSet$ is: applying $\UniWeakCtr$ backwards, or $\UniWeakWkn$ forward, over a formula $\UniFmA$ with multiplicity $\UniValA$ on the left of a sequent, we get a new sequent where the only difference is that $\UniFmA$ has multiplicity $\UniValB$.
We generalize it to tuples over $\UniNaturalSet$ to reflect applications on more than a single formula, and we consider their corresponding power set extensions.

For example, we have $\UniValA^3 \UniExtModRel{6}{3} \UniValA^6$ and by multiplying by this inequality by $\UniValA^2, \UniValA^5, \UniValA^8$ and $\UniValA^{11}$ we get the sequence of inequalities: $\UniValA^5 \UniExtModRel{6}{3} \UniValA^8 \UniExtModRel{6}{3} \UniValA^{11}\UniExtModRel{6}{3} \UniValA^{14}$. 
Likewise, we can get the inequalities $\UniValA^4 \UniExtModRel{6}{3} \UniValA^7 \UniExtModRel{6}{3} \UniValA^{10}\UniExtModRel{6}{3} \UniValA^{13}$, as well as $\UniValA^3 \UniExtModRel{6}{3} \UniValA^6 \UniExtModRel{6}{3} \UniValA^{9}\UniExtModRel{6}{3} \UniValA^{12}$. 
The pattern that emerges is  $\UniValA^k \UniExtModRel{6}{3} \UniValA^{k+3l}$, where $l \in \mathbb{N}$; on the other hand, it is impossible to obtain $\UniValA^2 \UniExtModRel{6}{3} \UniValA^{5}$ or $\UniValA \UniExtModRel{6}{3} \UniValA^{3}$, so $k \geq 3$. 
Focusing on the exponents, this suggests the following complete description of all non-reflexive cases: $k \UniExtModRel{6}{3} {k+3l}$, where $k, l \in \mathbb{N}$, $k \geq 3$. 
More generally, when $n<m$,  all the comparability relations are given by
\begin{center}
    $k \UniExtModRel{m}{n} {k+(m-n)l}$, where $k, l \in \mathbb{N}$, $k \geq n$,
\end{center}
together with the reflexive pairs.

In this direction, the formulation of $\UniExtModRel{m}{n}$ has to take into consideration some important aspects of backwards successive applications of $\UniWeakCtr$ or forward applications of  $\UniWeakWkn$.
We illustrate them using the specific case of $\UniWeakCRule{6}{3}$.
First of all, we might consider zero applications of this rule, so we need to have $\UniValA \UniExtModRel{6}{3} \UniValA$, for all $\UniValA \in \UniNaturalSet$ (which gives us the reflexivity of this ordering).
Also, when we start applying this rule to, say,
$\UniSequent{\UniFmA^{14}}{\UniFmB}$, 
one application leads to
$\UniSequent{\UniFmA^{11}}{\UniFmB}$, 
a second one produces to
$\UniSequent{\UniFmA^{8}}{\UniFmB}$,
and a third produces $\UniSequent{\UniFmA^{5}}{\UniFmB}$. 
Note that the multiplicities have decreased, but remained in the
same equivalence class modulo $6 - 3 = 3$ (the remainder of 14, 11, 8 and 5 in a division by 3 is 2).
If we experiment a bit more, we will see that the same holds for the other equivalence classes modulo 3.
In general, we should expect that, when 
$\UniValA \UniExtModRel{6}{3} \UniValB$, 
we need that $\UniValA \leq \UniValB$ and that $\UniValA$ and $\UniValB$ should be in the same class modulo 3:
$\UniValA \UniEquivMod{3} \UniValB$. 
Finally, there are some limit cases that we need to observe. For example, from
$\UniSequent{\UniFmA^{5}}{\UniFmB}$ 
we cannot apply the rule again, since we do not have at least 6 copies of the formula, as required by the rule, so we require that $\UniValB \geq 6$. 
Also, we cannot have $2 \UniExtModRel{6}{3} 8$, even though the previous conditions hold, so we also require $\UniValA \geq 3$.
Note that even though 2 is outside one of these boundaries, we still need to ensure that $2 \UniExtModRel{6}{3} 2$.
Generalizing from this example, we can replace $6$ for $m$ and $3$ for $n$, with $0 \leq n < m$, obtaining the following  definition.

\begin{definition}\label{d: knottedorder}
For $0 \leq n < m$, let
$\UniExtModRel{m}{n} \, \subseteq \, \UniNaturalSet \times \UniNaturalSet$ be such that, for all $\UniValA, \UniValB \in \UniNaturalSet$,
    \begin{gather*}
		\UniValA \UniExtModRel{m}{n} \UniValB
		\text{ iff }
		\text{either }
		\UniValA = \UniValB,\\
		\text{ or }
		\UniValA \geq n,
		\UniValB \geq m,
		\UniValA < \UniValB \text{ and } {\UniValA}\UniEquivMod{m-n}{\UniValB}.
    \end{gather*}

	\noindent Also, we define  $\UniWqoExtModRel{m}{n} \UniSymbDef \UniStruct{\UniNaturalSet, 
                \UniExtModRel{m}{n},
                \lambda \UniValA. \UniValA}$.
\end{definition}

The following simpler and equivalent formulation will be more convenient to work with.

\begin{proposition}
\label{fact:equiv-characterization-mn}
For all $0 \leq n < m$ and all $\UniValA, \UniValB \in \UniNaturalSet$,
	\begin{gather*}
		\UniValA \UniExtModRel{m}{n} \UniValB
		\text{ iff }
		\text{either }
		\UniValA = \UniValB
        \text{ and } \UniValA,\UniValB \in \UniSet{0,\ldots,n-1}\\
		\text{ or }
		\UniValA,\UniValB \not\in \UniSet{0,\ldots,n-1},
		\UniValA \leq \UniValB \text{ and } {\UniValA}\UniEquivMod{m-n}{\UniValB}.
	\end{gather*}
\end{proposition}
\begin{proof}
We will show that the relation $\UniExtModRel{m}{n}^\prime$ defined in the statement is equal to the relation $\UniExtModRel{m}{n}$ given in Definition~\ref{d: knottedorder}.
It is clear that $\UniExtModRel{m}{n} \subseteq \UniExtModRel{m}{n}^\prime$.
For the other direction, we proceed by cases under the assumption that $\UniValA \UniExtModRel{m}{n}^\prime \UniValB$.
If $\UniValA = \UniValB \in \UniSet{0,\ldots,n-1}$, then clearly $\UniValA \UniExtModRel{m}{n} \UniValB$.
If $\UniValA,\UniValB \geq n$, $\UniValA \leq \UniValB$ and (a): $\UniValA \UniEquivMod{m-n} \UniValB$, we have that the case $\UniValA = \UniValB$ is obvious and if $\UniValA \neq \UniValB$ we just need to prove that and $\UniValB \geq m$. 
Assume for the sake of contradiction that $\UniValB < m$.
Since $\UniValA < \UniValB$, we also have $\UniValA < m$.
Thus $n \leq \UniValA < \UniValB < m$.
Note that $\UniValA$ and $\UniValB$ can only assume values among $n, n+1, \ldots, m-1$, and all of them must be in different classes modulo $m-n$, contradicting (a). 
Therefore, $\UniValB \geq m$, so we are done.
\end{proof}

Below we show the Hasse diagram of this ordering for the case $n<m$ (if we allowed $m<n$, we would get the dual poset).

\begin{center}
    \begin{tikzpicture}
        \node (n0) 
        [
            circle, 
            fill, 
            inner sep=1.5pt,
            label={[right]$0$}
        ]{};
        \node (n1) 
        [
            right of=n0,
            circle, 
            fill, 
            inner sep=1.5pt,
            label={[right]$1$}
        ]{};
        \node (ldots) 
        [
            right of=n1,
        ]{$\cdots$};
        \node (nnmone) 
        [
            right of=ldots,
            circle, 
            fill, 
            inner sep=1.5pt,
            label={[right]$n-1$}
        ]{};
        
        \node (fstclass0) 
        [
            right of=nnmone,
            circle, 
            fill, 
            xshift=20pt,
            inner sep=1.5pt,
            label={[right]$n$}
        ]{};
        \node (fstclass1) 
        [
            above of=fstclass0,
            circle, 
            fill, 
            inner sep=1.5pt,
            label={[right]$m$}
        ]{};
        \node (fstclass2) 
        [
            above of=fstclass1,
            circle, 
            fill, 
            inner sep=1.5pt,
            label={[right]$2m-n$}
        ]{};
        \node (fstclassdots) 
        [
            above of=fstclass2,
            label={$\vdots$}
        ]{};
        \draw (fstclass0) -- (fstclass1);
        \draw (fstclass1) -- (fstclass2);
        \draw (fstclass2) -- (fstclassdots);

        %%%
        \node (sfstclass0) 
        [
            right of=fstclass0,
            circle, 
            fill, 
            inner sep=1.5pt,
            xshift=35pt,
            label={[right]$n+1$}
        ]{};
        \node (sfstclass1) 
        [
            above of=sfstclass0,
            circle, 
            fill, 
            inner sep=1.5pt,
            label={[right]$m+1$}
        ]{};
        \node (sfstclass2) 
        [
            above of=sfstclass1,
            circle, 
            fill, 
            inner sep=1.5pt,
            label={[right]$2m-n+1$}
        ]{};
        \node (sfstclassdots) 
        [
            above of=sfstclass2,
            label={$\vdots$}
        ]{};
        \draw (sfstclass0) -- (sfstclass1);
        \draw (sfstclass1) -- (sfstclass2);
        \draw (sfstclass2) -- (sfstclassdots);
        %%%

        \node (ldotsclass)
        [
            right of = sfstclass0,
            xshift=35pt,
        ]{$\cdots$};

        \node (sndclass0) 
        [
            right of=ldotsclass,
            circle, 
            fill, 
            inner sep=1.5pt,
            xshift=10pt,
            label={[right]$m-1$}
        ]{};
        \node (sndclass1) 
        [
            above of=sndclass0,
            circle, 
            fill, 
            inner sep=1.5pt,
            label={[right]$2m-n-1$}
        ]{};
        \node (sndclass2) 
        [
            above of=sndclass1,
            circle, 
            fill, 
            inner sep=1.5pt,
            label={[right]$3m-2n-1$}
        ]{};
        \node (sndclassdots) 
        [
            above of=sndclass2,
            label={$\vdots$}
        ]{};
        \draw (sndclass0) -- (sndclass1);
        \draw (sndclass1) -- (sndclass2);
        \draw (sndclass2) -- (sndclassdots);
    \end{tikzpicture}
\end{center}

$\UniWqoExtModRel{m}{n}$ was proved to be a  nwqo in~\cite[\S 7]{vanalten2005} via an \emph{ad hoc} argument; we refer to this nwqo as a \emph{basic knotted nwqo}.
We will provide a more general proof, using the framework of \emph{strong reflections}, which will also be important in obtaining length theorems (that is, upper bounds on length of bad sequences) for $k$-powers and power set extensions of various other nwqos, which we collectively call \emph{knotted nwqos}.

\section{Knotted (n)qos are (n)wqos, via strong reflections}

\emph{Strong reflections}~\cite{figueira2011,bala2020} are mappings between nwqos that allow for the transferring of results such as the wqo property (finiteness of bad sequences) and length theorems (to be precisely defined in the next section) among the relevant nwqos.
Since we will extensively use such tools in our study of knotted wqos, we define them in detail.

\begin{definition}
    \label{def:strong-reflections}
	A \emph{reflection} (or \emph{order-reflecting map}) between two qos  $\UniWqoA_1 \UniSymbDef \UniStruct{\UniWqoSet_1, \UniWqoRel{1}}$ and $\UniWqoA_2 \UniSymbDef \UniStruct{\UniWqoSet_2, \UniWqoRel{2}}$ is a mapping $\UniReflectionA :  \UniWqoSet_1 \to \UniWqoSet_2$ such that, for all $\UniValA, \UniValB \in \UniSetA_1$,
\[
\text{if } \UniReflectionA(\UniValA) \UniWqoRel{2} \UniReflectionA(\UniValB), \text{then } \UniValA \UniWqoRel{1} \UniValB;
\]
in symbols $\UniWqoA_1 \UniReflArrow{g} \UniWqoA_2$.
We say that $\UniWqoA_2$ \emph{reflects} $\UniWqoA_1$ when such a map exists, and write $\UniWqoA_1 \UniReflArrow{} \UniWqoA_2$.
When these qos are normed and $\UniNorm{\UniReflectionA(\UniValA)}{\UniWqoA_2} \leq \UniNorm{\UniValA}{\UniWqoA_1}$ for all $\UniValA \in \UniSetA_1$ (i.e., $\UniReflectionA$ is norm-decreasing), we say that $\UniReflectionA$ is a \emph{strong reflection} and write $\UniWqoA_1 \UniStrongReflArrow{\UniReflectionA} \UniWqoA_2$; if such a $g$ exists we write $\UniWqoA_1 \UniStrongReflArrow{} \UniWqoA_2$.
\end{definition}

Note that when $\UniWqoRel{2}$ is an order (i.e., it is anti-symmetric) then the reflection condition for $g$ above implies that $g$ is an injective map. 
This is the reason that we used the symbol $\UniReflArrow{}$, which is usually reserved for embeddings.
Moreover, observe that an nwqo isomorphism is always a strong reflection, and the same applies to an order isomorphism that is norm-decreasing. 

\begin{example}
    For any $\UniWqoA$ and $r, s \in \UniNaturalSetNN$, we have $\UniWqoA^r \UniStrongReflArrow{g} \UniWqoA^{r+s}$, where $g$ is given by $g((a_1,\ldots,a_r)) := (a_1,\ldots,a_r,\underbrace{a_r\ldots,a_r}_{s \text{ times}})$.
\end{example}

The following transfer result of the wqo property is easy to prove.

\begin{lemma}[\cite{schmitz2012notes}]
\label{fact:refl-pres-wellness}
    If $\UniWqoA_1,\UniWqoA_2$ are qos, $\UniWqoA_1 \UniReflArrow{} \UniWqoA_2$ and $\UniWqoA_2$ is a wqo, then $\UniWqoA_1$ is a wqo as well.
\end{lemma}
\begin{proof}
Note that the assumed reflection, call it $f$, guarantees that a bad sequence over $\UniWqoA_1$ can be transformed into a bad sequence over $\UniWqoA_2$ of same length.
Indeed, consider a bad sequence $a_0, a_1, \ldots$ over $\UniWqoA_1$ and the corresponding sequence $f(a_0),f(a_1),\ldots$ over $\UniWqoA_2$.
Since $a_i \not\UniWqoRel{\UniWqoA_1} a_j$ for all $i < j$, we must have by the reflection condition $f(a_i) \not\UniWqoRel{\UniWqoA_2} f(a_j)$ for all $i < j$, thus  $f(a_0),f(a_1),\ldots$ is a bad sequence of the same length. 
Hence, an infinite bad sequence over $\UniWqoA_1$ would translate via $f$ to an infinite bad sequence over $\UniWqoA_2$, which would contradict the assumption that $\UniWqoA_2$ is a wqo.
Therefore, $\UniWqoA_1$ must be wqo as well.
\end{proof}

Here we will provide many reflections between knotted qos and well-known nwqos involving $\UniWqoModNatural$ (including product and power set extensions); this will allow us to transfer results from the latter to the former.
We begin with the simplest case of $\UniWqoExtModRelProd{m}{n}{}$ itself.

\begin{lemma}
\label{fact:refl-nm-n-single}
For all $0 \leq n < m$,
$\UniWqoExtModRelProd{m}{n}{} \UniStrongReflArrow{} \; \UniFlatWqo{m} \cdot \UniWqoModNatural.$ 
Thus, $\UniWqoExtModRelProd{m}{n}{}$ is an nwqo.
\end{lemma}
\begin{proof}
For $\UniValA \in \UniNaturalSet$, let $g(\UniValA) \UniSymbDef \UniValA$ if $\UniValA < n$ and $g(\UniValA) \UniSymbDef n + \UniValA \% (m-n)$ otherwise. 
Here we write $a\% k$ for the element of $\mathbb{Z}_k$ that is equivalent to $a$ modulo $k$.
Note that the range of $g$ is $\UniSet{0, \ldots, m-1}$.
The candidate reflection, denoted by $(\UniEncodingVecMod{\cdot}{n}{m})$, is given by $\UniEncodingVecMod{\UniValA}{n}{m} \UniSymbDef \UniTuple{g(a),\UniValA}$.
For the reflection property, suppose that $\UniEncodingVecMod{\UniValA}{n}{m} \leq \UniEncodingVecMod{\UniValB}{n}{m}$.
Then (i): $g(a) = g(b)$ and (ii): $a \leq b$.
We proceed by cases on the value of $\UniValA$.
If $\UniValA < n$, we have {$ g(\UniValB)= g(\UniValA)=\UniValA  < n$}, thus $\UniValB=g(\UniValB) = g(\UniValA)=\UniValA $, hence $\UniValA = \UniValB$ and $\UniValA, \UniValB \in \UniSet{0,\ldots,n-1}$, and thus $\UniValA \UniExtModRel{m}{n} \UniValB$, as desired.
Otherwise, if $\UniValA \geq n$, we have $n + \UniValA\%(m-n) {= g(a)}= g(b)$, meaning that $g(b) \geq n$, thus $b \geq n$ by the definition of $g$.
Therefore, $n + \UniValA\%(m-n) = n + \UniValB\%(m-n)$, which gives $\UniValA \UniEquivMod{m-n} \UniValB$.
From this, (ii) and the fact that $\UniValA,\UniValB \not\in \UniSet{0,\ldots,n-1}$, we obtain $\UniValA \UniExtModRel{m}{n} \UniValB$ as desired.
This reflection is strong since $\UniNorm{(g(\UniValA), \UniValA)}{ \UniFlatWqo{m}\cdot\UniWqoModNatural} = \UniValA = \UniNorm{\UniValA}{\UniWqoExtModRelProd{m}{n}{}}$.
Finally, Lemma~\ref{fact:refl-pres-wellness} and Proposition~\ref{fact:operations_preserve_wqo} entail that  $\UniWqoExtModRelProd{m}{n}{}$ is an nwqo.
\end{proof}

The above can be used to investigate $\UniWqoExtModRelProd{m}{n}{k}$ and its power set extensions, as we show now by means of further useful properties of strong reflections.

\begin{lemma}
	\label{fact:products-preserve-reflections}
	If $\UniWqoA_1, \UniWqoA_2$ are nqos and 
    $\UniWqoA_1 \UniStrongReflArrow{} \UniWqoA_2$, then 
    $\UniWqoA^k_1 \UniStrongReflArrow{} \UniWqoA^k_2$ and 
    $k \cdot \UniWqoA_1 \UniStrongReflArrow{} k \cdot \UniWqoA_2$,  for all $k \in \UniNaturalSetNN$.
\end{lemma}
\begin{proof} If  $\UniWqoA_1 \UniStrongReflArrow{g} \UniWqoA_2$, then 
$\UniReflectionA'(\UniTuple{\UniValA_1,\ldots,\UniValA_k}) \UniSymbDef \UniTuple{\UniReflectionA(\UniValA_1),\ldots,\UniReflectionA(\UniValA_k)}$
defines a suitable strong reflection witnessing $\UniWqoA^k_1 \UniStrongReflArrow{} \UniWqoA^k_2$. 
For $k \cdot \UniWqoA_1 \UniStrongReflArrow{g} k \cdot \UniWqoA_2$, we take the mapping to be $g''((i,a)) \UniSymbDef (i, g(a))$.
\end{proof}

Using this lemma, we obtain a strong reflection for $\UniWqoExtModRelProd{m}{n}{k}$.

\begin{lemma}
\label{fact:refl-mn-n-k} \label{fact:majoring-nm-wqo}
For all $0 \leq n < m$ and $k \in \UniNaturalSetNN$,
$\UniWqoExtModRelProd{m}{n}{k} 
\UniStrongReflArrow{}
\; \UniFlatWqo{m^k} \cdot \UniWqoModNatural^k.$
Therefore $\UniWqoExtModRelProd{n}{m}{k}$ 
and  
$\UniMajoringFntWqo{\UniWqoExtModRelProd{n}{m}{k}}$ 
are nwqos.
\end{lemma}
\begin{proof}
By Lemma~\ref{fact:refl-nm-n-single} we have
$\UniWqoExtModRelProd{m}{n}{} 
\UniStrongReflArrow{}\;\UniFlatWqo{m} \cdot
\UniWqoModNatural$, 
hence 
$\UniWqoExtModRelProd{m}{n}{k} \UniStrongReflArrow{}\;(\UniFlatWqo{m} \cdot \UniWqoModNatural)^k$ 
by Lemma~\ref{fact:products-preserve-reflections}. 
By Proposition~\ref{fact:order-isomorph-powers},
$(\UniFlatWqo{m} \cdot \UniWqoModNatural)^k$
is nwqo-isomorphic to $\UniFlatWqo{m^k} \cdot \UniWqoModNatural^k,$
thus 
$\UniWqoExtModRelProd{m}{n}{k} \UniStrongReflArrow{}\;\UniFlatWqo{m^k} \cdot \UniWqoModNatural^k$ 
and, by Lemma~\ref{fact:refl-pres-wellness}, 
$\UniWqoExtModRelProd{n}{m}{k}$ is a nwqo.
Finally, by Proposition~\ref{fact:maj-is-nwqo}, we get that   
$\UniMajoringFntWqo{\UniWqoExtModRelProd{n}{m}{k}}$ 
is a (n)wqo.
\end{proof}

From the  Lemmas~\ref{fact:products-preserve-reflections} and~\ref{fact:refl-mn-n-k}  it follows that:

\begin{corollary}
\label{coro:disj-sums-mn-refle-nat}
For all $0 \leq n < m$ and $k, d \in \UniNaturalSetNN$,
$d \cdot \UniWqoExtModRelProd{m}{n}{k} \UniStrongReflArrow{}\;
(d \cdot \UniFlatWqo{m^k}) \cdot \UniWqoModNatural^k.$
\end{corollary}

Now, regarding reflections involving
power set extensions, we
have the following result:

\begin{lemma}
\label{fact:min-maj-refl-preserv}
If  $\UniWqoA_1,  \UniWqoA_2$ are nqos   and $\UniWqoA_1 \UniStrongReflArrow{\UniReflectionA} \UniWqoA_2$, then
    \begin{enumerate}
        \item $\UniMajoringFntWqo{\UniWqoA_1} \UniStrongReflArrow{\UniReflectionA[\cdot]} \UniMajoringFntWqo{\UniWqoA_2}$, and
        \item 
         $\UniMinoringFntWqo{\UniWqoA_1} \UniStrongReflArrow{\UniReflectionA[\cdot]} \UniMinoringFntWqo{\UniWqoA_2}$,
    \end{enumerate}
where $\UniReflectionA[\UniSetB] \UniSymbDef \UniSet{\UniReflectionA(\UniValB) \mid \UniValB \in \UniSetB}$ for all $\UniSetB \subseteq \UniSetA_1$.
\end{lemma}
\begin{proof}
For (1), we first show that $\UniReflectionA[\cdot]$ is a reflection.
Suppose that $\UniSetB \not\UniMajoringWqoRel{\UniWqoA_1}{}\UniSetC$.
Then there is $\UniValB \in \UniSetB$ such that $\UniValB \not\UniWqoRel{\UniWqoA_1} \UniValC$ for all $\UniValC \in \UniSetC$.
Because $\UniReflectionA$ is a reflection, we have 
$\UniReflectionA(\UniValB) \not\UniWqoRel{\UniWqoA_2} \UniReflectionA(\UniValC)$ 
for all $\UniValC \in \UniSetC$, thus 
$\UniReflectionA[\UniSetB]\not\UniMajoringWqoRel{\UniWqoA_2}{}\UniReflectionA[\UniSetC]$. 
To see that $\UniReflectionA[\cdot]$ is strong, note that 
$\UniPowerFntWqoNorm{\UniReflectionA[\UniSetB]}{\UniWqoA_2} = \max\left(\UniSet{\UniSetCard{\UniReflectionA[\UniSetB]}} \cup \UniSet{\UniNorm{\UniReflectionA(\UniValB)}{\UniWqoA_2} : \UniReflectionA(\UniValB) \in \UniReflectionA[\UniSetB]}\right) \leq \max\left(\UniSet{\UniSetCard{\UniSetB}} \cup \UniSet{\UniNorm{\UniValB}{\UniWqoA_2} : \UniValB \in \UniSetB}\right) = \UniPowerFntWqoNorm{\UniSetB}{\UniWqoA_1}$, since 
$\UniSetCard{\UniReflectionA[\UniSetB]} \leq \UniSetCard{\UniSetB}$ and $\UniNorm{\UniReflectionA(\UniValB)}{\UniWqoA_2} \leq \UniNorm{\UniValB}{\UniWqoA_1}$ 
for all $\UniValB \in \UniSetB$, because $\UniReflectionA$ is a strong reflection. 
For (2), the proof is analogous (the proof of the reflection part can also be found in~\cite[Theorem V.3]{abriola2014}).
\end{proof}

From the above result and the reflections we obtained before, we easily produce strong reflections for 
$\UniPowerMaj{\UniMajoringFntWqo{\UniWqoExtModRelProd{m}{n}{k}}}{d}$ and 
$\UniPowerMaj{\UniMinoringFntWqo{\UniWqoExtModRelProd{m}{n}{k}}}{d}$.

\begin{lemma}
\label{fact:refl-power-set-ext}
For all $0 \leq n < m$ and $k,d\in \UniNaturalSetNN$,
\begin{enumerate}
    \item $\UniPowerMaj{\UniMajoringFntWqo{\UniWqoExtModRelProd{m}{n}{k}}}{d} \UniStrongReflArrow{}\UniPowerMaj{\UniMajoringFntWqo{\UniWqoModNatural^k}}{d \cdot m^k}$ 
    and
    \item 
    $\UniPowerMaj{\UniMinoringFntWqo{\UniWqoExtModRelProd{m}{n}{k}}}{d} \UniStrongReflArrow{} \UniPowerMaj{\UniMinoringFntWqo{\UniWqoModNatural^k}}{d \cdot m^k}$.
\end{enumerate}
Therefore, $\UniPowerMaj{\UniMinoringFntWqo{\UniWqoExtModRelProd{m}{n}{k}}}{d}$ is a nwqo.
\end{lemma}
\begin{proof}
\sloppy
By using Lemma~\ref{fact:refl-mn-n-k} and Lemma~\ref{fact:min-maj-refl-preserv}, we obtain the strong reflections 
$\UniMajoringFntWqo{\UniWqoExtModRelProd{m}{n}{k}} \UniStrongReflArrow{} \UniMajoringFntWqo{ \UniFlatWqo{m^k} \cdot \UniWqoModNatural^k}$ 
and 
$\UniMinoringFntWqo{\UniWqoExtModRelProd{m}{n}{k}} \UniStrongReflArrow{}\UniMinoringFntWqo{ \UniFlatWqo{m^k} \cdot 
\UniWqoModNatural^k}$.
Then, by Proposition~\ref{fact:order-isomorph-powerset-extensions}, $\UniMajoringFntWqo{ \UniFlatWqo{m^k} \cdot \UniWqoModNatural^k}$
and
$\UniMinoringFntWqo{ \UniFlatWqo{m^k} \cdot \UniWqoModNatural^k}$ are order isomorphic to 
$\left (\UniMajoringFntWqo{\UniWqoModNatural^k} \right)^{m^k}$
and
$\left (\UniMinoringFntWqo{\UniWqoModNatural^k} \right)^{m^k}$, respectively, and it is not hard to see that these isomorphisms induce strong reflections
$\UniMajoringFntWqo{\UniWqoExtModRelProd{m}{n}{k}}\UniStrongReflArrow{} \UniPowerMaj{\UniMajoringFntWqo{\UniWqoModNatural^k}}{m^k}
$ 
and
$\UniMinoringFntWqo{\UniWqoExtModRelProd{m}{n}{k}} \UniStrongReflArrow{}
\UniPowerMaj{\UniMinoringFntWqo{\UniWqoModNatural^k}}{m^k}
$.
The desired reflections result from
Lemma~\ref{fact:products-preserve-reflections} and the fact that,  by Proposition~\ref{fact:order-isomorph-powers}, $(\UniWqoA^k)^d$ is nwqo-isomorphic to $\UniWqoA^{kd}$ for any nwqo $\UniWqoA$.

That $\UniPowerMaj{\UniMinoringFntWqo{\UniWqoExtModRelProd{m}{n}{k}}}{d}$ is a nwqo follows from (2) and Proposition~\ref{fact:min-preserved-products-sums}.
\end{proof}

\section{Complexity bounds via length theorems}
\label{s: lenght theorems}

Algorithms relying on wqos for termination are usually amenable to complexity analysis based on the length of bad sequences.
Results that provide upper bounds for the length of bad sequences are known as \emph{length theorems}.
As we mentioned before, unfortunately there is no upper bound for the length of arbitrary bad sequences in an arbitrary wqo (that is, length theorems are not available in general).
However, by imposing restrictions on the growth rate of the norms of the elements in the bad sequence we obtain such bounds; these restrictions are formalized using the notion of  \emph{controlled bad sequences}.
In this setting, we may define a so-called \emph{length function}that takes a number $t$ and outputs the maximum length of a controlled bad sequence where the norm of the first element is bounded by $t$.

\begin{example}
In the nwqo $\UniWqoModNatural$, the underlying order relation is total, so bad sequences are always strictly decreasing sequences (and every strictly decreasing sequence is bad).
Therefore, in particular $a, a-1,\ldots, 1, 0$ is a  bad sequence for every $a \in \UniNaturalSet$, hence there is no maximum length of a bad sequence. 
However, if we restrict the norm of the initial term to be no more than $t$ then there is a maximal length, $t+1$, for all such bad sequences (there are finitely many initial elements $a$ with $\UniNorm{\UniValA}{} \leq t$ and  the norms of the elements in a bad sequence of $\UniWqoModNatural$ strictly decrease, thus producing finitely many elements as possible second coordinates etc.). 
We can think of $L[\UniWqoModNatural]: t\mapsto t+1$ as a length function for the nwqo $\UniWqoModNatural$. 
Unfortunately, the norms of the elements of a bad sequence do not decrease in every  nwqo, so $\UniWqoModNatural$ is special. 
For  $\UniWqoModNatural^2$, for example, we can start with an element of norm no bigger than $t=3$. 
But every sequence of the form $(0,3), (a,2), (a-1, 2), \ldots, (1,2)$ is bad and its initial element has norm at most $3$, so there is no length function $L[\UniWqoModNatural^2]: t\mapsto L[\UniWqoModNatural^2](t)$ for $\UniWqoModNatural^2$. 
\end{example}

In this section, we define controlled bad sequences in detail and present length theorems for the knotted nwqos.

\subsection{Controlled bad sequences and length functions}

Considering bad sequences where the norms of the elements strictly decrease would result in a length function for every nwqo, but the applicability of this notion is too limited. 
Instead, we allow the norms to grow, but we control this growth rate. 
To maximize the applicability of the notion we take the growth rate itself to be exponential/compositional in terms of an auxiliary base function $f$. 
In other words, $f$ aims at indicating how the norm of a term $a_i$ of the sequence will restrict the norm of the next term: {$\UniNorm{\UniValA_{i+1}}{} \leq \UniControlFunctionA(\UniNorm{\UniValA_i}{})$.} 
By iterating this inequality we then get $\UniNorm{\UniValA_i}{} \leq \UniControlFunctionA^i(\UniControlParam)$ for all $i$. 
Length functions $\UniLengFunc{\UniWqoA}{\UniControlFunctionA}$ will then be defined in terms of both the nwqo $\m Q$ and the control function $f$.

\begin{definition}
A \emph{control function} is a mapping $\UniControlFunctionA : \UniNaturalSet \to \UniNaturalSet$ that is strictly increasing (for all $\UniValA\in \UniNaturalSet$, $\UniControlFunctionA(\UniValA+1) > \UniControlFunctionA(\UniValA)$) and expansive ($\UniControlFunctionA(\UniValA) \geq \UniValA$ for all $\UniValA \in \UniNaturalSet$).
\end{definition}

\begin{definition}{\cite{figueira2011}}
\label{def:controlfunction}
Given an nwqo $\UniWqoA$, a control function 
$\UniControlFunctionA$ and 
$\UniControlParam \in \UniNaturalSet$, an \emph{$(\UniControlFunctionA, \UniControlParam)$-controlled bad sequence over $\UniWqoA$} is a bad sequence 
$\UniValA_0,\ldots,\UniValA_{\UniLengthSequence-1}$ 
over $\UniWqoA$ where 
$\UniNorm{\UniValA_i}{} \leq \UniControlFunctionA^i(\UniControlParam)$ for all $0 \leq i < \UniLengthSequence$.
\end{definition}

For example, in $\UniWqoModNatural^2$, for $t=3$ and $f(x)=x+1$, the following are $(\UniControlFunctionA, \UniControlParam)$-controlled bad sequences: $(0,3), (1,2), (0, 2), (1,1), (0,1), (1,0), (0,0)$ and $(0,3), (1,2), (2, 1), \allowbreak (3,0), (2,0),\allowbreak (1,0),\allowbreak (0,0)$. If $f(x)=x^2$ instead, we have that $(0,3), (7,2), (80, 1), \allowbreak (81^2,0), \allowbreak (81^2-1,0),\ldots, (2,0),\allowbreak(1,0),\allowbreak (0,0)$ is a $(f,t)$-controlled bad sequence.

Now, unlike the general case of bad sequences, $(\UniControlFunctionA, \UniControlParam)$-controlled bad sequences have a maximum length~\cite{figueira2011}.
The proof is instructive, as it is a direct application of K\"onig's Lemma; we briefly reproduce it here for completeness.
Let $\UniWqoA$ be a nwqo and $\UniControlFunctionA$ be a control function. 
The idea is to construct a labelled tree in which a sequence of elements of $\UniWqoA$ is a branch if, and only if, it is an $(f,t)$-controlled bad sequence.
We start from an unlabelled node, the root (level 0), and add as children nodes labelled with the elements $a \in Q$ such that $\UniNorm{a}{\UniWqoA} \leq t$ (level 1). 
There are only finitely many of such elements, since we have a proper norm. 
For each of these new nodes, we add, as children, nodes labelled with the elements $b \in Q$ such that $\UniNorm{b}{\UniWqoA} \leq \UniControlFunctionA(t)$ in such a way that no increasing pair appears in a branch  (level 2). 
We keep doing this, adding as children of a node at level $i$ the elements with norm at most $f^{i-1}(t)$, while also guaranteeing that the wqo property is satisfied per branch.
This tree is finitely branching and all branches are finite, as they are bad sequences over $\UniWqoA$ (ignoring the root).
By K\"onig's Lemma, the tree must be finite, thus there are finitely many $(\UniControlFunctionA, t)$-controlled bad sequences over $\UniWqoA$; in particular, one of maximum length exists.
As we mentioned before, we usually express the maximum length as a function of the \emph{initial parameter} $\UniControlParam$.

The \emph{length function of an nwqo $\UniWqoA$ for a control function $\UniControlFunctionA$} is the mapping $\UniLengFunc{\UniWqoA}{\UniControlFunctionA} : \UniNaturalSet \to \UniNaturalSet$ that assigns to each $\UniControlParam \in \UniNaturalSet$ the maximum length of $(\UniControlFunctionA, \UniControlParam)$-controlled bad sequences over $\UniWqoA$. 

\begin{example}
In the nwqo $\UniWqoModNatural$ the underlying order relation is total, so bad sequences are always decreasing sequences.
For any control function $\UniControlFunctionA$ and $t \in \UniNaturalSet$, we have that $t, t-1,\ldots, 1, 0$ is a $(\UniControlFunctionA, t)$-controlled bad sequence over $\UniWqoModNatural$ of maximum length.
In other words, 
$\UniLengFunc{\UniWqoA}{\UniControlFunctionA}(t) = t+1$.
\end{example}

\begin{example} 
Arguably $\UniWqoModNatural^2$  is the next simplest nwqo where we could witness more complex length functions.
In fact, even for linear control functions one can already find non-elementary length functions, as the example in~\cite[p.~26]{schmitz2012notes} demonstrates.
This points to the fact that deriving length theorems is usually a nontrivial task.
Intuitively, for $\UniWqoModNatural^2$ one could start building a longest $(f,t)$-controlled bad sequence with the element $(t,t)$ (since it is $t$-controlled).
The space for the next element is thus all elements in $\UniWqoModNatural^2/(t,t) := \{ (a,b) \mid (t,t) \not\UniWqoRel{\UniWqoModNatural^2} (a,b) \}$ with norm at most $f(t)$---the space $\UniWqoModNatural^2/(t,t)$ is called the \emph{residual induced by $(t,t)$}, which induces a simpler nwqo (a substructure of $\UniWqoModNatural^2$)---but which one to choose is the hard part.
Residuals however are at the heart of the usual methods to derive length theorems for nwqos~\cite[Sec.~2.3]{schmitz2012notes}, since the length function of any nwqo $\UniWqoA$ is easily expressed recursively in terms of residuals via the so-called \emph{descent equation}:
$\UniLengFunc{\UniWqoA}{\UniControlFunctionA}(t) = \max_{x \in \UniWqoA_{\leq t}} \{ 1 + \UniLengFunc{\UniWqoA/x}{\UniControlFunctionA}(\UniControlFunctionA(t)) \}$,
where $\UniWqoA_{\leq t}$ is the set of all elements of $\UniWqoA$ of norm at most $t$.
\end{example}

Regarding length functions, it is well-known~\cite[Remark 2.2]{schmitz2012notes} that $\UniLengFunc{\UniWqoA}{\UniControlFunctionA}$ is increasing and if $\UniControlFunctionA_1 \leq \UniControlFunctionA_2$ then $\UniLengFunc{\UniWqoA}{\UniControlFunctionA_1} \leq \UniLengFunc{\UniWqoA}{\UniControlFunctionA_2}$.
    
In many of our applications, the upper bound $t$ for the norm of the initial element in a bad sequence will be the size $\langle h \rangle$ of the input hypersequent $h$ for proof-search algorithms we will design. 
The algorithm will need to decide whether $h$ is provable in a given system and we will be interested in its running time as a function of $t=\langle h \rangle$. 
Therefore, it is very convenient that we have defined  $\UniLengFunc{\UniWqoA}{\UniControlFunctionA}$ as a function of a single argument, since, as we will soon detail, understanding the complexity class of $\UniLengFunc{\UniWqoA}{\UniControlFunctionA}$ essentially gives the complexity of provability as a function of the size of its input.

The already signaled combinatorial complexity of controlled bad sequences leads to length functions that grow extremely fast.
Because of that, in order to prove the length theorems of interest, we first need some basic definitions and results regarding the theory of subrecursive and fast-growing hierarchies, which we present in the next subsection.

\subsection{Subrecursive hierarchies}
\label{sec:subrec-hierarchies}

As usual a function $f : \UniNaturalSet^k \to \UniNaturalSet$ is \emph{increasing} if $\vec x \leq \vec y$ (i.e., $x_i \leq y_i$ for all $1 \leq i \leq k$) implies $f(\vec x) \leq f(\vec y)$ for all $\vec x, \vec y \in \UniNaturalSet^k$.
Also, $f$ is \emph{expansive} if $\max_i x_i \leq f(\vec x)$ for all $\vec x \in \UniNaturalSet^k$.
Given $h_1,\ldots,h_k : \UniNaturalSet^v \to \UniNaturalSet$, we denote by $f \circ (h_1,\ldots,h_k)$ the function such that 
$f \circ (h_1,\ldots,h_k)(\vec y) = f(h_1(\vec y),\ldots,h_k(\vec y))$ for all $\vec y \in \UniNaturalSet^v$.
Moreover, when there is no risk of confusion, we will denote the single-variable version of $f$ by the same symbol, i.e., we define $f(x) \UniSymbDef f(x,\ldots,x)$ whenever $f$ accepts multiple inputs.
We denote also by $\circ$ the usual function composition.

Given $f,g : \UniNaturalSet^k \to \UniNaturalSet$, we say that $g$ \emph{eventually upper bounds} $f$ when there is $M \in \UniNaturalSet$ such that for all $\vec x \in \UniNaturalSet^k$ with $\max_i x_i \geq M$, $f(\vec x) \leq g(\vec x)$.
We also allow for single-variable functions to upper bound multiple-variable functions in the following, standard, way.
Given $f:\UniNaturalSet^k\to\UniNaturalSet$ and $g : \UniNaturalSet \to \UniNaturalSet$, we say that $g$ eventually upper bounds $f$ when there is $M \in \UniNaturalSet$ such that, for all $\vec x$ with $\max_i x_i \geq M$, we have $f(x_1,\ldots,x_k) \leq g(\max_i x_i)$. 

We denote by $\EpsZero$  the smallest epsilon number in Cantor's terminology: the smallest ordinal number $\varepsilon$ such that $\varepsilon=\omega^\varepsilon$.
We hereby assume familiarity with basic arithmetic operations over ordinals up to $\EpsZero$.
Recall that every $\alpha < \varepsilon_0$ has a unique Cantor Normal Form (CNF) representation, consisting of an  expression of the form $\omega^{\beta_m} + \cdots + \omega^{\beta_1}$, where $m \in \UniNaturalSet$, $\beta_i < \varepsilon_0$ for all $1 \leq i \leq m$, and $\beta_m \geq \beta_{m-1} \geq \ldots \geq \beta_1 \geq 0$.
We note that if consecutive $\beta$'s are equal then the corresponding part $\omega^{\beta} + \cdots + \omega^{\beta}$ of the CNF can be alternatively written as $\omega^{\beta}\cdot c$, where $c$ is the number of summands. 
Also, in the following we write $\gamma+ \omega^{\beta_1}$, where $\gamma=\omega^{\beta_m} + \cdots + \omega^{\beta_2}$, when we want to focus on the last summand of the CNF of an ordinal.

Subrecursive hierarchies are hierarchies of computable functions that do not demand the full power of Turing machines to be computed (justifying the name `subrecursive').
The subrecursive hierarchies we will work with are recursively defined in terms of ordinals. For that, we need a way to obtain a predecessor for limit ordinals (for successor ordinals the choice of predecessor is obvious), which relies on the notion of fundamental sequences.
A \emph{fundamental sequence} for a limit ordinal $\lambda < \varepsilon_0$ is a sequence $\{ \lambda_x \}_{x < \omega}$ of ordinals with supremum~$\lambda$. 
We fix a fundamental sequence (as in~\cite{schmitzschnoebelen2011}) for each such $\lambda$ as follows:
$\lambda_x\UniSymbDef \gamma + \omega^\beta \cdot (x+1)$ if $\lambda=\gamma + \omega^{\beta+1}$
and 
$\lambda_x\UniSymbDef\gamma + \omega^{\lambda_x'}$ if
$\lambda=\gamma + \omega^{\lambda'}$
(note that here $\lambda'$ is a limit ordinal).

Let $\UniControlFunctionA : \UniNaturalSet \to \UniNaturalSet$ be a control function.
The \emph{Hardy hierarchy} for $\UniControlFunctionA$ is given by $\{ \UniControlFunctionA^\alpha \}_{\alpha < \varepsilon_0}$, where
$\UniControlFunctionA^0(x) \UniSymbDef x$,
$\UniControlFunctionA^{\alpha+1}(x) \UniSymbDef f^{\alpha}(f(x))$
and
$\UniControlFunctionA^{\lambda}(x) \UniSymbDef
\UniControlFunctionA^{\lambda_x}(x)$---here the superscripts are part of the notation and do not denote $n$-fold composition.
The \emph{Cichoń hierarchy} (or \emph{length hierarchy}) for $\UniControlFunctionA$ is given by $\{ \UniControlFunctionA_\alpha \}_{\alpha < \varepsilon_0}$, where
$\UniControlFunctionA_0(x) \UniSymbDef 0$,
$\UniControlFunctionA_{\alpha+1}(x) \UniSymbDef 1 + \UniControlFunctionA_{\alpha}(\UniControlFunctionA(x))$
and
$\UniControlFunctionA_{\lambda}(x) \UniSymbDef \UniControlFunctionA_{\lambda_x}(x)$.
The \emph{fast-growing hierarchy} for $\UniControlFunctionA$,
$\{ \UniFastGFunc{\UniControlFunctionA}{\alpha} \}_{\alpha < \varepsilon_0}$, is defined by
\begin{center}
$\UniFastGFunc{\UniControlFunctionA}{0}(x) \UniSymbDef \UniControlFunctionA(x),
\UniFastGFunc{\UniControlFunctionA}{\alpha+1}(x) \UniSymbDef \mathsf f^{x+1}_{\UniControlFunctionA,\alpha}(x)$
and
$\UniFastGFunc{\UniControlFunctionA}{\lambda}(x) \UniSymbDef \mathsf f_{\UniControlFunctionA,\lambda_x}(x)$,
\end{center}
where $\lambda$ is a limit ordinal; here, and in the displayed formulas below, the exponent $x+1$ denotes the multiple composition of the function with itself $(x+1)$-many times.
A typical choice for $\UniControlFunctionA$ is $H(x) \UniSymbDef x+1$ (the successor function).
This leads to the fast-growing hierarchy for $H$, written
$\{ F_\alpha \}_{\alpha<\varepsilon_0}$:
\begin{center}
$F_0(x) := x + 1,
F_{\alpha + 1}(x) := F_\alpha^{x+1}(x),$
and
$F_{\lambda}(x) := F_{\lambda_x}(x)$.   
\end{center}
We have that $F_1(x) = 2x+1$ and $F_2(x) = 2^{x+1}(x+1)-1$, and that $F_3$ is non-elementary (grows faster than the tetration function).
Also, each primitive recursive function is eventually upper bounded by $F_k$ for some $k < \omega$.

In Lemma~\ref{lem:fgh-facts} below, we show some fundamental relationships between these hierarchies.

\begin{lemma}[{\cite[App. C.3]{schmitzschnoebelen2011}}]
    \label{lem:fgh-facts}
    For all $\alpha < \varepsilon_0$
    and control functions $\UniControlFunctionA$,
    \begin{enumerate}
        \item $\UniControlFunctionA_\alpha(t) \leq \UniControlFunctionA^{\alpha}(t)-t$ {for all $t \in \UniNaturalSet$}; equality holds if $f$ is the successor function;
        \item $\UniControlFunctionA^{\omega^\alpha \cdot r}(t) = \UniFastGFunc{\UniControlFunctionA}{\alpha}^r(t)$ for all $t \in \UniNaturalSet$ and $r < \omega$;
        \item if $\UniControlFunctionA$ is eventually upper bounded by $F_\gamma$, then $\UniFastGFunc{\UniControlFunctionA}{\alpha}$ is eventually upper bounded by $F_{\gamma + \alpha}$.
    \end{enumerate}
\end{lemma}

A common problem with the above hierarchies is that they are not monotonic with respect to the ordinal index under the usual well order on ordinals~(e.g., in the Cichoń hierarchy,
$\UniControlFunctionA_{\omega}(x) = 
\UniControlFunctionA_{\omega_x}(x) = 
\UniControlFunctionA_{x+1}(x)
= 1 + \UniControlFunctionA_x(f(x)) = x+1 < x + 2 = f_{x+2}(x)$,
even though $x+2 < \omega$).
In order to solve this, we  consider the so-called \emph{pointwise ordering}.

For each $x \in \UniNaturalSet$, the \emph{pointwise-at-$x$ ordering} $\UniPointWiseOrd{x}$ on ordinals is the smallest transitive relation such that, for all ordinals $\alpha$ and limit ordinals $\lambda$, $\alpha \UniPointWiseOrd{x} \alpha + 1$ and $\lambda_x \UniPointWiseOrd{x} \lambda$.
The above counter-example to monotonicity of $\{ h_\alpha \}_{\alpha < \epsilon_0}$ is no longer problematic as $x+2 \not\UniPointWiseOrd{x} \omega$.
In Lemma~\ref{lem:point-ord-facts} below we present some facts regarding the pointwise ordering, including the desired monotonicity property.

\begin{lemma}
    \label{lem:point-ord-facts}
    The following holds:
    \begin{enumerate}
        \item $\UniPointWiseOrd{0} \subseteq 
        \ldots
        \subseteq
        \UniPointWiseOrd{x}
        \subseteq
        \UniPointWiseOrd{x+1}
        \subseteq 
        \ldots
        \subseteq
        \bigcup_{x \in \UniNaturalSet} \UniPointWiseOrd{x} \; = \; <$.
        \item 
        for any control function $\UniControlFunctionA$, it is the case that
        $\alpha' \UniPointWiseOrd{x} \alpha$
        implies $\UniControlFunctionA_{\alpha'}(x) \leq \UniControlFunctionA_\alpha(x)$.
    \end{enumerate}
\end{lemma}

In order to relate this new ordering with the usual one, we use the notion of \emph{leanness}~\cite[{App. B.3}]{schmitzschnoebelen2011}.
Recall that the strict CNF of an ordinal $\alpha$ is obtained by grouping summands of its CNF with same exponent:
$\alpha = \omega^{\gamma_1} \cdot c_1 + \cdots + \omega^{\gamma_m} \cdot c_m$,
where $\gamma_1 > \ldots > \gamma_m$ and $c_i > 0$ for all $1 \leq i \leq m$.
Then we recursively define a norm $\UniOrdNorm{\cdot}$ on ordinals by setting
$\UniOrdNorm{0} \UniSymbDef 0$
and 
$\UniOrdNorm{\omega^{\gamma_1} \cdot c_1 + \ldots + \omega^{\gamma_m} \cdot c_m}
\UniSymbDef \max\{ c_1,\ldots,c_m, \UniOrdNorm{\gamma_1},\ldots,\UniOrdNorm{\gamma_m} \}$.
The leanness of an ordinal then tells us, via $\UniOrdNorm{\cdot}$, an upper bound for its coefficients.

\begin{definition}
    An ordinal $\alpha < \epsilon_0$ is $k$-lean, for $k \in \UniNaturalSet$, whenever $\UniOrdNorm{\alpha} \leq k$.
\end{definition}

Then, by the following lemma, we have the connection between the two orderings.

\begin{lemma}[{\cite[Lem.~B.1]{schmitzschnoebelen2011}}]\label{lem:point-ord-usual-ord-rel}
    If $\alpha$ is $k$-lean,
    $\alpha < \gamma 
    \text{ iff }
    \alpha \UniPointWiseOrd{k} \gamma$, 
    for all $\gamma$.
\end{lemma}

By putting all the above concepts together, we get a way to obtain monotonicity of the Cichoń hierarchy with respect to $<$:

\begin{corollary}
Let $\UniControlFunctionA$ be a control function.
If $\alpha < \gamma$, then for all 
$x \geq \UniOrdNorm{\alpha}$, 
we have 
$f_{\alpha}(x) \leq f_{\gamma}(x)$.
\end{corollary}
\begin{proof}
Observe that $\alpha$ is $\UniOrdNorm{\alpha}$-lean.
Thus by Lemma~\ref{lem:point-ord-usual-ord-rel}, we have
$\alpha \UniPointWiseOrd{\UniOrdNorm{\alpha}} \gamma$.
Let $x \geq \UniOrdNorm{\alpha}$.
By Lemma~\ref{lem:point-ord-facts} (1), we have 
$\alpha \UniPointWiseOrd{x} \gamma$.
Thus by Lemma~\ref{lem:point-ord-facts} (2), we have 
$f_{\alpha}(x) \leq f_{\gamma}(x)$ as desired.
\end{proof}

We have defined hierarchies of functions starting from a base function.
Now, based on them, we define hierarchies of classes of functions and classes of problems, which are useful to uniformly characterize functions and decision problems with respect to their complexity.

The \emph{extended Grzegorczyk hierarchy} is a family  
$\{ \UniGrzHLevel{\UniOrdinalA} \}_{\UniOrdinalA<\EpsZero}$ 
of collections of functions of type {$\UniNaturalSet^k \to \UniNaturalSet$ ($k \geq 1$)}, also called the \emph{levels} of the hierarchy (as usual, $\UniOrdinalA < \EpsZero$ means that the index $\UniOrdinalA$ runs over all ordinals less than $\EpsZero$).
The definition of the levels of this hierarchy relies essentially on the family
$\{ \UniFastGrowingFunction{\UniOrdinalA}\}_{\UniOrdinalA<\EpsZero}$, 
and is given in detail below.

    \begin{definition}
        \label{def:precise-fgh}
        For $\alpha < \varepsilon_0$, the class 
        $\UniFGHLevel{\alpha}$
        is defined as the closure of the set
        $\{ 
            \lambda x. 0,
            \lambda (x, y).x + y,
            \lambda (x_1,\ldots,x_n).x_i
            (\text{all } n \geq 1, 1 \leq i \leq n),
            F_\alpha
        \}$%
        ---that is, the zero constant, sum, projections
        and $F_\alpha$---under the following operations:
        \begin{itemize}
            \item substitution:
            if 
            $h_0,\ldots,h_p \in \UniFGHLevel{\alpha}$, such that $h_0$ has arity $p$, and $h_1,\ldots,h_p$ have arity $n$,
            \[
            \lambda (x_1,\ldots,x_n).
            h_0(h_1(x_1,\ldots,x_n),\ldots,h_p(x_1,\ldots,x_n)) \in \UniFGHLevel{\alpha}
            \]
            \item limited primitive recursion:
            if
            $h_0,h_1,g \in \UniFGHLevel{\alpha}$,
            then also 
            $f \in \UniFGHLevel{\alpha}$,
            in case $h_0$ has arity $n$, $h_1$ has arity $n+2$, $g$ has arity $1$, $f$ has arity $n+1$ and
            \begin{align*}
                f(0,x_1,\ldots,x_n)
                &= h_0(x_1,\ldots,x_n)\\
                \qquad f(y+1,x_1,\ldots,x_n)
                &= h_1(y,x_1,\ldots,x_n,f(y,x_1,\ldots,x_n))\\
                f(y,x_1,\ldots,x_n)
                &\leq g(\max\{ y,x_1,\ldots,x_n \})
                \text{ for all }
                y,x_1,\ldots,x_n \in \UniNaturalSet
            \end{align*}
        \end{itemize}
    \end{definition}
\noindent Note that by removing the third clause in the above form of limited recursion we get the usual notion of primitive recursion.

An important feature of these levels is that $\UniGrzHLevel{\UniOrdinalA}$, for $\UniOrdinalA \geq 2$, is the class of functions computable by deterministic Turing machines in time bounded by a function in $O(\UniFastGrowingFunction{\UniOrdinalA}^c)$, for some constant $c \in \UniNaturalSet$ (where $\UniFastGrowingFunction{\UniOrdinalA}^c$ denotes $c$-fold composition of $F_\alpha$). 
We let $\UniFGHLevel{<\alpha} \UniSymbDef \bigcup_{\beta < \alpha} \UniFGHLevel{\beta}$.
The following lemmas  about this hierarchy will be important to us.

\begin{lemma}[{\cite[Exercise 2.3 (4)]{schmitz2012notes}}]
\label{lem:bounds-in-fgh}
    For $\alpha > 0$, if  $f \in \UniFGHLevel{\alpha}$ has arity $k$, then there is $c \in \UniNaturalSet$ such that
    $f(x_1,\ldots,x_k) < F_\alpha^c(\max_i x_i + 1)$
    for every
    $x_1,\ldots,x_k \in \UniNaturalSet$.
\end{lemma}

\begin{lemma}[{\cite{schmitz2016hierar}}]
\label{lem:funcs-bound-by-fs}
Every function in $\UniFGHLevel{\beta}$ is eventually upper bounded by $F_\alpha$ for every $\alpha > \beta$.
\end{lemma}

\begin{lemma}\label{lem:max-preserve-level}
If $\alpha \geq 2$ is an ordinal and $h_1,\ldots,h_k : \UniNaturalSet^v \to \UniNaturalSet \in \UniFGHLevel{<\alpha}$, then the function
$\max_{1,\ldots,k}$ is in $\UniFGHLevel{<\alpha}$, 
where
$\max_{1,\ldots,k}(\vec x) \UniSymbDef \max(h_1(\vec x),\ldots,h_k(\vec x))$.
\end{lemma}
\begin{proof}
Since
$h_i \in \UniFGHLevel{\beta_i}$,
where $\beta_i < \alpha$, for all $1 \leq i \leq k$, we have 
$\beta \UniSymbDef \max_i\{ \beta_i \} < \alpha$, 
hence $h_i \in \UniFGHLevel{\beta}$ for all $1 \leq i \leq k$.
Moreover, the $k$-ary function $\max$ is in $\UniFGHLevel{1}$, therefore $\max_{1,\ldots,k}$ is in $\UniFGHLevel{\beta}$ as a composition of function in $\UniFGHLevel{\beta}$.
Since $\beta < \alpha$, we have $\max_{1,\ldots,k} \in \UniFGHLevel{<\alpha}$.
\end{proof}

\begin{lemma}\label{lem:it-is-prim-rec}
    If $f : \UniNaturalSet \to \UniNaturalSet$ is primitive recursive, then so is the function
    $\mathsf{it}_f$, defined by 
    $\mathsf{it}_f(0, x) \UniSymbDef x$
    and $\mathsf{it}_f(i+1, x) \UniSymbDef f(\mathsf{it}_f(i, x))$.
\end{lemma}
\begin{proof}
    Directly from the fact that $\mathsf{it}_f$ was defined by primitive recursion from a primitive-recursive function.
\end{proof}

\begin{lemma}[{\cite[Lem.~4.6]{schmitz2016hierar}}]\label{lem:comp-before-after}
    If $f, f' \in \UniFGHLevel{<\alpha}$, then 
    there is $p \in \UniFGHLevel{<\alpha}$ such that
    $f \circ F_\alpha \circ f'$
    is eventually upper bounded by $F_\alpha \circ p$.
\end{lemma}

\subsection{The fast growing complexity classes}\label{fast-growing-cc}

Schmitz~\cite{schmitz2016hierar} presented a hierarchy 
$\{ \UniFGHProbOneAppLevel{\UniOrdinalA} \}_{\UniOrdinalA<\EpsZero}$ 
of decision problems (known as \emph{fast-growing complexity classes}) based on Grzegorczyk's in order to provide a more fine-grained characterization of non-elementary complexities. 
In this hierarchy, $\UniFGHProbOneAppLevel{\UniOrdinalA}$ is the class of problems solvable by a deterministic Turing machine in time bounded by some function in $O(F_{\alpha} \circ h)$, for some
$h : \UniNaturalSet \to \UniNaturalSet \in \UniFGHLevel{<\alpha}$.
In order to facilitate our analysis, we let  
$\UniFGHOneAppLevel{\UniOrdinalA}$ be the set comprising the functions $\UniFastGrowingFunction{\UniOrdinalA} \circ h : \UniNaturalSet \to \UniNaturalSet$, for an $h : \UniNaturalSet \to \UniNaturalSet \in \UniFGHLevel{<\alpha}$.
Note that if a problem has complexity eventually upper bounded by a function in $\UniFGHOneAppLevel{\UniOrdinalA}$, then this problem is in
$\UniFGHProbOneAppLevel{\UniOrdinalA}$.

In this hierarchy, the classes
$\UniFGHProbOneAppLevel{\omega}$ (or \ACK)
of Ackermannian problems and $\UniFGHProbOneAppLevel{\omega^{\omega}}$ (or \HACK) of hyper-Ackermannian problems are of primary interest to us.
The former contains those problems solvable in time bounded by a function consisting of compositions of primitive recursive functions and  {then} a single application of an Ackermann-like function (the function $\UniFastGrowingFunction{\omega}$).
The latter contains the problems solvable in time bounded by a function consisting of compositions of multiply-recursive functions (that is, functions defined by nested or $k$-fold recursion~\cite[\S 10]{roszapeter1967}; the Ackermann function, for example, is defined by 2-fold recursion) and a single application of a hyper-Ackermannian function (the function $\UniFastGrowingFunction{\omega^\omega}$, which grows faster than any multiply-recursive function).

We close with a lemma expressing that composing functions of lower complexity with functions bounded in $\UniFGHOneAppLevel{\alpha}$ results in a function whose single-variable version is still bounded in $\UniFGHOneAppLevel{\alpha}$.

\begin{lemma}\label{lem:ev-ub-multvariable}
    Let $\alpha \geq 2$ be a limit ordinal,
    $E : \UniNaturalSet^k \to \UniNaturalSet$
    be an increasing function in
    $\UniFGHLevel{<\alpha}$
    and
    $h_i : \UniNaturalSet^v \to \UniNaturalSet$ be eventually upper bounded by a function in $\UniFGHOneAppLevel{\alpha}$ for every $1 \leq i \leq k$.
    Then 
    $\lambda x.E(h_1(x,\ldots,x),\ldots, h_k(x,\ldots,x))$
    is eventually upper bounded by a function in $\UniFGHOneAppLevel{\alpha}$.
\end{lemma}
\begin{proof}
First of all, for each $1 \leq i \leq k$, let $h'_i \UniSymbDef F_\alpha \circ p_i$ be a function in $\UniFGHOneAppLevel{\alpha}$ that eventually upper bounds $h_i$, with $p_i \in \UniFGHLevel{<\alpha}$, and let $N_i$ be the constant such that for all $\vec x$ with $\max_j x_j \geq N_i$ we have $h_i(\vec x) \leq h'_i(\vec x)$.
Then for $p'_i(\vec x) \UniSymbDef p_i(\vec x)+ \max_i x_i$ we have that $p_i(\vec x) \leq p'_i(\vec x)$ for all $\vec{x} \in \UniNaturalSet^v$, and each $p'_i$ is expansive and  in $\UniFGHLevel{<\alpha}$ (as a sum of such functions).

Recall by Lemma~\ref{lem:funcs-bound-by-fs} that for some $M_1$
and $\beta < \alpha$,
$E(\vec x) \leq F_\beta(\max_j x_j)$
for all $\vec x$ with $\max_j x_j \geq M_1$.

Moreover, for
$\max_{1,\ldots,k}(\vec z) \UniSymbDef \max(p'_1(\vec z),\ldots,p'_k(\vec z))$,
by Lemma~\ref{lem:max-preserve-level} we have  {$\max_{1,\ldots,k} \in \UniFGHLevel{<\alpha}$}. 
So, by Lemma~\ref{lem:comp-before-after}, we have 
$F_\beta \circ F_\alpha \circ \max_{1,\ldots,k}(\vec z) \leq F_\alpha \circ q (\vec z)$ for all $\vec z$ with $\max_i z_i \geq M_2$, for some $M_2$ and some $q \in \UniFGHLevel{<\alpha}$.

Also note that if $\max_j z_j \geq M_1$, we have $\max_i p'_i(\vec z) \geq M_1$ since $\max_j z_j \leq p'_i(\vec z) \leq \max_i p'_i(\vec z)$.
Since $F_\alpha$ is expansive,
$\max_i p'_i(\vec z) \leq F_\alpha(\max_i p'_i(\vec z))$, thus
(a): $M_1 \leq F_\alpha(\max_i p'_i(\vec z))$ as well.

Now, let $M \UniSymbDef \max\{ \max_i N_i, M_1,M_2\}$.
Then, for all $\vec y$ with $\max_i y_i \geq M$,
\begin{align*}
&E(h_1(\vec y),\ldots, h_k(\vec y))\\
&\leq
    E(F_\alpha(p_1(\vec y)),\ldots, F_\alpha(p_k(\vec y))) &E \text{ and } F_\alpha \text{ are increasing}\\
    &\leq 
    E(F_\alpha(p'_1(\vec y)),\ldots, F_\alpha(p'_k(\vec y)))&E \text{ and } F_\alpha \text{ are increasing}\\
    &\leq
    E(F_\alpha(\max_i p'_i(\vec y)),\ldots, F_\alpha(\max_i p'_i(\vec y))) & E \text{ and } F_\alpha \text{ are increasing}\\
    &\leq F_\beta(\max\{F_\alpha(\max_i p'_i(\vec y)),\ldots, F_\alpha(\max_i p'_i(\vec y))\})
    &
    \text{by (a)}
    \\
    &=
    F_\beta(F_\alpha(\max_i p'_i(\vec y)))\\
    &=
    F_\beta \circ F_\alpha \circ \max_{1,\ldots,k}(\vec y)\\
    &\leq
    F_\alpha \circ q(\vec y)
    & \max_i y_i \geq M \geq M_2
\end{align*}
In particular, for every $x \geq M$, we have
$$E(h_1(x,\ldots,x),\ldots, h_k(x,\ldots,x))
\allowbreak\leq F_\alpha(q(x,\ldots,x)),$$ 
hence 
$E'(x) \UniSymbDef E(h_1(x,\ldots,x),\ldots, h_k(x,\ldots,x))$ is eventually bounded by $q'(x) \UniSymbDef F_\alpha(q(x,\allowbreak\ldots,x))$.
Let $\UniId$ be the identity function on $\UniNaturalSet$.
Note that $q''(x) \UniSymbDef q(x,\ldots,x) \allowbreak= q(\UniId(x),\ldots,\UniId(x))$.
Since $q$ and $\UniId$ are in $\UniFGHLevel{<\alpha}$, we get $q'' \in \UniFGHLevel{<\alpha}$ (because the definition of $q''$ matches the substitution scheme in Definition~\ref{def:precise-fgh}).
Therefore, by the definition of $\UniFGHOneAppLevel{\alpha}$, we have $q' = F_\alpha \circ q'' \in \UniFGHOneAppLevel{\alpha}$, and $E'$ is eventually upper bounded by $q'$.
\end{proof}

\subsection{Length theorems for knotted nwqos}

In this section, we provide length theorems (upper bounds to length functions) for the knotted nwqos.
As we will see, the particular control functions $f$ we need for our applications will always be primitive recursive; so we are interested in theorems that prove
$\UniLengFunc{\UniWqoA}{\UniControlFunctionA}$ 
to be in a given fast-growing complexity class whenever $f$ is primitive recursive. 
We will define a binary relation $\UniFGBoundedNwqosIn$ between nwqos and the complexity classes that captures exactly this situation.

\begin{definition}
For a nwqo $\UniWqoA$ and an ordinal $\UniOrdinalA < \varepsilon_0$, we  write $\UniWqoA \UniFGBoundedNwqosIn \UniFGHOneAppLevel{\UniOrdinalA}$ when $\UniLengFunc{\UniWqoA}{\UniControlFunctionA}$ is eventually upper bounded by a function in $\UniFGHOneAppLevel{\UniOrdinalA}$, for all primitive recursive control functions $f$.
Let $\mathcal{Q}$ be a class of nwqos parameterized by a finite vector $\vec p$ of parameters, i.e., there is $v \in\UniNaturalSetNN$ such that $\mathcal{Q}=\{\UniWqoA_{\vec p}: \vec{p} \in \mathbb{N}^v\}$.
We denote by $\UniLengFunc{\mathcal{Q}}{f}$ the function $\lambda (\vec p, x).\UniLengFunc{\UniWqoA_{\vec p}}{f}(x)$.
Then $\mathcal{Q} \UniFGBoundedNwqosIn \UniFGHOneAppLevel{\UniOrdinalA}$ means that $ \UniLengFunc{\mathcal{Q}}{ \UniControlFunctionA}$ is eventually upper bounded by a function in 
$\UniFGHOneAppLevel{\UniOrdinalA}$ for all primitive recursive control functions $\UniControlFunctionA$.
We write $\UniFGBoundedNwqosIn_\gamma$, with $\gamma < \omega$, for restricting the above definitions to control functions at the level $\UniFGHLevel{\gamma}$.
\end{definition}

\begin{remark}
Upper bounds of the form $\mathcal{Q} \UniFGBoundedNwqosIn \UniFGHOneAppLevel{\UniOrdinalA}$ (that is, for parameterized classes of nwqos) are employed when the parameters themselves are inputs (or derived from the inputs) of the algorithm whose complexity is being analysed via these nwqos.
Still, it is useful to know upper bounds for the case in which parameters are fixed (i.e., not part of the input), since in most situations the length functions will then fall in lower levels of the fast-growing hierarchy.
This is known as \emph{parameterized complexity analysis}.
\end{remark}

Given a parameterized class of nwqos $\mathcal{Q}$, it will be useful for us to consider its \emph{diagonal subfamily}
$\UniDiagSubf{Q} \UniSymbDef \{ \UniWqoA_x \UniSymbDef \UniWqoA_{(x,\ldots,x)} \}_{x \in \UniNaturalSet}$, for which $\UniLengFunc{\UniDiagSubf{Q}}{f}$ is defined as $\lambda x.\UniLengFunc{\UniWqoA_x}{f}(x)$.
That is because we will be developing algorithms that depend on a single input (from which parameters are derived), and therefore being able to pick an appropriate nwqo based on this single input (instead of multiple parameters) will be important.
It is not hard to see that if $\mathcal{Q} \UniFGBoundedNwqosIn \UniFGHOneAppLevel{\UniOrdinalA}$, then 
$\UniDiagSubf{Q} \UniFGBoundedNwqosIn \UniFGHOneAppLevel{\UniOrdinalA}$.

We denote by $\UniNormLenFunc{\UniWqoA}{f}(t)$ the maximum norm of an element in a $(f,t)$-controlled bad sequence of $\UniWqoA$, and also extend this notation to parameterized classes of nwqos $\mathcal{Q}$ as we did before, i.e., $\UniNormLenFunc{\mathcal{Q}}{f} \UniSymbDef \lambda(\vec p, x).\UniNormLenFunc{\UniWqoA_{\vec p}}{f}(x)$ and $\UniNormLenFunc{\UniDiagSubf{Q}}{f} \UniSymbDef \lambda x.\UniNormLenFunc{\UniWqoA_{(x,\ldots,x)}}{f}(x)$. 
Observe that, by definition of controlled bad sequence, we must have $\UniNormLenFunc{\mathcal{Q}}{f}(\vec p, x) = \UniNormLenFunc{\UniWqoA_{\vec p}}{f}(x) \leq f^{\UniLengFunc{\UniWqoA_{\vec p}}{f}(x)}(x)$.
Therefore, we have $\UniNormLenFunc{\UniDiagSubf{Q}}{f}(x) \leq f^{\UniLengFunc{\UniDiagSubf{Q}}{f}(x)}(x)$. (Here, $f^k$ denotes the $k$-fold iteration of $f$, as in Lemma~\ref{lem:it-is-prim-rec}.)

The following lemma allows us to derive upper bounds for the maximum norm of an element in a controlled bad sequence given upper bounds for the length of the sequence.

\begin{lemma}\label{lem:bound-norm-bad-seq}
For a limit ordinal $\alpha \geq 2$, if $\mathcal{Q} \UniFGBoundedNwqosIn \UniFGHOneAppLevel{\UniOrdinalA}$, then $\UniNormLenFunc{\UniDiagSubf{Q}}{f}$ is upper bounded by a function in $\UniFGHOneAppLevel{\UniOrdinalA}$ for any primitive recursive control function $\UniControlFunctionA$.
\end{lemma}
\begin{proof}
Since $f$ is primitive recursive, consider the function 
$\mathsf{it}_f$ from Lemma~\ref{lem:it-is-prim-rec}.
Note that $\mathsf{it}_f$ is increasing in its first argument, i.e., $i_1 \leq i_2$ implies $\mathsf{it}_f(i_1,x) \leq \mathsf{it}_f(i_2,x)$, because $f$ satisfies $f(a) \geq a$ for all $a \in \UniNaturalSet$. 
Moreover, because $f$ is increasing, $\mathsf{it}_f$ is also increasing in its second argument.

Then, since $\mathcal{Q} \UniFGBoundedNwqosIn \UniFGHOneAppLevel{\UniOrdinalA}$, we have $\UniDiagSubf{Q} \UniFGBoundedNwqosIn \UniFGHOneAppLevel{\UniOrdinalA}$, and thus there is $x_0 \in \UniNaturalSet$ and $r \in \UniFGHLevel{<\alpha}$ such that, for all $x \geq x_0$, we have $\UniLengFunc{\UniDiagSubf{Q}}{f}(x) \leq F_{\alpha}(r(x))$.
Therefore, for all $x \geq x_0$, we have 
$\UniNormLenFunc{\UniDiagSubf{Q}}{f}(x) \leq f^{\UniLengFunc{\UniDiagSubf{Q}}{f}(x)}(x) \leq f^{F_{\alpha}(r(x))}(x) = \mathsf{it}_f(F_{\alpha}(r(x)), x) \leq \mathsf{it}_f(F_{\alpha}(r(x)), F_\alpha(x))$, 
and the latter is eventually upper bounded by a function in $\UniFGHOneAppLevel{\UniOrdinalA}$ in view of Lemma~\ref{lem:ev-ub-multvariable} (as mentioned before, $ f^{\UniLengFunc{\UniDiagSubf{Q}}{f}(x)}(x)$ denotes exponentiation with respect to functional composition).%
\qedhere
\end{proof}

In what follows, we will show results that upper bound the length functions of specific (classes of) nwqos (most of them based on knotted nwqos) in terms of the 
$\{ \UniFGHOneAppLevel{\UniOrdinalA} \}_{\UniOrdinalA<\EpsZero}$ hierarchy. 

First of all, if we consider the nwqos based on
$\UniWqoModNatural^k$
(employed, for example, in the complexity analysis of decision problems in counter machines~\cite{schmitz2016hierar}), we have the following length theorem.

\begin{theorem}
\label{fact:length_theorem_naturalset}
For $k, r \in \UniNaturalSetNN$, we have
${r \cdot \UniWqoModNatural^k} \UniFGBoundedNwqosIn_\gamma \UniFGHOneAppLevel{ \gamma+k+1}$.
Moreover,
$$\{r \cdot \UniWqoModNatural^k \}_{(r,k)\in\UniNaturalSet^2}\UniFGBoundedNwqosIn
\UniFGHOneAppLevel{\omega}.$$
\end{theorem}
\begin{proof}
Let $f$ be a primitive recursive function in $\UniFGHLevel{\gamma}$, where we assume $\gamma \geq 2$ without loss of generality.
By \cite[Thm.~3.15]{schmitzthesis}, we have that, for all $k, r \in \UniNaturalSetNN$,
$\UniLengFunc{r \cdot \UniWqoModNatural^k}{\UniControlFunctionA}(x) \leq h_{\omega^k \cdot r}(kx)$ for all $x \in \UniNaturalSet$, where
$h(x) := k\cdot\UniControlFunctionA(x)$.
For the first statement, note that
$h_{\omega^k \cdot r}(kx) \leq h^{w^k\cdot r}(kx) - kx = \mathsf{f}_{h,k}^r(kx)-kx \leq \mathsf{f}_{h,k}^r(kx)$
by Lemma~\ref{lem:fgh-facts}, items (1) and (2).
Since $\mathsf{f}_{h,k}$ is eventually upper bounded by $F_{\gamma + 1 + k}$ by Lemma~\ref{lem:fgh-facts} (3), we are done.
For the second statement, note that $\omega^k \cdot r \UniPointWiseOrd{kr} \omega^\omega$ by Lemma~\ref{lem:point-ord-usual-ord-rel} since $\omega^k \cdot r$ is $(kr)$-lean.
Then, since $r \geq 1$, for all $x \geq 1$ we have
$h_{\omega^k \cdot r}(kx) \leq 
h_{\omega^k \cdot r}(rkx)
\leq h_{\omega^\omega}(rkx)
\leq h^{\omega^\omega}(rkx) - rkx
= \mathsf{f}_{h,\omega}(rkx) - rkx \leq \mathsf{f}_{h,\omega}(rkx)$
by Lemma~\ref{lem:point-ord-facts}(2) and Lemma~\ref{lem:fgh-facts} (1),(2).
Also, for some $x_0 \in \UniNaturalSet$ and all
$rkx \geq x_0$
we have
$\mathsf{f}_{h,\omega}(rkx) \leq F_{\gamma+\omega}(rkx) = F_\omega(rkx)$
by Lemma~\ref{lem:fgh-facts} (3).
Since $\lambda (r,k,x).rkx \in \UniFGHLevel{1}$, we are done by Lemma~\ref{lem:bounds-in-fgh}.
\end{proof}

Balasubramanian~\cite{bala2020} established upper bounds for
$(\UniControlFunctionA, \UniControlParam)$-controlled bad sequences over
$\UniPowerMaj{\UniMajoringFntWqo{\UniWqoModNatural^k}}{d}$ and 
$\UniPowerMaj{\UniMinoringFntWqo{\UniWqoModNatural^k}}{d}$, 
which we employ to prove the following.

\begin{theorem}
\label{fact:length-theorem-power-set-wqo-nat}
For every $k,d \in \UniNaturalSet$, we have
$\UniPowerMaj{\UniMajoringFntWqo{\UniWqoModNatural^k}}{d},\UniPowerMaj{\UniMinoringFntWqo{\UniWqoModNatural^k}}{d}\UniFGBoundedNwqosIn \UniFGHOneAppLevel{\omega^{k}}$.
Moreover,
$$\left\{{\UniPowerMaj{\UniMajoringFntWqo{\UniWqoModNatural^k}}{d}}\right\}_{(k,d)\in\UniNaturalSet^2}, \left\{{\UniPowerMaj{\UniMinoringFntWqo{\UniWqoModNatural^k}}{d}}\right\}_{(k,d)\in\UniNaturalSet^2} \UniFGBoundedNwqosIn {\UniFGHOneAppLevel{\omega^\omega}}.$$
\end{theorem}
\begin{proof}
This proof is similar to the one of Theorem~\ref{fact:length_theorem_naturalset}.
Let $f$ be a primitive recursive function at $\UniFGHLevel{\gamma}$, where $\gamma \geq 2$ without loss of generality.

We first work on the majoring case.
By~\cite[{Thm.~4.5}]{bala2020}, we have that, for all $x \geq 1$,
$\UniLengFunc{\UniPowerMaj{\UniMajoringFntWqo{\UniWqoModNatural^k}}{d}}{\UniControlFunctionA}(x) \leq h_{\alpha}(4kdx)$,
where $\alpha = \omega^{\omega^{k-1}\cdot d}$ and $h(y) = 4y \cdot \UniControlFunctionA(y)$ (note that $h$ is also in  $\UniFGHLevel{\gamma}$).
Then we have $h_\alpha(4kdx) = h^{\alpha}(4kdx) - 4kdx 
= \mathsf{f}_{h,\omega^{k-1}\cdot d}(4kdx)-4kdx
\leq
\mathsf{f}_{h,\omega^{k-1}\cdot d}(4kdx)$; since $\mathsf{f}_{h,\omega^{k-1}\cdot d}$ is eventually upper bounded by
$F_{\omega^{k}}$, we are done.
Moreover, $\omega^{\omega^{k-1}\cdot d} \UniPointWiseOrd{kd} \omega^{\omega^\omega}$, thus for all $x \geq 1$,
$h_\alpha(4kdx) \leq h_{\omega^{\omega^{\omega}}}(4kdx) = h^{\omega^{\omega^{\omega}}}(4kdx) - 4kdx 
= \mathsf{f}_{h,\omega^\omega}(4kdx)-4kdx
\leq
\mathsf{f}_{h,\omega^\omega}(4kdx)$; and for sufficiently large
$4kdx$ (i.e., $4kdx \geq x_0$ for some $x_0 \in \UniNaturalSet$) we will have $\mathsf{f}_{h,\omega^\omega}(4kdx) \leq F_{\gamma + \omega^\omega}(4kdx) = F_{\omega^\omega}(4kdx)$ and we are essentially done.

We now move to the minoring case.
By~\cite[Thm.~6.3]{bala2020},
we have that
$\UniLengFunc{\UniPowerMaj{\UniMinoringFntWqo{\UniWqoModNatural^k}}{d}}{\UniControlFunctionA}(x) \leq h_{\alpha}(c(k,d) \cdot \UniControlFunctionA(x)^{2k})$ for sufficiently large $x$,
where $c$ is primitive recursive,
$\alpha = \omega^{\omega^{k-1}\cdot (2^k \cdot d)}$
and 
$h(y) = 4dy \cdot q(\UniControlFunctionA(y))$, with $q(y) = (y+1)^k$.
The rest of the proof is then very similar to the one above.
\end{proof}

In the previous section, using strong reflections we showed that all the knotted qos, their products and power set extensions are nwqos. 
Using the same reflections, we will now establish length theorems for them.
Below we present transfer results  of length theorems via strong reflections (recall the proof of Lemma~\ref{fact:refl-pres-wellness}).

\begin{lemma}[\cite{schmitz2012notes}]
\label{fact:strong-refl-preserv-bad-length}
If $\UniWqoA_1,\UniWqoA_2$ are nqos and $\UniWqoA_1 \UniStrongReflArrow{} \UniWqoA_2$, then 
$\UniLengFunc{\UniWqoA_1}{\UniControlFunctionA} \leq \UniLengFunc{\UniWqoA_2}{\UniControlFunctionA}$ for all control functions $\UniControlFunctionA$.
\end{lemma}

\begin{corollary}
\label{coro:strong-refl-f-hierar}
If $\UniWqoA_1\UniStrongReflArrow{} \UniWqoA_2$,
and $\UniWqoA_2 \UniFGBoundedNwqosIn \UniFGHOneAppLevel{\UniOrdinalA}$, then  $\UniWqoA_1 \UniFGBoundedNwqosIn \UniFGHOneAppLevel{\UniOrdinalA}$.
\end{corollary}
\begin{proof}
By assumption, we have that there is $x_0 \in \UniNaturalSet$
such that, for all 
$x \geq x_0$,
$\UniLengFunc{\UniWqoA_2}{f}(x) \leq F_\alpha(p(x))$ for some $p \in \UniFGHLevel{<\alpha}$.
Then, by Lemma~\ref{fact:strong-refl-preserv-bad-length}, we have 
$\UniLengFunc{\UniWqoA_1}{f}(x) \leq \UniLengFunc{\UniWqoA_2}{f}(x)$, 
so we are done.
\end{proof}

We now prove a similar result, this time for parameterized classes of nwqos.
Let $\mathcal{Q}_1 \UniSymbDef \{ \UniWqoA_{\vec p} : \vec p \in \mathbb{N}^{v_1} \}$ and
$\mathcal{Q}_2 \UniSymbDef \{ \UniWqoB_{\vec q} : \vec q \in \mathbb{N}^{v_2} \}$.
Given expansive $r_1,\ldots,r_{v_2} : \UniNaturalSet^{v_1} \to \UniNaturalSet$, we write
$\mathcal{Q}_1 \UniStrongReflArrow{\vec r} \mathcal{Q}_2$
in case for every 
$\vec p \in \UniNaturalSet^{v_1}$
we have
$\UniWqoA_{\vec p} \UniStrongReflArrow{} \UniWqoB_{(r_1(\vec p),\ldots,r_{v_2}(\vec p))}$.
Then we obtain the following transfer result.

\begin{lemma}
\label{fact:strong-refl-preserv-nwqo-param}
If $\mathcal{Q}_1\UniStrongReflArrow{\vec r} \mathcal{Q}_2$,
$r_1,\ldots,r_{v_2} \in \UniFGHLevel{<\alpha}$
and 
$\mathcal{Q}_2 \UniFGBoundedNwqosIn \UniFGHOneAppLevel{\UniOrdinalA}$, 
then  
$\mathcal{Q}_1 \UniFGBoundedNwqosIn \UniFGHOneAppLevel{\UniOrdinalA}$.
\end{lemma}
\begin{proof}
By Lemma~\ref{fact:strong-refl-preserv-bad-length} we have that
\begin{align*}
\UniLengFunc{\mathcal{Q}_1}{f}(\vec p,x) &= \UniLengFunc{\UniWqoA_{\vec p}}{f}(x)\\ 
&\leq \UniLengFunc{\UniWqoB_{r_1(\vec p),\ldots,r_{v_2}(\vec p)}}{f}(x)\\
&= \UniLengFunc{\mathcal{Q}_2}{f}(r_1(\vec p),\ldots,r_{v_2}(\vec p),x).
\end{align*}
By assumption, there is $x_0 \in \UniNaturalSet$ and $q \in \UniFGHLevel{<\alpha}$ such that for all $(\vec p, x)$ with 
$\max \{ \max_i p_i, \allowbreak x \} \geq x_0$, we have 
\[\UniLengFunc{\UniWqoB_{\vec p}}{f}(x) \leq F_\alpha(q(\max\{ p_1,\ldots,p_{v_2},x\}))
\leq 
F_\alpha(q(\max\{ \max_i r_i(\vec p),x\}))\]
since each $r_i$ is expansive.
The result follows by Lemma~\ref{lem:max-preserve-level} and Lemma~\ref{lem:bounds-in-fgh}.
\end{proof}

We are finally ready to prove length theorems for knotted nwqos.

\begin{theorem}
\label{fact:length-theo-nm}
For all $r,k \in \UniNaturalSetNN$,
${r \cdot}\UniWqoExtModRelProd{m}{n}{k} \UniFGBoundedNwqosIn_\gamma \UniFGHOneAppLevel{\gamma + k+1}$.
Also,
$$\left\{{r \cdot}\UniWqoExtModRelProd{m}{n}{k}\right\}_{(r,k)\in\UniNaturalSet^2} \UniFGBoundedNwqosIn {\UniFGHOneAppLevel{\omega}}.$$
\end{theorem}
\begin{proof}
Let $f$ be a primitive recursive function at $\UniFGHLevel{\gamma}$, with $\gamma \geq 2$ without loss of generality.
For the first part, from Corollary~\ref{coro:disj-sums-mn-refle-nat} and Theorem~\ref{fact:length_theorem_naturalset} we have 
$r \cdot \UniWqoExtModRelProd{m}{n}{k} \UniStrongReflArrow{}\;
(r \cdot \UniFlatWqo{m^k}) \cdot \UniWqoModNatural^k$
and
$(r\cdot m^k) \cdot \UniWqoModNatural^{k}\UniFGBoundedNwqosIn_\gamma \UniFGHOneAppLevel{\gamma + k+1}$,  thus
${r \cdot}\UniWqoExtModRelProd{m}{n}{k} \UniFGBoundedNwqosIn_\gamma \UniFGHOneAppLevel{\gamma + k+1}$ by Corollary~\ref{coro:strong-refl-f-hierar}.

For the second part, let $q_1(r, k) = r \cdot m^k$ and $q_2(r, k) = k$.
Both are expansive functions in $\UniFGHLevel{2}$---recall that $m$ is a constant here.
Hence, the reflection mentioned above implies that
$\left\{{r \cdot}\UniWqoExtModRelProd{m}{n}{k}\right\}_{(r,k)\in\UniNaturalSet^2} \UniStrongReflArrow{\vec q}
\{r \cdot \UniWqoModNatural^k\}_{(r,k)\in\UniNaturalSet^2}$
and since
$\{r \cdot \UniWqoModNatural^k \}_{(r,k)\in\UniNaturalSet^2}\UniFGBoundedNwqosIn
\UniFGHOneAppLevel{\omega}$, the result follows by Corollary~\ref{fact:strong-refl-preserv-nwqo-param}.%
\qedhere
\end{proof}

Similarly, the following length theorems for majoring and minoring extensions of knotted nwqos hold.
    
\begin{theorem}
\label{fact:length-theorem-power-set-wqo-maj-mn}
For every $r, d, k \in \UniNaturalSet$,
 ${\UniPowerMaj{\UniMajoringFntWqo{
{r \cdot} \UniWqoExtModRelProd{m}{n}{k}}}{d}},{\UniPowerMaj{\UniMinoringFntWqo{
{r \cdot} \UniWqoExtModRelProd{m}{n}{k}}}{d}}
\UniFGBoundedNwqosIn
{\UniFGHOneAppLevel{\omega^k}}$.
 Moreover,
 $$\left\{{\UniPowerMaj{\UniMajoringFntWqo{
{r \cdot} \UniWqoExtModRelProd{m}{n}{k}}}{d}}\right\}_{(r,k,d)\in\UniNaturalSet^3},\left\{{\UniPowerMaj{\UniMinoringFntWqo{
{r \cdot} \UniWqoExtModRelProd{m}{n}{k}}}{d}}\right\}_{(r,k,d)\in\UniNaturalSet^3}
\UniFGBoundedNwqosIn
{\UniFGHOneAppLevel{\omega^\omega}}.$$
\end{theorem}
\begin{proof}
Recall that
$\UniPowerMaj{\UniMajoringFntWqo{s \cdot \UniWqoA}}{d} \UniIsomorph \UniPowerMaj{\UniMajoringFntWqo{\UniWqoA}}{d\cdot s}$ by Proposition~\ref{fact:order-isomorph-powerset-extensions}.
From this fact and Lemma~\ref{fact:refl-power-set-ext}, 
$$
\UniPowerMaj{\UniMajoringFntWqo{r\cdot\UniWqoExtModRelProd{m}{n}{k}}}{d}
\UniStrongReflArrow{}
\UniPowerMaj{\UniMajoringFntWqo{\UniWqoExtModRelProd{m}{n}{k}}}{dr} \UniStrongReflArrow{}
\UniPowerMaj{\UniMajoringFntWqo{\UniWqoModNatural^k}}{dr \cdot m^k}
$$
and by Theorem~\ref{fact:length-theorem-power-set-wqo-nat} we know that
$\UniPowerMaj{\UniMinoringFntWqo{\UniWqoModNatural^k}}{dr\cdot m^k}\UniFGBoundedNwqosIn \UniFGHOneAppLevel{\omega^{k}}$.
The first statement (for the majoring nwqos) then holds by  Corollary~\ref{coro:strong-refl-f-hierar}.
The same statement for the minoring nwqos is proved analogously.

Regarding the second statement, let $q_1(r, k, d) := k $ and $q_2(r, k, d) := dr \cdot m^k$.
Thus, by the above reflection, we have
\[\left\{{\UniPowerMaj{\UniMajoringFntWqo{
{r \cdot} \UniWqoExtModRelProd{m}{n}{k}}}{d}}\right\}_{(r,k,d)\in\UniNaturalSet^3} \UniStrongReflArrow{\vec q}
\left\{{\UniPowerMaj{\UniMajoringFntWqo{\UniWqoModNatural^k}}{d}}\right\}_{(k,d)\in\UniNaturalSet^2}
\]
and since
$\left\{{\UniPowerMaj{\UniMajoringFntWqo{\UniWqoModNatural^k}}{d}}\right\}_{(k,d)\in\UniNaturalSet^2} \UniFGBoundedNwqosIn {\UniFGHOneAppLevel{\omega^\omega}}$
by Theorem~\ref{fact:length-theorem-power-set-wqo-nat}, the result follows by Corollary~\ref{fact:strong-refl-preserv-nwqo-param}.
The proof for the minoring ordering is similar.
\qedhere
\end{proof}

We close with a length theorem that we will use  in subsequent sections for lowering hyper-Ackermannian upper bounds to Ackermannian for some of the logics we consider.
In what follows, let $\UniPowerMaj{\UniMajoringFntSingWqo{\UniWqoExtModRelProd{m}{n}{k}}}{d}$
and $\UniPowerMaj{\UniMinoringFntSingWqo{\UniWqoExtModRelProd{m}{n}{k}}}{d}$
be, respectively, like
$\UniPowerMaj{\UniMajoringFntWqo{\UniWqoExtModRelProd{m}{n}{k}}}{d}$
and
$\UniPowerMaj{\UniMinoringFntWqo{\UniWqoExtModRelProd{m}{n}{k}}}{d}$
but with domains consisting only of those $d$-tuples in which all components are empty sets except for one, which must be a singleton containing a $k$-tuple from $\UniNaturalSet^k$.

\begin{theorem} The following holds.
\label{fact:length-theo-nm-not-fixed-maj-min-sing}
$$\left\{\UniPowerMaj{\UniMajoringFntSingWqo{{r \cdot}\UniWqoExtModRelProd{m}{n}{k}}}{d}\right\}_{(r,k,d)\in\UniNaturalSet^3},\left\{\UniPowerMaj{\UniMinoringFntSingWqo{{r \cdot}\UniWqoExtModRelProd{m}{n}{k}}}{d}\right\}_{(r,k,d)\in\UniNaturalSet^3}
\UniFGBoundedNwqosIn 
{\UniFGHOneAppLevel{\omega}}.$$
\end{theorem}
\begin{proof}
We consider first the case $r=1$ and show that
$\UniPowerMaj{\UniMajoringFntSingWqo{\UniWqoExtModRelProd{m}{n}{k}}}{d} \allowbreak\UniStrongReflArrow{\UniReflectionA}
d \cdot \UniWqoExtModRelProd{m}{n}{k}$
and 
$\UniPowerMaj{\UniMinoringFntSingWqo{\UniWqoExtModRelProd{m}{n}{k}}}{d}
\UniStrongReflArrow{\UniReflectionA} d \cdot \UniWqoExtModRelProd{m}{n}{k}
$,
where $\UniReflectionA(\UniTuple{X_1,\ldots,X_d}) \UniSymbDef (i, \vec a)$
and where $i$ is the unique index and $\vec a$ the unique tuple such that $X_i = \{ \vec a \}$.
The reflection property is immediate.
To see that these are strong reflections, let $\vec X \UniSymbDef \UniTuple{\varnothing,\ldots,\varnothing,\{ \vec a \}, \varnothing,\ldots,\varnothing}$
and note that
$\UniNorm{\vec X}{\left(\UniMajoringFntSingWqo{\UniWqoExtModRelProd{m}{n}{k}}\right)^d} = \max \{ 1, \UniNorm{\vec a}{\UniWqoExtModRelProd{m}{n}{k}}\}$.
From this, it follows that
$\UniNorm{\vec a}{\UniWqoExtModRelProd{m}{n}{k}} \leq \UniNorm{\vec X}{\left(\UniMajoringFntSingWqo{\UniWqoExtModRelProd{m}{n}{k}}\right)^d}$, as desired.
The same argument works for 
$\UniPowerMaj{\UniMinoringFntSingWqo{\UniWqoExtModRelProd{m}{n}{k}}}{d}$.
Then use Theorem~\ref{fact:length-theo-nm} and Lemma~\ref{fact:strong-refl-preserv-nwqo-param} as in the above proofs.
The result generalizes to any $r \geq 1$ in view of Proposition~\ref{fact:order-isomorph-powerset-extensions}.
\end{proof}

\subsection{Fast-growing proof-search algorithms}
\label{sec:fast-growing-algo}
In this section, we describe in general terms the two main approaches we will employ to decide provability and deducibility for the logics of interest. 
As we saw in Chapter~\ref{sec:preliminaries}, all these logics have a well-behaved hypersequent calculus, and we will use these calculi to devise decision algorithms for the associated decision problems.

Indeed, considering the provability problem for a hypersequent calculus $\UniHyperCalcA$, we will be interested in determining, given a hypersequent $h$ (of size $\langle h \rangle$),  whether $h$ is provable in $\UniHyperCalcA$, i.e., if the set $P(h)$  of proofs of $h$ in $\UniHyperCalcA$ is non-empty. 
For the deducibility problem, the input consists of a hypersequent $h$ and  a finite set of assumptions $\UniSetFmA$, and the question is whether the set $P(\Gamma, h)$  of deductions of $h$ from $\Gamma$ in $\UniHyperCalcA$ is non-empty. 
We will show in Chapter~\ref{sec:ub-ww} how to uniformly reduce this problem to an instance of provability by updating the calculus $\UniHyperCalcA$ with new rules that incorporate $\UniSetFmA$ in the  derivations. 
        
In what follows, we will write $a$ for the input $h$ in the provability setting and for the pair $(\Gamma, h)$ in the deducibility setting, and say that $a$ is provable (in the deducibility setting) if $h$ is provable from $\Gamma$. 
Also, $\langle a \rangle$ will denote $\langle h \rangle$ or $\langle \Gamma \rangle + \langle h \rangle$, respectively.
        
We will establish that  the required time (as a function of {$\langle a \rangle$}) is eventually upper bounded by a function in a suitable class $\UniFGHOneAppLevel{\alpha}$; this then guarantees that the problem is in the fast-growing complexity class $\UniFGHProbOneAppLevel{\alpha}$. 
Note that for levels of the fast-growing hierarchy that are beyond the second one, the distinction between space and time, as well as between determinism and nondeterminism, vanishes~\cite[Sec. 4.2.1]{schmitz2016hierar}, a fact that simplifies our strategies below. 
Recall that the calculi $\UniHyperCalcA$ we consider will either involve knotted contraction or knotted weakening and we will take a different approach for each of the two cases.
The next remark details how we deal with the case of knotted contraction.

\begin{remark}
\label{r: spacetimeC} 
In case the calculus $\UniHyperCalcA$ has a knotted contraction rule we claim that to guarantee that the required time to check $P(a) \neq \varnothing$ (as a function of $\langle a \rangle$) is upper bounded by a function in  $\UniFGHOneAppLevel{\alpha}$, it suffices to have a finite set $T(a)$ of finite trees for every $a$, such that 

\begin{enumerate}[(a)]
\item if $P(a)$ is non-empty then $P(a) \cap T(a)$ is non-empty 
\item the space (as a function of $\langle a \rangle$)  occupied by each element of $T(a)$ is upper bounded uniformly on $\UniSizeHyper{a}$ by a function in $\UniFGHOneAppLevel{\alpha}$ and
\item given a tree $t$, the time to check whether $t \in P(a)$ is primitive recursive on the size of $t$.
\end{enumerate}

Indeed, checking that $P(a)$ is non-empty is equivalent to checking that $P(a) \cap T(a)$ is non-empty, by (a). 
To check the latter, we first (non-deterministically) guess an element $t$ of $T(a)$ and then check if $t \in P(a)$, a process that by (c) takes primitive recursive time in terms of the space occupied by $t$, whose space is itself uniformly upper bounded by a function in  $\UniFGHOneAppLevel{\alpha}$ by (b); so the total process takes time that is uniformly upper bounded by a function in  $\UniFGHOneAppLevel{\alpha}$.
        
We mention that determining if a particular tree is in $P(a)$ relies on a local check of each label and its children (ensuring that they indeed correspond to applications of the rules of $\UniHyperCalcA$ that label the nodes) and that this takes linear time on the size of the particular tree; so (c) will automatically hold in our applications.        
The question is what is a suitable process $a \mapsto T(a)$ that satisfies (a) and (b). 
We will obtain $T(a)$ as candidates for minimal proofs (where minimality is defined via well-quasi-orders) in a variant of $\UniHyperCalcA$; these trees will satisfy the required space restrictions of (b). 
In order to guarantee (a), we will show that looking for minimal proofs in this variant calculus is enough to decide whether $a$ is provable in the original calculus or not.
         
In more detail,  $T(a)$ will contain trees
labelled with hypersequents. 
We will prove that it is enough that the trees in $T(a)$ have a maximum branching degree determined by $\UniHyperCalcA$ and their branches are controlled bad sequences in a suitable nwqo, the control being in terms of $\UniSizeHyper{a}$. 
Hence there is a bound for their length determined by $\UniSizeHyper{a}$. 
Also, we will show that there is a bound for the size of each hypersequent appearing in such a branch. 
These results will allow us to establish (b) by using length theorems for the nwqo. 
We will further argue that if $a$ is provable in $\UniHyperCalcA$ then there is a proof of it in $T(a)$, thus establishing (a).
\end{remark}

We now provide a fairly general lemma that will give us the required bounds for condition (b) of Remark~\ref{r: spacetimeC}.
In what follows, we define the \emph{size} of a labelled tree to be the sum of the sizes of its labels.

\begin{lemma}\label{lem:bound-trees}
If a labelled tree has branching degree at most $K \in \UniNaturalSet$, the length of each branch is at most $N \in \UniNaturalSet$, and the labels have size at most
$S \in\UniNaturalSet$, then the size of the labelled tree
is at most $K^{N+1} \cdot S$.
Moreover, if there are only finitely many labels of size at most $S$, then the set of such labelled trees is finite.
\end{lemma}
\newcommand{\UniTreeA}{\mathsf{t}}
\begin{proof}
At level $i$ of the tree, there are at most $K^i$ nodes, each of size at most $S$.
The number of levels is at most $N$, since every branch has length at most $N$. Thus the number of nodes in the tree is upper bounded by
\[
\sum_{i=0}^N K^i \leq K^{N+1}
\] 
and the size of the labelled tree is upper bounded by {$K^{N+1} \cdot S$}.

The number of (unlabelled) trees with at most $K^{N+1}$-many nodes is finite, and by assigning a label to each node among the finitely many labels of size $S$ we get a finite set of labelled trees satisfying the conditions in the statement.%
\qedhere
\end{proof}

Let $A$ be a set equipped with a map $\UniSizeHyper{\cdot} : A \to \UniNaturalSet$ and $\mathfrak{L}(a)$ a set of labels for each $a \in A$ also equipped with a map $\UniSizeHyper{\cdot} : \mathfrak{L}(a) \to \UniNaturalSet$ such that there are finitely many labels of the same {\emph{size}, i.e., 
$\UniSizeHyper{\cdot}$-value,} in $\mathfrak{L}(a)$.
For  $K \in \UniNaturalSet$, for $a \in A$, and for functions $L,S : \UniNaturalSet \to \UniNaturalSet$, let $T[\mathfrak{L}(a),K,L,S](\UniSizeHyper{a})$ be the (finite) set of all labeled trees with labels from $\mathfrak{L}(a)$, having branching degree at most $K$, branch length bounded by $L(\UniSizeHyper{a})$ and size of labels bounded by $S(\UniSizeHyper{a})$.

As mentioned above, in our applications, the set $A$ will consist either of all hypersequents $h$ or of all pairs $( \Gamma, h)$, where $\Gamma$ is a finite set of formulas and $h$ is a hypersequent; also the set of labels $\mathfrak{L}(a)$ will be the set of all $\Omega$-hypersequents, where $\Omega$ is the set of subformulas of the formulas appearing in $a$.
In this setting, we obtain the following results.

\begin{corollary}
Let $A$ be a set equipped with a labeling $\mathfrak{L}: A \to \mathfrak{L}[A]$ and a map $\UniSizeHyper{\cdot} : A \cup \mathfrak{L}[A] \to \UniNaturalSet$.
If {$K \in \UniNaturalSet$ and} $L,S : \UniNaturalSet \to \UniNaturalSet$ are functions eventually upper bounded by functions in $\UniFGHOneAppLevel{\alpha}$, then there is a function $f$ in $\UniFGHOneAppLevel{\alpha}$ and $M \in \UniNaturalSet$ such that, for all $a \in A$ with $\UniSizeHyper{a} \geq M$, the size of each tree in $T[\mathfrak{L}(a),K,L,S](\UniSizeHyper{a})$ is upper bounded by $f(\UniSizeHyper{a})$.
\end{corollary}
\begin{proof}
By assumption, there are $p_L,p_S \in \UniFGHLevel{<\alpha}$ such that for some $x_L, x_S \in \UniNaturalSet$, $L(x) \leq F_\alpha(p_L(x))$ for all $x \geq x_L$ and $S(x) \leq F_\alpha(p_S(x))$ for all $x \geq x_S$.
Note also that, for $E(x,y) \UniSymbDef K^{x+1} \cdot y$ and $E'(x) \UniSymbDef E(F_\alpha(p_L(x)),F_\alpha(p_S(x)))$, we have by Lemma~\ref{lem:ev-ub-multvariable} that for some $x_0 \in \UniNaturalSet$ and $p \in \UniFGHLevel{<\alpha}$ it holds that $E'(x) \leq F_\alpha(p(x))$ for all $x \geq x_0$. Observe that $E$ is increasing.

Therefore, by Lemma~\ref{lem:bound-trees}, for all $a \in A$ with $\UniSizeHyper{a} \geq \max\{x_L,x_S,x_0\}$ (this is the required $M$), the maximum size of a tree in $T[\mathfrak{L}(a),K,L,S](\UniSizeHyper{a})$ is at most
\begin{align*}
K^{L(\UniSizeHyper{a})+1} \cdot S(\UniSizeHyper{a}) &= E(L(\UniSizeHyper{a}),S(\UniSizeHyper{a}))\\
&\leq E(F_\alpha(p_L(\UniSizeHyper{a})),F_\alpha(p_S(\UniSizeHyper{a})))\\ 
&= E'(\UniSizeHyper{a})\\
&\leq  F_\alpha(p(\UniSizeHyper{a}))
\end{align*}
and $F_\alpha \circ p \in \UniFGHOneAppLevel{\alpha}$.
Thus take the latter function as $f$ and $M$ as above, and we are done.%
\qedhere
\end{proof}

\begin{lemma}
\label{lem:size-of-minimal-proofs} 
Let $\alpha > 2$ be a limit ordinal and $\mathcal{Q} \UniSymbDef \{\UniWqoA_{x}: x \in \UniNaturalSet\}$ be a parameterized class of nwqos with $\mathcal{Q} \sqsubseteq \UniFGHOneAppLevel{\alpha}$, where the underlying set of $\UniWqoA_{\UniSizeHyper{a}}$ is bijective to $\mathfrak{L}(a)$. 
Moreover, for each $a \in A$, let $T(a)$ be a finite set of finite $\mathfrak{L}(a)$-labelled trees with branching degree at most $K$.
Also, let $f$ be a primitive recursive control function, and let $S \in \UniFGHLevel{<\alpha}$ and
$S' \in \UniFGHOneAppLevel{\alpha}$ 
be increasing binary functions. 
If, for all $a \in A$, every label $g \in \mathfrak{L}(a)$ in the trees of $T(a)$ satisfies 
$\UniSizeHyper{g} \leq S(\UniNorm{g}{\UniWqoA_{\UniSizeHyper{a}}},\UniSizeHyper{a})$ and in every tree of $T(a)$ the branches have the form
\[
g_0, s^0_1,\ldots,s^0_{k_0}, \UniHypersequentB_1, s^1_1,\ldots,s^1_{k_1},
\ldots,
g_p, s^p_1,\ldots,s^p_{k_p},
\]
where $g_0,g_1,\ldots,g_p$ corresponds to a $(f,\UniSizeHyper{a})$-controlled bad sequence over $\UniWqoA_{\UniSizeHyper{a}}$, each $k_j \leq S'(j,\UniSizeHyper{a})$ and $\UniNorm{s^j_i}{\UniWqoA_{\UniSizeHyper{a}}} = \UniNorm{g_j}{\UniWqoA_{\UniSizeHyper{a}}}$, then for some $\tilde{L}, \tilde{S}:\UniNaturalSet\to\UniNaturalSet$ eventually upper bounded by a function in $\UniFGHOneAppLevel{\alpha}$, we have $T(a) \subseteq T[\mathfrak{L}(a),K,\tilde{L}, \tilde{S}](\UniSizeHyper{a})$ for all $a \in A$.
\end{lemma}
\begin{proof}
Let $L' := \UniLengFunc{\UniWqoA_{\UniSizeHyper{a}}}{f}(\UniSizeHyper{a})$, where $f$ is a primitive recursive control function.
By assumption, the branches of the trees in $T(a)$ have length at most 
\[L' + \sum_{i=0}^{L'} S'(i,\UniSizeHyper{a})
\leq
2L' \cdot S'(L',\UniSizeHyper{a}).\]
Let $\tilde{L}(\UniSizeHyper{a}) := 2 \cdot \UniLengFunc{\UniWqoA_{\UniSizeHyper{a}}}{f}(\UniSizeHyper{a}) \cdot S'(\UniLengFunc{\UniWqoA_{\UniSizeHyper{a}}}{f}(\UniSizeHyper{a}),\UniSizeHyper{a})$.
By Lemma~\ref{lem:ev-ub-multvariable} and the assumption that $\mathcal{Q} \sqsubseteq \UniFGHOneAppLevel{\alpha}$, we have that {$\tilde{L}$} is eventually upper bounded by a function in $\UniFGHOneAppLevel{\alpha}$.

Also, each label $g_i$ in the tree has norm $\UniNorm{g}{\UniWqoA_{\UniSizeHyper{a}}}$ at most
$\UniNormLenFunc{\UniWqoA_{\UniSizeHyper{a}}}{f}({\UniSizeHyper{a}})$.
This is also true for each $s^i_j$ as $\UniNorm{s^i_j}{\UniWqoA_{\UniSizeHyper{a}}} = \UniNorm{g_i}{\UniWqoA_{\UniSizeHyper{a}}}$, thus for $\tilde{S}(\UniSizeHyper{a}) \UniSymbDef S(\UniNormLenFunc{\UniWqoA_{\UniSizeHyper{a}}}{f}({\UniSizeHyper{a}}),\UniSizeHyper{a})$, the result follows by Lemmas~\ref{lem:bound-norm-bad-seq} and~\ref{lem:ev-ub-multvariable}. 
Indeed, by Lemma~\ref{lem:bound-norm-bad-seq}, there is a function $U \in \UniFGHOneAppLevel{\alpha}$ that eventually upper bounds 
$\lambda x.\UniNormLenFunc{\UniWqoA_{x}}{f}({x})$,
thus $\tilde{S}(x) \leq S(U(x),x)$ for sufficiently large $x$. 
Then by Lemma~\ref{lem:ev-ub-multvariable}, $\lambda x.S(U(x),x)$ is eventually upper bounded by a function in $\UniFGHOneAppLevel{\alpha}$, thus $\tilde{S}$ is upper bounded by a function in $\UniFGHOneAppLevel{\alpha}$, as desired. 
\end{proof}

Now, in the next remark we detail the strategy for the calculi with knotted weakening rules.

\begin{remark}
\label{r: spacetimeW} 
In case the system contains knotted weakening rules, our strategy will be to explore successively larger and larger sets of provable hypersequents following specific conditions at each stage, starting from the set of axiomatic sequents, and then checking whether $h$ is in one of the sets. 
We will focus on deducibility, since no deduction theorem is available for these logics (so deductions cannot be reduced to theorems), and provability is a particular case of deducibility; hence $a$ will be a pair $(\UniSetFmA,\UniHypersequentA)$ and $A$ will denote the set of all such pairs.
     
In such cases, our first step will be to expand $\UniHyperCalcA$ to a suitable calculus $\UniHyperCalcA_\UniSetFmA$ with rules that mimic the usage of the assumptions in $\UniSetFmA$.
Now, given a calculus $\UniHyperCalcA$, we claim that to guarantee that the required time to check $P(a) \neq \varnothing$ (as a function of $\langle a \rangle$) is upper bounded by a function in  $\UniFGHOneAppLevel{\alpha}$, it is enough to have, for every $a \in A$, a set $\textit{Seq}(a)$ of sequences $(D_0, D_1, \ldots, D_N$) of sets of hypersequents, such that
\begin{enumerate}[(a)]
 \item $P(a)$ is non-empty if and only if there is a sequence in $\textit{Seq}(a)$ containing $h$ in one of its elements and  built uniformly by forward applications of rules of $\UniHyperCalcA_\UniSetFmA$;
 \item the space (as a function of $\langle a \rangle$)  occupied by each element of $\textit{Seq}(a)$ is uniformly upper bounded by a function in  $\UniFGHOneAppLevel{\alpha}$ and
 \item the time to check whether a 
 sequence in $\textit{Seq}(a)$ satisfies the conditions in item (a) is primitive recursive on $\UniSizeHyper{a}$.
 \end{enumerate}

Thus, by item (a), it is enough to guess an element of $\textit{Seq}(a)$, which occupies $\UniFGHOneAppLevel{\alpha}$-space by (b), and check in primitive recursive time whether it meets the conditions in (a), by item (c).
The space to store each element dominates the space complexity of this non-deterministic algorithm; thus it decides in time $\UniFGHOneAppLevel{\alpha}$ whether $a$ is provable or not in the calculus.
\end{remark}

We establish a lemma that will help us to prove item (b) of the above remark whenever needed in the next sections.
As above, let $A$ be a set equipped with a map
$\UniSizeHyper{\cdot} : A \to \UniNaturalSet$
and $\mathfrak{L}(a)$ a set of labels for each $a \in A$ also equipped with a map $\UniSizeHyper{\cdot} : \mathfrak{L}(a) \to \UniNaturalSet$ such that there are finitely many labels of the same size in $\mathfrak{L}(a)$.
For all $a \in A$, let $\textit{Seq}[\mathfrak{L}(a), L,S](\UniSizeHyper{a})$  be the set of all sequences $D_0,\ldots,D_{N}$ of finite subsets of $\mathfrak{L}(a)$, each having size of representation at most $S(\UniSizeHyper{a})$, and where $N \leq L(\UniSizeHyper{a})$. 
Recall that the size of the representation of a subset $D$ of $\mathfrak{L}(a)$, denoted by $\UniSizeHyper{D},$ is the sum of the sizes of its elements.

\begin{lemma}\label{lem:seq-sets-ub} 
Let $A$ be a set equipped with a labeling $\mathfrak{L}: A \to \mathfrak{L}[A]$ and a map $\UniSizeHyper{\cdot} : A \cup \mathfrak{L}[A] \to \UniNaturalSet$, and let $L,S : \UniNaturalSet \to \UniNaturalSet$ be functions eventually upper bounded by functions in $\UniFGHOneAppLevel{\alpha}$.
Then there is a function $f$ in $\UniFGHOneAppLevel{\alpha}$ and $M \in \UniNaturalSet$ such that, for all $a \in A$ with $\UniSizeHyper{a} \geq M$, the size of each sequence in $\textit{Seq}[\mathfrak{L}(a), L,S](\UniSizeHyper{a})$ is upper bounded by $f(\UniSizeHyper{a})$.
\end{lemma}
\begin{proof}
By assumption, there are $p_L,p_S \in \UniFGHLevel{<\alpha}$ such that for some $x_L, x_S \in \UniNaturalSet$, $L(x) \leq F_\alpha(p_L(x))$ for all $x \geq x_L$, and $S(x) \leq F_\alpha(p_S(x))$ for all $x \geq x_S$.
Note also that, for $E(x,y) = (x+1)y$ and $E'(x) = E(F_\alpha(p_L(x)),F_\alpha(p_S(x)))$, we have by Lemma~\ref{lem:ev-ub-multvariable} that, for some $x_0 \in \UniNaturalSet$ and $p \in \UniFGHLevel{<\alpha}$, $E'(x) \leq F_\alpha(p(x))$ for all $x \geq x_0$. 

Take $a \in A$ with
$\UniSizeHyper{a} \geq \max\{x_L,x_S,x_0\}$ (take the latter as $M$).
Then the size of a sequence in $\textit{Seq}[\mathfrak{L}(a), L,S](\UniSizeHyper{a})$ is given by
$\sum_{i=0}^{N} \UniSizeHyper{D_i} \leq
(N+1)\cdot S(\UniSizeHyper{a})
\leq 
(L(\UniSizeHyper{a})+1)\cdot S(\UniSizeHyper{a}) \leq (F_\alpha(p_L(\UniSizeHyper{a}))+1) \cdot (F_\alpha(p_S(\UniSizeHyper{a}))) = E'(\UniSizeHyper{a}) \leq F_\alpha(p(\UniSizeHyper{a}))$.
Thus take $f$ as the latter function $F_\alpha \circ p$ and $M$ as above.
\end{proof}

In the next chapters, we
will employ the above proof-search strategies to analyze the complexity of knotted substructural logics.

%% file: tex/weak-contraction-ub.tex
In this section, we extend the known proof-search procedure and upper-bound argument for $\UniCalcExt{\UniFLeExtLogic{\UniCRule}}{ \UniAxiomSetA}$~\cite{revantha2020,BalLanRam21LICS} (a generalization of Kripke's argument for $\UniFLeExtLogic{\UniCProp}$), where $\UniAxiomSetA$ is a set of acyclic axioms in $\mathcal{P}_3^\flat$, to the context of all extensions 
$\UniCalcExt{\UniFLeExtLogic{\UniWeakCProp{m}{n}}}{ \UniAxiomSetA}$ 
and we obtain complexity bounds for provability of these logics. 
Actually, by using the deduction theorem that holds for these logics (see Section~\ref{sec:ded-theorem-provability}, and also~\cite{gavin2019,galatos2022}), we also obtain the same complexity bounds for the deducibility of these logics, as well. 
Note that all of these logics contain the exchange rule, but in Chapter~\ref{sec:noncom-ub} we will show how to extend the results even to logics where exchange is replaced by the (weaker) generalized versions of it that we saw in Section~\ref{sec:wck-substructural-logics}.

Kripke's argument for decidability of $\UniFLeExtLogic{\UniCProp}$ uses essentially three ingredients.
First, it replaces the existing sequent calculus by an equivalent one,
$\UniHRuleAbsorb{\UniFLeExtSCalc{\UniCProp}}$, that does not contain the cut rule and the contraction rule, and whose logical rules absorb a fixed number of applications of contraction.
Second, it considers an nwqo,  
$\UniStruct{
\UniOmegaSequentsSet{\UniSubfmlaHyperseqSet},
\UniWqoRel{\UniCProp},
\lambda (\UniSequent{\UniMSetFmA}{\UniMSetSucA}). 
\max_{\UniFmA\in\UniMSetFmA}\UniListCountElem{\UniMSetFmA}{\UniFmA}}$,
over $\UniSubfmlaHyperseqSet$-sequents (recall Definition~\ref{omega-hyper}), such that $s_1 \UniWqoRel{\UniCProp} s_2$ iff $s_1$ can be obtained from $s_2$ by successive applications of contraction.
Third, it establishes the existence of 
\emph{$\UniWqoRel{\UniCProp}$-minimal} or 
\emph{$\UniWqoRel{\UniCProp}$-irredundant} proofs: proofs where no  branch contains an increasing pair of sequents, i.e., every branch is a bad sequence over $\UniWqoRel{\UniCProp}$.
The latter ingredient is usually presented as a corollary of the so-called \emph{Curry's Lemma}.

\begin{lemma}[Curry's Lemma]
    If $\UniSequentA$ has a proof in 
    $\UniHRuleAbsorb{\UniFLeExtSCalc{\UniCProp}}$
    of height $k$ and $\UniSequentA' \UniWqoRel{\UniCProp} \UniSequentA$, then $\UniSequentA'$ has a proof in $\UniHRuleAbsorb{\UniFLeExtSCalc{\UniCProp}}$ of height at most $k$.
\end{lemma}

The proof-search procedure then boils down to considering all of the (finitely many) possible minimal cut-free derivations and checking whether one of them constitutes a proof of the target sequent.
The complexity of this procedure is dominated by the lengths of the branches in the underlying search tree, which are controlled by the known length theorem for 
$\UniStruct{
\UniOmegaSequentsSet{\UniSubfmlaHyperseqSet},
\UniWqoRel{\UniCProp},
\lambda (\UniSequent{\UniMSetFmA}{\UniMSetSucA}). 
\max_{\UniFmA\in\UniMSetFmA}\UniListCountElem{\UniMSetFmA}{\UniFmA}}$.

We will follow here the same approach for logics of the form $\UniCalcExt{\UniFLeExtLogic{\UniWeakCProp{m}{n}}}{ \UniAxiomSetA}$.
Actually, we will consider extensions of $\UniFLeExtHCalc{\UniWeakCProp{m}{n}}$ by a finite set $\UniAnaRuleSet$ of hypersequent analytic structural rules (denoted by $\UniHFLecR$).
Since by Section~\ref{s: P3} each $\UniAxiomSetA$ corresponds to such an $\UniAnaRuleSet$, we will get the desired results for each $\UniCalcExt{\UniFLeExtLogic{\UniWeakCProp{m}{n}}}{ \UniAxiomSetA}$.

First, we define a modified hypersequent calculus, $\UniHFLecRAbsorb$, that absorbs {a fixed} finite amount of applications of $\UniWeakCRule{m}{n}$ inside the other rules.
Then we introduce knotted nwqos over hypersequents and establish length theorems for them, via a strong reflection/encoding into the knotted nwqos over natural numbers developed in Chapter~\ref{sec:knotted-wqos}.
Finally, we prove the corresponding version of Curry's Lemma via \emph{height-preserving admissibility} or \emph{hp-admissibility} of knotted contraction, (EC) and (EW), a concept we define below.

A rule schema is \emph{hp-admissible} if whenever the premises of an instance are derivable there is a derivation of the conclusion with no greater height.
 
Then, working in the modified calculus, we search backwards in such a way that the branches of the underlying search tree are bad sequences over the knotted nwqo and this yields termination of the search procedure.
Finally, we apply the length theorems to derive complexity upper bounds for this proof-search algorithm (and thus for provability in the logics
$\UniCalcExt{\UniFLeExtLogic{\UniWeakCProp{m}{n}}}{ \UniAxiomSetA}$).

\section{Formula multiplicity and active component number} 

In this section we introduce two concepts associated with a hypersequent calculus that will allow us to abstract away unimportant details of rule schemas and focus on the relevant aspects in the proof of the desired hp-admissibility result.
In the following, recall that any component in the conclusion of a rule schema that is not a hypersequent-variable is an active component in the conclusion (see Section~\ref{sec:prelims-proof-theory}).

The \emph{active component number} of a rule schema is the number of active components in the conclusion.

The \emph{formula multiplicity} of a rule schema is~$1$ more than the maximum number of elements in the antecedents of the active components in the conclusion.

For example, the conclusion~$\UniHyperMSetA \VL \UniSequent{\UniMSetFmA}{\UniMSetSucA}$ of~(EW) has a single active component and its antecedent is $\UniMSetFmA$. 
Since there is just a single element in this list, the formula multiplicity is~$2$.

The conclusion~$\UniHyperMSetA \VL \UniSequent{\UniMSetFmA,\UniMSetFmB_1,\ldots,\UniMSetFmB_n}{\UniMSetSucA}$ of~$\UniWeakCAnaRule{m}{n}$ has one active component. 
Its antecedent is $\UniMSetFmA,\UniMSetFmB_1,\ldots,\UniMSetFmB_n$. 
Since there are $n+1$ elements in this list, the formula multiplicity is~$n+2$.

For further examples, see Figure~\ref{fig-str-rules-app}.

\begin{figure*}
\footnotesize
\centering
\begin{tabular}{cc}
\AxiomC{$\UniHyperMSetA \VL  \UniSequent{\UniMSetFmA_{1},\UniMSetFmB_{1}}{\UniMSetSucA_{1}}$}
\AxiomC{$\UniHyperMSetA \VL  \UniSequent{\UniMSetFmA_{2},\UniMSetFmB_{2}}{\UniMSetSucA_{2}}$}
\RightLabel{(com)}
\BinaryInfC{$\UniHyperMSetA \VL \UniSequent{\UniMSetFmA_{1},\UniMSetFmB_{2}}{\UniMSetSucA_{1}} \VL \UniSequent{\UniMSetFmA_{2},\UniMSetFmB_{1}}{\UniMSetSucA_{2}}$}
\noLine
\UnaryInfC{$\UniLandone{\UniPropA\imp \UniPropB}\lor \UniLandone{\UniPropB\imp \UniPropA}$}
\noLine
\UnaryInfC{$(3,2)$}
\DisplayProof
&
\AxiomC{$\UniHyperMSetA \VL \UniSequent{\UniMSetFmA, \UniMSetFmB}{}$}
\RightLabel{(wem)}
\UnaryInfC{$\UniHyperMSetA \VL \UniSequent{\UniMSetFmA}{} \VL \UniSequent{\UniMSetFmB}{}$}
\noLine
\UnaryInfC{$\UniLandone{\UniPropA\imp 0}\lor \UniLandone{(\UniPropA\imp 0)\imp 0}$}
\noLine
\UnaryInfC{$(2,2)$}
\DisplayProof
\\[3.5em]
\AxiomC{$\UniHyperMSetA \VL 
\UniSequent{\UniMSetFmA_{i}, \UniMSetFmA_{j}}{ \UniMSetSucA_{i}} (0\leq i,j\leq k; i\neq j)$}
\RightLabel{($Bwk$)}
\UnaryInfC{$\UniHyperMSetA \VL \UniSequent{\UniMSetFmA_{0}}{\UniMSetSucA_{0}} \VL \ldots \VL \UniSequent{\UniMSetFmA_{k}}{\UniMSetSucA_{k}}$}
\noLine
\UnaryInfC{$\lor_{i=0}^{k}\UniLandone{\UniPropA_{i}\imp (\lor_{j\neq i} \, \UniPropA_{j})}$}
\noLine
\UnaryInfC{$(2,k+1)$}
\DisplayProof
&
\AxiomC{$\UniHyperMSetA \VL 
\UniSequent{\UniMSetFmB,\UniMSetFmA_{1}}{\UniMSetSucA}$}
\AxiomC{$\UniHyperMSetA \VL 
\UniSequent{\UniMSetFmB,\UniMSetFmA_{2}}{\UniMSetSucA}$}
\RightLabel{(mingle)}
\BinaryInfC{$\UniHyperMSetA \VL 
\UniSequent{\UniMSetFmB,\UniMSetFmA_{1},\UniMSetFmA_{2}}{\UniMSetSucA}$}
\noLine
\UnaryInfC{$\UniPropA\cdot \UniPropA\imp \UniPropA$}
\noLine
\UnaryInfC{$(4,1)$}
\DisplayProof\\[1em]
\end{tabular}
%%%%%%%%%%
\\[1.2em]
\AxiomC{$\{\UniHyperMSetA\VL \UniSequent{\UniMSetFmA, \UniMSetFmB_{1}^{r_1}, \ldots, \UniMSetFmB_{n}^{r_n},\UniMSetFmC}{\UniMSetSucA}\}_{\sum r_i = m}$}
\RightLabel{$\UniWeakKAnaRule{m}{n}$}
\UnaryInfC{$\UniHyperMSetA \VL \UniSequent{\UniMSetFmA, \UniMSetFmB_1, \ldots, \UniMSetFmB_n, \UniMSetFmC}{\UniMSetSucA}$}
\noLine
\UnaryInfC{$\UniPropA^{n}\imp \UniPropA^{m}\text{ ($n,m\geq 0$)}$}
\noLine
\UnaryInfC{$(n+2,1)$}
\DisplayProof{}\\[1.2em]
\AxiomC{$\UniHyperMSetA \VL 
\UniSequent{\UniMSetFmA_{i}, \UniMSetFmA_{j}}{\UniMSetSucA_{i}} (0\leq i\leq k-1; i+1\leq j\leq k)$}
\RightLabel{($Bck$)}
\UnaryInfC{$\UniHyperMSetA \VL 
\UniSequent{\UniMSetFmA_{0}}{\UniMSetSucA_{0}} \VL
\ldots \VL 
\UniSequent{\UniMSetFmA_{k-1}}{\UniMSetSucA_{k-1}}
\VL \UniSequent{\UniMSetFmA_{k}}{}$}
\noLine
\UnaryInfC{$\UniLandone{\UniPropA_{0}}\lor \UniLandone{\UniPropA_{0}\imp \UniPropA_{1}}\lor \ldots \lor \UniLandone{(\UniPropA_{0}\land\ldots \land \UniPropA_{k-1})\imp \UniPropA_{k}}$}
\noLine
\UnaryInfC{$(2,k+1)$}
\DisplayProof
\caption{Some hypersequent analytic structural rule schemas in presence of exchange, including the rules $\UniWeakKAnaRule{m}{n}$ that realize the knotted extensions. 
Below each rule is the corresponding axiom and below that is its pair $(\text{formula multiplicity},\text{active component number})$. 
Note: $\UniLandone{\UniFmA}$ is notation for~$(\UniFmA\land 1)$.
}
\label{fig-str-rules-app}
\end{figure*}

The \emph{formula multiplicity of the calculus $\UniHFLecR$}, where $\UniAnaRuleSet$ is a finite set of hypersequent analytic structural rules, is the maximum of the formula multiplicities of its rule schemas.
Its \emph{active component number} is the maximum of the active component numbers of its rule schemas.

For instance, the formula multiplicity of
$\UniFLeExtHCalc{\UniWeakCProp{m}{n}}$ is~$\max\{4,n+2\}$ and its active component number is~$1$. 
Keeping track of these numbers is important, as we will present a calculus where applications of contraction, up to a value defined with such numbers, appearing directly below logical rules are absorbed inside the logical rules themselves; other applications of contraction, in excess of this value, will not be absorbed, but rather permuted upward above the logical rules.

\section{The refined calculus $\UniHFLecRAbsorb$} 

In order to facilitate the formulation of the modified/refined calculus and the proof that it satisfies the desired height-preserving admissibility, we define the binary relations
$\UniIntCtrRel{k}{\UniWeakCtr}$,
$\UniExtCtrRel{k}{\text{(EC)}}$
and
$\UniIntExtCtrRel{k}{l}$ on hypersequents (for each $k, l \geq 0$) such that:

\begin{itemize}
\item $\UniHypersequentA_1 \UniIntCtrRel{k}{\UniWeakCtr} \UniHypersequentA_{2}$ if, and only if, $\UniHypersequentA_{2}$ can be obtained from~$\UniHypersequentA_1$ by applying some number of instances of $\UniWeakCtr$ such that every formula in a component in~$\UniHypersequentA_{2}$ occurs   up to~$k$ times less in that component than in the corresponding component in~$\UniHypersequentA_1$.

\item $\UniHypersequentA_1 \UniExtCtrRel{l}{\text{(EC)}} \UniHypersequentA_{2}$ if, and only if, $\UniHypersequentA_{2}$ can be obtained from~$\UniHypersequentA_1$ by applying some number of instances of~$\text{(EC)}$ such that every component of $\UniHypersequentA_{2}$ occurs up to~$l$ times less in~$\UniHypersequentA_{2}$ than in~$\UniHypersequentA_1$.

\item $\UniHypersequentA_1\UniIntExtCtrRel{k}{l}\UniHypersequentA_{2}$ if, and only if, there exists $\UniHypersequentA'$ such that~$\UniHypersequentA_1 \UniIntCtrRel{k}{\UniWeakCtr} \UniHypersequentA'$ and $\UniHypersequentA' \UniExtCtrRel{l}{\text{(EC)}} \UniHypersequentA_{2}$.
\end{itemize}

\begin{example}
\label{ex:ctr-relation-deriv-app}
Consider $\UniWeakCRule{5}{3}$ and the hypersequents that follow:
\begin{center}
\AxiomC{$\UniHyperMSetA \VL \UniSequent{\UniMSetFmA,\UniFmA^5}{\UniFmC}$}
\RightLabel{$\UniWeakCRule{5}{3}$}
\UnaryInfC{$\UniHyperMSetA \VL \UniSequent{\UniMSetFmA,\UniFmA^3}{\UniFmC}$}
\DisplayProof
\end{center}
\begin{align*}
& \UniHypersequentA_{0}\,:=\,	
\UniSequent{\UniPropA^7,\UniPropB^2}{\UniPropC} \VL 
\UniSequent{\UniPropA^{5}, \UniPropB}{\UniPropC} \VL 
\UniSequent{\UniPropA^3,\UniPropB}{\UniPropC} \VL 
\UniSequent{}{\UniPropD}\\
& \UniHypersequentA_{1}\,:=\,	
\UniSequent{\UniPropA^5,\UniPropB^2}{\UniPropC} \VL 
\UniSequent{\UniPropA^{3},\UniPropB}{\UniPropC} \VL 
\UniSequent{\UniPropA^3,\UniPropB}{\UniPropC} \VL 
\UniSequent{}{\UniPropD}\\
& \UniHypersequentA_{2}\,:=\,	
\UniSequent{\UniPropA^3,\UniPropB}{\UniPropC} \VL 
\UniSequent{\UniPropA^3,\UniPropB}{\UniPropC} \VL 
\UniSequent{\UniPropA^3,\UniPropB}{\UniPropC} \VL 
\UniSequent{}{\UniPropD}\\
& \UniHypersequentA_{3}\,:=\,	
\UniSequent{\UniPropA^3,\UniPropB}{\UniPropC} \VL 
\UniSequent{}{\UniPropD}\\
& \UniHypersequentA_{4}\,:=\,	
\UniSequent{\UniPropA^4,\UniPropB}{\UniPropC} \VL 
\UniSequent{\UniPropA^5,\UniPropB}{\UniPropC} \VL 
\UniSequent{\UniPropA^3,\UniPropB}{\UniPropC} \VL 
\UniSequent{}{\UniPropD}\\
& \UniHypersequentA_{5}\,:=\,	
\UniSequent{\phantom{a}}{\UniPropD}\\
& \UniHypersequentA_{6}\,:=\,	
\UniSequent{\UniPropA,  \UniPropB}{\UniPropC} \VL 
\UniSequent{\UniPropA^3,\UniPropB}{\UniPropC} \VL 
\UniSequent{\UniPropA^3,\UniPropB}{\UniPropC} \VL 
\UniSequent{}{\UniPropD}
\end{align*}

\noindent Then~$\UniHypersequentA_{0}\UniIntCtrRel{2}{\UniWeakCRule{5}{3}} \UniHypersequentA_{1}$.
Also, $\UniHypersequentA_{0} \UniIntCtrRel{4}{\UniWeakCRule{5}{3}} \UniHypersequentA_{2}$ and $\UniHypersequentA_{2} \UniExtCtrRel{2}{\text{(EC)}} \UniHypersequentA_{3}$. Thus $\UniHypersequentA_{0}\UniIntExtCtrRel{4}{2}\UniHypersequentA_{3}$.
However, $\UniHypersequentA_{0}\not\UniIntCtrRel{4}{\UniWeakCRule{5}{3}} \UniHypersequentA_{4}$ since $\UniHypersequentA_4$ \textit{cannot} be obtained from $\UniHypersequentA_0$ by~$\UniWeakCRule{5}{3}$,
and
$\UniHypersequentA_{3}\not\UniExtCtrRel{2}{\text{(EC)}} \UniHypersequentA_{5}$ since~$\UniHypersequentA_{5}$ \textit{cannot} be obtained from~$\UniHypersequentA_{3}$ by~(EC). 

Also, $\UniHypersequentA_{2}\not\UniExtCtrRel{1}{\text{(EC)}} \UniHypersequentA_{3}$ since~$\UniHypersequentA_{3}$ has two 
fewer~$\UniSequent{\UniPropA^3,\UniPropB}{\UniPropC}$ than~$\UniHypersequentA_{2}$ (one fewer would have been fine). 

Finally, $\UniHypersequentA_{2}\not\UniIntCtrRel{2}{\UniWeakCRule{5}{3} } \UniHypersequentA_{6}$ since at least five copies of $\UniPropA$ are needed in the antecedent to apply $\UniWeakCRule{5}{3}$.
\end{example}

The following is a fact that comes easily from the definition of $\UniIntExtCtrRel{k}{l}$:

\begin{proposition}
\label{prop:descendands-in-c}
    If
    $\UniHypersequentA_1\UniIntExtCtrRel{k}{l}
    \UniHypersequentA_2$,
    then, for each component
    $\UniSequentA_1 \in \UniHypersequentA_1$,
    there is
    $\UniSequentA_2 \in \UniHypersequentA_2$
    such that
    $\UniSequentA_1 \UniIntCtrRel{k}{\UniWeakCtr} \UniSequentA_{2}$.
    We call $\UniSequentA_2$ a \emph{descendant} of $\UniSequentA_1$
    in $\UniHypersequentA_2$.
\end{proposition}
\begin{proof}
By definition,
$\UniHypersequentA_1\UniIntExtCtrRel{k}{l}
    \UniHypersequentA_2$
means that there exists
$\UniHypersequentA'$ such that~$\UniHypersequentA_1 \UniIntCtrRel{k}{\UniWeakCtr} \UniHypersequentA'$ and $\UniHypersequentA' \UniExtCtrRel{l}{\text{(EC)}} \UniHypersequentA_{2}$.
We fix a $\UniSequentA_1 \in \UniHypersequentA_1$ and note that it is enough to show that there is a descendant of $\UniSequentA_1$ in $\UniHypersequentA'$, since $\UniEC$ does not make any component disappear in derivations (it only reduces multiplicities).
We pick a derivation witnessing
$\UniHypersequentA_1 \UniIntCtrRel{k}{\UniWeakCtr} \UniHypersequentA'$ and that contains only applications of $\UniWeakCtr$.
Looking at it top-down, a component at node $i$ has an \emph{immediate descendant} at node $i+1$, which is either the same component (in case of being in the instantiation of the hypersequent variable) or a component coming from it via an application $\UniWeakCtr$ (that is, the component is active in the rule application).
In any case, a component and its immediate descendant are related under $\UniIntCtrRel{k}{\UniWeakCtr}$.
Starting from $s_1$, then, we can get a trace of descendants ending in a component
$\UniSequentA' \in \UniHypersequentA_1$ and the transitivity of $\UniIntCtrRel{k}{\UniWeakCtr}$ guarantees that $\UniSequentA_1 \UniIntCtrRel{k}{\UniWeakCtr} \UniSequentA'$.
\end{proof}

\begin{definition} 
\label{def:acn-fm}
Let $\UniFmlaMultFixed$ denote the formula multiplicity of $\UniHFLecR$, where $\UniAnaRuleSet$ is a finite set of hypersequent analytic structural rules, and let $\UniActiveCompFixed$ denote its active component number. 
\end{definition}

The refined calculus $\UniHFLecRAbsorb$ is obtained by removing the cut rule and replacing each rule schema~$\UniRuleSchemaA$ in $\UniHFLecR$ with the rule schema
$\UniHRuleAbsorb{\UniRuleSchemaA}$ displayed below, where $g$ ranges over all hypersequents with $\UniHypersequentA_{0} \UniIntExtCtrRel{\UniAbsCtrValue}{\UniActiveCompFixed} \UniHypersequentB$:
\begin{center}
\begin{tabular}{c@{\hspace{1cm}}c}
\AxiomC{$\UniHypersequentA_{1}\quad\cdots\quad \UniHypersequentA_{n}$}
\RightLabel{$\UniRuleSchemaA$}
\UnaryInfC{$\UniHypersequentA_{0}$}
\DisplayProof
&
\AxiomC{$\UniHypersequentA_{1}\quad\ldots\quad \UniHypersequentA_{n}$}
\RightLabel{$\UniHRuleAbsorb{\UniRuleSchemaA}$}
\UnaryInfC{$\UniHypersequentB$}
\DisplayProof
\end{tabular}
\end{center}
The choice of the value $\UniAbsCtrValue$ will be justified later, inside the proof of the result.

The \textit{base instance} of~$\UniHRuleAbsorb{\UniRuleSchemaA}$ is the particular instance of~$\UniHRuleAbsorb{\UniRuleSchemaA}$ whose conclusion is~$\UniHypersequentA_{0}$, i.e., no knotted contractions are applied.
Clearly, the conclusion~$\UniHypersequentB$ of a non-base instance is obtained from the conclusion~$\UniHypersequentA_{0}$ of the base instance by performing a restricted number of applications of $\UniWeakCtr$ (via $\UniIntCtrRel{\UniAbsCtrValue}{\UniWeakCtr}$) and then a restricted number of applications of (EC) (via $\UniExtCtrRel{\UniActiveCompFixed}{\text{(EC)}}$).

\begin{lemma}\label{lem-equiv-cal}
If $\UniAnaRuleSet$ is a finite set of hypersequent analytic structural rules, 
the calculi
$\UniHFLecR$ and~$\UniHFLecRAbsorb$ derive the same hypersequents.
\end{lemma}
\begin{proof}
First of all, recall that $\UniHFLecR$ satisfies cut elimination, so removing $\UniCutRule$ from it is harmless for provability.
Second, note that every instance of a rule schema~$\UniHRuleAbsorb{\UniRuleSchemaA}\in \UniHFLecRAbsorb$ can be simulated in~$\UniHFLecR$ by a rule instance of~$\UniRuleSchemaA$ followed by some applications of~$\UniWeakCtr$ and~(EC).
Conversely, every instance of a rule schema~$\UniRuleSchemaA$ of $\UniHFLecR$ can be simulated in~$\UniHFLecRAbsorb$ by a base instance of~$\UniHRuleAbsorb{\UniRuleSchemaA}$.
\end{proof}

\section{Height-preserving admissibility of $\UniWeakCRule{m}{n}$}\label{sec-internalising}

We now have all the notions needed for the height-preserving admissibility result.
We stress that height-preserving admissibility is proved for the rule $\UniWeakCtr$ rather than for its linearized version $\UniWeakCAnaRule{m}{n}$, which is still retained fully in the system (since its presence is necessary for making sure that the calculus is cut free). In the case of $\m {FL_{ec}}$ the corresponding refined system does not contain contraction, because the linearized version of contraction is equal to contraction; therefore, our case is more complex than $\m {FL_{ec}}$ in how to handle height-preserving admissibility.

\begin{lemma}\label{lem-hp-admissible}
If $\UniAnaRuleSet$ is a finite set of hypersequent analytic structural rules, then \textnormal{(EW)}, $\UniWeakCtr$ and \textnormal{(EC)} are hp-admissible in~$\UniHFLecRAbsorb$.
\end{lemma}

\begin{proof}
We prove hp-admissibility for each of the mentioned rules.

The rule \textbf{(EW)}: Given a derivation~$\UniDerivationA$ in~$\UniHFLecRAbsorb$ of~$\UniHypersequentA$, we will obtain a derivation of~$\UniHypersequentA \VL \UniSequent{\UniMSetFmA}{\UniMSetSucA}$ of no greater height. 
We proceed by induction on the height of~$\UniDerivationA$. 

\textit{Base case}: $\UniHypersequentA$ is an instance of an initial hypersequent, hence so is~$\UniHypersequentA \VL \UniSequent{\UniMSetFmA}{\UniMSetSucA}$, since it is enough to include $\UniSequent{\UniMSetFmA}{\UniMSetSucA}$ in the instantiation of the hypersequent variable (which always occurs in axiom instances, and only once; see Section~\ref{s: FEPvarieties}).

\textit{Inductive step}: Let~$\UniHRuleAbsorb{\UniRuleSchemaA}$ be the last rule instance applied in~$\UniDerivationA$ and assume the premises are~$\UniHypersequentB \VL \UniHypersequentB_{1}$ and~$\UniHypersequentB \VL \UniHypersequentB_{2}$ and the conclusion is~$\UniHypersequentB|\UniHypersequentB_{0}$, where~$\UniHypersequentB$ is the hypersequent that instantiates the hypersequent-variable of the rule (every rule schema in~$\UniHFLecRAbsorb$ has exactly one hypersequent-variable); here we  illustrate the argument for rules with two premises as the general case is analogous.
By the induction hypothesis (IH) we obtain derivations for~$\UniHypersequentB \VL \UniSequent{\UniMSetFmA}{\UniMSetSucA} \VL \UniHypersequentB_{1}$ and~$\UniHypersequentB \VL \UniSequent{\UniMSetFmA}{\UniMSetSucA} \VL \UniHypersequentB_{2}$ of height less than the height of~$\UniDerivationA$. 
Applying~$\UniHRuleAbsorb{\UniRuleSchemaA}$ to these hypersequents with the multiset-variable now instantiated by~$\UniHypersequentB \VL \UniSequent{\UniMSetFmA}{\UniMSetSucA}$, we obtain~$\UniHypersequentB \VL \UniSequent{\UniMSetFmA}{\UniMSetSucA} \VL \UniHypersequentB_{0}$. 
This is a derivation of $\UniHypersequentA  \VL \UniSequent{\UniMSetFmA}{\UniMSetSucA}$ whose height is no greater than that of~$\UniDerivationA$.

The rule {$\UniWeakCtr$:} Given a derivation~$\UniDerivationA$ of $\UniHypersequentA \VL \UniFmA^m, \UniSequent{\UniMSetFmA}{\UniMSetSucA}$, we will obtain a derivation of $\UniHypersequentA \VL \UniSequent{\UniFmA^n,\UniMSetFmA}{\UniMSetSucA}$ of no greater height. 
We proceed again by induction on the height of~$\UniDerivationA$. 

\textit{Base case}: If $\UniHypersequentA \VL \UniSequent{\UniFmA^m,\UniMSetFmA}{\UniMSetSucA}$ is an instance of an initial hypersequent, then $h$ contains an instance of the active component of the initial hypersequent (see the calculus in Figure~\ref{figure-HFLec}), hence 
$\UniHypersequentA \VL \UniSequent{\UniFmA^n,\UniMSetFmA}{\UniMSetSucA}$ is also an instance of an initial hypersequent. 

\textit{Inductive case}:
For what follows, recall Definition~\ref{def:acn-fm}.
If~$\UniHRuleAbsorb{\UniRuleSchemaA}$ is the last rule applied in~$\UniDerivationA$, then there must be a base instance~$\UniHypersequentB_{0}$ of~$\UniHRuleAbsorb{\UniRuleSchemaA}$ and some~$\UniHypersequentB_{1}$ such that
\[
\UniHypersequentB_{0} 
\UniIntCtrRel{\UniAbsCtrValue}{\UniWeakCtr} 
\UniHypersequentB_{1} 
\UniExtCtrRel{\UniActiveCompFixed}{\text{(EC)}} 
\UniHypersequentA \VL \UniSequent{\UniFmA^m,\UniMSetFmA}{\UniMSetSucA}
\]

\noindent
If $k$ is the number of copies of~$\UniFmA$ in~$\UniMSetFmA$, then  $k\geq 0$ and $\UniMSetFmA=\UniFmA^{k},\UniFmMSetMinusFm{\UniMSetFmA}$ for some $\UniFmMSetMinusFm{\UniMSetFmA}$ with $\UniFmA\not\in \UniFmMSetMinusFm{\UniMSetFmA}$.
Hence, $\UniHypersequentA \VL \UniSequent{\UniFmA^m,\UniMSetFmA}{\UniMSetSucA}$ is equal to $\UniHypersequentA \VL \UniSequent{\UniFmA^{k+m},\UniFmMSetMinusFm{\UniMSetFmA}}{\UniMSetSucA}$.
Also, if~$l$ is the number components in ~$\UniHypersequentA$ equal to $\UniSequent{\UniFmA^m,\UniMSetFmA}{\UniMSetSucA}$ (i.e. $\UniSequent{\UniFmA^{k+m},\UniFmMSetMinusFm{\UniMSetFmA}}{\UniMSetSucA}$), where $l\geq 0$, then there is a hypersequent $\UniFmMSetMinusFm{\UniHypersequentA}$ that does not contain $\UniSequent{\UniFmA^{k+m},\UniFmMSetMinusFm{\UniMSetFmA}}{\UniMSetSucA}$ such that $\UniHypersequentA$ is equal to $\UniFmMSetMinusFm{\UniHypersequentA}$  together with $l$ components equal to $\UniSequent{\UniFmA^{k+m},\UniFmMSetMinusFm{\UniMSetFmA}}{\UniMSetSucA}$. 
Thus
$\UniHypersequentA \VL \UniSequent{\UniFmA^m,\UniMSetFmA}{\UniMSetSucA}$ can be
rewritten as
\[
\underbrace{\UniFmMSetMinusFm{\UniHypersequentA} 
\VL 
\overbrace{\UniSequent{\UniFmA^{k+m},\UniFmMSetMinusFm{\UniMSetFmA}}{\UniMSetSucA} 
\VL 
\ldots 
\VL 
\UniSequent{\UniFmA^{k+m},\UniFmMSetMinusFm{\UniMSetFmA}}{\UniMSetSucA}}^{\text{$l$ components}}}_{\text{$\UniHypersequentA$}} 
\VL 
\UniSequent{\UniFmA^{k+m},\UniFmMSetMinusFm{\UniMSetFmA}}{\UniMSetSucA}
\]

Since $\UniHypersequentB_{1} \UniExtCtrRel{\UniActiveCompFixed}{\text{(EC)}}  \UniHypersequentA \VL \UniSequent{\UniFmA^{k+m},\UniFmMSetMinusFm{\UniMSetFmA}}{\UniMSetSucA}$, 
we can partition~$\UniHypersequentB_{1}$ into the portion~$\UniHypersequentB_{1}^{-}$ that externally contracts to~$\UniFmMSetMinusFm{\UniHypersequentA}$ (i.e. $\UniHypersequentB_{1}^{-}\UniExtCtrRel{\UniActiveCompFixed}{\text{(EC)}} \UniFmMSetMinusFm{\UniHypersequentA}$), and the portion that externally contracts to the remainder. 
Specifically, there exists~$\alpha \geq 1$ with $(l + \alpha)-(l+1) \leq \UniActiveCompFixed$, or, more simply, ~$1\leq\alpha\leq\UniActiveCompFixed+1$, such that
\[
\UniHypersequentB_{1} = 
\UniHypersequentB_{1}^{-} \VL 
\overbrace{\UniSequent{\UniFmA^{k+m},\UniFmMSetMinusFm{\UniMSetFmA}}{\UniMSetSucA} 
\VL 
\ldots
\VL 
\UniSequent{\UniFmA^{k+m},\UniFmMSetMinusFm{\UniMSetFmA}}{\UniMSetSucA}}^{\text{$l+\alpha$ components}}
\]

Moreover, for each one of these $l+\alpha$ components, say the $i^{\text{th}}$ component, there is a~$\beta_{i}$ with~$m\leq \beta_{i}\leq m+\UniAbsCtrValue$ and a multiset~$\UniMSetFmA_{i}$ with no occurrences of $\UniFmA$ such that
\[
\UniHypersequentB_{0} = \UniHypersequentB_{0}^{-} \VL \overbrace{
\UniSequent{\UniFmA^{k+\beta_{1}},\UniMSetFmA_{1}}{\UniMSetSucA} 
\VL 
\ldots 
\VL 
\UniSequent{\UniFmA^{k+\beta_{l+\alpha}},\UniMSetFmA_{l+\alpha}}{\UniMSetSucA}
}^{\text{$l+\alpha$ components}}
\]
for some hypersequent $\UniHypersequentB_{0}^{-}$ with 
$\UniHypersequentB_{0}^{-} 
\UniIntCtrRel{\UniAbsCtrValue}{\UniWeakCtr} 
\UniHypersequentB_{1}^{-}$ 
and
\begin{multline}
\label{eq:betas-to-ms}
\overbrace{
\UniSequent{\UniFmA^{k+\beta_{1}},\UniMSetFmA_{1}}{\UniMSetSucA}
\VL
\ldots
\VL
\UniSequent{\UniFmA^{k+\beta_{l+\alpha}},\UniMSetFmA_{l+\alpha}}{\UniMSetSucA}}^{\text{$l+\alpha$ components}}
\quad
\UniIntCtrRel{\UniAbsCtrValue}{\UniWeakCtr}\\
\underbrace{
\UniSequent{\UniFmA^{k+m},\UniFmMSetMinusFm{\UniMSetFmA}}{\UniMSetSucA}
\VL
\ldots
\VL
\UniSequent{\UniFmA^{k+m},\UniFmMSetMinusFm{\UniMSetFmA}}{\UniMSetSucA}}_{\text{$l+\alpha$ components}}
\end{multline}

The scheme below summarizes the above observations:

\begin{figure}[H]
    \centering
    \begin{tikzpicture}
        \node[] (g0) at (0,6.0) {$\UniHypersequentB_{0} \quad = \quad  \UniHypersequentB_{0}^{-} \VL \overbrace{
\UniSequent{\UniFmA^{k+\beta_{1}},\UniMSetFmA_{1}}{\UniMSetSucA} 
\VL 
\ldots 
\VL 
\UniSequent{\UniFmA^{k+\beta_{l+\alpha}},\UniMSetFmA_{l+\alpha}}{\UniMSetSucA}
}^{\text{$l+\alpha$ components}}$};
        \node[] (g1) at (-.3,3) {$\UniHypersequentB_{1} \quad = \quad  
\UniHypersequentB_{1}^{-} \VL 
\overbrace{\UniSequent{\UniFmA^{k+m},\UniFmMSetMinusFm{\UniMSetFmA}}{\UniMSetSucA} 
\VL 
\ldots
\VL 
\UniSequent{\UniFmA^{k+m},\UniFmMSetMinusFm{\UniMSetFmA}}{\UniMSetSucA}}^{\text{$l+\alpha$ components}}$};
        \node[] (H) at (.1,.2) {$h \VL \UniSequent{\UniFmA^m,\UniMSetFmA}{\UniMSetSucA} \quad = \quad 
        {\UniFmMSetMinusFm{\UniHypersequentA} 
\VL 
\overbrace{\UniSequent{\UniFmA^{k+m},\UniFmMSetMinusFm{\UniMSetFmA}}{\UniMSetSucA} 
\VL 
\ldots 
\VL 
\UniSequent{\UniFmA^{k+m},\UniFmMSetMinusFm{\UniMSetFmA}}{\UniMSetSucA}}^{\text{$l$ components}}}
\VL 
\UniSequent{\UniFmA^{k+m},\UniFmMSetMinusFm{\UniMSetFmA}}{\UniMSetSucA}$};
        %%% arrows
        \node[] at (1.3, 1.5) {$\UniVertSquig{\UniActiveCompFixed}{\UniEC}$};
        \node[] at (-2.5, 1.5) {$\UniVertSquig{\UniActiveCompFixed}{\UniEC}$};
        \node[] at (1.8, 4.5) {$\UniVertSquig{\UniAbsCtrValue}{\UniWeakCtr}$};
        \node[] at (-1.9, 4.5) {$\UniVertSquig{\UniAbsCtrValue}{\UniWeakCtr}$};
    \end{tikzpicture}
\end{figure}

Our goal now is to show that the multiplicities $k+\beta_i$ of $\UniFmA$ in the $\alpha$ last components in $g_0$ (displayed above) can be reduced to $k + n$ without increasing the height of the derivation.
Without loss of generality, in the list $\beta_{l+1},\ldots,\beta_{l+\alpha}$ we can assume that there is an initial segment $\beta_{l+1},\ldots,\beta_{l+N}$ (possibly empty) such that each of these values is in between $1+\UniAbsCtrValue$ and $m+\UniAbsCtrValue$, and that the rest in this list take values strictly less than~$(\UniFmlaMultFixed-1)(m-1)+1$. 

By the definition of formula multiplicity, the number of items in the antecedent of every component in the conclusion of the rule schema $\UniRuleSchemaA$ has at most~$\UniFmlaMultFixed-1$ elements.
Therefore, for every~$i$ such that $l+1\leq i\leq l+N$, either the component~$\UniSequent{\UniFmA^{k+\beta_i},\UniMSetFmA_{i}}{\UniMSetSucA}$ is in the $\UniRuleSchemaA$-instantiation of the hypersequent-variable, or at least~$m$ copies of~$\UniFmA$ occur in the $\UniRuleSchemaA$-instantiation of some multiset-variable $\Theta_i$ from that component.
Indeed, assume that the component of the conclusion of $\UniRuleSchemaA$ that instantiates to $\UniSequent{\UniFmA^{k+\beta_i},\UniFmMSetMinusFm{\UniMSetFmA}}{\UniMSetSucA}$ has the form $\Delta_1,\ldots,\Delta_{\rho_1},\UniFmB_1,\ldots,\UniFmB_{\rho_2}  \UniSequent{}{\UniMSetSucA}$ and that there are less than $m$ copies of $\UniFmA$ in the instantiation of each $\Delta_j$, i.e., $|\Delta_j|_{\UniFmA} \leq m-1$. 
Since each metavariable is instantiated with either a single formula or a multiset of formulas, by focusing on the number of copies of $\UniFmA$, we get that $k + \beta_i = \sum_j |\Delta_j|_{\UniFmA} + \sum_l [\UniFmB_l \mapsto \UniFmA]$,
where $[\UniFmB_l \mapsto \UniFmA]$ evaluates
to $1$ if $\UniFmB_l$ instantiates to $\UniFmA$
and $0$ otherwise. 
By the definition of $\UniFmlaMultFixed$, we have $\rho_1 + \rho_2\leq \UniFmlaMultFixed-1$, so using
$m-1 > 0$ we obtain $k + \beta_i \leq \rho_1 \cdot (m-1) + \rho_2 \leq \rho_1 (m-1) + \rho_2 (m-1) = (\rho_1 + \rho_2) (m-1) \leq (\UniFmlaMultFixed-1)(m-1)$, which contradicts the assumption 
$1 + (\UniFmlaMultFixed-1)(m-1)\leq \beta_i$.
  
In either case, applying the induction hypothesis (apply it over each premise in which  $\Theta_i$ occurs, for each $l+1 \leq i \leq l+N$; note that there might be  successive applications over the same premise, for $\beta_i$ and $\beta_j$ in different components $\Theta_i$ and $\Theta_j$)\footnote{The multiset-variable is not required to occur in the premises (e.g., if $\UniRuleSchemaA$~was the standard weakening rule). In that case, no use of the induction hypothesis on the premises is required; we can simply use~$\UniFmA^n$ instead of~$\UniFmA^m$ in the instantiation.}
ultimately results in the same premises, where occurrences of~$\UniFmA^m$ have been converted into~$\UniFmA^n$. 
Then applying~$\UniRuleSchemaA$ yields:
{
\small
\begin{multline}
\label{eq:new-base-instance}
\UniHypersequentB_{0}^{-} \VL
\overbrace{
\UniSequent{\UniFmA^{k+m},\UniMSetFmA_{1}}{\UniMSetSucA} 
\VL
\ldots
\VL
\UniSequent{\UniFmA^{k+m},\UniMSetFmA_{l}}{\UniMSetSucA}
}^{\text{$l$ components}} \VL \\
\overbrace{
\UniSequent{\UniFmA^{k+\beta_{l+1}-m+n},\UniMSetFmA_{l+1}}{\UniMSetSucA} \VL
\ldots \VL
\UniSequent{\UniFmA^{k+\beta_{l+N}-m+n},\UniMSetFmA_{l+N}}{\UniMSetSucA}}^{\text{$\UniFmA^m\mapsto \UniFmA^n$ via IH on premises}} \VL \\ 
\UniSequent{\UniFmA^{k+\beta_{l+N+1}},\UniMSetFmA_{l+N+1}}{\UniMSetSucA} \VL
\ldots \VL
\UniSequent{\UniFmA^{k+\beta_{l+\alpha}},\UniMSetFmA_{l+\alpha}}{\UniMSetSucA}
\end{multline}
}

\noindent Observe that the process $\UniFmA^m\mapsto\UniFmA^n$ that we described does not change anything else in the conclusion because the multiset-variables in the conclusion occur exactly once (by the property of linear conclusion), and the hypersequent-variable occurs exactly once.
It remains to show that this new $\UniRuleSchemaA$-instance is related under
$\UniIntExtCtrRel{\UniAbsCtrValue}{\UniActiveCompFixed}$
to the target sequent $\UniHypersequentA \VL \UniSequent{\UniFmA^n,\UniMSetFmA}{\UniMSetSucA}$, and thus their combination constitutes an instance of $\UniHRuleAbsorb{\UniRuleSchemaA}$.

We need first to show that each $k+\beta_i-m+n$, for $l+1 \leq i \leq l+N$, and each $k+\beta_j$, for $l+N+1 \leq j \leq l + \alpha$, can be reduced to $k + n$.
This demands showing that the difference between these values and $k+n$ is at most $(\UniFmlaMultFixed-1)(m-1)$, and that these values are in the same equivalence class of $k+n$ modulo $m-n$.
Note that, for $l + 1 \leq i \leq l+N$, we have
$(k+\beta_i-m+n)-(k+n) = \beta_i-m$. Since
$m \leq \beta_i \leq m + (\UniFmlaMultFixed-1)(m-1)$, we have 
$0 \leq \beta_i - m \leq (\UniFmlaMultFixed-1)(m-1)$.
Moreover, Equation~\ref{eq:betas-to-ms} tell us that $\beta_i \UniEquivMod{m-n} m$, thus $k + \beta_i - m + n \UniEquivMod{m-n} k + n$.
Now, for $l + N + 1 \leq j \leq l + \alpha$, we have $(k + \beta_j) - (k + n) = \beta_j - n$.
Since $\beta_j \leq (\UniFmlaMultFixed-1)(m-1)$, we have $\beta_j - n < (\UniFmlaMultFixed-1)(m-1)$.
Again, since $\beta_j \UniEquivMod{m-n} m$, we have $k + \beta_j \UniEquivMod{m-n} k + m$, and $k + m \UniEquivMod{m-n} k + m - (m - n) = k + n$, thus $k + \beta_j \UniEquivMod{m-n} k + n$.

We have thus established that the multiplicity of~$\UniFmA$ in each of the last $\alpha$ components listed in the above hypersequent (Equation~\ref{eq:new-base-instance}) can be decreased to $k+n$  using $\UniWeakCtr$  and the quantity of decrease is at most $(\UniFmlaMultFixed-1)(m-1)$. 
In other words, the above is related under~$\UniIntCtrRel{(\UniFmlaMultFixed-1)(m-1)}{\UniWeakCtr}$ to

\[
\UniHypersequentB_{1}^{-} \VL 
\overbrace{
\UniSequent{\UniFmA^{k+m},\UniFmMSetMinusFm{\UniMSetFmA}}{\UniMSetSucA} \VL
\ldots \VL
\UniSequent{\UniFmA^{k+m},\UniFmMSetMinusFm{\UniMSetFmA}}{\UniMSetSucA}
}^{\text{$l$ components}} \VL
\overbrace{
\UniSequent{\UniFmA^{k+n},\UniFmMSetMinusFm{\UniMSetFmA}}{\UniMSetSucA} 
\VL
\ldots 
\VL
\UniSequent{\UniFmA^{k+n},\UniFmMSetMinusFm{\UniMSetFmA}}{\UniMSetSucA}}
^{\text{$\alpha$ components}}
\]

Since $1\leq\alpha\leq\UniActiveCompFixed+1$ and $\UniHypersequentB_{1}^{-}\UniExtCtrRel{\UniActiveCompFixed}{\text{(EC)}} \UniFmMSetMinusFm{\UniHypersequentA}$, the above is related under~$\UniExtCtrRel{\UniActiveCompFixed}{\text{(EC)}}$ to
\[
\UniFmMSetMinusFm{\UniHypersequentA} \VL
\overbrace{
\UniSequent{\UniFmA^{k+m},\UniFmMSetMinusFm{\UniMSetFmA}}{\UniMSetSucA} 
\VL
\ldots 
\VL
\UniSequent{\UniFmA^{k+m},\UniFmMSetMinusFm{\UniMSetFmA}}{\UniMSetSucA}}
^{\text{$l$ components}} \VL 
\UniSequent{\UniFmA^{k+n},\UniFmMSetMinusFm{\UniMSetFmA}}{\UniMSetSucA}
\]
As~$\UniFmA^{k+n},\UniFmMSetMinusFm{\UniMSetFmA}$ is~$\UniFmA^n,\UniMSetFmA$, this is exactly $\UniHypersequentA \VL \UniSequent{\UniFmA^n,\UniMSetFmA}{\UniMSetSucA}$
as desired.

The rule \textbf{(EC)}: Given a derivation~$\UniDerivationA$ of~$\UniHypersequentA \VL s \VL s$, we will obtain a derivation of~$\UniHypersequentA \VL s$ of the same height, by induction on the height of~$\UniDerivationA$.

\textit{Base case}: If~$\UniHypersequentA \VL s \VL s$ is an instance of an initial hypersequent, then so is~$\UniHypersequentA \VL s$: if the metavariable $H$ is instantiated to $\UniHypersequentA \VL s$, then by instantiating $H$ to $h$ instead, we get that $\UniHypersequentA \VL s$ is an initial hypersequent. 
If $s \VL s$ is part of the instantiation of $H$, changing that instantiation to contain a single $s$ gives the initial hypersequent  $\UniHypersequentA \VL s$. 

\textit{Inductive case}: If~$\UniHRuleAbsorb{\UniRuleSchemaA}$ is the last rule applied in~$\UniDerivationA$, then there must be a base instance~$\UniHypersequentB_{0}$ of~$\UniHRuleAbsorb{\UniRuleSchemaA}$ and some~$\UniHypersequentB_{1}$ such that
\[
\UniHypersequentB_{0}\UniIntCtrRel{(\UniFmlaMultFixed-1)(m-1)}{\UniWeakCtr} 
\UniHypersequentB_{1} \UniExtCtrRel{\UniActiveCompFixed}{\text{(EC)}} 
\UniHypersequentA \VL s \VL s
\]

If~$l$ is the number of components of~$\UniHypersequentA$ that are equal to~$s$, then $l\geq 0$ and there is $\UniFmMSetMinusFm{\UniHypersequentA}$ such that~$s\not\in\UniFmMSetMinusFm{\UniHypersequentA}$ and 
\[
\UniHypersequentA \VL s \VL s = \UniFmMSetMinusFm{\UniHypersequentA} \VL 
\overbrace{s \VL \ldots \VL s}^{\text{$l$}} \VL s \VL s
\]

Since~$\UniHypersequentB_{1} \UniExtCtrRel{\UniActiveCompFixed}{\text{(EC)}} \UniHypersequentA \VL s \VL s$, there exists~$\alpha$ with~$2\leq\alpha\leq\UniActiveCompFixed+2$, and~$\UniHypersequentB_{1}'$ with~$\UniHypersequentB_{1}' \UniExtCtrRel{\UniActiveCompFixed}{\text{(EC)}} \UniFmMSetMinusFm{\UniHypersequentA}$, such that
\[
\UniHypersequentB_{1} =\UniHypersequentB_{1}' \VL \overbrace{s \VL \ldots \VL s}^{\text{$l+\alpha$}}
\]

In case~$\alpha\leq\UniActiveCompFixed+1$, we have~$\UniHypersequentB_{1} \UniExtCtrRel{\UniActiveCompFixed}{\text{(EC)}} \UniHypersequentA \VL s$, so~$\UniHypersequentB_{0}\UniIntExtCtrRel{(\UniFmlaMultFixed-1)(m-1)}{\UniActiveCompFixed} \UniHypersequentA \VL s$ as required. 
If, on the other hand,~$\alpha=\UniActiveCompFixed+2$, then $\UniHypersequentB_{0}$ has the following form for some~$\UniHypersequentB_{0}'$ with~$\UniHypersequentB_{0}'\UniIntCtrRel{(\UniFmlaMultFixed-1)(m-1)}{\UniWeakCtr} \UniHypersequentB_{1}'$, and sequents~$s_{i}$ ($1\leq i\leq l+\UniActiveCompFixed+2$) such that $s_{i} \UniIntCtrRel{(\UniFmlaMultFixed-1)(m-1)}{\UniWeakCtr} s$:
\[
\UniHypersequentB_{0} =\UniHypersequentB_{0}' \VL \overbrace{s_{1}
\VL
\ldots
\VL
s_{l+\UniActiveCompFixed+2}}^{\text{$l+\UniActiveCompFixed+2$}}
\]

Since the number of active components in a rule is at most~$\UniActiveCompFixed$, we may assume without loss of generality that the last two components~$s_{l+\UniActiveCompFixed+1}$ and~$s_{l+\UniActiveCompFixed+2}$ are in the {$\UniHRuleAbsorb{\UniRuleSchemaA}$-instantiation} of the hypersequent-variable. 
Using the hp-admissibility of~$\UniWeakCtr$, proved above, in each premise we can replace $s_{l+\UniActiveCompFixed+1} \VL s_{l+\UniActiveCompFixed+2}$ with $s \VL s$, and we  can replace~$s \VL s$ with~$s$ using the induction hypothesis. 
By applying {$\UniHRuleAbsorb{\UniRuleSchemaA}$ to these premises instead} we obtain the base instance
\[
\UniHypersequentB_{0}' \VL
\overbrace{
s_{1}\VL
\ldots \VL
s_{l+\UniActiveCompFixed}}^{l+\UniActiveCompFixed} \VL 
s
\] 

Since~$\UniHypersequentB_{0}'\UniIntCtrRel{(\UniFmlaMultFixed-1)(m-1)}{\UniWeakCtr}\UniHypersequentB_{1}'$ and each~$s_{i}\UniIntCtrRel{(\UniFmlaMultFixed-1)(m-1)}{\UniWeakCtr}s$, the above is related under~$\UniIntCtrRel{(\UniFmlaMultFixed-1)(m-1)}{\UniWeakCtr}$ to
\[
\UniHypersequentB_{1}' \VL \overbrace{s \VL \ldots|s}^{l+\UniActiveCompFixed} \VL s
\]
Because we had~$\UniHypersequentB_{1}'\UniExtCtrRel{\UniActiveCompFixed}{\text{(EC)}} \UniFmMSetMinusFm{\UniHypersequentA}$ it follows that the above is related under~$\UniExtCtrRel{\UniActiveCompFixed}{\text{(EC)}}$ to
\[
\UniFmMSetMinusFm{\UniHypersequentA}\VL\overbrace{s\VL \ldots \VL s}^{l}\VL s,
\]
which is equal to~$\UniHypersequentA \VL s$.
\end{proof}

We show next that the above hp-admissibility result  allows us to prove a version of Curry's lemma for the knotted contraction logics.
For that, we first need a quasi-order on hypersequents based on applications of knotted contraction (here, we simply need to consider the non-analytic version $\UniWeakCtr$), $\UniEW$ and $\UniEC$---thus matching the rules considered in the hp-admissibility result.
This preorder will then be shown in the next section to yield a nwqo with an associated length theorem.
For that we will use the developments on knotted nwqos from Chapter~\ref{sec:knotted-wqos}.

\begin{definition}\label{def-der-hone-htwo}
    For a finite set of formulas $\UniSubfmlaHyperseqSet$ and for $\UniSubfmlaHyperseqSet$-hypersequents $\UniHypersequentA_1$ and $\UniHypersequentA_2$, we define $\UniHypersequentA_1\UniHyperseqCtrWqo{m}{n}{\UniSubfmlaHyperseqSet}~\UniHypersequentA_2$ to mean that $\UniHypersequentA_1$ is derivable from $\UniHypersequentA_2$ using the rules $\UniEW$, $\UniWeakCtr$ and $\UniEC$.
\end{definition}

\begin{lemma}[Curry's Lemma for knotted contraction logics]\label{lem:curry-knotted}
If $\UniHypersequentA$ has a proof in $\UniHFLecRAbsorb$ of height $k \geq 1$ and $\UniHypersequentA' \UniHyperseqCtrWqo{m}{n}{\UniSubfmlaHyperseqSet} \UniHypersequentA$,
then $\UniHypersequentA'$ has a proof in $\UniHFLecRAbsorb$ of height at most $k$.
\end{lemma}
\begin{proof}
Suppose that $\UniHypersequentA$ has a proof in $\UniHFLecRAbsorb$ of height $k$ and $\UniHypersequentA' \UniHyperseqCtrWqo{m}{n}{\UniSubfmlaHyperseqSet} \UniHypersequentA$.
From the latter, there is a proof $\UniDerivationA$ of $\UniHypersequentA'$ starting from $\UniHypersequentA$ using only rules (EW), $\UniWeakCtr$ and (EC).
We proceed by induction on the height $d$ of $\UniDerivationA$.
If $d=1$, then $\UniHypersequentA=\UniHypersequentA'$ and we are done.
Suppose that the statement holds for all  derivations of height less than $d$.
The induction hypothesis applies to the premise of the last rule applied, thus there is a derivation of it with height at most $k$.
The last rule applied was one of (EW), $\UniWeakCtr$ and (EC), thus by hp-admissibility of them (Lemma~\ref{lem-hp-admissible}) we obtain the derivation of $\UniHypersequentA'$ with height no more than $k$.
\end{proof}

\section{Knotted nwqos over hypersequents}

In this section, we introduce certain knotted nwqos over hypersequents, and establish length theorems for them.

To obtain a normed wqo using the quasi order relation $\UniHyperseqCtrWqo{m}{n}{\UniSubfmlaHyperseqSet}$ given in Definition~\ref{def-der-hone-htwo}, we introduce norms for $\UniSubfmlaHyperseqSet$-hypersequents.
We define $$\UniNormSSeq{\UniSequent{\UniListFmA}{\UniMSetSucA}} \UniSymbDef 
\max_{\UniFmA \in \UniListFmA} 
\UniListCountElem{\UniListFmA}{\UniFmA}$$ and 
\[\UniNormHSeq{\UniSequentA_1 \VL \cdots \VL 
\UniSequentA_k
} \UniSymbDef 
\max\{ k^\ast, \max_{1 \leq i \leq k} \UniNormSSeq{\UniSequentA_i} \},\]
where $k^\ast \leq k$ is obtained by partitioning the components of the hypersequent per succedent and taking the size of a maximal partition.

We observe that the norm of a hypersequent is always smaller than or equal its size:

\begin{proposition}\label{prop:norm-size-hseqe}
For all hypersequents $\UniHypersequentA$, $\UniNormHSeq{\UniHypersequentA} \leq \UniSizeHyper{\UniHypersequentA}$.
\end{proposition}
\begin{proof}
We first observe that $\UniSizeHyper{\UniFmB} \geq 1$, for every formula $\UniFmB$. 
So we have $\UniNormSSeq{\UniSequent{\UniFmA_1,\ldots,\UniFmA_p}{\UniMSetSucA}} 
= \max\{|\UniFmA_1,\ldots,\UniFmA_p|_{\UniFmA_i}:1 \leq i \leq p\}
\leq  p 
\leq \sum_{i=1}^p \UniSizeHyper{\UniFmA_i} 
\leq \UniSizeHyper{\UniSequent{\UniFmA_1,\ldots,\UniFmA_p}{\UniMSetSucA}}$.

For $h=\UniSequentA_1 \VL \cdots \VL 
\UniSequentA_k$ we have $1 \leq \UniSizeHyper{\UniSequentA_i} $, for all $i$, hence 
$ k^\ast \leq k \leq \sum_{s \in \UniHypersequentA} \UniSizeHyper{s} \leq \UniSizeHyper{\UniHypersequentA}$. 
Moreover, 
$\UniNormSSeq{\UniSequentA_i}\leq \UniSizeHyper{\UniSequentA_i} \leq  \UniSizeHyper{\UniHypersequentA}$, 
for all $i$, by the preceding paragraph. Therefore, $\UniSizeHyper{\UniHypersequentA}$ is at least as big as the maximum of all of these values.\qedhere
\end{proof}

We denote by $\UniOmegaHypersequentsSet{\UniSubfmlaHyperseqSet}$ the collection of all $\UniSubfmlaHyperseqSet$-hypersequents and we observe that it supports the following nqo.
\begin{definition}
\label{def:nwqo-ec-mn}
$\UniNwqoHSeqName{m}{n} \UniSymbDef 
\UniStruct{\UniOmegaHypersequentsSet{\UniSubfmlaHyperseqSet},
\UniHyperseqCtrWqo{m}{n}{\UniSubfmlaHyperseqSet},
\UniNormHSeq{\cdot}}$.
\end{definition}

We will provide a strong reflection from  $\UniNwqoHSeqName{m}{n}$ to $\UniMinoringFntWqo{\UniWqoExtModRelProd{m}{n}{\UniSetCard{\UniSubfmlaHyperseqSet}}}^{\UniSetCard{\UniSubfmlaHyperseqSet} + 1}$ and use the results of Chapter~\ref{sec:knotted-wqos} to conclude that $\UniNwqoHSeqName{m}{n}$ is a nwqo and  to obtain length theorems for it.

\begin{definition}
\label{def:encoding-hypersequents}
If $\UniSubfmlaHyperseqSet \UniSymbDef \UniSet{\UniFmA_1,\ldots,\UniFmA_d}$ is an ordered set of formulas, where $d \in \UniNaturalSet$, and $\UniHypersequentA$ is an $\UniSubfmlaHyperseqSet$-hypersequent, we define
\[
\UniHyperTransNat{\UniHypersequentA}
\UniSymbDef
\UniTuple{X_0,\ldots,X_{d}}
\] 
to be the element of $(\UniPowerSetFin{\UniNaturalSet^d})^{d+1}$, where
\[
X_0 \UniSymbDef
\UniSet{
    \UniTuple{k_1,\ldots,k_d} \in \UniNaturalSet^d
    \mid
    (\UniSequent{\UniFmA_1^{k_1},\ldots,\UniFmA_d^{k_d}}{})
    \in \UniHypersequentA
}
\]
and, for each $1 \leq i \leq d$,
\[
X_i \UniSymbDef
\UniSet{
    \UniTuple{k_1,\ldots,k_d} \in \UniNaturalSet^d
    \mid
    (\UniSequent{\UniFmA_1^{k_1},\ldots,\UniFmA_d^{k_d}}{\UniFmA_i})
    \in \UniHypersequentA
}.
\]
\end{definition}

\begin{example}
Consider the ordered set of formulas $\UniSubfmlaHyperseqSet \UniSymbDef \UniSet{\UniPropA,\UniPropB,\UniPropA\imp\UniPropB}$  (in the listed order) and the $\UniSubfmlaHyperseqSet$-hypersequent
\[
\UniHypersequentA \UniSymbDef \UniSequent{}{\UniPropA} \VL
\UniSequent{\UniPropA\imp\UniPropB,\UniPropA,\UniPropA}{\UniPropA} \VL 
\UniSequent{\UniPropB}{} \VL 
\UniSequent{\UniPropB}{} 
\]
Then 
$
\UniHyperTransNat{\UniHypersequentA}
= 
\UniTuple{\UniSet{\UniTuple{0,1,0}}, 
          \UniSet{\UniTuple{0,0,0},\UniTuple{2,0,1}}, 
          \UniEmptySet, 
          \UniEmptySet}.
$
\end{example}

By the following lemma, the map $(\cdot)^\#: \UniNwqoHSeqName{m}{n} \to \UniPowerMaj{\UniMinoringFntWqo{\UniWqoExtModRelProd{m}{n}{\UniSetCard{\UniSubfmlaHyperseqSet}}}}{\UniSetCard{\UniSubfmlaHyperseqSet} + 1}$  provides a strong connection between the two nqos.

\begin{lemma}
\label{lem:hyperseq-nwqo-nat-corresp}
The following holds for all $\UniSubfmlaHyperseqSet$-hypersequents $\UniHypersequentA_1$, $\UniHypersequentA_2$, $\UniHypersequentA$:
\begin{enumerate}
\item $\UniHypersequentA_1\UniHyperseqCtrWqo{m}{n}{\UniSubfmlaHyperseqSet}\UniHypersequentA_2
\text{ if, and only if, }
\UniHyperTransNat{\UniHypersequentA_1}
\UniWqoRel{\UniMinoringFntWqo{{\UniWqoExtModRelProd{m}{n}{\UniSetCard{\UniSubfmlaHyperseqSet}}}}}^{\UniSetCard{\UniSubfmlaHyperseqSet}+1}
\UniHyperTransNat{\UniHypersequentA_2}
$; and
\item $\UniNorm{\UniHyperTransNat{\UniHypersequentA}}{\UniPowerMaj{\UniMinoringFntWqo{\UniWqoExtModRelProd{m}{n}{\UniSetCard{\UniSubfmlaHyperseqSet}}}}{\UniSetCard{\UniSubfmlaHyperseqSet} + 1}}
\leq \UniNormHSeq{\UniHypersequentA}$.
\end{enumerate}
\end{lemma}
\begin{proof}
For item (1), first note that (the restriction of) $(\cdot)^\#$  is  a bijection between sequences, $\varphi^k$, of the same formula, $\varphi$, and ${\UniWqoExtModRelProd{m}{n}{}}$. 
Also, the applications of $\UniWeakCRule{m}{n}$ on such formulas are preserved and reflected accurately by $(\cdot)^\#$ in the ordering of ${\UniWqoExtModRelProd{m}{n}{}}$, as this is exactly how the ordering was constructed. 
Moreover, $(\cdot)^\#$  is  a bijection also between sequents with a fixed succedent and $\UniWqoExtModRelProd{m}{n}{\UniSetCard{\UniSubfmlaHyperseqSet}}$, and  the applications of $\UniWeakCRule{m}{n}$ on individual components are preserved and reflected accurately by $(\cdot)^\#$ in the ordering of $\UniWqoExtModRelProd{m}{n}{\UniSetCard{\UniSubfmlaHyperseqSet}}$. 
Now, $(\cdot)^\#$ is not injective between  $\UniNwqoHSeqName{m}{n}$ and $\UniPowerMaj{\UniMinoringFntWqo{\UniWqoExtModRelProd{m}{n}{\UniSetCard{\UniSubfmlaHyperseqSet}}}}{\UniSetCard{\UniSubfmlaHyperseqSet} + 1}$, since it records identical components as a single tuple: each $X_i$ is a set, so multiplicity is lost. 
In that respect applications of (EC) are not registered at all by $(\cdot)^\#$ and the same holds for applications of (EW) that happen to be inverse-of-(EC) applications. 
An application of (EW) that is not of this form does yield different results via $(\cdot)^\#$, as it results in one new tuple being added to one of the ($|\Omega|+1$)-many sets $X_i$. 
Such a tuple, via a further application of $\UniWeakCRule{m}{n}$, may result in a $\UniWqoExtModRelProd{m}{n}{\UniSetCard{\UniSubfmlaHyperseqSet}}$-smaller tuple. 
Such situations are exactly what is captured by the ordering of $\UniMinoringFntWqo{\UniWqoExtModRelProd{m}{n}{\UniSetCard{\UniSubfmlaHyperseqSet}}}^{\UniSetCard{\UniSubfmlaHyperseqSet} + 1}$.
Item (2) is immediate from the definition of the involved norms.
In fact, note that
$\UniNorm{X_i}{\UniMinoringFntWqo{\UniWqoExtModRelProd{m}{n}{\UniSetCard{\UniSubfmlaHyperseqSet}}}} \leq \max\{ \UniSetCard{X_i}, \UniNormHSeq{\UniHypersequentA} \} = \UniNormHSeq{\UniHypersequentA}$.
\qedhere
\end{proof}

\begin{example}
Let
$\UniHypersequentA_1 \UniSymbDef
\UniSequent{\UniPropC^3, \UniPropA^2}{\UniPropB} \VL 
\UniSequent{\UniPropA^7, \UniPropC^6}{\UniPropB} \VL
\UniSequent{\UniPropA,\UniPropB}{\UniPropC}$
and
$\UniHypersequentA_2 \UniSymbDef
\UniSequent{\UniPropC^6,\UniPropA^5}{\UniPropB} \VL
\UniSequent{\UniPropC^9, \UniPropA^8}{\UniPropB} \VL
\UniSequent{\UniPropC^3,\UniPropA^2}{\UniPropB}$
be $\UniSubfmlaHyperseqSet$-hypersequents
for $\UniSubfmlaHyperseqSet =\{ \UniPropA,\UniPropB,\UniPropC \}$ (assume the elements are ordered in this way). 
Here is a derivation showing that 
$\UniHypersequentA_1\UniHyperseqCtrWqo{5}{2}{\UniSubfmlaHyperseqSet}\UniHypersequentA_2$
(we omit successive applications of the same rule):

\begin{center}
\AxiomC{$\UniSequent{\UniPropC^6,\UniPropA^5}{\UniPropB} \VL
\UniSequent{\UniPropC^9, \UniPropA^8}{\UniPropB} \VL
\UniSequent{\UniPropC^3,\UniPropA^2}{\UniPropB}$}
\RightLabel{$\UniWeakCRule{5}{2}$}
\UnaryInfC{$\UniSequent{\UniPropC^3, \UniPropA^2}{\UniPropB} \VL \UniSequent{\UniPropC^3, \UniPropA^2}{\UniPropB}
\VL \UniSequent{\UniPropC^3, \UniPropA^2}{\UniPropB}$}
\RightLabel{\UniEC}
\UnaryInfC{$\UniSequent{\UniPropC^3, \UniPropA^2}{\UniPropB}$}
\RightLabel{\UniEW}
\UnaryInfC{$
\UniSequent{\UniPropC^3, \UniPropA^2}{\UniPropB} \VL 
\UniSequent{\UniPropA^7, \UniPropC^6}{\UniPropB} \VL
\UniSequent{\UniPropA,\UniPropB}{\UniPropC}$}
\DisplayProof
\end{center}

\noindent
Call $\UniHypersequentB_1$ and $\UniHypersequentB_2$ the hypersequents appearing between $\UniHypersequentA_1$ and $\UniHypersequentA_2$ in a top-down reading of the above proof.
Regarding the encoding of these hypersequents, note that
\begin{align*}
\UniHyperTransNat{\UniHypersequentA_2} =& 
\UniTuple{
\{ \},
\{ \},
\{ \UniTuple{5, 0, 6}, 
\UniTuple{8, 0, 9}, 
\UniTuple{2, 0, 3} \},
\{ \}
}
\\
\UniHyperTransNat{\UniHypersequentB_1} =& 
\UniTuple{
\{ \},
\{ \},
\{ \UniTuple{2, 0, 3}, 
\UniTuple{2, 0, 3}, 
\UniTuple{2, 0, 3} \},
\{ \}
}
\\
\UniHyperTransNat{\UniHypersequentB_2} =& 
\UniTuple{
\{ \},
\{ \},
\{ \UniTuple{2, 0, 3} \},
\{ \}
}
\\
\UniHyperTransNat{\UniHypersequentA_1} =& 
\UniTuple{
\{ \},
\{ \},
\{ \UniTuple{2, 0, 3}, \UniTuple{7, 0, 6} \},
\{ \UniTuple{1, 1, 0} \}
}
\end{align*}
The reader may easily check that each component of the above tuples are correspondingly related under $\UniMinoringFntWqo{{\UniWqoExtModRelProd{5}{2}{\UniSetCard{\UniSubfmlaHyperseqSet}}}}$.
Moreover, 
$\UniNormHSeq{\UniHypersequentA_1} = \UniNorm{\UniHyperTransNat{\UniHypersequentA_1}}{\UniPowerMaj{\UniMinoringFntWqo{\UniWqoExtModRelProd{5}{2}{\UniSetCard{\UniSubfmlaHyperseqSet}}}}{\UniSetCard{\UniSubfmlaHyperseqSet} + 1}} = 7$
and
$\UniNormHSeq{\UniHypersequentA_2} = \UniNorm{\UniHyperTransNat{\UniHypersequentA_2}}{\UniPowerMaj{\UniMinoringFntWqo{\UniWqoExtModRelProd{5}{2}{\UniSetCard{\UniSubfmlaHyperseqSet}}}}{\UniSetCard{\UniSubfmlaHyperseqSet} + 1}} = 9$.
\end{example}

Interestingly, it is easy to see that $\UniNwqoHSeqName{m}{n}$ does not depend on the exact elements of $\UniSubfmlaHyperseqSet$, but only on the cardinality thereof.
For that reason, for each $k \geq 0$, if we fix $\UniSubfmlaHyperseqSet_k \UniSymbDef \{ \UniPropA_1,\ldots, \UniPropA_k\}$, then whenever $\UniSetCard{\UniSubfmlaHyperseqSet} = k$ we will have
$
\UniNwqoHSeqName{m}{n}
\cong
\UniNwqoHSeqNameVar{m}{n}{\UniSubfmlaHyperseqSet_k}$.
This observation allows us to consider the parameterized family of nwqos
$\{ \UniNwqoHSeqNameVar{m}{n}{\UniSubfmlaHyperseqSet_k}\}_{k \in \UniNaturalSet}$
to subsume all possible $\UniNwqoHSeqName{m}{n}$.
We then have the following result.

\begin{proposition}
\label{fact:hnwqo-length-theorem}
For all finite sets of formulas $\UniSubfmlaHyperseqSet$, $\UniNwqoHSeqName{m}{n}$ is a nwqo and $\UniNwqoHSeqName{m}{n} \sqsubseteq \UniFGHOneAppLevel{\omega^{\UniSetCard{\UniSubfmlaHyperseqSet}}}$.
Moreover, $\{ \UniNwqoHSeqNameVar{m}{n}{\UniSubfmlaHyperseqSet_k}\}_{k \in \UniNaturalSet} \sqsubseteq \UniFGHOneAppLevel{\omega^\omega}$.
\end{proposition}
\begin{proof}
For the first claim, by Lemma~\ref{lem:hyperseq-nwqo-nat-corresp} the function $(\cdot)^\#$, described in Definition~\ref{def:encoding-hypersequents}, is a strong reflection, i.e.,
$\UniNwqoHSeqName{m}{n} \UniStrongReflArrow{(\cdot)^\#} \UniPowerMaj{\UniMinoringFntWqo{\UniWqoExtModRelProd{m}{n}{\UniSetCard{\UniSubfmlaHyperseqSet}}}}{\UniSetCard{\UniSubfmlaHyperseqSet} + 1}
$.
Since $\UniPowerMaj{\UniMinoringFntWqo{\UniWqoExtModRelProd{m}{n}{\UniSetCard{\UniSubfmlaHyperseqSet}}}}{\UniSetCard{\UniSubfmlaHyperseqSet} + 1}\UniFGBoundedNwqosIn{\UniFGHOneAppLevel{\omega^{\UniSetCard{\UniSubfmlaHyperseqSet}}}}$ 
by Theorem~\ref{fact:length-theorem-power-set-wqo-maj-mn}, we are done in view of Corollary~\ref{coro:strong-refl-f-hierar}.

For the second statement, note that the above reflection implies that
\[\UniNwqoHSeqNameVar{m}{n}{\UniSubfmlaHyperseqSet_k} \UniStrongReflArrow{} \UniPowerMaj{\UniMinoringFntWqo{(k+1)\cdot\UniWqoExtModRelProd{m}{n}{k+1}}}{k+1}.\]
Thus
$\{ \UniNwqoHSeqNameVar{m}{n}{\UniSubfmlaHyperseqSet_k}\}_{k \in \UniNaturalSet} \UniStrongReflArrow{\lambda x.x+1}
\UniPowerMaj{\UniMinoringFntWqo{(k+1)\cdot\UniWqoExtModRelProd{m}{n}{k+1}}}{k+1}
$,
and thus by Theorem~\ref{fact:length-theorem-power-set-wqo-maj-mn} and Lemma~\ref{fact:strong-refl-preserv-nwqo-param}, we have
$\{ \UniNwqoHSeqNameVar{m}{n}{\UniSubfmlaHyperseqSet_k}\}_{k \in \UniNaturalSet} \sqsubseteq \UniFGHOneAppLevel{\omega^\omega}$.
\end{proof}

\section{Backward proof search and upper bounds}
\label{sec:wcontraction-proof-search}

By Lemma~\ref{lem-equiv-cal}, the calculi $\UniHFLecR$ and~$\UniHFLecRAbsorb$ derive the same hypersequents (and both of them are cut-free). 
So, when searching for a proof of a hypersequent in $\UniHFLecR$, we can  employ an algorithm that performs the  procedure of backward proof search (with some modifications) in $\UniHFLecRAbsorb$, instead.
From now on, we abbreviate $\UniHyperCalcA\UniSymbDef\UniHFLecRAbsorb$.

We note that a hypersequent $\UniHypersequentA$ is provable iff $\UniHypersequentA$ has a minimal proof (i.e., every subproof has minimal height).
Such a minimal proof cannot have a branch that contains a node labelled with $\UniHypersequentB$ whenever $\UniHypersequentB$ can be transformed via (EW), $\UniWeakCtr$, (EC) to some $\UniHypersequentB'$ that appears earlier on the branch (i.e., closer to the root): otherwise, by Lemma~\ref{lem:curry-knotted}, there would exist a proof of $\UniHypersequentB'$ of smaller height, violating minimality.
That is, every minimal proof is \emph{$\UniHyperseqCtrWqo{m}{n}{\UniSubfmlaHyperseqSet}$-minimal}.
Therefore, it is enough to search for $\UniHyperseqCtrWqo{m}{n}{\UniSubfmlaHyperseqSet}$-minimal proofs; in other words, we expand a branch with premises of a rule instance only when they do not violate $\UniHyperseqCtrWqo{m}{n}{\UniSubfmlaHyperseqSet}$-minimality (i.e., when they do not introduce increasing pairs of hypersequents in a branch).
We denote by $\BackPSComCtr(\UniHypersequentA)$ the proof search tree obtained by starting with $\UniHypersequentA$ and recursively expanding  the (existing stage of the) tree by independently adding to each leaf of the  tree, as children, the premises of all rule instances that have that leaf as conclusion if furthermore the resulting branches remain $\UniHyperseqCtrWqo{m}{n}{\UniSubfmlaHyperseqSet}$-bad sequences.
Note that if multiple current leaves are conclusions of rule instances that share the same premise sequents, these sequents are added (i.e., repeated) as children to each of the leaves.
We say that $\BackPSComCtr(\UniHypersequentA)$ produces a proof of  $\UniHypersequentA$ if there is a downward-closed (with respect to the tree ordering) subtree of $\BackPSComCtr(\UniHypersequentA)$ that is a proof of $\UniHypersequentA$.

\begin{lemma}\label{l: backsearch}
A sequent $\UniHypersequentA$ is provable in  $\UniHFLecRAbsorb$ iff it has a minimal proof in  $\UniHFLecRAbsorb$ iff $\BackPSComCtr(\UniHypersequentA)$ produces a proof of  $\UniHypersequentA$.
\end{lemma}

Therefore, every branch in $\BackPSComCtr(\UniHypersequentA)$ is a bad sequence over $\UniNwqoHSeqName{m}{n}$ (recall Definition~\ref{def:nwqo-ec-mn}), and thus the procedure of constructing $\BackPSComCtr(\UniHypersequentA)$ will terminate, since $\UniNwqoHSeqName{m}{n}$ is a nwqo by Proposition~\ref{fact:hnwqo-length-theorem}.

\begin{proposition}
\label{prop:back-ps-ek-control-bad-seq}
The branches of $\BackPSComCtr(\UniHypersequentA)$ are $(f,t)$-controlled bad sequences  over the nwqo $\UniNwqoHSeqName{m}{n}$, where
$\UniControlFunctionA(x) \UniSymbDef Mx$,  $M:=2\UniSizeHyper{\UniHyperCalcA}\UniAbsCtrValue$, 
and $t \UniSymbDef \UniSizeHyper{\UniHypersequentA}$.
\end{proposition}
\begin{proof}
Observe that $f^i(x) = M^i x$ for all $i \geq 0$.
The constant $M$ captures how the norm of a  premise $\UniHypersequentA'$ of a rule instance grows with respect to its conclusion $\UniHypersequentA$; namely, as we will see, $\UniNormHSeq{\UniHypersequentA'}{} \leq M\cdot \UniNormHSeq{\UniHypersequentA}{}$.
Also, recall that, in the backward proof-search procedure described above, $\UniHypersequentA'$ would appear right after $\UniHypersequentA$ in a branch of the underlying search tree.
With these observations in hand, the fact that such a branch is an $(\UniControlFunctionA, t)$-controlled bad sequence over $\UniNwqoHSeqName{m}{n}$, i.e., that $\UniNormHSeq{\UniHypersequentA_r}{} \leq f^r(t)$ for all $r$, is proved by induction on the length of the branches.
Indeed, the base case is given by the fact that $\UniNormHSeq{\UniHypersequentA} \leq \UniSizeHyper{\UniHypersequentA}= t = f^0(t)$
(recall Proposition~\ref{prop:norm-size-hseqe}). 
Furthermore, if $h=h_0,\ldots,h_{L-1}$ is a branch in this tree, the induction step is established as follows:
\begin{align*}
\UniNormHSeq{\UniHypersequentA_{i+1}} & \leq
    M\cdot \UniNormHSeq{\UniHypersequentA_i}\\
    &\leq M \cdot f^i(t) \qquad \text{(IH)}\\
    &= M \cdot M^i \cdot t\\
    &= M^{i+1} \cdot t\\
    &= f^{i+1}(t).
\end{align*}

The value of $M$ is defined to be $2\UniSizeHyper{\UniHyperCalcA}\alpha$, where $\alpha \UniSymbDef \UniAbsCtrValue$ and $\UniHyperCalcA \UniSymbDef \UniHFLecRAbsorb$ (recall Definition~\ref{def:size-hypersequent} for the meaning of $\UniSizeHyper{\UniHyperCalcA}$).
In order to see why this is enough for establishing $\UniNormHSeq{\UniHypersequentA'}{} \leq M\cdot \UniNormHSeq{\UniHypersequentA}{}$, we consider cases on the value of $\UniNormHSeq{\UniHypersequentA'}{}$ (where $\UniHypersequentA'$ is a premise and $\UniHypersequentA$ the conclusion of a rule $\UniHRuleAbsorb{\UniRuleSchemaA}$), having in mind that $\UniHypersequentA_0 \UniIntExtCtrRel{\UniAbsCtrValue}{\UniActiveCompFixed} \UniHypersequentA$, where $\UniHypersequentA_0$ is the conclusion of a base instance of $\UniHRuleAbsorb{\UniRuleSchemaA}$.
This is a picture of the whole setup, where $\rho$ is the number of premises of $\UniHRuleAbsorb{\UniRuleSchemaA}$ and $\UniIntExtCtrRel{}{}$ is $\UniIntExtCtrRel{\UniAbsCtrValue}{\UniActiveCompFixed}$ for simplicity:

\begin{center}
\AxiomC{$\UniHypersequentA_{1}
\quad\ldots\quad 
\UniHypersequentA'
\quad\ldots\quad 
\UniHypersequentA_{\rho}$
}
\RightLabel{$\UniHRuleAbsorb{\UniRuleSchemaA}$}
\UnaryInfC{\rotatebox[origin=c]{-90}{$
\text{\rotatebox[origin=c]{90}{$\UniHypersequentA_0$}}
\UniIntExtCtrRel{}{}
\text{\rotatebox[origin=c]{90}{$\UniHypersequentA$}}
$}
}
\DisplayProof
\end{center}

\noindent Therefore, we want to show $\UniNormHSeq{\UniHypersequentA'} \leq  2\UniSizeHyper{\UniHyperCalcA}\alpha\UniNormHSeq{\UniHypersequentA}$.
In what follows, given a sequent $\UniSequentB$ and a hypersequent $\UniHypersequentB$, we let $\UniMultAnt{\UniSequentB}{\UniFmA}$ be the multiplicity of $\UniFmA$ in the antecedent of $\UniSequentB$, and $\UniMultComp{\UniHypersequentB}{\UniSequentB}$ be the multiplicity of $\UniSequentB$ in $\UniHypersequentB$.
Note moreover that if $\UniNormHSeq{\UniHypersequentA'}{} > 0$, then $\UniNormHSeq{\UniHypersequentA}{} > 0$, thus we will assume $\UniNormHSeq{\UniHypersequentA}{} > 0$.
By definition (see after Definition~\ref{def-der-hone-htwo}), $\UniNormHSeq{\UniHypersequentA'}{}$ is the maximum of: the largest number of components with the same succedent, and the largest multiplicity of a formula occurring in the antecedent of a component; we examine these two cases separately. 

\begin{enumerate}
    \item  $\UniNormHSeq{\UniHypersequentA'}{}$ is the largest multiplicity of a formula occurring in the antecedent of a component; say that the formula in $h'$ with greatest multiplicity $k$ is $\UniFmA$ and occurs in component $s'$.
    If $s'$ is in the instantiation of the hypersequent variable (recall there is only one and the same in all rule schemas), then $s'$ also appears in $h_0$ and after applying  $\UniIntExtCtrRel{\alpha}{\UniActiveCompFixed}$, by Proposition~\ref{prop:descendands-in-c}, we find a corresponding component $s$ in $\UniHypersequentA$ (a descendant of $s'$) such that 
    $\UniMultAnt{s'}{\UniFmA} \leq \UniMultComp{s}{\UniFmA} + \alpha$,
    and thus 
    $\UniNormHSeq{\UniHypersequentA'}
    =
    \UniMultComp{s'}{\UniFmA}  \leq \UniMultComp{s}{\UniFmA} + \alpha 
    \leq 
    \UniNormHSeq{\UniHypersequentA}
    + \alpha
    \leq 
    2\UniSizeHyper{\UniHyperCalcA}\alpha\UniNormHSeq{\UniHypersequentA}$.
    Otherwise, $s'$ is an instantiation of a schematic component in a premise, whose form is 
    $\UniSequent{\Delta_1,\ldots,\Delta_{\mu_1},
    \UniFmC_1,\ldots,\UniFmC_{\mu_2}}{\UniMSetSucA}$.
    So,
    $k = \sum_{i=1}^{\mu_1} \Delta_i(\UniFmA) + \sum_{j=1}^{\mu_2} [\UniFmC_j \mapsto \UniFmA]$,
    with
    $\Delta_i(\UniFmA)$
    being the multiplicity of $\UniFmA$ in the instantiation of $\Delta_i$, $\mu_1 + \mu_2 \leq \UniSizeHyper{\UniHyperCalcA}$
    and $[\UniFmC_j \mapsto \UniFmA]$ evaluating to 1 if $\UniFmC_j$ is instantiated with $\UniFmA$ and 0 otherwise.
    Note that, for simplicity, we identify schematic variables with their instantiations.
    By the strong subformula and linear conclusion properties, there is a component $s_0$ in the conclusion $h_0$ of the rule in which $\Delta_{\mathsf{max}}$ occurs only once,  where 
    $\Delta_{\mathsf{max}} \in \{\Delta_1,\ldots,\Delta_{\mu_1}\}$ and
    $\Delta_{\mathsf{max}}(\varphi) \geq \Delta_1(\varphi),\ldots,\Delta_{\mu_1}(\varphi)$. Then:
    \begin{align*}
        k &=
        \sum_i \Delta_i(\UniFmA) + \sum_j [\UniFmC_j \mapsto \UniFmA]\\
        &\leq \sum_i \Delta_i(\UniFmA) + \UniSizeHyper{\UniHyperCalcA}\\
        &\leq \UniSizeHyper{\UniHyperCalcA}\Delta_{\mathsf{max}}(\UniFmA) + \UniSizeHyper{\UniHyperCalcA}\\
        &\leq \UniSizeHyper{\UniHyperCalcA}
        \UniMultComp{s_0}{\UniFmA} + \UniSizeHyper{\UniHyperCalcA}\\
        &\leq 2\UniSizeHyper{\UniHyperCalcA}
        \UniMultComp{s_0}{\UniFmA}
    \end{align*}
    In $h$, we must have a component $s$ such that $\UniMultComp{s_0}{\UniFmA} - \UniMultComp{s}{\UniFmA} \leq \alpha$, i.e., $\UniMultComp{s_0}{\UniFmA} \leq 
    \UniMultComp{s}{\UniFmA} + \alpha$, and thus
    $2\UniSizeHyper{\UniHyperCalcA}
    \UniMultComp{s_0}{\UniFmA} \leq 2\UniSizeHyper{\UniHyperCalcA}\UniMultComp{s}{\UniFmA} + 
    2\UniSizeHyper{\UniHyperCalcA}\alpha
    \leq 2\UniSizeHyper{\UniHyperCalcA}\alpha \UniMultComp{s}{\UniFmA} \leq 2\UniSizeHyper{\UniHyperCalcA}\alpha\UniNormHSeq{\UniHypersequentA}$.
    Therefore, $\UniNormHSeq{\UniHypersequentA'} = k \leq  2\UniSizeHyper{\UniHyperCalcA}\alpha\UniNormHSeq{\UniHypersequentA}$.
    \item 
    $\UniNormHSeq{\UniHypersequentA'}{}$ is the largest number of components with the same succedent; let $\UniMSetSucA$ be a  succedent with maximal number of components in the hypersequent.
    We know that the schematic hypersequent in $\UniRuleSchemaA$ that instantiates to $\UniHypersequentA'$ has the form $H \VL s^\ast$, where $s^\ast$ is a schematic sequent.
    We denote by $\UniHypersequentA'(\UniMSetSucA)$ the number of components of $\UniHypersequentA'$ with succedent $\UniMSetSucA$ (we use this notation for other hypersequents and succedents as well).
    Then $\UniHypersequentA'(\UniMSetSucA) \leq H(\UniMSetSucA) + [s^\ast \mapsto \UniMSetSucA] \leq H(\UniMSetSucA) + 1$, where $H(\UniMSetSucA)$ is the number of components with succedent $\UniMSetSucA$ in the instantiation of $H$, and $[s^\ast \mapsto \UniMSetSucA]$ evaluates to $1$ if $s^\ast$ instantiates to a component with succedent $\UniMSetSucA$ and to $0$ otherwise.
    We also know that $\UniHypersequentA_0$  is an instantiation of a schematic hypersequent of the form $H \VL s_1^\ast \VL \ldots \VL s_\rho^\ast$.
    Thus $H(\UniMSetSucA) \leq \UniHypersequentA_0(\UniMSetSucA)$ and therefore $\UniHypersequentA'(\UniMSetSucA) \leq H(\UniMSetSucA) + 1 \leq \UniHypersequentA_0(\UniMSetSucA) + 1$; i.e., $\UniNormHSeq{\UniHypersequentA'}{} \leq \UniNormHSeq{\UniHypersequentA_0}{} + 1$.
    
    We know that
    $\UniHypersequentA_0 \UniIntCtrRel{(\UniFmlaMultFixed-1)(m-1)}{\UniWeakCtr} \UniHypersequentB
    \UniExtCtrRel{\UniActiveCompFixed}{\text{(EC)}}
    \UniHypersequentA$
    for some hypersequent $\UniHypersequentB$. 
    Let
    $\UniDistinctComp{g}{\UniMSetSucA}$ be the set of those components of $g$ that have succedent $\UniMSetSucA$.
    By definition (see after Definition~\ref{def-der-hone-htwo}), $\UniNormHSeq{\UniHypersequentB}$ is the maximum of: the largest number of components with the same succedent, and the largest multiplicity of a formula occurring in the antecedent of a component. 

    \begin{enumerate}
        \item In case $\UniNormHSeq{\UniHypersequentB}$ is the
        largest number of
        components with the same succedent
        (say $\UniMSetSucA'$), since knotted contraction does not change the succedent of the involved components, and only decreases the multiplicities of formulas, we obtain 
        $\UniNormHSeq{\UniHypersequentA_0} = \UniNormHSeq{g}$.
        We know that
        $\UniMultAnt{g}{s} \leq \UniMultComp{h}{s}+\UniActiveCompFixed$
        for each $s \in g$. Thus
        \[
        \UniNormHSeq{g} =
        \sum_{s \in \UniDistinctComp{g}{\UniMSetSucA'}} 
        \UniMultComp{g}{s}
        \leq
        \left( \sum_{s \in \UniDistinctComp{g}{\UniMSetSucA'}
        } \UniMultComp{h}{s} \right)
        + 
        \UniSetCard{\UniDistinctComp{g}{\UniMSetSucA'}} \cdot \UniActiveCompFixed
        \]
        Note that
        $\UniDistinctComp{g}{\UniMSetSucA'} = \UniDistinctComp{h}{\UniMSetSucA'}$
        (since $\UniEC$ does not make a component disappear).
        Thus 
        $\sum_{s \in \UniDistinctComp{g}{\UniMSetSucA'}} 
        \UniMultComp{h}{s} =
        \sum_{s \in \UniDistinctComp{h}{\UniMSetSucA'}} 
        \UniMultComp{h}{s}
        \leq \UniNormHSeq{h}$.
        Also, $
        \UniSetCard{
        \UniDistinctComp{g}{\UniMSetSucA'}} = \UniSetCard{
        \UniDistinctComp{h}{\UniMSetSucA'}} \leq \UniNormHSeq{h}$.
        Therefore, we have
        $\UniNormHSeq{h_0}=\UniNormHSeq{g} \leq 2 \cdot \UniActiveCompFixed\cdot \UniNormHSeq{h}$.
        Thus
        $\UniNormHSeq{\UniHypersequentA'}{} \leq \UniNormHSeq{\UniHypersequentA_0}{} + 1 \leq  2 \cdot \UniActiveCompFixed\cdot \UniNormHSeq{h} + 1
        \leq 
        (2 \cdot \UniActiveCompFixed + 1)\cdot \UniNormHSeq{h}
        \leq 2\UniSizeHyper{\UniHyperCalcA}\alpha\UniNormHSeq{\UniHypersequentA}$.
        \item In case it is the largest multiplicity of a formula occurring in the antecedent of a component: there is some formula $\UniFmA$ occurring in some component $s \in g$ that witnesses the latter.
        Since 
        $\UniHypersequentA_0 \UniIntCtrRel{(\UniFmlaMultFixed-1)(m-1)}{\UniWeakCtr} \UniHypersequentB$
        and knotted contraction does not change the number of components nor the succedent of components,
        $\UniNormHSeq{h_0} = \UniMultAnt{s'_0}{\UniFmB}$
        for some $\UniFmB$ in the antecedent of a component $s'_0 \in h_0$.
        But then for some $s' \in g$ we have $\UniMultAnt{s_0'}{\UniFmB} 
        \leq 
        \UniMultAnt{s'}{\UniFmB} + \alpha$ by Proposition~\ref{prop:descendands-in-c}. Certainly $\UniMultAnt{s'}{\UniFmB} \leq \UniMultAnt{s}{\UniFmA}$, thus
        $\UniMultAnt{s_0'}{\UniFmB} \leq 
        \UniMultAnt{s}{\UniFmA} + \alpha$,
        and 
        $\UniNormHSeq{h_0} \leq \UniNormHSeq{g} + \alpha$.
        Since $\UniHypersequentB
    \UniExtCtrRel{\UniActiveCompFixed}{\text{(EC)}}
    \UniHypersequentA$ 
    and $\UniEC$ reduces the number of components without making any of them disappear, it must be that
    $\UniNormHSeq{g}=\UniNormHSeq{h}$, thus 
    $\UniNormHSeq{h_0} \leq \UniNormHSeq{h} + \alpha$.
    Finally,
    $\UniNormHSeq{\UniHypersequentA'}{} \leq \UniNormHSeq{\UniHypersequentA_0}{} + 1
    \leq 
    \UniNormHSeq{h} + \alpha + 1 \leq 
    (\alpha + 2)\UniNormHSeq{h}
    \leq 2\UniSizeHyper{\UniHyperCalcA}\alpha\UniNormHSeq{\UniHypersequentA}
    $.
    \end{enumerate}
\end{enumerate}

\noindent In both cases,
$\UniNormHSeq{\UniHypersequentA'} \leq  2\UniSizeHyper{\UniHyperCalcA}\alpha\UniNormHSeq{\UniHypersequentA}$ as desired.

In summary, we have shown that each branch of the backward proof search  in $\UniHyperCalcA \UniSymbDef \UniHFLecRAbsorb$ of a hypersequent $h$ is an $(\UniControlFunctionA, t)$-controlled bad sequence over $\UniNwqoHSeqName{m}{n}$, where $\UniControlFunctionA(x) \UniSymbDef Mx$,  $M:=2\UniSizeHyper{\UniHyperCalcA}\alpha$, $\alpha \UniSymbDef \UniAbsCtrValue$ and $t \UniSymbDef \UniSizeHyper{\UniHypersequentA}$. 
\end{proof}

In order to develop the complexity analysis of provability in the logics under consideration, we will need the following definition, which basically counts the number of sequents that exist under some particular constraints.

\begin{definition}
\label{def:counting-msets-seqs}
Let $\eta, \rho \geq 0$. We define
    \begin{enumerate}
    \item 
    $\UniNMSet{\eta}{\rho} := (\rho+1)^\eta$ as the number of multisets over a set of size $\eta$ with each element having maximum multiplicity $\rho$.
    \item $\UniNDistComp{\eta}{\rho} := (\eta + 1) \cdot \UniNMSet{\eta}{\rho}$ as the number of distinct sequents over $\eta$ distinct formulas where the maximum multiplicity of a formula in the antecedent is $\rho$.
    \end{enumerate}
\end{definition}

\begin{theorem}
\label{fact:flecmnr-ackermannian}
Let $0 < n < m$.
\begin{enumerate}
\item
If $\UniAnaRuleSet$ is a finite set of hypersequent analytic structural rules, then provability in $\UniHFLecRAbsorb$ is in $\UniFGHProbOneAppLevel{\omega^{\omega}}$.
\item 
If $\UniAnaRuleSet$ is a finite set of sequent analytic structural rules, then provability in $\UniSFLecRAbsorb$ is in $\UniFGHProbOneAppLevel{\omega}$.
\end{enumerate}
\end{theorem}
\begin{proof}
(1) By Remark~\ref{r: spacetimeC} it is enough to show that, given a hypersequent $\UniHypersequentA$, there exists a set $T(h)$, consisting of trees of $\UniSubfmlaHyperseqSet$-hypersequents, such that (a) $\UniHypersequentA$ is provable in $\UniHyperCalcA \UniSymbDef \UniHFLecRAbsorb$ iff $T(h)$ contains a proof of $\UniHypersequentA$ and (b) the space occupied by each tree in $T(h)$ is in  $\UniFGHOneAppLevel{\omega^{\omega}}$. 
We will now describe a suitable choice of $T(h)$.

By Lemma~\ref{l: backsearch}, in order to check provability of $\UniHypersequentA$ it is enough to search for a proof that is a subtree of $\BackPSComCtr(\UniHypersequentA)$.
By Proposition~\ref{prop:back-ps-ek-control-bad-seq} the branches in $\BackPSComCtr(\UniHypersequentA)$ are $(f,t)$-controlled bad sequences in the nwqo $\UniNwqoHSeqName{m}{n}$, where $f$ and $t$ are given in the statement of that proposition. 
Thus they correspond to $(f,t)$-controlled bad sequences over $\UniNwqoHSeqNameVar{m}{n}{\UniSubfmlaHyperseqSet_{\UniSizeHyper{\UniHypersequentA}}}$.
Therefore, (a) holds if we take $T(\UniHypersequentA)$ to be the set of trees whose branches are $(f,t)$-controlled bad sequences in the nwqo $\UniNwqoHSeqName{m}{n}$ and whose branching degree at any node is no bigger than $\UniSizeHyper{\UniHyperCalcA}$. 
Given that $\UniSizeHyper{\UniHyperCalcA}$ is constant, to establish (b) we will employ Lemma~\ref{lem:size-of-minimal-proofs}.
We check each of the assumptions of this lemma below.

\begin{itemize}
\item We instantiate the lemma by taking $\alpha = \omega^\omega$, $A$ the set of hypersequents, $\mathfrak{L}(a)$ the set of $\UniSubfmlaHyperseqSet$-hypersequents, where $\UniSubfmlaHyperseqSet$ is the set of all subformulas of $a \in A$, and $\mathcal{Q}=
\{ \UniNwqoHSeqNameVar{m}{n}{\UniSubfmlaHyperseqSet_k}\}_{k \in \UniNaturalSet}$.
    \item 
    By Proposition~\ref{fact:hnwqo-length-theorem},
    $\{ \UniNwqoHSeqNameVar{m}{n}{\UniSubfmlaHyperseqSet_k}\}_{k \in \UniNaturalSet} \sqsubseteq \UniFGHOneAppLevel{\omega^\omega}$.
    \item It is easy to see that, if $\UniSubfmlaHyperseqSet$ is the set of subformulas of $\UniHypersequentA$, we have $\UniSetCard{\UniSubfmlaHyperseqSet} < \UniSizeHyper{h}$, and thus bad sequences over $\UniNwqoHSeqName{m}{n}$ are bad sequences over $\UniNwqoHSeqNameVar{m}{n}{\UniSubfmlaHyperseqSet_{\UniSizeHyper{h}}}$.
    \item
    There is $S \in \UniFGHLevel{<\omega^\omega}$
    such that $\UniSizeHyper{g} \leq S(\UniNorm{g}{}, \UniSizeHyper{h})$ 
    for all $g$ in a tree of $T(h)$.
    Indeed, $g$ has at most 
    $A_1 \UniSymbDef {\UniNorm{g}{}}\cdot{\UniNDistComp{\UniSetCard{\UniSubfmlaHyperseqSet}}{\UniNorm{g}{}}} \leq \UniSizeHyper{h}(\UniNorm{g}{}+1)^{\UniSizeHyper{h}}$
    components (each distinct component has multiplicity at most $\UniNorm{g}{}$, which justifies the first factor, and each formula in an antecedent has at most multiplicity $\UniNorm{g}{}$, which justifies the second factor).
    Also, each component has size at most $A_2 \UniSymbDef c_1 \cdot \UniNorm{g}{} \cdot{\UniSetCard{\UniSubfmlaHyperseqSet}}
    \cdot \UniSizeHyper{\UniHypersequentA} \leq c_1 \cdot \UniNorm{g}{}\UniSizeHyper{\UniHypersequentA}^2$
    (the constant $c_1$ accounts for the succedent, commas and sequent symbol;
    $\UniNorm{g}{} \cdot{\UniSetCard{\UniSubfmlaHyperseqSet}}$
    is the number of formula occurrences in the antecedent, and each formula has size upper bounded by $\UniSizeHyper{\UniHypersequentA}$).
    Thus the size of the said hypersequent is $c_2 \cdot A_1 \cdot A_2 \leq S(\UniNorm{g}{},\UniSizeHyper{h})$, where the constant $c_2$ accounts for the $\VL$ symbols and $S(x,y) \UniSymbDef c_2 \cdot y(x+1)^x \cdot c_1 \cdot xy^2$.
    Clearly $S \in \UniFGHLevel{2}$ is increasing.
    \item Note that since the branches in the trees of $T(h)$ are bad sequences, the values of $k_j$ in the statement of Lemma~\ref{lem:size-of-minimal-proofs} are all $0$, and thus it is enough to take $S'(j,x) \UniSymbDef 0$.
\end{itemize}
Thus in view of Lemma~\ref{lem:size-of-minimal-proofs}, (b) is satisfied, and we are done.

    (2) The same proof-search procedure can be used over $\UniSFLecRAbsorb$ when it is analytic.
    Then the branches will contain only sequents, whose encodings will be elements of
    $\left(\UniMinoringFntSingWqo{\UniWqoExtModRelProd{n}{m}{\UniSetCard{\UniSubfmlaHyperseqSet}}}\right)^{\UniSetCard{\UniSubfmlaHyperseqSet}+1}$.
    Indeed, if a $\UniSubfmlaHyperseqSet$-hypersequent has a single component, it is just a $\UniSubfmlaHyperseqSet$-sequent $\UniSequent{\UniMSetFmA}{\UniMSetSucA}$.
    This will then be encoded (see Definition~\ref{def:encoding-hypersequents}) as the tuple $(\varnothing,\ldots,\varnothing,\{ \UniMSetFmA \}, \varnothing,\ldots,\varnothing)$, where the position of $\{ \UniMSetFmA \}$ depends on the index of $\UniMSetSucA$ in the assumed enumeration of the formulas in $\UniSubfmlaHyperseqSet$.
    So the bad sequences are over this nwqo, and the Ackermannian upper bound follows from Theorem~\ref{fact:length-theo-nm-not-fixed-maj-min-sing}.
\end{proof}

\begin{corollary}
\label{cor-ded-Fw}
Let $0 < n < m$.
\begin{enumerate}
    \item If $\UniAxiomSetA$ is a finite
set of acyclic $\mathcal{P}_3^\flat$ axioms,
then provability and deducibility in 
$\UniAxiomExt{\UniFLeExtLogic{\UniWeakCProp{m}{n}}}{ \UniAxiomSetA}$
are in $\UniFGHProbOneAppLevel{\omega^\omega}$.
    \item If $\UniAxiomSetA$ is a finite
set of $\mathcal{N}_2$ axioms, then provability and deducibility in $\UniAxiomExt{\UniFLeExtLogic{\UniWeakCProp{m}{n}}}{ \UniAxiomSetA}$ are in $\UniFGHProbOneAppLevel{\omega}$.
In particular, provability and deducibility in $\UniFLeExtLogic{\UniWeakCProp{m}{n}}$ are in $\UniFGHProbOneAppLevel{\omega}$.
\end{enumerate}
\end{corollary}
\begin{proof} 
We first consider the claims about provability.
For (1), by the discussion in Section~\ref{s: P3}, provability in
$\UniAxiomExt{\UniFLeExtLogic{\UniWeakCProp{m}{n}}}{ \UniAxiomSetA}$,
for $\UniAxiomSetA$ a finite set of acyclic $\mathcal{P}_3^\flat$ axioms, corresponds to provability in the calculus $\UniHFLecRAbsorb$ for some finite set of hypersequent analytic structural rules $\UniAnaRuleSet$. 
Thus the proof is a direct consequence of Theorem~\ref{fact:flecmnr-ackermannian} (1).
For (2), the result follows from the fact that $\UniAxiomExt{\UniFLeExtLogic{\UniWeakCProp{m}{n}}}{ \UniAxiomSetA}$, for $\UniAxiomSetA$ a (possibly empty) set of $\mathcal{N}_2$ axioms, has an analytic sequent calculus by Lemma~\ref{l: N_2^-0}, and hence the proof is already a direct use of Theorem~\ref{fact:flecmnr-ackermannian}(2).

The deducibility results follow from the provability results and the deduction theorem from~\cite{galatos2022,gavin2019}, which implies that, for any axiomatic extension $\UniLogicA$ of 
$\UniFLeExtLogic{\UniWeakCProp{m}{n}}$,
$\UniFmA_1,\ldots,\UniFmA_k \vdash_{\UniLogicA} \UniFmB$ if, and only if,
$\vdash_{\UniLogicA} (1 \land \UniFmA_1 \land \ldots \land \UniFmA_k)^n \to \UniFmB$
for all $k \geq 0$ and formulas $\UniFmA_1,\ldots,\UniFmA_k,\UniFmB$.
\end{proof}

In Chapter~\ref{sec:lowerbounds} we will establish lower bounds for these logics and we will be able to conclude that  provability and deducibility of $\UniFLeExtLogic{\UniWeakCProp{m}{n}}$ are \ACK-complete.

%% file: tex/weak-weakening-ub.tex
In this chapter, we show how to extend the decidability argument and complexity analysis that was developed in \cite{BalLanRam21LICS} for provability in
$\UniAxiomExt{\UniFLeExtLogic{\UniWProp}}{\UniAxiomSetA}$, 
where $\UniAxiomSetA$ is a set of acyclic $\mathcal{P}_3^\flat$ axioms.
Whenever not explicitly stated, we assume that $0 \leq m < n$ throughout this section.
Similar to what we did in the previous section, our approach will be to work with extensions of the hypersequent calculi $\UniFLeExtHCalc{\UniWeakWProp{m}{n}}$ by finite sets $\UniAnaRuleSet$ of hypersequent analytic structural rules (denoted by $\UniHFLewR$).
The results for $\UniAxiomExt{\UniFLeExtLogic{\UniWeakWProp{m}{n}}}{ \UniAxiomSetA}$ then follow because every $\UniAxiomSetA$ corresponds to such an $\UniAnaRuleSet$.
Additionally, we will not only replace weakening by knotted weakening, but also adapt the underlying algorithm to solve deducibility instead of only provability (recall that complexity upper bounds for the former yield upper bounds for the latter, but not the other way around).
We emphasize that, unlike the case of knotted contraction in the previous section, a suitable deduction theorem is \emph{not} available to transport upper bounds from provability to deducibility in the knotted weakening logics.
Thus, having a proof-search procedure for deducibility has great value for these logics.
The argument is based on the design and analysis of \emph{forward} proof-search over modified versions of $\UniHFLewR$. 
See Remark~\ref{r: spacetimeW} for an overview of this strategy.

The study of deducibility, however, brings an extra concern: what is the space of hypersequents to be explored in the proof search?
The subformula property that holds for provability does not extend in an obvious way to deducibility, since allowing for new initial sequents in derivations breaks cut elimination.
We now show how to adapt a hypersequent calculus that is cut-free (by usual cut-elimination argument) for provability in extensions of $\UniFLExtLogic{}$ into one that is also cut-free for instances of deducibility.

\section{Analytic hypersequent calculi for deducibility}
\label{sec:deducibility-adapts}

Let $\UniHyperCalcA$ be an extension of $\UniFLExtHCalc{}$ by hypersequent analytic structural rules.
Note that in this calculus applications of cut can be eliminated by the usual cut-elimination procedure when we are talking about provability, but, as already mentioned, this does not happen for deducibility.
We now show how to get around this problem.

Henceforth, we consider hypersequent rule schemas that may also contain fixed formulas (or \emph{parameters}) in their hypersequents.
When a rule schema has parameters, we will call it \emph{parametric}.
As expected, instantiation of parametric rule schemas works as in non-parametric schemas, with the difference that the parameters remain unchanged (they are always instantiated to themselves).

For  a finite set $\UniSetFmA$ of formulas, we denote by ${\UniRuleDedSet{\UniSetFmA}}$ the set of parametric rule schemas of the form
\begin{center}   
\AxiomC{$\UniHyperMSetA \VL 
\UniSequent{{\UniMSetFmB, }\UniFmB,\UniMSetFmC}{\UniMSetSucA}$}
\RightLabel{$\UniRuleDed{\UniFmB}$}
\UnaryInfC{$\UniHyperMSetA \VL   
\UniSequent{{\UniMSetFmB, }\UniMSetFmC}{\UniMSetSucA}$}
\DisplayProof
\end{center}
for each $\UniFmB \in \UniSetFmA$,  in which $\UniFmB$ is the only parameter. Extensions of $\UniHyperCalcA$ by $\UniRuleDedSet{\UniSetFmA}$ are of interest because cut-free provability for them precisely  captures $\UniConseqRel{\UniHyperCalcA}$ (that is, the consequence relation of $\UniHyperCalcA$) in the sense presented below.
Recall from Section~\ref{sec:prelims-proof-theory} that $\UniDelCut{(\cdot)}$ deletes the cut rule from the input calculus.

\begin{lemma}
\label{fact:cut-restrict-weak-weakening}
{
For all finite
$\UniSetFmA \subseteq\UniLangSet{\UniPropVars}$
and all hypersequents $\UniHypersequentA$,
\[
\UniSet{\UniSequent{}{\UniFmB} \mid \UniFmB \in \UniSetFmA} \UniHyperDerivRel{{\UniHyperCalcA}} \UniHypersequentA
\text{ if, and only if, }
\UniEmptySet
\UniHyperDerivRel{\UniDelCut{{\UniExtCalcDed{\UniHyperCalcA}{\UniRuleDedSet{\UniSetFmA}}}}}
\UniHypersequentA.\]
Therefore, for all formulas $\UniFmA$,
\[
\UniSetFmA
\UniConseqRel{\UniHyperCalcA}
\UniFmA
\text{ if, and only if, }
\UniEmptySet
\UniConseqRel{\UniDelCut{{\UniExtCalcDed{\UniHyperCalcA}{\UniRuleDedSet{\UniSetFmA}}}}}
\UniFmA.
\]}
\end{lemma}
\begin{proof}
For the right-to-left direction, we replace applications of $\UniRuleDed{\UniFmB}$ with applications of cut using as left premises sequents of the form $\UniSequent{}{\UniFmB}$. 
The other direction amounts to showing that we can eliminate applications of the cut rule when we have the rules $\UniRuleDed{\UniFmB}$ at our disposal.
Here we can follow the usual cut elimination procedure, but now we have to consider a new case: when cut is applied with $\UniSequent{}{\UniFmB}$ (some $\UniFmB \in \UniSetFmA$) as left premise (notice that such sequent cannot appear as right premise).
In that case, we replace the application of cut with the  appropriate application of $\UniRuleDed{\UniFmB}$.
Therefore, having eliminated cut, we can simulate every deduction $\UniSetFmA \UniConseqRel{\UniHyperCalcA} \UniFmA$ by rules in $\UniDelCut{{\UniExtCalcDed{\UniHyperCalcA}{\UniRuleDedSet{\UniSetFmA}}}}$ plus the initial sequents $\UniSequent{}{\UniFmB}$ (where $\UniFmB \in \UniSetFmA$). 
Moreover, each $\UniSequent{}{\UniFmB}$ can be obtained by applying $\UniRuleDed{\UniFmB}$ to the initial sequent $\UniFmB \UniSequent{}{\UniFmB}$, which is already provable in $\UniDelCut{{\UniExtCalcDed{\UniHyperCalcA}{\UniRuleDedSet{\UniSetFmA}}}}$.
\end{proof}

The above result shows that
$\UniDelCut{{\UniExtCalcDed{\UniHyperCalcA}{\UniRuleDedSet{\UniSetFmA}}}}$ expresses deducibility in ${{\UniHyperCalcA}}$ from a set $\UniSetFmA$ of assumptions; note that when $\UniSetFmA = \UniEmptySet$ this becomes provability in ${\UniHyperCalcA}$ (since cut is eliminable in $\UniHyperCalcA$).
Now we prove the desired analyticity result for this calculus, relative to a modified subformula property.

\begin{theorem}
    \label{the:subformula_property}
    For all finite $\UniSetFmA \subseteq \UniLangSet{\UniPropVars}$ and all hypersequents $\UniHypersequentA$, we have 
    $\UniEmptySet\UniHyperDerivRel{\UniDelCut{{\UniExtCalcDed{\UniHyperCalcA}{\UniRuleDedSet{\UniSetFmA}}}}}
    \UniHypersequentA$
    iff there is a proof of
    $\UniHypersequentA$
    in $\UniDelCut{{\UniExtCalcDed{\UniHyperCalcA}{\UniRuleDedSet{\UniSetFmA}}}}$
    in which only subformulas of the formulas in $\UniHypersequentA$ and in $\UniSetFmA$ appear.
\end{theorem}
\begin{proof}
Note that the right-to-left direction
is obvious from the definition of
$\UniHyperDerivRel{\UniDelCut{{\UniExtCalcDed{\UniHyperCalcA}{\UniRuleDedSet{\UniSetFmA}}}}}$,
so we proceed with the other direction.
Let $\UniSetFmA$ be a finite set of formulas and, for a hypersequent $\UniHypersequentA$, let $\Xi(\UniHypersequentA)$ be the set of subformulas of the formulas in $\UniHypersequentA$ and in $\UniSetFmA$. 
By induction on the structure of cut-free derivations in
$\UniDelCut{{\UniExtCalcDed{\UniHyperCalcA}{\UniRuleDedSet{\UniSetFmA}}}}$,
it is enough to prove $P(\ProofA) \UniSymbDef $ ``if $\ProofA$ is a proof of the hypersequent $\UniHypersequentA'$, then there is a derivation $\ProofA'$ that is also a proof of $\UniHypersequentA'$ and contains only formulas in $\Xi(\UniHypersequentA')$'', for all hypersequents $\UniHypersequentA'$.
When $\ProofA$ has a single node, we are obviously done.
For the inductive step, since all rules of $\UniHyperCalcA$ satisfy the strong subformula property, we only need to check the rules in 
$\UniRuleDedSet{\UniSetFmA}$.
If $\UniHypersequentB \VL \UniSequent{{\Delta }, \UniMSetFmA}{
\UniMSetSucA
}$ 
follows from an application of $\UniRuleDed{\UniFmB}$ on a proof of 
$\UniHypersequentB\VL\UniSequent{{\Delta }, \UniFmB, \UniMSetFmA}{
\UniMSetSucA
}$,
then by the induction hypothesis there is a proof of the latter using only formulas in
$\Xi(\UniHypersequentB \VL \UniSequent{{\Delta }, \UniFmB, \UniMSetFmA}{
\UniMSetSucA
})$. 
We apply to this proof the same rule $\UniRuleDed{\UniFmB}$ and we get  $\UniHypersequentB \VL \UniSequent{{\Delta }, \UniMSetFmA}{\UniMSetSucA}$.
\end{proof}

In particular, we obtain the following corollary for the logics we are studying in this work; we reference the correspondence between axioms and rules given in Section~\ref{s: P3}.

\begin{corollary}\label{cor:ded-weak-is-prov-mod-calc}
Let $\UniWEProp{\vec a}$ be a weak exchange axiom, $\UniWeakKProp{m}{n}$ be a knotted axiom, $\UniAxiomSetA$ be a finite set of  axioms in $\mathcal{P}_3^\flat$ and $\UniAnaRuleSet$ be the  corresponding set of hypersequent analytic structural rules.
Then
\begin{enumerate}
    \item $
\UniSetFmA
\UniConseqRel{\UniAxiomExt{\UniFLweExtLogic{\vec a}{\UniWeakKProp{m}{n}}}{\UniAxiomSetA}}
\UniFmA
\text{ if, and only if, }
\UniEmptySet
\UniConseqRel{\UniHFLwekRAug{\UniSetFmA}}
\UniFmA
$; and
\item 
$\UniSetFmA
\UniConseqRel{\UniAxiomExt{\UniFLweExtLogic{\vec a}{\UniWeakKProp{m}{n}}}{\UniAxiomSetA}}
\UniFmA$
if, and only if, there is a proof of $\UniSequent{}{\UniFmA}$ in the calculus $\UniHFLwekRAug{\UniSetFmA}$ using only subformulas of the formulas in $\UniSetFmA \cup \{\UniFmA\}$.
\end{enumerate}
\end{corollary}

\section[Upper bounds for commutative logics with knotted weakening]{Decidability---via proof search---and upper bounds for deducibility (and thus provability) in commutative logics with knotted weakening}
\label{sec:decid-ub-ww-proof}

The results from the previous section tell us, in particular, that in order to decide whether a formula $\UniFmA$ follows from a finite set of formulas $\UniSetFmA$ in $\UniAxiomExt{\UniFLeExtLogic{\UniWeakWProp{m}{n}}}{ \UniAxiomSetA}$ (that is, deducibility for this logic), it is enough to search for a proof of
$\UniSequent{}{\UniFmA}$ in $\UniHFLewRAug{\UniSetFmA}$ considering only subformulas of the formulas in $\UniSetFmA \cup \{ \UniFmA \}$
(by Corollary~\ref{cor:ded-weak-is-prov-mod-calc}).
With this in mind, we now describe a \emph{forward} proof-search algorithm for $\UniHFLewRAug{\UniSetFmA}$ that is uniform on $\UniSetFmA$ (i.e., the same algorithm works for any choice of $\UniSetFmA$), thus providing a decision procedure for deducibility in
$\UniAxiomExt{\UniFLeExtLogic{\UniWeakWProp{m}{n}}}{ \UniAxiomSetA}$.
We will prove its termination and provide upper bounds on its complexity using knotted wqos and the length theorems of Chapter~\ref{sec:knotted-wqos}.

Expanding on the above discussion, when searching for a derivation of a given conclusion from a given set of hypotheses, we can limit ourselves to sequents over formulas in the finite set $\UniSubfmlaHyperseqSet$ of subformulas of the conclusion and the hypotheses, which gives us a manageable search space. 
Since usual weakening (one of the rules we want to cover) can introduce arbitrary formulas in the forward search, resulting in an infinite search space, limiting the search space to $\UniSubfmlaHyperseqSet$-hypersequents is necessary, as we did in the previous section.
Moreover, it will be important to be able to express when a hypersequent follows from another via applications of (EW), $\UniWeakWkn$ and (EC).

\begin{definition}
    \label{def:ordering-hsqe}
    For $\UniSubfmlaHyperseqSet$-hypersequents 
    $\UniHypersequentA_1$ and $\UniHypersequentA_2$,
    we write $\UniHypersequentA_1\UniHyperseqWknWqo{m}{n}{\UniSubfmlaHyperseqSet}~\UniHypersequentA_2$
    if, and only if,
    $\UniHypersequentA_2$
    is derivable from $\UniHypersequentA_1$ by
    $\UniEW$, $\UniWeakWkn$ and $\UniEC$.
\end{definition}

Recalling Remark~\ref{r: spacetimeW}, the general idea of the forward proof-search algorithm is to start from the simplest derivable hypersequents (essentially, the instances of initial rule schemas), and then apply the proof rules of the calculus until reaching a predetermined stage where we can definitively tell whether the input hypersequent is derivable or not.
We see each stage of rule applications as expanding the initial set of hypersequents with new hypersequents, forming an increasing sequence of sets of hypersequents 
$\UniDerivSet_0, \UniDerivSet_1, \UniDerivSet_2, \ldots$ provable from the assumptions.
To guarantee termination (i.e., that the sequence is finite), we will be careful with how we apply the rules (EW), $\UniWeakWkn$ and (EC) when building each stage  $\UniDerivSet_i$.

\begin{definition}\label{def-derive-sets}
Let $\UniSubfmlaHyperseqSet$ be a finite set of formulas closed under subformulas, and set $\UniHyperCalcA \UniSymbDef \UniHFLewRAug{\UniSetFmA}$.
We define $\UniDerivSet_{0}$ as the set of all instances of initial rule schemas in~$\UniHyperCalcA$ such that
\begin{enumerate}[a)]
\item formula-variables are instantiated to elements of~$\UniSubfmlaHyperseqSet$;
\item succedent-variables are instantiated to an element in~$\UniSubfmlaHyperseqSet$ or as empty; 
\item multiset-variables are instantiated to $\UniSubfmlaHyperseqSet$-multisets with maximum multiplicity at most $n-1$ (i.e., multisets $M~:~\UniSubfmlaHyperseqSet \to \UniNaturalSet$ such that $M(\UniFmA) \leq n-1 \text{ for each } \UniFmA \in \UniSubfmlaHyperseqSet$);
\item hypersequent-variables are instantiated as empty.
\end{enumerate}

\noindent For $i > 0$, we define
$\UniDerivSet_{i+1}:=\UniDerivSet_i \cup \partial\UniDerivSet_i$,
where $\partial\UniDerivSet_i$ is the set of $\UniSubfmlaHyperseqSet$-hypersequents $\UniHypersequentA$ satisfying the following conditions:
\begin{enumerate}
    \item $\UniHypersequentA_1 \cdots \UniHypersequentA_p/h$ is a rule instance of $\UniHyperCalcA$ such that, for all $1 \leq k \leq p$, there is $\minus{\UniHypersequentA_k}\in \UniDerivSet_i$ with $\minus{\UniHypersequentA_k}\UniHyperseqWknWqo{m}{n}{\UniSubfmlaHyperseqSet}~{\UniHypersequentA_k}$;
    \item the multiplicity of
    every formula in the antecedent of a component of $\UniHypersequentA$ is at most $ 
    \UniNormHSeq{\UniDerivSet_i}
    \UniMulttNumbers \UniSizeHyper{\UniHyperCalcA} \UniMulttNumbers n$
    (where
    $\UniNormHSeq{\UniDerivSet_i} 
    \UniSymbDef
    \max_{\UniHypersequentA \in \UniDerivSet_i} \UniNormHSeq{\UniHypersequentA}$)
    \item the frequency of each component in $\UniHypersequentA$ is at most $\UniSizeHyper{\UniHyperCalcA}$; and
    \item there does not exist $\UniHypersequentA'\in \UniDerivSet_i$ such that $\UniHypersequentA'\UniHyperseqWknWqo{m}{n}{\UniSubfmlaHyperseqSet} \UniHypersequentA$.
\end{enumerate}
\end{definition}

The first condition states that {$\partial\UniDerivSet_i$} contains $\UniSubfmlaHyperseqSet$-hypersequents obtainable from the upward
$\UniHyperseqWknWqo{m}{n}{\UniSubfmlaHyperseqSet}$-closure of $\UniDerivSet_i$ by applying a single rule. 
Conditions two and three ensure that the conclusion of that rule instance is ``small'' and that each stage can be computed from the previous one in finite time. 
The final condition states that redundant elements are not added, i.e., an element of $\UniDerivSet_{i+1}{\setminus}\UniDerivSet_i$ cannot be obtained from $\UniDerivSet_i$ by applications of (EW), $\UniWeakWkn$ and (EC).

Since~$\UniSubfmlaHyperseqSet$, $\UniSetFmA$ and~$\UniHyperCalcA$ are finite, $\UniDerivSet_{0}$ is a finite set of elements like $(\UniSequent{\UniPropA}{\UniPropA})$
(for each propositional variable
$\UniPropA\in\UniSubfmlaHyperseqSet$), $(\UniSequent{}{1})$ and $(\UniSequent{0}{})$. 
Also, directly from the definition, we get
$\UniDerivSet_i \subseteq \UniDerivSet_{i+1}$ for every~$i \geq 0$.
We now prove a key result: computing the levels $\UniDerivSet_i$ is enough to reach any $\UniSubfmlaHyperseqSet$-hypersequent
derivable in $\UniHyperCalcA$.

\begin{lemma}
\label{lem:SN}
Let $0 \leq m < n$ and $\UniSubfmlaHyperseqSet$ be the set of formulas containing all subformulas of the formulas in $\UniSetFmA$ and $\UniHypersequentA$.
If
$\UniHyperDerivRel{\UniHFLewRAug{\UniSetFmA}} \UniHypersequentA$,
then there is~$N \in \UniNaturalSet$ and~$\UniHypersequentA' \in \UniDerivSet_N$ such that~$\UniHypersequentA' \UniHyperseqWknWqo{m}{n}{\UniSubfmlaHyperseqSet}  \UniHypersequentA$.
\end{lemma}
\begin{proof}
We will prove the statement  by induction on the height of a derivation $\UniDerivationA$ of~$\UniHypersequentA$ in $\UniHFLewRAug{\UniSetFmA}$.

\textit{Base case}: If $\UniDerivationA$ has height $1$, then $\UniHypersequentA$ is an instance of an initial rule schema. Though $\UniDerivSet_0$ consists of a proper subset of initial rule instances (hypersequent-variables were instantiated as empty), $\UniHypersequentA$ can be obtained by applying (EW) or $\UniWeakWkn$ to a suitable element of $\UniDerivSet_0$.
More precisely, if the instantiation of the hypersequent variable is not empty, thus not covered by restriction (d), we apply (EW) to obtain these additional components, and if the maximum multiplicity is restricted by (c), then we apply $\UniWeakWkn$ to obtain any multiplicity bigger than $n-1$.
We can do this because in between $m$ and $n-1$ (recall that $m \leq n-1$) there are $n-m$ numbers. 
Since there are $n-m$ equivalence classes modulo $n-m$, each such number must belong to a different
class. 
Thus, to obtain a multiplicity in equivalence class $0 \leq c \leq n-m-1$, we apply $\UniWeakWkn$ over the sequent with the amount of formulas in that equivalence class.

\textit{Inductive case}: If $\UniDerivationA$ has height greater than one and the last step of the proof is an instance of the rule schema~$\UniRuleSchemaA$, then, by the subformula property given in Theorem~\ref{the:subformula_property}, every subformula in each premise~$\UniHypersequentA_{k}$ in this $\UniRuleSchemaA$-instance is in~$\UniSubfmlaHyperseqSet$. 
By the induction hypothesis applied to the $k^{\text{th}}$ premise, there exists $N_{k}$ and a hypersequent $\minus{\UniHypersequentA_k} \in \UniDerivSet_{N_{k}}$ such that $\minus{\UniHypersequentA_k} \UniHyperseqWknWqo{m}{n}{\UniSubfmlaHyperseqSet}  \UniHypersequentA_k$. 
Since $\UniDerivSet_i \subseteq \UniDerivSet_{i+1}$ for every~$i$, we have  $\minus{\UniHypersequentA_k}\in \UniDerivSet_N$, for $N:=\max_k N_{k}$.
Let~$\UniDerivationA_{k}$ be the derivation witnessing $\minus{\UniHypersequentA_{k}} \UniHyperseqWknWqo{m}{n}{\UniSubfmlaHyperseqSet}  \UniHypersequentA_{k}$, i.e., a sequence of applications of (EW), $\UniWeakWkn$ and (EC) that takes $\minus{\UniHypersequentA_{k}}$ to $\UniHypersequentA_{k}$. 

By Definition~\ref{def-derive-sets}, if the antecedent of every component contains each formula with multiplicity $\leq \UniNormHSeq{\UniDerivSet_i} \UniMulttNumbers \UniSizeHyper{\UniHyperCalcA} \UniMulttNumbers n$ 
and every component has multiplicity $\leq\UniSizeHyper{\UniHyperCalcA}$, then either $\UniHypersequentA\in \UniDerivSet_{N+1}$ or there is some $\UniHypersequentA'\in \UniDerivSet_N$ such that $\UniHypersequentA'\UniHyperseqWknWqo{m}{n}{\UniSubfmlaHyperseqSet}\UniHypersequentA$, as desired. 
For the remaining cases, we proceed by a sub-induction on the value~$\UniSizeHyper{\UniHypersequentA}$:

$\bullet$
Some component~$s$ in~$\UniHypersequentA$ 
has a formula $\UniFmA$ in its antecedent whose multiplicity~$\alpha$ is 
$> 
\UniNormHSeq{\UniDerivSet_i} 
\UniMulttNumbers \UniSizeHyper{\UniHyperCalcA} \UniMulttNumbers n$. 
Then there are two possibilities:
\begin{enumerate}
\item \emph{$s$ is in the instantiation of the hypersequent-variable}: We call the corresponding component of~$s$ in each premise~$\UniHypersequentA_{k}$ a \textit{marked component}.
In every component of the (single) leaf $\minus{\UniHypersequentA_{k}}$ of $\UniDerivationA_k$ the multiplicity of $\UniFmA$ is $\leq\UniNormHSeq{\UniDerivSet_N}$ since $\minus{\UniHypersequentA_{k}}\in \UniDerivSet_N$, and hence $<\alpha$. 
This means that, tracing upwards from root towards the leaf, there is a `final ancestor' hypersequent that witnesses the first change in the multiplicity of $\UniFmA$. 
To be precise, if the marked component is the active component of the conclusion of $\UniWeakWkn$ then it is a final ancestor; if the marked component is the active component of the conclusion of (EW) then it is a final ancestor; if the marked component is the active component of the conclusion of (EC) then the corresponding two components in the premise are active components; if a marked-component is not active in a rule instance then the corresponding component in its premise is a marked component.

We obtain $\UniDerivationA_k'$ by replacing every final ancestor of type (EW) with a (EW) rule instance where $\UniFmA$ has multiplicity $\alpha-(n-m)$ instead of $\alpha$ (all else remains the same); we omit a $\UniWeakWkn$ rule whose conclusion is a final ancestor of type $\UniWeakWkn$. In this way we obtain a hypersequent $\UniHypersequentA_k'$ that is the same as~$\UniHypersequentA_k$ except that the number of occurrences of the formula~$\UniFmA$ in the antecedent of the marked component is $n-m$ fewer than before. 
Also $
\minus{\UniHypersequentA_{k}}
\UniHyperseqWknWqo{m}{n}{\UniSubfmlaHyperseqSet}
\UniHypersequentA_k'
\UniHyperseqWknWqo{m}{n}{\UniSubfmlaHyperseqSet} 
\UniHypersequentA_k$ 
(the latter application of $\UniWeakWkn$ is possible since $\alpha\geq n$ hence $\alpha-(n-m)\geq m$). 
By applying the same rule as before but now with the $\{\UniHypersequentA_k'\}_k$ as premises we obtain a hypersequent $\UniHypersequentA'$ such that 
$\UniHypersequentA' 
\UniHyperseqWknWqo{m}{n}{\UniSubfmlaHyperseqSet} 
\UniHypersequentA$ and $\hypsize{\UniHypersequentA'}<\hypsize{\UniHypersequentA}$. 
By the sub-induction hypothesis there is $M$ and $\UniHypersequentA''$ such that $\UniHypersequentA''\in \UniDerivSet_M$ and 
$\UniHypersequentA''
\UniHyperseqWknWqo{m}{n}{\UniSubfmlaHyperseqSet}
\UniHypersequentA'$. Since 
$\UniHypersequentA'
\UniHyperseqWknWqo{m}{n}{\UniSubfmlaHyperseqSet} 
\UniHypersequentA$ we are done.

\item 
\emph{$s$~is an active component---i.e., the component not in the instantiation of the hypersequent-variable---of the rule instance}: 
Since the number of occurrences of schematic-variables in any rule schema is $\leq \UniSizeHyper{\UniHyperCalcA}$ there must be some multiset-variable~$\UniMSetFmA$ (occurring exactly once in the conclusion, by linear conclusion) whose instantiation contains $\beta\geq 
\UniNormHSeq{\UniDerivSet_N}\UniMulttNumbers n$ 
occurrences of~$\UniFmA$.
In every component of the (single) leaf $\minus{\UniHypersequentA_{k}}$ of $\UniDerivationA_k$ the multiplicity of $\UniFmA$ is  
$\leq\UniNormHSeq{\UniDerivSet_N}$
(and thus $<\alpha$) since $\minus{\UniHypersequentA_{k}} \in \UniDerivSet_N$. 
This means that, as before, tracing upwards from root towards the leaf, there is a `final ancestor' hypersequent that witnesses the first change in the multiplicity of $\UniFmA$.

We call each component in each premise~$\UniHypersequentA_{k}$ that corresponds to a component containing~$\UniMSetFmA$ in the rule schema a \textit{marked component}. 
We trace the ancestor(s) of $\UniFmA^{\alpha}$ upwards in $\UniDerivationA_k$ from the root $\UniHypersequentA_k$ until we encounter final ancestor(s) that witnesses the change $\UniFmA^{\alpha}\mapsto \UniFmA^{\alpha-(n-m)}$ (due to $\UniWeakWkn$) or $\UniFmA^{\alpha}\mapsto \UniEmptySet$ (due to (EW)). 
We obtain a hypersequent $\UniDerivationA_k'$ by deleting the $\UniWeakWkn$ rule, and replacing it with a (EW) rule whose active component has $n-m$ fewer copies of $\UniFmA$. 
In this way we obtain $\UniHypersequentA_k'$ that is the same as~$\UniHypersequentA_k$ except that the number of occurrences of the formula~$\UniFmA$ in the antecedent of the marked component is $n-m$ fewer than before. Also $\minus{\UniHypersequentA_{k}}
\UniHyperseqWknWqo{m}{n}{\UniSubfmlaHyperseqSet} 
\UniHypersequentA_k'
\UniHyperseqWknWqo{m}{n}{\UniSubfmlaHyperseqSet} 
\UniHypersequentA_k$ 
(the latter application of $\UniWeakWkn$ is possible since $\alpha\geq n$ hence $\alpha-(n-m)\geq m$). 
By applying the same rule as before but this time to $\{\UniHypersequentA_k'\}_k$ as premises we obtain $\UniHypersequentA'$ such that $\UniHypersequentA'
\UniHyperseqWknWqo{m}{n}{\UniSubfmlaHyperseqSet} 
\UniHypersequentA$ and $\hypsize{\UniHypersequentA'}<\hypsize{h}$. 
By the sub-induction hypothesis there is $M > 0$ and $\UniHypersequentA''$ such that $\UniHypersequentA''\in \UniDerivSet_M$ and 
$\UniHypersequentA''
\UniHyperseqWknWqo{m}{n}{\UniSubfmlaHyperseqSet}  
\UniHypersequentA'$. 
Thus 
$\UniHypersequentA''\UniHyperseqWknWqo{m}{n}{\UniSubfmlaHyperseqSet} \UniHypersequentA$ 
and we are done.
\end{enumerate}

$\bullet$ Some component~$\UniSequentA$ in~$\UniHypersequentA$ has frequency strictly bigger than $\UniSizeHyper{\UniHyperCalcA}$: 
Since the number of components in the conclusion of every rule schema is at most $\UniSizeHyper{\UniHyperCalcA}$, the instantiation of the hypersequent-variable in the conclusion of the rule must have the form $s \VL s \VL \UniHypersequentB$. 
This hypersequent-variable occurs in every premise~$\UniHypersequentA_k$ of the rule schema. 
Since $\minus{\UniHypersequentA_k} 
\UniHyperseqWknWqo{m}{n}{\UniSubfmlaHyperseqSet}  
\UniHypersequentA_k$ for some $\minus{\UniHypersequentA_k}\in \UniDerivSet_{N_{k}}$, we get 
$\minus{\UniHypersequentA_k}
\UniHyperseqWknWqo{m}{n}{\UniSubfmlaHyperseqSet}  
\UniHypersequentA_k^*$ where $\UniHypersequentA_k^*$ is the same as $\UniHypersequentA_k$ except $\UniSequentA \VL \UniSequentA$ has been converted to $\UniSequentA$ using (EC). 
Applying the rule to the premises of the form $\UniHypersequentA_k^*$, we obtain a sequent $\UniHypersequentA^*$ which  is the same as $\UniHypersequentA$ except $\UniSequentA \VL \UniSequentA$ has been converted to $\UniSequentA$. 
Since $\UniSizeHyper{\UniHypersequentA^*}$ is smaller than $\UniSizeHyper{\UniHypersequentA}$, by the sub-induction hypothesis we obtain that there is $M > 0$ such that $\UniHypersequentA'\in \UniDerivSet_{M}$ and 
$\UniHypersequentA'
\UniHyperseqWknWqo{m}{n}{\UniSubfmlaHyperseqSet}   
\UniHypersequentA^*$. 
Since 
$\UniHypersequentA^*
\UniHyperseqWknWqo{m}{n}{\UniSubfmlaHyperseqSet}   
\UniHypersequentA$ 
we have 
$\UniHypersequentA'
\UniHyperseqWknWqo{m}{n}{\UniSubfmlaHyperseqSet}   
\UniHypersequentA$, as desired.
\qedhere
\end{proof}

A proof-search algorithm for checking the provability in $\UniHFLewRAug{\UniSetFmA}$ of a hypersequent $\UniHypersequentA$  proceeds as follows. 
We take $\UniSubfmlaHyperseqSet$ to be the set of subformulas of the formulas in $\UniSetFmA$ and in $h$, and relative to that we generate the sets $\UniDerivSet_0, \UniDerivSet_1, \UniDerivSet_2, \ldots$, according to Definition~\ref{def-derive-sets}: for each $i$, we include in $\UniDerivSet_{i+1}$  instantiations of rule schemas to hypersequents from the $\UniHyperseqWknWqo{m}{n}{\UniSubfmlaHyperseqSet}$-closure of $\UniDerivSet_{i}$, excluding the hypersequents of the closure itself; we note that this can be done in exponential time in the representation of $\UniDerivSet_i$.  
We will show right below that this process terminates  at some stage $N$, i.e., $\UniDerivSet_N=\UniDerivSet_{N+1}$, hence also $\UniDerivSet_N=\UniDerivSet_{N+k}$ for every $k$.
We will sometimes call $N$ the \emph{stabilizing point} of this sequence.
Finally, it is enough to check whether $\UniHypersequentA' \UniHyperseqWknWqo{m}{n}{\UniSubfmlaHyperseqSet}  \UniHypersequentA$ for some $\UniHypersequentA' \in \UniDerivSet_{N}$ to determine provability of $\UniHypersequentA$.
Below we provide the desired complexity bounds, but employing a nondeterministic version of this procedure following the general strategy of Remark~\ref{r: spacetimeW}.

We appeal again to wqo theory for proving termination of the procedure describe above.
First of all, it is not hard to check that the following is a nqo.

\begin{definition}
\label{def:nqo-weakening-def-hyper}
Given $0 \leq m < n$, we define $$\UniNwqoWknHSeqName{m}{n} \UniSymbDef \UniStruct{\UniOmegaHypersequentsSet{\UniSubfmlaHyperseqSet}, \UniHyperseqWknWqo{m}{n}{\UniSubfmlaHyperseqSet},
\UniNormHSeq{\cdot}}.$$
\end{definition}

Then we have the analogous of Lemma~\ref{lem:hyperseq-nwqo-nat-corresp} (recall what is $(\cdot)^\#$ in Definition~\ref{def:encoding-hypersequents}), where the only difference is the use of the majoring extension instead of the minoring extension.
The proof is also analogous, as we show.

\begin{lemma}
    \label{fact:weak-majoring-corresp}
    For all $0 \leq m < n$ and all $\UniSubfmlaHyperseqSet$-hypersequents $\UniHypersequentA_1$, $\UniHypersequentA_2$, $\UniHypersequentA$,
    \begin{enumerate}
        \item $\UniHypersequentA_1\UniHyperseqWknWqo{m}{n}{\UniSubfmlaHyperseqSet}\UniHypersequentA_2
    \text{ if, and only if, }
    \UniHyperTransNat{\UniHypersequentA_1}
\UniWqoRel{\UniMajoringFntWqo{{\UniWqoExtModRelProd{n}{m}{\UniSetCard{\UniSubfmlaHyperseqSet}}}}}^{\UniSetCard{\UniSubfmlaHyperseqSet}+1}
    \UniHyperTransNat{\UniHypersequentA_2}$.
    \item $
\UniNorm{\UniHyperTransNat{\UniHypersequentA}}{\UniMajoringFntWqo{\UniWqoExtModRelProd{n}{m}{\UniSetCard{\UniSubfmlaHyperseqSet}}}^{\UniSetCard{\UniSubfmlaHyperseqSet} + 1}}
    \leq
    \UniNormHSeq{\UniHypersequentA}
    $.
    \end{enumerate}
\end{lemma}
\begin{proof}
    For item (1), the observations in the proof of Lemma~\ref{lem:hyperseq-nwqo-nat-corresp} regarding applications of the knotted rule over a single formula and the ordering of $\UniWqoExtModRelProd{n}{m}{\UniSetCard{\UniSubfmlaHyperseqSet}}$ apply analogously here.
    For the hypersequent aspects of these orderings, it is enough to observe that when passing from $\UniHypersequentA_1$ to $\UniHypersequentA_2$ via $\UniEC$ or $\UniEW$, the components of $\UniHypersequentA_1$ are preserved up to a modification in multiplicity.
    The mapping $(\cdot)^\#$ only records applications of $\UniEW$ when a new component is introduced; but this introduction preserves the ordering of 
    $\UniMajoringFntWqo{\UniWqoExtModRelProd{n}{m}{\UniSetCard{\UniSubfmlaHyperseqSet}}}^{\UniSetCard{\UniSubfmlaHyperseqSet} + 1}$,
    since what matters for it is that each component in $\UniHypersequentA_1$ has a correspondent in $\UniHypersequentA_2$ (and not the opposite, say).
    The proof for item (2) is the same as for Lemma~\ref{lem:hyperseq-nwqo-nat-corresp} (2).
\end{proof}

As in Chapter~\ref{sec:ub-wc}, for each $k \geq 0$, if we fix $\UniSubfmlaHyperseqSet_k \UniSymbDef \{ \UniPropA_1,\ldots, \UniPropA_k\}$, then whenever $\UniSetCard{\UniSubfmlaHyperseqSet} = k$ we will have
$
\UniNwqoWknHSeqName{m}{n}
\cong
\UniNwqoWknHSeqNameVar{m}{n}{\UniSubfmlaHyperseqSet_k}$.
From this and the previous facts, we obtain the following result, analogous to Proposition~\ref{fact:hnwqo-length-theorem}.

\begin{proposition}
\label{fact:hnwqo-wkn-length-theorem}    
For all finite sets of formulas $\UniSubfmlaHyperseqSet$,
$\UniNwqoWknHSeqName{m}{n}$ is a nwqo and $\UniNwqoWknHSeqName{m}{n} \sqsubseteq \UniFGHOneAppLevel{\omega^{\UniSetCard{\UniSubfmlaHyperseqSet}}}$. Moreover, $\{ \UniNwqoWknHSeqNameVar{m}{n}{\UniSubfmlaHyperseqSet_k}\}_{k \in \UniNaturalSet} \sqsubseteq \UniFGHOneAppLevel{\omega^\omega}$.
\end{proposition}
\begin{proof}
Analogous to the proof of Proposition~\ref{fact:hnwqo-length-theorem}, this time using Lemma~\ref{fact:weak-majoring-corresp}.
\end{proof}

To show termination, as promised above, we assume towards a contradiction that no stabilizing $N$ exists, i.e., $\UniDerivSet_0\subset \UniDerivSet_1\subset \UniDerivSet_2 \subset \ldots$; so there exists an infinite sequence $\UniHypersequentA_0, \UniHypersequentA_1, \ldots $ such that $\UniHypersequentA_0 \in \UniDerivSet_0$ and $\UniHypersequentA_{i}\in \UniDerivSet_{i}\setminus \UniDerivSet_{i-1}$ for every $i \geq 1$. 
By condition 4 in Definition~\ref{def-derive-sets}, there is no pair $i<j$ with $\UniHypersequentA_i\UniHyperseqWknWqo{m}{n}{\UniSubfmlaHyperseqSet}\UniHypersequentA_j$. 
Thus $\UniHypersequentA_0, \UniHypersequentA_1, \ldots $ is an infinite bad sequence over $\UniNwqoWknHSeqName{m}{n}$, contradicting Proposition~\ref{fact:hnwqo-wkn-length-theorem}. 

Next, we extend Definition~\ref{def:counting-msets-seqs} with the following, in order to facilitate the complexity analysis that follows.

\begin{definition}
\label{def:counting-hseqs}
If $\eta, \rho, \mu \geq 0$, we define
$\UniNDistHSeq{\eta}{\rho}{\mu}
\UniSymbDef(\mu+1)^{\UniNDistComp{\eta}{\rho}}$
as the number of hypersequents over $\eta$ formulas such that the maximum multiplicity of a component is $\mu$, and the multiplicity of a formula in an antecedent is at most $\rho$.
\end{definition}

For the complexity analysis, we first observe that an upper bound for the length of the bad sequence
$\{{\UniHypersequentA_i}\}_{i\in\UniNaturalSetNN}$ 
applies to $N$. 
For our length theorem (Proposition~\ref{fact:hnwqo-wkn-length-theorem}) it is enough to show that this bad sequence is $(\UniControlFunctionA,\UniControlParam)$-controlled for some primitive recursive control function $\UniControlFunctionA$ and some $\UniControlParam \in \UniNaturalSet$.

For what follows, note that  
$\UniSetCard{\UniSubfmlaHyperseqSet} \leq \UniSizeHyper{\UniSetFmA}+\UniSizeHyper{\UniHypersequentA}$
and that, for any $\UniFmA \in \UniSubfmlaHyperseqSet$, we have
$\UniSizeHyper{\UniFmA} \leq \UniSizeHyper{\UniSetFmA}+\UniSizeHyper{\UniHypersequentA}$ as well.
Also,
$\UniSizeHyper{\UniHyperCalcA} \leq  2 \cdot \UniSizeHyper{\UniHFLewR} \cdot \UniSizeHyper{\UniSetFmA}$.

\begin{lemma}\label{lem:norm-sets-ub-by-iteration}
Let $0 \leq m < n$, $\UniSetFmA$ be a finite set of formulas and $\UniHypersequentA$ be a hypersequent. Let $f(x)=Mx$, 
$M = 2\UniSizeHyper{\UniHyperCalcA}n$,
and $t = M$. Also, let 
$\{D_i\}_{i \in \UniNaturalSetNN}$,
$N$ and $\{{\UniHypersequentA_i}\}_{i\in\UniNaturalSetNN}$ be as defined above, relative to the subformulas of the formulas in $\UniSetFmA$ and $\UniHypersequentA$. Then,
\begin{enumerate}
\item 
$\UniNormHSeq{D_{i}} \leq f^i(t)$
for all $0 \leq i \leq N$.
\item 
$\{{\UniHypersequentA_i}\}_{i\in\UniNaturalSetNN}$ is a $(f,t)$-controlled bad sequence over $\UniNwqoWknHSeqName{m}{n}$.\end{enumerate}
\end{lemma}
\begin{proof}

First, observe that
$f^i(x) = M^i x$ 
and that $M > 1$.
We now prove each item.

\begin{enumerate}
\item 
We proceed by induction to show that
$\UniNormHSeq{D_{i}} \leq f^i(t)$
for all $0 \leq i \leq N$. 
In the base case, we have
$
\UniNormHSeq{D_0} < n + \UniSizeHyper{\UniHyperCalcA} \leq 2 \UniSizeHyper{\UniHyperCalcA}n = M = f^0(t)$.
For the induction step, with $i > 0$, note that, by conditions (2) and (3) of Definition~\ref{def-derive-sets}, we have
$\UniNormHSeq{D_{i+1}} \leq \UniSizeHyper{\UniHyperCalcA} + \UniNormHSeq{D_i} \cdot \UniSizeHyper{\UniHyperCalcA} \cdot n \leq 2\UniSizeHyper{\UniHyperCalcA}n \cdot\UniNormHSeq{D_i} = M\UniNormHSeq{D_i}$.
By the induction hypothesis, then, we get
$\UniNormHSeq{D_{i+1}}\leq 
Mf^{i}(t) = f^{i+1}(t)$.
\item 
Because $\UniNormHSeq{\UniHypersequentA_i} \leq \UniNormHSeq{D_{i}}$, we have $\UniNormHSeq{\UniHypersequentA_i} \leq f^i(t)$ by item (1), as desired.
\qedhere
\end{enumerate}
\end{proof}

We denote by $\UniDerivSet_i^{\UniSetFmA,\UniHypersequentA}$ the $i$-th set in the sequence of Definition~\ref{def-derive-sets} built with respect to the subformulas of the formulas in $\UniSetFmA$ and $\UniHypersequentA$.
The next result provides a uniform upper bound for each $\UniSizeHyper{\UniDerivSet_i^{\UniSetFmA,\UniHypersequentA}}$.

\begin{lemma}
\label{lem:unif-bound-weak-sets}
If $0 \leq m < n$, there is an increasing and elementary function $S$ such that, for all finite sets of formulas $\UniSetFmA$ and hypersequents $\UniHypersequentA$,
$\UniSizeHyper{\UniDerivSet_i^{\UniSetFmA,\UniHypersequentA}} \leq S(i,\UniSizeHyper{\UniSetFmA}+\UniSizeHyper{\UniHypersequentA})$.
\end{lemma}
\begin{proof}
As usual, let $\UniSubfmlaHyperseqSet$ be the set of subformulas of $\UniHypersequentA$ and of the formulas in $\UniSetFmA$.
Note that, for each $\UniFmB \in \UniSetFmA$, the rule $\UniRuleDed{\UniFmB}$ has 15 symbols plus the symbols in $\UniFmB$, so $\UniSizeHyper{\UniRuleDedSet{\UniSetFmA}}\leq 15 \UniSizeHyper{\UniSetFmA}$;  for $M$ as in Lemma~\ref{lem:norm-sets-ub-by-iteration}, we get 
$M = 2n\UniSizeHyper{\UniHyperCalcA} = 2n ( \UniSizeHyper{\UniHFLewR} + \UniSizeHyper{\UniRuleDedSet{\UniSetFmA}}) \leq  
 2n \UniSizeHyper{\UniHFLewR} (1+ \UniSizeHyper{\UniSetFmA})
\leq  C \cdot (\UniSizeHyper{\UniSetFmA}+\UniSizeHyper{\UniHypersequentA})$, where $C=2n \UniSizeHyper{\UniHFLewR}$ is independent of  $\UniSetFmA$.
Note that any element in $D_0$ is an instantiation of an axiomatic hypersequent following Definition~\ref{def-derive-sets}.
Since hypersequent variables are instantiated with the empty hypersequent, the elements of $D_0$ are actually sequents.
There are at most $\UniSizeHyper{\UniHyperCalcA}$ variables to be instantiated in an axiomatic hypersequent.
Each instantiated variable is either a formula-variable (actually, a proposition-variable) or a multiset-variable.
Recall that $\UniSizeHyper{\UniFmA} \leq \UniSizeHyper{\UniSetFmA}+\UniSizeHyper{\UniHypersequentA}$ for all $\UniFmA \in \UniSubfmlaHyperseqSet$.
Moreover, multisets are limited to having each formula with multiplicity at most $M$. Thus their size is upper bounded by $\sum_{\UniFmA\in\UniSubfmlaHyperseqSet}M\UniSizeHyper{\UniFmA} = M\sum_{\UniFmA\in\UniSubfmlaHyperseqSet}\UniSizeHyper{\UniFmA} \leq M\UniSetCard{\UniSubfmlaHyperseqSet}(\UniSizeHyper{\UniSetFmA}+\UniSizeHyper{\UniHypersequentA}) \leq C \cdot (\UniSizeHyper{\UniSetFmA}+\UniSizeHyper{\UniHypersequentA})^3$.
From these observations, we have that an element in $D_0$ has size
$\leq C' \cdot (\UniSizeHyper{\UniSetFmA}+\UniSizeHyper{\UniHypersequentA})^2$, where $C' > C$ accounts for commas and the sequent symbol as well.
There can be at most
$\UniNDistComp{\UniSetCard{\UniSubfmlaHyperseqSet}}{M} \leq \UniNDistComp{\UniSizeHyper{\UniSetFmA}+\UniSizeHyper{\UniHypersequentA}}{C \cdot (\UniSizeHyper{\UniSetFmA}+\UniSizeHyper{\UniHypersequentA})}$
of such sequents.
Thus let $S_0(x) \UniSymbDef \UniNDistComp{x}{C \cdot x}\cdot C' \cdot x^2$, which is an increasing elementary function for which
$\UniSizeHyper{D_0} \leq S_0(\UniSizeHyper{\UniSetFmA}+\UniSizeHyper{\UniHypersequentA})$.

Observe now that the size of a hypersequent  meeting the conditions to be in $D_{j+1}$, for $0 \leq j < L$, is upper bounded by
$A_1 \UniSymbDef c_1 \UniSizeHyper{\UniHyperCalcA}
\UniNDistComp{\UniSetCard{\UniSubfmlaHyperseqSet}}
{\UniSizeHyper{\UniHyperCalcA}n\UniNormHSeq{D_j}}
\cdot \UniSizeHyper{\UniHyperCalcA}n\UniNormHSeq{D_j} \cdot \UniSetCard{\UniSubfmlaHyperseqSet} 
\cdot \UniSizeHyper{\UniHypersequentA}$,
where $c_1$ accounts for commas, $\VL$'s, sequent symbols and succedents.
Indeed,
$\UniSizeHyper{\UniHyperCalcA}
\UniNDistComp{\UniSetCard{\UniSubfmlaHyperseqSet}}
{\UniSizeHyper{\UniHyperCalcA}n\UniNormHSeq{D_j}}$ is the maximum number of components and 
$\UniSizeHyper{\UniHyperCalcA}n\UniNormHSeq{D_j} \cdot \UniSetCard{\UniSubfmlaHyperseqSet} 
\cdot (\UniSizeHyper{\UniSetFmA}+\UniSizeHyper{\UniHypersequentA})$
is the maximum size of a component.
From the observations in the above paragraph, we know that
$\UniNormHSeq{D_j} \leq M^{j+1}$.

Hence,
$A_1 \leq 
c_1 \UniSizeHyper{\UniHyperCalcA}
\UniNDistComp{\UniSetCard{\UniSubfmlaHyperseqSet}}
{\UniSizeHyper{\UniHyperCalcA}nM^{j+1}}
\cdot \UniSizeHyper{\UniHyperCalcA}nM^{j+1} \cdot \UniSetCard{\UniSubfmlaHyperseqSet} 
\cdot (\UniSizeHyper{\UniSetFmA}+\UniSizeHyper{\UniHypersequentA})
=
c_1 \UniSizeHyper{\UniHyperCalcA}
\cdot
{
(\UniSetCard{\UniSubfmlaHyperseqSet}+1)
\cdot
(\UniSizeHyper{\UniHyperCalcA}nM^{j+1}+1)^{\UniSetCard{\UniSubfmlaHyperseqSet}}
}
\cdot \UniSizeHyper{\UniHyperCalcA}nM^{j+1} 
\cdot \UniSetCard{\UniSubfmlaHyperseqSet} 
\cdot (\UniSizeHyper{\UniSetFmA}+\UniSizeHyper{\UniHypersequentA})$,
which is bounded by $E(j,
\UniSizeHyper{\UniSetFmA}+\UniSizeHyper{\UniHypersequentA})$,
where $E$ is an increasing elementary function.
(Recall that
$\UniSizeHyper{\UniHyperCalcA} \leq  c \cdot \UniSizeHyper{\UniSetFmA}$ for a constant $c$ that only depends on the base calculus ${\UniExtCalcDed{\UniDelCut{\UniFLeExtHCalc{\UniWeakWProp{m}{n}}}}{\UniAnaRuleSet }}$.)
This means that the size of $D_{j+1}$ is upper bounded by
\[
\UniNDistHSeq{\UniSetCard{\UniSubfmlaHyperseqSet}}
{\UniSizeHyper{\UniHyperCalcA}n\UniNormHSeq{D_{j}}}
{\UniSizeHyper{\UniHyperCalcA}} \cdot 
E(j,
\UniSizeHyper{\UniSetFmA}+\UniSizeHyper{\UniHypersequentA})
\leq 
E'(j,
\UniSizeHyper{\UniSetFmA}+\UniSizeHyper{\UniHypersequentA})
\]
for $E'$ an elementary function.
Thus set $S(j,
\UniSizeHyper{\UniSetFmA}+\UniSizeHyper{\UniHypersequentA}) \UniSymbDef
S_0(\UniSizeHyper{\UniSetFmA}+\UniSizeHyper{\UniHypersequentA})+E'(j,
\UniSizeHyper{\UniSetFmA}+\UniSizeHyper{\UniHypersequentA})$, 
and we are done.
\end{proof}

We are now ready to prove the desired complexity upper bounds.

\begin{theorem}
\label{fact:flewmnr-ackermannian}
Let $0 \leq m < n$ and $\UniSetFmA$ be a finite set of formulas.
\begin{enumerate}
\item \sloppy
If $\UniAnaRuleSet$ is a finite set of hypersequent analytic structural rules, then provability in $\UniHFLewRAug{\UniSetFmA}$ is in $\UniFGHProbOneAppLevel{\omega^{\omega}}$. 
\item 
If $\UniAnaRuleSet$ is a finite set of sequent analytic structural rules, then provability in $\UniSFLewRAug{\UniSetFmA}$ is in $\UniFGHProbOneAppLevel{\omega}$.
\end{enumerate}
Moreover, the underlying algorithms and upper bounds are uniform on ${\UniSetFmA}$.
\end{theorem}
\begin{proof}
Let $C \UniSymbDef (D_0,\ldots,D_N)$ be the sequence of sets of hypersequents described in Definition~\ref{def-derive-sets}, where $N$ is the stabilizing point provided by Lemma~\ref{lem:SN}.
We begin by employing Lemma~\ref{lem:seq-sets-ub} to obtain an upper bound on $\UniSizeHyper{C} = \sum_i \UniSizeHyper{D_i}$.

We start by providing an upper bound for $N$ in terms of the size of the input, $\UniSizeHyper{\UniSetFmA}+\UniSizeHyper{\UniHypersequentA}$.
Since
$\UniSetCard{\UniSubfmlaHyperseqSet} \leq \UniSizeHyper{\UniSetFmA}+\UniSizeHyper{\UniHypersequentA}$,
bad sequences over $\UniNwqoWknHSeqName{m}{n}$ are bad sequences over 
$\UniNwqoWknHSeqNameVar{m}{n}{\UniSubfmlaHyperseqSet_{\UniSizeHyper{\UniSetFmA}+\UniSizeHyper{\UniHypersequentA}}}$.
In view of Lemma~\ref{lem:norm-sets-ub-by-iteration}(2), we have that
$N \leq \UniLengFunc{\UniNwqoWknHSeqNameVar{m}{n}{\UniSubfmlaHyperseqSet_{\UniSizeHyper{\UniSetFmA}+\UniSizeHyper{\UniHypersequentA}}}}{\UniControlFunctionA}(t)
\leq 
\UniLengFunc{\UniNwqoWknHSeqNameVar{m}{n}{\UniSubfmlaHyperseqSet_{\UniSizeHyper{\UniSetFmA}+\UniSizeHyper{\UniHypersequentA}}}}{\UniControlFunctionA}(R(\UniSizeHyper{\UniSetFmA}+\UniSizeHyper{\UniHypersequentA}))$, 
where $t$ and $\UniControlFunctionA$ are given in Lemma~\ref{lem:norm-sets-ub-by-iteration} and $R$ is a linear function.
By Proposition~\ref{fact:hnwqo-wkn-length-theorem}, we know that
$\{ \UniNwqoWknHSeqNameVar{m}{n}{\UniSubfmlaHyperseqSet_k}\}_{k \in \UniNaturalSet} \sqsubseteq \UniFGHOneAppLevel{\omega^\omega}$, 
therefore $N$ is upper bounded by $\tilde{L}(\UniSizeHyper{\UniSetFmA}+\UniSizeHyper{\UniHypersequentA})$, 
for some $\tilde{L} \in \UniFGHOneAppLevel{\omega^\omega}$
(given that $\UniSizeHyper{\UniHypersequentA}$ is sufficiently large).

Now we provide an upper bound for the size of each element in the sequence. 
Note that since this sequence is actually an increasing chain, we must have
$\UniSizeHyper{\UniDerivSet_i} \leq \UniSizeHyper{\UniDerivSet_N}$
for all $0 \leq i \leq N$, so it is enough to upper bound $\UniSizeHyper{\UniDerivSet_N}$.
By Lemma~\ref{lem:unif-bound-weak-sets}, we have that $\UniSizeHyper{\UniDerivSet_N} \leq S(N,\UniSizeHyper{\UniSetFmA}+\UniSizeHyper{\UniHypersequentA})$,
where $S \in \UniFGHLevel{2}$ is increasing.
Thus $\UniSizeHyper{\UniDerivSet_N} \leq S(\tilde{L}(\UniSizeHyper{\UniSetFmA}+\UniSizeHyper{\UniHypersequentA}),
\UniSizeHyper{\UniSetFmA}+\UniSizeHyper{\UniHypersequentA})$
and by Lemma~\ref{lem:ev-ub-multvariable} we know that
$\lambda x. S(\tilde{L}(x),x) \in \UniFGHOneAppLevel{\omega^\omega}$.
We have thus established that the size of each element in the sequence is upper bounded by a function of 
$\UniSizeHyper{\UniSetFmA}+
\UniSizeHyper{\UniHypersequentA}$ in $\UniFGHOneAppLevel{\omega^\omega}$.

Thus, by Lemma~\ref{lem:seq-sets-ub}, we have that the chain is a member of $\textit{Seq}[\mathfrak{L}, \tilde{L},S](\UniSizeHyper{\UniSetFmA}+\UniSizeHyper{\UniHypersequentA})$,
where $\mathfrak{L}$ is the set of $\UniSubfmlaHyperseqSet$-hypersequents, and thus its size is bounded
(uniformly on $\UniSizeHyper{\UniSetFmA}+\UniSizeHyper{\UniHypersequentA}$) by a function in $\UniFGHOneAppLevel{\omega^\omega}$.
Because $\tilde{L}$ and $S$ are independent of $\UniSetFmA$, the algorithm and the above analysis are uniform on $\UniSetFmA$.

Now, by Remark~\ref{r: spacetimeW}, we define $\textit{Seq}(\UniSetFmA,h) \UniSymbDef \textit{Seq}[\mathfrak{L}, \tilde{L},S](\UniSizeHyper{\UniSetFmA}+\UniSizeHyper{\UniHypersequentA})$.
From the above discussion and by Lemma~\ref{lem:SN}, we have that $\UniHypersequentA$ is provable in the calculus
iff
$C \in \textit{Seq}(\UniSetFmA,h)$ and there is $\UniHypersequentA' \in \UniDerivSet_N$
with
$\UniHypersequentA' \UniHyperseqWknWqo{m}{n}{\UniSubfmlaHyperseqSet}  \UniHypersequentA$, which is testable in primitive recursive time.
Thus we obtain a nondeterministic algorithm running in space $\UniFGHOneAppLevel{\omega^\omega}$, as desired.

The proof of (2) is analogous to that of Theorem~\ref{fact:flecmnr-ackermannian} (2), that is, the same forward proof-search procedure can be used over the sequent calculus corresponding to
$\UniHFLewRAug{\UniSetFmA}$ 
when it is analytic.
The bad sequence extracted from the derivation sets will contain only sequents, whose encodings will be elements of
$\left(\UniMajoringFntSingWqo{\UniWqoExtModRelProd{n}{m}{\UniSetCard{\UniSubfmlaHyperseqSet}}}\right)^{\UniSetCard{\UniSubfmlaHyperseqSet+1}}$.
So the bad sequences are over this nwqo, and the Ackermannian upper bound follows from Theorem~\ref{fact:length-theo-nm-not-fixed-maj-min-sing}.
\end{proof}

From the above and Lemma~\ref{fact:cut-restrict-weak-weakening}, we obtain the following.

\begin{corollary}
\label{c: FLew(m,n) complexity}
Let $0 \leq m < n$.
\begin{enumerate}
    \item If $\UniAxiomSetA$ is a finite
set of acyclic $\mathcal{P}_3^\flat$ axioms,
then provability and deducibility in 
$\UniAxiomExt{\UniFLeExtLogic{\UniWeakWProp{m}{n}}}{ \UniAxiomSetA}$
are in $\UniFGHProbOneAppLevel{\omega^\omega}$.
    \item If $\UniAxiomSetA$ is a finite
set of $\mathcal{N}_2$ axioms, then provability and deducibility in $\UniAxiomExt{\UniFLeExtLogic{\UniWeakWProp{m}{n}}}{ \UniAxiomSetA}$
are in $\UniFGHProbOneAppLevel{\omega}$.
In particular, provability and deducibility in 
$\UniFLeExtLogic{\UniWeakWProp{m}{n}}$
are in $\UniFGHProbOneAppLevel{\omega}$.
\end{enumerate}
\end{corollary}
\begin{proof}
As we know, provability is a particular case of deducibility, so it is enough to prove the claim for the latter.
For (1), by the discussion in Section~\ref{s: P3}, deducibility in $\UniAxiomExt{\UniFLeExtLogic{\UniWeakWProp{m}{n}}}{ \UniAxiomSetA}$
corresponds to deducibility in $\UniHFLewR$, where $\UniAnaRuleSet$ is a finite set of analytic structural hypersequent rules.
Given a finite set of formulas $\UniSetFmA$ and a hypersequent $\UniHypersequentA$, by Corollary~\ref{cor:ded-weak-is-prov-mod-calc} deciding whether $\UniHypersequentA$ follows from $\UniSetFmA$ in $\UniAxiomExt{\UniFLeExtLogic{\UniWeakWProp{m}{n}}}{ \UniAxiomSetA}$ is the same as deciding whether $\UniHypersequentA$ is provable in $\UniHFLewRAug{\UniSetFmA}$.
Thus, upon taking $\UniSetFmA$ and $\UniHypersequentA$ as input, one just has to write down the rules in $\UniRuleDedSet{\UniSetFmA}$ (which takes space $c \cdot \UniSizeHyper{\UniSetFmA}$ for a constant $c$ that only depends on the base calculus $\UniHFLewR$) and employ the algorithm for provability given in Theorem~\ref{fact:flewmnr-ackermannian} (1)---this is possible because the algorithm is uniform on $\UniSetFmA$.
By this theorem, this provability problem is in
$\UniFGHProbOneAppLevel{\omega^\omega}$, and this bound is uniform on 
$\UniSizeHyper{\UniSetFmA}+\UniSizeHyper{\UniHypersequentA}$.
Thus the deducibility problem in 
$\UniAxiomExt{\UniFLeExtLogic{\UniWeakWProp{m}{n}}}{ \UniAxiomSetA}$
is in $\UniFGHProbOneAppLevel{\omega^\omega}$ as well, as desired.

For (2), the result follows from the fact that $\UniAxiomExt{\UniFLeExtLogic{\UniWeakWProp{m}{n}}}{ \UniAxiomSetA}$,
for $\UniAxiomSetA$ a (possibly empty) set
of $\mathcal{N}_2$ axioms,
has an analytic sequent calculus by Lemma~\ref{l: N_2^-0}; thus the proof is like the one above, but using item (2) of Theorem~\ref{fact:flewmnr-ackermannian}.
\qedhere
\end{proof}

\begin{remark}\label{r: shrinking} 
\label{remark:problem-horcik-terui}
A proof rule is called \emph{shrinking} if removing from it all metavariables contained in a given  set of metavariables results either in a premise identical with the conclusion, or each premise contains strictly fewer occurrences of symbols than the conclusion (this has to hold for all given sets of metavariables).
In \cite{horcik2011}, extensions of $\UniFLeExtLogic{}$ having an analytic sequent calculus and only \emph{shrinking} proof rules were shown to be in \PSPACE.
The authors state that 
$\UniFLeExtLogic{\UniWeakWProp{m}{n}}$
is in \PSPACE (much lower than our $\UniFGHProbOneAppLevel{\omega}$ upper bound), based on the claim that $\UniWeakWAnaRule{2}{3}$ is a shrinking rule. 
Unfortunately, this claim is not true, thus presenting a serious gap in the  \PSPACE claim. 
Indeed, 

\begin{center}
\AxiomC{
$\{\UniSequent{\UniMSetFmA, \UniMSetFmB_{1}^{r_1}, \UniMSetFmB_2^{r_2}, \UniMSetFmB_3^{r_3}}
{\UniMSetSucA}\}_{\sum r_i = 2}$
}
\RightLabel{
$\UniWeakWAnaRule{2}{3}$
}
\UnaryInfC{$
\UniSequent{\UniMSetFmA,\UniMSetFmB_1,\UniMSetFmB_2, \UniMSetFmB_3}{\UniMSetSucA}$}
\DisplayProof
\end{center}

is not a shrinking rule, since after removing the symbols of the set $S \UniSymbDef \UniSet{\UniMSetFmA,\UniMSetFmB_2,\UniMSetFmB_3}$ from the rule, we get the conclusion 
$\UniSequent{\UniMSetFmB_1}{\UniMSetSucA}$
and the premise
$\UniSequent{\UniMSetFmB_1,\UniMSetFmB_1}{\UniMSetSucA}$.
We believe that the authors were considering the rule $\UniWeakWRule{m}{n}$ instead, which is shrinking and it is equivalent to (interderivable with) $\UniWeakWAnaRule{m}{n}$ in the presence of cut. 
However, the calculus with $\UniWeakWRule{m}{n}$ fails cut elimination and thus its cut-free version has a different set of provable sequents than the (cut-free) calculus with $\UniWeakWAnaRule{m}{n}$; so the set of theorems of the cut-free calculus with $\UniWeakWRule{m}{n}$ does not correspond to the logic $\UniFLeExtLogic{\UniWeakWProp{m}{n}}$, and the correct rule to consider is $\UniWeakWAnaRule{m}{n}$, which is unfortunately not shrinking. 
It seems that at the moment Corollary~\ref{c: FLew(m,n) complexity} gives the best known upper bound.
\end{remark}

%% file: tex/noncommutative-upper-bounds.tex
In the previous two sections, we developed proof-search procedures and complexity upper-bound results for knotted commutative extensions of $\UniFLExtLogic{}$.
In this section, we show how to adapt these results to weakly commutative logics (for both deducibility and provability); these are extensions of  $\UniFLExtLogic{}$ by weakly commutative equations, defined in Section~\ref{s: P3/N2}.
In general, the lack of commutativity does not allow for the representation of hypersequents to be given only by the multiplicities of formulas in the antecedents of components---now the order of the formulas also matters.
This difference is so drastic that $\UniFLExtLogic{\UniCProp}$ has undecidable  provability and deducibility, as shown in \cite{horcik2016}.
However, we will see that weak commutativity allows us to define an appropriate nwqo, which can then be encoded in a knotted nwqo over the natural numbers and thus lead to similar results as in the commutative case. 
As we will see, there are many technical issues that come up and the analysis is much more intricate than in the commutative case.

Given the lack of commutativity, we need to work with lists in the antecedents of sequents instead of multisets.
In this section, for a finite set $A$, $\UniListClass{A}$ denotes the collection of all finite lists/sequences of elements in $A$.
As usual, we write lists by simply juxtaposing their elements.
The empty list is denoted by $\UniEmptyList$ and the operation of list concatenation is denoted by $\UniListConcat$.
Also, we denote by
$\UniListCountElem{\UniWeakComListA}{a}$ 
the number of occurrences of $a$ in a list $\UniWeakComListA$ and by $\mathsf{del}^a(\UniWeakComListA)$ the list obtained from $\UniWeakComListA$ by deleting all occurrences of $a \in A$.

\section{Weakly commutative Full Lambek calculi}

In what follows, for $1 \leq i \leq j$ and variables $x, y_i, \ldots,  y_{j-1}$, we consider the words/lists $w_{i,j} \UniSymbDef x y_i x y_{i+1} \ldots y_{j-1} x$.
For $r \in \mathbb{Z}^+$ and $\vec{a} \UniSymbDef \UniTuple{a_0,\ldots,a_r} \in \UniNaturalSet^{r+1}$ with $\sum_{i=0}^{r} a_i = r+1$ and $\prod_{i=0}^{r} a_i = 0$, we denote by $e(\vec a)^\star$ the weak commutativity equation 
\[ 
x y_1 x y_2 \cdots y_r x = x^{a_0} y_1 x^{a_1} y_2 \cdots y_r x^{a_r}
\]
that was defined in Section~\ref{sec:wck-substructural-logics} (note that the left-hand side is just $w_{1,r+1}$), and by $e(\vec a)$ the set of the corresponding (linearized) simple equations. 
For example, the linearizations of $xyx=xxy$ were given explicitly in Section~\ref{sec:wck-substructural-logics}.  
We denote by $\mathcal{K}(\vec a)$ the variety of monoids that satisfy the equations in $e(\vec a)$.
For simplicity, we will treat $e(\vec a)$ as a single equation when no ambiguity happens.
As observed in~\cite[Sec. 3.4]{Cardona2015} each equation of the form $e(\vec a)$ implies a set of equations (defined right below) that allow us to manipulate expressions much more effectively than direct single applications of $e(\vec a)$ do.  
We opt for working with that set and then observe that our results apply immediately to the (stronger) equation $e(\vec a)$.

Given $r$ and $\vec{a}$ as above, based on \cite[Sec. 3.4]{Cardona2015}  we define
$\UniEllParam \UniSymbDef 2r + 1$, 
$$\UniPParam \UniSymbDef \max\{ j : 0 \leq j \leq r \text{ and } a_i = 1 \text{ for all } i < j \}$$ and 
$$\UniQParam \UniSymbDef \max\{ j : 0 \leq j \leq r \text{ and } a_i = 1 \text{ for all } i > r-j \}.$$ 
(Here, $\mathsf{fw}$ stands for `front wall' and $\mathsf{bw}$ for `back wall'; see Definition~\ref{def:walls}.)
Without loss of generality, we may assume that $\UniPParam > 0$, since we can multiply the equation by $xy_0$ from the left to obtain an equation that satisfies $\UniPParam > 0$; since the former equation implies the latter, our results for the latter form will end up applying to the former, as well.
See Example~\ref{ex:wcsimple} for an illustration.
We denote by $\mathcal{K}(\UniEllParam, \UniPParam, \UniQParam)$ the variety of monoids that satisfy the  equations 
\[w_1 w w_2 = w_1 w' w_2, \]
where $w_1 \UniSymbDef w_{1,\UniPParam}$, $w_2 \UniSymbDef w_{\UniEllParam-\UniQParam+1, \UniEllParam}$
and $\mathsf{del}^x(w) = \mathsf{del}^{x}(w') = y_{\UniPParam} y_{\UniPParam+1}\cdots y_{\UniEllParam-\UniQParam}$,
with $|w|_x = |w'|_x = \UniEllParam - \UniPParam - \UniQParam$.
We denote the above equation by $e(w,w')$ 
and the collection of  all such equations by $E(\UniEllParam,\UniPParam,\UniQParam)$. 
By overloading notation we will use the same symbols for the simple-equation versions of them (recall Section~\ref{s: N2}); note that this collection is finite. 
These are powerful equations that will allow us to control the number of different types of sequents within hypersequents.

It is shown in~\cite[Sec. 3.4]{Cardona2015} that $\mathcal{K}(\vec a)$ is a subvariety of $\mathcal{K}(\UniEllParam,\UniPParam,\UniQParam)$, therefore any weak commutativity equation will allow us to classify $\Omega$-sequents, for a finite set of formulas $\Omega$, into finitely many types. 
Given a set $E$ of equations over the signature of $\mathsf{FL}$, we denote by $\mathsf{FL}_{E}$ the subvariety of $\mathsf{FL}$ axiomatized (relative to $\mathsf{FL}$) by the equations in $E$; we omit curly braces in set notation when there is no chance of confusion.     
Thus $\mathsf{FL}_{e(\vec a)}$ is a subvariety of $\mathsf{FL}_{E(\UniEllParam,\UniPParam,\UniQParam)}$ and we obtain the following result.

\begin{proposition}
\label{fact:variety-equality-wc}
$\mathsf{FL}_{ e(\vec a) } = \mathsf{FL}_{\{ e(\vec a) \} \cup E(\UniEllParam,\UniPParam,\UniQParam)}$.
\end{proposition}

By the results in Section~\ref{s: P3/N2}, all equations in $\{ e(\vec a) \} \cup E(\UniEllParam,\UniPParam,\UniQParam)$ belong to $\mathcal{N}_2^{-0}$ and they induce analytic structural (hyper)sequent rules. 
We denote  by $\mathsf e(\vec a)$ the rule corresponding to $e(\vec a)$, by $\mathsf e(W,W')$ the rule corresponding to $e(w,w')$, where $W, W'$ are the sequent-variable sequences corresponding to $w,w'$, and by $\mathsf E(\UniEllParam,\UniPParam,\UniQParam)$ the collection of all $\mathsf e(W,W')$.
We now define the weakly commutative base calculus that will be central in this section.

\begin{definition}
We denote by
$\UniFLExtHCalc{\UniWEProp{\vec{a}}}^\star$
the hypersequent calculus $\UniFLExtHCalc{}$ augmented with the analytic structural rules in $\{ \mathsf e(\vec a) \} \cup \mathsf E(\UniEllParam,\UniPParam,\UniQParam)$.
\end{definition}

The next proposition establishes that adding the rules in $\mathsf E(\UniEllParam,\UniPParam,\UniQParam)$ to analytic structural extensions of $\UniFLExtHCalc{\UniWEProp{\vec{a}}}$ does not change the set of provable sequents.

\begin{proposition}
Let $\mathcal{H}$ be a set of hypersequents, $\UniHypersequentA$ be a hypersequent and $\UniAnaRuleSet$ be a finite set of hypersequent analytic structural rules. Then
$\mathcal{H}\UniHyperDerivRel{\UniCalcExt{\UniFLExtHCalc{\UniWEProp{\vec{a}}}}{\mathcal R}} \UniHypersequentA$
iff
$\mathcal{H}\UniHyperDerivRel{
\UniCalcExt{\UniFLExtHCalc{\UniWEProp{\vec{a}}}^\star}{\mathcal{R}}} \UniHypersequentA$.
\end{proposition}

\begin{proof}
    This follows directly from the fact that the extra rules correspond to $\mathcal{N}_2^{-0}$-formulas that are already implied by the existing ones.
\end{proof}

\begin{proposition}
\label{prop:rules-wc}
For each $\mathsf e(W,W') \in \mathsf E(\UniEllParam,\UniPParam,\UniQParam)$, the following are particular instances thereof\footnote{To be more precise, the instance would have multiple premise-hypersequents, but all of them would be equal, so it is simpler (and safe) to consider this as the rule instance of interest.}:
\begin{center}
    \small
        \AxiomC{$
            \UniHyperMSetA \VL
            \UniSequent{\Theta_1,
            W_{1,\UniPParam},
                W,
                W_{(\UniEllParam-\UniQParam+1),
                \UniEllParam},\Theta_2
                }{\UniMSetSucA}
            $
            }
        \RightLabel{$\mathsf{e}[W,W']$}
        \UnaryInfC{$\UniHyperMSetA \VL
        \UniSequent{\Theta_1,
            W_{1,\UniPParam},
                W',
                W_{(\UniEllParam-\UniQParam+1),
                \UniEllParam},\Theta_2}{\UniMSetSucA}$}
        \DisplayProof
\end{center}
where $W_{i,j}$ is the segment of the list $\UniMSetFmB,\UniMSetFmA_1,
\UniMSetFmB,\UniMSetFmA_2\ldots$ of sequence-variables that starts at the $i$-th occurrence of $\UniMSetFmB$ and ends at the $j$-th occurrence, and
$W$, $W'$ are such that $\mathsf{del}^\UniMSetFmB(W) = \mathsf{del}^\UniMSetFmB(W') = \UniMSetFmA_{\UniPParam}, \UniMSetFmA_{\UniPParam+1}, \ldots, \UniMSetFmA_{\UniEllParam-\UniQParam}$
and $\UniListCountElem{W}{\UniMSetFmB} = \UniListCountElem{W'}{\UniMSetFmB} = \UniEllParam-\UniPParam-\UniQParam$.
\end{proposition}
\begin{proof}
It follows from the definition of the rules and the discussion in Section~\ref{s: P3/N2}.
\end{proof}

\begin{example}
\label{ex:wcsimple}
For the equation $xyx = xxy$, we have $r = 1$, $\vec a = \UniTuple{2,0}$, $\UniEllParam = 3$, $\UniPParam=0$ and $\UniQParam=0$. 
Since $\UniPParam=0$, we consider an equation that is implied by  $xyx = xxy$, namely $xzxyx = xzxxy$, for which
$r = 2$, $\vec a = \UniTuple{1,2,0}$, $\UniEllParam = 5$, $\UniPParam=1$ and $\UniQParam=0$.
For these values, the following are instances of rules in $\mathsf E(\UniEllParam,\UniPParam,\UniQParam)$:
\begin{center}
        \AxiomC{$
            \UniHyperMSetA \VL
            \UniSequent{\Theta_1,\UniMSetFmB,
                W,\Theta_2}{\UniMSetSucA}
            $}
        \RightLabel{$\mathsf{e}[W,W']$}
        \UnaryInfC{$\UniHyperMSetA \VL
        \UniSequent{\Theta_1,\UniMSetFmB,
            W',\Theta_2}{\UniMSetSucA}$}
        \DisplayProof
\end{center}
\noindent where $W$ and $W'$ are such that $\mathsf{del}^\UniMSetFmB(W) = \mathsf{del}^\UniMSetFmB(W') = \UniMSetFmA_1, \UniMSetFmA_2,\UniMSetFmA_3,\UniMSetFmA_4$
and $\UniListCountElem{W}{\UniMSetFmB} = \UniListCountElem{W'}{\UniMSetFmB} = 4$.
\end{example}

We now present definitions that will ultimately allow us to prove a normal form theorem for the provable hypersequents in the calculi defined above.

\begin{definition}
An instance of $\mathsf{e}(W,W') \in \mathsf E(\UniEllParam,\UniPParam,\UniQParam)$ is called \emph{basic} if it is of the form $\mathsf{e}[W,W']$ given in Proposition~\ref{prop:rules-wc} and $\Delta$ is instantiated with a single formula.
\end{definition}

\begin{definition}\label{def:nc-height}
The \emph{nc-height}\footnote{Here, `nc' stands for `non-commutativity'} of a derivation $\UniDerivationA$ in 
$\UniFLExtHCalc{\UniWEProp{\vec{a}}}^\star$
(and any rule extension thereof) is defined recursively by: if the last rule of $\UniDerivationA$ was $\UniRuleSchemaA$, then the \emph{nc-height} of $\UniDerivationA$ is
\begin{itemize}
    \item 1, if $\UniRuleSchemaA$ is axiomatic;
    \item $n$, if the instance of $\UniRuleSchemaA$ is a basic instance of $\mathsf{e}(W,W')$, where $n$ is the nc-height of the derivation of the premise  sequent;
    \item $n_1 + \ldots + n_p + 1$ in any other case, where $n_i$ is the nc-height of the derivation of the $i$-th premise sequent, for $1 \leq i \leq p$.
\end{itemize}
\end{definition}

\section{A normal form for the derivable hypersequents}
\label{sec:wc-normal-form}

It turns out that weak commutativity provides a way to classify the antecedents of sequents into a finite amount of types, and this will allow us to obtain a well-quasi-ordering on hypersequents in a way similar to the one employed in the commutative case. 
This classification will also provide us with a normal form for the derivable hypersequents in the extensions of $\UniFLExtHCalc{\UniWEProp{\vec{a}}}^\star$.
We hereby develop such ideas in detail and, as mentioned before, we work with equations of the form $e(w,w')$ instead of using $e(\vec a)$ directly.

Motivated by the discussion in the previous section, we begin by fixing $\UniEllParam, \UniPParam, \UniQParam \in \UniNaturalSet$ such that $\UniPParam + \UniQParam < \UniEllParam$ and $\UniPParam > 0$.
The following definitions will be instrumental to us; they consist of some operations on $\UniListClass{A}$ implicitly parameterized by $\UniEllParam,\UniPParam$ and $\UniQParam$.

\begin{definition}\label{def:walls}
If $A$ is a set, $\UniWeakComListA \in \UniListClass{A}$, $a \in A$ and $\UniListCountElem{\UniWeakComListA}{a} \geq \UniEllParam$, then
    \begin{enumerate}
        \item the \emph{front wall  $\UniFrontWall{a}(\UniWeakComListA)$ of $\UniWeakComListA$ for $a$} is the list corresponding to the initial segment of $\UniWeakComListA$ ending at the $\UniPParam$-th occurrence of $a$;
        \item the \emph{back wall $\UniBackWall{a}(\UniWeakComListA)$ of $\UniWeakComListA$ for $a$} is the final segment of $\UniWeakComListA$ that starts at the $(\UniListCountElem{\UniWeakComListA}{a}-\UniQParam+1)$-th occurrence of $a$ (or simply the $\UniQParam$-th occurrence of $a$ when reading $\UniWeakComListA$ from the end).
        \item the \emph{middle part $\UniMid{a}(\UniWeakComListA)$ of $\UniWeakComListA$ for $a$} is the list obtained from $\UniWeakComListA$ by deleting the segments $\UniFrontWall{a}(\UniWeakComListA)$ and $\UniBackWall{a}(\UniWeakComListA)$.
    \end{enumerate}
\end{definition}

\begin{example}
\label{ex:operationsonlists}
For $\UniEllParam \UniSymbDef 4$, $\UniPParam \UniSymbDef 2$, $\UniQParam \UniSymbDef 1$, $A \UniSymbDef \{ \mathtt{a}, \mathtt{b}, \mathtt{c}, \mathtt{d} \}$ and $\UniWeakComListA \UniSymbDef \mathtt{caabcababadbc}$, we have
$\UniFrontWall{\mathtt a}(\UniWeakComListA)
= \mathtt{caa}$;
$\UniBackWall{\mathtt a}(\UniWeakComListA)
= \mathtt{adbc}$;
and 
$\UniMid{\mathtt a}(\UniWeakComListA)
= \mathtt{bcabab}$.
\end{example}

Note that for every $\UniWeakComListA \in \UniListClass{A}$ and $a \in A$ with $\UniListCountElem{\UniWeakComListA}{a} \geq \UniEllParam$, we have
$$\UniWeakComListA = \UniFrontWall{a}(\UniWeakComListA) \UniListConcat \UniMid{a}(\UniWeakComListA) \UniListConcat \UniBackWall{a}(\UniWeakComListA).$$
Now we describe the operations $\UniAlphaN{}$ and $\UniAlphaD{}$ on $\UniListClass{A}$, which
$(\mathsf{N})$ormalize a list and provide a $(\mathsf{T})$ype for a list, respectively,
all with respect to the fixed parameters 
$\UniEllParam, \UniPParam$ and $\UniQParam$.
Intuitively, given a letter $a$ and a list $l$, these operations move all occurrences of $a$ inside the middle part of $l$ to a specific location (the front of the middle part); $\UniAlphaD{}$ further truncates the amount of $a$'s in that location.

\begin{definition}[{\cite[p. 372]{Cardona2015}}]
\label{def:alpha_defs}
For a set $A$,
$\UniWeakComListA \in \UniListClass{A}$,
$a \in A$
and
$\star \in \{ \mathsf{N}, \mathsf{T} \}$, we define
\begin{equation*}
    \UniAlphaStar{a}(\UniWeakComListA) \UniSymbDef 
    \begin{cases}
        \UniFrontWall{a}(\UniWeakComListA) 
        \UniListConcat 
        a^{K}
        \UniListConcat
        \mathsf{del}^a(\UniMid{a}(\UniWeakComListA))
        \UniListConcat 
        \UniBackWall{a}(\UniWeakComListA) & \text{if } 
        \UniListCountElem{\UniWeakComListA}{a} \geq \UniEllParam,\\
        \UniWeakComListA & \text{otherwise.}\\
    \end{cases}
\end{equation*}
where $K \UniSymbDef \UniListCountElem{\UniWeakComListA}{a} - \UniPParam - \UniQParam$ 
if $\star = \mathsf{N}$ and $K \UniSymbDef \UniEllParam - \UniPParam - \UniQParam$ if $\star = \mathsf T$.
Also, if  $A = \{ a_1,\ldots,a_k \}$ is finite, we set
$\UniAlphaStar{}(\UniWeakComListA) \UniSymbDef \UniAlphaStar{a_k}(\ldots(\UniAlphaStar{a_2}(\UniAlphaStar{a_1}(\UniWeakComListA))))$.
\end{definition}

\begin{remark}
\label{rem:wc-type-order-indep}
As argued in \cite[p.~25]{Cardona2015}, the way we enumerate the elements of $A$ does not
affect the result of the operations $\UniAlphaN{}$ and $\UniAlphaD{}$.
\end{remark}

\begin{example}
We continue Example~\ref{ex:operationsonlists}, where  $\UniEllParam \UniSymbDef 4$, $\UniPParam \UniSymbDef 2$ and $\UniQParam \UniSymbDef 1$, and $A \UniSymbDef \{ \mathtt{a}, \mathtt{b}, \mathtt{c}, \mathtt{d} \}$, by taking $\UniWeakComListA \UniSymbDef \mathtt{bccababacacbaadcdd}$.
Then:
\begin{enumerate}
    \item $\UniAlphaN{\mathtt a}(\UniWeakComListA) = 
        \mathtt b \mathtt c \mathtt c \mathtt a \mathtt b \mathtt a \mathtt a^3
    \mathtt b \mathtt c \mathtt c \mathtt b
    \mathtt a \mathtt d \mathtt c \mathtt d \mathtt d$
    \item $\UniAlphaD{\mathtt a}(\UniWeakComListA) = 
        \mathtt b \mathtt c \mathtt c \mathtt a \mathtt b \mathtt a \mathtt a
    \mathtt b \mathtt c \mathtt c \mathtt b
    \mathtt a \mathtt d \mathtt c \mathtt d \mathtt d$
    \item $\UniAlphaN{}(\UniWeakComListA) = 
        \mathtt b \mathtt c \mathtt c \mathtt c \mathtt c 
        \mathtt a \mathtt b \mathtt b  \mathtt a \mathtt a^3
        \mathtt b
        \mathtt a \mathtt d \mathtt c \mathtt d \mathtt d$
    \item $\UniAlphaD{}(\UniWeakComListA) = 
        \mathtt b \mathtt c \mathtt c \mathtt c 
        \mathtt a \mathtt b \mathtt b \mathtt a
        \mathtt a 
        \mathtt b
        \mathtt a \mathtt d \mathtt c \mathtt d \mathtt d$\qedhere
\end{enumerate} 
\end{example}

The next proposition shows that each list in $\UniListClass{A}$ is classified into one among finitely many types and it will be essential for obtaining the desired nwqo in Section~\ref{sec:nwqo-wc}.

\begin{proposition}
\label{fact:finitely-many-types}
If $A$ is a finite set, then $\UniAlphaD{}[\UniListClass{A}]$ is finite.
\end{proposition}
\begin{proof}
Indeed, each type is a list in which each element appears at most $\UniEllParam$ times, and there are finitely many of such lists.
\end{proof}

We now observe that basic instances of $\mathsf{e}(W,W')$ are type-preserving.

\begin{lemma}\label{lem:basic-inst-e-type-preserv}
If $(g \VL \UniSequent{\UniMSetFmA_1}{\UniMSetSucA})/ (g \VL \UniSequent{\UniMSetFmA_2}{\UniMSetSucA})$ is a basic instance of $\mathsf{e}(W,W')$, then $\UniAlphaD{}(\UniMSetFmA_1) = \UniAlphaD{}(\UniMSetFmA_2)$.
\end{lemma}
\begin{proof}
    Assume that the metavariable $\UniMSetFmB$ was instantiated with the formula $\UniFmA$ in this instance of $\mathsf{e}(W,W')$.
    This already means that 
    $\UniListCountElem{\UniMSetFmA_1}{\UniFmA} = \UniListCountElem{\UniMSetFmA_2}{\UniFmA} \geq \UniEllParam$, since there are $\UniEllParam$ occurrences of $\UniMSetFmB$ in $\mathsf{e}(W,W')$.
    Thus by Definition~\ref{def:alpha_defs}, we have $\UniAlphaD{\UniFmA}(\UniMSetFmA_1) = \UniAlphaD{\UniFmA}(\UniMSetFmA_2)$, which implies
    $\UniAlphaD{}(\UniMSetFmA_1) = \UniAlphaD{}(\UniMSetFmA_2)$
    (recall Remark~\ref{rem:wc-type-order-indep}).
\end{proof}

Finally, the following normal form for derivable hypersequents will be useful for the decidability argument we will soon develop.

\begin{proposition}
    \label{fact:normalization-wc}
    A hypersequent $\UniSequent{\UniListFmA_1}{\UniMSetSucA_1} \VL \cdots \VL \UniSequent{\UniListFmA_k}{\UniMSetSucA_k}$ has a proof in a rule extension of $\UniFLExtHCalc{\UniWEProp{\vec{a}}}^\star$ with nc-height $\eta$
    iff the hypersequent $\UniSequent{\UniAlphaN{}(\UniListFmA_1)}{\UniMSetSucA_1} \VL \cdots \VL \UniSequent{\UniAlphaN{}(\UniListFmA_k)}{\UniMSetSucA_k}$ has a proof in the same calculus using only basic instances of $\mathsf{e}(W,W')$, with nc-height $\eta$.
\end{proposition}
\begin{proof}
    We provide a sketch of the proof. We will work only with sequents, since the case of hypersequents follows easily.
    Given a sequent $\UniSequent{\UniMSetFmA}{\Pi}$ and $\UniFmA \in \UniMSetFmA$, we argue that we can derive $\UniSequent{\UniMSetFmA}{\Pi}$ with nc-height $\eta$ if we can derive 
    $\UniSequent{\UniAlphaN{\UniFmA}(\UniMSetFmA)}{\Pi}$
    with the same nc-height $\eta$; the converse will be analogous.
    We will show how to get from one sequent to the other using only applications of  basic instances of the rules $\mathsf{e}(W,W')$ given in Proposition~\ref{prop:rules-wc}.
    Note that, if
    $\UniListCountElem{\UniMSetFmA}{\UniFmA} < \UniEllParam$,
    then the two sequents are equal as
    $\UniAlphaN{\UniFmA}(\UniMSetFmA) = \UniMSetFmA$.
    Otherwise,
     $\UniSequent{\UniMSetFmA}{\Pi}$
    equals
    $\UniSequent{
        \UniFrontWall{\UniFmA}(\UniMSetFmA),
        \UniFmA^\rho,
        \Theta,
        \UniBackWall{\UniFmA}(\UniMSetFmA)
    }{\Pi}$,
    where $\UniListCountElem{\Theta}{\UniFmA} = \UniListCountElem{\UniMSetFmA}{\UniFmA}-\UniPParam-\UniQParam-\rho$.
    We show that we can turn this sequent into 
    $\UniSequent{
        \UniFrontWall{\UniFmA}(\UniMSetFmA),
        \UniFmA^{\rho+\beta},
        \Theta',
        \UniBackWall{\UniFmA}(\UniMSetFmA)
    }{\Pi}$
    via a basic instance of $\mathsf{e}(W,W')$ for any $\beta \leq \UniEllParam-\UniPParam-\UniQParam$, where $\Theta'$ is like $\Theta$ but without the first $\beta$ occurrences of $\UniFmA$.
    Then a repeated application of this argument gives us the desired result.
    
    Note first that all antecedents of premises of $\mathsf{e}(W,W')$ have the form (recall we are instantiating the hypersequent variable with empty):
    \begin{align*}
    \underbrace{\Theta_1,\UniMSetFmB,\UniMSetFmA_1,\UniMSetFmB,\ldots,
    \UniMSetFmB,\UniMSetFmA_{\UniPParam-1},\UniMSetFmB}_{\text{front wall, $\UniPParam$ copies of $\UniMSetFmB$}},
    \underbrace{
    \UniMSetFmA_{\UniPParam},\UniMSetFmB,
    \ldots,
    \UniMSetFmB,
    \UniMSetFmA_{\UniEllParam-\UniQParam}
    }_{\text{$\UniEllParam-\UniPParam-\UniQParam$ copies of $\UniMSetFmB$}},\\
    \underbrace{
    \UniMSetFmB,
    \UniMSetFmA_{\UniEllParam-\UniQParam+1},
    \UniMSetFmB,\ldots,
    \UniMSetFmB,
    \UniMSetFmA_{\UniEllParam-1},
    \UniMSetFmB,
    \Theta_2}_{\text{back wall, $\UniQParam$ copies of $\UniMSetFmB$}}
    \end{align*}
    We instantiate $\UniMSetFmB$ with only $\UniFmA$. It is clear that we can instantiate the variables 
    $\UniMSetFmA_1,\ldots,\UniMSetFmA_{\UniPParam-1}$
    so that the above front wall matches $\UniFrontWall{\UniFmA}(\UniMSetFmA)$,
    as it has necessarily $\UniPParam$ copies of $\UniFmA$ and before each of them there is one sequence of other formulas.
    A similar reasoning applies to the back wall.
    Now we instantiate the middle section
    $\UniMSetFmA_{\UniPParam},\UniMSetFmB,
    \ldots,
    \UniMSetFmB,
    \UniMSetFmA_{\UniEllParam-\UniQParam}$, by first instantiating 
    $\UniMSetFmA_{\UniPParam}$ with the list starting right after the front wall and ending right before the 
    $(\UniPParam+\rho-(\UniEllParam-\UniPParam-\UniQParam-\beta)+1)$-th occurrence of $\UniFmA$.
    Then we instantiate $\UniMSetFmA_{\UniEllParam-\UniQParam}$ with the list starting right after the ($\UniPParam+\rho+\beta$)-th occurrence of $\UniFmA$ and ending right before the starting of $\UniBackWall{\UniFmA}(\UniMSetFmA)$.
    The remaining variables need to be instantiated to result in a list with $\beta$ occurrences of $\UniFmA$ that starts and ends with single occurrences of $\UniFmA$; this can clearly be done.
    The particular $\mathsf{e}(W,W')$ we take is the one with conclusion having as antecedent
    \begin{align*}
    \Theta_1,\UniMSetFmB,\UniMSetFmA_1,\UniMSetFmB,\ldots,
    \UniMSetFmB,\UniMSetFmA_{\UniPParam-1},\UniMSetFmB,
    \underbrace{
    \UniMSetFmB,
    \ldots,
    \UniMSetFmB,
    \UniMSetFmA_{\UniPParam},
    \ldots,
    \UniMSetFmA_{\UniEllParam-\UniQParam}
    }_{\text{all $\UniEllParam-\UniPParam-\UniQParam$ copies of $\UniMSetFmB$
    in front}}
    ,\\
    \UniMSetFmB,
    \UniMSetFmA_{\UniEllParam-\UniQParam+1},
    \UniMSetFmB,\ldots,
    \UniMSetFmB,
    \UniMSetFmA_{\UniEllParam-1},
    \UniMSetFmB,
    \Theta_2
    \end{align*}
    which is exactly what we want, since the middle section starts with
    $\UniEllParam-\UniPParam-\UniQParam$ copies of $\UniFmA$ from the $\UniMSetFmB$'s plus
    $\rho-(\UniEllParam-\UniPParam-\UniQParam-\beta)$ copies coming from the beginning of the instantiation of $\UniMSetFmA_{\UniPParam}$ ($\rho+\beta$ in total).

    Observe that we can further get to
    $\UniSequent{\UniAlphaN{}(\UniListFmA)}{\UniMSetSucA}$ by successively applying the same reasoning for each formula in $\UniListFmA$, thus matching the definition of $\UniAlphaN{}(\cdot)$ in Definition~\ref{def:alpha_defs}.
\end{proof}

 We present a simple example to illustrate the above result.
    
\begin{example}
    We take
    $\UniSequent{\UniMSetFmA}{\Pi}$
    to be the following sequent:
    \[
    \UniSequent{
    \UniFmB_1,\UniFmB_2,
    \UniFmA,
    \UniFmB_2,\UniFmB_1,\UniFmB_2,
    \UniFmA,
    \UniFmB_1,
    \UniFmA,
    \UniFmA,
    \UniFmB_2,\UniFmB_1,
    \UniFmA,
    \UniFmB_2,
    \UniFmA,
    \UniFmB_1,
    \UniFmA,
    \UniFmB_2, \UniFmB_2,
    \UniFmA,
    \UniFmB_1
    }{\Pi}
    \]
    We will employ the rule depicted below (obtained by taking $\UniEllParam=6, \UniPParam=2, \UniQParam=1$) to derive
    $\UniSequent{\UniAlphaN{\UniFmA}(\UniMSetFmA)}{\Pi}$.

    \begin{center}
            \AxiomC{$
                \UniHyperMSetA \VL
                \UniSequent{
                    \Theta_1,
                    \UniMSetFmB,
                    \UniMSetFmA_1,
                    \UniMSetFmB,
                    \UniMSetFmA_2,
                    \UniMSetFmB,
                    \UniMSetFmA_3,
                    \UniMSetFmB,
                    \UniMSetFmA_4,
                    \UniMSetFmB,
                    \UniMSetFmA_5,
                    \UniMSetFmB,
                    \Theta_2}{\UniMSetSucA}
                $}
            \UnaryInfC{$\UniHyperMSetA \VL
            \UniSequent{
                    \Theta_1,
                    \UniMSetFmB,
                    \UniMSetFmA_1,
                    \UniMSetFmB,
                    \UniMSetFmA_2,
                    \UniMSetFmB,
                    \UniMSetFmB,
                    \UniMSetFmB,
                    \UniMSetFmA_3,
                    \UniMSetFmA_4,
                    \UniMSetFmA_5,
                    \UniMSetFmB,
                    \Theta_2
                }
                {\UniMSetSucA}$}
            \DisplayProof
    \end{center}
    
    We consider the instantiation indicated below (we take $\UniMSetFmB$ instantiated to $\UniFmA$ and $\UniMSetFmA_3$ to the empty list):
    \[
        \UniSequent{
        \underbrace{\UniFmB_1,\UniFmB_2}_{\Theta_1},
        \UniFmA,
        \underbrace{\UniFmB_2,\UniFmB_1,\UniFmB_2}_{\UniMSetFmA_1},
        \UniFmA,
        \underbrace{\UniFmB_1}_{\UniMSetFmA_2},
        \UniFmA,
        \UniFmA,
        \underbrace{\UniFmB_2,\UniFmB_1}_{\UniMSetFmA_4},
        \UniFmA,
        \underbrace{\UniFmB_2,
        \UniFmA,
        \UniFmB_1,
        \UniFmA,
        \UniFmB_2, \UniFmB_2}_{\UniMSetFmA_5},
        \UniFmA,
        \underbrace{\UniFmB_1}_{\Theta_2}
        }{\Pi}
    \]
    
    By applying this particular instance, we get the sequent
    \[
        \UniSequent{
        \underbrace{\UniFmB_1,\UniFmB_2}_{\Theta_1},
        \UniFmA,
        \underbrace{\UniFmB_2,\UniFmB_1,\UniFmB_2}_{\UniMSetFmA_1},
        \UniFmA,
        \UniFmA,
        \UniFmA,
        \UniFmA,
        \underbrace{\UniFmB_1}_{\UniMSetFmA_2},
        \underbrace{\UniFmB_2,\UniFmB_1}_{\UniMSetFmA_4},
        \underbrace{\UniFmB_2,
        \UniFmA,
        \UniFmB_1,
        \UniFmA,
        \UniFmB_2, \UniFmB_2}_{\UniMSetFmA_5},
        \UniFmA,
        \underbrace{\UniFmB_1}_{\Theta_2}
        }{\Pi}
    \]
    
    Now we consider the {instance of the rule with premise}
    \[
        \UniSequent{
        \underbrace{\UniFmB_1,\UniFmB_2}_{\Theta_1},
        \UniFmA,
        \underbrace{\UniFmB_2,\UniFmB_1,\UniFmB_2}_{\UniMSetFmA_1},
        \UniFmA,
        \underbrace{
        \UniFmA,
        \UniFmA}_{\UniMSetFmA_2},
        \UniFmA,
        \underbrace{\UniFmB_1,
        \UniFmB_2,\UniFmB_1,
        \UniFmB_2}_{\UniMSetFmA_3},
        \UniFmA,
        \underbrace{\UniFmB_1}_{\UniMSetFmA_4},
        \UniFmA,
        \underbrace{\UniFmB_2, \UniFmB_2}_{\UniMSetFmA_5},
        \UniFmA,
        \underbrace{\UniFmB_1}_{\Theta_2}
        }{\Pi}
    \]
    
    The application of the instance produces the sequent
        \[
        \UniSequent{
        \underbrace{\UniFmB_1,\UniFmB_2}_{\Theta_1},
        \UniFmA,
        \underbrace{\UniFmB_2,\UniFmB_1,\UniFmB_2}_{\UniMSetFmA_1},
        \UniFmA,
        \UniFmA,
        \UniFmA,
        \UniFmA,
        \underbrace{
        \UniFmA,
        \UniFmA}_{\UniMSetFmA_2},
        \underbrace{\UniFmB_1,
        \UniFmB_2,\UniFmB_1,
        \UniFmB_2}_{\UniMSetFmA_3},
        \underbrace{\UniFmB_1}_{\UniMSetFmA_4},
        \underbrace{\UniFmB_2, \UniFmB_2}_{\UniMSetFmA_5},
        \UniFmA,
        \underbrace{\UniFmB_1}_{\Theta_2}
        }{\Pi}
    \]
    which is precisely the sequent $\UniSequent{\UniAlphaN{\UniFmA}(\UniMSetFmA)}{\Pi}$ we wanted. By starting from it and applying the inverse process we get back to $\UniSequent{\UniMSetFmA}{\Pi}$.
\end{example}

In order to cover deducibility in the logics of interest, we will follow the strategy developed in Section~\ref{sec:deducibility-adapts} and consider, for each finite set of formulas $\UniSetFmA$, a finite collection of rules $\UniRuleDedSet{\UniSetFmA}$ (cf.~Lemma~\ref{fact:cut-restrict-weak-weakening}).
Thus we will always work with calculi extended also with these rules. 
As it is the case in  Section~\ref{sec:decid-ub-ww-proof}, we will see that the proofs are not affected by this modification.

\section[Wqos for knotted rules with weak commutativity]{Well-quasi-orderings for handling knotted rules in the presence of weak commutativity}
\label{sec:nwqo-wc}

Our goal here is to adapt the proof search and upper-bound analysis of previous sections to the weakly commutative setting.
The first obstacle is finding a suitable nwqo on hypersequents from which we can define a suitable notion of minimal proof or irredundancy. In the presence of commutativity, we were able to represent a hypersequent using only the vectors of multiplicities of formulas in the antecedents of its components.
The nwqo, then, was essentially mirroring the effect of applying a knotted rule, making it possible to prove, for example, the hp-admissibility of such rule with respect to this ordering, thus establishing the existence of minimal proofs (with respect to such ordering). 
Without commutativity, we cannot do the same, as the position of the formulas is important and is needed in order to characterize a hypersequent, while the rigidity that this creates does not allow the application of a knotted rule.

We will show that using weak commutativity and a knotted rule we are able to define a suitable nwqo.
Before moving on, we provide some intuition for the definition. 
The key idea is that, in presence of weak commutativity, we may still consider only the multiplicities of the formulas in the components of the hypersequents insofar as we also include the type of the antecedents of the components in the definition of the ordering. 
Since the set $\UniAlphaD{}[\UniListClass{\UniSubfmlaHyperseqSet}]$ of all concerned types (where $\UniSubfmlaHyperseqSet$ is a finite set of formulas closed under subformulas) is finite by Proposition~\ref{fact:finitely-many-types}, we will be able to prove that such ordering is indeed a nwqo by providing a strong reflection to minoring and majoring ordering relations over tuples of natural numbers, from which we will also inherit length theorems.
The information of the type of the antecedent of a component will allow us also to handle the application of knotted rules and reach the desired notion of irredundant proofs.

For simplicity, we begin by defining an ordering on $\UniSubfmlaHyperseqSet$-sequents and then we lift it to $\UniSubfmlaHyperseqSet$-hypersequents using essentially the majoring and the minoring orderings.
Note that we use the same symbols for the orderings on sequents and hypersequents to avoid complicating the notation.

\begin{definition}\label{def:rel-weak-com-ctr}
Let $0 < n < m$ and let $\UniWEProp{\vec a}$ be a weak commutativity axiom. We define $\UniSequent{\UniListFmA_1}{\UniMSetSucA_1}\UniHyperseqWeCtrWqo{m}{n}{\UniSubfmlaHyperseqSet}~\UniSequent{\UniListFmA_2}{\UniMSetSucA_2}$ by: 
\begin{enumerate}
    \item $\UniMSetSucA_1 = \UniMSetSucA_2$;
    \item $\UniAlphaD{}(\UniListFmA_1) = \UniAlphaD{}(\UniListFmA_2)$; and
    \item 
    for all $\UniFmA \in \UniSubfmlaHyperseqSet$,
    $\UniListCountElem{\UniListFmA_1}{\UniFmA} \UniExtModRel{m'}{n'} 
    \UniListCountElem{\UniListFmA_2}{\UniFmA}$
    where $m' \UniSymbDef \UniEllParam + m$
    and $n' \UniSymbDef \UniEllParam+n$.
\end{enumerate}
\end{definition}

\noindent Observe that item (3) above is equivalent to the following property
($\ast$): for all $\UniFmA \in \UniSubfmlaHyperseqSet$, either
$\UniListCountElem{\UniListFmA_1}{\UniFmA} = \UniListCountElem{\UniListFmA_2}{\UniFmA}$,
or
$\UniListCountElem{\UniListFmA_1}{\UniFmA} \geq \UniEllParam + n, \UniListCountElem{\UniListFmA_2}{\UniFmA} \geq \UniEllParam+m$ %}
and
$\UniListCountElem{\UniMid{\UniFmA}(\UniListFmA_1)}{\UniFmA} \UniExtModRel{m}{n} 
\UniListCountElem{\UniMid{\UniFmA}(\UniListFmA_2)}{\UniFmA}$.

Indeed, to prove $(\ast)$ from $(3)$ assume that 
$\UniListCountElem{\UniListFmA_1}{\UniFmA} \UniExtModRel{m'}{n'} \UniListCountElem{\UniListFmA_2}{\UniFmA}$. By Definition~\ref{d: knottedorder}, we get that
$\UniListCountElem{\UniListFmA_1}{\UniFmA} =\UniListCountElem{\UniListFmA_2}{\UniFmA}$
or
$\UniListCountElem{\UniListFmA_1}{\UniFmA} \geq \UniEllParam + n, \UniListCountElem{\UniListFmA_2}{\UniFmA} \geq \UniEllParam+m$,
$\UniListCountElem{\UniListFmA_1}{\UniFmA}
< \UniListCountElem{\UniListFmA_2}{\UniFmA}$
and
$\UniListCountElem{\UniListFmA_1}{\UniFmA}
\UniEquivMod{m'-n'} \UniListCountElem{\UniListFmA_2}{\UniFmA}$.
In the first case, we are done.
Otherwise, note that $\UniListCountElem{\UniListFmA_1}{\UniFmA} < \UniListCountElem{\UniListFmA_2}{\UniFmA}$
implies that
$
\UniListCountElem{\UniMid{\UniFmA}(\UniListFmA_1)}{\UniFmA}=\UniListCountElem{\UniListFmA_1}{\UniFmA}-(\UniPParam+\UniQParam)
< \UniListCountElem{\UniListFmA_2}{\UniFmA}
-
(\UniPParam+\UniQParam)
=
\UniListCountElem{\UniMid{\UniFmA}(\UniListFmA_2)}{\UniFmA}$,
since
$\UniListCountElem{\UniListFmA_1}{\UniFmA}, \UniListCountElem{\UniListFmA_2}{\UniFmA} \geq \UniEllParam > \UniPParam+\UniQParam$.
Moreover, $m'-n' = (m+\UniEllParam)-(n+\UniEllParam) = m - n$, thus 
$\UniListCountElem{\UniListFmA_1}{\UniFmA}
\UniEquivMod{m-n} \UniListCountElem{\UniListFmA_2}{\UniFmA}$ and thus 
$\UniListCountElem{\UniMid{\UniFmA}(\UniListFmA_1)}{\UniFmA}=\UniListCountElem{\UniListFmA_1}{\UniFmA}-
(\UniPParam+\UniQParam)
\UniEquivMod{m-n} \UniListCountElem{\UniListFmA_2}{\UniFmA}-
(\UniPParam+\UniQParam)=\UniListCountElem{\UniMid{\UniFmA}(\UniListFmA_2)}{\UniFmA}$.
Finally,
$\UniListCountElem{\UniMid{\UniFmA}(\UniListFmA_1)}{\UniFmA}=\UniListCountElem{\UniListFmA_1}{\UniFmA}-
(\UniPParam+\UniQParam)
\geq 
(\UniEllParam+n)-(\UniPParam+\UniQParam) \geq n$.
Similarly, we get
$\UniListCountElem{\UniMid{\UniFmA}(\UniListFmA_2)}{\UniFmA} \geq m$.
Therefore, by Definition~\ref{d: knottedorder}, we have 
$\UniListCountElem{\UniMid{\UniFmA}(\UniListFmA_1)}{\UniFmA} \UniExtModRel{m}{n} 
\UniListCountElem{\UniMid{\UniFmA}(\UniListFmA_2)}{\UniFmA}$
as desired.

Getting $(3)$ from $(\ast)$ is similar.
Indeed, assume that
$\UniListCountElem{\UniListFmA_1}{\UniFmA} = \UniListCountElem{\UniListFmA_2}{\UniFmA}$,
or
$\UniListCountElem{\UniListFmA_1}{\UniFmA} \geq \UniEllParam + n, \UniListCountElem{\UniListFmA_2}{\UniFmA} \geq \UniEllParam+m$
and
$\UniListCountElem{\UniMid{\UniFmA}(\UniListFmA_1)}{\UniFmA} \UniExtModRel{m}{n} \UniListCountElem{\UniMid{\UniFmA}(\UniListFmA_2)}{\UniFmA}$.
From the latter, we have 
$\UniListCountElem{\UniMid{\UniFmA}(\UniListFmA_1)}{\UniFmA} < \UniListCountElem{\UniMid{\UniFmA}(\UniListFmA_2)}{\UniFmA}$
and
$\UniListCountElem{\UniMid{\UniFmA}(\UniListFmA_1)}{\UniFmA} \UniEquivMod{m-n} \UniListCountElem{\UniMid{\UniFmA}(\UniListFmA_2)}{\UniFmA}$,
and reasoning as above we obtain
$\UniListCountElem{\UniMSetFmA_1}{\UniFmA} < \UniListCountElem{\UniMSetFmA_2}{\UniFmA}$
and
$\UniListCountElem{\UniMSetFmA_1}{\UniFmA} \UniEquivMod{m'-n'} \UniListCountElem{\UniMSetFmA_2}{\UniFmA}$
as desired.

\begin{definition}\label{def:rel-weak-com-ctr-hyper}
For all $\UniSubfmlaHyperseqSet$-hypersequents
$\UniHypersequentA_1$, $\UniHypersequentA_2$,
we define $\UniHypersequentA_1 \UniHyperseqWeCtrWqo{m}{n}{\UniSubfmlaHyperseqSet}~\UniHypersequentA_2$
by: for each component $\UniSequentA_2 \in \UniHypersequentA_2$, there is a component $\UniSequentA_1 \in \UniHypersequentA_1$ such that
$\UniSequentA_1 \UniHyperseqWeCtrWqo{m}{n}{\UniSubfmlaHyperseqSet}~\UniSequentA_2$.
Also, we set
$\UniNwqoWeCtnHSeqName{m}{n} \UniSymbDef 
\UniStruct{\UniOmegaHypersequentsSet{\UniSubfmlaHyperseqSet},
\UniHyperseqWeCtrWqo{m}{n}{\UniSubfmlaHyperseqSet},
\UniNormHSeq{\cdot}}$.
\end{definition}

\begin{definition}\label{def:rel-weak-com-wkn-hyper}
For all $\UniSubfmlaHyperseqSet$-hypersequents
$\UniHypersequentA_1$, $\UniHypersequentA_2$,
we define $\UniHypersequentA_1 \UniHyperseqWeWknWqo{m}{n}{\UniSubfmlaHyperseqSet}~\UniHypersequentA_2$
by: for each component $\UniSequentA_1 \in \UniHypersequentA_1$,
there is a component $\UniSequentA_2 \in \UniHypersequentA_2$
such that
$\UniSequentA_1 \UniHyperseqWeCtrWqo{m}{n}{\UniSubfmlaHyperseqSet}~\UniSequentA_2$.
Also, set
$\UniNwqoWeWknHSeqName{m}{n} \UniSymbDef 
\UniStruct{\UniOmegaHypersequentsSet{\UniSubfmlaHyperseqSet},
\UniHyperseqWeWknWqo{m}{n}{\UniSubfmlaHyperseqSet},
\UniNormHSeq{\cdot}}$.
\end{definition}

In order to prove that the above are nwqos and to provide a length theorem for them, we present an encoding into the nwqo
$(\UniPowerSetFin{\UniFlatWqo{\UniSetCard{\UniAlphaD{}(\UniListClass{\UniSubfmlaHyperseqSet})}} 
\cdot \UniWqoExtModRelProd{m'}{n'}{d}})^{d+1}$.
The definition is essentially the one in Definition~\ref{def:encoding-hypersequents}, with the difference that now we include the possible types of the antecedents of the $\UniSubfmlaHyperseqSet$-sequents.

\begin{definition}
    \label{def:encoding-hypersequents-types}
    Let $0 < n < m$, let $\UniWEProp{\vec a}$ be a weak commutativity axiom and let
    $\UniSubfmlaHyperseqSet \UniSymbDef
    \UniSet{\UniFmA_1,\ldots,\UniFmA_d}$ be an ordered set of formulas with $d \in \UniNaturalSet$; we write $\UniAlphaD{}[\UniListClass{\UniSubfmlaHyperseqSet}]$ for the nwqo $\UniStruct{\UniAlphaD{}[\UniListClass{\UniSubfmlaHyperseqSet}], =, \lambda a. 0}$.
    For an $\UniSubfmlaHyperseqSet$-hypersequent $\UniHypersequentA$, we define 
    \[
    \UniHyperTransNat{\UniHypersequentA}
    \UniSymbDef
    \UniTuple{X_0,\ldots,X_{d}} \in
    (\UniPowerSetFin{\UniFlatWqo{\UniAlphaD{}[\UniListClass{\UniSubfmlaHyperseqSet}]} 
        \times \UniNaturalSet^d})^{d+1},
    \]
    where
    \[
    X_0 \UniSymbDef
        \UniSet{
            \UniTuple{\UniAlphaD{}(\UniListFmA), \UniTuple{\UniListCountElem{\UniListFmA}{\UniFmA_1},\ldots,\UniListCountElem{\UniListFmA}{\UniFmA_d}}} \in  
            (\UniFlatWqo{{\UniAlphaD{}[\UniListClass{\UniSubfmlaHyperseqSet}]}} \times  \UniNaturalSet^d)
            \mid
            \UniSequent{\UniListFmA}{}
            \in \UniHypersequentA
        }
    \]
    and, for each $1 \leq i \leq d$,
    \[
    X_i \UniSymbDef
    \UniSet{
        \UniTuple{\UniAlphaD{}(\UniListFmA), \UniTuple{\UniListCountElem{\UniListFmA}{\UniFmA_1},\ldots,\UniListCountElem{\UniListFmA}{\UniFmA_d}}} \in  
        (\UniFlatWqo{\UniAlphaD{}[\UniListClass{\UniSubfmlaHyperseqSet}]} \times  \UniNaturalSet^d)
        \mid
        \UniSequent{\UniListFmA}{\UniFmA_i}
        \in \UniHypersequentA
    }.
    \]
\end{definition}

With the above encoding, we can prove the following length theorem.

\begin{proposition}
\label{fact:wc-length-theorem}
Let $\UniWEProp{\vec a}$ be a weak commutativity axiom.
For all finite sets of formulas $\UniSubfmlaHyperseqSet$, $\UniNwqoWeCtnHSeqName{m}{n}$ (for {$0 < n < m$}) and $\UniNwqoWeWknHSeqName{m}{n}$
(for {$0 \leq m < n$}) are nwqos and $\UniNwqoWeCtnHSeqName{m}{n}, \UniNwqoWeWknHSeqName{m}{n} \sqsubseteq \UniFGHOneAppLevel{\omega^{\UniSetCard{\UniSubfmlaHyperseqSet}}}$.
Moreover,
$$\{ \UniNwqoWeCtnHSeqNameVar{m}{n}{\UniSubfmlaHyperseqSet_k}\}_{k \in \UniNaturalSet},
\{ \UniNwqoWeWknHSeqNameVar{m}{n}{\UniSubfmlaHyperseqSet_k}\}_{k \in \UniNaturalSet}\sqsubseteq \UniFGHOneAppLevel{\omega^\omega}.$$
\end{proposition}
\begin{proof}
Clearly, $(\cdot)^\#$, described in Definition~\ref{def:encoding-hypersequents-types}, is a strong reflection, respectively, into
$\UniPowerMaj{\UniMinoringFntWqo{{{\UniAlphaD{}[\UniListClass{\UniSubfmlaHyperseqSet}]} \times}\UniWqoExtModRelProd{m'}{n'}{k}}}{d}$
and
$\UniPowerMaj{\UniMajoringFntWqo{{{\UniAlphaD{}[\UniListClass{\UniSubfmlaHyperseqSet}]} \times}\UniWqoExtModRelProd{n'}{m'}{k}}}{d}$,
which are isomorphic to
$\UniPowerMaj{\UniMinoringFntWqo{{\UniSetCard{\UniAlphaD{}[\UniListClass{\UniSubfmlaHyperseqSet}]} \cdot}\UniWqoExtModRelProd{m'}{n'}{k}}}{d}$
and
$\UniPowerMaj{\UniMajoringFntWqo{{\UniSetCard{\UniAlphaD{}[\UniListClass{\UniSubfmlaHyperseqSet}]} \cdot}\UniWqoExtModRelProd{n'}{m'}{k}}}{d}$ respectively.
The result then follows from Theorem~\ref{fact:length-theorem-power-set-wqo-maj-mn}, reasoning as in the proof of Proposition~\ref{fact:hnwqo-length-theorem}.
\end{proof}

We now proceed with the application of these nwqos to proof search and to an upper-bound analysis of the logics of interest.

\section[Upper bounds: weakly commutativity with knotted contraction]{Upper bounds for weakly commutative logics with knotted contraction}

We fix a finite set of formulas $\UniSetFmA$ that will serve as the set of assumptions in an instance of deducibility.
We begin by defining a version of 
$\UniCalcExt{\UniFLExtHCalc{\UniWEProp{\vec{a}}\UniWeakCProp{m}{n}}^\star}{\UniAnaRuleSet\cup \UniRuleDedSet{\UniSetFmA}}$ 
(recall Section~\ref{sec:deducibility-adapts}), where $\UniAnaRuleSet$ is a finite set of analytic structural rules, that absorbs applications of knotted contraction in the middle part of antecedents of components of hypersequents.

\begin{definition}
For $0 < n < m$, $k \geq 0$, and weak commutativity axiom $\UniWEProp{\vec a}$, we define the following binary relations on hypersequents:

\begin{itemize}
    \item $\UniHypersequentA_1 \UniIntCtrRel{k}{\UniWeakCtr} \UniHypersequentA_{2}$ iff
        for each $\UniSequentA_2 \UniSymbDef (\UniSequent{\UniListFmA_2}{\UniMSetSucA}) \in \UniHypersequentA_2$ there is
            $\UniSequentA_1 \UniSymbDef (\UniSequent{\UniListFmA_1}{\UniMSetSucA}) \in \UniHypersequentA_1$
            such that $\UniSequentA_2 \UniHyperseqWeCtrWqo{m}{n}{\UniSubfmlaHyperseqSet} \UniSequentA_1$
            and for each formula $\UniFmA$ in the antecedent of $\UniSequentA_2$,  
        $0 \leq \UniListCountElem{\UniMid{\UniFmA}(\UniListFmA_1)}{\UniFmA} - \UniListCountElem{\UniMid{\UniFmA}(\UniListFmA_2)}{\UniFmA} \leq k$.
    \item $\UniHypersequentA_1 \UniExtCtrRel{k}{\text{(EC)}} \UniHypersequentA_{2}$ iff~$\UniHypersequentA_{2}$ can be obtained from~$\UniHypersequentA_1$ by applying some number of~$\text{(EC)}$ such that every component is up to~$k$ occurrences fewer in~$\UniHypersequentA_{2}$ than in~$\UniHypersequentA_1$. 
    \item $\UniHypersequentA\UniIntExtCtrRel{k}{l}\UniHypersequentA_{1}$ iff there is $\UniHypersequentA'$ such that~$h \UniIntCtrRel{k}{\UniWeakCtr} \UniHypersequentA'$ and $\UniHypersequentA' \UniExtCtrRel{l}{\text{(EC)}} \UniHypersequentA_{1}$.
    \end{itemize}
\end{definition}

\begin{definition}\label{def:absorb-calc-weak-com}
    Let $\UniFmlaMultFixed$ and $\UniActiveCompFixed$ be as in Definition~\ref{def:acn-fm} and set
    $\UniFixedCtrAmount \UniSymbDef (\UniFmlaMultFixed - 1)(\UniEllParam + m-1)$.
    Then $\UniHFLwecRAbsorb$ is obtained from
    $\UniCalcExt{\UniFLExtHCalc{\UniWEProp{\vec{a}}\UniWeakCProp{m}{n}}^\star}{\UniAnaRuleSet\cup \UniRuleDedSet{\UniSetFmA}}$
    by removing the cut rule and the knotted contraction rule, and replacing each rule schema~$\UniRuleSchemaA$ with the rule
    $\UniHRuleAbsorb{\UniRuleSchemaA}$
    displayed below, in which $\UniHypersequentA_{0} \UniIntExtCtrRel{\UniFixedCtrAmount}{\UniActiveCompFixed} \UniHypersequentB$:
    \begin{center}
    \begin{tabular}{c@{\hspace{1cm}}c}
    \AxiomC{$\UniHypersequentA_{1}\quad\cdots\quad \UniHypersequentA_{n}$}
    \RightLabel{$\UniRuleSchemaA$}
    \UnaryInfC{$\UniHypersequentA_{0}$}
    \DisplayProof
    &
    \AxiomC{$\UniHypersequentA_{1}\quad\cdots\quad \UniHypersequentA_{n}$}
    \RightLabel{$\UniHRuleAbsorb{\UniRuleSchemaA}$}
    \UnaryInfC{$\UniHypersequentB$}
    \DisplayProof
    \end{tabular}
    \end{center}
    The \textit{base instance} of~$\UniHRuleAbsorb{\UniRuleSchemaA}$ is the particular instance of~$\UniHRuleAbsorb{\UniRuleSchemaA}$ whose conclusion is~$\UniHypersequentA_{0}$, i.e., no contractions are applied.
\end{definition}

\begin{proposition}\label{prop:wc-same-ded-absord}
Let $0 < n < m$, $\UniWEProp{\vec a}$ be a weak commutativity axiom, $\UniAnaRuleSet$ be
a finite set of analytic structural rules
and
$\UniHyperCalcA\UniSymbDef\UniCalcExt{\UniFLExtHCalc{\UniWEProp{\vec{a}}\UniWeakCProp{m}{n}}^\star}{\UniAnaRuleSet\cup \UniRuleDedSet{\UniSetFmA}}$.
Then $\UniDelCut{\UniHyperCalcA}$ and $\UniHFLwecRAbsorb$ induce the same deducibility relation on hypersequents, that is,
$\UniHyperDerivRel{\UniDelCut{\UniHyperCalcA}} \;=\; \UniHyperDerivRel{\UniHFLwecRAbsorb}$.
\end{proposition}
\begin{proof}
The inclusion
$\UniHyperDerivRel{\UniDelCut{\UniHyperCalcA}} \;\subseteq\; \UniHyperDerivRel{\UniHFLwecRAbsorb}$
is clear, as instances of rules of $\UniDelCut{\UniHyperCalcA}$ are base instances of the same rules in $\UniHFLwecRAbsorb$.
The converse inclusion is proved by showing that every instance of
$\UniHRuleAbsorb{\UniRuleSchemaA}$ 
is derivable in $\UniDelCut{\UniHyperCalcA}$.
In fact, if
$\UniHypersequentA_{0} \UniIntExtCtrRel{\UniFixedCtrAmount}{\UniActiveCompFixed} \UniHypersequentB$,
we have that there is
$\UniHypersequentA'$ such that~$\UniHypersequentA \UniIntCtrRel{\UniFixedCtrAmount}{\UniWeakCtr} \UniHypersequentA'$ and $\UniHypersequentA' \UniExtCtrRel{\UniActiveCompFixed}{\text{(EC)}} \UniHypersequentA_{1}$.
We can then obtain $\UniHypersequentA'$ from $\UniHypersequentA$ by basic instances of $\mathsf{e}(W,W')$ 
(check Proposition~\ref{fact:normalization-wc}) and applications of
$\UniWeakCRule{m}{n}$, and then get from $\UniHypersequentA'$ to $\UniHypersequentA_1$ by applications of $\mathrm{(EC)}$.
\end{proof}

Now, we prove Curry's Lemma for 
$\UniHFLwecRAbsorb$,
which is key in showing the existence of irredundant proofs with respect to the nwqo
$\UniNwqoWeCtnHSeqName{m}{n}$.
For what follows, recall the notion of nc-height given in Definition~\ref{def:nc-height}.
 
\begin{lemma}[Curry's Lemma for $\UniHFLwecRAbsorb$]
\label{lem:curry-lemma-wc}
\sloppy
Let $0 < n < m$ and $\UniWEProp{\vec a}$ be a weak commutativity axiom.
If $\UniHypersequentA_1$ has a proof in $\UniHFLwecRAbsorb$ with nc-height $\eta$ and 
$\UniHypersequentA_2 \UniHyperseqWeCtrWqo{m}{n}{\UniSubfmlaHyperseqSet} \UniHypersequentA_1$, then $\UniHypersequentA_2$ has a proof in $\UniHFLwecRAbsorb$ with nc-height at most $\eta$.
\end{lemma}
\begin{proof}
    Let $\UniHypersequentA_1 \UniSymbDef \UniSequentA_1 \VL \cdots \VL \UniSequentA_l$ and assume 
    $\UniHypersequentA_2 \UniHyperseqWeCtrWqo{m}{n}{\UniSubfmlaHyperseqSet} \UniHypersequentA_1$,
    for some $\UniHypersequentA_2 \UniSymbDef \UniSequentB_1 \VL \cdots \VL \UniSequentB_{l'}$, with $l, l' \geq 0$. 
    This means, by Definition~\ref{def:rel-weak-com-ctr-hyper}, that for each $1 \leq i \leq l$ there is $1 \leq j_i \leq l'$ such that $\UniSequentB_{j_i} \UniHyperseqWeCtrWqo{m}{n}{\UniSubfmlaHyperseqSet} \UniSequentA_i$.
    We proceed by induction on the nc-height of a proof $\UniDerivationA$ of $\UniHypersequentA_1$ in $\UniHFLwecRAbsorb$.

    In the base case, if $\UniDerivationA$ has nc-height $1$, then $\UniHypersequentA_1$ must result from an initial rule schema followed by applications of basic instances of $\mathsf{e}(W,W')$, and either for no distinct hypersequent $\UniHypersequentA_2$ we have
    $\UniHypersequentA_2 \UniHyperseqWeCtrWqo{m}{n}{\UniSubfmlaHyperseqSet} \UniHypersequentA_1$
    or we can derive $\UniHypersequentA_2$ by picking an appropriate instance of the same initial rule (such that the multiplicity of the formulas in the middle part of the sequents is adjusted to match the multiplicities in $\UniHypersequentA_2$).
    
    In the inductive step, suppose that (IH): the lemma holds for every hypersequent provable with nc-height less than $\eta$, where $\eta > 1$.
    We will prove that it holds for hypersequents provable with nc-height $\eta$.
    First, we show how to prove the hypersequent $\UniHypersequentA_2' \UniSymbDef \UniSequentB_{j_1} \VL \cdots \VL \UniSequentB_{j_l}$
    obtained from $\UniHypersequentA_1$ by replacing $\UniSequentA_i$ with its corresponding $\UniSequentB_{j_i}$, using  a derivation of height at most $\eta$. 
    To get from $\UniHypersequentA_2'$ to $\UniHypersequentA_2$, then, we will only need to consider $\mathrm{(EC)}$ and $\mathrm{(EW)}$, but these cases will be essentially the same as in the proof of admissibility of these rules in Lemma~\ref{lem-hp-admissible}.
    
    We work with one sequent from $\UniHypersequentA_1$ at a time. We pick $\UniSequentA_1$ and assume, without loss of generality, that 
    $\UniSequentB_{1} \UniHyperseqWeCtrWqo{m}{n}{\UniSubfmlaHyperseqSet} \UniSequentA_1$; we write $\UniHypersequentA_1$ as $\UniHypersequentB \VL \UniSequentA_1$.
    We will show how to derive $\UniHypersequentA_1' \UniSymbDef \UniHypersequentB \VL \UniSequentB_1$
    with nc-height no more than $\eta$.
    The definition of $\UniHyperseqWeCtrWqo{m}{n}{\UniSubfmlaHyperseqSet}$ allows us to work on one formula $\UniFmA$ at a time. 
    The idea is to reduce the multiplicity of each $\UniFmA$ in the middle part (for $\UniFmA$), obtaining intermediate hypersequents provable with nc-height at most $\eta$, until reaching the desired $\UniSequentB_1$. 
    
    Let us consider the last rule instance applied in $\UniDerivationA$, say it was
    $\UniHRuleAbsorb{\UniRuleSchemaA}$; we write $\UniSequentA_1$ as $\UniSequent{\UniListFmA_1}{\UniMSetSucA}$
    and
    $\UniSequentB_1$ as $\UniSequent{\UniListFmA_2}{\UniMSetSucA}$
    (succedents are the same because $\UniSequentB_{1} \UniHyperseqWeCtrWqo{m}{n}{\UniSubfmlaHyperseqSet} \UniSequentA_1$).
    Then there must be a base instance of~$\UniHRuleAbsorb{\UniRuleSchemaA}$ with conclusion $\UniHypersequentB_0$ such that
        \[
        \UniHypersequentB_{0} 
        \UniIntCtrRel{\UniFixedCtrAmount}{\UniWeakCtr} 
        \UniHypersequentB_{1} 
        \UniExtCtrRel{\UniActiveCompFixed}{\text{(EC)}} 
        \UniHypersequentB \VL \UniSequentA_1.
        \]

    If $\UniFmA$ is a formula in the antecedent of $\UniSequentB_1$ and $\UniListCountElem{\UniListFmA_1}{\UniFmA} = \UniListCountElem{\UniListFmA_2}{\UniFmA}$, we simply use basic instances of $\mathsf{e}[W,W']$ and note that the nc-height is not affected by such applications.
    Otherwise, we have
    $\UniListCountElem{\UniListFmA_1}{\UniFmA} \geq \UniEllParam + m, \UniListCountElem{\UniListFmA_2}{\UniFmA} \geq \UniEllParam+n$ and
    $\UniListCountElem{\UniMid{\UniFmA}(\UniListFmA_1)}{\UniFmA} \UniExtModRel{m}{n} 
    \UniListCountElem{\UniMid{\UniFmA}(\UniListFmA_2)}{\UniFmA}$.
    If we define $\UniSequentA_1^\UniFmA$ to be the sequent obtained from $\UniSequentA_1$ by removing $m-n$ occurrences of $\UniFmA$ in $\UniMid{\UniFmA}(\UniListFmA_1)$, we show now how to obtain $\UniHypersequentB \VL \UniSequentA_1^\UniFmA$ via a derivation with nc-height at most $\eta$.

    Note that we can write 
    $\UniSequentA_1$ as 
    $\UniSequent{\UniFrontWall{\UniFmA}(\UniListFmA_1),\Theta,\UniBackWall{\UniFmA}(\UniListFmA_1)}{\UniMSetSucA}$ and
    $\UniHypersequentB \VL \UniSequentA_1$ as:

    {\footnotesize
    \[
\underbrace{\UniFmMSetMinusFm{\UniHypersequentB}
    \VL
    \overbrace{
    \UniSequent{\UniFrontWall{\UniFmA}(\UniListFmA_1),\Theta,\UniBackWall{\UniFmA}(\UniListFmA_1)}{\UniMSetSucA}
    \VL
    \ldots
    \VL
    \UniSequent{\UniFrontWall{\UniFmA}(\UniListFmA_1),\Theta,\UniBackWall{\UniFmA}(\UniListFmA_1)}{\UniMSetSucA}
    }^{\text{$k$ components}}}_{\text{$\UniHypersequentB$}}
    \VL
    \UniSequent{\UniFrontWall{\UniFmA}(\UniListFmA_1),\Theta,\UniBackWall{\UniFmA}(\UniListFmA_1)}{\UniMSetSucA}
    \]
    }%
    where $k \geq 0$ and $\UniListCountElem{\Theta}{\UniFmA} = \gamma + m$,
    for $\gamma \geq 1$
    (recall that $\UniPParam + \UniQParam < \UniEllParam$) %}
    and $\UniFmMSetMinusFm{\UniHypersequentB}$
    having no occurrences of $\UniSequentA_1$.

    Since~
    $\UniHypersequentB_{1} \UniExtCtrRel{\UniActiveCompFixed}{\text{(EC)}}  \UniHypersequentB
    \VL
    \UniSequentA_1$,
    we can partition~$\UniHypersequentB_{1}$ into the portion~$\UniHypersequentB_{1}'$ that externally contracts to~$\UniFmMSetMinusFm{\UniHypersequentB}$ (i.e.,
    $\UniHypersequentB_{1}'\UniExtCtrRel{\UniActiveCompFixed}{\text{(EC)}} \UniFmMSetMinusFm{\UniHypersequentB}$), and the portion that externally contracts to the remainder, which consists of occurrences of $\UniSequentA_1$. 
    Specifically, by the definition of $\UniExtCtrRel{\UniActiveCompFixed}{\text{(EC)}}$,
    there exists~$\alpha \geq 1$ with $(k + \alpha)-(k+1) \leq \UniActiveCompFixed$,
    or, more simply, ~$1\leq\alpha\leq\UniActiveCompFixed+1$, such that
    \[
    \UniHypersequentB_{1} =
    \UniHypersequentB_{1}' \VL
    \overbrace{
    \UniSequent{\UniFrontWall{\UniFmA}(\UniListFmA_1),\Theta,\UniBackWall{\UniFmA}(\UniListFmA_1)}{\UniMSetSucA}
    \VL
    \ldots
    \VL
    \UniSequent{\UniFrontWall{\UniFmA}(\UniListFmA_1),\Theta,\UniBackWall{\UniFmA}(\UniListFmA_1)}{\UniMSetSucA}
    }^{\text{$k+\alpha$ components}}
    \]

    From this, let us open up the structure of the hypersequent $\UniHypersequentB_0$ introduced above (recall that $\UniHypersequentB_{0} 
    \UniIntCtrRel{\UniFixedCtrAmount}{\UniWeakCtr} \UniHypersequentB_{1}$). 
    For the $i^{\text{th}}$ the above $k+\alpha$ components, there is~$\beta_{i}$ with~$0 \leq \beta_{i}\leq \UniFixedCtrAmount$ and lists~$\Theta_i, \UniMSetFmA'_{i}$ such that
    \[
    \UniHypersequentB_{0} = \UniHypersequentB_{0}' \VL \overbrace{
    \UniSequent{\UniFrontWall{\UniFmA}(\UniListFmA'_1),\Theta_{1},\UniBackWall{\UniFmA}(\UniListFmA'_1)}{\UniMSetSucA}
    \VL 
    \ldots 
    \VL 
    \UniSequent{\UniFrontWall{\UniFmA}(\UniListFmA'_{k+\alpha}),\Theta_{k+\alpha},\UniBackWall{\UniFmA}(\UniListFmA'_{k+\alpha})}{\UniMSetSucA}
    }^{\text{$k+\alpha$ components}}
    \]
    with 
    $|\Theta_{i}|_{\UniFmA} = \gamma + m + \beta_i$,
    $\UniHypersequentB_{0}' 
    \UniIntCtrRel{\UniFixedCtrAmount}{\UniWeakCtr} 
    \UniHypersequentB_{1}'$ and
    \begin{multline*}
    \overbrace{
        \UniSequent{\UniFrontWall{\UniFmA}(\UniListFmA'_1),\Theta_{1},\UniBackWall{\UniFmA}(\UniListFmA'_1)}{\UniMSetSucA}
    \VL 
    \ldots 
    \VL 
    \UniSequent{\UniFrontWall{\UniFmA}(\UniListFmA'_{k+\alpha}),\Theta_{k+\alpha},\UniBackWall{\UniFmA}(\UniListFmA'_{k+\alpha})}{\UniMSetSucA}
    }^{\text{$k+\alpha$ components}}
    \quad 
    \UniIntCtrRel{\UniFixedCtrAmount}{\UniWeakCtr}\\
    \underbrace{
    \UniSequent{\UniFrontWall{\UniFmA}(\UniListFmA_1),\Theta,\UniBackWall{\UniFmA}(\UniListFmA_1)}{\UniMSetSucA}
    \VL
    \ldots
    \VL
    \UniSequent{\UniFrontWall{\UniFmA}(\UniListFmA_1),\Theta,\UniBackWall{\UniFmA}(\UniListFmA_1)}{\UniMSetSucA}
    }_{\text{$k+\alpha$ components}}
    \end{multline*}

    We easily obtain that
    $m + \beta_i \leq  \UniFixedCtrAmount + m$
    (by the definition of $\UniIntCtrRel{\UniFixedCtrAmount}{\UniWeakCtr}$).
    We now want to be able to use an application of the same $\UniHRuleAbsorb{\UniRuleSchemaA}$ to obtain 
    \begin{multline}\label{eq:target-hypersequent-wc}
        g_1' \VL
\underbrace{\UniSequent{\UniFrontWall{\UniFmA}(\UniListFmA_1),\Theta,\UniBackWall{\UniFmA}(\UniListFmA_1)}{\UniMSetSucA}
        \VL
        \cdots
        \VL
        \UniSequent{\UniFrontWall{\UniFmA}(\UniListFmA_1),\Theta,\UniBackWall{\UniFmA}(\UniListFmA_1)}{\UniMSetSucA}
    }_{k}
    \VL\\
    \underbrace{
        \UniSequent{\UniFrontWall{\UniFmA}(\UniListFmA_1),\Theta',\UniBackWall{\UniFmA}(\UniListFmA_1)}{\UniMSetSucA}
        \VL
        \cdots
        \VL
        \UniSequent{\UniFrontWall{\UniFmA}(\UniListFmA_1),\Theta',\UniBackWall{\UniFmA}(\UniListFmA_1)}{\UniMSetSucA}
    }_{\alpha}
    \end{multline}
    instead, where 
    $\UniListCountElem{\Theta'}{\UniFmA} = \UniListCountElem{\Theta}{\UniFmA} - (m-n) = \gamma + n$.
    
    Since $m - n < m$, we have $\beta_i + (m-n) < m + \beta_i$ and thus
    $\beta_i + (m - n) < \UniFixedCtrAmount + m$.
    Without loss of generality we can assume that there is some $N \geq 0$ such that
    $\beta_j + (m-n)$, for all $k+1 \leq j \leq k+N$, is between $\UniFixedCtrAmount$ and $\UniFixedCtrAmount + m$,
    and $\beta_{j'} + (m-n)$ is less than $\UniFixedCtrAmount$ for all $j' > k+N$.
    By the definition of formula multiplicity, the number of items in the antecedent of every component in the conclusion of the rule schema has at most~$\UniFmlaMultFixed-1$ elements. 
    Therefore, for every~$i$ such that $k+1\leq i\leq k+N$, either the component
    $\UniSequent{\UniFrontWall{\UniFmA}(\UniListFmA'_i),\Theta_{i},\UniBackWall{\UniFmA}(\UniListFmA'_i)}{\UniMSetSucA}$ is in the instantiation of the hypersequent-variable, or by the pigeonhole principle at least~$m+\UniEllParam$ copies (recall the value of $\UniFixedCtrAmount$ in Definition~\ref{def:absorb-calc-weak-com}) of~$\UniFmA$ occur in the instantiation of some multiset-variable from that component. 
    In either case we apply the induction hypothesis to those instantiations in the premise(s) to make $m$ copies of $\UniFmA$ into $n$ copies in the middle part (for $\UniFmA$).
    Note that this is possible because there are more than $\UniEllParam$ copies of the formula, thus this reduction in multiplicity does not change the type of the sequent; in particular, the front and back walls remain the same. 
    Then we apply~$\UniHRuleAbsorb{\UniRuleSchemaA}$ to get the following base instance:
        {
        \small
        \begin{multline*}
        \UniHypersequentB_{0}' \VL
        \overbrace{
        \UniSequent{\UniFrontWall{\UniFmA}(\UniListFmA_1),\Theta,\UniBackWall{\UniFmA}(\UniListFmA_1)}{\UniMSetSucA}
        \VL
        \cdots
        \VL
        \UniSequent{\UniFrontWall{\UniFmA}(\UniListFmA_1),\Theta,\UniBackWall{\UniFmA}(\UniListFmA_1)}{\UniMSetSucA}
        }^{\text{$k$ components}} \VL \\
        \overbrace{
        \UniSequent{\UniFrontWall{\UniFmA}(\UniListFmA_1),\Theta'_{k+1},\UniBackWall{\UniFmA}(\UniListFmA_1)}{\UniMSetSucA}
        \VL
        \cdots
        \VL
        \UniSequent{\UniFrontWall{\UniFmA}(\UniListFmA_1),\Theta'_{k+N},\UniBackWall{\UniFmA}(\UniListFmA_1)}{\UniMSetSucA}
        }^{\text{$m$ copies of $\UniFmA$ to $n$ copies in the middle part via IH on premises}} \VL \\ 
        \UniSequent{\UniFrontWall{\UniFmA}(\UniListFmA_1),\Theta_{k+N+1},\UniBackWall{\UniFmA}(\UniListFmA_1)}{\UniMSetSucA} \VL
        \ldots \VL
        \UniSequent{\UniFrontWall{\UniFmA}(\UniListFmA_1),\Theta_{k+\alpha},\UniBackWall{\UniFmA}(\UniListFmA_1)}{\UniMSetSucA}
        \end{multline*}
        }%
where $\UniListCountElem{\Theta'_{i}}{\UniFmA} = \gamma + n + \beta_i$.
    The above is then related to the hypersequent displayed in Equation \ref{eq:target-hypersequent-wc} under $\UniIntCtrRel{\UniFixedCtrAmount}{\UniWeakCtr}$. The latter is then related via $\UniExtCtrRel{\UniActiveCompFixed}{\text{(EC)}}$ to 
    \begin{multline*}
    \label{eq:final-target-hypersequent-wc}
    g^{-} \VL
    \underbrace{
        \UniSequent{\UniFrontWall{\UniFmA}(\UniListFmA_1),\Theta,\UniBackWall{\UniFmA}(\UniListFmA_1)}{\UniMSetSucA}
        \VL
        \cdots
        \VL
        \UniSequent{\UniFrontWall{\UniFmA}(\UniListFmA_1),\Theta,\UniBackWall{\UniFmA}(\UniListFmA_1)}{\UniMSetSucA}
    }_{k}
    \VL\\
        \UniSequent{\UniFrontWall{\UniFmA}(\UniListFmA_1),\Theta',\UniBackWall{\UniFmA}(\UniListFmA_1)}{\UniMSetSucA}
    \end{multline*}
    which is just $g \VL \UniSequentA_1^\UniFmA$, as desired.
    The nc-height of this derivation is at most $\eta$.
    We repeat the same argument for $\UniFmA$ starting from the obtained hypersequent, until the desired multiplicity (the one observed in $\UniSequentB_1$) is obtained.
    Then we work on the other formulas in the same way. We do this also for the remaining sequents from $\UniHypersequentA$, until $\UniSequentB_{j_1} \VL \cdots \VL \UniSequentB_{j_{l'}}$ is reached, via a derivation of nc-height $\eta$, as desired.
    As explained before, to get to $\UniHypersequentA_2$, it only remains to consider the hp-admissibility of rules $\mathrm{(EC)}$ and $\mathrm{(EW)}$, as done in Lemma~\ref{lem-hp-admissible}.
\end{proof}

\begin{proposition}
    \label{prop:minimal-proofs-wc}
    Let $0 < n < m$ and $\UniWEProp{\vec a}$ be a weak commutativity axiom.
    If $\UniHypersequentA$ has a proof $\UniDerivationA$ in $\UniHFLwecRAbsorb$, then it has a proof $\UniDerivationA'$ in $\UniHFLwecRAbsorb$ such that, for every hypersequents $\UniHypersequentB_1,\UniHypersequentB_2$ appearing in $\UniDerivationA'$, if $\UniHypersequentB_1$ is below $\UniHypersequentB_2$
    (or, equivalently, if we can reach $\UniHypersequentB_2$ from $\UniHypersequentB_1$ following backward rule applications),
    then
    $\UniHypersequentB_1 \UniHyperseqWeCtrWqo{m}{n}{\UniSubfmlaHyperseqSet} \UniHypersequentB_2$
    iff the path from $\UniHypersequentB_1$ to $\UniHypersequentB_2$ has only applications of basic instances of rules $\mathsf{e}(W,W')$.
\end{proposition}
\begin{proof}
Suppose that
$\UniHypersequentA$ has a proof $\UniDerivationA$ in $\UniHFLwecRAbsorb$.
Note that if the path from hypersequents 
$\UniHypersequentB_1$ to $\UniHypersequentB_2$  in the desired $\UniDerivationA'$
has only basic applications of rules $\mathsf{e}[W,W']$, then by Lemma~\ref{lem:basic-inst-e-type-preserv} it is clear that $\UniHypersequentB_1 \UniHyperseqWeCtrWqo{m}{n}{\UniSubfmlaHyperseqSet} \UniHypersequentB_2$, so the only direction that can fail in the desired property is the left-to-right.
We will find $\UniDerivationA'$ by induction on the nc-height of $\UniDerivationA$.
The base case follows vacuously, as a proof of nc-height 1 has an axiomatic hypersequent in the leaf and is followed by applications of basic instances of rules $\mathsf{e}(W,W')$; i.e., $\UniDerivationA$ itself will do. 
We assume the statement holds for proofs of nc-height less than $\eta$. 
Assume that $\UniDerivationA$ fails the desired condition.
Pick a hypersequent $\UniHypersequentB_1$ below $\UniHypersequentB_2$ in $\UniDerivationA$ for which there is at least one application of a rule instance that is not a basic instance of a rule $\mathsf{e}(W,W')$ in the path between them.
Then the nc-height of $\UniHypersequentB_2$, call it $\eta_2$, is less than the one of $\UniHypersequentB_1$, say $\eta_1$; i.e., $\eta_2 < \eta_1$.
Then, by Lemma~\ref{lem:curry-lemma-wc}, there is a proof of $\UniHypersequentB_1$ of height at most $\eta_2$.
Take $\UniDerivationA'$ as $\UniDerivationA$ but with this proof of $\UniHypersequentB_1$ replaced by this proof with height at most $\eta_2$.
Then $\UniDerivationA'$ has nc-height smaller than $\eta$, and we are done by the (IH).
\end{proof}

\begin{remark}\label{rem:not-repeated-in-branch-is-safe}
Observe that in any derivation $\UniDerivationA$ it is safe to assume that every branch contains no repeated hypersequents.
Indeed, if such repetition happens on a branch, one can safely remove it by picking the highest occurrence (i.e., the farthest occurrence from the root) of the repeated hypersequent and use its proof as the proof of the lowest occurrence. 
By repeatedly doing this in every branch and for every repeated hypersequent, the result is a derivation of the same hypersequent but without the repetitions.
\end{remark}

We define now a proof-search procedure for $\UniHFLwecRAbsorb$.

\begin{definition}
\label{def-ps-procedure}
Let $0 < n < m$ and $\UniWEProp{\vec a}$ be a weak commutativity axiom. 
We consider a proof-search tree $\mathsf{t}$ built by applying rule instances backwards in the following way. 
For a node $\mathsf{n}$ labelled with a hypersequent $\UniHypersequentB$, we define the following expansion procedure: for each rule instance (recall that $\UniCutRule$ is not in the calculus) with conclusion $\UniHypersequentB$, we add its premises as children of $\mathsf{n}$ in case
    \begin{enumerate}
        \item this is a basic instance of $\mathsf{e}(W,W')$ and this premise does not occur in the path from the root of $\mathsf{t}$ to $\mathsf{n}$; or else
        \item none of the premises $h'$ is such that $h'' \UniHyperseqWeCtrWqo{m}{n}{\UniSubfmlaHyperseqSet} h'$ for some $h''$ in the path from the root to $\mathsf{n}$.
    \end{enumerate}
    The proof-search procedure starts with a single node labelled with $\UniHypersequentA$, the input hypersequent, and invokes the above procedure recursively, expanding each node.
    If a subtree of the resulting tree is a proof of $\UniHypersequentA$ we output  `provable'; if not, we output `unprovable'.
\end{definition}

Of course,  the definition above does not guarantee that the procedure terminates and is correct, i.e., that this is indeed a decision procedure for provability in $\UniHFLwecRAbsorb$. 
We will prove this now.

\begin{theorem}
    \label{fact:wc-proof-search-contraction}
    For all $0 < n < m$, all weak commutativity axioms $\UniWEProp{\vec a}$, and all finite sets of formulas $\UniSetFmA$, provability of a hypersequent in $\UniHFLwecRAbsorb$ is decidable via the proof-search procedure described in Definition~\ref{def-ps-procedure}.
\end{theorem}
\begin{proof}
    First of all, we show that the procedure terminates. It is enough to guarantee that every branch of the proof-search tree is finite, as the number of rule instances applicable at each expansion is finite.
    By construction, if we pick the sequence of hypersequents on a branch, we have the following picture:
    $\UniHypersequentA = g_0, s^0_1,\ldots,s^0_{k_0}, \UniHypersequentB_1, s^1_1,\ldots,s^1_{k_1},\UniHypersequentB_2,\ldots$, 
    for some hypersequents $s_{k_q}^{p}$, where $g_i \not\UniHyperseqWeCtrWqo{m}{n}{\UniSubfmlaHyperseqSet} g_j$ for every $i < j$.
    Here the hypersequent $s^i_1$ is obtained from $g_i$ by applying $\mathsf{e}(W,W')$, as is $s^i_{j+1}$ from $s^i_j$ ($1\leq j<k_i$), and $g_{i+1}$ is obtained from $s^i_{k_i}$ by some other rule.
    Observe that the sequence $g_0,g_1,\ldots$ is a bad sequence over $\UniHyperseqWeCtrWqo{m}{n}{\UniSubfmlaHyperseqSet}$,
    and therefore must be finite, thus the whole branch must be finite. 

    Now we show that the procedure is a decision procedure for $\UniHFLwecRAbsorb$.
    The key observation is that every derivation with root labelled with $\UniHypersequentA$ satisfying the conditions in Proposition~\ref{prop:minimal-proofs-wc} (and also considering Remark~\ref{rem:not-repeated-in-branch-is-safe}) is a subtree of the proof-search tree, and vice-versa.
    Then, from the mentioned proposition, decidability follows because to look for a proof of a hypersequent $\UniHypersequentA$ in $\UniHFLwecRAbsorb$ it is
    enough to exhaustively search for all
    possible derivations satisfying the conditions listed there.
\end{proof}

\begin{theorem}
\label{fact:flewmnr-weak-com-ackermannian}
Let $0 < n < m$, $\UniWEProp{\vec a}$ be a weak exchange axiom and $\UniSetFmA$ be a finite set of formulas.
\begin{enumerate}
    \item \sloppy
    If $\UniAnaRuleSet$ is a finite set of hypersequent analytic structural rules, then provability in $\UniHFLwecRAbsorb$ is in $\UniFGHProbOneAppLevel{\omega^{\omega}}$. 
    \item If 
    $\UniAnaRuleSet$ is a finite set of sequent analytic structural rules, then provability in $\UniSFLwecRAbsorb$ is in $\UniFGHProbOneAppLevel{\omega}$.
\end{enumerate}
Moreover, the underlying algorithms and upper bounds are uniform on $\UniMSetFmA$.
\end{theorem}
\begin{proof}
We only prove in detail item (1), as the proof of item (2) is analogous to the proof of Theorem~\ref{fact:flecmnr-ackermannian} (2).
By Remark~\ref{r: spacetimeC}, it is enough to show that, given a hypersequent $\UniHypersequentA$, there exists a set $T(\UniSetFmA,h)$, consisting of trees of $\UniSubfmlaHyperseqSet$-hypersequents (for $\UniSubfmlaHyperseqSet$ as usual being the set of subformulas of the formulas in $\UniSetFmA$ and $\UniHypersequentA$), such that (a) $\UniHypersequentA$ is provable in $\UniHFLwecRAbsorb$ iff $T(\UniSetFmA,h)$ contains a proof of $\UniHypersequentA$ and (b) the space occupied by each tree in $T(h)$ is in  $\UniFGHOneAppLevel{\omega^{\omega}}$. 
In order to resolve (a) and (b), we will take inspiration from Theorem~\ref{fact:wc-proof-search-contraction} and Lemma~\ref{lem:size-of-minimal-proofs}.

From Proposition~\ref{prop:minimal-proofs-wc}, in order to check provability of $\UniHypersequentA$ it is enough to search for a proof whose branches are sequences of the form
\[
h=g_0, s^0_1,\ldots,s^0_{k_0}, \UniHypersequentB_1, s^1_1,\ldots,s^1_{k_1},
\ldots,
g_p, s^p_1,\ldots,s^p_{k_p},
\]
where $g_0,\ldots,g_p$ is a bad sequence over $\UniNwqoWeCtnHSeqName{m}{n}$, which corresponds to a bad sequence over $\UniNwqoWeCtnHSeqNameVar{m}{n}{\UniSubfmlaHyperseqSet_{\UniSizeHyper{\UniSetFmA}+\UniSizeHyper{\UniHypersequentA}}}$ since $\UniSetCard{\UniSubfmlaHyperseqSet} \leq \UniSizeHyper{\UniSetFmA}+\UniSizeHyper{\UniHypersequentA}$.
In fact, by reasoning
similarly as in Proposition~\ref{prop:back-ps-ek-control-bad-seq}, we can see that this sequence is $(f,\UniSizeHyper{\UniMSetFmA}+\UniSizeHyper{\UniHypersequentA})$-controlled, for some primitive recursive control function $f$---note that the term $\UniSizeHyper{\UniMSetFmA}$ will appear everywhere in our analysis since we want the upper bounds to be uniform on $\UniMSetFmA$. 
In addition, the branching of such tree is bounded by a constant $K$ determined by the base calculus
$\UniCalcExt{\UniFLExtHCalc{\UniWEProp{\vec{a}}\UniWeakCProp{m}{n}}^\star}{\UniAnaRuleSet}$, i.e., it is independent of $\UniSetFmA$, since the rules
in $\UniRuleDedSet{\UniSetFmA}$
all have a single premise.
Together with the above discussion, the next items show that we have all the assumptions of Lemma~\ref{lem:size-of-minimal-proofs} satisfied:
\begin{itemize}
    \item
    We instantiate the lemma by taking $\alpha = \omega^\omega$, $A$ as the set of pairs $(\UniSetFmA,\UniHypersequentA)$ of finite sets of formulas and hypersequents, $\mathfrak{L}(\UniSetFmA,\UniHypersequentA)$ as the set of $\UniSubfmlaHyperseqSet$-hypersequents, where $\UniSubfmlaHyperseqSet$ is the set of all subformulas of the formulas in $(\UniSetFmA,\UniHypersequentA) \in A$, and 
    $\mathcal{Q}=
    \{ \UniNwqoWeCtnHSeqNameVar{m}{n}{\UniSubfmlaHyperseqSet_k}\}_{k \in \UniNaturalSet}$.
    \item We have $\{ \UniNwqoWeCtnHSeqNameVar{m}{n}{\UniSubfmlaHyperseqSet_k}\}_{k \in \UniNaturalSet}\sqsubseteq \UniFGHOneAppLevel{\omega^\omega}$ by Proposition~\ref{fact:wc-length-theorem}.
    \item There is an increasing $S \in \UniFGHLevel{<\omega^\omega}$ such that $\UniSizeHyper{g} \leq S(\UniNorm{g}{}, \UniSizeHyper{\UniSetFmA}+\UniSizeHyper{h})$ for all $g$ in the tree (this follows by a reasoning that is similar to the one developed in the proof of Theorem~\ref{fact:flecmnr-ackermannian}).
    \item 
    There is $S'$ primitive recursive such that $k_j \leq S'(j,\UniSizeHyper{\UniMSetFmA}+\UniSizeHyper{\UniHypersequentA})$ for every $1 \leq j \leq p$.
    Indeed, note that the hypersequents $s_j^1,\ldots,s^j_{k_j}$ must be all distinct by construction.
    Moreover, they can be seen as being like $g_j$ up to a permutation of formulas in the antecedent.
    We know that in such antecedent there can be at most ${\UniSetCard{\UniSubfmlaHyperseqSet}}\UniNormHSeq{g_j}$
    formula occurrences, and thus an upper bound for the possible permutations is
    $({\UniSetCard{\UniSubfmlaHyperseqSet}}\UniNormHSeq{g_j})!$.
    Hence, by the control property of the bad sequence,
    $k_j \leq ({\UniSetCard{\UniSubfmlaHyperseqSet}}\UniNormHSeq{g_j})! \leq ({\UniSetCard{\UniSubfmlaHyperseqSet}}f^j(\UniSizeHyper{\UniMSetFmA}+\UniSizeHyper{\UniHypersequentA}))! \leq ({(\UniSizeHyper{\UniMSetFmA}+\UniSizeHyper{\UniHypersequentA})}f^j(\UniSizeHyper{\UniMSetFmA}+\UniSizeHyper{\UniHypersequentA}))!$.
    We take $S'$ as the latter expression, which is primitive recursive in view of Lemma~\ref{lem:it-is-prim-rec}.
    \item From the above, we also have that the norm of each $s^j_i$ is the same as the norm of $g_j$ for all $1 \leq j \leq p$.
    \end{itemize}
    By Lemma~\ref{lem:size-of-minimal-proofs}, then, we have that such a tree must be a member of
    $T[\mathfrak{L}(\UniSetFmA,\UniHypersequentA),K,\tilde{L},\allowbreak \tilde{S}](\UniSizeHyper{\UniSetFmA}+\UniSizeHyper{h})$.
    So take this set to be $T(\UniSetFmA,h)$, thus obtaining (a), and each element there is upper bounded by a function in $\UniFGHOneAppLevel{\omega^\omega}$. 
    Checking whether this is a proof is performed in exponential time/space, so we are done.
    Observe that this whole analysis is uniform on $\UniSetFmA$.\qedhere
\end{proof}

\begin{corollary}
\label{c: FLknotcontr+wc<Ack}
Let $0 < n < m$ and $\UniWEProp{\vec a}$ be a weak commutativity axiom.
\begin{enumerate}
\item
If $\UniAxiomSetA$ is a finite set of acyclic $\mathcal{P}_3^\flat$ axioms, then provability and deducibility of $\UniAxiomExt{\UniFLweExtLogic{\vec a}{\UniWeakCProp{m}{n}}}{\UniAxiomSetA}$ is in $\UniFGHProbOneAppLevel{\omega^{\omega}}$.
\item 
If $\UniAxiomSetA$ is a finite set of $\mathcal{N}_2$ axioms, then provability and deducibility of $\UniAxiomExt{\UniFLweExtLogic{\vec a}{\UniWeakCProp{m}{n}}}{\UniAxiomSetA}$ is in $\UniFGHProbOneAppLevel{\omega}$.
\end{enumerate}
\end{corollary}
\begin{proof}
Similar to the proofs of Corollaries~\ref{cor-ded-Fw} and~\ref{c: FLew(m,n) complexity}.
\end{proof}

By Corollary~\ref{c: FLknotcontr+wc<Ack} and Lemma~\ref{l: alg_tanslation}, we obtain the following.

\begin{corollary}\label{c: knotcontr+wc<Ack}\
\begin{enumerate}
    \item The complexity of the quasiequational (hence also of the equational) theory of a variety of (pointed) residuated lattices axiomatized by any {finite} set of $\mathcal{P}_3^\flat$ equations containing a knotted contraction equation and a weak commutativity equation is in $\UniFGHProbOneAppLevel{\omega^{\omega}}$.
    \item The complexity of the quasiequational (hence also of the equational) theory of a variety of (pointed) residuated lattices axiomatized by any {finite} set of $\mathcal{N}_2$ equations containing a knotted contraction equation and a weak commutativity equation  is in $\UniFGHProbOneAppLevel{\omega}$, i.e., at most Ackermannian.
\end{enumerate}
\end{corollary}

\section{Upper bounds for weakly commutative logics with knotted weakening}

We continue the approach of modifying previous procedures in order to cover the weakly commutative logics.
For the logics with knotted weakening, we begin by generalizing Definition \ref{def-derive-sets}.
Even though the modifications are small, for convenience we write below the whole definition.

\begin{definition}\label{def-derive-sets-wc}
Let $0 \leq m < n$ and $\UniWEProp{\vec a}$ be a weak commutativity axiom.
For  a finite set of formulas $\UniSetFmA$, we abbreviate $\UniHyperCalcA \UniSymbDef \UniDelCut{\UniCalcExt{\UniFLExtHCalc{\UniWEProp{\vec{a}}\UniWeakWProp{m}{n}}^\star}{\UniAnaRuleSet\cup \UniRuleDedSet{\UniSetFmA}}}$ and we define $\UniDerivSet_{0}$ as the set of all instances of initial rule schemas in~$\UniHyperCalcA$ such that
\begin{enumerate}[a)]
    \item formula-variables are instantiated using elements from~$\UniSubfmlaHyperseqSet$;
    \item succedent-variables are instantiated by an element in~$\UniSubfmlaHyperseqSet$ or empty; 
    \item list-variables are instantiated with lists $\UniListFmA$ such that $\UniListCountElem{\UniListFmA}{\UniFmA} \leq {\UniEllParam + (n-1)} \text{ for each } \UniFmA \allowbreak\in \UniSubfmlaHyperseqSet$;
    \item hypersequent-variables are instantiated as empty.
\end{enumerate}

\noindent For $i \geq 0$, we define
$\UniDerivSet_{i+1}:=\UniDerivSet_i \cup \partial\UniDerivSet_i$,
where $\partial\UniDerivSet_i$ is the set of $\UniSubfmlaHyperseqSet$-hypersequents $\UniHypersequentA$ satisfying the following conditions:
\begin{enumerate}
    \item $\UniHypersequentA_1 \cdots \UniHypersequentA_p/h$ is a rule instance of $\UniHyperCalcA$ (not of $\UniCutRule$) such that, for all $1 \leq k \leq p$, there is $\minus{\UniHypersequentA_k}\in \UniDerivSet_i$ with $\minus{\UniHypersequentA_k}\UniHyperseqWknWqo{m}{n}{\UniSubfmlaHyperseqSet}~{\UniHypersequentA_k}$;
   \item the multiplicity of every formula in the antecedent of a component of $\UniHypersequentA$
    is at most $\UniNormHSeq{\UniDerivSet_i} \UniMulttNumbers \UniSizeHyper{\UniHyperCalcA} \UniMulttNumbers (\UniEllParam + n)$;
    \item the frequency of each component in $\UniHypersequentA$ is at most $\UniSizeHyper{\UniHyperCalcA}$; and
    \item there does not exist $\UniHypersequentA'\in \UniDerivSet_i$ such that $\UniHypersequentA'\UniHyperseqWeWknWqo{m}{n}{\UniSubfmlaHyperseqSet} \UniHypersequentA$.
\end{enumerate}
\end{definition}

Note that the differences with respect to Definition~\ref{def-derive-sets} are in items (c), (2) and (4). 
Recall, in addition, that the nwqo is the one introduced in Definition~\ref{def:rel-weak-com-wkn-hyper}.
With these modifications the analogue of Theorem~\ref{lem:SN} follows.
The reader may also check that the algorithm and complexity analysis of Section~\ref{sec:decid-ub-ww-proof} apply analogously to the weakly commutative setting.
Therefore, we obtain the following results.

\begin{theorem}
\label{fact:flewwmnr-weak-com-ackermannian}
Let $0 < m < n$, $\UniWEProp{\vec a}$ be a weak commutativity axiom and $\UniSetFmA$ be a finite set of formulas.
\begin{enumerate}
\item \sloppy
If 
$\UniAnaRuleSet$
is a finite set of hypersequent analytic structural rules,
then provability in
$\UniDelCut{\UniCalcExt{\UniFLExtHCalc{\UniWEProp{\vec{a}}\UniWeakWProp{m}{n}}^\star}{\UniAnaRuleSet\cup \UniRuleDedSet{\UniSetFmA}}}$ 
is in $\UniFGHProbOneAppLevel{\omega^{\omega}}$. 
\item If 
$\UniAnaRuleSet$ is a finite set of sequent analytic structural rules, then provability in 
$\UniDelCut{\UniCalcExt{\UniFLExtSCalc{\UniWEProp{\vec{a}}\UniWeakWProp{m}{n}}^\star}{\UniAnaRuleSet\cup \UniRuleDedSet{\UniSetFmA}}}$
is in $\UniFGHProbOneAppLevel{\omega}$.
\end{enumerate}
\end{theorem}

\begin{corollary}\label{cor:ub-flwewmn-logic}
Let $0 < m < n$ and $\UniWEProp{\vec a}$ be a weak commutativity axiom.
\begin{enumerate}
    \item If $\UniAxiomSetA$ is a finite set of acyclic $\mathcal{P}_3^\flat$ axioms, then provability and deducibility of $\UniAxiomExt{\UniFLweExtLogic{\vec a}{\UniWeakWProp{m}{n}}}{\UniAxiomSetA}$ is in $\UniFGHProbOneAppLevel{\omega^{\omega}}$.
    \item If $\UniAxiomSetA$ is a finite set of $\mathcal{N}_2$ axioms, then provability and deducibility of $\UniAxiomExt{\UniFLweExtLogic{\vec a}{\UniWeakWProp{m}{n}}}{\UniAxiomSetA}$
    is in $\UniFGHProbOneAppLevel{\omega}$.
\end{enumerate}
\end{corollary}

By Corollary~\ref{cor:ub-flwewmn-logic} and Lemma~\ref{l: alg_tanslation}, we obtain the following.

\begin{corollary}\label{c: knotweak+wc<Ack}\
\begin{enumerate}
    \item The complexity of the quasiequational (hence also of the equational) theory of a variety of (pointed) residuated lattices axiomatized by any finite set of $\mathcal{P}_3^\flat$ equations containing a knotted weakening equation and a weak commutativity equation is in $\UniFGHProbOneAppLevel{\omega^{\omega}}$.
    \item The complexity of the quasiequational (hence also of the equational) theory of a variety of (pointed) residuated lattices axiomatized by any {finite} set of $\mathcal{N}_2$ equations containing a knotted weakening equation and a weak commutativity equation is in $\UniFGHProbOneAppLevel{\omega}$;. i.e., at most Ackermannian.
\end{enumerate}
\end{corollary}

%% file: tex/noncommutative-integral-ub.tex
So far we have presented upper bounds for deducibility and provability in all the logics corresponding to the classes of algebras covered by the decidability results in Theorem~\ref{t: FEP}(1), namely commutative or weakly commutative knotted extensions of $\UniFLExtLogic{}$ extended by axioms corresponding to analytical structural (hyper)sequent rules.
Item (2) of that theorem, however, has not been contemplated yet, and this is what we intend to do in the present section.
We will show that the techniques employed for the commutative and weakly commutative logics with knotted weakening can be applied to obtain complexity results for deducibility and provability in 
$\UniAxiomExt{\UniFLExtLogic{\UniWeakWProp{0}{1}}}{\UniAxiomSetA}
=
\UniAxiomExt{\UniFLExtLogic{\UniIProp}}{\UniAxiomSetA}$,
where $\UniAxiomSetA$ is a finite set of axioms in $\mathcal{P}_3^\flat$ and $(\UniIProp)$ corresponds to the axiom $x \leq 1$ of integrality (or, in terms of structural rules, the rule of left weakening).
We want to stress that these logics do not satisfy commutativity, nor any form of weak commutativity, so the results are in some sense surprising, but they can be attributed to the strength of the integrality equation among the knotted rules.
We emphasize that it is well-known that provability in $\UniFLExtLogic{\UniIProp}$ is decidable and \PSPACE-complete~\cite{horcik2011}, and it is 
$\UniFGHProbOneAppLevel{\omega^\omega}$-complete for deducibility~\cite{GreatiRamanayake2024}
but no result was known until now for the infinitely many axiomatic extensions of the form
$\UniAxiomExt{\UniFLExtLogic{\UniIProp}}{\UniAxiomSetA}$
at the level of generality we are presenting here. 
For example, we will obtain a complexity upper bound for the noncommutative version of \textbf{MTL} (monoidal t-norm logic); we note that an upper bound for usual (commutative) \textbf{MTL} was only recently established by \cite{BalLanRam21LICS}.

We will see that the forward proof-search procedure developed in Chapter~\ref{sec:ub-ww} will  adapt to the logics studied in this section with some important modifications due to the full lack of commutativity.
Besides that, a crucial step is showing that the ordering given in Definition~\ref{def:ordering-hsqe} gives rise to an nwqo.
We first define as candidate nwqo essentially the noncommutative version of Definition~\ref{def:nqo-weakening-def-hyper}, where the norm is also slightly adapted to consider the whole length of antecedents instead of only the maximum multiplicity of a formula; the reason for this modification will be clear below, when we present the notion of \emph{Higman's extensions}.

\begin{definition}
Let
$\UniNwqoWknHSeqNCName{m}{n} \UniSymbDef 
\UniStruct{\UniOmegaNCHypersequentsSet{\UniSubfmlaHyperseqSet},
\UniHyperseqWknWqo{m}{n}{\UniSubfmlaHyperseqSet},
\UniNormHNCSeq{\cdot}}$, where
$\UniHyperseqWknWqo{m}{n}{\UniSubfmlaHyperseqSet}$
is as in Definition~\ref{def:ordering-hsqe} but now defined over noncommutative hypersequents and noncommutative rules, and
\[\UniNormHNCSeq{\UniSequentA_1 \VL \cdots \VL 
\UniSequentA_k
} \UniSymbDef 
\max\{ k^\ast, \max_{1 \leq i \leq k} \UniNormSNCSeq{\UniSequentA_i} \},\]
where
$\UniNormSNCSeq{\UniSequent{\UniListFmA}{\UniMSetSucA}}\UniSymbDef 
\UniListSize{\UniListFmA}$ (the length of the list $\UniListFmA$) and $k^\ast \leq k$ is obtained by partitioning the components of the hypersequent per succedent and taking the size of
a maximal partition.
\end{definition}

Interestingly, for most of the noncommutative logics with  knotted weakening, the above is not a nwqo, as the next example shows.

\begin{example}
    The following is an infinite bad sequence over $\UniNwqoWknHSeqNCName{m}{n}$ whenever $m > 0$ or $n > 1$:
    \[
    \UniSequent{\UniPropA}{\UniPropC} \qquad
    \UniSequent{\UniPropA,\UniPropB}{\UniPropC} \qquad
    \UniSequent{\UniPropA,\UniPropB,\UniPropA}{\UniPropC} \qquad
    \UniSequent{\UniPropA,\UniPropB,\UniPropA,\UniPropB}{\UniPropC} \qquad
    \UniSequent{\UniPropA,\UniPropB,\UniPropA,\UniPropB,\UniPropA}{\UniPropC} \qquad
    \ldots\qedhere
    \]
\end{example}

Integrality, however, is much stronger than any other knotted weakening rule. 
It immediately blocks the above example, because to be applied it does not require formulas to be adjacent (neither in the premise nor in the conclusion of the knotted weakening rule). 
In fact, it is immediate to see that all pairs of successive elements in the above sequence are increasing if we see it as a sequence over $\UniNwqoWknHSeqNCName{0}{1}$ (so it is not a bad sequence). 
We prove now that the corresponding nqo is in fact a nwqo and derive a length theorem for it. For that, we need Higman's ordering on finite lists/sequences over a given set.

\begin{definition}
    Given an nqo $\UniWqoA$, we define the following ordering on $\UniListClass{\UniWqoSet}$:
    \begin{gather*}
        a_1 \cdots a_\eta \UniHigmanOrd{\UniWqoA}b_1 \cdots b_\rho
        \text{ iff }
        \text{ for some $1 \leq i_1 < i_2 < \ldots < i_\eta \leq \rho$,}\\
        \text{we have $a_{j} \UniWqoRel{\UniWqoA} b_{i_j}$
        for all $1 \leq j \leq \eta$}.
    \end{gather*}
    We denote by $\UniHigmanWqo{\UniWqoA}$ the nqo $\UniStruct{\UniListClass{\UniWqoSet}, \UniHigmanOrd{\UniWqoA}, \UniNorm{\cdot}{\UniHigmanWqo{\UniWqoA}}}$, where the norm is given by 
    $\UniNorm{a_1 \cdots a_\eta}{\UniHigmanWqo{\UniWqoA}} \allowbreak\UniSymbDef
    \max\{ \eta, \UniNorm{a_1}{{\UniWqoA}},
    \ldots, \UniNorm{a_\eta}{{\UniWqoA}}\}$ and refer to it as the \emph{Higman extension} of $\UniWqoA$.
\end{definition}

The following comes from a well-known result in wqo-theory~\cite{Higman1952}.

\begin{lemma}[Higman's Lemma]
    If $\UniWqoA$ is a nwqo, then 
    $\UniHigmanWqo{\UniWqoA}$ is a nwqo.
\end{lemma}

The following simple nqwos will be
important to us:

\begin{definition}
Given a finite set $A$,  $\UniPlainWqo{A}$ denotes the nwqo
$\UniStruct{A, =, \lambda a. 0}$, called the \emph{flat} nwqo over $A$.
\end{definition}

Regarding length theorems, \cite{schmitzschnoebelen2011} proved a tight length theorem for the so-called \emph{exponential nwqos} (the collection of nwqos containing all the flat nwqos up to isomorphism and their Higman's extensions, and closed under direct products and disjoint sums).
In~\cite{hasse2013}, a generalization of the above-mentioned theorem was employed to also cover nwqos with nested applications of Higman's extensions (adding this new closure condition gives rise to the so-called \emph{elementary nwqos}). 
We specialize these results to the nwqos of interest to us.

As in previous sections, we define, for each $k \geq 0$, $\UniSubfmlaHyperseqSet_k \UniSymbDef \{ \UniPropA_1,\ldots, \UniPropA_k\}$.

\begin{theorem}[{\cite{schmitzschnoebelen2011}}, Theorem 5.3, adapted]
\label{the:len-the-higman}
For all $k \geq 0$ and $d,r \geq 1$, we have
${r \cdot \UniHigmanWqo{\UniPlainWqo{\UniSubfmlaHyperseqSet_k}}} \sqsubseteq \UniFGHOneAppLevel{{\omega^{k}}}$
and ${
(\UniHigmanWqo{
(\UniHigmanWqo{\UniPlainWqo{\UniSubfmlaHyperseqSet_k}})
})^d
}\sqsubseteq\UniFGHOneAppLevel{{\omega^{\omega^{\omega^{k}}}}}$.
Moreover,
$\{ r \cdot \UniHigmanWqo{\UniPlainWqo{\UniSubfmlaHyperseqSet_k}}\}_{(r,k) \in \UniNaturalSet^2} \sqsubseteq \UniFGHOneAppLevel{{\omega^{\omega}}}$ and
$\{ (\UniHigmanWqo{
    (\UniHigmanWqo{\UniPlainWqo{\UniSubfmlaHyperseqSet_k}})
    })^d\}_{(d,k)\in\UniNaturalSet^2} \sqsubseteq\UniFGHOneAppLevel{{\omega^{\omega^{\omega^\omega}}}}$.
\end{theorem}
\begin{proof}
In \cite[Sec. 4.2.3]{hasse2013}, each elementary nwqo $\UniWqoA$ is associated with an ordinal $o(\UniWqoA)$, and this ordinal is used to compute the level of the fast-growing hierarchy to which the corresponding length function belongs.
We first compute the ordinals corresponding to these nwqos.
Following the definitions presented in the mentioned work, we obtain that
$$o(r \cdot \UniHigmanWqo{\UniPlainWqo{\UniSubfmlaHyperseqSet_k}})
= \omega^{\omega^{k-1}} \cdot r
< \omega^{\omega^{k} + 1} \quad
\text{ and } \quad
o((\UniHigmanWqo{
(\UniHigmanWqo{\UniPlainWqo{\UniSubfmlaHyperseqSet_k}})
        })^d) = \omega^{\omega^{\omega^{\omega^{k-1}}} \cdot d}
        < \omega^{\omega^{\omega^{\omega^{k}}}+1}.$$
The result then follows directly from (the generalization to elementary nwqos of)~\cite[Prop.~5.2]{schmitzschnoebelen2011}, reasoning as in the proofs of Theorems~\ref{fact:length_theorem_naturalset} and \ref{fact:length-theorem-power-set-wqo-nat} to obtain the uniformity claims.
\end{proof}

We now show how to strongly reflect the nqo on hypersequents into an nwqo based on nested Higman's extensions of flat nwqos.

\begin{definition}
    For an $\UniSubfmlaHyperseqSet$-hypersequent $\UniHypersequentA$, let $\UniHypersequentA[\UniMSetSucA]$ be the hypersequent containing only the components of $\UniHypersequentA$ having succedent $\UniMSetSucA$.
    Note that $\UniHypersequentA$ is the multiset union of all
    $\{\UniHypersequentA[\UniMSetSucA]\}_{\UniMSetSucA \in (\UniSubfmlaHyperseqSet\cup\{ \varnothing \})}$ (where $\varnothing$ represents the empty succedent).
\end{definition}

\begin{definition}
    \label{def:encoding-hypersequents-integral}
    Assume that, in a hypersequent, components with same succedent are sorted according to some total order on their antecedents.
    If $\UniSubfmlaHyperseqSet \UniSymbDef \UniSet{\UniFmA_1,\ldots,\UniFmA_d}$ is an ordered set of formulas, where $d \in \UniNaturalSet$, and $\UniHypersequentA$ is an $\UniSubfmlaHyperseqSet$-hypersequent, we define
    \[    \UniHyperTransNat{\UniHypersequentA}
    \UniSymbDef
    \UniTuple{X_0,\ldots,X_{d}} \] 
   to be the element of $ (\UniHigmanWqo{(\UniHigmanWqo{\UniSubfmlaHyperseqSet})})^{d+1}$
    where
    \[
    X_0 \UniSymbDef
(\UniMSetFmA_1)\cdots(\UniMSetFmA_\eta),
    \text{ for }
    h[\varnothing] = (
    \UniSequent{\Gamma_1}{}
    \VL \cdots
    \VL 
    \UniSequent{\Gamma_\eta}{})
    \]
    and, for each $1 \leq i \leq d$,
    \[
    X_i \UniSymbDef
(\UniMSetFmA_1)\cdots(\UniMSetFmA_\eta),
    \text{ for }
    h[\UniFmA_i] =
    (\UniSequent{\Gamma_1}{\UniFmA_i}
    \VL \cdots
    \VL 
    \UniSequent{\Gamma_\eta}{\UniFmA_i})
    \]
    Notice that, when $\eta$ above is 0, the generated sequence is $\UniEmptyList$ (the empty sequence).
\end{definition}

\begin{example}
    Let $\UniSubfmlaHyperseqSet \UniSymbDef \{ \UniPropA,\UniPropB,\UniPropC,\UniPropA\to\UniPropB \}$
    and 
    $$\UniHypersequentA \UniSymbDef
    \UniSequent{\UniPropA,\UniPropA\to\UniPropB}{\UniPropC} \VL 
    \UniSequent{}{\UniPropA} \VL 
    \UniSequent{\UniPropC,\UniPropA}{} \VL 
    \UniSequent{\UniPropA,\UniPropB}{\UniPropC} \VL
    \UniSequent{\UniPropB}{}
    .$$
    Then
    $\UniHyperTransNat{\UniHypersequentA}
    = \UniTuple{
        (\UniPropC\UniPropA)(\UniPropB),
        (\UniEmptyList{}),
        \UniEmptyList{},
        (\UniPropA\UniPropA\to\UniPropB)(\UniPropA\UniPropB),
        \UniEmptyList{}
    }$.
    Note that $\UniEmptyList{}$ is an empty sequence, while $(\UniEmptyList{})$ is a sequence with one element: the empty sequence itself.
\end{example}

\begin{lemma}
For all finite sets of formulas
$\UniSubfmlaHyperseqSet$,
\[
\UniNwqoWknHSeqNCName{0}{1}
\UniStrongReflArrow{\UniHyperTransNat{(\cdot)}}
\UniPowerMaj{\UniHigmanWqo{
    (\UniHigmanWqo{\UniPlainWqo{\UniSubfmlaHyperseqSet}})
}}{\UniSetCard{\UniSubfmlaHyperseqSet}+1}.
\]
\end{lemma}
\begin{proof}
    Let $\UniSubfmlaHyperseqSet$ be a finite set of formulas, $\UniHypersequentA_1$
    and $\UniHypersequentA_2$ be  
$\UniSubfmlaHyperseqSet$-hypersequents, and $d \UniSymbDef \UniSetCard{\UniSubfmlaHyperseqSet} + 1$.
For simplicity, we denote by $\UniWqoRel{}$ the order relation of
    $\UniPowerMaj{\UniHigmanWqo{
        (\UniHigmanWqo{\UniPlainWqo{\UniSubfmlaHyperseqSet}})
    }}{\UniSetCard{\UniSubfmlaHyperseqSet}+1}$.
    Let
    $\UniHyperTransNat{\UniHypersequentA_1} \UniSymbDef \UniTuple{X_0,\ldots,X_d}$
    and
    $\UniHyperTransNat{\UniHypersequentA_2} \UniSymbDef\UniTuple{X'_0,\ldots,X'_d}$,
    and suppose that
    $\UniHyperTransNat{\UniHypersequentA_1}
    \UniWqoRel{}
    \UniHyperTransNat{\UniHypersequentA_2}$.
    We pick an arbitrary $0 \leq i \leq d$
    and assume it corresponds to the succedent $\UniMSetSucA$.
    Then $X_i = l_1\cdots l_\eta$ and $X'_i = l'_1\cdots l'_{\eta'}$ (elements of $\UniHigmanWqo{
(\UniHigmanWqo{\UniPlainWqo{\UniSubfmlaHyperseqSet}})
    }$)
    are such that
    $l_p \UniWqoRel{\ast} l'_{p_j}$
    for some $1 \leq {p_1} < \ldots < p_\eta \leq \eta' $.
    Note that since $l_1$ is in $X_i$, the component $\UniSequent{l_1}{\UniMSetSucA}$ is in $\UniHypersequentA_1$ and likewise
    $\UniSequent{l_{p_1}}{\UniMSetSucA} \in \UniHypersequentA_2$.
    Since $l_1 \UniWqoRel{\ast} l'_{p_1}$,
    $l_1$ is obtained from $l'_{p_1}$
    by deleting some formulas, hence
    $\UniSequent{l_{p_1}}{\UniMSetSucA}$
    is obtained from $\UniSequent{l_1}{\UniMSetSucA}$
    by left-weakening.
    As  the same reasoning applies to the other indices, we have that the components of
    $\UniHypersequentA_2[\UniMSetSucA]$
    corresponding to these indices (say a hypersequent $g$) are obtained by weakening from $\UniHypersequentA_1[\UniMSetSucA]$.
    The other components that might be in
    $\UniHypersequentA_2$ all follow then from $g$ by (EW).
    Since $0 \leq i \leq d$ was arbitrary, $\UniHypersequentA_2$ is derivable using only left-weakening and $\UniEW$ from $\UniHypersequentA_1$, and thus $\UniHypersequentA_1 \UniHyperseqWknWqo{0}{1}{\UniSubfmlaHyperseqSet}\UniHypersequentA_2$.
    It is clear
    that the reflection
    is strong by the definition of
    the involved norms.
\end{proof}

Observe that $\UniNwqoWknHSeqNCName{0}{1} \cong \UniNwqoWknHSeqNCNameVar{0}{1}{\UniSubfmlaHyperseqSet_k}$ whenever $\UniSetCard{\UniSubfmlaHyperseqSet} = k$.
Then the above result together with Theorem~\ref{the:len-the-higman} entail the following, by reasoning as in the proof of Proposition~\ref{fact:hnwqo-length-theorem}.

\begin{proposition}
    \label{prop:length-the-hyper-nc}
    For all $k \geq 0$,  ${\UniNwqoWknHSeqNCNameVar{0}{1}{\UniSubfmlaHyperseqSet_k}}
\sqsubseteq\UniFGHOneAppLevel{{\omega^{\omega^{\omega^{k}}}}}$. 
Moreover,
$$\{\UniNwqoWknHSeqNCNameVar{0}{1}{\UniSubfmlaHyperseqSet_k} \}_{k \in \UniNaturalSet} \sqsubseteq \UniFGHOneAppLevel{{\omega^{\omega^{\omega^{\omega}}}}}.$$
\end{proposition}

Actually, in the context of sequent calculi, the domain of
$\UniNwqoWknHSeqNCName{0}{1}$
changes to 
$\UniSubfmlaHyperseqSet$-sequents 
and with the obvious adaptations of the order relation and the norm we obtain an nwqo denoted by $\UniNwqoWknSSeqNCName{0}{1}$.
Again, observe that
$\UniNwqoWknSSeqNCName{0}{1} \cong \UniNwqoWknSSeqNCNameVar{0}{1}{\UniSubfmlaHyperseqSet_k}$
whenever $\UniSetCard{\UniSubfmlaHyperseqSet} = k$.
In this case, we obtain a smaller upper bound.

\begin{proposition}
    \label{prop:length-the-sequents-nc}
    For all $k \geq 0$, 
    ${\UniNwqoWknSSeqNCNameVar{0}{1}{\UniSubfmlaHyperseqSet_k}}{\UniControlFunctionA}\sqsubseteq\UniFGHOneAppLevel{{\omega^{k}}}$.
    Moreover,
    $$\{ \UniNwqoWknSSeqNCNameVar{0}{1}{\UniSubfmlaHyperseqSet_k}\}_{k \in \UniNaturalSet} \sqsubseteq \UniFGHOneAppLevel{{\omega^{\omega}}}.$$
\end{proposition}
\begin{proof}
It is not hard to see that
$\UniNwqoWknSSeqNCName{0}{1}
\UniStrongReflArrow{}
(\UniSetCard{\UniSubfmlaHyperseqSet}+1) \cdot
{{
\UniHigmanWqo{\UniPlainWqo{\UniSubfmlaHyperseqSet}}
}}$, 
so we can reason as in the proof of Proposition~\ref{fact:hnwqo-length-theorem}, now using Theorem~\ref{the:len-the-higman}.
\end{proof}

We now turn to the hypersequent calculi of the form
$\UniHFLiRAug{\UniSetFmA}$
(for $\UniAnaRuleSet$ a finite set of analytic hypersequent rules and $\UniRuleDedSet{\UniSetFmA}$ as defined in Section~\ref{sec:deducibility-adapts}), adapting the forward proof-search procedure described in Chapter~\ref{sec:ub-ww} to the noncommutative setting in presence of integrality.
Denote by $\UniCountLists{\rho}{\eta}$ the number of distinct lists over $\rho$ elements having length at most $\eta$ and note that $\UniCountLists{\rho}{\eta} \leq \rho^{\eta+1}$.

\begin{definition}\label{def-derive-sets-hfli}
Let $\UniSubfmlaHyperseqSet$ be a finite set of formulas closed under subformulas.
We define $\UniDerivSet_{0}$ as the set of all instances of initial rule schemas in~$\UniHyperCalcA \UniSymbDef \UniHFLiRAug{\UniSetFmA}$ such that

\begin{enumerate}[a)]
\item formula-variables are instantiated to elements of~$\UniSubfmlaHyperseqSet$;
\item succedent-variables are instantiated to an element in~$\UniSubfmlaHyperseqSet$ or as empty; 
\item sequence-variables are instantiated to 
the empty sequence;
\item hypersequent-variables are instantiated as empty.
\end{enumerate}

\noindent For $i > 0$, we define
$\UniDerivSet_{i+1}:=\UniDerivSet_i \cup \partial\UniDerivSet_i$,
where $\partial\UniDerivSet_i$ is the set of $\UniSubfmlaHyperseqSet$-hypersequents $\UniHypersequentA$ satisfying the following conditions:
\begin{enumerate}
    \item $\UniHypersequentA_1 \cdots \UniHypersequentA_p/h$ is a rule instance of $\UniHyperCalcA$ such that, for all $1 \leq k \leq p$, there is $\minus{\UniHypersequentA_k}\in \UniDerivSet_i$
    with $\minus{\UniHypersequentA_k}\UniHyperseqWknWqo{0}{1}{\UniSubfmlaHyperseqSet}~{\UniHypersequentA_k}$;
    \item the length of the antecedent of each component of $\UniHypersequentA$ is at most 
    $\UniAlgLimitNC{i}$
    (where
    $\UniNormHNCSeq{\UniDerivSet_i} 
    \UniSymbDef
    \max_{\UniHypersequentA \in \UniDerivSet_i} \UniNormHNCSeq{\UniHypersequentA}$
    )
    \item the frequency of each component in $\UniHypersequentA$ is at most $\UniSizeHyper{\UniHyperCalcA}$; and
    \item there is no $\UniHypersequentA'\in \UniDerivSet_i$ such that $\UniHypersequentA'\UniHyperseqWknWqo{0}{1}{\UniSubfmlaHyperseqSet} \UniHypersequentA$.
\end{enumerate}
\end{definition}

We now proceed to prove the analogous of Lemma~\ref{lem:SN}.
Before that, we need a couple of technical lemmas.
In what follows, if $L_1,\ldots,L_m$ are lists, $L_1\cdots L_m$ denotes their concatenation.

\begin{lemma}
\label{lem:break-list-higman}
If a sequent ${\UniSequentA'}$ yields
$\UniSequentA : = (\UniSequent{L_1\cdots L_m}{\UniMSetSucA})$
by successive applications of $\UniWeakWRule{0}{1}$, then there are sequences $L'_1,\ldots,L'_m$ such that
$\UniSequentA' = (\UniSequent{L'_1\cdots L'_m}{\UniMSetSucA})$
and
$L'_i \UniWqoRel{\ast} L_i$ for all $1 \leq i \leq m$. 
\end{lemma}
\begin{proof}
We proceed by induction on a deduction containing only applications of $\UniWeakWRule{0}{1}$ witnessing that from ${\UniSequentA'}$ we derive
$\UniSequentA : = (\UniSequent{L_1\cdots L_m}{\UniMSetSucA})$.
If it has a single node, the involved sequents are the same, so take $L'_i \UniSymbDef L_i$.
In the inductive step, assume
$s = \UniSequent{L_1 \cdots L^1_i \UniFmA L^2_i \cdots L_m}{\UniMSetSucA}$ was obtained by weakening from 
$s'' = \UniSequent{L_1 \cdots L^1_i L^2_i \cdots L_m}{\UniMSetSucA}$, where $L_i = L^1_i L^2_i$.
Then, by the (IH), 
$s' = \UniSequent{L'_1 \cdots L'_i \cdots L'_m}{\UniMSetSucA}$
with $L'_j \UniWqoRel{\ast} L_j$ for all $1 \leq j \leq m$.
In particular, $L'_i \UniWqoRel{\ast} L_i^1L_i^2 \UniWqoRel{\ast} L_i^1\UniFmA L_i^2$, 
and we are done.
\end{proof}

\begin{lemma}\label{lem:aux-deriv-nc}
    Let $\UniHypersequentA,\UniHypersequentB$ be $\UniSubfmlaHyperseqSet$-hypersequents.
    If $\UniHypersequentA\UniHyperseqWknWqo{0}{1}{\UniSubfmlaHyperseqSet}
    \UniHypersequentB$,
    then for every component $\UniSequentA \in \UniHypersequentB$, either
    (1) there is $\UniSequentA' \in \UniHypersequentA$ such that $\UniSequentA$ is obtained from $\UniSequentA'$ by successive applications of $\UniWeakWRule{0}{1}$,
    or (2) there is a hypersequent $\UniHypersequentA'$ and a sequent $\UniSequentA'$ such that
    $\UniHypersequentA \UniHyperseqWknWqo{0}{1}{\UniSubfmlaHyperseqSet}
    \UniHypersequentA' \to_{\UniEW} \UniHypersequentA' \VL \UniSequentA'
    \UniHyperseqWknWqo{0}{1}{\UniSubfmlaHyperseqSet}
    \UniHypersequentB$
    and $\UniSequentA$ is obtained from $\UniSequentA'$ by successive applications of $\UniWeakWRule{0}{1}$.
\end{lemma}
\begin{proof}
    We proceed by induction on the structure of a derivation containing only applications of $\UniEW$, $\UniWeakWRule{0}{1}$ and $\UniEC$ witnessing 
    $\UniHypersequentA\UniHyperseqWknWqo{0}{1}{\UniSubfmlaHyperseqSet}
    \UniHypersequentB$.
    If it has a single node, the claim is obvious.
    Otherwise, let $g$ be obtained from $g'$ by one of those three rules.
    Note that $\UniHypersequentA
    \UniHyperseqWknWqo{0}{1}{\UniSubfmlaHyperseqSet}
    \UniHypersequentB'$
    and that the (IH) applies to $g'$.
    Let $s \in g$.
    The first case we consider is if $s$ is in the instantiation of the hypersequent variable in the rule.
    In this case, the claim for $s$ follows by the induction hypothesis, since $s \in g'$ as well.
    Also, if the last rule applied was $\UniEC$ and $s$ was an active component, then $s \in g'$ as well, as the claim follows again for it.
    Otherwise, we reason by the kind of rule applied (by the above reasoning, $\UniEC$ does not need consideration).
    \begin{enumerate}
        \item If $\UniWeakWRule{0}{1}$ and $s$ was an active component, then there is $s'' \in g'$ that yields $s$ by $\UniWeakWRule{0}{1}$.
        By (IH), we consider two cases.
        First, there is $s' \in \UniHypersequentA$ that yields $s''$ by $\UniWeakWRule{0}{1}$.
        Since $s''$ yields $s$ by this same rule, we are done.
        In the other case, there is a hypersequent $\UniHypersequentA'$ and a sequent $\UniSequentA'$ such that
        $\UniHypersequentA \UniHyperseqWknWqo{0}{1}{\UniSubfmlaHyperseqSet}
        \UniHypersequentA' \to_{\UniEW} \UniHypersequentA' \VL \UniSequentA'
        \UniHyperseqWknWqo{0}{1}{\UniSubfmlaHyperseqSet}
        \UniHypersequentB'$
        and $\UniSequentA''$ is obtained from $\UniSequentA'$ by successive  applications of $\UniWeakWRule{0}{1}$. But then $\UniHypersequentA \UniHyperseqWknWqo{0}{1}{\UniSubfmlaHyperseqSet}
        \UniHypersequentA' \to_{\UniEW} \UniHypersequentA' \VL \UniSequentA'
        \UniHyperseqWknWqo{0}{1}{\UniSubfmlaHyperseqSet}
        \UniHypersequentB$
        since $g' \UniHyperseqWknWqo{0}{1}{\UniSubfmlaHyperseqSet} g$.
        Also, $s'$ yields $s$ by successive applications of $\UniWeakWRule{0}{1}$, $s''$ yields $s$ by one application of such rule, and we are done.
        \item If $\UniEW$ and $s$ was introduced, we have $g = g' \VL s$.
        Then 
        $\UniHypersequentA \UniHyperseqWknWqo{0}{1}{\UniSubfmlaHyperseqSet}
        \UniHypersequentB' \to_{\UniEW} \UniHypersequentB' \VL \UniSequentA
        \UniHyperseqWknWqo{0}{1}{\UniSubfmlaHyperseqSet}
        \UniHypersequentB$, and $s$
        obviously follow from $s$ by zero applications of $\UniWeakWRule{0}{1}$.\qedhere
    \end{enumerate}  
\end{proof}

\begin{lemma}
\label{lem:NonComSN}
Let $\UniSubfmlaHyperseqSet$ be the set of all subformulas of the formulas in a hypersequent $\UniHypersequentA$ and in a set of formulas $\UniSetFmA$.
If $\UniHyperDerivRel{\UniHFLiRAug{\UniSetFmA}} \UniHypersequentA$, then there is~$N \in \UniNaturalSet$ and~$\UniHypersequentA' \in \UniDerivSet_N$ such that~$\UniHypersequentA' \UniHyperseqWknWqo{0}{1}{\UniSubfmlaHyperseqSet}\UniHypersequentA$.
\end{lemma}
\begin{proof}
The proof strategy is similar to the one of Lemma~\ref{lem:SN}, but there are some key changes due to the absence of commutativity.

We will prove the statement  by induction on the height of a derivation $\UniDerivationA$ of~$\UniHypersequentA$ in $\UniHFLiRAug{\UniSetFmA}$.

\textit{Base case}: If $\UniDerivationA$ has height $1$, then $\UniHypersequentA$ is an instance of an initial rule schema. 
Though $\UniDerivSet_0$ consists of a proper subset of initial rule instances, $\UniHypersequentA$ can be obtained by applying (EW) or $\UniWeakWRule{0}{1}$ to a suitable element of $\UniDerivSet_0$.

\textit{Inductive case}: If $\UniDerivationA$ has height greater than one and the last step of the proof is an instance of the rule schema~$\UniRuleSchemaA$ with premises $h_1,\ldots,h_k$, then, by the subformula property given in Theorem~\ref{the:subformula_property}, every subformula in each premise~$\UniHypersequentA_{j}$ in this $\UniRuleSchemaA$-instance is in~$\UniSubfmlaHyperseqSet$. 
By the induction hypothesis applied to the $j^{\text{th}}$ premise, there exists $N_{j}$ and a hypersequent $\minus{\UniHypersequentA_j} \in \UniDerivSet_{N_{j}}$ such that $\minus{\UniHypersequentA_j} \UniHyperseqWknWqo{0}{1}{\UniSubfmlaHyperseqSet}  \UniHypersequentA_j$. 
Since $\UniDerivSet_i \subseteq \UniDerivSet_{i+1}$ for every~$i$, we have every $\minus{\UniHypersequentA_j}\in \UniDerivSet_N$, for $N:=\max_i N_{j}$.
Let~$\UniDerivationA_{j}$ be the derivation witnessing $\minus{\UniHypersequentA_{j}} \UniHyperseqWknWqo{0}{1}{\UniSubfmlaHyperseqSet} \UniHypersequentA_{j}$, i.e., a sequence of applications of (EW), $\UniWeakWRule{0}{1}$ and (EC) that takes $\minus{\UniHypersequentA_{j}}$ to $\UniHypersequentA_{j}$. 

By Definition~\ref{def-derive-sets-hfli}, if the antecedent of every component has length at most ${\UniAlgLimitNC{N}}$ and every component has multiplicity at most 
$\UniSizeHyper{\UniHyperCalcA}$, 
then either $\UniHypersequentA\in \UniDerivSet_{N+1}$ or there is some $\UniHypersequentA'\in \UniDerivSet_N$ such that $\UniHypersequentA'\UniHyperseqWknWqo{0}{1}{\UniSubfmlaHyperseqSet}\UniHypersequentA$, as desired. 
For the remaining cases, we proceed by a sub-induction on the value~$\UniSizeHyper{\UniHypersequentA}$:

$\bullet$ 
Some component~$s$ in~$\UniHypersequentA$ has an antecedent whose length~$\alpha$ is strictly greater than ${\UniAlgLimitNC{N}}$. 
Then there are two possibilities:

\begin{enumerate}
    \item \emph{$s$ is in the instantiation of the hypersequent-variable}:
    In a derivation $\UniDerivationA_i$, we use the term \emph{marked component} to refer to components related to $s$ through applications of $\UniEC$.
    Specifically, call this particular occurrence of $s$ in the root (i.e., in $h_i$) a marked component, and tracing upwards in $\UniDerivationA_i$, whenever a marked component $s$ follows by $\UniEC$ from $s \VL s$, then call each of these occurrences of $s$ a marked component.
    Let $\UniDerivationA_i'$ be like $\UniDerivationA_i$ but with all applications of $\UniEC$ involving marked components deleted.
    We call its root $h'_i$.
    Note that $\UniDerivationA_i'$ witnesses $\minus{\UniHypersequentA_i} \UniHyperseqWknWqo{0}{1}{\UniSubfmlaHyperseqSet} \UniHypersequentA_i'$
    and observe that the only difference between $h_i$ and $h'_i$ is that the marked components that were `absent' from the former (as they were removed by $\UniEC$ applications in $\UniDerivationA_i$) now appear in the latter.
    
    We write $\UniSequentA$ as $\UniSequent{\UniMSetFmA}{\UniMSetSucA}$, we assume item (1) of Lemma~\ref{lem:aux-deriv-nc} holds for $h'_i$ (each $1 \leq i \leq k$) with respect to the marked components $s_{i1},\ldots,s_{ik_i}$ ($k_i \geq 0$) and we denote by $s_{ij}' = \UniSequent{\UniMSetFmA_{ij}}{\UniMSetSucA}$ the component in $\minus\UniHypersequentA_i$ that leads to the $j$th marked component $s_{ij}$ in $h'_i$ by successive applications of $\UniWeakWRule{0}{1}$.
    We take the sequence $\UniMSetFmA$ and align the elements of the sequences $\UniMSetFmA_{ij}$ with it (there must be an alignment since we are assuming item (1) of Lemma~\ref{lem:aux-deriv-nc}), always taking the leftmost alignment, and cross all those elements off $\UniMSetFmA$; what remains are those elements of $\UniMSetFmA$ that can be seen as resulting from weakened in every marked component.
    Note that if $\UniMSetFmA_{ij}=\UniMSetFmA_{il}$ ($j \neq l$) the crossed off subsequence they produce is the same since we always consider the leftmost match in $\UniMSetFmA$.
    We do this for every $1 \leq i \leq k$.
    We denote by $\UniMSetFmA'$ the subsequence of $\UniMSetFmA$ of crossed-off formulas and consider the sequent $s' = \UniSequent{\UniMSetFmA'}{\UniMSetSucA}$
    (in particular, if all $k_i = 0$, $\UniMSetFmA'$ is the empty list).
    Since the length of each $\UniMSetFmA_{ij}$ is at most $\UniNormHNCSeq{\UniDerivSet_N}$ (because $s_{ij}' \in \minus{h_i}$
    and $\minus{h_i} \in D_N$), $\UniMSetFmA'$ has length at most $\UniNormHNCSeq{\UniDerivSet_N} \cdot {\UniCountLists{\UniSetCard{\UniSubfmlaHyperseqSet}}{\UniNormHNCSeq{\UniDerivSet_N}}}
    < \alpha$. 
    We consider the hypersequent $h''_i$ that is as $h'_i$ but with each marked component replaced by $s'$ (for all $1 \leq i \leq k$).
    Then $\minus{\UniHypersequentA}_i\UniHyperseqWknWqo{0}{1}{\UniSubfmlaHyperseqSet} \UniHypersequentA''_i$
    (for those $s'$ replacing marked components that satisfy item (2) of Lemma~\ref{lem:aux-deriv-nc} instead of item (1), we make the application of $\UniEW$ that introduced them introduce the component $s'$ instead).
    We apply $\UniEC$ successively over $s'$ occurrences starting from $h''_j$ and obtain $h'''_j$ for each $1 \leq j \leq k$, which is like the original $h_j$ but with the component $s$ replaced by $s'$.
    Thus $\minus{\UniHypersequentA_i} \UniHyperseqWknWqo{0}{1}{\UniSubfmlaHyperseqSet} h'''_i$ for every $1 \leq i \leq k$.
    
    Then we apply the same rule $\UniRuleSchemaA$ to the premises $h'''_1,\ldots,h'''_k$, and the resulting hypersequent $h'$ is as $h$ but with $s$ replaced by $s'$ in the instantiation of the hypersequent variable.
    We clearly have that $h' \UniHyperseqWknWqo{0}{1}{\UniSubfmlaHyperseqSet} h$, but since we are not sure if $h'$ is in $D_N$ (it might have other components that break condition (2)), we need to use the subinduction hypothesis: since we saw that the size of $s'$ is strictly smaller than the size of $s$, we have $\UniSizeHyper{h'} < \UniSizeHyper{h}$, and the subinduction hypothesis applies, meaning that there are $M$ and $h''$ such that $h'' \in D_M$ and $h'' \UniHyperseqWknWqo{0}{1}{\UniSubfmlaHyperseqSet} h'$, and we are done for this case.
    \item \emph{$s$~is an active component---i.e., the component not in the instantiation of the hypersequent-variable---of the rule instance}:
    Since the number of occurrences of schematic-variables in any rule schema is $\leq \UniSizeHyper{\UniHyperCalcA}$ there must be some sequence-variable~$\UniMSetFmA$ (occurring exactly once in the conclusion, by linear conclusion) whose instantiation has antecedent with length $\beta > 
    \UniNormHNCSeq{\UniDerivSet_N}
    \cdot {\UniCountLists{\UniSetCard{\UniSubfmlaHyperseqSet}}{\UniNormHNCSeq{\UniDerivSet_N}}}$. 
    Recall that in every component of the (single) leaf $\minus{\UniHypersequentA_{k}}$ of $\UniDerivationA_k$ the length of the antecedent is
    $\leq\UniNormHNCSeq{\UniDerivSet_N}$
    (and thus $<\alpha$) since $\minus{\UniHypersequentA_{k}} \in \UniDerivSet_N$.
    Our goal is to modify (only) the instantiation of
    $\UniMSetFmA$ in $\UniRuleSchemaA$
    to obtain a smaller hypersequent and apply the sub-induction hypothesis as we did in the previous item.
    We write each $h_i$ as $g \VL s_i$, and  $h$ as $g \VL s$.
    As in the previous item, change each $h_i$ to $h_i'$ by considering the marked components associated to $s_i$; again we  have that
    $\minus{\UniHypersequentA_i} \UniHyperseqWknWqo{0}{1}{\UniSubfmlaHyperseqSet} \UniHypersequentA_i'$
    and the only difference between $h_i$ and $h'_i$ is that now all marked components
    appear in $h'_i$.

    From $h'_i$ we obtain $h''_i$ as in the previous item, with some key changes.
    First, $\UniMSetFmA$ is not the whole antecedent of $s$, but only the instantiation of a schematic variable (we call it $\UniMSetFmA$ too, counting on the context to disambiguation) in $\UniRuleSchemaA$.
    The $\UniMSetFmA_{ij}s$ are now lists in the components $s_{ij}$ that lead to $\UniMSetFmA$ by applications of $\UniWeakWRule{0}{1}$, which exist by Lemma~\ref{lem:break-list-higman}, while $\UniMSetFmA'$ is defined as expected: the crossed-off list in $\UniMSetFmA$ after crossing off the lists
    $\UniMSetFmA_{ij}$ in $\UniMSetFmA$.
    We then obtain $s_i'$ by replacing $\UniMSetFmA$ in the antecedent with $\UniMSetFmA'$.
    Then, as in the previous item, we obtain $h'''_i$ by applying $\UniEC$ over the corresponding substitutes of marked components, such that the only difference with respect to $h_i$ is that the highlighted occurrence of $s_i$ was replaced by $s'_i$.
    Then we consider an instantiation of $\UniRuleSchemaA$ that only differs from the applied one by instantiating the schematic variable $\UniMSetFmA$ with $\UniMSetFmA'$ instead.
    By applying this instance over the premises $h'''_i$, we get $h'$, which differs from $h$ only in that $\UniMSetFmA$ is instantiated with $\UniMSetFmA'$.
    As in the previous item, the subinduction hypothesis applies, and we are done.
\end{enumerate}

$\bullet$ Some component~$\UniSequentA$ in~$\UniHypersequentA$ has frequency strictly bigger than $\UniSizeHyper{\UniHyperCalcA}$:
The situation is essentially the same as the corresponding case in the proof of Lemma~\ref{lem:SN}.
\end{proof}

Note that the sets $D_i$ are guaranteed to be finite, so the forward proof-search procedure is essentially the same as the one described in Chapter~\ref{sec:ub-ww} and terminates due to the same reason: one can extract a bad sequence of hypersequents $\{ h_i \}_{i=0}^{N-1}$ from the sets $D_0 \subset D_1 \subset \ldots \subset D_{N}$ ($N \geq 0$), where $h_i \in D_{i}$ for all $0 \leq i \leq N-1$.
This time, this bad sequence is over $\UniNwqoWknHSeqNCName{0}{1}$.

\begin{lemma}\label{lem:control-hfli}
Let $f(x) := 2\UniSizeHyper{\UniHyperCalcA}(\UniSetCard{\UniSubfmlaHyperseqSet}+2)^{2(x+1)}$ and $t \UniSymbDef \UniSizeHyper{\UniHyperCalcA}$.
Also, let $N$ and $\{{\UniHypersequentA_i}\}_{i\in\UniNaturalSetNN}$ be as defined above.
Then,
\begin{enumerate}
    \item $\UniNormHNCSeq{D_{i}} \leq f^i(t)$ for all $0 \leq i < N$.
    \item 
    $\{{\UniHypersequentA_i}\}_{i\in\UniNaturalSetNN}$ is a $(f,t)$-controlled bad sequence.
\end{enumerate}
\end{lemma}
\begin{proof}
For item (1), we show that
$\UniNormHNCSeq{D_{i}} \leq f^i(t)$
by induction on $i$.
First of all, Definition~\ref{def-derive-sets-hfli} implies that
$\UniNormHNCSeq{D_{i+1}}
\leq
\UniSizeHyper{\UniHyperCalcA}
+
\UniSizeHyper{\UniHyperCalcA}
\UniSetCard{\UniSubfmlaHyperseqSet}^{\UniNormHNCSeq{D_{i}}+1}
\UniNormHNCSeq{D_{i}}
\leq
2\UniSizeHyper{\UniHyperCalcA}
(\UniSetCard{\UniSubfmlaHyperseqSet}+2)^{\UniNormHNCSeq{D_{i}}+1}
\UniNormHNCSeq{D_{i}}
\leq
2\UniSizeHyper{\UniHyperCalcA}
(\UniSetCard{\UniSubfmlaHyperseqSet}+2)^{2(\UniNormHNCSeq{D_{i}}+1)}
= f(\UniNormHNCSeq{D_{i}})$
(here we use that
$y \leq K^y$ for all $y \in \UniNaturalSet$
whenever $K > 1$, and that is why we have `$\UniSetCard{\UniSubfmlaHyperseqSet}+2$' in the above expression).
Note that
$\UniNormHNCSeq{D_0}
\leq \UniSizeHyper{\UniHyperCalcA}
= t = f^0(t)$,
so the base case follows.
For the induction step, we have
$\UniNormHNCSeq{D_{i+1}} 
\leq 
f(\UniNormHNCSeq{D_{i}})
\leq f(f^i(t)) = f^{i+1}(t)$.

Item (2) follows from the fact that $\UniNormHNCSeq{h_i} \leq \UniNormHNCSeq{D_{i}}$.
\end{proof}

Before proceeding with the analysis, we adapt Definitions~\ref{def:counting-msets-seqs} and~\ref{def:counting-hseqs} to the noncommutative setting.

\begin{definition}
    For $\eta,\rho,\mu \geq 0$, we define
    \begin{enumerate}
        \item $\UniNDistCompNC{\eta}{\rho} 
        \UniSymbDef
        (\sum_{i=0}^{\rho} \eta^i) \cdot (\eta+1)
        \leq \eta^{\rho+1} \cdot (\eta+1)
        \leq (\eta+1)^{\rho+1}$
        as the maximum number of distinct (noncommutative) sequents over $\eta$ formulas, each having antecedent with length at most $\rho$.
        \item 
        $\UniNDistHSeqNC{\eta}{\rho}{\mu}
        \UniSymbDef
        \mu^{\UniNDistCompNC{\eta}{\rho}}$
        as the number of (noncommutative) hypersequents over $\eta$ formulas such that the maximum multiplicity of a component is $\mu$, and the length of the antecedent in a component is at most $\rho$.
    \end{enumerate}
\end{definition}

Similar to our approach in Chapter~\ref{sec:ub-ww}, we denote by $\UniDerivSet_i^{\UniSetFmA,\UniHypersequentA}$ the $i$-th set in the sequence of Definition~\ref{def-derive-sets-hfli} built with respect to the subformulas  of the formulas in $\UniSetFmA$ and $\UniHypersequentA$.
The next result provides an uniform upper bound on the size of each $\UniDerivSet_i^{\UniSetFmA,\UniHypersequentA}$.

\begin{lemma}\label{lem:unif-bound-weak-sets-hfli}
    There is an increasing and elementary function $S$ such that, for all finite sets of formulas $\UniSetFmA$ and hypersequents $\UniHypersequentA$,
    $\UniSizeHyper{\UniDerivSet_i^{\UniSetFmA,\UniHypersequentA}} \leq S(i,\UniSizeHyper{\UniSetFmA}+\UniSizeHyper{\UniHypersequentA})$.
\end{lemma}
\begin{proof}
    We essentially follow the proof of Lemma~\ref{lem:unif-bound-weak-sets}.

    Note that any element in $D_0$ is an instantiation of an axiomatic hypersequent following Definition~\ref{def-derive-sets-hfli}. 
    There are at most $\UniSizeHyper{\UniHyperCalcA}$ variables to be instantiated
    in an axiomatic hypersequent (recall that
$\UniSizeHyper{\UniHyperCalcA} \leq  2  \cdot \UniCalcExt{\UniDelCut{\UniFLExtHCalc{\UniIProp}}}{\UniAnaRuleSet}\cdot    \UniSizeHyper{\UniSetFmA}$).
    Each instantiated variable is either a formula-variable (actually, a proposition-variable) or a list-variable. 
    Hypersequent variables are instantiated with the empty hypersequent, so elements of $D_0$ are actually sequents.
    Recall that $\UniSizeHyper{\UniFmA} < \UniSizeHyper{\UniSetFmA}+\UniSizeHyper{\UniHypersequentA}$
    for all $\UniFmA \in \UniSubfmlaHyperseqSet$.
    Moreover, list variables are instantiated with the empty list.
    From these observations, we have that the size of a sequent in $D_0$ is
    $\leq C \cdot (\UniSizeHyper{\UniSetFmA}+\UniSizeHyper{\UniHypersequentA})$,
    where $C \geq \UniSizeHyper{\UniHyperCalcA}$ accounts for commas and the sequent symbol as well.
    By inspection, we have that the length of the antecedent of these sequents is at most 1. Thus there can be at most
    $\UniNDistCompNC{\UniSetCard{\UniSubfmlaHyperseqSet}}{1}$ sequents in $D_0$.
    Therefore, take
    $S_0(x) \UniSymbDef \UniNDistCompNC{x}{1}\cdot C \cdot x$, which is an increasing elementary function, and by the above analysis we have
    $\UniSizeHyper{\UniDerivSet_0^{\UniSetFmA,\UniHypersequentA}} \leq S_0(\UniSizeHyper{\UniSetFmA}+\UniSizeHyper{\UniHypersequentA})$.

    The size of a hypersequent in $D_{j+1}$ is upper bounded by 
    \begin{multline*}
    A_1 \UniSymbDef c_1 \cdot\UniSizeHyper{\UniHyperCalcA}
    \UniNDistCompNC{\UniSetCard{\UniSubfmlaHyperseqSet}}
    {\UniAlgLimitNC{j}}\\
\cdot\allowbreak\UniAlgLimitNC{j} \allowbreak\cdot \UniSetCard{\UniSubfmlaHyperseqSet} 
    \cdot \UniSizeHyper{\UniHypersequentA}
    \end{multline*}
    where $c_1$ accounts for commas, $\VL$'s, sequent symbols and succedents.
    
    From Lemma~\ref{lem:control-hfli}, we know that $\UniNormHNCSeq{D_j} \leq f^j(t)$; also, we know that
    $\UniSetCard{\UniSubfmlaHyperseqSet} < \UniSizeHyper{\UniSetFmA}+\UniSizeHyper{\UniHypersequentA}$.
    Hence the above expression is upper bounded by an increasing elementary function $E$ on $j,\UniSizeHyper{\UniSetFmA}+\UniSizeHyper{\UniHypersequentA}$.
    This means that the size of $D_{j+1}$ is upper bounded by
    \begin{align*}
\UniNDistHSeqNC{\UniSetCard{\UniSubfmlaHyperseqSet}}
    {\UniAlgLimitNC{j}}
    {\UniSizeHyper{\UniHyperCalcA}} \cdot 
    E(j,\UniSizeHyper{\UniSetFmA}+\UniSizeHyper{\UniHypersequentA})
    &\leq 
    E'(j,\UniSizeHyper{\UniSetFmA}+\UniSizeHyper{\UniHypersequentA})\\ 
    &\leq 
    E'(j+1,\UniSizeHyper{\UniSetFmA}+\UniSizeHyper{\UniHypersequentA})
    \end{align*}
    for $E'$ an increasing elementary function. 
    Thus set $S(j,
    \UniSizeHyper{\UniSetFmA},\UniSizeHyper{\UniHypersequentA}) \UniSymbDef
    S_0(\UniSizeHyper{\UniSetFmA},\UniSizeHyper{\UniHypersequentA})+E'(j,
    \UniSizeHyper{\UniSetFmA},\UniSizeHyper{\UniHypersequentA})$, 
    and we are done.
\end{proof}

From the above, we finally obtain the main results of this section.

\begin{theorem}
\label{the:calc-compl-results}
Let $\UniSetFmA$ be a finite set of formulas.
\begin{enumerate}
\item If $\UniAnaRuleSet$ is a finite set of hypersequent analytic structural rules, then provability in
$\UniCalcExt{\UniDelCut{\UniFLExtHCalc{\UniIProp}}}{\UniAnaRuleSet\cup\UniRuleDedSet{\UniSetFmA}}$
is in 
$\UniFGHProbOneAppLevel{\omega^{\omega^{\omega^\omega}}}$.
\item If $\UniAnaRuleSet$ is a finite set of sequent analytic structural rules, then provability in 
$\UniCalcExt{\UniDelCut{\UniFLExtSCalc{\UniIProp}}}{\UniAnaRuleSet\cup \UniRuleDedSet{\UniSetFmA}}$ is in $\UniFGHProbOneAppLevel{\omega^\omega}$.
\end{enumerate}
The underlying algorithms and upper bounds are uniform on $\UniSetFmA$.
\end{theorem}
\begin{proof}
    The proofs of both items follow the same strategy employed in Theorem~\ref{fact:flewmnr-ackermannian}, but this time using Lemmas~\ref{lem:control-hfli}, \ref{lem:unif-bound-weak-sets-hfli}, \ref{prop:length-the-hyper-nc} and \ref{prop:length-the-sequents-nc}.
\end{proof}

\begin{corollary}\
\begin{enumerate}
    \item If $\UniAxiomSetA$ is a finite set of acyclic $\mathcal{P}_3^\flat$ axioms, then provability and deducibility of $\UniAxiomExt{\UniFLExtLogic{\UniIProp}}{\UniAxiomSetA}$ is in $\UniFGHProbOneAppLevel{\omega^{\omega^{\omega^\omega}}}$.
    \item If $\UniAxiomSetA$ is a finite set of $\mathcal{N}_2$ axioms, then provability and deducibility of $\UniAxiomExt{\UniFLExtLogic{\UniIProp}}{\UniAxiomSetA}$ is in $\UniFGHProbOneAppLevel{\omega^\omega}$.
\end{enumerate}
\end{corollary}
\begin{proof}
    Analogous to the proof of Corollary~\ref{c: FLew(m,n) complexity}.
\end{proof}

Note that the above result directly applies to the analytic extensions of ${\UniFLExtLogic{\UniWProp}}$ (where, recall, $\UniWProp$ refers to the combination  of integrality, i.e., left weakening, and right weakening).

\begin{remark}
Recall that the above upper bound at
$\UniFGHProbOneAppLevel{\omega^{\omega^{\omega^\omega}}}$
comes from strongly reflecting the nwqo underlying the proof-search procedure (i.e., $\UniNwqoWknHSeqNCName{0}{1}$) into nested Higman extensions over a finite alphabet.
It is not hard to see that this nwqo also reflects into
$\UniMajoringFntWqo{\UniPlainWqo{\UniSubfmlaHyperseqSet}}$,
for $\UniSubfmlaHyperseqSet$ the set of subformulas of the input hypersequent.
We could thus potentially derive lower upper bounds for this same procedure if a length theorem for such nwqo places its length function in lower levels of the fast-growing hierarchy.
Tight length theorems for this nwqo, however, have not been investigated yet in the literature.
\end{remark}

\begin{corollary}\label{c: intComplexity}\
\begin{enumerate}
    \item The complexity of the quasiequational (hence also of the equational) theory of a variety of (pointed) residuated lattices axiomatized by any {finite} set of $\mathcal{P}_3^\flat$ equations containing integrality is in $\UniFGHProbOneAppLevel{\omega^{\omega^{\omega^\omega}}}$.
    \item The complexity of the quasiequational (hence also of the equational) theory of a variety of (pointed) residuated lattices axiomatized by any {finite} set of $\mathcal{N}_2$ equations containing integrality  is in $\UniFGHProbOneAppLevel{\omega^\omega}$.
\end{enumerate}
\end{corollary}

\begin{corollary}
The provability and deducibility of any extension of  noncommutative $\m {MTL}$  with a finite set of acyclic $\mathcal{P}_3^\flat$ axioms is in
$\UniFGHProbOneAppLevel{\omega^{\omega^{\omega^\omega}}}$.
\end{corollary}

%% file: tex/lb-algebraic-machines.tex
In the following sections, we will establish complexity lower bounds for a broad class of extensions of $\UniFLExtLogic{}$. 
The idea is to adapt to our more general setting the arguments and constructions from Urquhart~\cite{urquhart1999}, where it is shown that the provability problem for $\UniFLeExtLogic{\UniCProp}$ is Ackermannian. %
While Urquhart carried out this argument syntactically for $\UniFLeExtLogic{\UniCProp}$, it is more suitable for the general extensions under consideration here to employ a modular approach using semantic techniques developed in \cite{galatos2022} and based on the residuated frames of Section~\ref{s: RelSem}. 
In particular, using residuated frames we will reduce the reachability problem for counter machines used by Urquhart~\cite{urquhart1999} to the theory of certain varieties of residuated lattices, thus establishing lower bounds; 
then we use the fact that (pointed) residuated lattices are the equivalent algebraic semantics of $\m {FL}$, as mentioned in Section~\ref{s: AlgSem}, to transfer these results to the corresponding logics. 
In particular, we will consider and-branching counter machines and introduce the notion of a computation forest, which will allow us to visualize and argue about computations in these machines. 
Then we will define specific complexity problems on these machines and encode them in residuated lattices via quasiequations, establishing their soundness due to the idempotent semiring reducts. 
The particular equations we will be considering will be represented by computational glitches and our proof of the completeness of the encoding via residuated frames will amount to showing that the computations of the machines are impervious to these glitches.

We start by defining some types of simple equations that will be important for what follows. 
In Theorem~\ref{thm:branchincLB} and Theorem~\ref{t: ackhardJD} we will establish lower bounds for the quasiequational theory/deducibility of quasi-varieties/logics axiomatized by the following types of equations/rules.
Recall that an equation $x_1 x_2 \cdots x_\ell \leq t_1 \jn  \cdots \jn t_k$ is simple if the variables $x_1,\ldots,x_\ell$ are pairwise distinct, each joinand $t_i$ is a product of variables, and every variable in the equation occurs on both sides; 
by exception, integrality is also included.
 \begin{definition}\label{def:TypesOfSimpEq} 
 A simple equation is:
 \begin{itemize}
 \item 
 \emph{{\joinincreasing}}, if there is a joinand on the right-hand side in which each %\NOTE{N}{of the $n$-many variables occur} 
 variable occurring in the equation appears at least once;
 \item 
 \emph{{\joindecreasing}}, if there is a joinand on the right-hand side in which each variable occurring in the equation appears at most once (i.e., a linear joinand). 
 \end{itemize}
 \end{definition}

 We say that a $\{\vee,\cdot,1\}$-equation (not necessarily simple) is \emph{\joinincreasing}, if the simple equation(s) obtained via the (linearization) procedure of Lemma~\ref{l: N_2^-0} is \joinincreasing; 
 since this procedure leaves simple equations unchanged, this is an extension of the definition for simple equations. 
 Similarly, we say that a $\{\vee,\cdot,1\}$-equation is \emph{\joindecreasing}, if the simple equation(s) obtained via the linearization procedure is \joindecreasing.  
 As Lemma~\ref{l: N_2^-0} does not provide a linearization process for trivializing equations (those with only the one-element---trivial---model), e.g., the equation $\mathsf{t}:1\leq x$, we explicitly define trivializing $\{\vee,\cdot,1\}$-equations to be \joinincreasing, but not  \joindecreasing. 
 Later on, in Section~\ref{sec: eqs_geometric}, we give a more ``geometric'' description of these general, not necessarily simple, {\joinincreasing} and {\joindecreasing} equations.

For example, in the linearization, below, of the equation $x^2\leq x^3$ of Example~\ref{ex: lin}:
$$uv\leq u^3\vee u^2v\vee uvu \vee uv^2\vee vu^2\vee vuv\vee v^2 u\vee v^3,$$
any one of the joinands $u^2v, uvu , uv^2, vu^2, vuv, v^2 u$ witnesses that the equation is {\joinincreasing}. 
However, it is not {\joindecreasing}: every joinand has a variable that occurs more than once.
Likewise, the linearization of $x^3\leq x^2$:
$$uvw
\leq
u^2 \vee uv \vee uw \vee vu \vee v^2 \vee vw \vee wu \vee wv \vee w^2 $$
is  {\joindecreasing}, as witnessed by any one of the joinands $uv$, $uw$, $vu$, $vw$, $wu$, $wv$, but it is not {\joinincreasing}.

Note that an equation can be both {\joinincreasing} and {\joindecreasing}: in the simple equation $x\leq x^2\vee 1$ the joinand $x^2$ witnesses the fact that the equation is {\joinincreasing} and $1$ witnesses that it is {\joindecreasing}. 
On the other hand, in the simple equation $xy\leq x^2\vee y^2$ neither of the joinands contain both variables (so the equation is not \joinincreasing) and each joinand contains more than one instance of some variable (so it is not \joindecreasing).

We now prove that the observations above for $x^2\leq x^3$ and $x^3\leq x^2$ hold in general, thus showing that {\joinincreasing} and {\joindecreasing} equations are generalizations of knotted equations. 
We call an equation $\simpeq$ \emph{non-trivializing} if it does not entail (modulo the theory of residuated lattices) the trivializing equation $\mathsf{t}: 1\leq x$; i.e., if $\UniRLV+\simpeq\not\models \mathsf{t}$. 
We denote the set of all non-trivial {\joinincreasing} equations by $\eqset_\mathrm{inc}$; in the proposition below, we verify that the variety of residuated lattices satisfying $\eqset_\mathrm{inc}$ coincides with the variety of contractive commutative residuated lattices. 
Similarly, we call an equation $\simpeq$ \emph{non-integral} if it does not entail (modulo the theory of residuated lattices) the integrality equation $\mathsf{i}: x\leq 1$; i.e., if $\UniRLV+\simpeq\not\models \mathsf{i}$.
We denote the set of all non-integral {\joindecreasing} equations by $\eqset_\mathrm{dec}$; in the proposition below we show that the associated variety contains all integral commutative residuated lattices (i.e., each equation in $\eqset_\mathrm{dec}$ is entailed by integrality and commutativity). 
Later on, in Section~\ref{sec: CRLa}, we show the variety is finitely based and we call its algebras  \emph{almost integral} residuated lattices.

In what follows, we denote by $\mathsf{CRL}$ the variety of commutative residuated lattices.
Moreover, we adopt the following notation: if $\mathcal{V}$ is a variety and $\mathcal{E}$ is a set of equations, we denote by $\mathcal{V}_\mathcal{E}$ the subvariety of $\mathcal{V}$ axiomatized by $\mathcal{E}$.
\begin{proposition}\label{prop:AlterEqs}\ 
\begin{enumerate}
\item\label{altereqs1}
% For $m,n$ positive integers, the rules $\UniWeakCAnaRule mn$ and $\UniWeakWAnaRule mn$ correspond to non-integral equations that are, respectively, {\joinincreasing} and {\joindecreasing}.
For $m,n$ positive integers, the rules $\UniWeakCAnaRule mn$ and $\UniWeakWAnaRule mn$ correspond to non-integral equations that are {\joinincreasing} and {\joindecreasing}, respectively.
\item\label{altereqs2} 
Every non-trivializing {\joinincreasing} equation holds in $\UniCRLV_\mathsf{c}$ and every 
{\joindecreasing} equation holds in $\UniCRLV_\mathsf{i}$,
i.e., $\UniCRLV_\mathsf{c} \subseteq \mathsf{RL} +\varepsilon$, for every  {\joinincreasing} equation $\varepsilon$, and $\UniCRLV_\mathsf{i} \subseteq \mathsf{RL} +\varepsilon$, for every 
{\joindecreasing} equation $\varepsilon$.
Moreover, $\UniCRLV_\mathsf{c}=\UniRLV_{\eqset_\mathrm{inc}}$ and $\UniCRLV_\mathsf{i}=\UniRLV_{\eqset_\mathrm{dec}}\cup\{\mathsf{i}\}$.
\end{enumerate}
\end{proposition}
\begin{proof}
For (1), by Section~\ref{s: N2} we know that the rules $\UniWeakCRule mn$ and $\UniWeakWRule mn$ correspond to the inequality $x^n \leq x^m$, for $n<m$ and $n>m$, respectively.
Via the linearization process of Lemma~\ref{l: N_2^-0}, the simple equation $\varepsilon$ corresponding to the inequality $x^n \leq x^m$ is 
$$x_1 x_2 \cdots x_n \leq \bigvee \{x_{i_1}\cdots x_{i_m}: i_1, \ldots, i_m \in \{1, \ldots, n\}\}.$$
If $n<m$ then $x_1x_2 \cdots x_n\cdots x_n$ appears in the join, hence $\simpeq$ is {\joinincreasing}. If $n>m$, then the joinand $x_1x_2 \cdots x_m$ appears in the join, hence $\simpeq$ is {\joindecreasing}.

For (2), first recall that an equation is {\joinincreasing}/{\joindecreasing} if its equivalent simple equation (modulo the theory of RLs) obtained via the linearization of Lemma~\ref{l: N_2^-0} is {\joinincreasing}/{\joindecreasing}.
So it suffices to prove the claim for simple equations. 
Let $\simpeq: x_1 \cdots x_\ell \leq t_1 \jn \cdots \jn t_k$ be a simple equation and let $t_\simpeq$ denote the joinand witnessing the {\joinincreasing} or the  {\joindecreasing} property. 
Note that, modulo commutativity, the term $t_\simpeq$ is equivalent to $x_1^{m_1}\cdots x_\ell^{m_\ell}$, for some $m_1, \ldots, m_\ell \in \mathbb{N}$. 
If $\simpeq$ is {\joindecreasing} then $m_i\leq 1$, for all $i$, so integrality implies $x_1 \cdots x_\ell \leq x_1^{m_1}\cdots x_\ell^{m_\ell}$; hence modulo commutativity we have $x_1 \cdots x_\ell \leq t_\simpeq \leq t_1 \jn \cdots \jn t_k$. 
Consequently, $\UniCRLV_\mathsf{i}\subseteq\UniRLV_{\eqset_\mathrm{dec}}$, while clearly $\UniRLV_{\eqset_\mathrm{dec}} + \{\mathsf{i}\} \subseteq \UniCRLV_\mathsf{i}$ since integrality is assumed and commutativity ($xy\leq yx$) is a {\joindecreasing} equation. 
Likewise, if $\simpeq$ is  {\joinincreasing} then $m_i\geq 1$ for each $i$, so contraction implies $x_1 \cdots x_\ell \leq x_1^{m_1}\cdots x_\ell^{m_\ell}$; hence modulo commutativity we have $x_1 \cdots x_\ell \leq t_\simpeq \leq t_1 \jn \cdots \jn t_k$.
Consequently, $\UniCRLV_\mathsf{c}\subseteq\UniRLV_{\eqset_\mathrm{inc}}$ and the reverse inclusion holds since contraction and commutativity are both {\joinincreasing} equations.
\end{proof}

Our results apply to these general types of rules, hence they will specialize nicely to the special cases of interest: $\UniWeakCAnaRule mn$ and $\UniWeakWAnaRule mn$.

\section{And-branching counter machines}
The encoding of complexity problems in the (quasi)equational theories of interest will be done via abstract models of computation, known as \emph{And-branching counter machines} (ACMs), that have proven suitable tools for establishing complexity, and even undecidability, results (see \cite{lincoln1992,urquhart1999,horcik2015,horcik2016,galatos2022}). 
ACMs are equivalent to a type of \emph{Vector Addition System with States} (VASS) known as \emph{Alternating} VASS (AVASS). Our formalization of these machines will have an ``algebraic'' flavor, as is done in \cite{galatos2022}, and, in contrast to \cite{lincoln1992,urquhart1999}, our formulation makes the machines more amenable to the logics we consider. 
We refer to our formulation as algebraic, and use $\leq$ to denote the computation relation, because, as we will see, this relation is taken to be compatible with the (algebraic) operations of multiplication and join.

An ACM is a tuple ${\acm} \UniSymbDef (\Reg, {\State},{\Inst})$ representing a type of (parallel-computing) counter machine, where 
\begin{itemize}
	\item $\Reg$ is a finite set of \textit{registers}, each able to store a nonnegative integer which corresponds to the number of \emph{tokens} in the register,      
	\item ${\State}$ is a finite set of \textit{states} with a designated \emph{initial state} $\qin$ and \textit{final state} $\qfin$, and
	\item ${\Inst}$ is a finite set of \textit{instructions} that allow for \emph{incrementing} or \emph{decrementing} by adding or removing (if possible) a single token to a given register and transitioning from one state to another; they also include \textit{forking} instructions that duplicate register contents into parallel computations (or branches).
\end{itemize}
For a positive integer $\kreg$, a \emph{$\kreg$-ACM} is an ACM with $\kreg$-many registers, i.e., $\kreg=|\Reg|$; in such a case we will often represent the set of registers by $\Reg_\kreg:=\{\reg_1,...,\reg_\kreg\}$ and we refer to $\kreg$ as the \emph{dimension} of the ACM.

A \emph{configuration} of an ACM $\acm$ consists of a single state and, for each register, a nonnegative integer indicating the contents of that register (if the integer is $0$, we say that register is \emph{empty}); we denote by $\conf{\acm}$ the set of all configurations.  
As the machine is intended to capture parallel computation (due to forking the machine can be at multiple configurations), the status of the machine is given by a finite multiset of configurations, known as an \emph{instantaneous description} (ID); we denote by $\indes{\acm}$ the set of all possible instantaneous descriptions. 

We denote by $\mathbf{A_{\acm}}\UniSymbDef(A_{\acm},\vee,\cdot, 1)$ the commutative semiring freely generated by the finite set $\State\cup\Reg$ and we represent its elements concretely as multisets (written as joins) of monoid words over $\State\cup\Reg$. 
Recall that $(A,\vee,\cdot,  1)$ is called a \emph{commutative semiring} if $(A, \jn)$ is a commutative semigroup, $(A, \cdot, 1)$ is a commutative monoid and $\cdot$ distributes over $\jn$. 
In this way, the sets $\conf{\acm}$ and $\indes{\acm}$ may be seen as special subsets of $A_{\acm}$: a configuration, usually denoted by $\cf$, is formally a multiplicative term $\qstate \reg_1^{n_1} \reg_2^{n_2}\cdots \reg_\kreg^{n_\kreg}$, where $d$ is the dimension of $\acm$ (note that the contents of the registers are captured by the exponents) and an ID, usually denoted by $\uid$, is a join $\cf_1\vee\cdots\vee \cf_\ell$ of configurations (as we will see, this will represent the leaves in the branching tree at some point of the run of the machine), where $\ell$ is a positive integer.
For example, if $\State=\{\qin, \qfin, \qstate\}$ and $\Reg=\{\reg_1, \reg_2, \reg_3\}$, then $\qstate\reg_1^2\reg_3$ is a configuration and $\qstate\reg_1^2\reg_3 \jn \qin \reg_2\reg_3\jn \qfin$ is an ID indicating that at that moment the machine is simultaneously in three different configurations via the branching.

Note that we do not assume that the symbol $\jn$ of  $\mathbf{A_{\acm}}$ is idempotent, as that provides more control when working with the computation relation of the machine; nevertheless, since $\jn$ will be mapped to/interpreted as the (idempotent) join operation of a residuated lattice, the lack of idempotency is merely a technical convenience and it can safely be thought as idempotent by the reader. 
Due to the distributivity of $\cdot$ over $\jn$ (the utility of which will become apparent later on), the terms of $\mathbf{A_{\acm}}$ admit a normal form of joins of monoid words, up to the commutativity of join; we will work with the understanding that all terms are fully distributed like that, even when we write $\uid x$ where $\uid$ is an ID and $x \in (\State\cup\Reg)^*$; here, and for the remainder of the monograph, we denote by $X^*$ the free \emph{commutative} monoid of the set $X$; please note that this deviates from the usage of this notation in previous sections, where it was denoting the free monoid (without commutativity). 
Also, even though $\mathbf{A}_\acm$ has no unit/identity for $\lor$, it embeds in $\mathbf{A}_\acm^\bot=\mathbf{A}_\acm\cup\{\bot\}$, the commutative semiring with additive identity $\bot$ (and where $\bot$ is an absorptive element for multiplication) freely generated by $\State\cup\Reg$. 

The set $\Inst$ of \emph{instructions} of an ACM consists of pairs of a configuration and an ID (where we denote $p\UniSymbDef(a,b)$ by $p: a \leq b$) of the following forms for states $\qstate,\qstate',\qstate''\in \State$ and register $\reg\in \Reg$. 
In particular, we represent
\begin{center}
``increment-$\reg$ at $\qstate$" as $p: \qstate\leq \qstate'\reg$\\ 
``decrement-$\reg$ at $\qstate$" as $p: \qstate\reg\leq \qstate'$\\ 
``fork at $\qstate$" as $p: \qstate\leq \qstate'\vee \qstate''$
\end{center}
(Note that these instructions are of the form $\cf\leq \uid$, where $\cf$ is a configuration and $\uid$ is an ID.) 
As we see below, the intended meaning of the instruction $ \qstate\leq \qstate'\reg$, for example, is not  merely that the machine can transition from the configuration $ \qstate$ to the configuration $\qstate'\reg$.

The \emph{computation relation} $\leq_\acm$ of $\acm$ is defined as the smallest $\{\cdot,\vee \}$-compatible preorder over $\mathbf{A}_\acm$ containing $\Inst$. 
It will be useful to have representations of pairs in $\leq_\acm$ in a way that witnesses the fact that they are indeed generated by pairs of $\Inst$; we call such a representation a \emph{computation} as we detail below.
For each $p\in \Inst$, we define $\leq^p$ to be the smallest relation over $\mathbf{A}_\acm$ containing $p$ and closed under the inference rules 
$$\infer[{[\cdot]}] {ux\leq^p vx }{u\leq^p v} 
\quad\begin{array}{c}\mathsf{and}\\~\end{array}\quad
\infer[{[\vee]}]{ u\vee w \leq^p v\vee w}{u\leq^p v},$$
for any terms $u,v,w\in A_{\acm}$ and word $x\in (\State\cup\Reg)^*$.  
It can be easily shown that, during the closure process, these inference rules  can be safely applied in a particular order: all the multiplication rule instances are applied before all the join instances.  
Consequently, if $p: {\cf}\leq {\uid}$, then: $u\leq^p v$ iff $u=_{\mathbf{A}_\acm}{\cf}x\vee w$ and $v=_{\mathbf{A}_\acm}\uid x\vee w$, for some word $x\in (\State\cup\Reg)^*$ and term $w \in A^\bot_{\acm}$; allowing for $w$ to be $\bot$ is a convenient way for us to include the case where $u$ and $v$ are IDs comprising of a single configuration, while still making sure that $u$ and $v$ are never themselves equal to $\bot$, hence they remain elements of $A_{\acm}$. 
There could be many occurrences of ${\cf}x$ in $u$ and, in order to have control over which occurrence we refer to, we consider this information as part of the instance of $p$: an \emph{instance} of $p$ consists of the above (i.e., $\cf, x, \uid, w$) together with the particular \emph{occurrence} of ${\cf}x$ in $u$. 
The pair $(\cf x, \uid x)$ is called the \emph{active component} of the instance $u\leq^p v$; note that $\cf x$ is a monoid word, but $\uid x$ could be a join of monoid words.
Also, the term $w$ above is called the \emph{passive component}, if $w \in A_\acm$; if $w=\bot$, then the instance has no passive component. We illustrate some of these definitions in Example~\ref{ex:acmleq} below.

It turns out that the computation relation $\leq_\acm$ is equal to the smallest preorder containing $\bigcup \UniSet{{\leq^p}:p\in \Inst}$.
So, $u\leq_{\acm} v$ iff there exist $\complength \in \mathbb{N}$, 
a sequence of $\mathbf{A}_{\acm}$-terms $u_0, \ldots, u_\complength$ and a sequence of \emph{instances} of instructions $p_1, \ldots, p_\complength$ from ${\Inst}$---collectively called a \textit{computation in ${\acm}$} of \emph{length} $\complength$ witnessing $u\leq_{\acm} v$---such that
\begin{equation}\label{eq:ACMcomps}
u=_{\mathbf{A}_\acm} u_0 \leq^{p_1} u_1 \leq^{p_2}\cdots \leq^{p_\complength} u_\complength=_{\mathbf{A}_\acm} v.
\tag{$\comp$}
\end{equation}
Given such a computation $\comp$ we will define the labeled forest ${\labeledforest}_\comp$ of it as follows. 
We will see that these labeled forests capture computations and provide a visual rendering of them that is amenable to more intuitive and direct arguments in the proofs below. 
We consider the set of all \emph{occurrences} of monoid terms in the computation (in all of the $u_i$'s) and  we define an equivalence relation on it: we identify two occurrences if their terms are equal and they appear on two sides, $\cf x \jn w$ and $\uid x \jn w$, of some instance $\cf x \jn w\leq ^p \uid x \jn w$ of an instruction $p$ within the passive component $w$ of that instance, and then we take the equivalence-relation closure. 

\begin{example}\label{e: computation}
For example, in the computation 
\begin{equation}\label{eq:compforestEXeq}
c_1 \jn c_2 \jn c_1 \leq^{p_1} c_4 \jn c_1 \jn c_3 \jn c_2 \jn c_1 \leq^{p_2}  c_4 \jn c_1 \jn c_5 \jn c_6 \jn c_2 \jn c_1
\end{equation}
which, for easy reference of the occurrences, we also write as
 $$c_1^{1} \jn c_2^{1} \jn c_1^{2} \leq^{p_1} c_4^{1} \jn c_1^{3} \jn c_3^{1} \jn c_2^{2} \jn c_1^{4} \leq^{p_2}  c_4^{2} \jn c_1^{5} \jn c_5^{1} \jn c_6^{1} \jn c_2^{3} \jn c_1^{6}$$
and where the active component of $p_1$ is $(c_1^{1}, c_4^{1} \jn c_1^{3} \jn c_3^{1})$ and the active component of $p_2$ is $(c_3^{1}, c_5^{1} \jn c_6^{1})$, the set of occurrences contains all 14 occurrences. 
Note that we wrote $c_1^{3}$ for the third occurrence of $c_1$, instead of the more standard $(c_1, 3)$. Since the passive component of the first instance is $c_2 \jn c_1 (=c_2^{1} \jn c_1^2=c_2^{2} \jn c_1^{4})$, the terms $c_1^{2}$ and $c_1^{4}$ are identified (and $c_2^{1}$ is identified with $c_2^{2}$). 
These are further identified with $c_1^{6}$ as the passive component of the last instance is $c_4^{1} \jn c_1^{3} \jn c_2^{2} \jn c_1^{4}=c_4^{2} \jn c_1^{5} \jn c_2^{3} \jn c_1^{6}$. 
We define $F_\comp$ to be the resulting set of equivalence classes of occurrences. 
In the above example, all occurrences of $c_2$ form a single equivalence class in $F_\comp$ and the same holds for all other terms, except for $c_1$ whose occurrences split into the three equivalences classes $\{c_1^{1}\},  \{c_1^{2}, c_1^{4}, c_1^{6}\},  \{c_1^{3}, c_1^{5}\}$; so  $F_\comp=\{[c_1^{1}], [c_1^{2}], [c_1^{3}], [c_2^{1}],[c_3^{1}],[c_4^{1}],[c_5^{1}],[c_6^{1}]\}$, where we chose the representative of each class as the first occurrence of each term in the computation. 
\end{example}

We define the ordering relation $\preceq$ on $F_\comp$ to be the transitive and reflexive closure of the relation $\prec$ given by: for $\tnode,\tnode' \in F_\comp$ we write $\tnode \prec \tnode'$ if there exist occurrences $c\in \tnode$ and $d \in \tnode'$ such that $(c, v)$ is the active component of an instance of an instruction in the computation and $d$ is a joinand of $v$. 
In Example~\ref{e: computation}, $[c_1^{1}] \prec [c_4^{1}]$,
$[c_1^{1}] \prec [c_1^{3}]$ and $[c_1^{1}] \prec [c_3^{1}]$, since $(c_1^{1}, c_4^{1} \jn c_1^{3} \jn c_3^{1})$ is the active component of $p_1$.
Since we never identify instances across active components, $\preceq$ is antisymmetric, hence a partial order. 
We denote the resulting poset by $(F_\comp, \preceq)$, and we refer to the minimal elements as \emph{roots} and the maximal elements as \emph{leaves}. 

We now assign \emph{labels} to the elements of $F_\comp$, as follows. 
If $\tnode \in F_\comp$ contains occurrences of the monoidal term $c$ and if $u_i$ is the first term in $\comp$ where an element of $\tnode$ appears, then we label $\tnode$ by $(c, i,p_i)$; if $i=0$, since no $p_0$ exists, we take $p_0$ to be the empty set. 
In Example~\ref{e: computation}, $u_1= c_4^1 \jn c_1^3 \jn c_3^1 \jn c_2^2 \jn c_1^4$ is the first term of the computation where an element of $[c_1^3]$ appears, so the label of $[c_1^{3}]$ is $(c_1, 1,p_1)$, encoding the step of the computation where $c_1^{3}$ first appeared and the instruction that produced it.
We denote by $\ell$ the labeling function described above and we write $\labeledforest_\comp=(F_\comp, \preceq, \ell)$ for the resulting labeled poset.  

Note that the labeled posets of computations are finite forests (every principal downset is a chain) and have labels that are triples $(c, i,p)$, where $c$ is a monoid term, $i \in \mathbb{N}$ and $p\in \Inst \cup \{\emptyset \}$, such that the labels of roots are of the form $(q, 0, \emptyset)$ and for every non-leaf node $\tnode$ with set of covers $\{\tnode_1, \ldots, \tnode_k\}$, the labels of $\tnode, \tnode_1, \ldots, \tnode_k$ are of the form  $(c,j,q)$, $ (c_1, i,p), \ldots, (c_k, i,p)$, for some $0\leq j<i$, and $c \leq^{p} c_1 \jn \cdots \jn c_k$ is an instance of the instruction $p$.
We call such labeled posets \emph{computation forests}. 

Below is the forest and the computation forest for the computation $\comp$ from \eqref{eq:compforestEXeq} of Example~\ref{e: computation}; in the latter we present the labels at each node. 
\begin{equation*}
\begin{array}{c | c}
\text{The forest $(F_\comp,\preceq)$} & \text{The computation forest $\labeledforest_\comp \UniSymbDef(F_\comp,\preceq,\ell)$}\\
\hline
\scalebox{.75}{
\begin{tikzpicture}[node distance = .5cm, every node/.style={minimum height=0cm,minimum width=0cm}]
%left tree
\node (1) {$\{c_1^{1}\}$};
\node[above = of 1] (2) {$\{c_1^3,c_1^5\}$};
\node[left = of 2] (l2) {$\{c_4^1,c_4^2\}$};
\node[right = of 2] (r2) {$\{c_3^1\}$};
\node[above = of r2] (3) {$\{c_5^1\}$};
\node[right = of 3] (r3) {$\{c_6^1\}$};
%right roots
\node[right = of 1] (1r) {$\{c_2^{1},c_2^{2},c_2^{3}\}$};
\node[right = of 1r] (1rr) {$\{c_1^{2},c_1^{4},c_1^{6}\}$};
%arrows
\draw (1)--(2);
\draw (1)--(l2);
\draw (1)--(r2);
\draw (r2)--(3);
\draw (r2)--(r3);
\end{tikzpicture}
}
&
\scalebox{.75}{
\begin{tikzpicture}[node distance = .5cm, every node/.style={minimum height=0cm,minimum width=0cm,draw}]
%left tree
\node (1) {$c_1,0,\emptyset$};
\node[above = of 1] (2) {$c_1,1,p_1$};
\node[left = of 2] (l2) {$c_4,1,p_1$};
\node[right = of 2] (r2) {$c_3,1,p_1$};
\node[above = of r2] (3) {$c_5, 2, p_2$};
\node[right = of 3] (r3) {$c_6, 2, p_2$};
%right roots
\node[right = of 1] (1r) {$c_2,0,\emptyset$};
\node[right = of 1r] (1rr) {$c_1,0,\emptyset$};
%arrows
\draw (1)--(2);
\draw (1)--(l2);
\draw (1)--(r2);
\draw (r2)--(3);
\draw (r2)--(r3);
\end{tikzpicture}
}
\end{array}
\end{equation*}

Given a computation forest $\labeledforest \UniSymbDef (F, \preceq, \ell)$, for all $i$ that appear as second coordinates of a label, $F_{i}:=\{\tnode: \ell(\tnode)=(c,j,p), j \leq i\}$ is subforest of $\labeledforest$; we define the term $u_i$ to be the join of all the monoid terms of the leaves of  $F_{i}$.
It is easy to see that then $u_0 \leq^{p_1} u_1 \leq^{p_2}\cdots \leq^{p_\complength} u_\complength$ is a computation; for example, from the labeled forest above we obtain the initial computation (\ref{eq:compforestEXeq}).

The correspondence between computations and computation forests is a bijection modulo renaming the nodes: labeled forests are special among computation forests only in that their nodes are not arbitrary elements, but rather equivalence classes of occurrences of monoid terms. In other words, the composition one way (computation $\rightarrow$ forest $\rightarrow$ computation) is the identity and the composition the other way (forest $\rightarrow$ computation $\rightarrow$ forest) is not the identity, but it is an isomorphism of labeled forests.

A number of useful properties become apparent using this perspective, as we see in the following lemma. 
Items 2, 3($a\Leftrightarrow b$), and 4 can also be found in \cite[Lemmas~3.1 and 3.4]{galatos2022}).
\begin{lemma}\label{lem:basicACMfacts}
    If $\acm$ is an $ACM$ and $u,v,w\in A_\acm$, then the following hold.
    \begin{enumerate}
        \item\label{i1:basicACMfacts} 
        $u\leq_\acm v$ iff there is a computation forest with root nodes labeled by the joinands of $u$ and leaf nodes labeled by the joinands of $v$.
        \item\label{i2:basicACMfacts}
        Any computation starting from a join of $\m{A}_\acm$-terms can be partitioned into two disjoint computations. 
        Conversely, any two computations may be combined to form a single computation. 
        Formally, $u\vee v\leq_\acm w$ iff $ u\leq_\acm w'$ and $v\leq_\acm w''$, for some $w',w''\in A_\acm$ with $w=w'\vee w''$.
        \item\label{i3:IDs2IDs} 
        The following are equivalent whenever $u\leq_\acm v$ holds:
        \begin{enumerate}
            \item $u\in\indes{\acm}$
            \item $v\in\indes{\acm}$
            \item Every node in a computation forest witnessing $u\leq_\acm v$ contains some configuration $\cf\in \conf{\acm}$ as its label-term.
        \end{enumerate}
    \end{enumerate}
\end{lemma}
\begin{proof}
    (1) is justified by the discussion above. 
    For the forward direction of (2), by (1) there exists a computation forest $\labeledforest$ witnessing  $u\vee v\leq_\acm w$. 
    Since $\labeledforest$ is a forest, the upsets $F_u$ and $F_v$ of the root nodes whose labels consist of the joinands in $u$ and $v$, respectively, are subforests of $\labeledforest$; let $w'$ and $w''$ be the join of the monoidal terms in the leaves of $F_u$ and $F_v$, respectively. 
    By relabeling (with different consecutive numbers) the center index the labels of nodes in $F_u$ and $F_v$,  we obtain computation forests $\labeledforest_u$ and $\labeledforest_v$, respectively, witnessing  $u\leq_\acm w'$ and $v\leq w''$.    
    Conversely, if $u\leq_\acm w'$ and $v\leq_\acm w''$, then using the fact that  $\leq_\acm$ is a preorder compatible with the operations, in particular $\vee$, we get $u\vee v \leq_\acm w'\vee v \leq_\acm w'\vee w'' $. 

    For (3), let $\labeledforest$ be a computation forest witnessing $u\leq_\acm v$. 
    As each instruction of the ACM contains precisely one state variable on the left-hand side and one state variable (in each joinand of) the right-hand side, the number of state variables appearing in the labels of the nodes of every maximal convex subchain $C$ of $\labeledforest$ is constant. 
    Consequently, the term-label of the root of $C$ is a configuration iff the leaf label is a configuration.
    Since every node appears in some maximal convex subchain, the equivalences follow. 
\end{proof}

\begin{example}\label{ex:acmleq}
The following $2$-ACM $\acm_\leq=(\Reg_2,\State_\leq,\Inst_\leq)$ verifies that, when starting with state $\qin$ and reaching an ID consisting only of the configuration $\qfin$, the contents of register $\reg_1$ are no greater than that of $\reg_2$. 
The instructions and states are given as follows:
\begin{align*}
    \Dec\reg_1 &: 
    \qin\reg_1\leq \qstate_1 
    &
    \Dec\reg_2 &: 
    \qstate_1\reg_2 \leq \qin 
    &\\ 
    \reg_1\ztr &:
    \qin \leq \qstate_2 \vee \zstate_1 
    & 
    \reg_2{\to}\text{0} &: \qstate_2\reg_2\leq \qstate_2
    &
    \reg_2\ztr &:
    \qstate_2\leq \qfin \vee \zstate_2
    \label{acmleq}\tag{$\acm_\leq$} \\
    \text{Also, for}&~ \{i,j \}=\{1,2\}:
    &
    {\zinst}^i_j &:
    \zstate_i\reg_j \leq \zstate_i 
    &
    {\zinst}^i_\FIN &:
    \zstate_i \leq \qfin \vee \qfin
\end{align*}

The following are examples of $\acm_\leq$-computations: 
% $$
% \begin{array}{r ll ll ll l }
%     \comp_0 :&
%     \qin \reg_1\reg_2^2 
%     & \leq^{\Dec\reg_1}&
%     \qstate_1 \reg_2^2
%     &\leq^{\Dec\reg_2}&
%     \qin\reg_2
%     &\leq^{\reg_1\ztr}&
%     \qstate_2\reg_2\vee\zstate_1 \reg_2
%     ;\\    
%     \comp_1 :&
%     \qstate_2 \reg_2 
%     & \leq^{\reg_2{\to}\text{0}}&
%     \qstate_2
%     &\leq^{\reg_2\ztr}&
%     \qfin \vee \zstate_2
%     &\leq^{{\zinst}^2_\FIN}&
%     \qfin \vee \qfin \vee \qfin
%     ;\\    
%     \comp_2: &
%     \zstate_1\reg_2 
%     &\leq^{{\zinst}^1_2}&
%     \zstate_1 
%     &\leq^{{\zinst}^1_\FIN}&
%     \qfin\vee\qfin;    
%     &\text{and}~ \comp'_2:&    
%     \zstate_1\reg_2 
%     \leq^{{\zinst}^1_\FIN}
%     \qfin\reg_2\vee\qfin\reg_2
%     .\qedhere
% \end{array}
% $$
$$
\arraycolsep=1pt
\begin{array}{r| ll  ll ll l}
    \comp_0 ~&~
    \qin \reg_1\reg_2^2 
     &\leq^{\Dec\reg_1}&
    \qstate_1 \reg_2^2
    &\leq^{\Dec\reg_2}&
    \qin\reg_2
    &\leq^{\reg_1\ztr}&
    \qstate_2\reg_2\vee\zstate_1 \reg_2
    ;\\ 
    \hline
    \comp_1 ~&~
    \qstate_2 \reg_2 
     &\leq^{\reg_2{\to}\text{0}}&
    \qstate_2
    &\leq^{\reg_2\ztr}&
    \qfin \vee \zstate_2
    &\leq^{{\zinst}^2_\FIN}&
    \qfin \vee \qfin \vee \qfin
    ;\\
    \hline
    \comp_2 ~&~
    \zstate_1\reg_2 
    &\leq^{{\zinst}^1_2}&
    \zstate_1 
    &\leq^{{\zinst}^1_\FIN}&
    \qfin\vee\qfin; 
    \\
    \hline
    \comp'_2~&~    
    \zstate_1\reg_2 
    &\leq^{{\zinst}^1_\FIN}&
    \qfin\reg_2&\vee\hspace{1pt}\qfin\reg_2
    .
\end{array}
$$
While it may not be the most succinct machine with this property, it is easily verified that $\acm_\leq$ performs the desired task. 
\end{example}
For two $\acm$-computations $\comp: u_0\leq^{p_1}\cdots\leq^{p_m} u_m$ and $\comp':v_0\leq^{p'_1}\cdots \leq^{p'_n} v_n$, we define the $\acm$-computation $\comp\jn \comp': u_0 \jn v_0\leq^{p_1}\cdots\leq^{p_m} u_m\vee v_0 \leq^{p'_1} u_m\vee v_1 \leq^{p'_2}\cdots \leq^{p'_n} u_m\vee v_n$. 
Also, for two `composable' $\acm$-computations $\comp: u_0\leq^{p_1}\cdots\leq^{p_m} u\vee v_0$, and $\comp': v_0\leq^{p'_1}\cdots \leq^{p'_n} v_n$, we define the $\acm$-computation $\comp\circ\comp': u_0\leq^{p_1}\cdots\leq^{p_m} u\vee v_0 \leq^{p'_1} u\vee v_1 \leq^{p'_2}\cdots \leq^{p'_n} u\vee v_n$. 

We consider now the following computations of \ref{acmleq}: 
$\comp_{12}:= \comp_1\jn \comp_2$, $\comp_{21}:= \comp_2\jn \comp_1$, $\comp_{012}:= \comp_0\circ \comp_{12}$, $\comp_{021}:= \comp_0\circ \comp_{21}$; 
and $\comp'_{012}:= \comp_0 \circ (\comp_1\circ \comp'_2)$.  
Note that  $\comp_{012}$ witnesses that $\qin\reg_1\reg_2^2$ is accepted in $\acm$. 
The computation $\comp'_{012}$ also starts with $\qin\reg_1\reg_2^2$, but it does not reach an accepted ID, even though it is maximal, i.e., no further instruction can be applied. 
Therefore, performing a maximal computation that fails to accept a configuration does not imply that the configuration is not accepted by the machine.

Also, notice that the computations $\comp_{012}$ and $\comp_{021}$ are distinct $\acm_\leq$-computations, with the same initial and final terms, since $\comp_{12}$ and $\comp_{21}$ are distinct computations with the same property.

\begin{example}\label{ex:ACMleqCompForest}
Consider the posets $(F_{12},\preceq_1)$ and $(F_{21},\preceq_2)$ corresponding to the \ref{acmleq}-computations $\comp_{12}$ and $\comp_{21}$, respectively. 
These forests are isomorphic:
\begin{equation*}
\begin{array}{c | c | c}
    \text{\scriptsize$\acm_\leq$-computation $\comp_{12}$:} 
    & \text{\scriptsize Forest 
    $(F_{12},\preceq_1) \cong (F_{21},\preceq_2)$:} 
    & \text{\scriptsize $\acm_\leq$-computation $\comp_{21}$:} 
    \\ 
    \hline
        \scalebox{.8}{$
        \arraycolsep=1pt
        \begin{array}{c l}
            \qfin \vee \qfin \vee \qfin\vee \qfin \vee \qfin
            &\geq^{p_5} \\
            \qfin \vee \qfin \vee \qfin \vee \zstate_1
            &\geq^{p_4} \\
            \qfin \vee \qfin \vee \qfin \vee \zstate_1 \reg_2
            &\geq^{p_3} \\
            \qfin \vee \zstate_2 \vee \zstate_1\reg_2
            &\geq^{p_2} \\
            \qstate_2 \vee \zstate_1\reg_2
            &\geq^{p_1} \\
            \qstate_2\reg_2 \vee \zstate_1\reg_2 
        \end{array}
        $}   
    &    
    \scalebox{.7}{
    \begin{tikzpicture}[baseline={([yshift=-.8ex]current bounding box.center)}, node distance = .5cm, every node/.style={minimum height=0cm,minimum width=0cm,draw}]
        %left tree
        \node (1) {$\qstate_2\reg_2$};
        \node[above = of 1] (2) {$\qstate_2$};
        \node[above = of 2] (3) {$\zstate_2$};
        \node[left = of 3] (l3) {$\qfin$};
        \node[above = of 3] (4) {$\qfin$};
        \node[left  = of 4] (l4) {$\qfin$};
        \draw (1)--(2);
        \draw (2)--(3);
        \draw (3)--(4);
        \draw (2)--(l3);
        \draw (3)--(l4);
        %right tre
        \node[right = of 1] (1') {$\zstate_1\reg_2$};
        \node[above = of 1'] (2') {$\zstate_1$};
        \node[above = of 2'] (3') {$\qfin$};
        \node[right = of 3'] (r3') {$\qfin$};
        \draw (1')--(2');
        \draw (2')--(3');
        \draw (2')--(r3');
    \end{tikzpicture}
    }
    & 
        \scalebox{.8}{$
        \arraycolsep=1pt
        \begin{array}{c l}
            \qfin \vee \qfin \vee \qfin\vee \qfin \vee \qfin 
            & \geq^{p_5'} \\
            \qfin \vee \zstate_2 \vee \qfin \vee \qfin
            & \geq^{p_4'} \\
            \qstate_2 \vee \qfin \vee \qfin
            & \geq^{p_3'} \\
            \qstate_2\reg_2 \vee \qfin \vee \qfin
            & \geq^{p_2'} \\
            \qstate_2\reg_2 \vee \zstate_1
            & \geq^{p_1'} \\
            \qstate_2\reg_2 \vee \zstate_1\reg_2
        \end{array}
        $}
\end{array}
\end{equation*}
In the middle column above, we chose to give representatives of the equivalence classes that serve as nodes (instead of putting the whole equivalence class) and, furthermore, in order to show the isomorphism between the two (distinct) forests, we suppressed the particular occurrences of the monoid terms. 

As can be seen from the table above, the forest does not fully capture the computation; in particular, it does not have the information of which instruction is used when.
\footnote{
    In this particular case, all \ref{acmleq}-computations witnessing the acceptance of $\qstate_2\reg_2\vee\zstate_1\reg_2$ actually have the same forest, but $\qin\reg_2$, for example, has two accepting computations with different forests; so forests distinguish some computations but not all.
    } 
However, we see below that the labeling distinguishes the computation forests $\labeledforest_{12}=(F_{12},\preceq_1,\ell_1)$ and $\labeledforest_{22}=(F_{21},\preceq_2,\ell_2)$, displayed with the label affixed at each node; their corresponding computations can be recovered according to our definition. 
\begin{equation*}
\begin{array}{c|c}
\scalebox{.65}{
    \begin{tikzpicture}[node distance = .5cm, every node/.style={minimum height=0cm,minimum width=0cm,draw}]
    %left tree
    \node (1) {$\qstate_2\reg_2,0,\emptyset$};
    \node[above = of 1] (2) {$\qstate_2,1,\reg_2{\to}\text{0}$};
    \node[above = of 2] (3) {$\zstate_2, 2, \reg_2\ztr$};
    \node[left = of 3] (l3) {$\qfin, 2, \reg_2\ztr$};
    \node[above = of 3] (4) {$\qfin, 3,{\zinst}^2_\FIN$};
    \node[left = of 4] (l4) {$\qfin, 3,{\zinst}^2_\FIN$};
    \draw (1)--(2);
    \draw (2)--(3);
    \draw (3)--(4);
    \draw (2)--(l3);
    \draw (3)--(l4);
    %right tre
    \node[right = of 1] (1') {$\zstate_1\reg_2, 0,\emptyset$};
    \node[above = of 1'] (2') {$\zstate_1, 4, {\zinst}^1_2$};
    \node[above = of 2'] (3') {$\qfin, 5, {\zinst}^1_\FIN$};
    \node[right = of 3'] (r3') {$\qfin, 5, {\zinst}^1_\FIN$};
    \draw (1')--(2');
    \draw (2')--(3');
    \draw (2')--(r3');
    \end{tikzpicture}
    }
&
\scalebox{.65}{
    \begin{tikzpicture}[node distance = .5cm, every node/.style={minimum height=0cm,minimum width=0cm,draw}]
    %left tree
    \node (1) {$\qstate_2\reg_2,0,\emptyset$};
    \node[above = of 1] (2) {$\qstate_2,3,\reg_2{\to}\text{0}$};
    \node[above = of 2] (3) {$\zstate_2, 4, \reg_2\ztr$};
    \node[left = of 3] (l3) {$\qfin, 4, \reg_2\ztr$};
    \node[above = of 3] (4) {$\qfin, 5,{\zinst}^2_\FIN$};
    \node[left = of 4] (l4) {$\qfin, 5,{\zinst}^2_\FIN$};
    \draw (1)--(2);
    \draw (2)--(3);
    \draw (3)--(4);
    \draw (2)--(l3);
    \draw (3)--(l4);
    %right tree
    \node[right = of 1] (1') {$\zstate_1\reg_2, 0,\emptyset$};
    \node[above = of 1'] (2') {$\zstate_1, 1, {\zinst}^1_2$};
    \node[above = of 2'] (3') {$\qfin, 2, {\zinst}^1_\FIN$};
    \node[right = of 3'] (r3') {$\qfin, 2, {\zinst}^1_\FIN$};
    \draw (1')--(2');
    \draw (2')--(3');
    \draw (2')--(r3');
    \end{tikzpicture}
    }
\\
\text{The computation forest } \labeledforest_{12} & \text{The computation forest } \labeledforest_{21}\qedhere
\end{array}
\end{equation*}
\end{example}

\section[Encoding ACMs in residuated lattices: soundness]{Encoding complexity problems of ACMs in residuated lattices: soundness}
Of the vast landscape of complexity problems arising from the computation relation of ACMs, here we will use problems based on the notions of \emph{acceptance} and \emph{termination} in ACMs. The \emph{initial} configuration of an ACM $\acm$ is defined to be $\qin$ and the \emph{final} configuration is $\qfin$; note that each of these configurations consists of a single state and all registers are empty. 
We say that a term $u\in A_\acm$ is \emph{accepted in $\acm$} if $u\leq_\acm \ufin$ for some ID $\ufin$ consisting only of final configurations, i.e., $\ufin$ is a non-empty join of $\qfin$'s; such a $\ufin$ is called a \emph{final} ID and the set of accepted terms is denoted $\Acc(\acm)$. The machine $\acm$ is called \emph{terminating} if the initial configuration is accepted in $\acm$; i.e., $\qin\in \Acc(\acm)$. 

The following proposition can be found in \cite[Lemma~3.4(1)]{galatos2022}, but is also an immediate consequence of Lemma~\ref{lem:basicACMfacts}(3) and the fact that the accepted terms are the ones that compute a final ID.
\begin{proposition}
    For any ACM $\acm$, any accepted term must be an ID, i.e., $\Acc(\acm)\subseteq \indes{\acm}$.
\end{proposition}

\begin{example}
    Continuing with our Example~\ref{ex:acmleq} above, a complete description of $\Acc(\acm_\leq)\cap \conf{\acm_\leq}$ is given, for $x=\reg_1^{n_1}\reg_2^{n_2}$, by: 
    \begin{align*}
        \text{For $i=1,2$, }\zstate_i x\in \Acc(\acm_\leq)
        & \iff
        n_i=0 
        &;&&
        \qstate_2x\in \Acc(\acm_\leq)
        & \iff
        n_1 = 0 
        ;\\
        \qstate_1x\in \Acc(\acm_\leq)
        & \iff
        n_1 < n_2
        &;&&
        \qin x\in \Acc(\acm_\leq)
        & \iff 
        n_1\leq n_2
        .
    \end{align*}
    The complete description of $\Acc(\acm_\leq)$ can be garnered from the above, the fact that $\Acc(\acm_\leq)\subseteq \indes{\acm_\leq}$, and the forthcoming Lemma~\ref{lem:computationtree}(ii).
    Moreover, $\acm_\leq$ is terminating as $\qin = \qin\reg_1^0\reg_2^0 \in \Acc(\acm_\leq)$.
\end{example}

In \cite{galatos2022}, the acceptance problem for a given ACM $\acm =(\Reg,\State,\Inst)$ and a $u\in A_\acm$ is encoded in the theory of residuated lattices by the quasiequation $\qeacc_{\acm}(u)$: 
\begin{equation}\label{eq:acc(u)}\tag{$\qeacc_{\acm}(u)$}
    \amper \Inst_\mathrm{com} \implies {u}\leq \qfin ,
\end{equation}
where the antecedent is the finite conjunction of inequations $\Inst_\mathrm{com}:=\Inst \cup\{xy\leq yx: x,y\in \State\cup \Reg \}$, i.e., it consists of the instructions of $\acm$ and a finite set of \emph{commuting instructions}, which can be understood as enough ``commutativity'' to implement the instructions of $\Inst$.  
The presence of the commuting instructions should not be seen as an indication that the results hold only in the commutative setting; on the contrary, the results hold \emph{even} in the (more challenging) commutative setting, as well as the noncommutative setting.

Note that the quasiequation $\qeacc_{\acm}(u)$ is over the signature $\{\vee,\cdot,1\}$.
In \cite[\S 4]{galatos2022} it is shown that via the above encoding the computations of a machine $\acm$ can be carried out in the theory of residuated lattices, since their $\{\vee,\cdot,1\}$-reduct is an idempotent semiring, resulting in the following soundness result. This soundness result automatically holds for every subquasivariety of $\UniRLV$ (since quasiequations that hold in $\UniRLV$, also hold for its subquasivarieties).

\begin{lemma}[{\cite[Lemma~4.1]{galatos2022}}]\label{FramesForward}
If $\acm$ is an ACM and  $u\in \Acc(\acm)$ then $\UniRLV$ satisfies $\qeacc_\acm(u)$. 
\end{lemma}

The converse (completeness of the encoding) is guaranteed only for certain (quasi)varieties, however.
In particular, for every machine $\acm$ we will define/construct an algebra  $\mathbf{W}_{\acm}^{+}$ using residuated frames \cite{GalJip13}. 
The completeness of the encoding relative to a  sub(quasi)variety  $\quasivar$ of $\UniRLV$ is guaranteed when $\quasivar$ contains $\mathbf{W}_{\acm}^{+}$, because accepting computations in $\acm$ can be extracted from $\mathbf{W}_{\acm}^{+}$. 
In particular, in \cite[Theorem~4.5]{galatos2022} it is shown that the complexity of deciding quasiequations in any (quasi)variety of residuated lattices which contains $\mathbf{W}_{\acm}^{+}$ is at least as hard as deciding membership in $\Acc(\acm)$.  
This is done by focusing on quasiequations of the special form \ref{eq:acc(u)} seen above: they have a fixed antecedent $\Inst_\mathrm{com}$, corresponding to the instructions of a single machine associated to $\quasivar$, and the succedent varies depending on the input $u$ given to the machine. 
Quasiequations with fixed antecedent (corresponding to a presentation of an algebra with generators and relators) with varying succedent (corresponding to the equation involving generators, to be checked for validity in the presented algebra) are the ones constituting what is known as the \emph{(local) word problem} (of the given class of algebras). 
Lower bounds for the word problem imply lower bounds for the quasiequational theory, since if deciding quasiequations with fixed antecedent is hard then deciding arbitrary quasiequations is at least as hard.

The general scheme mentioned above is used to encode \emph{undecidable} problems into such theories, but for our purposes this will be too strong. 
Indeed, the quasiequational theory for knotted subvarieties of commutative residuated lattices is decidable, as a consequence of the FEP \cite{vanalten2005}, or, alternatively, as the results of Chapter~\ref{sec:fepP3} show. 
As such strong lower-bounds simply do not hold, in what follows we establish weaker (and correct for our situation) lower-bounds by encoding weaker complexity problems. 

Another interesting feature of our results is that in order to establish the hardness of arbitrary quasiequations (yielding the hardness of the quasiequational theory) we will not focus on quasiequations with fixed antecedent and varying succedent, mentioned above, but rather on quasiequations with a fixed succedent and a varying antecedent. 
As the antecedent corresponds to the instructions of a machine, this means that instead of a fixed machine we will be varying the machines. 
In more detail, this will be achieved by encoding the \emph{termination problem} for ACMs ($\qin \leq_{\acm} \uid_\FIN$, where $\uid_\FIN$ is a finite join of $\qfin$'s) into quasiequations of the form $\qeter(\acm):=\qeacc_{\acm}(\qin)$:
\begin{equation}\label{eq:termination}\tag{$\qeter(\acm)$}
  \amper \Inst_\mathrm{com} \implies {\qin}\leq \qfin 
\end{equation}
which range over the antecedent (i.e., over \emph{machines}) rather than the consequent; recall that $\Inst_\mathrm{com}$ depends on $\acm$, as it contains $\Inst$. 
We will prove that there is a class $\acmclass$ of ACMs, whose membership problem is \ACK-hard, and that for each (quasi)variety of residuated lattices containing all algebras $\mathbf{W}_{\acm}^+$, where $\acm \in \acmclass$, the complexity of deciding quasiequations  in this (quasi)variety is at least as hard as deciding membership in $\acmclass$, i.e., \ACK-hard.

To that aim, we will adapt the arguments and constructions of \cite{galatos2022} in order to handle (quasi)varieties of residuated lattices (and their corresponding logics) axiomatized by certain simple equations (rules, respectively); these will include in particular all commutative residuated lattice varieties (and $\UniFLeExtLogic{}$ extensions) axiomatized by a (non-integral) knotted equation (rule, respectively). 

\section{Simple equations and ACMs suitable for them}
Urquhart introduces the notion of \emph{expansive} (And-branching) counter machines in \cite{urquhart1999}, which are (And-branching) counter machines in which computations may contain ``glitches" that spontaneously increase the value of any nonempty register.
Allowing for such glitches results in computations that are not sound in arbitrary commutative residuated lattices, but they are sound in contractive CRLs (in the logic $\UniFLeExtLogic{\UniCProp}$), and conversely they incorporate enough of the contraction rule into the computation relation to guarantee that acceptance in the machine fully reflects the corresponding quasiequations of the theory of $\UniCRLV_{\mathsf{c}}$ (of the logic $\UniFLeExtLogic{\UniCProp}$). 
These glitches are formally given by \emph{expansive instructions}  of the form ``if in state $\mathtt{q}$ and the register $\reg{}$ is nonempty, add one token to $\reg{}$ and remain in state $\mathtt{q}$'', where $\mathtt{q}$ and $\reg{}$ range over all possible states and registers; %
as ACMs work non-deterministically (any applicable instruction could be applied, starting a branch of a new computation) these instructions can be applied spontaneously and any number of times. %
To represent these instructions in our particular formalism of ACMs, we define the \emph{$\mathsf{c}$-computation relation $\leq_{\mathsf{c}\acm}$} of an ACM $\acm$ to be the smallest $\UniSet{\lor,\cdot}$-compatible preorder containing $\Inst$ and every instance $w\leq w^2$, where $w\in (\State\cup\Reg)^*$. 
Recall that machines were defined to have a finite set of instructions, so adding all of the (infinitely many) glitches as instructions would not result in a machine. 
That we do not get a machine is a good thing, as it would not make sense to have such an infinite machine as input to an algorithm. 
However, it does make sense to ask for the existence of an algorithm that takes as input an ACM $\acm$ and decides whether a pair of IDs is in the $\mathsf{c}$-computation relation $\leq_{\mathsf{c}\acm}$ of $\acm$.
Therefore, and because it is tedious to refer to the computation of $\acm$ and the $\mathsf{c}$-computation  of $\acm$ for distinguishing them, we do allow ourselves to talk about the (infinitary) machine $\mathsf{c}\acm$ obtained from $\acm$, by using this expanded set of instructions, whose computation is naturally the $\mathsf{c}$-computation relation $\leq_{\mathsf{c}\acm}$ of $\acm$. 
We will do this with the understanding that when we  talk about the complexity of $\mathsf{c}\acm$, we will mean it in the way above (where $\acm$ is the input, not  $\mathsf{c}\acm$, and the question is about $\leq_{\mathsf{c}\acm}$). 
The above definition of the $\mathsf{c}$-computation relation $\leq_{\mathsf{c}\acm}$ allows even for state variables to be duplicated (against the intended meaning of expansive-instructions) but it has the benefit that it is compatible with multiplication (as it applies to all $w$'s). 
Moreover, even though the two relations differ in that respect, they coincide to the extent that a configuration $\cf$ is accepted by an expansive ACM $\acm$ if and only if $\cf$ is accepted in $\mathsf{c}\acm$ (i.e., $\cf\leq_{\mathsf{c}\acm} \ufin$ for some final ID $\ufin$). 
The forward direction of this claim follows from the fact that every expansive instruction can be obtained by applying $\reg{} \leq \reg{}^2$  to the appropriate register and then multiplying both sides by the rest of the configuration. 
The reverse direction follows from the fact that, while  instances of contraction $w \leq w^2$ may add state-variables in a computation (from left to right) to existing state-variables, there is no instruction in $\Inst$ nor an instance of contraction that can completely remove state variables (they can only convert one state to another). Thus, any $\mathsf{c}$-computation resulting in an ID must have started from an ID: $u\leq_{\mathsf{c\acm}} v\in \indes{\acm}$ implies $u\in \indes{\acm}$. 
Hence, instances of contraction, at least in computations between IDs, can be restricted to register-variables only (in \cite{galatos2022} this property is called \emph{state-admissibility}). Since this is the only difference between Urquhart's expansive version of a machine $\acm$ and our $\mathsf{c\acm}$, the results of \cite{urquhart1999} yield the following. 

\begin{proposition}\label{thm:expansiveterm}
There is a class $\acmclass$ of ACMs for which the following hold:
\begin{enumerate}
\item The termination problem for $\acmclass$ is \ACK-hard.
\item $\acmclass$ is \emph{$\mathsf{c}$-termination admissible}, i.e., $\mathsf{c}\acm$ terminates iff $\acm$ terminates, for each $\acm \in\acmclass$.
\end{enumerate}
\end{proposition}

Note that $\mathsf{c}$-termination admissibility means that the termination of the machine is unaffected by (it is impervious to) the presence of $\mathsf{c}$-glitches.
Urquhart obtains the lower bounds for the quasiequational theory of contractive CRLs by combining Proposition~\ref{thm:expansiveterm} above with  an argument that expansive ACMs represent exactly the inequalities of (the $\{\jn, \cdot, 1\}$-fragment of) contractive CRLs.
He proves this fact using a proof-theoretic analysis, but we will opt for an argument based on residuated frames. In other words, an alternative path to the one Urquhart takes is to prove that: 
(i) computations in expansive ACMs can be implemented by computations in $\mathsf{c}\acm$s (explained above); 
(ii) computations in $\mathsf{c}\acm$ can be implemented by $\{\jn, \cdot, 1\}$-inequalities in contractive CRLs (a result of the definition of computations); 
and (iii) that $\{\jn, \cdot, 1\}$-inequalities in contractive CRLs can be implemented in the computation of expansive ACMs (using residuated frames). 
This would close the circle and establish all implications as equivalences. 

We will follow this reasoning path for studying simple equations $\simpeq$ that are more general than contraction. 
In more detail, we will specify an extension $\leq_{\varepsilon \acm}$ of the computation relation of an ACM $\acm$ that will incorporate (a minimal amount of instances of) a given simple equation $\simpeq$ so as to ensure the algebras that we will construct via residuated frames  will be members of the variety $\UniCRLV+\simpeq$.
%}%end note

Given an ACM $\acm=(\Reg,\State,\Inst)$ and a simple equation $\simpeq: t_0\leq t_1 \jn \cdots \vee t_k$, we define  $\leq^\simpeq$ to be the smallest relation containing all evaluations $\sigma\simpeq$ of $\simpeq$ in the free commutative monoid $(\State\cup\Reg)^*$:  
\begin{equation}\label{eq:glitchrel}
\sigma \simpeq:\sigma t_0\leq \sigma t_1 \jn \cdots \vee \sigma t_k, \mbox{ for assignments }\sigma: \Xvar\to(\State\cup\Reg)^*,
\end{equation}
and closed under the inference rules $[\cdot]$ and $[\vee]$; this relation is the formalization of $\simpeq$-glitches.

Given a (possibly infinite) set $\eqset$ of simple equations, we denote by $\eqset\acm$ the (infinitary) ACM that is obtained from $\acm$ and whose computation relation $\leq_{ {\eqset\acm}}$ is the smallest compatible preorder containing ${\Inst}$ and $ {\leq}^\simpeq$, for each $\simpeq\in\eqset$; 
as explained above for the case of contraction, $\eqset\acm$ is an infinitary ACM, but when we consider algorithms about its computation relation the input will be the ACM $\acm$.  
As with instructions, a glitch instance $u\leq^\simpeq v$ holds iff $u=x\sigma t_0 \vee w$ and $v= \bigvee_{i=1}^k x\sigma t_i \vee w$, for some word $x\in (\State\cup\Reg)^*$, term $w\in A_\acm^\bot$, and assignment $\sigma$. It follows that $u\leq_{\eqset\acm} v$ holds iff there is a \emph{computation} in $\eqset\acm$ as in Equation~\eqref{eq:ACMcomps}, but where the labels $p_1,\ldots,p_n$ may include any $\simpeq\in \eqset$. 
We refer to (infinitary) machines of the form $\eqset\acm$, where $\acm$ is an ACM, as $\eqset$ACMs. 
Accepted terms in $\eqset\acm$ are defined similarly, where $u\in {\Acc}(\eqset{\acm})$ iff $u\leq_{ {\simpeq\acm}} \ufin$ for some final ID $\ufin$, and we say that $\eqset\acm$ is \emph{terminating} iff $\qin$ is accepted. 
We refer the reader to \cite[\S 5.1]{galatos2022} for a more detailed heuristic description of the computations in $\eqset\acm$,  and to \cite[\S 6]{galatos2022} for more details of the formal description.

As with Lemma~\ref{lem:basicACMfacts}, the following lemma follows easily from the definitions and the observations above. 
Item 3 can also be found in \cite[Lemma 6.1]{galatos2022}.

\begin{lemma}\label{lem:computationtree}
Let $\acm$ be an ACM and $\eqset$ a set of simple equations. Then for all $u,v,w\in A_\acm$:
\begin{enumerate}
\item $u\leq_{\eqset\acm} v$ iff there is a computation forest in which $u$ is a join of its root-labels and $v$ is a join of its leaf-labels.
\item\label{i2:computationtree}
$u\vee v\leq_{\eqset\acm} w$ iff $ u\leq_{\eqset\acm} w'$ and $v\leq_{\eqset\acm} w''$, for some $w',w''\in  A_{\acm}$ with $w=w'\vee w''$.
\item\label{lem:EqCompIDS} ${u}\vee {v}\in {\Acc}({\eqset{\acm}})$ iff ${u},v\in{\Acc}(\eqset{\acm})$.
\end{enumerate}
\end{lemma}

Let $\labeledforest$ be the computation forest of an $\eqset\acm$-computation 
in some ACM $\acm$ and $\tnode_1\prec \cdots\prec \tnode_k$ in $\labeledforest$; i.e., these nodes form a convex subchain of $\labeledforest$.
For each $1\leq j\leq k$, let $(c_j, i_j, p_j)$ be the label of $\tnode_j$.
The convex subchain is called \emph{decreasing} (resp, \emph{increasing}) if for all $1\leq j<k$, $p_j\in \eqset$ implies $c_{j}$ contains (resp., is contained in) $c_{j+1}$ as a subterm in $(\State\cup\Reg)^*$; i.e., up to commutativity of multiplication. 

\begin{example}
    Continuing Example~\ref{ex:acmleq}, we consider the simple equation $\mathsf{d}: x\leq x^2 \vee 1$ and the following $\mathsf{d}\acm_{\leq}$-computation forest:
$$
\scalebox{.68}{
\begin{tikzpicture}[node distance = .5cm, every node/.style={minimum height=0cm,minimum width=0cm,draw}]
%left tree
\node (0) {$\qin\reg_1\reg_2^2,0,\emptyset$};
\node[above left = of 0] (1l) {$\qin\reg_1\reg_2^3,1,\mathsf{d}$};
\node[above right = of 0] (1r) {$\qin\reg_1\reg_2,1,\mathsf{d}$};
\node[above = of 1l] (2) {$\qstate_1\reg_2^3,2,\Dec\reg_1$};
\node[above = of 2] (3r) {$\qstate_1\reg_2,3,\mathsf{d}$};
\node[ left = of 3r] (3l) {$\qstate_1\reg_2^5,3,\mathsf{d}$};
\node[above left = of 3l] (4) {$\qin\reg_2^4,4,\Dec\reg_2$};
\node[above left = of 3r] (5) {$\qin,5,\Dec\reg_2$};

\node[above =  of 1r] (6l) {$\qstate_2\reg_1\reg_2,6,\reg_1\ztr$};
\node[ right =  of 6l] (6r) {$\zstate_1\reg_1\reg_2,6,\reg_1\ztr$};
\node[above = of 6r] (9l) {$\zstate_1^2\reg_1^2\reg_2,9,\mathsf{d}$};
\node[ right = of 9l] (9r) {$\reg_2,9,\mathsf{d}$};
\node[above = of 9l] (10) {$\qfin\zstate_1\reg_1^2\reg_2,10,{\zinst}^1_\FIN$};
\node[right = of 10] (10') {$\qfin\zstate_1\reg_1^2\reg_2,10,{\zinst}^1_\FIN$};

\node[above = of 6l] (7) {$\qstate_2\reg_1,7,\reg_2{\to}\text{0}$};
\node[above = of 7] (8r) {$\zstate_2\reg_1,8,\reg_2\ztr$};
\node[left = of 8r] (8l) {$\qfin\reg_1,8,\reg_2\ztr$};
\draw (0)--(1l);
\draw (0)--(1r);

\draw (1l)--(2);
\draw (2)--(3l);
\draw (2)--(3r);
\draw (3l)--(4);
\draw (3r)--(5);
\draw (1r)--(6l);

\draw (1r)--(6r);
\draw (6r)--(9l);
\draw (6r)--(9r);
\draw (9l)--(10);
\draw (9l)--(10');
\draw (6l)--(7);
\draw (7)--(8l);
\draw (7)--(8r);
\end{tikzpicture}
}
$$
Here, the convex subchain ${\downarrow}(\qin\reg_2^4,4,\Dec\reg_2)$ is increasing; each of the convex subchains ${\downarrow}(\qfin\reg_1,8,\reg_2\ztr)$, ${\downarrow}(\zstate_2\reg_1,8,\reg_2\ztr)$, ${\downarrow}(\reg_2,9,\mathsf{d})$ are decreasing; finally the convex subchain ${\downarrow}(\qfin\zstate_1\reg_1^2\reg_2,10,{\zinst}^1_\FIN)$ is neither increasing nor decreasing. Note that the first coordinate in nodes $(\reg_2,9,\mathsf{d})$ and $(\qfin\zstate_1\reg_1^2\reg_2,10,{\zinst}^1_\FIN)$ are not configurations.
\end{example}

\begin{lemma}\label{lem:branchconfigurations}
For any decreasing or increasing convex subchain $\tnode_1\prec \cdots\prec \tnode_\ell$ of an $\eqset\acm$-computation forest, the following are equivalent.
\begin{enumerate}
\item The term-labels for the nodes $\tnode_1$ and $\tnode_\ell$ are configurations in $\acm$. \item For each $j\in \{1,\ldots, \ell\}$, the term-label of the node $\tnode_j$ is a configuration in $\acm$.
\end{enumerate}
\end{lemma}
\begin{proof}
Each instruction from $\Inst$ preserves the total number of states in each joinand, from left-to-right. Hence, by the definition of a {\joindecreasing} (resp., \joinincreasing) convex subchain, the total number of states can never increase (resp., decrease) going from left-to-right. 
Since configurations have precisely one state variable, the number of states in the labels must be invariant and equal to one. Therefore all term-labels are configurations in $\acm$.
\end{proof}

\section{$\eqset$ACMs and residuated frames: completeness of the encoding}
By Equation~\ref{eq:glitchrel} and Lemma~\ref{lem:computationtree}\eqref{lem:EqCompIDS}, we get that if $\eqset$ is a set of simple equations and $\acm$ an ACM, then for each $\simpeq: t_0\leq t_1 \jn \cdots \jn t_k$ in $\eqset$ the following implication (written vertically) holds:
\begin{equation}\label{eq:simpinfrule}
\infer[]
{u\cdot\sigma t_0 \in \Acc(\eqset\acm)}{
u\cdot\sigma t_1\in \Acc(\eqset\acm) & \cdots & u\cdot\sigma t_k\in \Acc(\eqset\acm)}
\end{equation}
for each $u\in A_\acm$ and assignment $\sigma:\Xvar\to (\State\cup\Reg_k)^*$. 

To establish the completeness of the encoding into ACMs, we use residuated frames (as defined in Section~\ref{s: RelSem}). %
For an ACM ${\acm}=(\Reg, {\State},{\Inst})$ and a set $\eqset$ of simple equations, we define the tuple $\mathbf{W}_{\eqset\acm}=(W_{\acm},W_{\acm}, {\Nuc}_{\eqset\acm}, \cdot, 1)$, where $W_{\acm}:=({\State}\cup \Reg)^*$ and $x~\Nuc_{\eqset\acm}~y$ iff $xy\in {\Acc}({\eqset\acm})$, for all $x,y\in W_{\acm}$. 
For ease of notation we will drop the subscripts from $W_\acm$, $\Nuc_{\eqset\acm}$, and $\mathbf{W}_{\eqset\acm}$, when they are clear from context.
It is easily seen that the relation $\Nuc_{\eqset\acm}$ is nuclear for $z \sslash y = yz$ (see \cite[Lemma~4.2]{galatos2022} for the details), so $\mathbf{W}_{\eqset\acm}$ is a commutative residuated frame, as defined in Section~\ref{s: RelSem}. 
Moreover, by Equation~\ref{eq:simpinfrule}, the following inference rule holds in $\mathbf{W}_{\eqset\acm}$:
$$
\infer[]
{t_0^{\mathbf{W}_{\acm}}(w_1,\ldots,w_\ell)~ \Nuc_{\eqset\acm} ~ u}{
t_1^{\mathbf{W}_{\acm}}(w_1,\ldots,w_\ell)~ \Nuc_{\eqset\acm} ~ u \quad  & \cdots & \quad t_k^{\mathbf{W}_{\acm}}(w_1,\ldots,w_\ell)~ \Nuc_{\eqset\acm} ~ u}
$$
for all $u,w_1,\ldots,w_\ell\in W_{\acm}$. 
By using general properties of residuated frames (see \cite[Theorem~3.10]{GalJip13}), we obtain the following lemma (see also \cite{galatos2022}).

\begin{lemma}[{\cite[Lemma~6.2]{galatos2022}}]\label{WmInclus}
If $\acm$ is an ACM and $\eqset$ is a set of simple equations, then $\mathbf{W}_{\eqset\acm}^+$ is a member of $\UniCRLV+\eqset$.
\end{lemma}

\begin{lemma}[cf. {{\cite[Lemma~{4.3}]{galatos2022}}}]\label{valuation}
Let ${\acm}=(\Reg,{\State},{\Inst})$ be an ACM, $\eqset$ a set of simple equations,  $e:{\State}\cup \Reg_k\to W_{\eqset\acm}^+$ the assignment given by $e(a):=\{a\}^{\triangleright\triangleleft}$, and $\bar e$ the homomorphic extension of $e$ from the term algebra over ${\State}\cup \Reg_k$ and the language of residuated lattices into $\mathbf{W}_{\eqset\acm}^+$. 
Then $\mathbf{W}_{\eqset\acm}^+,\bar e\models{\Inst_\mathrm{com}}$ and $\bar{e} (u) = \{w_1,\ldots, w_\ell\}^{\triangleright\triangleleft}$ for all $w_1,\ldots,w_\ell \in W_{\acm}$ with $u=_{\mathbf{A}_\acm} w_1\vee\cdots \vee w_\ell$. 
\end{lemma}

\begin{proof}
The lemma is proved in {{\cite[Lemma~{4.3}]{galatos2022}}} for the special case where $\eqset$ is a singleton. 
The general proof for arbitrary sets of simple equations is the obvious variant of that proof, where in the computation of the machine we allow instructions corresponding to arbitrary equations of  $\eqset$.
\end{proof}

The proof of the converse of Lemma~\ref{FramesForward} is a very small variation of \cite[Lemma~4.4]{galatos2022}.

\begin{lemma}\label{FramesBack} 
Let ${\acm}$ be an ACM, $\eqset$ a set simple equations, and $\quasivar$ a quasivariety of residuated lattices containing $\mathbf{W}_{\eqset\acm}^+$. For any ${u}\in A_{\acm}$, if $\quasivar$ satisfies $\qeacc_{\acm}({u})$ then $u$ is accepted in $\eqset\acm$. 
\end{lemma}
\begin{proof} 
Since $\quasivar$ satisfies the quasiequation $\qeacc_{\acm}({u})$ and $\mathbf{W}^+:=\mathbf{W}_{\eqset\acm}^+$ is a member of $\quasivar$, it follows that $\mathbf{W}^+$ also satisfies $\qeacc_{\acm}({u})$. 
By Lemma~\ref{valuation}, $\mathbf{W}^+,\bar e\models {\Inst_\mathrm{com}}$ and so $\mathbf{W}^+,\bar e \models {u}\leq \qfin$; i.e, $\bar{e} (u) \subseteq \bar{e}(\qfin)$. 
Since ${u}\in A_{\acm}$, there exist $w_1,\ldots,w_\ell\in (\State\cup\Reg)^*$ with ${u}=_{\mathbf{A}_\acm}{w}_1\vee\cdots\vee {w}_\ell$, so $\bar{e} ({w}_1\vee\cdots\vee {w}_\ell) \subseteq \bar{e}(\qfin)$. 
By Lemma~\ref{valuation}, we get $\{{w}_1, \ldots, {w}_\ell\}^{\triangleright\triangleleft} \subseteq \UniSet{\qfin}^{\triangleright\triangleleft}$, which is equivalent to ${\{\qfin\}^\triangleright} \subseteq \{{w}_1,\ldots,{w}_\ell \}^\triangleright$.  
Since $\leq_{\eqset\acm}$ is reflexive, $\qfin\in {\Acc}({\eqset\acm})$ and thus $\qfin\Nuc ~1,$ i.e., $1\in\{\qfin\}^\triangleright$. 
Therefore, $1\in \{{w}_1,\ldots,{w}_\ell \}^\triangleright$, i.e., $\{{w}_1,\ldots,{w}_\ell \} \Nuc  1$, so $w_1 \Nuc  1, \ldots ,w_\ell \Nuc  1$. 
Hence ${w}_1,\ldots,{w}_\ell\in{\Acc}({\eqset\acm})$ and, by Lemma~\ref{lem:computationtree}\eqref{lem:EqCompIDS}, we conclude $u\in{\Acc}({\eqset\acm})$.
\end{proof}

A class $\acmclass$ of machines is called \emph{$\eqset$-termination admissible} if, for each $\acm \in\acmclass$, we have:  $\eqset\acm$ terminates iff $\acm$ terminates. 
In other words, computations of machines in  $\acmclass$ are impervious to $\eqset$-glitches when termination is concerned.

\begin{theorem}\label{thm:qeLB}
If $\eqset$ a set of simple equations, $\acmclass$ is $\eqset$-termination admissible a class of ACMs, and  $\quasivar$ a quasivariety of residuated lattices that contains $\mathbf{W}^+_{\eqset\acm}$ for each $\acm \in \acmclass$, then the complexity of the quasiequational theory of $\quasivar$ is at least as hard as the termination problem for $\acmclass$. 
\end{theorem}

\begin{proof}
We first show that, for every machine $\acm\in\acmclass $, $\acm$ terminates iff $\quasivar$ satisfies $\qeter(\acm)$. 
The forward direction follows from Lemma~\ref{FramesForward} since $\quasivar$ is a subclass of $\UniRLV$. Conversely, if $\quasivar$ satisfies $\qeter(\acm)$, then by Lemma~\ref{FramesBack} it follows that $\eqset\acm$ terminates. 
Since $\acm$ is $\eqset$-termination admissible, $\acm$ terminates as well, establishing our claim. Thus, any decision procedure for determining the satisfaction of quasiequations in $\quasivar$ must also decide whether members of $\acmclass$ terminate. 
\end{proof}

By Lemma~\ref{WmInclus} and Theorem~\ref{thm:qeLB} we obtain the following result.

\begin{corollary}\label{cor:qeLB}
If $\eqset$ a set of simple equations, $\acmclass$ is $\eqset$-termination admissible a class of ACMs, and  $\quasivar$ a quasivariety of residuated lattices that contains $\UniCRLV+\eqset$, then the complexity of the quasiequational theory of $\quasivar$ is at least as hard as the termination problem for $\acmclass$. 
\end{corollary}

%% file: tex/lb-joinand-increasing.tex
We are ready to generalize Urquhart's results to a broader class of simple extensions, which includes all {\joinincreasing} simple equations, hence also all knotted-contraction rules.

\begin{theorem}\label{thm:branchincLB}
If a quasivariety of residuated lattices contains all contractive commutative residuated lattices, then its  quasiequational theory is \ACK-hard. 
In particular, this holds for any quasivariety of residuated lattices that contains $\UniCRLV+\eqset$ (i.e., $\UniCRLV+\eqset\subseteq \quasivar \subseteq \UniRLV$), for some set $\eqset$ of {\joinincreasing} simple equations (hence also for any quasivariety of residuated lattices that contains $\UniCRLV+\simpeq$, for some {\joinincreasing} $\simpeq$).
\end{theorem}
\begin{proof}
Since $\quasivar$ includes $\UniCRLV_\mathsf{c}$, it contains $\mathbf{W}^+_{\mathsf{c}\acm}$ for every ACM $\acm$. 
By applying Theorem~\ref{thm:qeLB} for the class $\acmclass$ provided in Proposition~\ref{thm:expansiveterm}, we get that deciding quasiequations in $\quasivar$ is \ACK-hard. 
The second claim follows from Proposition~\ref{prop:AlterEqs}(2). 
\end{proof}

\begin{corollary}\label{c:ackharJI_logic}
If $\mathbf{L}$ is any logic in the interval of axiomatic extensions between $\UniFLExtLogic{}$ and $\UniFLeExtLogic{\UniCProp}$ (for example $\mathbf{L}=\UniFLeExtLogic{\UniWeakCProp{m}{n}}$), then the deducibility of $\mathbf{L}$ is \ACK-hard. 
Moreover, this complexity lower bound is already witnessed in its $\{\lor,\cdot,1 \}$-fragment.
\end{corollary}

Theorem~\ref{thm:branchincLB} and Corollary~\ref{c: knotcontr+wc<Ack}(2) yield a  tight complexity bound for the quasiequational theory of many varieties of (pointed) residuated lattices.

\begin{corollary}\label{c: knotcontr+wc=Ack}
The complexity of the quasiequational theory is exactly Ackermannian (i.e, $\UniFGHProbOneAppLevel{\omega}$) for every variety of (pointed) residuated lattices axiomatized by any set of $\mathcal{N}_2$ equations (in particular, by any set of $\{\jn, \cdot, 1\}$-equations) that (a) contains a knotted-contraction equation and a weak-commutativity equation  and (b) every equation in the set is a consequence of the conjunction of contraction and commutativity.

In particular, this holds for every variety axiomatized by a knotted contraction equation and a non-empty set of weak commutativity equations.
\end{corollary}

\begin{corollary}\label{c: knotcontr+wc=Ack Logics}
The complexity of deducibility is exactly Ackermannian (i.e, $\UniFGHProbOneAppLevel{\omega}$) for every extension of $\m {FL}$ axiomatized by any set of $\mathcal{N}_2$ equations that (a) contains a knotted contraction formula and a weak commutativity formula and (b) every formula  in the set is a consequence of the conjunction of contraction and exchange.

In particular, this holds for every extension axiomatized by a knotted contraction formula and a non-empty set of weak commutativity formulas.
\end{corollary}

In Chapter~\ref{s: joinand-decreasing}, we will prove an extension of Proposition~\ref{thm:expansiveterm} (also providing an alternative proof of Proposition~\ref{thm:expansiveterm}, as our proof is not based on Urquhart's paper) that will allow us to
establish an analogous theorem for {\joindecreasing} simple equations, as well.

\section{Complexity of the equational theory; i.e., of provability.}
\label{sec:ded-theorem-provability}

Here we show how the above results on the quasiequational theory/deducibility can be exported to lower bounds for the equational theory/provability, by making use of a deduction theorem that holds for certain logics.

A variety of commutative residuated lattices is called \emph{negatively $k$-potent}, where $k \in \mathbb{Z}^+$, if it satisfies the identity $(\UniPropAlgA\land 1)^{k+1}= (\UniPropAlgA\land 1)^k$; it is called \emph{negatively potent} if it is negatively $k$-potent for some $k$. 
We note that the seemingly more general condition of $(\UniPropAlgA\land 1)^{l}= (\UniPropAlgA\land 1)^k$, for positive integers $k< l$, is actually equivalent to negative $k$-potency.
Indeed, negative $k$-potency follows since $(\UniPropAlgA\land 1)^{l}\leq (\UniPropAlgA\land 1)^{l-1}\leq \cdots\leq(\UniPropAlgA\land 1)^{k+1} \leq (\UniPropAlgA\land 1)^k$ always holds, and thus the assumption $(\UniPropAlgA\land 1)^{k}= (\UniPropAlgA\land 1)^l$ collapses each inequality to an equality. Meanwhile,
that negative $k$-potency implies $(\UniPropAlgA\land 1)^{l}= (\UniPropAlgA\land 1)^k$ is immediate by transitivity since $(\UniPropAlgA\land 1)^{k+1} = (\UniPropAlgA\land 1)^k$ implies $(\UniPropAlgA\land 1)^{n+k+1} = (\UniPropAlgA\land 1)^{n+k}$ for every $n\geq 0$, so $(\UniPropAlgA\land 1)^{l}=(\UniPropAlgA\land 1)^{l-1}= \cdots  = (\UniPropAlgA\land 1)^k$. 
It turns out that such varieties enjoy a deduction theorem.

\begin{lemma}[{\cite[Lem.~5.6]{galatos2022}}]\label{ExpDedTh}
Given a  negatively $k$-potent subvariety  $\mathcal{V}$ of $\UniFLeV$, for each quasiequation $q$ there is an equation $\varepsilon_k(q)$ such that: $\mathcal{V}$ satisfies $q$ iff it satisfies  $\varepsilon_k(q)$. 
\end{lemma}

For completeness, we mention that, in terms of the corresponding logics, if negative $k$-potency is provable in an extension $\m L$ of $\m {FL_e}$, then $\m L$ has a deduction theorem of the following concrete form: for all sets of formulas $\Gamma \cup\{\phi\}$, we have 
\begin{center}
   $\Gamma \vdash_{\m L} \phi$ {\sf iff} there exist $\gamma_1, \ldots, \gamma_\ell \in \Gamma$ with $\vdash_{\m L} ((\gamma_1 \mt 1) \cdots (\gamma_\ell \mt 1))^k \rightarrow \phi$. 
\end{center}
Equivalently: for all sets of formulas $\Gamma \cup\{\phi\}$, we have 
\begin{center}
   $\Gamma \vdash_{\m L} \phi$ {\sf iff} there exist $\gamma_1, \ldots, \gamma_\ell \in \Gamma$ with $\vdash_{\m L} (\gamma_1 \mt  \cdots \mt \gamma_\ell \mt 1)^k \rightarrow \phi$. 
\end{center}

A sufficient condition (expressed in the $\{\vee,\cdot,1 \}$-reduct even) for  negative potency is any identity of the form
\begin{equation}\label{eq:eqexpansive}x^k\leq x^{k+c_1} \lor\cdots\lor x^{k+c_n}
\tag{$\mathsf{c}[k; k+c_1, \ldots, k+c_n]$}
\end{equation}
where $k, c_1,\ldots,c_n$ are positive integers; we denote this equation by $\mathsf{c}[k; k+c_1, \ldots, k+c_n]$ and call it \emph{multi-contraction}.
That it implies negative $k$-potency is seen by applying the substitution $x\mapsto x\land 1$ and using the fact that $\bigvee_{i\leq n}(x\land 1)^{k+p_i}\leq (x \land 1)^{k+1}\leq (x \land 1)^k$ holds in $\UniRLV$.  
A simple equation $\simpeq$ is called \emph{{\multicontractible}} if it has a multi-contraction equation \ref{eq:eqexpansive} as a consequence. 
Note that every multi-contraction equation is joinand-increasing, but this is not true for {\multicontractible} equations (for example, $xy\leq x^3 \vee y^3$ is {\multicontractible}, as it entails  $x^2\leq x^3$; the latter is joinand-increasing but the former is not). 
Also, every $(n,m)$-contraction is a multi-contraction equation. 
Therefore, a deduction theorem of the form above holds for every extension of $\m {FL}$ that contains $(n,m)$-contraction and exchange/commutativity.

\medskip

The proof of Lemma~\ref{ExpDedTh} makes use of commutativity. 
We will now show how to weaken the commutativity assumption, while still ensuring we have a \emph{local deduction theorem} (a weaker version of a deduction theorem), which in the presence of a {\multicontractible} equation yields a deduction theorem. 
A weak commutativity equation (see Section~\ref{sec:wck-substructural-logics}) is called \emph{initial} if it is of the form  
$xy_1x\cdots xy_{s}x=x^{a_0}y_1 \cdots x^{a_{s}}y_{s}$, for some $s \in \mathbb{Z}^+$, i.e., the last coordinate $a_{s+1}$ of the vector $\vec{a}$ is zero. 
Likewise, a \emph{final} weak commutativity equation is one of the form $xy_1x\cdots xy_{s}x=y_1 \cdots x^{a_{s}}y_{s}x^{a_s}$, for some $s \in \mathbb{Z}^+$, i.e., the first coordinate $a_{0}$ of the vector $\vec{a}$ is zero. 
We will show that  a pair of an initial and of a final weak commutativity equation can replace the need for commutativity in Lemma~\ref{ExpDedTh}, allowing also for the possibility of a single equation that is both initial and final.
For example $xyxzx=x^2yxz$, $xyxzx=x^3yz$, $xyxzx=xyx^2z$ and $xyxzx=yx^3z$  are initial weak commutativity equations, while the last one is a final weak commutativity equation, as well.

In fact, we will show that a weaker form of commutativity, dubbed `subcommutativity', will be enough for our purposes.
Given $q \in \mathbb{Z}^+$, a variety of residuated lattices is called \emph{$q$-subcommutative} \cite{GOR2008} if it satisfies $(x \mt 1)^qy=y(x \mt 1)^q$; it is called \emph{subcommutative} if it is $q$-subcommutative for some positive integer $q$. 
We will also make use of the stronger condition of \emph{$q$-commutativity}: $x^qy=yx^q$. 

Recall that every equation $s=t$ is equivalent to $1 \leq (s \ld t) \mt (t \ld s)$ and to $1 = 1 \mt [ (s \ld t) \mt (t \ld s)]$, over the theory of residuated lattices. 
In view of that, in the next lemma we prove the (local) deduction theorem for equations of that form. 
After that, in Theorem~\ref{thm:subcomDedThm}, we prove the result for equations of the general form $s=t$ making direct use of that lemma. 
Also, note that the implications below are not restricted to finite antecedents (i.e., they are not restricted to quasiequations) so as to make the connection to the logical form of the deduction theorem more direct. 

\begin{lemma}\label{lem:subcommQE}
Let $S \cup \{s_0\}$ be a  set of terms, $\mathcal{V}$ a subcommutative variety of residuated lattices and $\mathcal{W}$ a subcommutative and  negatively $k$-potent variety. 
\begin{enumerate}
\item\label{item:subcommQE1} 
In $\mathcal{V}$ we have: $\bigwedge_{s \in S} (s=1) \Rightarrow (s_0=1)$ iff there exist $r, p \in \mathbb{N}$ and $s_1, \ldots, s_r \in S$ with $(s_1\mt \cdots \mt s_r \mt 1)^p \leq s_0$ [iff  there exist $p \in \mathbb{N}$ with $(1 \mt \bigwedge S)^p \leq s_0$ (when $S$ is finite)].
\item\label{item:subcommQE2} 
In $\mathcal{W}$ we have: $\bigwedge_{s \in S} (s=1) \Rightarrow (s_0=1)$ iff there exists $r\in \mathbb{N}$ and $s_1, \ldots, s_r \in S$ with $(s_1\mt \cdots \mt s_r \mt 1)^k \leq s_0$ [iff $(1 \mt \bigwedge S)^k \leq s_0$ (when $S$ is finite)].
\item\label{item:subcommQE3} 
The conjunction of the two equations $x^by=x^{b-c}yx^c$ and $yx^d=x^eyx^{d-e}$, where $b\geq c$ and $d \geq e$ are elements of $\mathbb{Z}^+$, implies $q$-commutativity for some $q$. %}
\item\label{item:subcommQE4} 
The conjunction of any initial and any final weak commutativity equation implies $q$-commutativity for some $q$.
\end{enumerate}
\end{lemma}

\begin{proof}
(1) For all elements $x$ and $a$ of a $q$-subcommutative residuated lattice $\m A$, where $x \leq 1$, we have $ax^q=x^qa\leq xa$ and $x^qa=ax^q \leq ax$, so $x^q \leq a \ld xa$ and $x^q \leq ax \rd a$; since $x \leq 1$ we also have $x^q \leq 1$, thus  $x^q \leq  a \ld xa \land 1=\lambda_a(x)$ and $x^q \leq ax \rd a \land 1=\rho_a(x)$ (see Section~\ref{s: P3} for the definitions related to conjugates $\lambda_a(x)$ and $\rho_a(x)$). 
Therefore, if $a_1, \ldots, a_r \in A$ and $\delta_i\in \{\lambda_{a_i}, \rho_{a_i}\}$, for all $i \in \{1, \ldots, r\}$, then we get a lower bound for the corresponding iterated conjugate for any $x\leq 1$: 
    \begin{align*}
    x^{(q^r)} &= x^{(q^{r-1}){q}} 
    \leq \delta_r(x^{(q^{r-1})})
    \leq \ldots  
    \leq \delta_r\circ\cdots \circ \delta_3 ((x^q)^q) 
    \\
    &\leq \delta_r\circ\cdots \circ \delta_3 \circ \delta_2 (x^q)
    \leq \delta_r\circ\cdots \circ \delta_3 \circ \delta_2  \circ \delta_1(x).
    \end{align*}
Therefore, for every $x \leq 1$ and iterated conjugate $\gamma$, there exists $r \in \mathbb{N}$ such that $x^r \leq \gamma(x)$; also, conversely, for every $x \leq 1$ and $r \in \mathbb{N}$ there is an iterated conjugate $\gamma(x)=a \ld x1 \mt 1= x \mt 1=x$ with $\gamma(x) \leq x$. 
    
By general results in \cite[Theorem~3.47]{GalJipKowOno07} (or as explicitly argued in \cite[Section~5.3]{galatos2022}), we have that the implication $\bigwedge_{s \in S} (s=1) \Rightarrow s_0=1$ holds in $\mathcal{V}$ iff $s_0$ is above some product of iterated conjugates of some elements of $S$ in the free algebra of $\mathcal{V}$. 
By the above reasoning, iterated conjugates can be replaced by powers, so this is in turn equivalent to:  $s_0$ is above some product of powers of some elements of $S_{\mt 1}:=\{s \mt 1 : s \in S\}$, which is equivalent to:  $s_0$ is above a product of some elements of $S_{\mt 1}$, i.e., to: there exist $r \in \mathbb{N}$ and (not-necessarily distinct) $s_1, \ldots, s_r \in S$ with $(s_1 \mt 1)\cdots (s_r \mt 1) \leq s_0$. 
This is also equivalent to: there exist $r, p \in \mathbb{N}$ and $s_1, \ldots, s_r \in S$ with $(s_1\mt \cdots \mt s_r \mt 1)^p \leq s_0$.

(2) In the presence of negative $k$-potency, the statement $(\exists  p \in \mathbb{N})((s_1\mt \cdots \mt s_r \mt 1)^p \leq s)$ is equivalent to $(s_1\mt \cdots \mt s_r \mt 1)^k \leq s$, since for each negative element $a$ (i.e., $a \leq 1$), we have $a^p=a^k$, for all $p\geq k$.

(3)  We consider $x^by=x^{b-c}yx^c$ and $yx^d=x^eyx^{d-e}$, where $b\geq c$ and $d\geq e$ are elements of $\mathbb{Z}^+$. 
From the first equation, we get that $x^{b'}y=x^{b'-b}x^by=x^{b'-b}x^{b-c}yx^c=x^{b'-c}yx^c$, for all $b'\geq b$, and from the second equation, we get that $yx^{d'}=yx^{d}x^{d'-d}=x^eyx^{d-e}x^{d'-d}=x^eyx^{d'-e}$, for all $d' \geq d$, which are more general facts. 
Therefore, for $i, j \in \mathbb{Z}^+$ with $b-c \leq je$ and $d-e \leq ic$ (for example we can take the values $j:=\lceil \frac{b-c}{e}\rceil$ and $i:=\lceil \frac{d-e}{c}\rceil$), we have $je+c \geq b$ and $ic+e \geq d$, so the above two general facts apply and by repeated instantiations of them we obtain
    \begin{align*}
        x^{je+ic}y&
        =x^{je+(i-1)c}yx^c
        =x^{je+(i-2)c}yx^{2c}
        =\ldots
        =x^{je+c}yx^{(i-1)c}
        %=x^{je}yx^{ic}=
        =x^{je}yx^{ic}\\
        &=x^{(j-1)e}yx^{ic+e}
        =\ldots
        =x^{2e}yx^{ic+(j-2)e}
        =x^{e}yx^{ic+(j-1)e}
        =yx^{ic+je}
    \end{align*}
As a result, we obtain $q$-commutativity for $q=ic + je$.
    
(4) Setting $y_q=y$ and $y_1= \ldots =y_{q-1}=1$ in the initial weak commutativity equation $xy_1x\cdots xy_{q}x=x^{a_0}y_1 \cdots x^{a_{q}}y_{q}$ yields $x^{q+1}y=x^qyx$. 
Likewise, from the final weak commutativity equation we get $yx^{q+1}=xyx^q$. 
These two equations are of the form of item (3), hence they imply $q$-commutativity.
\end{proof}

We obtain the following deduction theorem for certain subcommutative varieties of residuated lattices.
\begin{theorem}\label{thm:subcomDedThm}
    If $\mathcal{V}$ is a subcommutative and negatively potent variety of residuated lattices, then the equational theory for $\mathcal{V}$ is just as hard as its quasiequational theory.
\end{theorem}
\begin{proof}
Recall that any quasiequation can be equivalently written, modulo the theory of residuated lattices, to one of the form $\bigwedge_{s \in S} (s=1) \Rightarrow (s_0=1)$ for finite $S$. Since $\mathcal{V}$ is negatively $k$-potent for some $k\geq 1$, Lemma~\ref{lem:subcommQE}\eqref{item:subcommQE2} yields that the satisfaction of such a quasiequation in $\mathcal{V}$ is equivalent to the satisfaction of the equation $(1\land \bigwedge S)^k \leq s_0$. 
So any procedure for deciding equations satisfied in $\mathcal{V}$  also decides the satisfiability of quasiequations.
\end{proof}

So, as we mention in the results below, if in a variety of residuated lattices  a multi-contraction equation  and  two weak commutativity equations---one initial and one final---hold then we obtain a deduction theorem and  the Ackermann hardness of the quasiequational theory of the variety is also transferred to the equational theory. Alternatively, we can have any single weak commutativity equation together with two equations as in item 3 above.
Alternatively, we can have a single weak commutativity equation that is both initial and final, such as $xyxzxwx=yx^2zx^2w$, or $xyxzxwx=yx^3zxw$, or $xyxzxwx=yzx^4w$.

Recall from Definition~\ref{def:TypesOfSimpEq} that a $\{\vee,\cdot,1\}$-equation is called {\joinincreasing} or {\joindecreasing}, if the simple equation obtained via the linearization procedure in Lemma~\ref{l: N_2^-0} is {\joinincreasing} or {\joindecreasing}, respectively.  
As we will see in the following lemma, the types of equations just defined above are all both {\joinincreasing} and {\joindecreasing}. 
For example, we consider the $2$-commutativity equation $x^2y = yx^2$, which is equivalent to the conjunction of $x^2y\leq yx^2$ and $yx^2\leq x^2y$. 
If $\sigma$ denotes the substitution where $x\mapsto u\vee v$ and $y\mapsto w$, then $\sigma[x^2y\leq yx^2]$ is
    $$
     uvw\leq u^2w\vee uvw \vee vuw \vee v^2 w=\sigma(x^2y)\leq \sigma(yx^2) = wu^2\vee wuv \vee wvu \vee wv^2 
    $$
and similarly, for $\sigma':x\mapsto u'\vee v'$ and $\sigma' :y\mapsto w'$, the equation $\sigma'[yx^2\leq yx^2]$ is 
    $$ w'u'v' \leq u'^2w'\vee u'v'w' \vee v'u'w' \vee v'^2 w' .$$
Note that in both cases each of the simple equations obtained has a linear joinand on the right-hand side containing all the variables (resp., $wuv$ and $u'v'w'$) of the left-hand side. 
In view of Definition~\ref{def:TypesOfSimpEq}, each inequation satisfies the property of being both {\joinincreasing} and {\joindecreasing}. 
We also mention that the standard way of obtaining a single simple equation from two simple equations (first disjointify the variables and then multiply the equations) preserves the properties of being {\joinincreasing}/{\joindecreasing}.
Therefore $2$-commutativity is {\joinincreasing} and {\joindecreasing}.

\begin{lemma}\label{lemma:wkcommJincJdec}
Every  $\{\vee,\cdot,1\}$-(in)equation where every variable occurs the same number of times on the two sides of the equality is {\joinincreasing} and {\joindecreasing}. 
In particular, every (initial or final) weak commutativity, as well as every $q$-commutativity equation satisfies the property of being both {\joinincreasing} and {\joindecreasing}. 
Similarly, this holds for both equations $x^by=x^{b-c}yx^c$ and $yx^d=x^eyx^{d-e}$, where $b\geq c$ and $d \geq e$.
\end{lemma}
\begin{proof}
Note that, in each of these equations, every variable occurs the same number of times on either side of the equality. 
Therefore, the simple equation obtained via the linearization process of Lemma~\ref{l: N_2^-0} produces a joinand on the right-hand side of the equation that contains every variable precisely once. 
Therefore it satisfies both properties in Definition~\ref{def:TypesOfSimpEq}. 
\end{proof}

By the Lemma~\ref{ExpDedTh} and Theorem~\ref{thm:branchincLB}, we obtain the following result.

\begin{corollary}\label{cor:ackhardEQtheory}
If a variety of residuated lattices  is subcommutative, negatively potent, and contains all contractive CRLs, then its equational theory is \ACK-hard.

In particular, deciding equations in a variety of any of the following types is \ACK-hard:
\begin{enumerate}
    \item A knotted-contractive commutative variety $\UniCRLV + (x^n\leq x^m)$, where $0<n<m$.    
    \item $\UniRLV +\eqset$, where $\eqset$ is a set of {\joinincreasing} $\{\vee,\cdot,1\}$-(in)equations containing at least one {\multicontractible} equation and satisfying an initial and a final weak commutativity equation.
    \item $\UniRLV +\eqset$, where $\eqset$ is a set of {\joinincreasing} $\{\vee,\cdot,1\}$-(in)equations containing at least one {\multicontractible} equation and satisfying the equations $x^by=x^{b-c}yx^c$ and $yx^d=x^eyx^{d-e}$, for some $b\geq c$ and $d\geq e$.
\end{enumerate}
\end{corollary}
\begin{proof}  
Since the variety is subcommutative and negatively potent, its equational theory is as hard as its quasiequational theory by Theorem~\ref{thm:subcomDedThm}. 
By Theorem~\ref{thm:branchincLB} its quasiequational theory is \ACK-hard, since the variety contains $\UniCRLV_\mathsf{c}$.
Therefore, the equational theory of the variety is \ACK-hard.

For the remaining claims, note that knotted-contractive equations $x^n\leq x^m$, where $0<n<m$, are {\joinincreasing} equations by Proposition~\ref{prop:AlterEqs}(\ref{altereqs1}). 
Also, since $\eqset$ consists \emph{only} of {\joinincreasing} equations, by Proposition~\ref{prop:AlterEqs}(\ref{altereqs2}) the variety contains $\UniCRLV_\mathsf{c}$. Since $\eqset$ is a \emph{non-empty} set of {\joinincreasing} equations, in all three cases the variety contains $\UniCRLV_\mathsf{c}$.

For (1) note that the knotted-contractive equation $x^n\leq x^m$, where $0<n<m$, implies negative $n$-potency, while subcommutativity follows by commutativity.

For the remaining two claims, subcommutativity follows from Lemma~\ref{lem:subcommQE}(\ref{item:subcommQE3},\ref{item:subcommQE4}). 
Finally,  $\eqset$ contains a {\multicontractible} equation which, by definition, has a multi-contraction equation $x^k\leq x^{k+c_1} \lor\cdots\lor x^{k+c_\ell}$ as a substitution instance; this, in turn, implies negative $k$-potency, as we have already proved. 
\end{proof}

Corollary~\ref{cor:ackhardEQtheory} and Lemma~\ref{l: alg_tanslation} yield the following result for the logics corresponding to the varieties of the corollary.

\begin{corollary}\label{cor:ackhardEQtheoryforLogics}
If an axiomatic extension of $\m {FL}$ is subcommutative, negatively potent, and is contained in 
$\UniFLeExtLogic{\UniCProp}$, then its provability is \ACK-hard.

In particular, deciding provability in a logic of any of the following types is \ACK-hard:
\begin{enumerate}
    \item $\UniFLeExtLogic{\UniWeakCProp{m}{n}}$, where $0<n<m$.   
    \item $\UniFLeExtLogic{}+\Sigma$, where $\Sigma$ is a set of {\joinincreasing} $\{\vee,\cdot,1\}$-formulas 
    containing at least one {\multicontractible} formula and satisfying an initial and a final weak commutativity formula. 
    \item $\UniFLeExtLogic{}+\Sigma$, where $\Sigma$ is a set of {\joinincreasing} $\{\vee,\cdot,1\}$-formulas containing at least one {\multicontractible} formula and satisfying the formulas corresponding to the equations $x^by=x^{b-c}yx^c$ and $yx^d=x^eyx^{d-e}$,  for some $b\geq c$ and $d\geq e$.
\end{enumerate}
\end{corollary}

As every knotted contraction equation implies negative potency, together Corollary~\ref{c: knotcontr+wc<Ack}(2) and Corollary~\ref{cor:ackhardEQtheory} yield a tight complexity bound for the equational theory of many varieties of (pointed) residuated lattices.
We stress that subcommutativity is enough for Corollary~\ref{cor:ackhardEQtheory}, but it is not an $\mathcal{N}_2$ equation so Corollary~\ref{c: knotcontr+wc<Ack}(2) does not apply to it. 
Nevertheless, the stronger condition of $q$-commutativity (for any $q$) is  an $\mathcal{N}_2$ equation, so both results apply in this context. 
Also, in condition (a) below, weak commutativity is needed for the upper complexity bound ($q$ commutativity is not enough for that result) and $q$-commutativity is needed for the deduction theorem (weak commutativity is not enough), so we include both conditions. 
The final sentence of the corollary explains a situation where both conditions hold.

\begin{corollary}\label{c: knotcontr-eq-theory+wc=Ack}
The complexity of the equational theory is exactly Ackermannian (i.e., $\UniFGHProbOneAppLevel{\omega}$) for every variety of (pointed) residuated lattices axiomatized by any finite set of $\mathcal{N}_2$ equations (in particular, by any set of $\{\jn, \cdot, 1\}$-equations) that (a) entails a knotted contraction equation, a weak commutativity equation and a $q$-commutativity equation, and (b) every equation in the set is a consequence of the conjunction of contraction and commutativity.

In particular, this holds for every variety axiomatized by any knotted contraction equation and any set of weak commutativity equations that contains an initial and a final weak commutativity equation.
\end{corollary}
\begin{proof}
Note that condition (b) below is a restatement of the demand that the variety contains the variety of contractive CRLs, which is needed to invoke Corollary~\ref{cor:ackhardEQtheory}. 

For the last part of the statement, we note that every weak commutativity or $q$-commutativity equation is a consequence of  commutativity, and any contractive knotted (or more generally, {\joinincreasing}) equation is a consequence of contraction and commutativity, by Proposition~\ref{prop:AlterEqs}\eqref{altereqs2}.
\end{proof}

Similarly, we obtain tight bounds for the corresponding logics via Corollary~\ref{c: FLknotcontr+wc<Ack} and Corollary~\ref{cor:ackhardEQtheoryforLogics}.

\begin{corollary}\label{c: knotcontr+wc=Ack Logic}
The complexity of provability is exactly Ackermannian (i.e., $\UniFGHProbOneAppLevel{\omega}$) for every extension of $\m {FL}$ axiomatized by any finite set of $\mathcal{N}_2$ formulas that (a) entails a knotted contraction formula, a weak commutativity formula and a $q$-commutativity formula, and (b) every formula in the set is a consequence of the conjunction of contraction and exchange.

In particular, this holds for every extension axiomatized by any a knotted contraction formula, a weak commutativity formula and a $q$-commutativity formula, or by a knotted contraction formula and any set of weak commutativity formulas that contains an initial and a final weak commutativity formula.
\end{corollary}

\section{Alternative argument via a proof-theoretic reduction}
We observe that the class of machines used to prove that deducibility in $\UniFLeExtLogic{\UniCProp{}}$ is \ACK-hard was used to prove the same result for joinand-increasing extensions of $\UniFLeExtLogic{}$.
Below we show that this has a proof-theoretic counterpart: provability in $\UniFLeExtLogic{\UniCProp{}}$ reduces to deducibility in $\UniFLExtLogic{}+\UniAnaRuleSet$ for any finite set $\UniAnaRuleSet$ of hypersequent rules derivable in $\UniFLeExtLogic{\UniCProp{}}$; this includes in particular the sequent rules corresponding to {\joinincreasing} equations.
A key observation here is that $\UniFLeExtLogic{\UniCProp{}}$ already satisfies cut elimination (provided we use the \emph{generalized mix rule} instead of the usual cut rule; see~\cite[p.~217]{GalJipKowOno07}).

\begin{theorem}\label{t: PTreduction}
Let $\UniAnaRuleSet$ be a set of hypersequent rules that are derivable in the cut-free hypersequent calculus $\UniFLeExtHCalc{\UniCProp}$ having 
$\frac{H \VL \UniSequent{\UniMSetFmA,\UniFmC,\UniFmC, \UniMSetFmB}{\UniFmD}}{H \VL \UniSequent{\UniMSetFmA,\UniFmC, \UniMSetFmB}{\UniFmD}}$ (c)
as the contraction rule and 
$\frac{H \VL \UniSequent{\UniMSetFmA,\UniFmB,\UniFmC, \UniMSetFmB}{\UniFmD}}{H \VL \UniSequent{\UniMSetFmA,\UniFmC,\UniFmB, \UniMSetFmB}{\UniFmD}}$ (e)
as the exchange rule.
Then, for all formulas $\UniFmA$, we have
    $$
    \vdash_{\UniFLeExtHCalc{\UniCProp}} \UniFmA
    \text{ iff }
    \mathcal{T}(\UniFmA) \vdash_{\UniFLExtHCalc{}+\UniAnaRuleSet}\UniFmA,
    $$
where 
$\mathcal{T}(\UniFmA) \UniSymbDef 
    \{ \UniFmB \ld (\UniFmB \fus \UniFmB) \mid \UniFmB \in \mathsf{subf}(\UniFmA) \}
    \cup
    \{ (\UniFmB \cdot \UniFmC) \ld (\UniFmC \fus \UniFmB) \mid \UniFmB,\UniFmC \in \mathsf{subf}(\UniFmA) \}$.
\end{theorem}
\begin{proof}
In the right-to-left direction, we assume we have a deduction witnessing $\mathcal{T}(\UniFmA) \vdash_{\UniFLExtHCalc{}+\UniAnaRuleSet} \UniFmA$ and work to get rid of applications of rules in $\UniAnaRuleSet$ and of the leaves in $\mathcal{T}(\UniFmA)$ with the help of contraction, thus obtaining a proof in $\UniFLeExtHCalc{\UniCProp}$ of the same formula $\UniFmA$.
The first task is easy from the assumption that the rules in $\UniAnaRuleSet$ are derivable in $\UniFLeExtHCalc{\UniCProp}$. 
The leaves, in turn, can be easily derived with the help of contraction and exchange:
    \begin{center}
        \AxiomC{$\UniSequent{\UniFmB}{\UniFmB}$}
        \AxiomC{$\UniSequent{\UniFmB}{\UniFmB}$}
        \BinaryInfC{$\UniSequent{\UniFmB,\UniFmB}{\UniFmB \fus \UniFmB}$}
        \RightLabel{(c)}
        \UnaryInfC{$\UniSequent{\UniFmB}{\UniFmB \fus \UniFmB}$}
        \UnaryInfC{$\UniSequent{}{\UniFmB \ld (\UniFmB \fus \UniFmB)}$}
        \DisplayProof
        \qquad
        \AxiomC{$\UniSequent{\UniFmC}{\UniFmC}$}
        \AxiomC{$\UniSequent{\UniFmB}{\UniFmB}$}
        \BinaryInfC{$\UniSequent{\UniFmC, \UniFmB}{\UniFmC \fus \UniFmB}$}
        \RightLabel{(e)}
        \UnaryInfC{$\UniSequent{\UniFmB, \UniFmC}{\UniFmC \fus \UniFmB}$}
        \UnaryInfC{$\UniSequent{\UniFmB \cdot \UniFmC}{\UniFmC \fus \UniFmB}$}
        \UnaryInfC{$\UniSequent{}{(\UniFmB \cdot \UniFmC) \ld (\UniFmC \fus \UniFmB)}$}
        \DisplayProof
    \end{center}

For the left-to-right direction, note first that by assumption we have a cut-free proof of $\UniFmA$, therefore in this proof only subformulas of $\UniFmA$ appear (in particular, only subformulas of $\UniFmA$ appear in applications of contraction). 
Because the only important differences between $\UniFLeExtHCalc{\UniCProp}$ and $\UniFLExtHCalc{}+\UniAnaRuleSet$ (for this direction) is the presence of contraction and exchange in the former, we need to show that we can simulate the contraction and exchange applications inside $\UniFLExtHCalc{}+\UniAnaRuleSet$, and for that we use the axioms in $\mathcal{T}(\UniFmA)$.
The proof is by structural induction.
Note that all rules except contraction and exchange are present in $\UniFLExtHCalc{}+\UniAnaRuleSet$, so there is nothing to be done in their corresponding cases; for contraction, the situation looks like the following (where $\mathcal{D}$ is a derivation of the premise sequent):
    \begin{center}
        \AxiomC{$\mathcal{D}$}
        \noLine
        \UnaryInfC{$g \VL \UniSequent{\UniMSetFmA,\UniFmC,\UniFmC, \UniMSetFmB}{\UniFmD}$}
        \RightLabel{(c)}
        \UnaryInfC{$g \VL \UniSequent{\UniMSetFmA,\UniFmC, \UniMSetFmB}{\UniFmD}$}
        \DisplayProof
    \end{center}

From the (IH) applied to the premise, we get that $g \VL \UniSequent{\UniMSetFmA,\UniFmC,\UniFmC, \UniMSetFmB}{\UniFmD}$ is deducible in $\UniFLExtHCalc{}+\UniAnaRuleSet$, from which we get $g \VL \UniSequent{\UniMSetFmA,\UniFmC \cdot \UniFmC, \UniMSetFmB}{\UniFmD}$ by the left rule for $\fus$. 
Then, by a cut with $\UniSequent{\UniFmC}{\UniFmC \cdot \UniFmC}$ (see a deduction of it below) we get $g \VL \UniSequent{\UniMSetFmA, \UniFmC, \UniMSetFmB}{\UniFmD}$ as desired.

    \begin{center}
        \AxiomC{$\UniSequent{}{\UniFmC \ld \UniFmC \fus \UniFmC}$}
        \AxiomC{$\UniSequent{\UniFmC}{\UniFmC}$}
        \AxiomC{$\UniSequent{\UniFmC}{\UniFmC}$}
        \AxiomC{$\UniSequent{\UniFmC}{\UniFmC}$}
        \BinaryInfC{$\UniSequent{\UniFmC , \UniFmC}{\UniFmC \cdot \UniFmC}$}
        \UnaryInfC{$\UniSequent{\UniFmC \cdot \UniFmC}{\UniFmC \cdot \UniFmC}$}
        \BinaryInfC{$\UniSequent{\UniFmC,\UniFmC \ld \UniFmC \fus \UniFmC}{\UniFmC\fus\UniFmC}$}
        \BinaryInfC{$\UniSequent{\UniFmC}{\UniFmC \cdot \UniFmC}$}
        \DisplayProof
    \end{center}
The case of exchange is similar and left to the reader.\qedhere
\end{proof}
    
\begin{remark}\label{rem:proof-the-lb-sequents}
The above result can be straightforwardly proved for the corresponding sequent systems (instead of hypersequent systems) when the extension by $\UniAnaRuleSet$ preserves cut-freeness.
\end{remark}

\begin{corollary}
Let $\UniAnaRuleSet$ be a finite set of analytic structural sequent rules corresponding to linearizations of joinand-increasing equations and weak commutativity equations.
Then deducibility in  $\UniFLExtSCalc{}+\UniAnaRuleSet$ is \ACK-hard.
In particular, deducibility in $\UniFLExtLogic{\UniWEProp{\vec a}\UniWeakCProp{m}{n}}$ is \ACK-hard.
\end{corollary}
\begin{proof}
Note first that any analytic structural sequent rule corresponding to the linearization of a joinand-increasing equation is derivable from 
$\frac{ \UniSequent{\UniMSetFmA,\UniFmC,\UniFmC, \UniMSetFmB}{\UniFmD}}{ \UniSequent{\UniMSetFmA,\UniFmC, \UniMSetFmB}{\UniFmD}}$ (c)
in presence of exchange.
In fact, consider an application of $\mathsf r \in \UniAnaRuleSet$.
Because it corresponds to a simple joinand-increasing equation, it must have a premise of the form 
$\UniSequent{\UniMSetFmA,\UniMSetFmB_1^{m_1},\ldots,\UniMSetFmB_k^{m_k}}{\UniFmC}$, 
where $m_i \geq 1$ for each $1 \leq i \leq k$. 
We argue that with (c) we can get from this premise to
$\UniSequent{\UniMSetFmA,\UniMSetFmB_1,\ldots,\UniMSetFmB_k}{\UniFmC}$.
Take a $\UniMSetFmB_i$ with $m_i > 1$ instantiated with the multiset $\UniFmA_1,\ldots,\UniFmA_r$. 
Then, since we have exchange, we can see $\UniMSetFmB_i^{m_i}$ as the multiset $\UniFmA_1^{m_i},\ldots,\UniFmA_r^{m_i}$, and we can clearly, using (c), get to $\UniMSetFmB_i$ from the latter. 
Apply this reasoning to each $1 \leq i \leq k$, and we are done. 
In the case of the rules corresponding to weak commutativity equations, it is clear that having the power (due to (e)) to move any formula around in the antecedent of a sequent is enough to derive any weak commutativity rule. 
Therefore, in view of Remark~\ref{rem:proof-the-lb-sequents}, the set $\UniAnaRuleSet$ satisfies the hypothesis of Theorem~\ref{t: PTreduction}, so deciding deducibility in $\UniFLExtSCalc{}+\UniAnaRuleSet$ is at least as hard as deciding provability in  $\UniFLeExtSCalc{\UniCProp}$, which is \ACK-hard.
\end{proof}

In the following section, we will obtain Ackermannian lower bounds also for logics of the form $\UniFLExtLogic{\UniWEProp{\vec a}\UniWeakWProp{m}{n}}$.
Could this have been obtained directly using the above approach, using $\UniFLeExtLogic{\UniWProp}$ instead as the source of hardness?
We could adapt the above proof-theoretic argument and use $\UniFLeExtLogic{\UniWProp}$ as a master problem but the result we would obtain is not sharp enough: provability and deducibility in $\UniFLeExtLogic{\UniWProp}$ is known to be primitive recursive (\PSPACE~\cite{horcik2011} and \TOWER~\cite{tanaka2022}, respectively), far from the Ackermannian lower bound that we will obtain.

%% file: tex/lb-joinand-decreasing.tex
We have established lower bounds for logics based on joinand-increasing rules using Proposition~\ref{thm:expansiveterm}. We will now provide a generalization of Proposition~\ref{thm:expansiveterm} that will also apply to {\joindecreasing} equations and use it to obtain complexity lower bounds for the latter. 
The proof that we will provide (and which will also serve as an alternative proof of Proposition~\ref{thm:expansiveterm}) will proceed via a reduction of \emph{Counter Machines with zero tests} (CM), also known as Minsky Machines or \emph{Vector Addition Systems with full zero tests} (VASS${}_0$), to ACMs (i.e., AVASS). 
We describe this reduction in this section.

A CM $\acm$ is the same as an ACM except forking instructions are not allowed and \emph{zero-test} instructions are allowed instead; the latter allow the machine to check whether a register is empty (has value $0$) or not, and update the state if emptiness is verified. %
The computation relation $\to_\acm$ of $\acm$ is the smallest preorder on $\mathbf{A}_\acm$ that contains 
the restrictions $\to^p$ of $\leq^p$ to multiplicative terms, where $p$ is an increment or decrement instruction of $\acm$, as well as all ordered pairs the form $\qstate x\leq \qstate'x$ where $x\in (\Reg\setminus\UniSet{\reg})^*$, for each \emph{zero-test $\reg$ instruction} $\qstate [\reg\ztr]\qstate'$ (if in state $\qstate$ and register $\reg$ is zero, goto state $\qstate'$). %
As with ACMs, computations in CMs correspond to computation forests; in fact these forests are unions of chains (actually a single chain when the computation starts with a single configuration), as there are no branching instructions.  
Unfortunately, the relation $\to_\acm$ is not compatible with multiplication (specifically with multiplication by register variables), and this is the reason why we do not use the symbol $\leq$ for it, which we reserve for compatible relations; if we would force it to be compatible then the zero test would lose its intended functionality of testing for absence of register tokens. 
As with ACMs, by a $\kreg$-CM we mean a CM with $\kreg$-many registers. 

The acceptance problem for CMs is undecidable, even when restricted to CM's with only two counters/registers, and this is the basis of the reduction in \cite{lincoln1992}, as well as in \cite{horcik2015,galatos2022}. 
The weaker problem that Urquhart utilizes has \ACK-hard complexity (even when restricting to the CMs that have three counters/registers, cf. \cite[Theorem~5.1]{urquhart1999}) and is called \emph{ACP} 
(Ackermann-bounded version of the CM halting problem):  
For a fixed non-primitive recursive function $A\in \UniFGHLevel{\omega}$, we ask
\begin{equation}\label{ACP}\tag{ACP}
\parbox{\dimexpr\linewidth-4em}{
    Does a given $3$-CM have a terminating computation in which the total register contents of any configuration in the computation do not exceed the value $A(n)$, where $n$ is the number of states of the machine?
    }
\end{equation} 
In \cite[Theorem~3.1]{FMR68}, Fischer, Meyer, and Rosenberg establish a (primitive recursive) space-reduction from Turing machines to CMs, based on an algorithm due to Fischer \cite{FISCHER66} for simulating a Turing machine by a $3$-CM. 
Using these results, Urquhart provides a (primitive recursive) space-reduction from the complexity class $\UniFGHProbOneAppLevel{\omega}$ to the ACP. 
While his reduction references a particular non-primitive recursive $A\in \UniFGHLevel{\omega}$, such a stipulation is immaterial for the non-primitive recursive problems in $\UniFGHProbOneAppLevel{\omega}$. 
Indeed, all such problems are reducible to each other in primitive recursive time (see Section~\ref{sec:subrec-hierarchies}), where the distinction between time and space disappear. 
Therefore, we have the following theorem for any non-primitive recursive member of $\UniFGHLevel{\omega}$. 
\begin{theorem}[{\cite[Theorem~5.1]{urquhart1999}}]\label{t: ACP is Ack-hard}
The \ref{ACP} is \ACK-hard.    
\end{theorem}

Since the computation relation $\to_\acm$ of CMs is not compatible with multiplication, it is not directly amenable to an algebraic formulation and, therefore, this causes a problem in having direct algebraic access to the \ref{ACP}. 
In \cite{lincoln1992} it is shown that the effect of zero-test instructions can actually be
simulated by certain forking instructions, if we are willing to restrict our attention to acceptance of configurations only---a small price to pay, given that we are only interested in acceptance of configurations anyway. 
In essence \cite{lincoln1992} marks the genesis of ACMs, which we have seen admit an algebraic description since their computation relations are compatible with multiplication and join. 
We will review this simulation procedure following \cite[\S7.1]{galatos2022}, as it will be necessary for what follows.

Let $\acm=(\Reg_\kreg,\State,\Inst\cup\Inst_0)$ be a $\kreg$-CM, where $\Inst$ is a set consisting of only increasing and decreasing instructions and $\Inst_0$ consists only of zero tests. 
We will construct an ACM $\acm_\lor$ from $\acm$ such that a configuration of $\acm$ is accepted in $\acm_\lor$ iff it is accepted in $\acm$. 
We first define the auxiliary ACM ${{\zinst}_\kreg}=(\Reg_\kreg,{\State}_{\zinst}\cup\{\qfin\}, {\Inst}_{\zinst})$, where ${\State}_{\zinst} = \{\zstate_1,\ldots,\zstate_\kreg\}$ is a fresh set of variables, $\qfin$ is the final state of $\acm$, and the instructions ${\Inst}_{\zinst}$ are given by: 
    $$
    \begin{array}{ l c r c l }
    {{\zinst}^i_j}&:&\zstate_i \reg_j&\leq& \zstate_i \\
    {{\zinst}^i_\FIN}&:& \zstate_i&\leq& \qfin\vee \qfin ,
    \end{array}
    $$
for each $i, j\leq \kreg$ with $i \not = j$, yielding a total of $\kreg^2$-many instructions. 
Then we define $\Inst_\lor$ to be the set consisting of the forking instructions $p_\lor:\qstate\leq \qstate'\vee \zstate_i$, for each zero-test instruction $p: \qstate[\reg_{i}\ztr] \qstate'$ in $\Inst_0$. 

The new ACM is defined to be $\acm_\lor:=(\Reg_\kreg,\State\cup{\State}_{\zinst},\Inst\cup{\Inst}_{\zinst}\cup\Inst_\lor)$. 
The proposition below establishes the simulation of $\acm$ in $\acm_\lor$, at least when restricting to computations between configurations in $\conf{\acm}$ (viewed as a subset of $\conf{\acm_\lor}$).  
The argument is essentially the one found in \cite[Lemma~3.4]{lincoln1992} but, due to our (slightly) different semantics for machines, we make precise the statements and arguments in our setting as they will be useful later on.

The \emph{size} of an (A)CM $\acm=(\Reg,\State,\Inst)$ is defined to be $|\Reg|+|\State|+|\Inst|$ and we will  denote it by $|\acm|$.

\begin{proposition}\label{prop:zerotest}
If $\acm$ is a $\kreg$-CM, $\cf,\cf'\in \conf{\acm}$ and $\zstate_ix\in\conf{{\zinst}_\kreg}$, then:
\begin{enumerate}
\item A configuration $\zstate_ix$ is accepted in ${\zinst}_\kreg$ iff $x$ contains no occurrence of $\reg_i$ \cite[Lemma~7.1]{galatos2022}. 
\item For each zero-test $p$, $\cf\to^p \cf'$ in $\acm$ iff $\cf\leq^{p_\lor} \cf' \lor \mathtt{Z}$ for some $\mathtt{Z}\in \Acc({\zinst}_\kreg)$ \cite[Lemma~7.2]{galatos2022}. 
\item $\cf$ is accepted in $\acm$ iff $\cf$ is accepted in $\acm_\lor$ \cite[Lemma~3.4]{lincoln1992}.
\end{enumerate}
 Moreover, $|\acm_\vee|=|\acm| + \kreg+\kreg^2$.
\end{proposition}

\begin{proof} 
Let $\acm=(\Reg_\kreg,\State,\Inst\cup \Inst_0)$ be a $\kreg$-CM, where $\Inst_0$ is the set of the zero test instructions. 
By construction, the size of $\acm_\vee$ is $|\Reg_\kreg|+|\State|+|\Inst|+|\Inst_\vee|+|\State_{\o}| +|\Inst_{\o}|=|\acm| + \kreg +\kreg^2$. 
The proof of (1) can be found in \cite[Lemma~7.1]{galatos2022}, but it is also an obvious consequence from the description of the ACM ${\zinst}_\kreg$. 
Similarly, item (2) can be found in \cite[Lemma~7.2]{galatos2022}, but it is also immediate from the description of ${\zinst}_\kreg$.
    
For item (3), suppose $\cf$ is accepted in $\acm$, i.e., $\cf\to_\acm \qfin$. 
The associated computation forest is actually a chain $\tnode_0\prec \cdots\prec \tnode_\complength$. 
From this chain we construct a tree by adding a fresh node $\tnode_i'$ for each instance $\cf_{i-1}\leq^{p} \cf_i$ of a zero test $p\in \Inst_0$, witnessed by the labeling of $\tnode_{i-1}\prec \tnode_i$, and add precisely one edge so that $\tnode'_i$ covers $\tnode_{i-1}$.  
We label this new node by $(\zf_i,i,p_\vee)$, where $\zf_i$ is the configuration obtained by replacing the state $\qstate'\in \State$ in $\cf_i$ by the  corresponding state $\zstate\in \State_{\o}$ (determined by $p$), while its sibling $\tnode_i$ is relabeled as $(\cf_i,i,p_\vee)$ (simply replacing $p$ with $p_\vee$); all other node labels remain unchanged. 
It is easy to see that this labeled tree is a computation forest for $\acm_\vee$ witnessing $\cf\leq_{\acm_\vee} \qfin \vee \bigvee_{i\in I} \zf_i$, where $I$ is the set of all indices where $p_i$ is a zero test. 
By item (2), it follows that $\zf_i\in \Acc(\zinst_\kreg)\subseteq \Acc(\acm_\vee)$, for all $i \in I$, so $\cf$ is accepted in $\acm_\vee$ by Lemma~\ref{lem:computationtree}(2).

Conversely, suppose $\cf$ is accepted in $\acm_\vee$ and let $\labeledforest$ be a computation forest witnessing this fact. 
As $\cf$ is a configuration, each node of the forest must be labeled by configurations (see Lemma~\ref{lem:basicACMfacts}\eqref{i3:IDs2IDs}), since its root $\tnode_0$ is labeled by a configuration, $\cf\in \conf{\acm}$, and its leaves are also labeled by the configuration $\qfin$.
We will specify a maximal subchain that naturally admits a relabeling witnessing the acceptance of $\cf$ in $\acm$.

We first define the \emph{next node} relation on $\labeledforest$, as follows. Recall that, given a non-leaf node $\tnode$ of $\labeledforest$, its covers have a common instruction label among them.  
We will consider cases on what that instruction is, so as to define which of these covers will be \emph{next nodes} of $\tnode$. 

If the instruction is in $\Inst$, then $\tnode$ has a single node, which we define to be the \emph{next node} of $\tnode$. 
If the instruction is in $\Inst_{\zinst}$, no node is a next node for $\tnode$. 
If the instruction is of the form $p_\vee$, where $p\in \Inst_0$ (a zero-test $\reg_j$-instruction), then $p_\lor:\qstate\leq \qstate'\vee \zstate_i$ and $\tnode$ has two covers corresponding to $\qstate'$ and to $\zstate_i$; we define the \emph{next node} of $\tnode$ to be the one corresponding to $\qstate'$.

Note that the next-node relation is a subrelation of the the covering relation of $\labeledforest$.  
Also, by the above definition, if $\tnode$ is a next-node (of some node) then the first coordinate of its label does not include any state variable of $\State_{\zinst}$, so its cover nodes (if any) do not have instructions in $\Inst_{\zinst}$. 
Therefore, if $\tnode$ is a next-node and it has covers (i.e., it is not a leaf), then it itself has a next node. 
Finally, since the state of $\tnode_0$ is not in $\State_{\zinst}$, we have that $\tnode_0$ has a next node, unless it is a leaf.
Therefore, the connected component of $\tnode_0$ under the reflexive and transitive closure of the next-node relation forms a subchain of $\labeledforest$ and its leaf is a leaf of $\labeledforest$; also all of the monoid-word labels are configurations in $\acm$. 

We will now argue this subchain witnesses a terminating computation in $\acm$.
We assume that $\tnode$, with monoid word label $c$, has a next node $\tnode'$, with monoid word label $c'$, and consider cases for the instruction of $\tnode'$. 
If the instruction $p$ is in $\Inst$, then $c \to^p c'$ is a valid computation step of $\acm$. 
If the instruction is of the form $p_\lor:\qstate\leq \qstate'\vee \zstate_i$, where $p\in \Inst_0$, then since every leaf above $\tnode$ has label $\qfin$, Lemma~\ref{lem:computationtree}(2) and item (2) imply that $c \to^p c'$ is a valid computation step of $\acm$. 
Therefore by replacing each instruction $p_\vee$ by $p\in \Inst_0$, and re-indexing, we obtain a computation (labeled chain) in $\acm$ witnessing the acceptance of $\cf$. 
\end{proof}
 
\begin{example}
The machine $\acm_\leq$ from Example~\ref{ex:acmleq} is precisely the machine $\acm_\vee$, where $\acm$ is the counter machine represented by the following state-flow diagram:
    \begin{equation*}
    \scalebox{.8}{
    \begin{tikzpicture}[node distance = 2cm]
    \node[state] (qin) {$~~\qin ~~$};
    \node[state, left = of qin] (q1) {$\qstate_1$};
    \node[state, right = of qin] (q2) {$\qstate_2$};
    \node[state, right = of q2] (qf) {$~~\qfin~~$};
    \draw[-{Latex[width=3mm]}, >=Latex]
    (qin) edge[bend right,below] node{$\Dec\reg_1$} (q1)
    (q1) edge[bend right, below] node{$\Dec\reg_2$} (qin)
    (qin) edge[above] node{$\reg_1\ztr$} (q2)
    (q2) edge[above] node{$\reg_2\ztr$} (qf)
    (q2) to[loop below, above right] node{~$\reg_2\rst$} (q2)
    ;
    \end{tikzpicture}
    }
    \end{equation*}
Here the labeled edges correspond to the following instructions in $\acm$: 
the edge with label $\Dec\reg_i$ denotes the same instruction $\Dec\reg_i$ in \ref{acmleq};
the edge labeled $\reg_2\rst$ corresponds to the instruction $\reg_2{\to}\text{0}$ and the edge labeled $\reg_i\ztr$ denotes the zero-test $p:\qstate[\reg_{i}\ztr] \qstate'$, so $p_\vee$ is the instruction $\reg_i\ztr$ in $\acm_\leq$.
\end{example}

\section{In the presence of $\eqset$-glitches}
In fact, testing whether a certain register is $0$ in this way can be implemented in $\eqset$ACMs as well, for any set $\eqset$ of simple equations that does not entail integrality. 
An ACM $\acm$ is called \emph{$\eqset$-admissible} if: for all $\cf\in \conf{\acm}$, $\cf$ is accepted in $\eqset\acm$ iff it is accepted in $\acm$. 
Note that this notion is stricter than asking that $\acm$ is $\eqset$-termination admissible.

\begin{lemma}\label{lem:nonintZT}
For every set $\eqset$ of simple equations and $\kreg \in \mathbb{Z}^+$, the ACM ${\zinst}_\kreg$ is $\eqset$-admissible iff $\eqset$ does not contain the integrality equation. 
\end{lemma}

\begin{proof}
If $\eqset$ contains integrality, $\simpeq: x \leq 1$, then by evaluating $x$ to $\reg_1$ we get
$$\zstate_1\reg_1\leq^{\simpeq} \zstate_1  \leq^{{\zinst}^1_\FIN} \qfin\vee\qfin \in \Acc({{\zinst}_\kreg})$$
so $\zstate_1\reg_1$ is accepted in $\eqset{\zinst}_\kreg$, but it is not accepted in ${\zinst}_\kreg$.

Conversely, suppose ${\zinst}_\kreg$ is not $\eqset$-admissible. 
As $\Acc({{\zinst}_\kreg})\subseteq \Acc(\eqset{{\zinst}_\kreg})$ by definition,  there is a configuration $\zf\in  \Acc(\eqset{{\zinst}_\kreg})\setminus \Acc({{\zinst}_\kreg})$. 
By Proposition~\ref{prop:zerotest}, $\zf=\zstate\reg_i x$ for some $\zstate\in\{\zstate_i,\qfin \}$ and $x\in\Reg_{\kreg}^*$. 
Let $\labeledforest=( F,\preceq,\ell)$ be the computation forest of some $\eqset{\zinst}_d$-computation witnessing the acceptance of $\zf$. 
Note that no instruction from $\Inst_{\zinst}$, nor instance of an equation from $\eqset$ (as it is simple), can introduce a variable distinct from those present in $x$ and the set $\{\zstate_i,\qfin \}$. 
So no member of any node from $\labeledforest$ has an occurrence of a symbol aside from these.  
Now, as $\reg_i$ does not appear in any final ID, and since the poset $F$ is finite, there must exist a non-leaf node $\tnode \in F$ with the property that the variable $\reg_i$ occurs in the monoid term label of $\tnode$ but it does not occur in  the monoid term of any cover of $\tnode$.
Let $p$ be the shared instruction appearing in the labels of the covers of $\tnode$, and $c_0\leq^p c_1\vee\cdots \vee c_k$ the corresponding instance of $p$. 
Note that $p$ is not a member of $\Inst_{\zinst}$, as no instruction with state $\zstate_i$ or $\qfin$ removes $\reg_i$. So $p=\sigma \simpeq$ for some simple equation $\simpeq: t_0\leq t_1\vee\cdots\vee t_k$ in $\eqset$. 
Therefore, there is a variable, say $x_1$, in $t_0$ such that $\reg_i$ occurs in $\sigma(x_1)$. As $\reg_i$ does not occur in any of $c_1,\ldots,c_k$, the variable $x_1$ cannot occur in any of $t_1,\ldots,t_k$. Therefore $\simpeq$ is integrality, by the definition of which equations are simple.
\end{proof}

\section{Lossy Counter Machines}
Our complexity reduction for {\joindecreasing} equations will be done relative to \emph{Lossy Counter Machines}.
More generally, given a set $\eqset$ of simple equations, $\eqset$CMs are defined by analogy to $\eqset$ACMs: 
Given a CM $\acm$, we define the infinitary CM $\eqset\acm$ to be the one obtained from $\acm$ by adding as instructions all instances of $\eqset$, so the computation relation $\to_{\eqset\acm}$ is the smallest preorder closed under $\to_\acm$ and $\leq^\eqset$. 
An $\eqset$CM is by definition an infinitary CM of the form $\eqset\acm$,  where $\acm$ is an CM.

In  particular, if $\acm$ is an CM and $\simpeq$ is an (in)equation, then $\{ \simpeq\}\acm$ is an $\{\simpeq\}$ACM; we will use the simplified notation $\simpeq\acm$. 
Of particular interest are the equations $\mathsf{i}: x\leq 1$ and $\mathsf{t}: 1\leq x$, where the former identity is \emph{integrality} and the latter is often referred to as \emph{triviality}---as a residuated lattice is trivial (i.e., it consists of a single element) iff it satisfies $\mathsf{t}$.
In order to match this with the terminology in the literature, given a CM  $\acm$, we also define the infinitary CM $\acm_\mathrm{lossy}$ whose computation relation differs from $\mathsf{i}\acm$ in that the compatibility is defined with respect to instances of $\leq^\eqset$ involving only register variables and no state variables; likewise we define the machine $\acm_\mathrm{inc}$ as a restricted variant of $\mathsf{t}\acm$. 
Machines of the form $\acm_\mathrm{lossy}$ and $\acm_\mathrm{inc}$ are called
\emph{lossy} and \emph{increasing}, respectively, and instead of $\to_{\acm_\mathrm{lossy}}$ and $\to_{\acm_\mathrm{inc}}$, we write $\lossycm$ and $\incm$.

The behavior of increasing CMs (variables can appear out of nowhere) is more radical than that of expansive CMs (variables can be duplicated), as glitches of the latter can be implemented by glitches of the former, but not the other way.

We note that  by inverting the set of instructions of any CM $\acm=(\Reg,\State,\Inst)$, we obtain a new CM $\acm^{-1}=(\Reg,\State,\Inst^{-1})$, where $(a,b)\in \Inst^{-1}$ iff $(b,a)\in \Inst$. 
As a result, the acceptance/termination problem for lossy CMs is intimately linked to the one of increasing CMs, as described below.

\begin{proposition}[{\cite[Exercise~3.3]{schmitz2012notes}}]\label{prop:invertedCM}
If $\acm=(\Reg,\State,\Inst)$ is a CM, then the following hold for all configurations $\cf,\cf'\in \conf{\acm}$:
\begin{enumerate}
\item $\cf\to_{\acm} \cf'$ iff $\cf'\to_{\acm^{-1}} \cf$;  
\item $\cf\to^\mathrm{lossy}_{\acm} \cf'$ iff $\cf'\to^\mathrm{inc}_{\acm^{-1}} \cf$. 
\end{enumerate}
Consequently, the complexities of the acceptance/termination problem for lossy and for increasing CMs coincide.
\end{proposition}

The following lemma shows that the differences between the two pairs of notions $\acm_\mathrm{lossy}$/$\mathsf{i}\acm$ and $\acm_\mathrm{inc}$/$\mathsf{t}\acm$ is immaterial for our purposes.

\begin{lemma}\label{lem:IntTrivACM}
If $\acm$ is a CM, then $\acm_\mathrm{lossy}$ terminates iff $\mathsf{i}\acm$ terminates. 
Similarly, $\acm_\mathrm{inc}$  terminates iff $\mathsf{t}\acm$ terminates. 
Consequently, the termination problem for lossy/increasing CMs coincides with that of $\mathsf{i}$CMs/$\mathsf{t}$CMs. 
The same result holds for ACMs. 
\end{lemma}

\begin{proof}
Clearly the forward direction of each claim holds as the computation relation for $\acm_{\mathrm{lossy}}$ (resp., $\acm_{\mathrm{inc}}$) is just a restriction of that for $\mathsf{i}\acm$ (resp., $\mathsf{t}\acm$). 
On the other hand, any computation in $\mathsf{i}\acm$ (resp, $\mathsf{t}\acm$) is a decreasing (resp., increasing)  convex subchain in the associated computation tree. 
By Lemma~\ref{lem:branchconfigurations}, every word appearing in the computation must be a configuration. So any instance of $\to^\mathsf{i}$ (resp, $\to^\mathsf{t}$) must be obtained by an assignment $\sigma$ in $\Reg^*$, which is a lossy (resp., an increasing) instance by definition. 
\end{proof}

The next lemma shows how an $\mathsf{i}$CM/$\mathsf{t}$CM computation can be extracted from an $\eqset$ACM computation, for certain sets of equations $\eqset$. 
It is the primary ingredient in the proof of Theorem~\ref{thm:lossyadmissmach}; the statement, and its proof, is analogous to the right-to-left implication of Proposition~\ref{prop:zerotest}(3).

\begin{lemma}\label{lem:branchinglem}
If $\acm$ is a CM, $\eqset$ is a set of non-integral {\joindecreasing}/increasing simple equations and $\eqset\acm_\lor$ terminates, then $\mathsf{i}\acm$/$\mathsf{t}\acm$ terminates. 
\end{lemma}

\begin{proof}
Let $\acm = (\Reg,\State,\Inst\cup\Inst_0)$ be a CM and recall that $\acm_\vee:=(\Reg,\State\cup\State_{\zinst},\Inst\cup \Inst_\vee\cup \Inst_{\zinst})$ is the ACM simulating $\acm$. 
Since $\eqset\acm_\vee$ terminates, there exists a witnessing computation $\comp: u_0\leq^{p_1} \cdots\leq ^{p_\complength} u_\complength$, i.e.,  $\qin=u_0$ and $u_\complength = \uid_\FIN$ for some final ID $\uid_\FIN$. 
Let $\labeledforest=(F, \preceq, \ell)$ be the computation forest of $\comp$; in particular, $F$ has a single root $\tnode_0$ with $\ell(\tnode_0) = (\qin,0,\emptyset)$, hence it is a tree.
Since $\comp$ is a terminating computation, the first coordinate of every leaf of $\labeledforest$ is $\qfin$. 

We first define the \emph{next node} relation on $\labeledforest$, as follows. 
Recall that given a non-leaf node $\tnode$ of $F$ its covers have a common instruction label among them. 
We will consider cases on what that instruction is, so as to define which of these covers will be \emph{next nodes} of $\tnode$. 

If the instruction is in $\Inst$, then $\tnode$ has a single node, which we define to be the \emph{next node} of $\tnode$. 
If the instruction is in $\Inst_{\zinst}$, no node is a next node for $\tnode$. 
If the instruction is of the form $p_\vee$, where $p\in \Inst_0$ (a zero-test $\reg_j$-instruction), then $p_\lor:\qstate\leq \qstate'\vee \zstate_i$ and $\tnode$ has two covers corresponding to $\qstate'$ and to $\zstate_i$; we define the \emph{next node} of $\tnode$ to be the one corresponding to $\qstate'$.
Finally, if the instruction is in $\eqset$, then, since $\eqset$ is {\joindecreasing}/{\joinincreasing}, 
there is at least one joinand witnessing this property; all of the covers of $\tnode$ corresponding to such a witnessing node are defined to be a \emph{next node} of $\tnode$.

Note that the next-node relation is a subrelation of the the covering relation of $\labeledforest$. 
Also, by the above definition, if $\tnode$ is a next-node (of some node) then the first coordinate of its label does not include any state variable of $\State_{\zinst}$, so its cover nodes (if any) do not have instructions in $\Inst_{\zinst}$. 
Therefore, if  $\tnode$ is a next-node and it has covers (i.e., it is not a leaf), then it  itself has a next node. 
Finally, since the state of $\tnode_0$ is not in $\State_{\zinst}$, we have that $\tnode_0$ has a next node, unless it is a leaf.
Therefore, the connected component of $\tnode_0$ under the reflexive and transitive closure of the next-node relation forms a subtree of $\labeledforest$ and all of its leaves are leaves of $\labeledforest$; also all of the monoid-word labels do not have states in $\State_{\zinst}$. 
(This subtree can be thought of as the computation tree of $\eqset\acm$).

We will now argue that every maximal subchain of this subtree witnesses a terminating computation in $\acm$.
We assume that $\tnode$, with monoid word label $c$, has a next node $\tnode'$, with monoid word label $c'$, and consider cases for the instruction of $\tnode'$.
If the instruction $p$ is in $\Inst$, then $c \to^p c'$ is a valid computation step of $\acm$. If the instruction is of the form $p_\lor:\qstate\leq \qstate'\vee \zstate_i$, where $p\in \Inst_0$, then since every leaf above $\tnode$ has label $\qfin$ and since  $\eqset$ is non-integral,  Lemma~\ref{lem:nonintZT}  implies that $c \to^p c'$ is a valid computation step of $\acm$. Finally, if the instruction is in $\eqset$, then, depending on whether $\eqset$ is joinand-increasing or joinand-decreasing, using the definition  we obtain $c\to^{\mathsf{i}} c'$ or $c \to^{\mathsf{t}}c'$.
\end{proof}

In contrast to CMs, the termination problem for lossy CMs is decidable \cite{mayr2003}, so the same holds for increasing CMs by Proposition~\ref{prop:invertedCM}. 
However, their complexities have been shown to be \ACK-complete---the hardness result is established via a reduction from \ref{ACP}. 
For lossy counter machines, this fact was first shown by Schnoebelen in \cite{schnoebelen2002}, and again in \cite{schnoebelen2010} with a simpler and more elegant presentation (see also \cite[Chapter~3]{schmitz2012notes}).
The construction we use is from \cite{schnoebelen2010}, where it is stated in terms of \emph{extended} counter machines, counter machines with some additional types of instructions. 
We define these machines within the proof of the theorem below, where we show how each new instruction type can be replaced by a collection of fresh registers/states/instructions hence expanded CMs can be simulated by CMs.

For a natural number $n$, we say that a CM $\acm$ \emph{terminates in space bounded by $n$}, if 
there is a computation witnessing the termination of $\acm$ in which the total register contents of any configuration in the computation does not exceed the value $n$.

\section{Complexity of lossy (or increasing) counter machines}\label{s: lossy}
The following theorem will sit at the basis of our reduction of the ACP. 
The proof is essentially given in \cite[Theorem~5.1]{schnoebelen2010}. 
However, the construction involves using additional types of instructions outside those allowed by our definition. 
Those machines are \emph{extended} CMs, which are CMs with some additional  instruction types, here called \emph{extended} instructions, of the form \emph{reset} (which empties some given register), \emph{copy} (place a copy of the contents of one register to another), \emph{equality test} (like zero test, a conditional instruction testing whether two registers have the same contents), or \emph{no-op} (transition between states performing no operation or test). 
Extended CMs are known to be equivalent to CMs, however it may not be immediately clear that the same correspondence holds when allowing lossy glitches. 
Instead of proving this more general result, we opt to adapt Schnoebelen's construction directly for CMs, as the basis of his construction is both elegant and edifying. 
This also allows us to obtain the result for increasing machines (item 4 below), a fact that is not stated in \cite{schnoebelen2010} but is readily implicit.

Let us recall from Section~\ref{sec:subrec-hierarchies} the fast-growing hierarchy $\{F_k:\mathbb{N}\to\mathbb{N} \}_{k<\omega}$ for the map $x\mapsto x+1$ defined by induction via the following scheme:
\begin{equation}\label{eq:FGFdef}
   F_0(x) := x+1 \qquad F_{k+1}(x) := F_k^{x+1}(x). 
\end{equation}
It is easily verified (by induction) that $F_1(x) = 2x+1$ and $F_2(x)=-1+(1+x)2^{1+x}$. On the other hand, the function $F_3$ is not elementary. In fact, every primitive recursive function is eventually dominated by some $F_k$. 
Therefore the function ${Ack:\mathbb{N}\to \mathbb{N}}$ defined via  
    $$
    Ack(n):= F_n(n) 
    $$
is a non-primitive recursive member of $\UniFGHLevel{\omega}$. 
The function $Ack$ is related, but not identical, to Ackermann's function in the sense that they are both non-primitive recursive in level $\UniFGHLevel{\omega}$. 
Using the above, and to give a sense of perspective for its growth-rate, the initial values of $Ack$ for $n=0,1,2$ are, respectively, $1,3,23$ (see also below for an alternative computation of $Ack(2)$), while 
$Ack(3)=F_3(3)$ and
    \begin{align*}
    F_3(3) 
    &=F_2^4(3)
    =F_2^3(-1+2^6)
    =F_3^2\left(-1+2^6\cdot 2^{2^6}\right) 
    \\
    &=F_3^1\left(-1+2^{70}\cdot 2^{2^{70}}\right)
    =-1 + 2^{70}\cdot 2^{2^{70}}\cdot 2^{2^{70}\cdot 2^{2^{70}}}     
    \end{align*}
noting that $(1+3)\cdot 2^{1+3}=2^6$ and $2^6\cdot 2^{2^6}=2^{70}$. 
Consequently, $Ack(3)\gtrsim 2^{2^{2^{70}}} >  10^{10^{10^{20}}}$.
The value of $Ack(4)$ is more cumbersome to describe exactly but, utilizing the fact that $F_3(m)$ is greater than $2^a$ where $a$ is power-tower of $2$'s of height $m$, we provide the following crude lower bound:
    $$
    Ack(4) = F_3^5(4) \geq 2^{2^{\cdot^{\cdot^{\cdot^2}}}}
    \phantom{2}^{\big\}
        \text{height $>$}
        2^{2^{\cdot^{\cdot^{\cdot^2}}}}
        \phantom{2}^{\big\}
        \text{height $>$}
        2^{2^{\cdot^{\cdot^{\cdot^2}}}}
        \phantom{2}^{\big\}
        \text{height $>$}
        2^{2^{\cdot^{\cdot^{\cdot^2}}}}
        \phantom{2}^{\big\}
        \text{height $>$} 2^{16}
        }
        }
        }
        }
    $$

For each $m\in \mathbb{N}$, there is a ``vectorial'' function $F:\mathbb{N}^{m+1}\times \mathbb{N}\to \mathbb{N}$ defined in \cite{schnoebelen2010} as a composition of fast-growing functions:
    \begin{equation}\label{eq:FGFdefVec}
    F(\vec{a}; n) =F(a_m,\ldots,a_0;n):= F_m^{a_m}(\ldots F_1^{a_1}(F_0^{a_0}(n))\ldots) .
    \end{equation}
In particular, we have that $F_k(n)=F(0^{m-k}, 1, 0^{k}; n)$, where $k \in \{0,\ldots,m\}$ and $0^a$ denotes a string of $a$-many zeros. 
Thus, the generative fast growing functions are coordinates of the vectorial function.

Actually, the function $F$ can be alternatively defined inductively via the  rules:
\begin{itemize}
    \item[(D0)]\label{rule:F0} $F(0^{m+1} ;n)= n$;
    \item[(D1)]\label{rule:F1} $F(a_m,\ldots,a_0+1;n)= F(a_m,\ldots,a_0;n+1)$;
    \item[(D2)]\label{rule:F2} $F(a_m,\ldots,a_{k+1} +1,0, 0^k; n) = F(a_m,\ldots,a_{k+1},n+1, 0^k; n)$, for each $k\in \{0,\ldots,m-1\}$;
\end{itemize}
we will denote by (D$2_k$) the rule (D2) restricted to a fixed $k\in \{0,\ldots,m-1\}$.  
To see that these two definitions coincide, note that if $F$ is defined by Eq.~\ref{eq:FGFdefVec}, then
(D0) and (D1) hold by the base case of Eq.~\ref{eq:FGFdef}, while (D2) holds by the inductive part of Eq.~\ref{eq:FGFdef}.
On the other hand, reading each rule from left to right we see that the vector gets lexicographically strictly smaller up to the point that only the right-most coordinate is present. 
So, via the inductive definition $F$ is defined on all vectors (i.e., it is a single-total relation). Finally, $F$ is not over-defined (i.e., it is a single-valued relation) as the domains/left-hand sides of each rule are disjoint. 
So, the inductive definition yields a unique $F$, and as the function defined by Eq.~\ref{eq:FGFdef} satisfies the inductive rules, the two functions coincide.

Furthermore, since using the inductive rules from left to right results in lexicographically strictly smaller vectors and since the lexicographic order on $\mathbb{N}^{m+1}$ is a well order, applying these one-directional rules results in a terminating procedure, which yields a number by a final application of (D0). 
In other words, this procedure allows us to compute/calculate the value of $F$ on any vector, and as we will see below this procedure can be implemented by a counter machine. 
In particular, given that $F(1,0^{m};m)=F_m(m) = Ack(m)$ by the equivalence of the two definitions of $F$, we can calculate the value of $Ack(m)$. 
For example, we can compute $Ack(2)$, as follows:
    $$ 
    \begin{array}{r l}
    Ack(2) &=
    F(1,0,0;2)
    \stackrel{\mathrm{D2}}{=}
    F(0,3,0;2)
    \stackrel{\mathrm{D2}}{=}
    F(0,2,3;2)
    \stackrel{\mathrm{D1}}{=}
    F(0,2,0;5)
    \\ &
    \stackrel{\mathrm{D2}}{=}
    F(0,1,6;5)
    \stackrel{\mathrm{D1}}{=}
    F(0,1,0;11)
    \stackrel{\mathrm{D2}}{=}
    F(0,0,12;11)
    \stackrel{\mathrm{D1}}{=}
    F(0,0,0;23)
    \stackrel{\mathrm{D0}}{=}
    23
    \end{array}
    $$

A key idea in the proof of the following theorem is to identify the vector $(a_m,\ldots,a_0;n)$ with the configuration $\qstate_\mathrm{evalF}\mathtt{a}_m^{a_m}\cdots \mathtt{a}_0^{a_0}\mathtt{n}^n$ in a given CM and to associate subCMs to the rules (D0), (D1) and (D2) that move from the input vector on the left of the rule to the input vector on the right of the rule, thus computing the value of $F$.

Moreover, as will be relevant to lossy/increasing glitches, the function $F$ is monotone with respect to the product order over $\mathbb{N}$ (c.f., \cite[Fact~3.1]{schnoebelen2010}); i.e., for all $\vec{a}_1, \vec{a}_2\in \mathbb{N}^{m+1}$ and $n_1,n_2 \in \mathbb{N}$, we have 
    \begin{equation}\label{eq:Fmono}
        \vec{a}_1 \leq \vec{a}_2 \text{ and }n_1\leq n_2 \implies F(\vec{a}_1, n_1)\leq F(\vec{a}_2,n_2).
    \end{equation}

\begin{theorem}[{cf. {\cite[Theorem~5.1]{schnoebelen2010}}}]\label{prop:AckCMs}
If $\acm$ is a CM and $m\in \mathbb{Z}^+$, then there exists a CM $\acm(m)$ with fresh states $\qstate_\mathrm{evalF}$ and $\qstate_\mathrm{backF}$ and fresh registers $\mathtt{A}_m=\{\mathtt{a}_m,\ldots,\mathtt{a}_0 \}$, such that the following are equivalent:
\begin{enumerate}
\item\label{item:AckCMs_ACP} $\acm$ terminates in space bounded by $Ack(m)$.
\item\label{item:AckCMs_CM} $\qstate_\mathrm{evalF} \mathtt{a}_m\mathtt{a}_{0}^m\to_{\acm(m)}\qstate_{\mathrm{backF}}\mathtt{a}_m\mathtt{a}_{0}^m$.
\item\label{item:AckCMs_iCM} $\qstate_\mathrm{evalF} \mathtt{a}_m\mathtt{a}_{0}^m\to_{\acm(m)}^\mathrm{lossy}\qstate_{\mathrm{backF}}\mathtt{a}_m\mathtt{a}_{0}^m$.
\item\label{item:AckCMs_tCM} $\qstate_\mathrm{evalF} \mathtt{a}_m\mathtt{a}_{0}^m\to_{\acm(m)}^\mathrm{inc}\qstate_{\mathrm{backF}}\mathtt{a}_m\mathtt{a}_{0}^m$.
\end{enumerate}
Moreover, the construction is such that $|\acm(m)|$ grows linearly with $|\acm|$ and $m$ as input.% is a 
\end{theorem}

\begin{proof}
We recall the important features in the construction from \cite{schnoebelen2010}. 
Let $\acm=(\Reg_d, \State,\Inst)$ be a $d$-CM, where $\Inst$ is the  disjoint union of the set $\Inst_{\mypm}$ of increment/decrement instructions and the set $\Inst_\mathtt{0}$ zero tests. 
The machine $\acm(m)$ contains the original states of $\acm$ together with two new special states $\qstate_{\mathrm{evalF}}$ and $\qstate_{\mathrm{backF}}$, among others, and is a composition of three submachines: 
    $$
    \scalebox{.6}{
    \begin{tikzpicture}[node distance = 5cm]
    \node[state] (0) {$\qstate_{\mathrm{evalF}}$};
    \node[state, right of = 0] (1) {$\qin^\mathrm{B}$};
    \node[state, right of = 1] (2) {$\qfin^\mathrm{B}$};
    \node[state, right of = 2] (3) {$\qstate_{\mathrm{backF}}$};
    \draw[-{Latex[width=3mm]}, >=Latex]
    (0) edge[dashed,above left] node{$\mathtt{evalF}$} (1)
    (1) edge[dashed,above] node{$\acm^\mathrm{B}$} (2)
    (2) edge[dashed,above right] node{$\mathtt{backF}$} (3)
    %(1) edge[above,dashed] node{$\acm(m)$} (2)
    (1) edge[ bend right,dashed,above] node{$\vdots$}  (2)
    (1) edge[ bend left,dashed] (2)
    ;
    \draw[dashed,>=Latex] 
    (0) to [in=60, out=30, loop] %node{$\mathtt{evalF}$} 
    ();
    \draw[dashed,>=Latex] 
    (3) to [in=120, out=150, loop] ()
    ;
    \end{tikzpicture}
    }
    $$

The central submachine $\acm^\mathrm{B}$, called the \emph{budget machine}, is a $(d+1)$-CM which simulates $\acm$ and at the same time keeps track of the total register contents with the help of an additional \emph{budget register} $\mathtt{n}$.
$\acm^\mathrm{B}$ acts only on registers from $\Reg_d\cup\{\mathtt{n}\}$ in such a way that for any instruction $p\in \Inst_{\mypm}$ coming from $\acm$, we add to $\acm^\mathrm{B}$ instructions (and a fresh intermediary state) given by the following program: 
    $$ 
    \scalebox{.6}{
    \begin{tikzpicture}[node distance = 2.5cm]
    \node[state] (0) {$\qstate$};
    \node[state, minimum size = .25,right of = 0] (1) {};
    \node[state, right of = 1] (2) {$\qstate'$};
    \draw[-{Latex[width=3mm]}] 
    (0) edge[above] node{$\Inc\reg$ ($\Dec\reg$)} (1)
    (1) edge[above] node{$\Dec\mathtt{n}$ ($\Inc\mathtt{n}$)} (2)
    ;
    \end{tikzpicture}
    }
    $$
In other words, if $p:\qstate \leq \qstate' \reg$ and $p_\mathrm{out}:\qstate^p$ denotes the fresh state, then we add the instructions $p_\mathrm{in}:\qstate \leq \qstate^p \reg$ and $\qstate^p\mathtt{n} \leq \qstate' $; if $p:\qstate \reg \leq \qstate'$, we add the instructions $p_\mathrm{in}:\qstate \reg\leq \qstate^p $ and $p_\mathrm{out}:\qstate^p \leq \qstate' \mathtt{n}$.
If the budget runs out then the machine may get stuck in this intermediary fresh state. 
As a result, the size of $\acm^\mathrm{B}$ is equal to $|\acm| + 2|\Inst_{\mypm}|$, since we add $|\Inst_{\mypm}|$-many fresh states and $|\Inst_{\mypm}|$-many new instructions.

Observe that the combined effect of the diagram above swaps a $\mathtt{n}$-token with a $\reg$-token, as if doing $\qstate \mathtt{n}\leq \qstate'\reg$, in case of $p$ is an increment instruction and \emph{vice versa}, as if doing $\qstate \reg\leq \qstate' \mathtt{n}$, in  case  $p$ a decrement instruction. 
Since zero test instructions do not alter register contents, in any computation between two configurations whose states are in $\State$ the total number of register tokens from $\Reg\cup \{\mathtt{n} \}$ in the configuration remains invariant. 
In particular, as explained in \cite[Section~4]{schnoebelen2010}, the original $\acm$ has terminating computation bounded by a value $N$ iff there is a computation in $\acm^\mathrm{B}$ from $\qin \mathtt{n}^N$ to $\qfin \mathtt{n}^N$. 
Indeed, if $\acm$ has such a bounded terminating computation, then whenever that computation is in a state from $\State$ (from $\acm$), the sum of all register contents is exactly $N$, so the budget machine does not get stuck. 
Conversely, if $\qin\mathtt{n}^N \to_{\acm^\mathrm{B}} \qfin\mathtt{n}^N$ then, from any witnessing computation of this relation, a terminating computation for $\acm$ may be extracted, which moreover is $N$-bounded. 

Note that if there is a nontrivial lossy computation  in $\acm^\mathrm{B}$  from $\qin \mathtt{n}^N$ to $\qfin\mathtt{n}^{N'}$, then (even if the lossy glitch(es) took place in a fresh state) the net effect of the computation ignoring the glitches keeps the number of register tokens from $\Reg\cup \{\mathtt{n} \}$ in the configurations invariant (as argued above). 
Factoring in the effect of the lossy glitches, we see that the  number of register tokens strictly decreases, hence $N>N' $ (cf., \cite[Cor.~4.5]{schnoebelen2010}). 
Likewise, in every nontrivial increasing computation in $\acm^\mathrm{B}$ from $\qin \mathtt{n}^N$ to $\qfin\mathtt{n}^{N'}$ we have $N<N' $.
These features will be important shortly. 

The purpose of the submachines $\mathtt{evalF}$ and $\mathtt{backF}$ are to, respectively, load and unload the budget register $\mathtt{n}$ with the value ${Ack}(m)$ before and after executing the budget machine above. 
The submachine $\mathtt{evalF}$ in \cite[Fig.~1]{schnoebelen2010} can be seen as a fusion of three machines, denoted here by $\mathtt{D0}$, $\mathtt{D1}$, and $\mathtt{D2}$. 
These machines are CMs with four additional types of instructions and are known as \emph{extended CMs}. 
These extra types of instructions can actually be implemented by usual CM instructions (under the addition of extra states, registers and instructions), so there is no need for considering extended CMs; i.e., extended CMs are equivalent to CMs. 
Instead of giving the proof of this general statement, below we will give the CM implementation of the extended CMs used in  \cite[Fig.~1]{schnoebelen2010}.
The machines $\mathtt{D1}$ and $\mathtt{D2}$ will implement the left-to-right direction of a single step of rules (D1) and (D2), respectively, with the latter machines being a fusion of submachines $\mathtt{D2}_k$ corresponding to (D$2_k$), for each $0\leq k<m$. 
The machine $\mathtt{D0}$ initializes the budget machine by testing that all registers $\mathtt{a}_m,\ldots,\mathtt{a}_0$ are empty (corresponding to $\vec{a}=0^{m+1}$ in (D0)), ensuring the process is completed with the resulting contents of $\mathtt{n}$ as budget for $\acm^\mathrm{B}$. 
The machine can be depicted as below
\begin{equation}\label{eq:evalFmach}\tag{$\mathtt{evalF}$}
\scalebox{.66}{
    \hspace{-0.5cm}
    \begin{tikzpicture}[node distance = 2cm,baseline=(current bounding box.center)]
    %D1 machine
    \node[state,minimum size = .6cm] (0) {$\qstate_\mathrm{evalF}$};
    \node[state,minimum size = .25cm, above of = 0] (d1) {};
    %D0 machine
    \node[state,minimum size = .25cm, right of = 0] (D0in) {$\qstate_1$};
    \begin{scope}[node distance = 1.6cm]
        \node[state,minimum size = .25cm,right of = D0in] (D1in) {};
    \end{scope}
    \node[state,minimum size = .25cm,right of = D1in] (Diin) {};
    \begin{scope}[node distance = 2.2cm]
        \node[state,minimum size = .25cm, right of = Diin] (Dkin){$\qstate_{k+1}$}; 
        \node[state,minimum size = .25cm,right of = Dkin] (Dk1in) {};
    \end{scope}
    \node[state,minimum size = .25cm, right of = Dk1in] (Dmin){$\qstate_{m}$};
    \begin{scope}[node distance = 2.2cm]
        \node[state, right of = Dmin] (output) {$\qin^\mathrm{B}$};
    \end{scope}
    %D2machine
    \begin{scope}[node distance = 3.5cm]
        \node[state,minimum size = .25cm, below of = D0in] (D0out) {$\qstate_1'$};
        \node[state,minimum size = .25cm,below of = D1in] (D1out) {};
        \node[state,minimum size = .25cm,below of = Diin] (Diout) {};
        \node[state,minimum size = .25cm, below of = Dkin] (Dkout){$\qstate_{k+1}'$};
        \node[state,minimum size = .25cm,below of = Dk1in] (Dk1out) {};
        \node[state,minimum size = .25cm, below of = Dmin] (Dmout){$\qstate_{m}'$};
    \end{scope}
    \draw[-{Latex[width=2mm]}] 
    %D1 edges
    (0) edge[bend right, below left] node{$\Dec \mathtt{a}_0$} (d1)
    (d1) edge[bend right, above] node{$\Inc \mathtt{n}\hspace{4mm}$}(0)
    %D0 edges
    (0) edge[below] node{\small$\mathtt{a}_0\ztr$} (D0in)
    (D0in) edge[below] node{\small$\mathtt{a}_1\ztr$} (D1in)
    (D1in) edge[dashed, ultra thick,above left] node{\large$\mathtt{D2}_k^\mathrm{in}$} (Diin)
    (Diin) edge[below] node{\small$\mathtt{a}_k\ztr$} (Dkin)
    (Dkin) edge[above] node{\small$~\mathtt{a}_{k+1}\ztr$} (Dk1in)
    (Dk1in) edge[dotted,above] node{\large$\cdots$} (Dmin)
    (Dmin) edge[above] node{$\mathtt{a}_m\ztr$} (output)
    %D2 edges down
    (D0in) edge[dotted,left] node{\small$\mathtt{D2}_0^\mathrm{step}$} (D0out)
    (D1in) edge[dotted,left] node{\small$\mathtt{D2}_1^\mathrm{step}$} (D1out)
    (Diin) edge[dotted,left] node{\Large$\cdots\hspace{.75cm}$} (Diout)
    (Dkin) edge[dashed,ultra thick,left] node{\large$\mathtt{D2}_k^\mathrm{step}$} (Dkout)
    (Dk1in) edge[dotted] 
    (Dk1out) %node{\small$\mathtt{D2}_{k+1}^\mathrm{step}$} (Dk1out)
    (Dk1in) edge[dotted,right] node{\Large$\cdots$} (Dk1out)
    (Dmin) edge[dotted,left] node{\small$\mathtt{D2}_{m-1}^\mathrm{step}$} (Dmout)
    %D2 edges left
    (Dmout) edge[dotted,above] node{\large$\cdots$} (Dk1out)
    (Dk1out) edge[above] node{\small$\mathtt{a}_k\ztr$} (Dkout)
    (Dkout) edge[below] node{\small$\mathtt{a}_{k-1}\ztr$} (Diout)
    (Diout) edge[dashed, ultra thick, above] node{\large$\mathtt{D2}_{k}^\mathrm{out}$} (D1out)
    (D1out) edge[below] node{\small$\mathtt{a}_0\ztr$} (D0out)
    (D0out) edge[bend left, below left] node{$\texttt{no}_\texttt{-}\texttt{op}$} (0)
    ;
    \draw[-{}]
    (0) edge[dashed, ultra thick] (D0in)
    (D0in) edge[dashed, ultra thick] (D1in)
    (Diin) edge[dashed, ultra thick] (Dkin)
    (D1out) edge[dashed, ultra thick] (D0out)
    (Dkout) edge[dashed, ultra thick] (Diout)
    ;
    %Boxes
    \node[draw=gray,rectangle,thick,fit= (0) (output)] {};
    \draw (output) node[above=1cm] {\Large$\mathtt{D0}\hspace{3cm}$};
    \draw (d1) node[right=1cm] {\Large$\mathtt{D1}$};
    \draw (Dmout) node[right=1cm] {\Large$\mathtt{D2}$};
    \node[draw=gray,rectangle,fit= (0) (d1)] {};
    \node[draw=gray,rectangle,fit= (0) (Dmout) (Dmin)] {};
    \end{tikzpicture}
    }
\end{equation}
with the machine $\mathtt{D2}_k$ for $0\leq k<m$ indicated by following the thick dashed lines.
The intention is that $\mathtt{evalF}$ will simulate the application of rules (D1) and (D2), read from left-to-right, in the sense that a completion of a single $\mathtt{D}_1$ loop, or respectively $\mathtt{D}_2$ loop, to and from the control state $\qstate_\mathrm{evalF}$ shall reflect a single application of (D1) or (D2), respectively, upon the correspondence
    $$
    \qstate_\mathrm{evalF}\mathtt{a}_m^{a_m}\cdots \mathtt{a}_0^{a_0}\mathtt{n}^n \equiv F(a_m,\ldots,a_0;n).
    $$

When $\qstate_\mathrm{eval}$ is initialized with register contents, it first transfers the contents of $\mathtt{a}_0$ to the budget $\mathtt{n}$, then enters a process of sequentially testing whether the register $\mathtt{a}_i$, for $i=0,\ldots,m$, is empty. 
In case all tests are positive, i.e., all registers but the budget are empty, $\mathtt{D0}$ is completed and the budget machine has been initialized. Otherwise, upon reaching the first state with index $k<m$ for which $\mathtt{a}_{k+1}$ is nonempty (i.e., completing subprogram $\mathtt{D2}_k^\mathrm{in}$) the machine executes $\mathtt{D2}^\mathrm{step}_k$. 
This, in turn, decrements $\mathtt{a}_{k+1}$ and begins a process of copying the contents of the budget $\mathtt{n}$ into $\mathtt{a}_k$, ending by placing an additional token to $\mathtt{a}_k$, then (re)verifying all registers $\mathtt{a}_i$, for $i<k$, are empty---marking the end of program $\mathtt{D2}_k$, and completing the loop by re-entering the control state $\qstate_\mathrm{evalF}$ via a ``no operation'' instruction $\texttt{no}_\texttt{-}\texttt{op}$; this is an extended instruction allowing the machine to transition from one state to another performing no operation or test.

Other than the single $\texttt{no}_\texttt{-}\texttt{op}$ and the submachines $\mathtt{D2}^\mathrm{step}_k$ for each $0\leq k<m$, every edge/instruction is represented using basic CM instructions. 
In \cite{schnoebelen2010}, the construction of  $\mathtt{D2}^\mathrm{step}_k$ (cf., \cite[Fig.~1]{schnoebelen2010}), depicted below on the left, utilizes a \emph{copy} instruction $\qstate[\mathtt{a}_k\cpy \mathtt{n}]\qstate'$, an extended instruction that allows the machine, while in state $\qstate$, to replace the contents stored in $\mathtt{a}_k$ by a copy of (all) the contents stored in $\mathtt{n}$, and transition to $\qstate'$. 
The inclusion of these extended instructions offer no major obstacle, and can be remedied by affixing to the machine a fresh register $\mathtt{t}$, used as temporary storage in the submachine $\mathtt{D2}_k^\mathrm{step}$. 
The submachine $\mathtt{D2}^\mathrm{step}_k$ for our purposes is depicted below on the right.
$$
\begin{array}{c | c}
\text{(D$2_k$) as an extended CM} & \text{(D$2_k$) as a CM} \\ \hline \\
\scalebox{0.65}{
    \begin{tikzpicture}[node distance = 2cm,baseline=(current bounding box.base)]   
        \node[state,minimum size =0.1cm] (0) {$\qstate_{k+1}\phantom{'}$};
        \node[ below of = 0] (0') {};
        \node[state,minimum size =0.1cm, right of =0] (1) {$\qstate\phantom{'}$};
        \node[state,minimum size =0.1cm, right of =1] (2) {$\qstate'$};
        \node[state,minimum size =0.1cm, right of= 2] (3)
        {$\qstate_{k+1}'$};
        \node[ below of = 3] (3') {};
        \draw[-{Latex[width=2mm]}] 
        (0') edge[dotted,thick,right] node{$\mathtt{a}_k\ztr$} (0)
        (0) edge[above] node{$\Dec \mathtt{a}_{k+1}$} (1)
        (1) edge[above] node{$\mathtt{a}_k\cpy \mathtt{n} $} (2)
        (2) edge[above] node{$\Inc \mathtt{a}_k$} (3)
        (3) edge[dotted,thick,left] node{$\mathtt{a}_{k-1}\ztr$} (3')
        ;
    \end{tikzpicture}
    }
&
\scalebox{0.65}{
    \begin{tikzpicture}[node distance = 2cm,baseline=(current bounding box.base)]
        \node[state,minimum size =0.1cm] (0) {$\qstate_{k+1}\phantom{'}$};
        \node[ below of = 0] (0') {};
        \node[state,minimum size =0.1cm, right of =0] (1) {$\qstate\phantom{'}$};
        \begin{scope}[node distance = 1.5cm]
        \node[state,minimum size =0.1cm, right of =1] (2) {};
        \node[state,minimum size =0.1cm, right of= 2] (3) {};
        \node[state,minimum size =0.1cm, right of= 3] (4) {{$\qstate'$}};
        \end{scope}
        \node[state,minimum size =0.1cm, right of= 4] (5){$\qstate_{k+1}'$};
        \node[below of = 5] (5') {};
    
        \node[state, minimum size = 0.25cm,below of = 2] (2') {};
        \node[state, minimum size = 0.25cm,below of = 3] (3'') {};
        \node[right of = 3''] (spot) {};
        \node[state, minimum size = 0.25cm, between = 3'' and spot] (3') {};
        %\node[state, minimum size = 0.25cm] at (12,-1.2) (3'') {};
        \draw[-{Latex[width=2mm]}] 
        (0') edge[dotted,thick,right] node{$\mathtt{a}_k\ztr$} (0)
        (0) edge[above] node{$\Dec \mathtt{a}_{k+1}$} (1)
        (1) edge[above] node{$\mathtt{t}\ztr $} (2)
        (2) edge[above] node{$\mathtt{n}\ztr $} (3)
        (3) edge[above] node{$\mathtt{t}\ztr $} (4)
        (4) edge[above] node{$\Inc \mathtt{a}_k$} (5)
        (2) edge[bend left,above left] node{$\Dec\mathtt{n}$} (2')
        (2') edge[bend left, below left] node{$\Inc\mathtt{t}$} (2)
        (3) edge[bend left, left] node{$\Dec\mathtt{t}$} (3')
        (3') edge[below] node{$\Inc\mathtt{a}_{k}$} (3'')
        (3'') edge[bend left,below left] node{$\Inc\mathtt{n}$} (3)
        (5) edge[dotted,thick,left] node{$\mathtt{a}_{k-1}\ztr$} (5')
        ;
    \end{tikzpicture}
    }
\end{array}
$$
Every $\texttt{no}_\texttt{-}\texttt{op}$ edge in \ref{eq:evalFmach} is simply replaced by a zero test instruction $\qstate'_1[\mathtt{t}\ztr]\qstate_\mathrm{evalF}$ of the temporary register, which must always be empty at any point outside the (sub)submachine from $\qstate$ to $\qstate'$ in any machine $\mathtt{D2}_k$ on any (glitchless) run.

As should be clear from the diagram, the CM  $\mathtt{D2}_k^\mathrm{step}$ performs the desired task.
Indeed, having entered state $\qstate_{k+1}$ with $\mathtt{a}_k$ verified as empty, if $\mathtt{a}_{k+1}$ is also nonempty the program begins by decrementing it. 
After a (redundant) test of $\mathtt{t}$ being empty, the program enters the first loop transferring the contents of $\mathtt{n}$ in the temporary storage $\mathtt{t}$. 
Once this is completed, verified by a zero test on $\mathtt{n}$, it enters a second loop reversing this process, but with an intermediate step of also incrementing $\mathtt{a}_k$, effectively restoring $\mathtt{n}$, and simultaneously placing its copy in $\mathtt{a}_k$, from the temporary storage $\mathtt{t}$. 
When this loop is completed with a zero test on $\mathtt{t}$, the program adds an additional token to $\mathtt{a}_k$.
In sum, this performs precisely the same operation, i.e., a step of rule (D$2_k$), as the extended machine, at least on glitchless runs. 
 
Towards establishing the (computational) equivalence of the extended and non-extended version of these submachines, and to use therefore the results of \cite{schnoebelen2010}, it will suffice to verify that for the extended instruction $p: \qstate[\mathtt{a}_k\cpy\mathtt{n}]\qstate'$, the following holds
    $$
    \begin{array}{r c l}
    \qstate\mathtt{n}^n [{\to^p} \circ  {\to^\mathrm{lossy}} ]\qstate' \mathtt{n}^{n'}\mathtt{a}_k^{n''} 
    &\iff&
    \qstate\mathtt{n}^n [ {\to_{\mathtt{D2}_k^\mathrm{step}}} \circ  {\to^\mathrm{lossy}}] \qstate'\mathtt{n}^{n'}\mathtt{a}_k^{n''} \\
    &\iff & 
    \qstate \mathtt{n}^n \to_{\mathtt{D2}_k^\mathrm{step}}^\mathrm{lossy} \qstate'\mathtt{n}^{n'}\mathtt{a}_k^{n''} 
    \end{array}
    $$

The first biconditional is obvious by the description above, as there are no lossy instances within internal states in the submachine by assumption. 
For the second biconditional, observe that the top-to-bottom direction is obvious as the latter relation subsumes the former. 
So suppose, when initialized with $n$ tokens in $\mathtt{n}$, there is a computation $\mathtt{D2}_k^\mathrm{step}$ reaching $\qstate'$. 
As any such program ends with a zero test of $\mathtt{t}$, and the instructions only involve registers $\mathtt{a_k},\mathtt{n},\mathtt{t}$, it must be that the computation results in a configuration $\qstate'\mathtt{n}^{n'}\mathtt{a}_k^{n''}$ for some $n',n''$. 
It is easy to see that any decrease due to lossy instances in registers $\mathtt{n}$ or $\mathtt{t}$ in the first loop yields a value of $t\leq n$ in $\mathtt{t}$ upon entering the second loop. 
Similarly, any loss in any register in the second loop yields $n'\leq t$ and $n''\leq t$. So $n',n''\leq t \leq n$. Thus $\qstate \mathtt{n}^n \to_{\mathtt{D2}_k^\mathrm{step}}\qstate'\mathtt{n}^n\mathtt{a}_k^n \to^\mathrm{lossy} \qstate'\mathtt{n}^{n'}\mathtt{a}_k^{n''}$ holds, completing the claim.

Similarly, for increasing glitches,
    $$
    \begin{array}{r c l}
    \qstate\mathtt{n}^n [{\to^p} \circ  {\to^\mathrm{inc}} ]\qstate' \mathtt{n}^{n'}\mathtt{a}_k^{n''} 
    &\iff&
    \qstate\mathtt{n}^n [ {\to_{\mathtt{D2}_k^\mathrm{step}}} \circ  {\to^\mathrm{inc}}] \qstate'\mathtt{n}^{n'}\mathtt{a}_k^{n''} \\
    &\iff & 
    \qstate \mathtt{n}^n \to_{\mathtt{D2}_k^\mathrm{step}}^\mathrm{inc} \qstate'\mathtt{n}^{n'}\mathtt{a}_k^{n''} 
    \end{array}
    $$
holds using similar reasoning but reversing the inequalities. 
In this way our construction of $\mathtt{evalF}$, and its lossy version, is computationally identical to that in \cite{schnoebelen2010}, at least when restricted to configurations with state $\qstate_\mathrm{evalF}$.

As is stated in \cite[Lemma~3.2]{schnoebelen2010}, if there is a (resp., lossy) $\mathtt{evalF}$-computation from $\qstate_\mathrm{evalF}\mathtt{a}_m^{a_m}\cdots \mathtt{a}_0^{a_0}  \mathtt{n}^n$ to $\qstate_\mathrm{evalF}\mathtt{a}_m^{a_m'}\cdots \mathtt{a}_0^{a_0'}  \mathtt{n}^{n'} $, then the value $F(a_m,\ldots,a_0,n)$ is (resp., greater than or) equal to $F(a_m',\ldots,a_0',n')$.

The analogous claim for increasing computations (where the inequality becomes $\leq$) is not stated in \cite{schnoebelen2010}, but does hold as the result of a similar argument. 
Indeed, suppose there is such an increasing computation and, without loss of generality, that it is a result of a single run of $\mathtt{D1}$ or some $\mathtt{D2}_k$. If it occurs while in $\mathtt{D1}$, the claim is immediate by the monotonicity of $F$ (Eq.~\ref{eq:Fmono}). 
Suppose then there is an increase within $\mathtt{D2}_k$. 
As it starts (and ends) with a sequence of zero tests, it must be that $a_0=\cdots=a_{k}=0$. 
By the biconditionals above, it must be that $ a'_{k},n'\geq n$. 
Since it must decrement $\mathtt{a}_{k+1}$, and this is the only instruction for this register, it must be that $a_{k+1}\leq a'_{k+1}+1$. 
As there are no instructions for the remaining registers $\mathtt{a}_{k+2},\ldots,\mathtt{a}_m$ in this submachine, it must be that $a_i\leq a'_i$ for each $k+1<i\leq m$. 
Hence
    $$
    \begin{array}{r c l l}
    F(a_m,\ldots,a_{0};n)
    &=& F(a_m,\ldots,\phantom{1+ {}}a_{k+1},0,0^{k};n) 
    \\
    &\leq&F(a'_m, \ldots,1+a'_{k+1},0,0^{k};n) 
    & \text{(Eq.\ref{eq:Fmono})} \\
    & =& 
    F(a'_m, \ldots,\phantom{1+ {}}a'_{k+1},n,0^{k};n) 
    & \text{(D$2_k$)}\\
    &\leq& 
    F(a'_m, \ldots,\phantom{1+ {}}a'_{k+1} ,a'_k,\ldots,a'_0;n') 
    & \text{(Eq.\ref{eq:Fmono})}
    \end{array}
    $$
and this completes the claim: if there is an increasing $\mathtt{evalF}$-computation from $\qstate_\mathrm{evalF}\mathtt{a}_m^{a_m}\cdots \mathtt{a}_0^{a_0}  \mathtt{n}^n$ to $\qstate_\mathrm{evalF}\mathtt{a}_m^{a_m'}\cdots \mathtt{a}_0^{a_0'}  \mathtt{n}^{n'} $, then the value $F(a_m,\ldots,a_0,n)$ is less than or equal to $F(a_m',\ldots,a_0',n')$.

The submachine $\mathtt{backF}$ is defined to be the inverse machine $\mathtt{evalF}^{-1}$, after having renamed fresh states appropriately. 
By Proposition~\ref{prop:invertedCM}(1), this machine is equivalent to running the reverse of $\mathtt{evalF}$ (in a non-deterministic way), effectively applying rules (D0)--(D2) read from right-to-left. 
By Proposition~\ref{prop:invertedCM}(2), the inequalities of $F$ for lossy computations are maintained (by the argument for increasing $\mathtt{evalF}$), and similarly for increasing computations the inequalities are maintained (by the argument for lossy $\mathtt{evalF}$).

We are now ready to prove the stated equivalences.
On the one hand, it is not difficult to see that items (1) and (2) are equivalent. Indeed, the machine $\acm$ terminates in space bounded by $Ack(m)$ iff the budget machine satisfies $\qin^\mathrm{B}\mathtt{n}^{Ack(m)} \to_{\acm^\mathrm{B}}\qfin^\mathrm{B}\mathtt{n}^{Ack(m)}$. 
The latter statement is equivalent to item (2) since, by the very construction of $\mathtt{evalF}$ and $\mathtt{backF}$, both $\qstate_\mathrm{evalF}\mathtt{a}_m\mathtt{a}_0^m \to_\mathtt{evalF} \qin^\mathrm{B}\mathtt{n}^{Ack(m)}$ and $\qfin^\mathrm{B}\mathtt{n}^{Ack(m)}\to_\mathtt{backF}
\qstate_\mathrm{backF}\mathtt{a}_m\mathtt{a}_0^m$ hold.

For the remaining equivalences, first note that item (2) implies both items (3) and (4) \emph{a fortiori} since the lossy/increasing machines are no less expressive. 
We will show each of (3) and (4), individually, imply item (1).

We assume that $\qstate_\mathrm{evalF}\mathtt{a}_m^{1}\mathtt{a}_0^m\to_{\acm(m)}^\simpeq \qstate_\mathrm{backF}\mathtt{a}_m^{1}\mathtt{a}_0^m$,
where $\simpeq$ denotes either lossy or increasing.
By construction, such a computation must pass through the budget submachine, i.e., there must exist numbers $n,n'\in \mathbb{N}$ such that
    \begin{equation}\label{eq:Ackglitchcomp}
    \qstate_\mathrm{evalF} \mathtt{a}_m^1 \mathtt{a}_0^m 
    \quad
    \to_\mathtt{evalF}^\simpeq 
    \quad
    \qin^\mathrm{B}\mathtt{n}^n
    \quad
    \to_{\acm^\mathrm{B}}^\simpeq
    \quad
    \qfin^\mathrm{B}\mathtt{n}^{n'}
    \quad
    \to_\mathtt{backF}^\simpeq
    \quad
    \qstate_\mathrm{backF} \mathtt{a}_m^1 \mathtt{a}_0^m.
    \end{equation}

Consider first the case when $\simpeq$ is lossy. 
As described above for the budget machine, it must be that $n\geq n'$. 
Recall also that, in the programs $\mathtt{evalF}$ and $\mathtt{backF}$, lossy glitches can only decrease the value of the function $F$.
Consequently, 
    $$
    {Ack}(m)=F(1,0^{m-1},m;0)\geq F(0^{m+1};n)\geq F(0^{m+1};n')\geq F(1,0^{m-1},m;0),
    $$
it therefore follows that $n=n'={Ack}(m)$. 
On the one hand, the fact that $n'=n$ implies that the central computation in Eq.~\eqref{eq:Ackglitchcomp} is witnessed with no lossy glitches, due to the behavior of the budget machine i.e., $\qin^\mathrm{B}\mathtt{n}^n
\to_{\acm^\mathrm{B}}\qfin^\mathrm{B}\mathtt{n}^{n}$.
As described above for the budget machine, this means that $\acm$ terminates in space bounded by $n=Ack(m)$. So item (1) holds.

The case where $\simpeq$ is increasing is similar. 
Indeed, if $\simpeq$ is increasing then it must be that $n\leq n'$. Again, by the properties of increasing $\mathtt{evalF}$ and $\mathtt{backF}$, and the behavior of $F$, we find $Ack(m)\leq n$ and $ n'\leq  Ack(m)$. Therefore, $n=n'=Ack(m)$ and item (1) follows by the same reasoning above. 
\end{proof}

We now recast this result in our framework. 
For a $3$-CM $\acm=(\Reg,\State,\Inst)$, we write $\acm\models \ref{ACP}$ to denote that $\acm$ satisfies the \ref{ACP}, i.e., $\acm$ terminates in space bounded by $Ack(|\State|)$.

\begin{corollary}\label{cor:ackmachines}
If $\acm$ is a $3$-CM, then there is a CM $\mathtt{Ack}(\acm)$ such that the following are equivalent:
    \begin{enumerate}
        \item\label{item:ackmachines_ACP} $\acm \models \ref{ACP}$.
        \item\label{item:ackmachines_CM} $\mathtt{Ack}(\acm)$ terminates.
        \item\label{item:ackmachines_iCM} $\mathsf{i}\mathtt{Ack}(\acm)$ terminates.
        \item\label{item:ackmachines_tCM} {
        $\mathsf{t}\mathtt{Ack}(\acm)$ terminates.
        }
    \end{enumerate}
Moreover, the construction is such that $|\mathtt{Ack}(\acm)|$ grows linearly with $|\acm|$ as input. 
\end{corollary}

\begin{proof}
Let $m$ be the total number of states of $\acm$, and let $\acm(m) = (\Reg,\State,\Inst)$ be the CM of Proposition~\ref{prop:AckCMs}.
The machine $\mathtt{Ack}(\acm)$ is obtained by simply fusing an initialing machine $\mathtt{In}$ and a finalizing machine $\mathtt{Fin}$, each with a fresh set of $2(m+2)$-many states and $2(m+1)$-many instructions, in order to compute and empty the input and output configurations of $\acm(m)$,  summarized by the following specification:
$$
\scalebox{.6}{
    \begin{tikzpicture}[node distance = 1cm]
    \node[state] (qin) {$~~\qin ~~$};
    \node[state, right = of qin] (q0) {$\qstate_0$};
    \node[right = of q0] (dots) {$\cdots$};
    \node[state, right = of dots] (qm) {$\qstate_{m}$};
    \node[state, right = of qm] (qm1) {$\qstate_\mathrm{evalF}$};
    \begin{scope}[node distance = 1.6cm]
        \node[state, right = of qm1] (qm2) {$\qstate_\mathrm{backF}$};  
    \end{scope}
    \node[state, right = of qm2] (qm') {$\qstate'_m$};
    \node[right = of qm'] (dots') {$\cdots$};
    \node[state, right = of dots'] (q0') {$\qstate'_0$};
    \node[state, right = of q0'] (qout) {$~~\qfin ~~$};
    \draw[-{Latex[width=3mm]}, >=Latex]
    (qin) edge[above] node{${+}\mathtt{a}_m$} (q0)
    (q0) edge[above] node{${+}\mathtt{a}_0$} (dots)
    (dots) edge[above] node{${+}\mathtt{a}_0$} (qm)
    (qm) edge[above] node{${+}\mathtt{a}_0$} (qm1)
    (qm2) edge[above] node{$\Dec\mathtt{a}_0$} (qm')
    (qm') edge[above] node{$\Dec\mathtt{a}_0$} (dots')
    (dots') edge[above] node{$\Dec\mathtt{a}_0$} (q0')
    (q0') edge[above] node{$\Dec\mathtt{a}_m$} (qout)
    (qm1) edge[above,dashed] node{$\acm(m)$} (qm2)
    (qm1) edge[ bend right,dashed,above] node{$\vdots$}  (qm2)
    (qm1) edge[ bend left,dashed] (qm2)
    ;
    \end{tikzpicture}
    }
$$
It is clear that $\qin \to_\mathtt{In} \qstate_\mathrm{evalF} \mathtt{a}_m\mathtt{a}_0^m$ and $\qstate_\mathrm{backF}\mathtt{a}_m\mathtt{a}_0^m\to_\mathtt{Fin}\qfin$, so:  $\mathtt{Ack}(\acm)$ terminates iff $\qstate_\mathrm{evalF} \mathtt{a}_m\mathtt{a}_0^m\to_{\acm(m)}\qstate_{\mathrm{backF}}\mathtt{a}_m\mathtt{a}_0^m$ iff  $\acm$ terminates in space bounded by $Ack(m)$ (by Proposition~\ref{prop:AckCMs}\eqref{item:AckCMs_ACP}). 

Moreover, it is easy to verify that the following hold for every $x\in \Reg^*$:
    $$
    \begin{array}{r c l c r c l}
    \qin 
    &\to_\mathtt{In}^\mathrm{lossy} &\qstate_\mathrm{evalF} x 
    &\iff& 
    \qstate_\mathrm{evalF}\mathtt{a}_m\mathtt{a}_0^m
    &\to^\mathrm{lossy}
    &\qstate_\mathrm{evalF} x; \quad \text{and} \\
    \qin 
    &\to_\mathtt{In}^\mathrm{inc} &\qstate_\mathrm{evalF} x 
    &\iff& 
    \qstate_\mathrm{evalF}\mathtt{a}_m\mathtt{a}_0^m
    &\to^\mathrm{inc}
    &\qstate_\mathrm{evalF} x \quad 
    \end{array}
    $$
as any lossy (or increasing) glitch in a computation witnessing $\qin\to_\mathtt{In}^\mathrm{lossy} \qstate_\mathrm{evalF}x$ can be permuted to occur later while the computation is in state $\qstate_\mathrm{evalF}$. 
Moreover, as $\mathtt{Fin}$ and $\mathtt{In}^{-1}$ define the same machine, up to the renaming of states ($\qstate_i' := \qstate_i^{-1}$ and $\qstate_{\mathrm{backF}}:= \qstate_{\mathrm{evalF}}^{-1}$), the following are immediate from Lemma~\ref{prop:invertedCM} for all $x\in \Reg^*$:
    $$
    \begin{array}{r c l c r c l}
    \qstate_\mathrm{backF}x 
     &\to_\mathtt{Fin}^\mathrm{inc}
     &\qfin
     &\iff&
     \qstate_\mathrm{backF}x
     &\to^\mathrm{lossy} 
     &\qstate_\mathrm{backF} \mathtt{a}_m\mathtt{a}_0^m; \quad \text{and} \\
     \qstate_\mathrm{backF}x 
     &\to_\mathtt{Fin}^\mathrm{lossy}
     &\qfin
     &\iff&
     \qstate_\mathrm{backF}x
     &\to^\mathrm{lossy} 
     &\qstate_\mathrm{backF} \mathtt{a}_m\mathtt{a}_0^m.
    \end{array}
    $$

By Lemma~\ref{lem:IntTrivACM}, $\mathsf{i}\mathtt{Ack}(\acm)$ terminates iff $\mathtt{Ack}(\acm)_\mathrm{lossy}$ terminates. 
By the observations above, this is equivalent to $\qstate_\mathrm{evalF}\mathtt{a}_m\mathtt{a}_0^m\to_{\mathtt{Ack(\acm)}}^\mathrm{lossy}  \qstate_\mathrm{backF}\mathtt{a}_m\mathtt{a}_0^m$.
By the construction of $\mathtt{Ack}(\acm)$, this is equivalent to 
$\qstate_\mathrm{evalF}\mathtt{a}_m\mathtt{a}_0^m\to_{\acm(m)}^\mathrm{lossy} \qstate_\mathrm{backF}\mathtt{a}_m\mathtt{a}_0^m$, which is in turn equivalent to item \eqref{item:ackmachines_ACP} by Proposition~\ref{prop:AckCMs}\eqref{item:AckCMs_iCM}. 
Similarly,  $\mathsf{t}\mathtt{Ack}(\acm)$ terminates iff $\mathtt{Ack}(\acm)_\mathrm{inc}$ terminates. 
By the observations above, this is equivalent to $\qstate_\mathrm{evalF}\mathtt{a}_m\mathtt{a}_0^m\to_{\mathtt{Ack(\acm)}}^\mathrm{inc}  \qstate_\mathrm{backF}\mathtt{a}_m\mathtt{a}_0^m$.
By the construction of $\mathtt{Ack}(\acm)$, this is equivalent to 
$\qstate_\mathrm{evalF}\mathtt{a}_m\mathtt{a}_0^m\to_{\acm(m)}^\mathrm{inc} \qstate_\mathrm{backF}\mathtt{a}_m\mathtt{a}_0^m$, which is in turn equivalent to item \eqref{item:ackmachines_ACP} by Proposition~\ref{prop:AckCMs}\eqref{item:AckCMs_tCM}.

Lastly, by Proposition~\ref{prop:AckCMs}, the (size of the) construction for $\acm(m)$ grows linearly with input $m$ and $|\acm|$. It is clear that also the size of $\mathtt{Ack}(\acm)$ is linear in  $|\acm|$ by our specification.
\end{proof}

\section{Lower (and tight) bounds for joinand-decreasing equations}
We are now able to offer an extension of Proposition~\ref{thm:expansiveterm}, which specializes to Proposition~\ref{thm:expansiveterm} as a different proof than that of Urquhart. 
The extension covers {\joindecreasing} equations, as well. 

\begin{theorem}\label{thm:lossyadmissmach}
There is a class $\acmclass$ of ACMs for which the following hold:
    \begin{enumerate}
    \item $\acmclass$ is $\eqset$-termination admissible for every set $\eqset$ of non-integral {\joindecreasing} simple equations and the termination problem for $\acmclass$ is \ACK-hard. 
    \item $\acmclass$ is $\eqset$-termination admissible for every set $\eqset$ of {\joinincreasing} simple equations and the termination problem for $\acmclass'$ is \ACK-hard.  
    \end{enumerate}
\end{theorem}
\begin{proof}
For a CM $\acm$, let $\mathtt{Ack}(\acm)$ be the CM from Corollary~\ref{cor:ackmachines}. 
Note that for every  CM $\acm$ and for every set $\eqset$ of simple equations:
    $$
    \arraycolsep=1pt
    \begin{array}{r c l l}
    \text{The CM }\mathtt{Ack}(\acm) \text{ terminates} 
    &\iff& \text{The ACM }\mathtt{Ack}(\acm)_\vee \text{ terminates} & \text{(Prop.~\ref{prop:zerotest}(3))}
    \\
    &\implies& \text{The $\eqset$ACM }\eqset\mathtt{Ack}(\acm)_\vee \text{ terminates}
    \end{array}
    $$
where the last implication follows from the fact every computation for $\mathtt{Ack}(\acm)_\vee$ is also a computation for $\eqset\mathtt{Ack}(\acm)_\vee$. 
By the same reasoning, if the CM $\mathtt{Ack}(\acm)^{-1}$ terminates, then the $\eqset$ACM $\eqset(\mathtt{Ack}(\acm)^{-1})_\vee$ terminates.

(1) If $\acm$ is a CM and $\eqset$ is a set of non-integral {\joindecreasing} simple equations, then
    $$
    \arraycolsep=1pt
    \begin{array}{r c l l}
        \text{The $\eqset$ACM }\eqset\mathtt{Ack}(\acm)_\vee \text{ terminates}
         &\implies&
         \text{The $\mathsf{i}$CM } \mathsf{i}\mathtt{Ack}(\acm) \text{ terminates}
         &
         \text{(Lem.~\ref{lem:branchinglem})}
         \\
         & \iff &
         \text{The CM }\mathtt{Ack}(\acm) \text{ terminates}
         & \text{(Cor.~\ref{cor:ackmachines})}
    \end{array}
    $$
Consequently, $\mathtt{Ack}(\acm)_\vee$ terminates iff $\eqset\mathtt{Ack}(\acm)_\vee$ terminates.  
Therefore, the class $\acmclass$  of ACMs $\mathtt{Ack}(\acm)_\vee$, where $\acm$ is a 3-CM,  is $\eqset$-termination admissible for any set $\eqset$ of non-integral {\joindecreasing} simple equations. 

(2) If $\acm$ is a CM and  $\eqset$ is a set of {\joinincreasing} simple equations, then 
    $$
    \arraycolsep=1pt
    \begin{array}{r c l l}
        \text{The $\eqset$ACM }\eqset\mathtt{Ack}(\acm)_\vee \text{ terminates}
         &\implies&
         \text{The $\mathsf{t}$CM } \mathsf{t}\mathtt{Ack}(\acm) \text{ terminates}
         &
         \text{(Lem.~\ref{lem:branchinglem})}
         \\
         & \iff &
         \text{The CM }\mathtt{Ack}(\acm) \text{ terminates}
         & \text{(Cor.~\ref{cor:ackmachines})}
    \end{array}
    $$
Consequently, $\mathtt{Ack}(\acm)^{-1}_\vee$ terminates iff $\eqset\mathtt{Ack}(\acm)^{-1}_\vee$ terminates. 
Therefore, the class $\acmclass$  of ACMs $\mathtt{Ack}(\acm)_\vee$, where $\acm$ is a 3-CM,  is $\eqset$-termination admissible for any set $\eqset$ of {\joinincreasing} simple equations. 
To show that the termination problem for $\acmclass$ is \ACK-hard, we argue that the \ref{ACP} can be reduced to it; this suffices, since the latter is  \ACK-hard by Theorem~\ref{t: ACP is Ack-hard}. 
We mention that the distinction between time and space complexity disappears at the level of Ackermann, i.e., Ackermann time is equal to Ackermann space, so we will not distinguish the two below.
First recall that by Corollary~\ref{cor:ackmachines}, $\acm\models\ref{ACP}$ iff
$\mathtt{Ack}(\acm)$ terminates and by Prop.~\ref{prop:zerotest}(3) this happens
iff $\mathtt{Ack}(\acm)_\vee$ terminates. Also, by the arguments above this happens iff $\eqset\mathtt{Ack}(\acm)_\vee$ terminates. 
Now we give the reduction of a 3-CM satisfying $\ref{ACP}$ to a member of  $\acmclass$ terminating.

Given a 3-CM $\acm$ we run a program that constructs the CM $\mathtt{Ack}(\acm)$. By Corollary~\ref{cor:ackmachines}, this process uses space that is linear relative to the the size of $\acm$. 
Then we  run a program that constructs the ACM $\mathtt{Ack}(\acm)_\vee$ from $\mathtt{Ack}(\acm)$, and by Proposition~\ref{prop:zerotest} this process uses space that is linear relative to the the size of $\mathtt{Ack}(\acm)$, hence linear relative to the the size of $\acm$. 
Then we run the machine $\mathtt{Ack}(\acm)_\vee$ from its initial state, allowing $\eqset$ instructions, and check for termination, i.e., we check whether $\eqset\mathtt{Ack}(\acm)_\vee$ terminates. 
Since, by the proceeding paragraph, checking termination for $\eqset\mathtt{Ack}(\acm)_\vee$ is equivalent to checking whether $\acm\models\ref{ACP}$, the overall process we described  checks whether $\acm\models\ref{ACP}$. 
If checking termination for members of  $\acmclass$ had complexity properly smaller than Ackermann, then the overall complexity of the process we described (for checking $\acm\models\ref{ACP}$) would be less than Ackermannian (since linear space is primitive recursive, hence properly smaller than Ackermannian); this would contradict that $\ref{ACP}$ is \ACK-hard.
Hence checking termination for members of $\acmclass$ is \ACK-hard.
\end{proof}

We are now finally ready to apply Theorem~\ref{thm:qeLB}. 
A residuated lattice is called \emph{almost integral} if it satisfies every non-integral {\joindecreasing} simple equation; we denote the set of these equations by $\eqset_\mathrm{dec}$. 
Note that commutativity is equivalent to the non-integral simple equation $xy\leq yx$, and is therefore contained in $\eqset_\mathrm{dec}$.
We will denote the variety of almost integral RLs by $\UniCRLV_\mathsf{a}$, so 
%$\UniCRLV_\mathsf{a}:=\UniRLV+ \eqset_\mathrm{dec}$; 
\begin{equation}\label{CRLa}
\UniCRLV_\mathsf{a}:=\UniRLV+ \eqset_\mathrm{dec}
\end{equation}
In Theorem~\ref{the:crla-char} we provide an  alternative characterization, which also explains the subscript `$\mathsf a$'.

By Theorem~\ref{thm:lossyadmissmach}, together with Theorem~\ref{thm:qeLB} and Corollary~\ref{cor:qeLB}, we get the main result of this section.

\begin{theorem}\label{t: ackhardJD}
Every quasivariety of residuated lattices that contains $\UniCRLV_\mathsf{a}$ has an \ACK-hard quasiequational theory. 
In particular, this holds for any quasivariety of residuated lattices that contains $\UniCRLV+\eqset$, for some $\eqset \subseteq \eqset_\mathrm{dec}$ (hence also for any quasivariety of residuated lattices that contains $\UniCRLV+\simpeq$, for some $\simpeq \in \eqset_\mathrm{dec}$).
\end{theorem}

Later on, we will show that the variety $\UniCRLV_\mathsf{a}$ of almost integral RLs is finitely based and contains (properly) all integral CRLs.

\begin{corollary}\label{c: ackhardJD_logic}
Let $\mathbf{L}$ be any logic in the interval of axiomatic extensions between $\UniFLExtLogic{}$ and the extension $\UniFLeExtLogic{}$ by all non-integral {\joindecreasing} simple rules. 
Then $\mathbf{L}$ has \ACK-hard deducibility, witnessed in its $\{\lor,\cdot,1 \}$-fragment. 
In particular, this holds for $\mathbf{L}=\UniFLeExtLogic{\UniWeakWProp{m}{n}}$.
\end{corollary}

Theorem~\ref{t: ackhardJD} and Corollary~\ref{c: knotweak+wc<Ack}(2) yield a  tight complexity bound for the quasiequational theories of many varieties of (pointed) residuated lattices.

\begin{corollary}\label{c: knotweak+wc=Ack}
The complexity of the quasiequational theory is exactly Ackermannian (i.e., $\UniFGHProbOneAppLevel{\omega}$) for every variety of (pointed) residuated lattices axiomatized by any set of $\mathcal{N}_2$ equations (in particular, by any set of $\{\jn, \cdot, 1\}$-equations) that (a) contains a knotted weakening equation and a weak commutativity equation and (b) every equation in the set is a consequence of the conjunction of commutativity and some set of equations from $\eqset_\mathrm{dec}$.

In particular, this holds for every variety axiomatized by a knotted weakening equation and a non-empty set of weak commutativity equations.
\end{corollary}

\begin{corollary}\label{c: knotweak+wc=Ack Logics}
The complexity of deducibility is exactly Ackermannian (i.e., $\UniFGHProbOneAppLevel{\omega}$) for every extension of $\m {FL}$ axiomatized by any set of $\mathcal{N}_2$ formulas that (a) contains a knotted weakening formula and a weak commutativity formula  and (b) every formula in the set is a consequence of the conjunction of exchange and some set of formulas from $\eqset_\mathrm{dec}$.

In particular, this holds for every extension axiomatized by a knotted weakening axiom and a non-empty set of weak commutativity axioms.
\end{corollary}

\begin{remark}
Note that even though the equation $\mathsf{t}$ is \joinincreasing, it is not simple, so it is not covered by Theorem~\ref{thm:lossyadmissmach}(2). 
Moreover, $\mathsf{i}$ is explicitly not covered by Theorem~\ref{thm:lossyadmissmach}(1). 

On the other hand, in view of  Proposition~\ref{prop:AlterEqs}, $\mathsf{CRL_c}$ is the smallest variety defined by {\joinincreasing} $\{\jn, \cdot, 1\}$-equations which Theorem~\ref{thm:lossyadmissmach}(2) covers. 
Also, by definition, $\mathsf{CRL_a}$ is the smallest variety defined by {\joindecreasing} $\{\jn, \cdot, 1\}$-equations  which the Theorem~\ref{thm:lossyadmissmach}(1) covers.

We now mention why for the cases of  $\mathsf{i}$ and  $\mathsf{t}$ the strategy in the proof of Theorem~\ref{thm:lossyadmissmach} does not work, i.e., why the following two series of reductions, indicated by $\rightsquigarrow$, break:
\begin{center}
    lossy/increasing CMs $\rightsquigarrow$ lossy/increasing ACMs $\rightsquigarrow$ integral/trivial residuated lattices.
\end{center}
Recall that by Lemma~\ref{lem:IntTrivACM}, the termination problem for lossy/increasing CMs coincides with that of $\mathsf{i}$CMs/$\mathsf{t}$CMs, and the same holds for ACMs, so we blur the distinction.
For the first series of reductions (i.e., the one relating lossy machines with integral RLs), in light of Lemma~\ref{lem:nonintZT} the analogue of Proposition~\ref{prop:zerotest}(3) for $\mathsf{i}$CMs and $\mathsf{i}$ACMs does not hold for lossy machines, (i.e., when taking $\eqset=\{\mathsf{i}: x\leq 1\}$ in the proposition), and this creates a potential issue with the first reduction above. 
In fact, this is an actual issue and the first reduction breaks, since the termination problem for lossy CMs is Ackermann hard, as we mentioned, but the termination problem for lossy ACMs is $2$-\EXPTIME-complete, as we explain in the proposition below; 
by the way the second reduction does not break, as deducibility in $\UniFLExtLogic{\mathbf{i}}$ is at least $2$-\EXPTIME-hard by \cite{GreatiRamanayake2024} (it was actually proved to be hyper-Ackermannian-complete).
(We also note that the commutative version of the second reduction does not break either, as deducibility $\UniFLeExtLogic{\mathbf{w}}$ is \TOWER-complete \cite{tanaka2022}.) 

For the second series of reductions, Lemma~\ref{lem:nonintZT} shows that $\eqset$-admissibility works for $\mathsf{t}: 1\leq x$ and actually the termination problem for increasing ACMs is \ACK-complete (see \cite[Theorem~7.2]{LazSchmitz2015}), so the first reduction holds. 
However, the equation $\mathsf{t}$ is not simple and therefore it is not supported by the frames construction (Lemma~\ref{WmInclus} is not applicable), and this creates a potential problem with the second reduction. 
Actually, the second reduction breaks: since every quasiequation holds in a trivial residuated lattice, the theory of trivial residuated lattices has constant-time complexity.

Therefore, not only do the reductions in the proof of Theorem~\ref{thm:lossyadmissmach} break for  $\mathsf{i}$ and  $\mathsf{t}$, but actually the complexities for these two cases (between machines and varieties) are not equal.
 
Actually, it is only in these boundary cases where the above sequence of reductions fails in the context of subvarieties of $\mathsf{CRL}$ axiomatized by {\joindecreasing} or {\joindecreasing} equations. 
In particular, we show in Lemma~\ref{lem:FLeiEqTh} that integrality (in the context of commutativity) implies all other joinand-decreasing equations and in Theorem~\ref{t: CRL_a} that almost integrality defines a cover $\mathsf{CRL_a}$ of integrality $\mathsf{CRL_i}$ among $\{\jn, \cdot, 1\}$-equations hence among {\joindecreasing} equations. 
In that respect, the `next' {\joindecreasing} equation outside the scope of Theorem~\ref{t: ackhardJD} is $\mathsf{i}$, and it is (up to equivalence) the only {\joindecreasing} equation for which the result fails. 
Likewise, as a consequence of Proposition~\ref{prop:AlterEqs}\eqref{altereqs2}, contraction defines a cover $\mathsf{CRL_c}$ of $\mathsf{t}$ in the context of {\joinincreasing} equations, since every {\joinincreasing} $\{\jn, \cdot, 1\}$-equation is either trivializing, or it is equivalent to a conjunction of simple ({\joinincreasing}) equations. 
In that respect, the `next' {\joinincreasing} equation outside the scope of Theorem~\ref{t: ackhardJD} is $\mathsf{t}$, and it is (up to equivalence) the only {\joinincreasing} equation for which the result fails.
\end{remark}

As promised, we now show that the termination problem for $\mathsf{i}$ACMs is $2$-\EXPTIME-complete. 
In \cite[\S 3.3.1]{LazSchmitz2015}, this result is attributed to \cite{CourtSchmitz2014}, even though it is neither stated explicitly nor proved. We provide the details below. 

\begin{proposition}[\cite{CourtSchmitz2014}]
\label{iACMbound}
The termination problem for $\mathsf{i}$ACMs is $2$-\EXPTIME-complete. 
\end{proposition}
\begin{proof}
In \cite[Theorem~3.1]{CourtSchmitz2014} it is shown that \emph{state reachability} in AVASS is in $2$-\EXPTIME. 
We note an AVASS is defined as an ACM without designated initial and final states.
State reachability (cf. \cite[\S2.2.2]{CourtSchmitz2014}), also known as \emph{coverability}, refers to the question of whether, given an AVASS $\acm$ and states $\qstate$ and $\qstate'$ as input, there a computation $\qstate\leq_\acm \qstate' x_1 \vee \cdots \vee \qstate' x_n$ for some $n\in \mathbb{Z}^+$. 
Any AVASS can be made into an ACM by designating two of its states to be initial and final (e.g., taking $\qin := \qstate$ and $\qfin := \qstate'$), respectively. 
So translated to our setting, coverability is the question: given an ACM $\acm$ as input, does $\qin\leq_\acm \qfin x_1 \vee \cdots \vee \qfin x_n$ hold for some $n\in \mathbb{Z}^+$ and register-words $x_1,\ldots,x_n\in \Reg^*$? 
We will show that it is equivalent to termination of $\mathsf{i}\acm$, i.e., $\qin \leq_{\mathtt{i}\acm} \ufin$, where $\ufin$ is a final ID.

For the forward direction (coverability of the ACM $\acm$ implies termination of $\mathsf{i}\acm$), first note that $\qfin x_1 \vee \cdots\vee \qfin x_n \leq^\mathtt{i} \ldots \leq^\mathtt{i} \qfin \vee\cdots\vee \qfin$, for all  $x_1,\ldots,x_n\in \Reg^*$. 
So, if $\qin\leq_\acm \qfin x_1 \vee \cdots\vee \qfin x_n$, we get $\qin\leq_{\mathsf{i}\acm} \qfin \vee \cdots\vee \qfin $, hence $\mathsf{i}\acm$ terminates. 
This short argument is mentioned in \cite[\S 3.2.2]{LazSchmitz2015} in terms of lossy termination.

Towards establishing the converse, we prove first that the computation relation $\leq_{\mathsf{i}\acm}$ for $\mathsf{i}\acm$ coincides with the composition ${\leq_\acm} \circ {\leq_\mathsf{i}}$, where $\leq_\mathsf{i}$ is the transitive closure of $\leq^\mathsf{i}$; we will show now the non-trivial inclusion:  $\leq_{\mathsf{i}\acm}$ is contained in ${\leq_\acm} \circ {\leq_\mathsf{i}}$.

Indeed, suppose $u\leq^\mathsf{i} u' \leq_\acm w$ for some terms $u,u',w\in A_\acm$.
Therefore, $u= cx \vee u''$, where $(cx,c)$ is the active component and $u''$ is the passive component of the instance $u\leq^\mathsf{i} u'$, for some  $c,x\in (\State\cup\Reg)^*$ and $u''\in A_\acm^\bot$; also, $u'= c \vee u''$.
Since $u'\leq_\acm w$, from Lemma~\ref{lem:computationtree}\eqref{i2:computationtree} we obtain that $c\leq_\acm w'$ and $u''\leq_\acm w'' $ for some terms with $w'\vee w'' = w$. 
We set $v= w'x \vee w''$. 
On the one hand, $v\leq_\mathsf{i} w$ is obtained from repeated applications of $\leq^\mathsf{i}$ by removing $x$ from each joinand of the (fully distributed) term $w'x$ to obtain $w'$. 
On the other hand, since $\leq_\acm$ is compatible with multiplication, we have $cx\leq_\acm w'x$, and since $\leq_\acm$ is transitive and also compatible with $\vee$, we conclude
    $$
    u= cx \vee u'' \leq_\acm w'x \vee u''\leq_\acm w'x\vee w''=v.
    $$
Thus $u\leq_\acm v \leq_\mathsf{i} w$, establishing this intermediary claim. 
By induction, it follows that ${\leq_\mathsf{i}}\circ{\leq_\acm}$ is contained in ${\leq_\acm}\circ{\leq_\mathsf{i}}$. 

Now, we verify that the computation relation for $\mathsf{i}\acm$ coincides with ${\leq_\acm} \circ {\leq_\mathsf{i}}$. 
Indeed, recall that $\leq_{\mathsf{i}\acm}$ is the transitive closure of ${\leq_\acm} \cup {\leq^\mathsf{i}}$, hence also the transitive closure of ${\leq_\acm} \cup {\leq_\mathsf{i}}$. 
From the definition of the transitive closure and ${\leq_\mathsf{i}}\circ{\leq_\acm} \subseteq {\leq_\acm}\circ{\leq_\mathsf{i}}$, it follows that $\leq_{\mathsf{i}\acm}$ is contained in ${\leq_\acm}\circ{\leq_\mathsf{i}}$. 

Finally, we show $\mathsf{i}\acm$ termination implies $\acm$ coverability. 
Suppose $\qin \leq_{\mathsf{i}\acm} \ufin$. 
Then there is a term $v\in A_\acm$ with $\qin\leq_\acm v$ and $v\leq_\mathsf{i}  \ufin$. 
The former fact implies $v$ is an ID by Lemma~\ref{lem:branchconfigurations}(1), and the latter fact, in addition, entails $v = \qfin x_1 \vee\cdots \qfin x_n$ for some register words $x_1,\ldots, x_n\in \Reg^*$.
Hence $\mathsf{i}\acm$ termination implies $\acm$ coverability.
\end{proof}

\section{Almost integrality}\label{sec: CRLa}
Recall from above that the variety of almost integral RLs is axiomatized, relative to RL, by the set of all non-integral {\joindecreasing} simple equations.
The following theorems establish that the variety of almost integral RLs is finitely based and a cover of the variety of integral CRLs in the subvariety lattice of RLs  restricted to varieties axiomatized by $\{\vee,\cdot,1\}$-equations.
\begin{theorem}\label{the:crla-char}
The variety of almost integral RLs is equivalent to the variety axiomatized, relative to $\UniCRLV$, by any of the items below:
\begin{enumerate}
    \item The (in)equation $\mathsf{a}: x^2 yz \leq xy \vee xy^2  z^2$.
    \item The simple equation $\mathsf{a}_\ell:wxyz\leq wy \vee wy^2z^2 \vee xy \vee xy^2z^2$.
    \item The simple equations $\mathsf{m}: xy\leq x\vee y$ and $\mathsf{b}:yz\leq y \vee y^2 z^2$.
\end{enumerate}
\end{theorem}
\begin{proof}
We first show that the equations in each item define the same subvariety of $\UniRLV$, and thus the same subvariety of $\UniCRLV$. 
On the one hand, $\mathsf{a}_\ell$ is simply the linearization of $\mathsf{a}$, as described in Lemma~\ref{l: N_2^-0}, and thus the two define the same subvariety. 
On the other hand, $\mathsf{a}_\ell$ can be `factored' as the `product' of $\mathsf{m}$ and $\mathsf{b}$:
    $$
    wxyz = wx\cdot yz \leq^{\mathsf{m}+\mathsf{b}} (w\vee x)(y\vee y^2z^2) =_\UniRLV wy \vee wy^2z^2 \vee xy \vee xy^2z^2 .
    $$
So $\UniRLV_{\mathsf{m}\mathsf{b}}\models \mathsf{a}_\ell$. 
Conversely, $\UniRLV_{\mathsf{a}_\ell}\models \mathsf{m}$ via the substitution $y,z\mapsto 1$ and $\UniRLV_{\mathsf{a}_\ell}\models \mathsf{b}$ via $w,x\mapsto 1$. 
Therefore, items 1--3 define the same subvariety of $\UniRLV$, which we denote by $\UniCRLV_\mathsf{a}$ (the next paragraphs explain why we gave the same name to the variety of almost integral RLs).

We now argue that $\UniCRLV_\mathsf{a}$ coincides with the variety of almost integral RLs. 
On the one hand, since the simple equations $\mathsf{a}$ and commutativity $xy\leq yx$ are non-integral {\joindecreasing} equations, $\UniCRLV_\mathsf{a}$ contains the variety of almost integral RLs by Proposition~\ref{prop:AlterEqs}\eqref{altereqs2}.
Towards establishing the converse, we first prove the following intermediary claim: for each positive integer $n$,
    \begin{enumerate}[(a)]
        \item $\UniCRLV_\mathsf{a}\models 
        %\NOTE{N}{
        x_1\cdots x_n \leq x_1 \jn \cdots \jn x_n 
        %\sout{\prod_{i=1}^n x_i \leq \bigvee_{i=1}^n x_i}}
        $, and
        \item $\UniCRLV_\mathsf{a}\models yz \leq y \vee (yz)^n$.
    \end{enumerate}    
The simple equation in item (a) is the linearization of the knotted equation $x^n\leq x$. 
For $n=1$, the equation always holds, while for $n>1$ it is a consequence, modulo the theory of $\UniRLV$, of the equation $x^2\leq x$; the linearization of $x^2\leq x$ is the mingle equation $\mathsf{m}$. 
Since $\mathsf{a}$ implies $\mathsf{m}$ in the presence of commutativity, Claim (a)  follows. 

For (b), let $k$ be such that $2^k\geq n$, and observe the following (modulo the theory of $\UniCRLV$)
    $$
    \arraycolsep=1pt
    \begin{array}{r l l}
    yz
    \leq^\mathsf{b}
    y \vee y^2z^2 
    \leq^\mathsf{b}
    y \vee y^2 \vee y^4z^4
    =^\mathsf{m}
    y \vee y^4z^4
    &\leq^\mathsf{b}& y \vee y^4 \vee y^8z^8\\
    &~\vdots&
    \quad\vdots\\
    &=^\mathsf{m}& 
    y \vee y^{2^k}z^{2^k}
    = y \vee (yz)^{2^k}
    \\
    &\leq^\mathsf{m}& 
    y \vee (yz)^n,
    \end{array}
    $$
thus establishing (b).    
    
Now, let $\simpeq: t_0 \leq t_1\vee \cdots  \vee t_k$ be {\joindecreasing} and non-integral; we will show that $\UniCRLV_\mathsf{a}\models \simpeq$. 
The former condition on $\simpeq$ (i.e., that it is {\joindecreasing}) entails there is a joinand $t_i$ which is a subword (up to commutativity) of the linear term $t_0$; we may assume, without loss of generality, that $t_1$ is such a witness. 
The latter condition (i.e., non-integrality) implies that every variable occurring in $t_0$ occurs in some joinand $t_i$; this is equivalent to the property that $t_0$ is a subword (up to commutativity) of the term $t:= t_1 t_2\cdots t_k$. 
Since $\simpeq$ is a simple equation, and therefore every variable occurring in some joinand $t_i$ occurs in $t_0$, it follows that $t$ is a subword (up to commutativity) of $t_0^m$ for some least positive integer $m$. 
    
Since each variable occurring in $t$ has multiplicity no greater than $m$, and $\UniCRLV_\mathsf{m}\models x^m\leq x^n$ for any $1\leq n\leq m$, it follows that $(\star):\UniCRLV_\mathsf{m}\models t_0^m\leq t$. Since $t_1$ is a subword of $t_0$, $t_0=t_1\cdot s$ for some $s$, modulo commutativity.   
Putting this all together, the following derivation holds in $\UniCRLV_\mathsf{a}$
    $$ 
    \begin{array}{r c l l}
    t_0 
    =
    t_1 s 
    &\leq& 
    t_1 \vee t_1^m{s}^m 
    & \text{(b)}
    \\
    &=& t_1 \vee t_0^m & \text{(commutativity)}
    \\
    &\leq& t_1 \vee t &{(\star)}\\
   &=& t_1 \vee \prod_{i=1}^k t_i&\text{(def. of $t$)}\\
   &\leq& t_1 \vee (t_1 \vee \cdots \vee t_k) & \text{(a)} \\
   &=& t_1 \vee \cdots \vee t_k &\text{(idempotency of $\vee$)}.
    \end{array}
    $$
Hence $\UniCRLV_\mathsf{a}\models \simpeq$. So $\UniCRLV_\mathsf{a}$ is a subvariety of almost integral RLs. 
Therefore the two varieties must coincide.
\end{proof}
We thus have that $\UniCRLV_\mathsf{a}$ is indeed the variety of almost integral residuated lattices, and, from Proposition~\ref{prop:AlterEqs}\eqref{altereqs2}, that
every {\joindecreasing} equation holds in $\UniCRLV_\mathsf{i}$; thus, $\UniCRLV_\mathsf{i}\subseteq \UniCRLV_\mathsf{a} $. 
We claim that this containment is proper and, moreover, that $\UniCRLV_\mathsf{a}$ is the smallest subvariety, axiomatized relative to $\UniRLV$ by $\{\vee,\cdot,1\}$ equations, with this property.

Towards establishing these claims, we first elucidate which simple equations are satisfied in $\UniCRLV_\mathsf{i}$ in the following lemma. 
The same characterization can be obtained via a semantic argument, utilizing the fact that residuated lattices are a conservative extension of idempotent semirings (cf. \cite[Thm.~2.3.4]{gavin2019}), and the variety of integral commutative idempotent semirings satisfies $\simpeq$ iff $\simpeq$ is {\joindecreasing} (cf. \cite[Cor.~4.7(2)]{SpadaStJohn2024}). 
In the interest of completeness, we provide a syntactic proof via algebraization (cf. Section~\ref{s: AlgSem}).

\begin{lemma}\label{lem:FLeiEqTh}
    For every simple equation $\simpeq$, we have  $\UniCRLV_\mathsf{i}\models \simpeq$ iff $\simpeq$ is {\joindecreasing}.
\end{lemma}
\begin{proof}
As mentioned above, Proposition~\ref{prop:AlterEqs}\eqref{altereqs2} verifies the right-to-left implication. 
Let $\simpeq: t_0\leq t_1\vee \cdots\vee t_k$ be a simple equation. 
We will show that $\simpeq$ is {\joindecreasing} via a syntactic argument for the logic $\UniFLeExtLogic{\mathbf{i}}$ (see Figure~\ref{figure-HFLec}).

First, we show that $\UniCRLV_\mathsf{i}\models \simpeq$ iff the sequent $\UniSequent{t_0}{t_1\vee \cdots \vee t_k}$ is derivable in $\UniFLeExtLogic{\mathbf{i}}$. 
Indeed, by residuation, in any given RL, $\simpeq$ is satisfied iff the equation $1\leq t_0\backslash \bigvee_{i=1}^k t_i$ is satisfied. 
By algebraization (Section~\ref{s: AlgSem}), $\UniCRLV_\mathsf{i} \models 1\leq t_0\backslash \bigvee_{i=1}^k t_i$ iff the sequent $\UniSequent{}{t_0\backslash \bigvee_{i=1}^k t_i}$ is derivable in $\UniFLeExtLogic{\mathbf{i}}$. 
As is well known for (extensions) of $\UniFLExtLogic{}$, this last statement is equivalent to the sequent $\UniSequent{t_0}{ \bigvee_{i=1}^k t_i}$ being derivable in $\UniFLeExtLogic{\mathbf{i}}$, which itself is well-known to be equivalent to the derivability of the sequent $\UniSequent{x_1,\ldots, x_n}{\bigvee_{i=1}^k t_i}$; where $n\geq 0$ is the total number of (propositional) variables occurring in $t_0$.
In what follows, we will call such a sequent \emph{linear} and denote it by $S(\simpeq)$, for any $\{\vee,\cdot,1\}$-inequation $\simpeq$ whose left-hand side is a linear monoid term and whose right-hand side is a join of monoid terms. 
Note that this condition is weaker than being simple, as it allows the RHS to contain variables not occurring on the left, as well as the LHS to be the constant-term $1$.

So, to establish the left-to-right implication, it suffices to show that a linear equation $\simpeq$ is a {\joindecreasing} simple equation if the linear sequent $S(\simpeq)$ is derivable in $\UniFLeExtLogic{\mathbf{i}}$. 
So suppose there is a derivation of $S(\simpeq)$ in the logic $\UniFLeExtLogic{\mathbf{i}}$, where $\simpeq$ has a linear left-hand side (note that this condition is weaker than being simple as describe above). 
As the cut rule is an admissible rule in $\UniFLeExtLogic{\mathbf{i}}$, there must be a cut-free derivation of this sequent (see Figure~\ref{figure-HFLec}). We proceed by induction on the height of proof trees of linear sequents. 
    
Suppose the height is $0$. 
Then $S(\simpeq)$ must be an initial sequent. 
There are only two applicable cases: $n=0$ and $k=1$ (corresponding to $\UniSequent{}{1}$), or $n=k=1$ (corresponding to $\UniSequent{p}{p}$). 
This means either $\simpeq$ is the equation $1\leq 1$ or $x\leq x$, both of which are {\joindecreasing} simple equations.

Now suppose the claim holds for any derivable linear sequent with a smaller proof height. 
As the left-hand side of $S(\simpeq)$ consists only of a list of (propositional) variables $x_1,\ldots, x_n$, and the right-hand side is a finite join of formulas consisting only of variables and the symbol $\cdot$, the only possible rules that may result in $S(\simpeq)$ as the conclusion is the left-weakening structural rule $(\mathsf{i})$ or $(\mathsf{e})$, or one of the logical rules $(\cdot\mathrm{R})$ or $({\vee}\mathrm{R})$. 
The particular rule-instances of these cases (cf., Figure~\ref{figure-HFLec}) are summarized, respectively, below:
    $$
    \begin{array}{ c c }
    \infer[(\mathsf{i})]{\UniSequent{\Gamma_1,\Psi,\Gamma_2}{\bigvee_{i=1}^k t_i}}{\UniSequent{\Gamma_1,\Gamma_2}{\bigvee_{i=1}^k t_i}}
    &
    \infer[(\mathsf{e})]{\UniSequent{\Delta_1,\Sigma_1,\Sigma_2,\Delta_2}{\bigvee_{i=1}^k t_i}}{\UniSequent{\Delta_1,\Sigma_2,\Sigma_1,\Delta_2}{\bigvee_{i=1}^k t_i}}
    \\
    \\
    \infer[(\cdot\mathrm{R})]{\UniSequent{x_1,\ldots,x_{m},x_{m+1},\ldots, x_n }{t'_1\cdot t''_1}}{\UniSequent{x_1,\ldots,x_{m} }{t'_1} &\quad  \UniSequent{x_{m+1},\ldots,x_{n} }{t''_1}}
    &
    \infer[(\vee\mathrm{R})]{\UniSequent{x_1,\ldots, x_n}{\varphi_1\vee \varphi_2}}{\UniSequent{x_1,\ldots, x_n}{\varphi_j}}
    \end{array}
    $$
where $\Gamma_1,\Psi,\Gamma_2 = \Delta_1,\Sigma_1,\Sigma_2,\Delta_2 = x_1,\ldots, x_n$; $t_1 = t_1'\cdot t_1''$; and $\bigvee_{i=1}^k t_i = \varphi_{1}\vee\varphi_2$, with $j\in \{1,2\}$.
We note that in each case, the premises of each inference rule instance above consist of linear sequent(s) $S(\simpeq')$ (and $S(\simpeq'')$ in the case of $(\cdot\mathrm{R})$) for some simple equation(s) $\simpeq'$ (and $\simpeq''$). 
By the inductive hypothesis, $\simpeq'$ and $\simpeq''$ are {\joindecreasing} simple equations. 
For the cases $(\mathsf{e})$ and $(\vee\mathrm{R})$, there must be a joinand on the right-hand side of $S(\simpeq')$ whose propositional variables occur at most once on the left-hand side---in either case it follows that this property is inherited by $S(\simpeq)$, which establishes the claim.
For $(\cdot\mathrm{R})$ (where it must be that $k=1$), this same property must hold for $S(\simpeq')$ and $S(\simpeq'')$, and since the set of variables on the left-hand sides of these sequents are disjoint, it must be that each variable occurring in $t_1$ occurs at most once in $x_1,\ldots, x_n$. 
Lastly for $(\mathsf{i})$, as the left-weakening only introduces variables on the left-hand side of $S(\simpeq)$, the same joinand witnessing $\simpeq'$ being {\joindecreasing} establishes the same for $\simpeq$.

In any case, it follows that $\simpeq$ is a {\joindecreasing} simple equation.
\end{proof}

\begin{theorem}\label{t: CRL_a}
    The variety $\UniCRLV_\mathsf{a}$ is a cover for $\UniCRLV_\mathsf{i}$ in the subvariety lattice of CRLs when restricted to extensions by $\{\vee,\cdot,1\}$-equations.
\end{theorem}
\begin{proof}
On the one hand, every integral CRL is a member of $\UniCRLV_\mathsf{a}$ by Proposition~\ref{prop:AlterEqs}\eqref{altereqs2}. 
This inclusion is proper as witnessed by the 3-element Sugihara monoid $\mathbf{A}$, which is based on the three-element chain $\bot < 1 < \top$, and whose multiplicative structure is idempotent with $\bot$ is an absorbing element, and is thus necessarily commutative. 
It is easily verified that $\m A$ is an idempotent semiring and therefore supports the structure of a CRL, as it is finite.  
$\m{A}$ is not integral, as $1<\top$. 
Moreover, $\mathbf{A}\in \UniCRLV_\mathsf{a}$ as it is idempotent (i.e., satisfies $x^2 = x$), and thus $\m{A}$ satisfies 
    $$
    x^2yz = xyz \leq xy \vee xyz = xy \vee xy^2z^2.
    $$
Consequently, $\m{A}$ satisfies the equation $\mathsf{a}$ and so $\m{A}\in \UniCRLV_\mathsf{a}\setminus \UniCRLV_\mathsf{i}.$

To show that $\UniCRLV_\mathsf{a}$ is a cover, let $\mathcal{V}$ be a variety of residuated lattices  axiomatized relative to $\UniCRLV$ by $\{\vee,\cdot,1\}$-equations such that  $\UniCRLV_\mathsf{i}\subset \mathcal{V}$; by Lemma~\ref{l: N_2^-0}, $\mathcal{V} = \UniCRLV +\eqset$ for some set $\eqset$ of simple equations. 
Since $\UniCRLV_\mathsf{i}\subseteq \mathcal{V}$, we have $\UniCRLV_\mathsf{i}\models \simpeq$ for each $\simpeq\in \eqset$, so $\simpeq$ is {\joindecreasing} by Lemma~\ref{lem:FLeiEqTh}; hence $\eqset$ must be a set of {\joindecreasing} equations.
On the other hand, $\mathcal{V}$ does not satisfy integrality as it properly contains $\UniCRLV_\mathsf{i}$. 
Hence $\eqset$ is a set of non-integral {\joindecreasing} simple equations. 
Therefore  $\UniCRLV_\mathsf{a}\subseteq\mathcal{V}$.
\end{proof}

Let $\m L$ be an axiomatic extension of $\UniFLExtLogic{}$. 
We call $\m L$ \emph{simple} if it extends $\UniFLExtLogic{}$ only by a set of simple rules.

\begin{corollary}\label{cor-FLemb}
Let $\UniFLeExtLogic{\mathbf{m}\mathbf{b}}$ be the extension of $\UniFLeExtLogic{}$ by the axioms corresponding to the (simple) mingle rule $\mathsf{m} = \hyperlink{link:knottedrule}{\UniWeakKAnaRule{1}{2}}$ and the simple rule:
    $$
    \infer[(\mathsf{b})]{\Gamma,\Psi, \Delta \Rightarrow \Pi}{\Gamma,\Psi \Rightarrow \Pi &\quad \Gamma,\Psi^2,\Delta^2\Rightarrow \Pi}
    $$
Then the variety of almost integral RLs is the equivalent algebraic semantics of the ($0$-free fragment of) $\UniFLeExtLogic{\mathbf{m}\mathbf{b}}$, and $\UniFLeExtLogic{\mathbf{i}}$ is the least proper simple extension of $\UniFLeExtLogic{\mathbf{m}\mathbf{b}}$.
\end{corollary}

%% file: tex/conclusion.tex
\section{A summary of results}
For easy reference, we combine some of the results from the previous sections. The problems under consideration primarily fall within the fast-growing complexity classes, which are defined in Section~\ref{fast-growing-cc}.
\begin{theorem}\label{t: AckHard/complete_containing_CRLc/CRLa}
    Let $\mathcal{V}$ be a variety of residuated lattices containing the variety $\mathsf{CRL_c}$ (defined above Proposition~\ref{prop:AlterEqs}) or the variety $\mathsf{CRL_a}$ (Equation~\eqref{CRLa}).
    \begin{enumerate}
        \item The quasiequational theory for $\mathcal{V}$ is \ACK-hard. 
        In particular, this holds for any variety $\mathsf{RL}+\eqset$ where $\eqset$ consists of (not necessarily simple) $\{\vee,\cdot,1\}$-equations that are all non-trivializing {\joinincreasing} or all non-integral {\joindecreasing} (Definition~\ref{def:TypesOfSimpEq}).
        
        \item If, in addition, $\mathcal{V}$ is knotted and weakly commutative and (relatively) axiomatized by only $\mathcal{N}_2$ equations, then the quasiequational theory for $\mathcal{V}$ is \ACK-complete. 
        In particular, this holds for any variety $\mathsf{CRL}+\eqset$ where $\eqset$ is a finite set of (not necessarily simple) $\{\vee,\cdot,1\}$-equations that are all non-trivializing {\joinincreasing} or all non-integral {\joindecreasing}.
    \end{enumerate}
\end{theorem}
\begin{proof}
    The first claim of (1) is the content of Theorem~\ref{thm:branchincLB} (when containing $\mathsf{CRL_c}$) and of Theorem~\ref{t: ackhardJD} (when containing $\mathsf{CRL_a}$). 
    For the second claim, let $\mathcal{V} = \mathsf{CRL}+\eqset$ for some set of (not necessarily simple) equations. 
    If $\eqset$ is entirely non-trivializing {\joinincreasing}, then $\mathsf{CRL_c}\subseteq \mathcal{V}$ by Proposition~\ref{prop:AlterEqs}\eqref{altereqs2}. 
    On the other hand, if $\eqset$ consists entirely of (non-integral) {\joindecreasing} equations, then $\mathsf{CRL_i}\subseteq \mathcal{V}$, and since they are non-integral, from Theorem~\ref{t: CRL_a} it follows that $\mathsf{CRL_a}\subseteq \mathcal{V}$. 
Furthermore, (2) is immediate from (1) and Corollaries~\ref{c: knotcontr+wc<Ack}(2) and \ref{c: knotweak+wc<Ack}(2) (see also Corollaries~\ref{c: knotcontr+wc=Ack Logic} and \ref{c: knotcontr+wc=Ack}).
\end{proof}
Via algebraization, we may translate the above in logical terms.
\begin{theorem}\label{t: AckHard/complete_containing_FLec/FLea}
    Let $\mathbf{L}$ be any logic in the interval of axiomatic extensions between $\mathbf{FL}$ and $\mathbf{FL_{ec}}$, or between $\mathbf{FL}$ and $\mathbf{FL_{emb}}$ (defined in Corollary~\ref{cor-FLemb}).
    \begin{enumerate}
        \item Deducibility in $\mathbf{L}$ is \ACK-hard. In particular, this holds for any logic $\mathbf{FL}+\Sigma$ where $\Sigma$ consists 
        of  $\{\vee,\cdot,1\}$-formulas that are all non-trivializing {\joinincreasing} or all non-integral {\joindecreasing}.
        \item If, in addition, $\mathbf{L}$ is knotted and weakly commutative, and (relatively) axiomatized by only $\mathcal{N}_2$ formulas, then deducibility in $\mathbf{L}$ is \ACK-complete. In particular, this holds for any logic $\mathbf{FL_e}+\Sigma$ where $\Sigma$ is a finite set of $\{\vee,\cdot,1\}$-formulas that are all non-trivializing {\joinincreasing} or all non-integral {\joindecreasing}.
    \end{enumerate}
\end{theorem}

\begin{remark}
\label{rem:fli}
As mentioned in previous sections, deducibility in $\mathbf{FL_w}$ is
$\UniFGHProbOneAppLevel{\omega^\omega}$-complete, in contrast to the much lower \TOWER complexity of deducibility in $\mathbf{FL_{ew}}$. 
This hyper-Ackermannian complexity was proved in \cite{GreatiRamanayake2024} and  the lower bounds come from the reachability problem in \emph{lossy channel systems}. 
These are models of computation in which processes communicate via first-in-first-out channels, meaning that the \emph{order} in which messages arrive and leave the channels is essential. 
The lossy behavior allows any message to be lost at any moment in any channel, and is responsible for making the reachability problem decidable.
The fact that order matters is exactly what brings more expressiveness to these systems when compared to counter machines, and it is thanks to the lack of exchange in $\mathbf{FL_w}$ that we can encode such order-sensitive message-passing mechanism as a deducibility instance in this logic.
\end{remark}

One of our main results regarding the complexity of knotted substructural logics/algebras are summarized by the following theorem as a consequence of the Theorems~\ref{t: AckHard/complete_containing_CRLc/CRLa} and \ref{t: AckHard/complete_containing_FLec/FLea} above (as well as Corollaries~\ref{c: knotcontr+wc<Ack}(1) and \ref{c: knotweak+wc<Ack}(1)).

\begin{theorem}\label{t:genknotted_ackhard}
The complexity of deducibility/quasiequational theory in an extension of
$\mathbf{FL}$/$\mathsf{FL}$
by a weak commutativity axiom/equation and a non-integral knotted axiom/equation is exactly Ackermannian.
Deducibility in extensions of these logics/varieties by
$\mathcal{N}_2$
axioms/equations has Ackermannian upper bound, and by $\mathcal{P}_3^\flat$ axioms/equations has hyper-Ackermannian upper bound.
\end{theorem}

For fully noncommutative logics/algebras, we obtain the following:

\begin{theorem}
The complexity of deducibility/quasiequational theory in an extension of
$\mathbf{FL_i}$/$\mathsf{FL_i}$
by $\mathcal{N}_2$ axioms/equations is hyper-Ackermannian; when the extension is by $\mathcal{P}_3^\flat$ axioms/equations, it is $\UniFGHProbOneAppLevel{\omega^{\omega^{\omega^\omega}}}$.
\end{theorem}

For a full overview of our results, the reader is referred to Table~\ref{tab:table-contrib}.
In the next subsections, we will present many open problems regarding logics and algebras in the vicinity of the those that we considered in this work.

\section{Rules beyond the present analysis}

We have been able to prove the FEP, obtain decidability of the quasiequational theory, and provide upper complexity bounds for a wide range of varieties of residuated lattices/substructural logics. 
These logics involve extensions by $\mathcal{N}_2$ and even $\mathcal{P}_3^\flat$ axioms and extend, past the commutative case, to the weakly commutative case. 
However, all of these logics contain the assumption that a knotted rule holds. 
This has been crucial in both the algebraic and proof-theoretic analysis (for both knotted weakening and knotted contraction), as all of our arguments rely on an underlying well-quasi-ordered set (and the associated length theorems for our complexity upper bounds). 
The knotted rules ensure that in the free algebra, for every variable $x$, the powers of $x$ form a wqo (actually a finite disjoint union of singletons and copies of the natural numbers). 
This is the driving force, which together with (weak) commutativity ensures that any infinite aspect of monoid words on a finite set of variables is pushed to the exponents of the variables, thus resulting in a direct product of the one-variable wqo's mentioned before. 
This infinite aspect can be tamed by an analysis of the wqo and an application of wqo theory, as we have seen.

Nevertheless, if the axiom  we are considering has the form $m_0\leq m_1\vee \cdots\vee m_k$, where $k\in \mathbb{Z}^+$ and $m_0,\ldots, m_k$ are products of variables, and where $k>2$, i.e., there is a disjunction on the right-hand side, then the resulting order on the powers of $x$ does not need to result in a wqo.
As we discuss below (undertaking an analysis of simple equations with a small number of variables), in many cases the logics actually have undecidable deducibility, so no complexity upper bound exists, and of course there is no expectation of an analysis similar to ours using wqos. 
At the moment, we do not know if this undecidability is a fundamental aspect of the lack of a wqo structure or coincidental, as we do not have any other tools for establishing decidability other than wqo theory.

\begin{openproblem}\label{op:any_nonknotted_dec?}
Is there a variety of residuated lattices that is axiomatized by $\{\jn, \cdot, 1\}$-equations, does not satisfy any knotted equation, and has decidable  quasiequational theory?  
\end{openproblem}

We now mention two main undecidability results (\cite{horcik2015} and \cite{galatos2022}) and outline some of their consequences that justify the inability to obtain decidability/complexity upper bounds for the given logics.

\subsection{Limitations due to undecidability: fully noncommutative case}

The results of \cite{horcik2015} involve the Galois algebra of a specific residuated frame.
Given a set (of \emph{letters)}, the \emph{alphabet} $\Sigma$, and a subset of $L \subseteq \Sigma^*$, we define $W_L:= \Sigma^*$, $W'_L: \Sigma^*\times \Sigma^*$, and the  relation: $x ~N_L~(u,v)$ iff $uxv\in L$. 
It is easily verified that $(W_L,W'_L, N_L)$ supports a residuated frame and therefore the associated Galois algebra $\m W_L^+$ is a residuated lattice \cite[Ex.~2.6]{horcik2015} and that $\varnothing$ is its least element. 

In \cite{horcik2015} it is shown that every variety of residuated lattices that contains  $\m W_L^+$, for a specific language $L$, has  undecidable word problem, hence also undecidable quasiequational theory. 
This is used in the paper to show, as a central result, that this is the case for the variety of residuated lattices axiomatized by the conjunction of $x^3\leq x^2$ and $x \leq x^2$, and for every bigger variety of residuated lattices (for example the one axiomatized by $x \leq x^2 \jn 1$). 
Below we obtain some further corollaries, using the results and ideas in \cite{horcik2015}. 

A \emph{hereditarily square equation} is an equation of the form  $m_0\leq m_1\vee \cdots\vee m_k$, where $k\in \mathbb{Z}^+$ and $m_0,\ldots, m_k$ are products of variables, each variable appearing on both sides, and every substitution that sends variables to themselves or to $1$ either results in a trivial instance (one of the joinands is the left-hand side) or one of the joinands contains a square; here we say that a monoid word contains a square, if it is of the form $uv^2w$, for some words $u,v,w$ and $v$ is not the monoid identity.  
We say a monoid word is \emph{square-free} if it does not contain a square.

The following lemma extends the ideas found in \cite[Lem.~2.7]{horcik2015}.

\begin{lemma}
\label{l:hered_square_satisfaction} 
    If a language $L$ contains only square-free words, then $\m W_L^+$ satisfies every hereditarily square equation.
\end{lemma}
\begin{proof} 
    First note that since $L$ has only square-free words, if $y \in Y\subseteq W_L$ and $y$ contains a square, then $Y^\triangleright =\varnothing$, so $\gamma(Y)=Y^{\triangleright\triangleleft}=\varnothing^\triangleleft = W_L$.
    
    Given a hereditarily square equation $\simpeq:m_0\leq m_1\vee\cdots\vee m_k$ in $n$-variables $\{x_1,\ldots,x_n\}$, and $X_1,\ldots, X_n\in W_L^+$, we will show that the instance $\simpeq(X_1,\ldots,X_n)$ holds in $\m W_L^+$.
    If $X_i=\varnothing$ for some $i\leq n$, then $m_0(X_1,\ldots,X_n) = \varnothing$ (as evaluated in $\m W_L^+$), since $x_i$ occurs in  $m_0$,  so  $\simpeq(X_1,\ldots,X_n)$ holds. 
    We now consider the case that all sets $X_1,\ldots, X_n$ are nonempty.
    If all of the sets $X_i$ are singletons and contain the empty word, then $\{1\}=1_{\m W_L^+}$ (since the $X_i$'s are closed sets) and $m_\ell(X_1,\ldots,X_n) = 1_{\m W_L^+}$ for all $\ell$, thus $\simpeq(X_1,\ldots,X_n)$ trivially holds. 
    Note that the condition $\simpeq(X_1,\ldots,X_n)$ is identical to $\simpeq'(Y_1,\ldots,Y_r)$, where the $Y_i$'s are those $X_i$'s (in the same order) where $X_i\neq \{1\}$ and $\simpeq'=\sigma(\simpeq)$ is the substitution instance of $\simpeq$ under the substitution $\sigma$ where $x_i$ is mapped to $1$ if $X_i=\{1\}$ and to $x_i$ otherwise; so there exist $1 \neq y_i \in Y_i$ for all $i$, and the collection of the $Y_i$'s is non-empty. 
    Since $\simpeq$ is hereditarily square, $\simpeq'$ either has a joinand equal to the left-hand side, in which case $\simpeq'(Y_1,\ldots,Y_r)$ holds, or it has a joinand $m_j$ that contains a square; in this case  $m_j(y_1,\ldots,y_r)$ contains a square and belongs to  $m_j(Y_1,\ldots,Y_r)$, so the latter is equal to $W_L$ and thus, by the comments above, the join on the right is equal to $W_L$, and $\simpeq'(Y_1,\ldots,Y_r)$ holds.
\end{proof}

\begin{corollary}\label{c:hered_square_und} \label{c: horcik2015}
    Every variety of residuated lattices axiomatized by hereditarily square equations has an undecidable word problem, and hence also undecidable quasiequational theory.
\end{corollary}

\begin{proof}
    In \cite[Lem.~4.3]{horcik2015} it is shown that a certain language $\mathsf{ACan}$ consists of square-free words and in \cite[Thm.~5.1]{horcik2015} it is shown that any variety of residuated lattices containing $\m W_\mathsf{ACan}^+$ has an undecidable word problem. Since $\mathsf{ACan}$ is square-free, $\m W_\mathsf{ACan}^+$ satisfies all hereditarily-square equations by Lemma~\ref{l:hered_square_satisfaction}.
\end{proof}

 For example, this applies to the variety axiomatized by the equation $xy \leq x^2y \jn xy^2$. Consider the relevant substitutions in the definition of hereditary square equation: setting $y=1$ results in $x \leq x^2 \jn 1$, which has a square on the right-hand side, setting $x=1$ also results in a square, setting $x=y=1$ results in a trivial equation, and setting none of the variables equal to $1$ also results in a square on the right.  However, it does not apply to the similar  equation $xy\leq xyx\vee yxy$ (obtained by permuting the variables), as it is not hereditarily-square: under the identity substitution  the equation is neither trivial nor is there a joinand with a square.

\subsection{Limitations due to undecidability: 
(weakly) commutative case}

Similar to Corollary~\ref{c: horcik2015}, \cite{galatos2022} provides sufficient conditions for equations (actually most simple equations) to axiomatize a variety with undecidable quasiequational theory (and word problem) even in the presence of commutativity. 
Of course the equations involved cannot be knotted, as knotted commutative varieties have a decidable quasiequational theory. 
Knotted equations are part of a larger class of  equations excluded from the purview of \cite{galatos2022}, called \emph{spinal} equations, and of the more general class of \emph{prespinal} equations: simple equations that have a spinal equation as a substitution instance. Everything else is covered: \cite{galatos2022} ensures that every equation that is not prespinal, also known as a \emph{spineless} equation, together with commutativity axiomatizes a variety with undecidable quasiequational theory; even more generally, the same holds for every supervariety of such a variety.
In detail, an equation is \emph{spinal} if, after commuting variables, it can take the form 
$$x_1^{n_1}\cdots x_k^{n_k} \leq 1 \jn  \; x_1^{n_{11}} \jn x_1^{n_{12}}x_2^{n_{22}}\jn \cdots \jn x_1^{n_{1k}}\cdots x_k^{n_{kk}},$$
where $n_{ii}\neq 0$ and no joinand is equal to the left-hand side, or is the same form but where the joinand $1$ is not included. 
Therefore, spinal equations include all knotted equations (for $k=1$), as well as the equations $x\leq 1 \vee x^2$ and $xy\leq x^2\vee y^2$. Also,   the equation  $xy\leq x^2\vee y^2\vee y^3$ 
is prespinal, as it yields the spinal equation $x\leq 1\vee x^2$ under the substitution $y\mapsto 1$.

\begin{lemma}[{\cite[Thm.~5.5]{galatos2022}}] \label{l: spineless}
Every variety axiomatized by a finite set of \emph{spineless} equations together with commutativity has undecidable word problem, hence also quasiequational theory; the same holds for every supervariety of such a variety.
\end{lemma}

\begin{theorem}\label{t: wmulticontrUND}
The equational theory is undecidable for every variety axiomatized by the conjunction of $q$-commutativity (defined above Lemma~\ref{lem:subcommQE}) and any finite set of spineless equations containing a multicontraction ($x^n\leq x^{n+c_1}\vee \cdots \vee x^{n+c_k}$, for $k\geq 2$ and positive $c_j$'s).
\end{theorem}

\begin{proof}
The variety is negatively potent (see below Lemma~\ref{ExpDedTh}), so it has the deduction theorem by Lemma~\ref{lem:subcommQE}. 
As the above equations are spineless equations (in particular, since $q$-commutativity is rendered trivial under commutativity), the variety has an undecidable quasiequational theory by Lemma~\ref{l: spineless}, hence also an undecidable  equational theory by the deduction theorem.
\end{proof}

We observe that
\cite[Section~8.3]{galatos2022} provides a general algorithm for determining prespinality using notions from positive linear algebra.
The spineless equations in $1$-variable are characterized easily.
\begin{lemma}[{\cite[Lem.~5.4]{galatos2022}}]\label{l:spineless_sufficient}
    An equation $x^{n}\leq x^{n_1}\vee\cdots \vee x^{n_k}$ is spineless iff it is non-trivializing (i.e., $n>0$) and has two distinct joinands  both different from $1$, i.e., there are $n_i>n_j>0$ for some $i,j\leq k$.
\end{lemma}
    
Therefore, the only (pre)spinal equations in a single variable are either trivializing (equivalent to $1\leq x$), integral (equivalent to $x\leq 1$), knotted equations $x^n\leq x^m$ or ones of the form $x^n\leq 1\vee x^m$, for distinct $n,m>0$. Regarding the latter, we note those of the form $x^n\leq  1\vee x$, where $n\geq 2$, define the same variety as $x^2\leq 1\vee x $. 

\begin{lemma}\label{l: x2<xv1}
  For any $n\geq 2$, a residuated lattice satisfies $x^n\leq 1\vee x$ iff it satisfies $x^2\leq 1\vee x $.
\end{lemma}
\begin{proof}
For the forward direction, observe the substitution $x\mapsto x\vee 1$  yields $(x\vee 1)^n \leq 1 \jn (x\vee 1)= 1 \jn x$, and since $x^2$ appears as a joinand upon fully distributing $(x\vee 1)^n$, it follows $x^2\leq (x\vee 1)^n \leq 1\vee x$. 
Conversely, $x^2\leq x \vee 1$ yields
$x^{n}\leq x^{n-1} \vee x^{n-2}\leq x^{n-2} \vee x^{n-2} \vee x^{n-3} = x^{n-2} \vee x^{n-3}\leq \ldots \leq x\vee 1$.
\end{proof}

\subsection{Limitations to the FEP: equations satisfied in the integers}\label{sec:FEPlimits}

Even though the FEP itself does not provide complexity upper bounds, it does entail the decidability of the universal, hence also of the quasiequational, theory; this is the reason why we are interested in it and why we proved FEP results in Chapter~\ref{sec:fepP3}. 
There are however certain limitations to establishing the FEP; here we will mention some varieties for which the FEP provably fails, utilizing the following result due to Blok and van Alten \cite{BvA02} (see also \cite[Theorem~6.56]{GalJipKowOno07}). 

\begin{lemma}
\label{l:noFEP_BvA}
    The FEP fails for any class of residuated lattices containing the abelian $\ell$-group of integers $\mathbf{Z}$.
\end{lemma}
Towards using the above lemma, we provide a characterization of the equations of the form we are considering that are satisfied in $\mathbf{Z}$.
Given an $n$-variable equation $\simpeq: m_0\leq m_1\vee\ldots \vee m_k$, we consider the terms obtained from the joinands by arranging the variables in numerical order: for each $j \in\{1, \ldots, k\}$, if the joinand $m_j$  is equivalent to $x_1^{m_{j1}}\cdots x_n^{m_{jn}}$  modulo commutativity, then we associate it to  the tuple/point $\vec{m}_j:=(m_{j 1},\ldots,m_{j n})\in \mathbb{N}^n$. Also, we consider the  associated  equation $\simpeq_{\mathrm{com}}$:
$$x_1^{m_{01}}\cdots x_n^{m_{0n}}  \leq \bigvee_{j=1}^k x_1^{m_{j1}}\cdots x_n^{m_{jn}}.$$
We will call the equation $m_0\leq m_1\vee\ldots \vee m_k$ a \emph{$\mathbf{Z}$-equation} if, when viewing $\vec{m}_0,\vec{m}_1,\ldots,\vec{m_k}$ as points in $\mathbb{N}^n$, the point $\vec{m}_0$ is contained in the convex hull of the set of points $\{\vec{m}_1,\ldots,\vec{m}_k\}$; explicitly this means
\begin{equation}
\label{eq:Zeq}
    \left(\sum_{j=1}^k c_j\right)\vec{m}_0 =\sum_{j=1}^k c_j\vec{m}_j,\quad \mbox{for some $c_1,...,c_k\in \mathbb{N}$ not all zero.}
\end{equation} 

The following lemma characterizes the $\{\vee,\cdot,1\}$-equations satisfied in the $\ell$-group $\mathbf{Z}$. 
The argument is inspired by a similar one found in \cite[Remark~2]{ColMet2017}, using a variant of Farka's Lemma in linear programming known as \emph{Gordan's Lemma}.
    
\begin{lemma}\label{l:Zeqs}
The equation $\simpeq: m_0\leq m_1\vee\cdots \vee m_k$ is satisfied in $\mathbf{Z}$ iff it is a $\mathbf{Z}$-equation.
\end{lemma}
\begin{proof}
We first recall Gordan's Lemma: For any integer-valued matrix $M\in \mathbb{Z}^{n\times k}$, exactly one of the following hold:
\begin{itemize}
    \item there exists $\mathbf{n}\in \mathbb{Z}^n$ such that $\mathbf{n}^\top M< \mathbf{0}$. 
    \item there exists $\mathbf{x}\in \mathbb{N}^k\setminus \{\mathbf{0}\} $ such that $M\mathbf{x}= \mathbf{0}$. 
\end{itemize}
    Here  $\mathbf{n}< \mathbf{0}$ means that every entry of  $\mathbf{n}$ is strictly negative.
    %Let $\simpeq$ denote the equation $m_0\leq m_1\vee\cdots \vee m_k$. 
    We view the joinands $m_1,\ldots, m_k$ of $\simpeq$ as columns of the $\mathbb{N}^{n\times k}$ matrix $M=[\vec{m}_1~ \cdots~ \vec{m}_k]$ and set $M_{0}:=[(\vec{m}_1-\vec{m}_0)~ \cdots~ (\vec{m}_k-\vec{m}_0)]$, the matrix in $ \mathbb{Z}^{n\times k}$ obtained by subtracting $\vec{m}_0$ from each column of $M$; here we view $\mathbf{x}$ and $\vec{m}_0$ as column vectors.
    
    Observe that
    $$
    \begin{array}{ r c r l }
    \mathbf{Z}\models\varepsilon &\iff& (\forall \sigma \mbox{ substitution})& \sigma(m_0) \leq \max\{\sigma(m_j):1\leq  j\leq k \}\\
    &\iff& (\forall \sigma \mbox{ substitution}) & 0 \leq \max\{\sigma(m_j)-\sigma(m_0):1\leq  j\leq k \}\\
    &\iff& (\forall \mathbf{n}\in \mathbb{Z}^n) & 0 \leq \max\{\mathbf{n}^\top \vec{m}_j-\mathbf{n}^\top\vec{m}_0:1\leq  j\leq k \}\\
    &\iff& (\forall \mathbf{n}\in \mathbb{Z}^n)&  \mathbf{n}^\top( \vec{m}_j-\vec{m}_0) \geq 0, \mbox{ for some }1\leq  j\leq k\\
    &\iff& \lnot (\exists  \mathbf{n}\in \mathbb{Z}^n) & \mathbf{n}^\top M_{0} < \mathbf{0}\\
    \end{array}
    $$
    By Gordan's Lemma, the last line is equivalent to $M_{0} \mathbf{x} =\mathbf{0}$ for some $\mathbf{x}\in \mathbb{N}^k\setminus\{\mathbf{0}\}$.
    Now, $M_{0}\mathbf{x}= (M - [\vec{m}_0 \cdots \vec{m}_0])\mathbf{x}=
    M\mathbf{x} - [\vec{m}_0 \cdots \vec{m}_0]\mathbf{x}$ and 
    $[\vec{m}_0 ~\cdots~ \vec{m}_0]\mathbf{x}
    = \sum_{j=1}^m \mathbf{x}(j)\vec{m}_0 = 
    (\mathbf{x}^\top \mathbf{1})\vec{m}_0 
    $, where  $\mathbf{1}$ is the column vector of all $1$'s; so the condition that $M_{0} \mathbf{x} =\mathbf{0}$ for some $\mathbf{x}\in \mathbb{N}^k\setminus\{\mathbf{0}\}$ is equivalent to $(\mathbf{x}^\top \mathbf{1})\vec{m}_0 = M\mathbf{x}$ 
    for some $\mathbf{x}\in \mathbb{N}^k\setminus\{\mathbf0\}$ 
    which is a direct reformulation of Equation~\ref{eq:Zeq}.
\end{proof}
    
Consequently, by Lemmas~\ref{l:noFEP_BvA} and \ref{l:Zeqs}, the following is immediate.
    
\begin{theorem}\label{t:noFEP_Zeq}
The FEP fails in every variety of residuated lattices axiomatized by $\mathbf{Z}$-equations.
\end{theorem}
    
Of course, we already know that varieties axiomatized by sets of spineless $\mathbf{Z}$-equations cannot satisfy the FEP, as they have undecidable quasiequational theory.
These include the previously mentioned $xyz\leq x^2y\vee y^2z\vee xz^2$ (whose closure of joinands is $1$-dimensional as in Figure~\ref{fig:preknotted}(c)) and the equation $xy \leq x\vee x^2y \vee y^2$ (whose closure is $2$-dimensional). 
For (combinations of) prespinal $\mathbf{Z}$-equations the above theorem ensures that the FEP fails, but no (un)decidability results are known in the (weakly) commutative case. The simplest example is given here:

\begin{openproblem}
Is the quasiequational theory decidable for the variety axiomatized by commutativity and $x\leq 1\vee x^2$?
\end{openproblem}

We note that the equation $x\leq 1\vee x^2$ is {\joinincreasing} (actually also {\joindecreasing}), and therefore has a \ACK-hard theory. 
The simplest example of a (pre)spinal $\mathbf{Z}$-equation which is neither {\joinincreasing} nor {\joindecreasing} requires at least $2$-variables.
    
\begin{openproblem}
Is the quasiequational theory \ACK-hard for the variety axiomatized by commutativity and $xy\leq x^2\vee y^2$? Is it  decidable? 
\end{openproblem}

Since both equations above are hereditarily-square, the corresponding (fully noncommutative) varieties of residuated lattices have undecidable word problem.
An example of a (pre)spinal $\mathbf{Z}$-equation that is neither {\joinincreasing}, nor {\joindecreasing}, nor hereditarily-square is $vwxyz\leq xvyvx \vee  ywxwy \vee zvzwz$. Indeed, it is not hereditarily-square since each joinand is square-free, and it is not trivial; it is prespinal since the substitution sending all variables but $z$ to $1$ gives the spinal equation $z\leq 1\vee z^3$; it is not {\joinincreasing} since no joinand contains all variables; it is not {\joindecreasing} since every joinand contains at least two copies of some variable; lastly, it is a $\mathbf{Z}$-equation since $3\vec{m}_0 = \vec{m}_1 + \vec{m}_2+\vec{m}_3$ (i.e., there are 3 joinands and each variable occurs 3 times on the right and once on the left).

\begin{openproblem}
Is the quasiequational theory \ACK-hard for the variety axiomatized by $vwxyz\leq xvyvx \vee  ywxwy \vee zvzwz$? Is it decidable? 
\end{openproblem} 

\subsection{Elementary upper bounds}

Before undertaking an analysis of equations with a small number of variables, we also prove a result that (in contrast to the undecidability results of the previous sections) provides an elementary upper bound for certain potent logics.

A variety is called \emph{$(n,m)$-potent} if it satisfies the equation  $x^{n}=x^m$, for some distinct $n,m \in \mathbb{N}$, and \emph{potent} if it is $(n,m)$-potent for some $n,m$; we say that it is \emph{$n$-potent} if $x^n=x^{n+1}$, for some $n$. 
Note that $(n,m)$-potency does not always imply $n$-potency, even though this is the case for negative $(n,m)$-potency and negative $n$-potency, as discussed right before Lemma~\ref{ExpDedTh}.

\begin{lemma}\label{l: FEPpotency}
If a variety of residuated lattices is axiomatized by finitely many $\{\jn, \cdot, 1\}$-equations and satisfies (weak) commutativity and potency, then it has the FEP. 
Also, deciding the universal theory is in \ccfont{3-expspace}.
\end{lemma}

\begin{proof}
Given an algebra $\m A$ in the variety and a finite subset $B$ of size $n$, say $B=\{b_1, \ldots, b_n\}$, the submonoid $C$ of $\m A$ generated by $B$ is finite. 
Indeed, by weak commutativity we get only finitely many types of monoid words with exponents and there are only finitely many choices of the exponents, if the variety is $(k,\ell)$-potent. 
Using the results in \cite{Cardona2015}, it can be computed that if $s$ is the length of the vector in the weak commutativity equation, $p,q$ are the lengths of the front and back wall and $t:=\max\{2s, m+p_a+q_a, n+p_a+q_a\}$, then $|C|\leq (n+1)^{nt}\leq {2^{M n\ln n}}$, for some constant $M$. 
In the commutative case, we have words $b_1^{k_1}\cdots b_n^{k_n}$, where $0\leq k_i\leq \max\{k, \ell\}$, i.e., $|C|\leq \max\{k, \ell\}^n \leq 2^{Mn}$, for some constant $M$.
So, if $\m D$ is the join subsemilattice of $\m A$ generated by $C$, we have $|D|\leq 2^{|C|}\leq 2^{2^{M n\ln n}}$, and  $|D|\leq 2^{|C|}\leq  2^{2^{Mn}}$ in the commutative case.
Then we can expand $\m D$ to a residuated lattice, which witnesses the FEP; see Theorem~5.8 of \cite{AG} for the exact calculations and details. 
Also, $\m D$ stays in the variety as it is a $\{\jn, \cdot, 1\}$-subalgebra of $\m A$. 

So, to falsify a universal sentence with $n$-variables, we employ an alternating algorithm. 
On an existential state, we guess an algebra $\m D$ of size up to $ 2^{2^{M n\ln n}}$ (i.e., an underlying set with that 2-exp size, and operation tables of 3-exp size $|D|^{|D\times D|}$); then on a universal state we consider/guess each evaluation $v$ of the variables in the equation in $D$ (of size $n^{|D|}$) and verify that $\m D$ satisfies the axiomatization under the evaluation $v$; then in an existential state we guess a valuation that falsifies the universal sentence.
In total the algorithm is in \ccfont{atime}($2^{2^{2^n}}$) $\subseteq$ \ccfont{dspace}($2^{2^{2^n}}$), so we obtain \ccfont{3-expspace}.
The proof for the commutative case is similar.
\end{proof}

\begin{lemma}
\label{l: EqThpotency}
If a variety of residuated lattices is axiomatized by finitely many $\{\jn, \cdot, 1\}$-equations and satisfies {(weak)} commutativity and potency, then the complexity of the quasiequational theory is in \EXPTIME. 
\end{lemma}
\begin{proof}
If the set of equations is inconsistent or has just the non-trivial model, then every formula is true. Hence it suffices to consider the case when the variety is axiomatised by finitely many non-trivializing $\{\jn, \cdot, 1\}$-equations.
We rephrase this problem in terms of deducibility in the corresponding logics.
From the discussions in Section~\ref{s: P3/N2} and Chapter~\ref{sec:noncom-ub}, to solve the deducibility of a weakly commutative potent substructural logic axiomatized by $\{ \lor,\fus,1 \}$-equations, we can solve the provability in the \emph{sequent version} of the calculus
$
\UniHyperCalcA_\UniSetFmA:= \UniCalcExt{\UniFLExtHCalc{\UniWEProp{\vec{a}}\UniWeakPtProp{k}{\ell}}^\star}{\UniAnaRuleSet\cup \UniRuleDedSet{\UniSetFmA}}$,
where $\UniSetFmA$ is the finite set of assumptions in the instance of deducibility,
$\UniWEProp{\vec a}$ is a weak commutativity axiom and $\UniWeakPtProp{k}{\ell}$ is the
$(k,\ell)$-potency axiom; without loss of generality we assume $k < \ell$.
We first devise an algorithm for provability in $\UniHyperCalcA_\UniSetFmA$ and then we make it uniform on $\UniSetFmA$.
We start by the commutative case and then give directions to extend the argument to the weakly commutative setting.

Let $M_{\UniHyperCalcA_{\UniSetFmA}}$ be the maximum number of metavariables in the antecedent of a schematic sequent in a rule schema of $\UniHyperCalcA_{\UniSetFmA}$.
Let $C(\UniSubfmlaHyperseqSet)$ be the set of $\UniSubfmlaHyperseqSet$-sequents where the multiplicities of the formulas are at most $\kappa \UniSymbDef M_{\UniHyperCalcA_{\UniSetFmA}} \cdot (\ell-1)$.
Note that the cardinality of this set is exponential on $\UniSetCard{\UniSubfmlaHyperseqSet} < \UniSizeHyper{\UniSubfmlaHyperseqSet}$, namely
$(\UniSetCard{\UniSubfmlaHyperseqSet}+1) \cdot \kappa^{\UniSetCard{\UniSubfmlaHyperseqSet}}$.
For this commutative setting, \cite[Sec.~3.4]{gavin2019} shows that a $\UniSubfmlaHyperseqSet$-sequent $s$ is provable if, and only if, its truncation $\text{tr}(s) \in C(\UniSubfmlaHyperseqSet)$ has a $C(\UniSubfmlaHyperseqSet)$-proof, where $\text{tr}(s)$ is obtained by truncating the multiplicities of the formulas in $s$ to $k + (v-k)\%(\ell-k)$ whenever they are  $\geq \ell$.
Using this fact, we now devise an exponential-time algorithm that produces all provable $\UniSubfmlaHyperseqSet$-sequents.

Let $ \UniSubfmlaHyperseqSet \supseteq \UniSetFmA$ be a finite set of formulas closed under subformulas. First, the algorithm writes down all possible rule instances, what can be done in exponential time on $\UniSizeHyper{\UniSubfmlaHyperseqSet}$ in view of the imposed multiplicity constraints.

Second, the algorithm writes down 
the set $C(\UniSubfmlaHyperseqSet)$.
This can be done in exponential time, because writing down the sequents themselves takes time polynomial on the (exponential) cardinality of $C(\UniSubfmlaHyperseqSet)$.

Third, the algorithm proceeds by forward applications of the rule instances, marking in the list above all the provable $\UniSubfmlaHyperseqSet$-sequents.
If at some stage nothing new is marked, the algorithm stops, and one can safely conclude that nothing else is provable. 
More specifically, at stage 0, we mark all instances of axiomatic sequents, which takes time polynomial in the length of the list.
Now, in a loop, we run through the rule instances, and, for each of them, we check whether the premises are marked; if this is the case, we mark the conclusion, otherwise we go to the next rule instance.
The loop can only run at most the length of the list of possible sequents.  
From this, we obtain that the algorithm runs in
exponential time.

Now, assume we have an instance of deducibility, i.e., a finite set $\UniSetFmA$ of formulas and a formula $\UniFmA$.
Let $\UniSubfmlaHyperseqSet$ be the set of subformulas of the formulas in $\UniSetFmA \cup \{ \UniFmA \}$.
Creating the rules $\UniRuleDedSet{\UniSetFmA}$ takes time polynomial in $\UniSizeHyper{\UniSetFmA}$ (see the complexity analysis in Sections~\ref{sec:ub-ww} and ~\ref{sec:noncom-ub}).
We run the algorithm described above in time that is exponential in $\UniSizeHyper{\UniSubfmlaHyperseqSet}$, which is polynomially related to $\UniSizeHyper{\UniSetFmA} + \UniSizeHyper{\UniFmA}$.
Finally, we check whether the input formula (i.e., the sequent $\UniSequent{}{\UniFmA}$) is marked, which takes polynomial time in the length of the list.
Thus the whole procedure takes exponential time in $\UniSizeHyper{\UniSetFmA}+\UniSizeHyper{\UniFmA}$.

For the weakly commutative case, the algorithm is the same, but now we adjust the value of $\kappa$.
We take $\lambda\in \UniNaturalSet$
as defined in Section~\ref{sec:wc-normal-form}.
We set $\kappa \UniSymbDef
M_{\UniHyperCalcA_{\UniSetFmA}} \cdot
(\lambda + \ell-1)$
and make the truncation such that it maps to the value
$\lambda \leq k + \lambda + (v-\lambda-k)\%(\ell-k) < \lambda+\ell$
without affecting the type of the sequent, i.e., the image of the antecedent under $\UniAlphaD{}(\cdot)$~(see Definition~\ref{def:alpha_defs}) remains the same---this is always possible as we have more than $\lambda$ copies of the formula before and after the truncation.
One can then follow a similar proof as that in \cite[Sec.~3.4]{gavin2019} to show that we again only need to consider $C(\UniSubfmlaHyperseqSet)$-proofs for this new value of $\kappa$; the idea is that, recall, when a formula has multiplicity $\geq \lambda$, the middle part of the sequent for that formula behaves as if full commutativity was present.
Since the cardinality of $C(\UniSubfmlaHyperseqSet)$ is upper bounded by
$(\UniSetCard{\UniSubfmlaHyperseqSet}+1) \cdot \sum_{l=0}^{\UniSetCard{\UniSubfmlaHyperseqSet}\kappa} l! \leq     (\UniSetCard{\UniSubfmlaHyperseqSet}+1)^2\kappa \cdot (\UniSetCard{\UniSubfmlaHyperseqSet}\kappa)! \leq 2^{r(\UniSetCard{\UniSubfmlaHyperseqSet})}$, for a polynomial $r$, we still get an exponential-time procedure in the weakly commutative case.\qedhere
\end{proof}

\begin{lemma}
\label{l: EqThpotencyhyper}
If a variety of residuated lattices is axiomatized by finitely many $\mathcal{P}_3^\flat$-equations and satisfies {(weak)} commutativity and potency, then the complexity of the quasiequational theory is in \ccfont{2-exptime}. 
\end{lemma}
\begin{proof}
The algorithm is essentially the same as the one of  Lemma~\ref{l: EqThpotency}, with the difference that now we have to control which hypersequents need to be considered in the forward proof search.

We consider the hypersequent calculus
$\UniHyperCalcA_\UniSetFmA:= \UniCalcExt{\UniFLExtHCalc{\UniWEProp{\vec{a}}\UniWeakPtProp{k}{\ell}}^\star}{\UniAnaRuleSet\cup \UniRuleDedSet{\UniSetFmA}}$.
Recall from the proof of Lemma~\ref{l: EqThpotency} the definition of
$M_{\UniHyperCalcA_{\UniSetFmA}}$
and note that it upper bounds the maximum number of active components appearing in a schematic hypersequent in a rule schema of
$\UniHyperCalcA_{\UniSetFmA}$.
Moreover, we take $C(\UniSubfmlaHyperseqSet)$ as defined in the same proof, and let $H(\UniSubfmlaHyperseqSet)$ be all the hypersequents with components in  $C(\UniSubfmlaHyperseqSet)$ having multiplicity at most $M_{\UniHyperCalcA_{\UniSetFmA}}$.
A $H(\UniSubfmlaHyperseqSet)$-proof is one in which only hypersequents from $H(\UniSubfmlaHyperseqSet)$ appear.
    
Given a $\UniSubfmlaHyperseqSet$-hypersequent $h$, we define the truncation of $h$, $\text{tr}(h) \in H(\UniSubfmlaHyperseqSet)$, as the hypersequent obtained by first truncating each of its components (using $\text{tr}(\cdot)$ from Lemma~\ref{l: EqThpotency}), and then truncating their multiplicities to $M_{\UniHyperCalcA_{\UniSetFmA}}$ (whenever they are greater than this value).
One can then prove that $h$ is provable iff $\text{tr}(h) \in H(\UniSubfmlaHyperseqSet)$
has a $H(\UniSubfmlaHyperseqSet)$-proof.

From here, the algorithm from Lemma~\ref{l: EqThpotency} adapts to the hypersequent setting in the obvious way.
The complexity jump from $\EXPTIME$ to \ccfont{2-exptime} is witnessed by the fact that $H(\UniSubfmlaHyperseqSet)$ has now size bounded by $M_{\UniHyperCalcA_{\UniSetFmA}}^{\UniSetCard{C(\UniSubfmlaHyperseqSet)}} \leq 2^{2^{q(\UniSizeHyper{\UniSubfmlaHyperseqSet})}}$
for a polynomial $q$.\qedhere
\end{proof}

\begin{remark}
We have seen in this work two distinct ways of translating deducibility to provability in a logic determined by an extension $\UniHyperCalcA$ of the calculus $\mathbf{HFL}$.
The first one is well-known: when the logic has a deduction theorem, we have
$\UniMSetFmA \vdash_{\UniHyperCalcA} \UniFmB$
iff
$ \vdash_{\UniHyperCalcA} f(\UniSetFmA,\UniFmB)$, where
$f(\UniSetFmA,\UniFmB)$ is a formula. 
The second one consists in updating $\UniHyperCalcA$ with rules incorporating the assumptions $\UniSetFmA$, forming the cut-free calculus $\UniDelCut{\UniHyperCalcA_{\UniSetFmA}}$---call it the $\UniSetFmA$-modified version of $\UniHyperCalcA$---with the property 
$\UniMSetFmA \vdash_{\UniHyperCalcA} \UniFmB$
iff
$ \vdash_{\UniDelCut{\UniHyperCalcA_{\UniSetFmA}}} \UniFmB$.
While the former usually applies to very particular classes of substructural logics (e.g., the contractive ones shown Section~\ref{sec:ded-theorem-provability}), the second one is quite pervasive (see Section~\ref{sec:deducibility-adapts}), essentially depending on cut elimination, which holds in all of the extensions of $\mathbf{HFL}$ we considered here.
Observe that the deduction theorem is very strong in the following sense: once provability is decided for $\UniHyperCalcA$, deducibility is automatically decided.
Moreover, if $f$ is efficiently computable, then the complexity upper bounds of provability automatically transfer to deducibility.
The second approach, however, is weaker in the sense that understanding provability for $\UniHyperCalcA$ might not help at all to understand deducibility, since we now have to investigate provability for the new calculus
$\UniDelCut{\UniHyperCalcA_{\UniSetFmA}}$, which may lead to completely different complexity results compared to provability in $\UniHyperCalcA$.
The best examples are $\mathbf{FL_{ew}}$ and $\mathbf{FL_{w}}$, whose complexities for provability and deducibility differ drastically (see Table~\ref{tab:table-contrib}).
Indeed, provability for $\mathbf{FL_{ew}}$ and $\mathbf{FL_{w}}$ is easily solved by standard methods leading to \PSPACE upper bounds, while their $\UniSetFmA$-modified versions require wqos and lead to non-elementary upper bounds.
Still, working with
$\UniDelCut{\UniHyperCalcA_{\UniSetFmA}}$ helps in the study of deducibility by abstracting away the set $\UniSetFmA$ of assumptions, representing it as well-behaved rules.
As an alternative to this approach, one may consider working with restricted cuts, as done in~\cite{GreatiRamanayake2024}.
\end{remark}

\begin{remark}\label{remark:combins of eqs}
    To capitalize on Lemma~\ref{l: FEPpotency}, it will be important to know exactly when a variety satisfies an $(n,m)$-potent, or even a knotted, equation. 
    Given an equation of the form  $m_0\leq m_1\vee \cdots\vee m_k$, where $k\in \mathbb{Z}^+$, $m_0,\ldots, m_k$ are products of variables and each variable appears on both sides, we say that it is \emph{preknotted} if 
    there is a direct ($1$-variable) substitution instance resulting in some knotted equation.
    If $\eqset$ is a set of inequalities of the form above, by \cite[Thm.~2.4.1]{gavin2019}
    the variety $\mathrm{Mod}(\eqset)$ of residuated lattices/FL-algebras axiomatized by $\eqset$  satisfies a knotted equation iff $\eqset$ contains a preknotted equation.
    
    Related facts hold for other types of $1$-variable equations,  such as the following below, again stated in \cite[Thm.~2.4.1]{gavin2019}.
    Recall, from Section~\ref{sec:ded-theorem-provability}, that an equation is called \emph{multi-contractible} if it entails some multi-contraction equation $x^k\leq x^{k+c_1}\vee\cdots x^{k+c_n}$, where $k,c_1,\ldots,c_n\geq 1$. 
    Similarly,  $\mathrm{Mod}(\eqset)$  satisfies a multi-contraction equation iff $\eqset$ contains a multi-contractible equation with a direct  ($1$-variable) substitution instance yielding the multi-contraction. 
    Dually, call an equation \emph{multi-weakening} if it is of the form $x^k\leq x^{c_1}\vee\cdots x^{c_n}$, where $0\leq c_1,\ldots,c_n<k$, and an equation \emph{multi-weak} if a multi-weakening equation is a direct ($1$-variable) substitution instance. 
    Then  $\mathrm{Mod}(\eqset)$  satisfies a multi-weakening equation iff $\eqset$ contains a multi-weak equation.

    Finally, by \cite[Cor.~2.4.3]{gavin2019}, the variety  $\mathrm{Mod}(\eqset)$  is potent 
    iff $\eqset$ contains both a preknotted weakening (resp., contraction) equation and a multi-contractible (resp., multi-weak) equation.
\end{remark}

We also note that \cite{horcik2011} establishes lower complexity bounds for many substructural logics. 
Specializing Corollary~4.6 of \cite{horcik2011} (and using the fact that $\mathcal{M}_2$ axioms in the corollary include all $\mathcal{N}_2$ equations) we get the following result that is useful to our discussion.

\begin{lemma}[\cite{horcik2011}]\label{l: HorcikTerui}
Every non-trivial variety of residuated lattices axiomatized by $\{\jn, \cdot, 1\}$-equations has \PSPACE-hard equational theory.    
\end{lemma}

Also, recall the definition of a \emph{shrinking} equation from Remark~\ref{r: shrinking} and the fact that \cite{horcik2011} proves that every extension of $\m{FL}$ (or of $\m{FL_e}$) with a finite set of shrinking analytic structural rules is \PSPACE-complete.

\begin{openproblem}
What is the exact complexity of the (quasi)equational theory of the various varieties axiomatized by finitely many $\{\jn, \cdot, 1\}$-equations that entail (weak)commutativity and potency, such as the conjunction of commutativity and $x^3=x^4$?
\end{openproblem}

\subsection{Preknotted equations: a geometric description and lack of lower bounds}\label{sec: eqs_geometric}
As mentioned earlier, all of 
our methods for establishing upper bounds (even the FEP), apply only to knotted varieties. 
By Remark~\ref{remark:combins of eqs}, this is equivalent to considering preknotted equations instead, so here we give a useful geometric description of such equations. 
 
Let $\simpeq$ be an $n$-variable equation $m_0\leq m_1\vee\ldots \vee m_k$, and recall (from Section~\ref{sec:FEPlimits}) that to the monomials $m_0,\ldots,m_k$ we associate points $\vec{m}_0,\ldots,\vec{m}_k\in \mathbb{R}^n$.
   
\begin{lemma}\label{l: preknotted}
    The equation $\simpeq$ is preknotted iff the following two conditions hold: 
    (i) the set of points $\{\vec{m}_1,\ldots,\vec{m}_k\}$ lie on an affine hyperplane $H:\mathbf{n}^\top\mathbf{x}=b$ (i.e., affine subspace of codimension 1) with $b\geq 0$ and nonnegative coefficients (i.e., a normal vector) $\mathbf{n}$ whose entries consist of necessarily integer values;
    and (ii) the point $\vec{m}_0$ is not contained on $H$. 
\end{lemma} 

\begin{proof}
    If these two conditions are met, then the substitution $\sigma$ defined via $x_i\mapsto x^{\mathbf{n}(i)}$ gives an instance $\sigma(\simpeq)$ equal to $ x^a\leq x^b$ 
    where $b\geq 0$ (as  $H$ is the solution set of the linear system $\mathbf{n}^{\top}\mathbf{x} = b\geq 0$)
    by (i), 
    and $0\leq a\neq b$ by (ii). 
    Conversely, any $1$-variable substitution $\sigma: x_i\mapsto x^{a_i}$ in which $\sigma(\simpeq)$ is knotted yields $\mathbf{n}:= (a_1,\ldots, a_n)$ for which conditions (i) and (ii) must hold. 
    In this way, $\simpeq$ is preknotted iff $\simpeq_{\mathrm{com}}$ is preknotted.   
\end{proof}

\begin{figure}[h]
    \centering
    $\begin{array}{ccc}
    \begin{tikzpicture}[scale=.8, every node/.style={scale=.7}]
        % Axes
        \draw[-] (-0.25,0) -- (4,0);
        \draw[->] (0,-0.25) -- (0,4);      
        % Dashed grid lines
        \draw[dashed,-] (-.25,2) -- (4,2);
        \draw[dashed,-] (2,-.25) -- (2,4);
        % Shaded regions
        \fill[pattern=north east lines, pattern color=gray!70] (0,0) -- (2,0) -- (2,2) -- (0,2) -- cycle;
        \fill[pattern=north east lines, pattern color=gray!70] (2,2) -- (4,2) -- (4,4) -- (2,4) -- cycle;
            % Labels for shaded regions (regular font size)
        \node at (1,1) {\scriptsize${\downarrow}\vec{m}_0$};
        \node at (3,3) {\scriptsize${\uparrow }\vec{m}_0$};    
        % Normal vector n
        \draw[->, gray] (3.5,1.083) -- (4,1.512) node[pos = .2,right] {\scriptsize$\mathbf{n}$};
        \draw[thick, <->] (1,4) -- (4,.5) node[pos=0.85,  left ] {$H$~};
        \filldraw (2,2) circle (2pt) node[above left] {$\vec{m}_0$};   
    \end{tikzpicture}
    
    &
    
    \begin{tikzpicture}[scale=.8, every node/.style={scale=.7}]
        % Axes
        \draw[-] (-0.25,0) -- (4,0);
        \draw[->] (0,-0.25) -- (0,4);
        % Dashed grid lines
        \draw[dashed,-] (-.25,2) -- (4,2);
        \draw[dashed,-] (2,-.25) -- (2,4);
        % Shaded regions
        \fill[pattern=north east lines, pattern color=gray!70] (0,0) -- (2,0) -- (2,2) -- (0,2) -- cycle;
        \fill[pattern=north east lines, pattern color=gray!70] (2,2) -- (4,2) -- (4,4) -- (2,4) -- cycle;    
        % Labels for shaded regions (regular font size)
        \node at (1,1) {\scriptsize${\downarrow}\vec{m}_0$};
        \node at (3,3) {\scriptsize${\uparrow }\vec{m}_0$};    
        % Normal vector n
        \draw[->, gray] (2.5, 0.5) -- (3,1) node[pos = .2,right]  {\scriptsize$\mathbf{n}$};
        \draw[thick, <->] (-.25,3.25) -- (3.25,-.25) node[pos=0.8, left ] {$H$~};   
        % Point \vec{m}_0
        \filldraw (2,2) circle (2pt) node[above left] {$\vec{m}_0$};
    \end{tikzpicture}
    
    &
    
    \begin{tikzpicture}[scale=.8, every node/.style={scale=.7}]
        % Axes
        \draw[-] (-0.25,0) -- (4,0);
        \draw[->] (0,-0.25) -- (0,4); 
        % Dashed grid lines
        \draw[dashed,-] (-.25,2) -- (4,2);
        \draw[dashed,-] (2,-.25) -- (2,4);
        % Shaded regions
        \fill[pattern=north east lines, pattern color=gray!70] (0,0) -- (2,0) -- (2,2) -- (0,2) -- cycle;
        \fill[pattern=north east lines, pattern color=gray!70] (2,2) -- (4,2) -- (4,4) -- (2,4) -- cycle;
        % Labels for shaded regions (regular font size)
        \node at (1,1) {\scriptsize${\downarrow}\vec{m}_0$};
        \node at (3,3) {\scriptsize${\uparrow }\vec{m}_0$};
        % Normal vector n 
        \draw[->, gray] (3, 1) -- (3.5,1.5) node[pos = .2,right]  {\scriptsize$\mathbf{n}$};
        \draw[thick, <->] (.25,3.75) -- (3.5,.5) node[pos=0.8, left ] {$H$~};    
        % Point \vec{m}_0
        \filldraw (2,2) circle (2pt) node[below left] {$\vec{m}_0$};
    \end{tikzpicture}\\
    \text{\scriptsize (a) {$\sigma(\simpeq)$ is a knotted contraction}} 
    &
    \text{\scriptsize (b) {$\sigma(\simpeq)$ is a knotted weakening}}
    &
    \text{\scriptsize (c) {$\sigma(\simpeq)$ is trivial}}
    \end{array}$
    \caption{\scriptsize In each picture, the set $H$ is a hyperplane containing the points $\vec{m}_1,\ldots,\vec{m}_k$ corresponding to the joinands of $\simpeq$ and the vector $\mathbf{n}$ has non-negative integer entries and lies normal to $H$; i.e., $H$ is the solution set to the linear system $\mathbf{n}^\top\mathbf{x} = b$ for some integer $b\geq 0$. 
    The plane $\mathbb{N}^n$ is a disjoint union of the sets $H: \mathbf{n}^\top \mathbf{x} = b$, $H^-: \mathbf{n}^\top \mathbf{x} < b$, and $H^+: \mathbf{n}^\top \mathbf{x} > b$. 
    The shaded regions represent the negative and positive cones with vertex $\vec{m}_0$, respectively denoted by sets ${\downarrow}{\vec{m}_0}$ and  ${\uparrow}{\vec{m}_0}$.
    For $\sigma$ the substitution $x_i\mapsto x^{\mathbf{n}(i)}$, $\sigma(\simpeq)$ is the equation $x^a\leq x^b$ where $a = \mathbf{n}^\top \vec{m}_0$. 
    The case (a) shows when $\vec{m}_0\in H^-$, in which case $a<b$ and $\sigma(\simpeq)$ is a knotted contractive equation; moreover, $\simpeq$ is {\joinincreasing} iff $\{\vec{m}_1,\ldots ,\vec{m}_k\}\cap {\uparrow}\vec{m}_0 \neq \varnothing$.  
    The case (b) shows when $\vec{m}_0\in H^+$, in which case $a>b$ and $\sigma(\simpeq)$ is a knotted weakening equation; moreover, $\simpeq$ is {\joindecreasing} iff $\{\vec{m}_1,\ldots ,\vec{m}_k\}\cap {\downarrow} \vec{m}_0 \neq \varnothing$.  
    The case (c) shows when $\vec{m}_0\in H$,  in which case $a=b$ and $\sigma(\simpeq)$ is trivial. 
    }\label{fig:preknotted}
\end{figure}
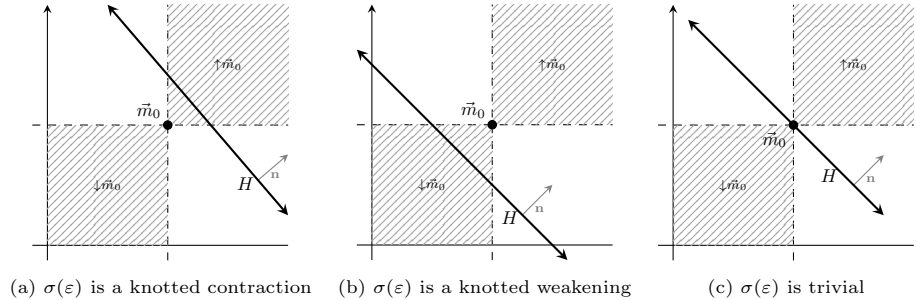
    
Note that the notions of an equation being {\joinincreasing} or {\joindecreasing} admits a simple geometric description; e.g., certain points are contained in the shaded regions depicted in Figure~\ref{fig:preknotted}.
Let ${\downarrow} \vec{m}_0$, called the \emph{negative cone} with vertex $\vec{m}_0$, denote the downset of $\vec{m}_0$ in $\mathbb{N}^n$.
Dually, let ${\uparrow} \vec{m}_0$, called the \emph{positive cone} with vertex $\vec{m}_0$, denote the upset of $\vec{m}_0$ in $\mathbb{N}^n$.
Then $\simpeq$ is {\joinincreasing} iff  $\{\vec{m}_1,\ldots ,\vec{m}_k\}\cap {\uparrow}\vec{m}_0 \neq \varnothing$, and  $\simpeq$ is {\joindecreasing} iff  $\{\vec{m}_1,\ldots ,\vec{m}_k\}\cap {\downarrow}\vec{m}_0 \neq \varnothing$.
For example,  the equation $xy \leq x \vee x^2y$ has $m_0=(1,1)$, $m_1=(1,0)$, and $m_2=(2,1)$, with $m_1 \in {\downarrow} m_0$ and $m_2\in {\uparrow}m_0$, so it is both {\joinincreasing} and {\joindecreasing}. 
We note that Figure~\ref{fig:preknotted} covers both cases of a non-trivial  preknotted equation: it is either joinand-increasing as in (a) or joinand-decreasing as in (b); it cannot be both according to condition (i) of Lemma~\ref{l: preknotted}, since the entries of the normal vector need to be positive. 
In the (weak)-commutative setting, every preknotted contractive (resp., weakening) equation $\simpeq$ specifies a variety with an Ackermannian upper bound, and this bound is tight  whenever that equation $\simpeq$ is {\joinincreasing} (resp., non-integral {\joindecreasing}). 
No lower bounds are known (aside from the $\PSPACE$-hard result from \cite{horcik2011}) for those failing such conditions. 
Below, we list two simple examples which are preknotted contraction (entails $x^2\leq x^3$) in the first, preknotted weakening (entails $x^3\leq x^2$) in the second, neither are {\joinincreasing} nor {\joindecreasing}, and so no tight bounds are known.
    
\begin{openproblem} 
Does the variety of residuated lattices axiomatized by commutativity and the equation $xy\leq x^3 \vee y^3$ admit a primitive recursive quasiequational theory? 
\end{openproblem}

\begin{openproblem} 
Does the variety of residuated lattices axiomatized by commutativity and the equation $xyz\leq x^2 \vee y^2\vee z^2$ admit a primitive recursive quasiequational theory?
\end{openproblem}

To understand potency geometrically, it suffices to find a description for multi-contractible and multi-weak equations. 
Indeed, mentioned in the above figure, a hyperplane $H\subseteq \mathbb{R}^n$ separates the space into three disjoint pieces: $\mathbb{R}^n=H^-\cup H\cup H^+$. 
It is then not difficult to see that $\simpeq$ is multi-contractible iff there exists an affine hyperplane $H:\mathbf{n}^\top \mathbf{x}= b $ for some $\mathbf{n}\in \mathbb{N}^n$ and $b>0$ such that $\vec{m}_0\in H^-$ while $\vec{m}_1,\ldots,\vec{m}_k\in H^+\cup H$. 
Similarly, $\simpeq$ is multi-weak iff there exists an affine hyperplane $H:\mathbf{n}^\top \mathbf{x}= b $ for some $\mathbf{n}\in \mathbb{N}^n$ and $b>0$ such that $\vec{m}_0\in H^+$ while $\vec{m}_1,\ldots,\vec{m}_k\in H^-\cup H$.

\begin{remark}
From a syntactic perspective, from all these geometric descriptions and from the use of Gordan's Lemma in Lemma~\ref{l:Zeqs}, it follows that no multi-contractible nor multi-weak (and hence no preknotted) equation can be a $\mathbf{Z}$-equation. 
This is also obvious from the semantical perspective since $\mathbf{Z}$ is a chain containing no idempotents (but $0$) in its positive and negative cone. 
\end{remark}

Of course, (at least individually) many such equations yield undecidable theories, even in the commutative case, such as the multi-contraction equation $x\leq x^2\vee x^3$ and the multi-weakening $x^3\leq x^2\vee x$, as they are spineless. 
In fact, as we will describe in detail later, all single variable equations, except for integrality, admit an \ACK-hard lower bound. Therefore, to identify equations not covered by any method, we need at least two variables. 
Such equations (for example multi-contractible) must be prespinal and of a particular form: $\{ m_1,\ldots,m_k\}$ must be contained in $H^+\cup H$ but not entirely in $H$ and be disjoint from ${\uparrow}m_0$, for some positively oriented hyperplane $H$ with non-negative coefficients. 
For instance, the following equation is multi-contractible ($x,y\mapsto x$), neither {\joinincreasing} nor {\joindecreasing}, not knotted (all of which can be seen geometrically) but prespinal ($x\mapsto x$ and $y\mapsto 1$).

\begin{openproblem}
Is the quasiequational theory \ACK-hard for the variety axiomatized by commutativity and $ xy\leq x^3\vee y^3 \vee y^4 $? Is it decidable? 
\end{openproblem} 

Similarly, the following equation is multi-weak ($x,y\mapsto x^2$ and $z\mapsto x$), neither {\joinincreasing} nor {\joindecreasing}, not knotted but prespinal ($x\mapsto x$ and $y,z\mapsto 1$).

\begin{openproblem}
Is the quasiequational theory \ACK-hard for the variety axiomatized by commutativity and $ xyz\leq x^2\vee y^2 \vee z^2 \vee z^3$? Is it decidable? 
\end{openproblem}

Recall that an equation $x_1 x_2 \cdots x_\ell \leq t_1 \jn  \cdots \jn t_k$ is simple if the variables
$x_1,\ldots,x_\ell$ are pairwise distinct, each joinand $t_i$ is a product of variables, and every variable in the equation occurs on both sides; by exception, integrality $x\leq 1$ is also considered a simple equation.
A simple equation is {\joinincreasing} ({\joindecreasing}) if there is a joinand on the right-hand side in which each variable occurring in the equation appears at least (resp. at most) once. 
Our results for lower bounds cover all the {\joinincreasing}  and all the {\joindecreasing} simple equations, thus vastly generalizing knotted equations of both types. 
There are still simple equations that are of neither form, such as $xy\leq x^2\vee y^2$.
Let us now consider in detail the simple equations over a single variable.

\subsection{One-variable $\{\vee,\cdot,1\}$-equations}
    
We will be considering $1$-variable inequalities of the form $x^n \leq x^{n_1} \jn \cdots \jn x^{n_k}$, as every $\{\vee,\cdot,1\}$-equation is equivalent to a conjunction of such equations. 
If one of the joinands on the right is $x^n$, 
then the equation is trivial and  defines the variety of all residuated lattices, which  has an undecidable quasiequational theory, even in the (weakly) commutative case (cf., \cite[Cor.~4.6]{galatos2022}); so we assume each $n_j$ is distinct from $n$. 
If $n = 0$, then the left-hand side is the term $1$ and the equation defines the trivial variety of 1-element algebras, so the quasiequational theory is decidable (with constant-time complexity). In the following we assume $0\neq n \neq n_j$, for all $j$.
    
Now, if $k\geq 2$ and there are at least two distinct non-identity joinands (i.e., two positive values $n_i,n_j>0$ with $i\neq j$), then the equation is spineless by Lemma~\ref{l:spineless_sufficient} and, by Lemma~\ref{l: spineless}, the corresponding variety (as well as every variety axiomatized by the equation and commutativity)  has an undecidable word problem.
Otherwise, $k\leq 2$ and the equation is of the form  $x^n\leq x^m$ or $x^n\leq x^m \vee 1$. 
If $m=0$ (i.e., the right-hand side only contains the term $1$), then both equations are of the form $x^n\leq  1$ which is equivalent to integrality and, as we have already discussed earlier, the quasiequational theory is hyper-Ackermannian-complete \cite{GreatiRamanayake2024} (\TOWER-complete in the commutative case \cite{tanaka2022}) and the equational theory is \PSPACE-complete (even in the commutative case) \cite{horcik2011}.
For the remaining cases ($k\leq 2$ and $m\geq 1$) we discuss separately the situation in the  fully noncommutative setting and in the (weakly) commutative setting.
    
If $m\geq 2$, then for each equation, $x^n\leq x^m$ or $x^n\leq x^m \vee 1$, the corresponding variety  has an undecidable word problem by Corollary~\ref{c: horcik2015}. 
If $m=1$, then each of the equations ($x^n\leq x$ and $x^n\leq x\vee 1$) is equivalent to a {\joindecreasing} equation, hence the corresponding variety has an \ACK-hard quasiequational theory. 
In particular, in the case of  mingle ($x^2 \leq x$) the variety has decidable universal theory (actually, the FEP) as shown in \cite{horcik2017}.
  
\begin{openproblem}\label{op:Minglenoncom}%
    What is the complexity  of the quasiequational theory of the variety axiomatized by $x^2\leq x$? Is it \ACK-complete?
\end{openproblem} 
    
However, decidability is open for all other knotted weakening equations ($x^{n}\leq x$ with $n \geq 3$) and for the equations $x^n\leq x\vee 1$, where $n\geq 2$.  
The equations $x^n\leq 1 \vee x$, for $n\geq2$, are all equivalent to each other by Lemma~\ref{l: x2<xv1} and the decidability of the quasiequational theory, even in the commutative setting, remains open. 

\begin{openproblem}\label{op:noncomm} ~
    \begin{enumerate}
        \item Is the quasiequational theory for the variety of residuated lattices axiomatized by $x^n  \leq x$ decidable, for $n\geq 3$? If so, is it $\ACK$-complete? Does it have the FEP?
        \item Is the quasiequational theory for the variety of residuated lattices axiomatized by $x^2  \leq x\vee 1$ decidable? If so, is it $\ACK$-complete? Does it have the FEP?
    \end{enumerate}
\end{openproblem}

In the (weakly) commutative setting (with $k\leq 2$ and $m\geq 1$), the equations $x^n\leq x^m$ are non-integral knotted equations and hence the corresponding varieties have \ACK-complete quasiequational theories.
The remaining equations, $x^n\leq x^m \vee 1$ with $n, m\geq 1$, are spinal but not preknotted, so the decidability of the quasiequational theory for the corresponding varieties is open; for $n<m$ as these are $\mathbf{Z}$-equations, so the FEP fails.

\begin{openproblem} 
    For any fixed positive integers $m,n$, we ask:
    Is the quasi\-equational theory for the variety of residuated lattices axiomatized by commutativity and $x^{m+n}\leq x^m\vee 1$  decidable (for example for $x^2\leq x \jn 1$)? If so, is it \ACK-complete? Does it have the FEP?
\end{openproblem} 
       
\begin{openproblem} 
    More generally, for any fixed positive integers $m,n$, we ask:
    For $n,m$ distinct,
    is the quasiequational theory or the variety axiomatized by commutativity and $ x^n\leq x^m \vee 1$ decidable (for example, $x \leq x^2\jn 1$)? If so, is it \ACK-complete? 
\end{openproblem} 

Finally, we consider varieties axiomatized by a set $\eqset$ of equations of the above types, focusing on the cases where this set is not a singleton.

In particular, we will assume $\eqset$ contains no trivial nor trivializing equations, i.e., for all equations in $\eqset$ the left-hand side does not occur as a joinand on the right (non-trivial), and the variable $x$ has at least one occurrence on the left (non-trivializing).

First we consider the case in which $\eqset$ entails integrality, modulo the theory of residuated lattices. By Remark~\ref{remark:combins of eqs}, $\eqset$ entails integrality iff it contains a \emph{pre-integral} equation (i.e., preknotted and the corresponding entailed knotted equation is integrality); such an equation has the form $x^n\leq 1$ with $n\geq 1$, which in turn is equivalent to integrality. 
It is therefore immediate from \cite{GreatiRamanayake2024} that any integral subvariety of residuated lattices, relatively axiomatized by $\mathcal{N}_2$ equations, has a hyper-Ackermannian upper bound---consequently this holds for the set $\eqset$ even in the presence of (weak) commutativity. Note that integrality entails all {\joindecreasing} equations.
If  $\eqset$ entails integrality and also contains some non-{\joindecreasing} equation, i.e., a multi-contraction equation $x^{n}\leq x^{n+c_1}\vee\cdots \vee x^{n+c_k}$ for some $n,c_1,\ldots, c_k\geq 1$, then $x^n \leq x^{n+c_1} \jn \cdots \jn x^{n+c_k} \leq x^{n+1}\leq x^n$, so $x \leq 1$ and $x^{n+1}=x^n$. 
Taking $n$ to be the minimal among all multi-contraction equations in $\eqset$, we get that $\eqset$  is equivalent to the conjunction of integrality and  $x^{n+1}=x^n$; indeed, $x\leq 1$ entails all {\joindecreasing} equations and
$x^n=x^{n+1}$ entails all other multi-contraction equations in $\eqset$ by the minimality of $n$ ($x^{m}\leq x^{m+d_1}\leq x^{m+d_1}\vee \cdots x^{m+d_\ell}$ for all $m\geq n$ and $d_1,\ldots, d_\ell\geq 1$).
In particular, if the minimal value is $n=1$, then $\eqset$ is equivalent to $x^2=x\leq 1$, so it axiomatizes the variety of Brouwerian algebras, which is \PSPACE-complete. 
On the other hand, if $n\geq 2$, then the exact complexity is unknown in the fully noncommutative case.
\begin{openproblem}
    What is the complexity of the (at most hyper-Ackerman\-nian) quasiequational theory of integral $n$-potent residuated lattices for $n\geq 2$? Is it primitive recursive?
\end{openproblem}
\noindent Restricting the question above to the (weak) commutative case, by Lemma~\ref{l: EqThpotency}, we know that the decision procedure is at most \EXPTIME.

We now assume that $\eqset$ is not pre-integral, i.e., every equation in $\eqset$ has some joinand different from $1$ on the right-hand side. We begin with the fully noncommutative case.   

If $\eqset$ consists entirely of hereditarily-square equations, then by  Corollary~\ref{c: horcik2015}  $\mathrm{Mod}(\eqset)$ has an undecidable quasiequational theory. 
Otherwise, $\eqset$ contains at least some nontrivial equation of the form $x^n\leq x$ or  $x^n\leq x\vee 1$, for $n \geq 2$ (we are excluding $x^n\leq  1$ as it is equivalent to integrality); i.e., at least some nontrivial equation of the form $x^n\leq x$, for $n \geq 2$, or  $x^2\leq x\vee 1$, by Lemma~\ref{l: x2<xv1}.
Note that these equations are shrinking so their equational theory is \PSPACE-complete. Also, all of these equations are non-integral joinand-decreasing, so if $\eqset$ consists entirely of such equations and possibly other joinand-decreasing hereditarily-square equations, then we obtain an Ackermaniann lower bound by  Theorem~\ref{t: AckHard/complete_containing_CRLc/CRLa}, but no upper bound is known.
In the remaining cases $\eqset$ contains equations of the form $x^n\leq x$, for $n \geq 2$, or  $x^2\leq x\vee 1$, and also of hereditarily-square equations not subsumed by these equations. 
Each equation of the form $x^n\leq x\vee 1$ is equivalent to $x^2\leq x\vee 1$.
Very little is known for these remaining cases; namely any set $\eqset$ containing either $x^{n}\leq x$ or $x^2\leq x\vee 1$.

\begin{openproblem}
    Is the quasiequational theory decidable for the variety axiomatized by  $x^3\leq x$ and $x^3 \leq x^4 \vee x^6$; and if so, is it primitive recursive? 
    What about $x=x^3$? How about the combination of $x^3\leq x$ and $x^2\leq x\vee 1$?
\end{openproblem}

Note that for the combination of $x^3\leq x$ and $x^3 \leq x^4 \vee x^6$ we have $x^3 \leq x^4 \vee x^3x^3\leq x^4 \vee xx^3=x^4$, so we have the combination of a knotted weakening and a knotted contraction equation, hence we obtain potency, but in the absence of weak commutativity, we do not have results that yield decidability of the quasiequational theory.

In the (weakly) commutative setting, if $\eqset$  is finite and consists entirely of spineless equations, then by Lemma~\ref{l: spineless} the corresponding variety has undecidable quasiequational theory (even word problem); recall, by Lemma~\ref{l:spineless_sufficient}, that an equation $x^n \leq x^{n_1} \jn \cdots \jn x^{n_k}$ with distinct $n_i$'s is spineless iff it contains two joinands $x^m$ and $x^k$, with $m>k>0$. If $\eqset$ is infinite little is known. 
Otherwise, $\eqset$ is finite and contains some prespinal equation, which has to be spinal (as it is a $1$-variable equation); i.e., it has the form $x^n \leq x^m$, or $x^n \leq 1 \jn x^m$, where $m>0$ (since we do not include integrality) and $0<n \neq m$  (since we assumed non-trivializing and nontrivial, respectively). 
We first assume that $\eqset$ contains a knotted equation $x^n \leq x^m$; in this case the quasiequational theory has an Ackermannian upper bound, by Theorem~\ref{t:genknotted_ackhard}. 
If, furthermore, $\eqset$ consists entirely of {\joinincreasing} or entirely of non-integral {\joindecreasing} equations, then this bound is tight by  Theorem~\ref{t: AckHard/complete_containing_CRLc/CRLa}.
Otherwise, $\eqset$ contains a knotted contraction and a non-{\joinincreasing} equation (i.e., some multi-weakening, as it is a one-variable equation)
or a knotted weakening equation  and a  non-{\joindecreasing} equation (i.e., a  multi-contraction); so by Remark~\ref{remark:combins of eqs}, the variety is potent, therefore it has \EXPTIME upper bound by Lemma~\ref{l: EqThpotency}. 

Finally, if $\eqset$ does not contain any knotted equation but does contain some  equation $x^n \leq 1 \jn x^m$, for distinct positive $m,n$ (note such equations are both {\joinincreasing} and {\joindecreasing}), there are no known upper bound, or even decidability, results  (see Open problem~\ref{op:any_nonknotted_dec?}).
If $\eqset$ consists entirely of {\joinincreasing} or entirely of {\joindecreasing} equations, then the quasiequational theory is \ACK-hard; otherwise, there are no known interesting lower bounds. 

\begin{openproblem}
    Is the quasiequational theory decidable for the variety axiomatized by commutativity, $x\leq x^2 \vee x^3$, and $x^2 \leq x \vee 1$; and if so, is it primitive recursive? 
\end{openproblem}

\subsection{Equational theory}

In terms of the equational theory, as we mentioned in Lemma~\ref{l: HorcikTerui}, \cite[Cor.~4.6]{horcik2011} establishes the \PSPACE-hardness for certain varieties; in particular for any variety of residuated lattices/FL-algebras axiomatized by $\{\vee,\cdot,1 \}$-equations. The main other method for establishing lower bounds use a deduction theorem, when it is available in the logic.

We now mention certain cases of $1$-variable equations of the form $x^n\leq x^{m_1}\vee\cdots\vee x^{m_k}$ that axiomatize varieties with known or unknown decidability of the equational theory. We note first that  integrality, $x^n\leq x$, and even the equations $x^n\leq x\vee 1$ (which are equivalent to $x^2\leq x\vee 1$, by Lemma~\ref{l: x2<xv1}) yield decidable equational theories with \PSPACE-complete complexity by \cite{horcik2011}, with or without (weak) commutativity.

\begin{remark}
    In fact, the equation $x^2\leq x\vee 1$ is the \emph{only} known example of a non-preknotted equation axiomatizing a decidable (in this case, equational) theory. We note that this equation is equivalent to the quasiequation $1 \leq x \Rightarrow x=x^2$, expressing that all elements in the positive cone are idempotent. Indeed, if an algebra satisfies $x^2\leq x\vee 1$ and $a$ is a positive element, then $a\leq a^2\leq a\vee 1=a$, so the quasiequation holds; conversely, if an algebra satisfies the quasiequation, then for  an arbitrary element $b$, $b\leq b^2 \vee b \vee 1 = (b\vee 1)^2 = b\vee 1$.
\end{remark}

In the noncommutative setting, 
by \cite[\S5.2]{horcik2016}, the equational theory is undecidable for every variety axiomatized by a (single)  knotted contraction equation; moreover, this holds for a (single) multi-contraction equation \ref{eq:eqexpansive}: $x^k\leq x^{k+c_1}\vee\cdots x^{k+c_n}$, where $k,c_1,\ldots,c_n\geq 1$, by arguing as in \cite[\S5.2]{horcik2016} and using the entailed $k$-potency.

In the  commutative setting, or more generally with the level of commutativity of Corollary~\ref{c: knotcontr+wc=Ack Logic}, the knotted contractive equations yield an \ACK-complete equational theory. 
For the knotted weakening varieties we have a decidable equational theory because  the FEP holds and the complexity of the quasiequational (hence also of the equational theory) is at most Ackermannian; however, tight bounds for the complexity are unknown (except for mingle, 
which has \PSPACE complexity as described above).

\begin{openproblem}
    Is the equational theory primitive recursive for the variety of commutative residuated lattices axiomatized by
    $x^{n+c}\leq x^n$ for $n\geq 2$ and $c >0$? 
\end{openproblem}

On the other hand, the conjunction of $q$-commutativity and any finite set of equations of the form $x^n\leq x^{n+c_1}\vee \cdots \vee x^{n+c_k}$, for $k\geq 2$ and positive $c_j$'s, yields a variety that has an undecidable  equational theory, by Theorem~\ref{t: wmulticontrUND}.
However, for the remaining cases, even decidability is unknown.

\begin{openproblem}
    Is the equational theory decidable for the variety of commutative residuated lattices axiomatized by $x^{n}\leq x^m \lor x^k \lor Y$, where    $0\leq m<n<k$ and $Y$ is a possibly empty join of powers of $x$?
\end{openproblem}

Let us now briefly mention some known results and open problems for commutative varieties $\mathsf{CRL}_{\eqset}$ axiomatized by a set $\eqset$ of 1-variable equations, not already discussed. 
If $\eqset$ consists entirely of shrinking equations, i.e., (the linearizations of) $x\leq 1$, $x^n\leq x$, and $x^n\leq x\vee 1$, then the equational theory of $\mathsf{CRL}_{\eqset}$ is \PSPACE-complete by \cite{horcik2011} (also see Remark~\ref{r: shrinking}).
If $\eqset$ entails a potent equation $x^n = x^{n+m}$, for $n,m\geq 1$, then the equational theory of $\mathsf{CRL}_{\eqset}$ has at most \EXPTIME complexity as a consequence of Lemma~\ref{l: EqThpotency} (and is \PSPACE-complete when $\eqset$ contains integrality and entails $x=x^2$); this occurs exactly when $\eqset$ contains a knotted contractive together with a multi-weakening equation, or a knotted weakening equation and a multi-contraction equation (cf., Remark~\ref{remark:combins of eqs}). 
If $\eqset$ consists only of non-trivializing {\joinincreasing} equations (i.e., $\eqset\subseteq \eqset_\mathrm{inc}$) and also some multi-contraction equation then the equational` theory of $\mathsf{CRL}_{\eqset}$ is \ACK-hard---it is \ACK-complete when $\eqset$ contains knotted contraction, undecidable if $\eqset$ contains only spineless equations (i.e., no equations of the form $x^n\leq x^m\vee 1$)---and otherwise the complexity is generally unknown. We state some simple examples of outstanding problems below.
\begin{openproblem}~
    \begin{enumerate}
        \item Is the (decidable) equational theory of commutative residuated lattices satisfying $x^2\leq x$ and $x\leq 1\vee x^2$ primitive recursive?
        \item Is the (\ACK-hard) equational theory of commutative residuated lattices satisfying $x\leq 1\vee x^2$ and $x\leq x^2\vee x^3$ decidable? If so, is it \ACK-complete?
    \end{enumerate}
\end{openproblem}

The situation is even worse in the noncommutative setting. As we have described above, any multi-contraction equation yields undecidable equational theory, but even the decidability of any other $1$-variable equation not already mentioned above remains open.

\section{Notable open problems in lower bounding}

The lower bound results that we obtained are restricted to $\mathcal{N}_2$-extensions. Given that the upper bounds for $\mathcal{P}^\flat_3$-extensions (related
to hypersequents) are worse (larger) than the ones for $\mathcal{N}_2$-extensions, it is expected that the lower bounds for some $\mathcal{P}^\flat_3$-extensions would be hyper-Ackermannian. However, it is unclear how the algebraic counter machine argument in this work should be extended.
We expect that a suitable modification of the machines and of the residuated frames (to the residuated hyperframe setting) is possible and remains to be fleshed out.

\begin{openproblem}
The monoidal t-norm based logic $\m{MTL}$ was introduced by Esteva and Godo~\cite{EstGod01} in 2001, and it is axiomatized by extending $\m{FL_{ew}}$ with prelinearity $(\UniFmA\imp \UniFmB)\lor(\UniFmB\imp \UniFmA)$. $\m{MTL}$ is prominent because of the importance of its features to fuzzy logics. Specifically, it describes the common behaviors of all fuzzy logics based on left-continuous t-norms. Its complexity is a ``long-standing problem within propositional fuzzy logics."~\cite{Han17}.
The only known~\cite{BalLanRam21LICS} upper bound is hyper-Ackermannian $\UniFGHProbOneAppLevel{\omega^\omega}$ (see Theorem~\ref{fact:flewwmnr-weak-com-ackermannian}); no non-trivial lower bound is known. What is a non-trivial lower bound for $\m{MTL}$?
\end{openproblem}

\section{On the possibility of unifying forward and backward proof search}

Our methods for obtaining complexity lower bounds for knotted contraction and for knotted weakening rules are essentially unified in view of Theorem~\ref{thm:lossyadmissmach}, which serves as a common basis for both cases. 
The same is true for our results on the finite embeddability property (hence also of decidability of the quasiequational theory/deducibility), which hold for knotted rules of both types.

On the other hand, our arguments---via proof calculi---for complexity upper bounds for the two cases are quite different. For the knotted contraction case, we consider backwards proof search focusing on trees of hypersequents, while for the knotted weakening case, we consider a forward proof search focusing on sequences of sets of hypersequents that are provable at each successive stage. 
Also, the proofs of the FEP that establish finiteness cannot be adapted in an obvious way to provide upper complexity bounds in a uniform way; as a result, at the moment there is no common approach to these two cases of knotted rules.

\section{Bounding the size of countermodels: identifying the obstacle}

\newcommand{\up}[1]{\text{up}(#1)}
\newcommand{\PS}{\mathcal{P}}

In Chapter~\ref{sec:fepP3}, we obtained finite algebraic countermodels for a large family of axiomatic extensions of $\m{FL}$ via the FEP. 
It is natural to ask if finiteness can be strengthened to an upper bound on the size of these countermodels as a function of the unprovable (input) formula. 
Here we present a finite countermodel construction for $\m{FL_{ec}}$ inspired by~\cite[Thm.~3.15]{GalJip13} (to our knowledge, the first for substructural logics that is based on a proof-theoretic analysis) in order to pinpoint the obstacle we encountered in extending finiteness to a size bound. 
%Incidentally, we are unaware of any result in the literature establishing non-primitive recursive upper bounds on the size of substructural countermodels.

Since $\m{FL_{ec}}$ has exchange, we write $\UniSequent{x}{\UniMSetSucA}$ to represent an $\UniSubfmlaHyperseqSet$-sequent $\UniSequent{\UniMSetFmA}{\UniMSetSucA}$, where $x \in \UniNaturalSet^{\UniSetCard{\UniSubfmlaHyperseqSet}}$
is the vector of multiplicities of the formulas in $\UniMSetFmA$ (assuming we fixed an enumeration of $\UniSubfmlaHyperseqSet$); also, we write $x_j$ for the $j$-th coordinate of $x$.
In the same spirit, we will write $x,y$ to mean $x+y$ (i.e., usual vector sum), hence
$\UniSequent{x,y}{\UniMSetSucA}$
represents
$\UniSequent{\UniMSetFmA,\UniMSetFmB}{\UniMSetSucA}$
when
$x$ and $y$ are respectively the vectors corresponding to the formula multiplicities of
$\UniMSetFmA$ and $\UniMSetFmB$.
% Moreover, to simplify the presentation, whenever we write
% `$\UniMSetSucA \in \UniSubfmlaHyperseqSet$',
% we mean either that $\UniMSetSucA$ is an element of $\UniSubfmlaHyperseqSet$ or it is the empty stoup.

Given  a non-provable sequent $s$, we define a residuated frame based on the naive proof search ~$\PS$ rooted at~$s$.
Take $W$ to be $\mathbb{N}^{\UniSetCard{\UniSubfmlaHyperseqSet}}$ and $W'$  to be $W \times \UniSubfmlaHyperseqSet \times W$ (here $\UniSubfmlaHyperseqSet$ is assumed also to contain the empty stoup). 
The frame relation is defined
\begin{center}
    $x \mathrel{N_s} (y, \UniMSetSucA, z)$ iff $\UniSequent{y,x,z}{\UniMSetSucA}$ is provable or it does not belong to $\PS$. 
\end{center}
To prove $\m W^+$ is a finite algebra (see Section~\ref{s: RelSem} for details), it suffices to show that there are finitely many basic closed sets $(y,\UniMSetSucA,z)^\triangleleft:=\{x \in W: x \mathrel{N_s} (y, \UniMSetSucA, z)\}$, where $(y,\UniMSetSucA,z) \in W'$. 
Equivalently, it suffices to show that the collection of complements of basic closed sets is finite, i.e.,
that the following set is finite
\[
\Big\{
\{
x \in \UniNaturalSet^{\UniSetCard{\UniSubfmlaHyperseqSet}} \mid \text{$\UniSequent{y,x,z}{\UniMSetSucA}$ unprovable and $\UniSequent{y,x,z}{\UniMSetSucA}\in\mathcal{P}$}
\}
\mid \UniMSetSucA\in \UniSubfmlaHyperseqSet, \, y,z\in\mathbb{N}^{\UniSetCard{\UniSubfmlaHyperseqSet}}
\Big\}.
\]
This is simply
\begin{multline}\label{eq-bounding-main}
\Big\{
\{
x \in 
\UniNaturalSet^{\UniSetCard{\UniSubfmlaHyperseqSet}} \mid \text{$\UniSequent{y,x,z}{\UniMSetSucA}$ unprovable}
\}
\;\cap\\
\{
x \in \UniNaturalSet^{\UniSetCard{\UniSubfmlaHyperseqSet}} \mid \UniSequent{y,x,z}{\UniMSetSucA}\in\mathcal{P}
\}
\mid
 \UniMSetSucA\in \UniSubfmlaHyperseqSet, y,z\in\mathbb{N}^{|\UniSubfmlaHyperseqSet|}
\Big\}.
\end{multline}
Let us obtain a finite handle on each of the two sets in the intersection.

Observe that $U(\UniMSetSucA) := \{ y \in  \UniNaturalSet^{\UniSetCard{\UniSubfmlaHyperseqSet}}\mid \text{$\UniSequent{y}{\UniMSetSucA}$ is unprovable}\}$, the set of unprovable sequents with succedent $\Pi$, is an upset with respect to the ordering of the contraction nwqo
$\UniWqoExtModRelProd{2}{1}{{\UniSetCard{\UniSubfmlaHyperseqSet}}}$, i.e., $\UniExtModRel{2}{1}^{\UniSetCard{\UniSubfmlaHyperseqSet}}$.
Recall that ${\uparrow} x\UniSymbDef\{y\in\UniNaturalSet^{\UniSetCard{\UniSubfmlaHyperseqSet}}\mid x\UniExtModRel{2}{1}^{\UniSetCard{\UniSubfmlaHyperseqSet}}y\}$ denotes the principal upset generated by $x$. 
In a wqo, every upset $V$ is generated by a finite set $\min(V)$ of minimal elements, i.e., $V = {\uparrow}{\min(V)}$~\cite{schmitz2012notes}.
Let $U_\UniMSetSucA \UniSymbDef \min(U(\UniMSetSucA))$
for each $\UniMSetSucA\in \UniSubfmlaHyperseqSet$.
Given $u \in \UniNaturalSet^{\UniSetCard{\UniSubfmlaHyperseqSet}}$, we define $u\dot{-} (y,z)$ as the set
$\{ x \in
\UniNaturalSet^{\UniSetCard{\UniSubfmlaHyperseqSet}}
\mid u\UniExtModRel{2}{1}^{\UniSetCard{\UniSubfmlaHyperseqSet}} y,x,z\}$ i.e., the elements that together with $y,z$ reach above $u$ in the contraction nwqo.
It is not hard to check that
$U(y,z,\UniMSetSucA) \UniSymbDef \{
x \in \UniNaturalSet^{\UniSetCard{\UniSubfmlaHyperseqSet}} \mid \text{$\UniSequent{y,x,z}{\UniMSetSucA}$ unprovable}
\}=\bigcup_{u \in U_\UniMSetSucA} (u\dot{-}(y,z))$, for any $\UniMSetSucA \in \UniSubfmlaHyperseqSet$ and $y,z\in\mathbb{N}^{\UniSetCard{\UniSubfmlaHyperseqSet}}$. 
Indeed, $x \in U(y,z,\UniMSetSucA)$ iff $\UniSequent{y,x,z}{\UniMSetSucA}$ is unprovable iff
$u\UniExtModRel{2}{1}^{\UniSetCard{\UniSubfmlaHyperseqSet}}
{y,x,z}$
for some $u \in U_\UniMSetSucA$
iff
$x \in \bigcup_{u \in U_\UniMSetSucA} (u\dot{-}(y,z))$.

To express the set $P(y,z,\UniMSetSucA) \UniSymbDef \{x \in \UniNaturalSet^{\UniSetCard{\UniSubfmlaHyperseqSet}} \mid \UniSequent{y,x,z}{\UniMSetSucA}\in\mathcal{P}\}$ in terms of a finite union, we prove in the following lemma that every sequent in $\mathcal{P}$ is $\UniExtModRel{2}{1}^{\UniSetCard{\UniSubfmlaHyperseqSet}}$-above some sequent in the finite proof search $\mathcal{F}$ that is conducted on the amended sequent calculus $\UniHRuleAbsorb{\UniFLeExtSCalc{\UniCProp}}$, whose logical rules absorb a fixed amount of contraction (see Chapter~\ref{sec:ub-wc}); in other words, we show that $\mathcal{P} \subseteq {\uparrow} \mathcal{F}$.
In what follows, we denote by $\UniWqoRel{}$ the wqo relation on (hyper)sequents $\UniHyperseqCtrWqo{2}{1}{\UniSubfmlaHyperseqSet}$ introduced in Definition~\ref{def:nwqo-ec-mn}.

\begin{lemma}\label{l: P0belowP}
Let $\PS$ be the naive proof search tree in $\m{FL_{ec}}$ rooted at sequent $s$, and let $\mathcal{F}$ be its finite proof-search tree in $\UniHRuleAbsorb{\UniFLeExtSCalc{\UniCProp}}$. 
For any node in $\PS$ labelled with the sequent $t$, there is a node in $\mathcal{F}$ labelled with the sequent $t'$ such that $t' \UniWqoRel{} t$.
\end{lemma}
\begin{proof}
We define the $\PS$-height of a node as the length of the unique path in $\PS$ from the root to that node.
We define the $\mathcal{F}$-height analogously. 
We prove the stronger claim that a node of $\PS$-height~$k$ with label~$t$ implies a node of $\mathcal{F}$-height $\leq k$ with label $t'$ such that $t' \UniWqoRel{} t$. 
This claim is established by induction on~$k$.

\textit{Base case.} If $k=1$, then $t$ is the root, so the claim clearly holds.

\textit{Inductive case ($k=K+1$)}. In that case, $\PS$ contains an instance of a rule~$r$ whose premise is the node labelled $t$, and whose parent is a node of $\PS$-height~$K$ labelled by some $t_0$. 
By induction hypothesis, there is a node in $\mathcal{F}$ of $\mathcal{F}$-height $\leq K$ labelled $t_0'$ such that 
$t_0' \UniWqoRel{} t_0$. 
If, for example, $r$ is contraction, then $t_0' \UniWqoRel{} t$ so we are done by transitivity of $\UniWqoRel{}$. 
To handle the general case, one can show  that there is a rule instance of $r$ with conclusion $t_0'$ whose premise is some $t' \UniWqoRel{} t$. 
Indeed, if it is not the case that $t' \UniWqoRel{} t$, this is because there are some coordinates in the antecedent at which $t'$ is $0$ and $t$ takes a positive value. 
This is precisely the situation that $\UniHRuleAbsorb{\UniFLeExtSCalc{\UniCProp}}$ was designed to handle: each rule receives new instances that incorporate a limited and fixed amount of contraction. 
As a consequence, the coordinates that were previously $0$ can take on the value $1$. Hence, there is a rule instance of $r$ in $\UniHRuleAbsorb{\UniFLeExtSCalc{\UniCProp}}$ 
with conclusion $t_0'$ and premise $t' \UniWqoRel{} t$. 
By construction of $\mathcal{F}$, either this $t'$ was written down, or it was not written down because there is some node labelled $t''$ below (i.e., closer to the root) the node labelled $t_0'$ in $\PS_0$ such that $t'' \UniWqoRel{} t'$. In either case, we have found a node of $\mathcal{F}$-height $\leq K+1$ whose label is $\UniWqoRel{} t$.
\end{proof}

Define $\mathcal{P}_\UniMSetSucA:=\{u: (\UniSequent{u}{\UniMSetSucA}) \in \mathcal{P}\}$ and $\mathcal{F}_\UniMSetSucA:=\{v: (\UniSequent{v}{\UniMSetSucA}) \in \mathcal{F}\}$.
The above lemma implies 
$\mathcal{P} \subseteq \bigcup_{\UniMSetSucA\in \UniSubfmlaHyperseqSet} {\uparrow}\mathcal{F}_\UniMSetSucA$,  and the reverse subset inequality holds since $\mathcal{F}$ is a subtree of $\PS$, i.e., $\mathcal{F}_\UniMSetSucA \subseteq\mathcal{P}_\UniMSetSucA $, and $(\UniSequent{y}{\UniMSetSucA})\in\PS$ implies $(\UniSequent{z}{\UniMSetSucA})\in\PS$ for $z\in{\uparrow}{y}$, i.e., $\mathcal{P}_\UniMSetSucA$ is an upset.
Moreover, observe that
$P(y,z,\UniMSetSucA)=
\bigcup_{v \in \mathcal{F}_\UniMSetSucA} (v\dot{-}(y,z))$ for any $\UniMSetSucA\in \UniSubfmlaHyperseqSet$ and $y,z\in\mathbb{N}^{\UniSetCard{\UniSubfmlaHyperseqSet}}$.
Indeed,
$x \in P(y,z,\UniMSetSucA)$
iff
$\UniSequent{y,x,z}{\UniMSetSucA} \in \PS$
iff
$y,x,z 
\UniExtModRel{2}{1}^{\UniSetCard{\UniSubfmlaHyperseqSet}}
v$
for some $v \in \mathcal{F}$
iff
$x \in v\dot{-}(y,z)$
for some $v \in \mathcal{F}$
iff
$x \in 
\bigcup_{v \in \mathcal{F}}(v\dot{-}(y,z))$.

Consequently, (\ref{eq-bounding-main}) can be written as
\begin{equation}\label{eq-bounding-second}
\left\{
\left(\bigcup_{u \in U_\UniMSetSucA} (u \dot{-}(y,z))\right)
\cap
\left(\bigcup_{v \in \mathcal{F}_\UniMSetSucA} (v\dot{-}(y,z))\right)
\mid
 \UniMSetSucA\in \UniSubfmlaHyperseqSet, y,z\in\mathbb{N} ^{\UniSetCard{\UniSubfmlaHyperseqSet}}
\right\}.
\end{equation}
For $y\in\mathbb{N}^{\UniSetCard{\UniSubfmlaHyperseqSet}}$ and $C\in\mathbb{N}$, the \emph{$C$-truncation} of $y$, denoted by
$y^C\in\mathbb{N}^{\UniSetCard{\UniSubfmlaHyperseqSet}}$, is defined at each coordinate~$j$ as follows: $y^C_j$ is $C$ if $y_j\geq C$, else it is $y_j$. 
We denote by $\UniNorm{u}{\infty}$ the norm of the contraction nwqo $\UniWqoExtModRelProd{2}{1}{{\UniSetCard{\UniSubfmlaHyperseqSet}}}$, i.e., $\UniNorm{u}{\infty} = \max_j u_j$.

We then have that 
$u\dot{-} (y,z)=u\dot{-} (y^C,z^C)$ for $C\geq
\UniNorm{u}{\infty}+1$.
Indeed, for the $\subseteq$-direction, if $x \in u\dot{-} (y,z)$, we have
$u_j \UniExtModRel{2}{1} y_j,x_j,z_j$ for all $1 \leq j \leq \UniSetCard{\UniSubfmlaHyperseqSet}$.
Then, for any such $j$, if $y_j, z_j < C$, we have 
$y_j^C = y_j$ and $z_j^C = z_j$, thus 
$u_j \UniExtModRel{2}{1} y_j^C,x_j,z_j^C$.
If $y_j \geq C$ or $z_j \geq C$,
we have $y_j^C + x_j + z_j^C \geq C > \UniNorm{u}{\infty} \geq u_j$ and, since $C > 0$, we get $u_j \UniExtModRel{2}{1} y_j^C,x_j,z_j^C$.
Thus, for any $1 \leq j \leq \UniSetCard{\UniSubfmlaHyperseqSet}$,
we have $u_j \UniExtModRel{2}{1} y_j^C,x_j,z_j^C$,
so $x \in u \dot{-} (y^C,z^C)$.
For the $\supseteq$-direction, assume that 
$u_j \UniExtModRel{2}{1} y_j^C,x_j,z_j^C$
for all $1 \leq j \leq \UniSetCard{\UniSubfmlaHyperseqSet}$, and the result follows from the fact that
$w^C \UniExtModRel{2}{1}^{\UniSetCard{\UniSubfmlaHyperseqSet}} w$ for any $w \in \UniNaturalSet^{\UniSetCard{\UniSubfmlaHyperseqSet}}$.

We define $B_U \UniSymbDef 
\max_{\UniMSetSucA \in \UniSubfmlaHyperseqSet} \max_{u \in U_\UniMSetSucA}
\UniNorm{u}{\infty}$
and 
$B_{\mathcal{F}} \UniSymbDef 
\max_{\UniMSetSucA \in \UniSubfmlaHyperseqSet} \max_{v \in \mathcal{F}_\UniMSetSucA}
\UniNorm{v}{\infty}$.
Now (\ref{eq-bounding-second}) can be written as follows, where $B := 1+\max\{ B_U, B_{\mathcal{F}} \}$,
\[
\left\{
\left(\bigcup_{u \in U_\UniMSetSucA} (u\dot{-}(y,z))\right)
\cap
\left(\bigcup_{v \in \mathcal{F}_\UniMSetSucA} (v\dot{-}(y,z))\right)
\mid
 \UniMSetSucA\in \UniSubfmlaHyperseqSet, y,z\in\mathbb{N}^{\UniSetCard{\UniSubfmlaHyperseqSet}}, %\text{ s.t. }
 \UniNorm{y}{\infty},\UniNorm{z}{\infty}\leq B
\right\}.
\]
Clearly this set is finite, hence finiteness of the countermodel is established. 

\begin{corollary}
If a {formula/equation} fails in $\m{FL_{ec}}/\mathsf{FL_{ec}}$, then there exists a finite countermodel for it that is constructed via backwards proof search.
\end{corollary}
In this approach, bounding the size of the countermodel would follow from a bound on $B_U$ and $B_{\mathcal{F}}$. 
We have an Ackermannian bound for $B_{\mathcal{F}}$, but how to bound $B_U$? 
Since $B_U$ is the maximum of the norms of minimal elements of $U(\UniMSetSucA) := \{ y \in  \UniNaturalSet^{\UniSetCard{\UniSubfmlaHyperseqSet}}\mid \text{$\UniSequent{y}{\UniMSetSucA}$ is unprovable}\}$ ($\UniMSetSucA\in \UniSubfmlaHyperseqSet$), bounding $B_U$ seems to require a refined understanding of \emph{un}provability
in $\mathbf{FL_{ec}}$.

%% file: tex/contrib-table-new.tex
    \begin{table}[!tbh]
        \centering
        
        \caption{Complexities of substructural logics that have 
        an analytic hypersequent calculus.
        Columns `LB' and `UB' indicate lower and upper bounds.
        A cell marked `open'
        indicates that the status is unknown for that class of logics. `FMP' and `FEP' indicate finite model property and finite embeddability property, respectively. 
        The citation beside `PS' is a reference to a terminating proof calculus.
        The contributions of this paper are marked in blue.
        }
        \label{tab:table-contrib}

        \scriptsize

%\Rotatebox{90}{%
        
         \begin{threeparttable}
        \begin{tabular}{@{}p{.4cm}llll|lll@{}}
             \toprule
             & Logic(s) & \multicolumn{3}{c}{Provability} & \multicolumn{3}{c}{Deducibility} \\
             \midrule
             & & Decidability & LB & UB & Decidability & LB & UB\\
             %& &\\
             \cmidrule{3-8}
              % & $\UniFLExtLogic{\UniCProp}$ & N\cite{horcik2016} & -- & -- & N\cite{horcik2015} & -- & --\\
             & $\UniFLeExtLogic{}$ & FMP\cite{OT99} PS\cite{Komori1986}   & \PSPACE\cite{horcik2011} & \PSPACE\cite{horcik2011} & undecidable~\cite{lincoln1992} & -- & --\\
              \parbox[t]{2mm}{\multirow{2}{*}[-28pt]{
             \rotatebox[origin=c]{90}{\makecell{Base logics}}
             }}
             & $\UniFLeExtLogic{\UniWProp}$ & FMP\cite{OT99} PS\cite{Komori1986}  & \PSPACE\cite{horcik2011} & \PSPACE\cite{horcik2011} & FEP\cite{vanalten2005} \UniHere{PS}(\ref{def-derive-sets})\tnote{b}   & \TOWER\cite{tanaka2022} & \TOWER\cite{tanaka2022}\\
             % & Any &  &  &  &  &  &  & \UniHere{$\UniFGHProbOneAppLevel{\omega}$} & \\
             & $\UniFLeExtLogic{\UniCProp}$ & FMP\cite{OT99} PS\cite{Kri59}   & $\UniFGHProbOneAppLevel{\omega}$\cite{urquhart1999} & $\UniFGHProbOneAppLevel{\omega}$\cite{urquhart1999} & FEP\cite{vanalten2005} PS\cite{Kri59}\tnote{a}   & $\UniFGHProbOneAppLevel{\omega}$\cite{urquhart1999} & $\UniFGHProbOneAppLevel{\omega}$\cite{urquhart1999}\tnote{a}\\
             & $\UniFLeExtLogic{\UniWeakCProp{m}{1}}, m > 2$ & FEP\cite{vanalten2005} PS\cite{hori1994}   & \UniHere{$\UniFGHProbOneAppLevel{\omega}$}(\ref{c: knotcontr+wc=Ack Logics})\tnote{a} & \UniHere{$\UniFGHProbOneAppLevel{\omega}$}(\ref{cor-ded-Fw}) & FEP\cite{vanalten2005} PS\cite{hori1994}   & \UniHere{$\UniFGHProbOneAppLevel{\omega}$}(\ref{c: knotcontr+wc=Ack Logics}) & \UniHere{$\UniFGHProbOneAppLevel{\omega}$}(\ref{cor-ded-Fw})\\
             & $\UniFLeExtLogic{\UniWeakCProp{m}{n}}, n \geq 2$ & FEP\cite{vanalten2005} \UniHere{PS}(\ref{fact:flecmnr-ackermannian})   & \UniHere{$\UniFGHProbOneAppLevel{\omega}$}(\ref{c: knotcontr+wc=Ack Logics})\tnote{a} & \UniHere{$\UniFGHProbOneAppLevel{\omega}$}(\ref{cor-ded-Fw}) & FEP\cite{vanalten2005} \UniHere{PS}(\ref{fact:flecmnr-ackermannian})   & \UniHere{$\UniFGHProbOneAppLevel{\omega}$}(\ref{c: knotcontr+wc=Ack Logics}) & \UniHere{$\UniFGHProbOneAppLevel{\omega}$}(\ref{cor-ded-Fw})\\
             & $\UniFLweExtLogic{\vec a}{\UniWeakCProp{m}{n}}$ & FEP\cite{Cardona2015} \UniHere{PS}(\ref{fact:flewmnr-weak-com-ackermannian})   & 
             \PSPACE\cite{horcik2011}
             %\UniHere{$\UniFGHProbOneAppLevel{\omega}$} (Cor.~\ref{c: knotcontr+wc=Ack Logic})\tnote{c}
             & \UniHere{$\UniFGHProbOneAppLevel{\omega}$}(\ref{c: FLknotcontr+wc<Ack}) & FEP\cite{Cardona2015} \UniHere{PS}(\ref{fact:flewmnr-weak-com-ackermannian})   & \UniHere{$\UniFGHProbOneAppLevel{\omega}$}(\ref{c: knotcontr+wc=Ack Logics}) & \UniHere{$\UniFGHProbOneAppLevel{\omega}$}(\ref{c: FLknotcontr+wc<Ack})\\
             & $\UniFLeExtLogic{\UniWeakWProp{1}{n}}, n \geq 2$ & FEP\cite{vanalten2005} PS\cite{hori1994}   & \PSPACE\cite{horcik2011} & \PSPACE\cite{horcik2011} & FEP\cite{vanalten2005} \UniHere{PS}(\ref{def-derive-sets})   & \UniHere{$\UniFGHProbOneAppLevel{\omega}$}(\ref{c: knotweak+wc=Ack Logics}) & \UniHere{$\UniFGHProbOneAppLevel{\omega}$}(\ref{c: FLew(m,n) complexity})\\
             & $\UniFLeExtLogic{\UniWeakWProp{m}{n}}, m\geq 2$ & FEP\cite{vanalten2005} \UniHere{PS}(\ref{def-derive-sets})   & \PSPACE\cite{horcik2011} & \UniHere{$\UniFGHProbOneAppLevel{\omega}$}(\ref{c: FLew(m,n) complexity}) & FEP\cite{vanalten2005} \UniHere{PS}(\ref{def-derive-sets})   & \UniHere{$\UniFGHProbOneAppLevel{\omega}$}(\ref{c: knotweak+wc=Ack Logics}) & \UniHere{$\UniFGHProbOneAppLevel{\omega}$}(\ref{c: FLew(m,n) complexity})\\
             & $\UniFLweExtLogic{\vec a}{\UniWeakWProp{m}{n}}$ & FEP\cite{Cardona2015} \UniHere{PS}(\ref{def-derive-sets-wc})   & \PSPACE\cite{horcik2011} & \UniHere{$\UniFGHProbOneAppLevel{\omega}$}(\ref{cor:ub-flwewmn-logic}) & FEP\cite{Cardona2015} \UniHere{PS}(\ref{def-derive-sets-wc})    & \UniHere{$\UniFGHProbOneAppLevel{\omega}$}(\ref{c: knotweak+wc=Ack Logics}) & \UniHere{$\UniFGHProbOneAppLevel{\omega}$}(\ref{cor:ub-flwewmn-logic})\\
             & $\UniFLExtLogic{\UniIProp}$ & FMP\cite{OT99}
             PS\cite{Komori1986}
               & \PSPACE\cite{horcik2011} & \PSPACE\cite{horcik2011} & 
             FEP\cite{BvA02} PS\cite{GreatiRamanayake2024}  & $\UniFGHProbOneAppLevel{\omega^\omega}$\cite{GreatiRamanayake2024} & $\UniFGHProbOneAppLevel{\omega^\omega}$\cite{GreatiRamanayake2024}\\
              & $\UniFLExtLogic{\UniWeakCProp{m}{n}}$ & undecidable~\cite{horcik2016} & -- & -- & undecidable~\cite{horcik2015} & -- & --\\
              & $\UniFLExtLogic{\UniWeakWProp{1}{2}}$ & FMP\cite{GalJip13} PS\cite{horcik2011}& \PSPACE\cite{horcik2011} & \PSPACE\cite{horcik2011} & FEP\cite{horcik2017} & open & open\\
              & $\UniFLExtLogic{\UniWeakWProp{1}{n}}$ & FMP\cite{GalJip13} PS\cite{horcik2011} & \PSPACE\cite{horcik2011} & \PSPACE\cite{horcik2011} & open & open & open\\
              & $\UniFLExtLogic{\UniWeakWProp{m}{n}}, m > 1$ & open & \PSPACE\cite{horcik2011} & open & undecidable~\cite{horcik2015} & -- & --\\
             % & $\UniFLeExtLogic{} + x^n \leq (1\lor)x^{p_0} \lor 
             % $&&&&&&&&\\  &$
             % x^{p_1} \lor \bigvee_{i=2}^k x^{p_i}, k\geq 0$ & ? & ? & ? & ? & N\cite{galatos2022} & -- & -- & --\\
             % & $\UniFLeExtLogic{} + x^n \leq   \bigvee_{i=0}^k x^{n+p_i},
             %  $&&&&&&&&\\  &$
             % k \geq 2$ & N\cite{galatos2022} & -- & -- & -- & N\cite{galatos2022} & -- & -- & --\\
             % & $\UniFLeExtLogic{} + x^n \leq x^m \lor 1 $ & ? & ? & ? & ? & ? & ? & \UniHere{$\UniFGHProbOneAppLevel{\omega}$} & ?\\
             \midrule
             \parbox[t]{2mm}{\multirow{2}{*}[-18pt]{
             \rotatebox[origin=c]{90}{\makecell{$\mathcal{A} \subseteq \mathcal{N}_2$}}
             }}
             & $\UniAxiomExt{\UniFLeExtLogic{\UniCProp}}{ \UniAxiomSetA}$ & FEP\cite{Cardona2015}  PS\cite{revantha2020}   &  \UniHere{$\UniFGHProbOneAppLevel{\omega}$}(\ref{c: knotcontr+wc=Ack Logic})\tnote{c} & $\UniFGHProbOneAppLevel{\omega}$\cite{BalLanRam21LICS} & FEP\cite{Cardona2015} PS\cite{BalLanRam21LICS}\tnote{a}   & \UniHere{$\UniFGHProbOneAppLevel{\omega}$}(\ref{c: knotcontr+wc=Ack Logics})\tnote{c} &$\UniFGHProbOneAppLevel{\omega}$\cite{BalLanRam21LICS}\tnote{a}\\
              & $\UniAxiomExt{\UniFLeExtLogic{\UniWeakCProp{m}{n}}}{ \UniAxiomSetA}$ & FEP\cite{Cardona2015} \UniHere{PS}(\ref{fact:flecmnr-ackermannian})   & \UniHere{$\UniFGHProbOneAppLevel{\omega}$}(\ref{c: knotcontr+wc=Ack Logic})\tnote{c} & \UniHere{$\UniFGHProbOneAppLevel{\omega}$}(\ref{cor-ded-Fw}) & FEP\cite{Cardona2015} \UniHere{PS}(\ref{fact:flecmnr-ackermannian})   & \UniHere{$\UniFGHProbOneAppLevel{\omega}$}(\ref{c: knotcontr+wc=Ack Logics})\tnote{c} & \UniHere{$\UniFGHProbOneAppLevel{\omega}$}(\ref{cor-ded-Fw})\\
               & $\UniAxiomExt{\UniFLweExtLogic{\vec a}{\UniWeakCProp{m}{n}}}{ \UniAxiomSetA}$ & FEP\cite{Cardona2015} \UniHere{PS}(\ref{fact:flewmnr-weak-com-ackermannian})   & \UniHere{$\UniFGHProbOneAppLevel{\omega}$}(\ref{c: knotcontr+wc=Ack Logic})\tnote{c} & \UniHere{$\UniFGHProbOneAppLevel{\omega}$}(\ref{c: FLknotcontr+wc<Ack}) & FEP\cite{Cardona2015} \UniHere{PS}(\ref{fact:flewmnr-weak-com-ackermannian})   & \UniHere{$\UniFGHProbOneAppLevel{\omega}$}(\ref{c: knotcontr+wc=Ack Logics})\tnote{c} & \UniHere{$\UniFGHProbOneAppLevel{\omega}$}(\ref{c: FLknotcontr+wc<Ack})\\
              & $\UniAxiomExt{\UniFLeExtLogic{\UniWProp}}{ \UniAxiomSetA}$ & FEP\cite{Cardona2015} PS\cite{BalLanRam21LICS}   &  \PSPACE\cite{horcik2011} & $\UniFGHProbOneAppLevel{\omega}$\cite{BalLanRam21LICS} & FEP\cite{Cardona2015} \UniHere{PS}(\ref{def-derive-sets})  & \UniHere{$\UniFGHProbOneAppLevel{\omega}$}(\ref{c: knotweak+wc=Ack Logics})\tnote{c} &$\UniFGHProbOneAppLevel{\omega}$\cite{BalLanRam21LICS}\tnote{a}\\
              & $\UniAxiomExt{\UniFLeExtLogic{\UniWeakWProp{m}{n}}}{ \UniAxiomSetA}$ & FEP\cite{Cardona2015} \UniHere{PS}(\ref{def-derive-sets})   & \PSPACE\cite{horcik2011} & \UniHere{$\UniFGHProbOneAppLevel{\omega}$}(\ref{c: FLew(m,n) complexity}) & FEP\cite{Cardona2015} \UniHere{PS}(\ref{def-derive-sets})   & \UniHere{$\UniFGHProbOneAppLevel{\omega}$}(\ref{c: knotweak+wc=Ack Logics})\tnote{c} & \UniHere{$\UniFGHProbOneAppLevel{\omega}$}(\ref{c: FLew(m,n) complexity})\\
               & $\UniAxiomExt{\UniFLweExtLogic{\vec a}{\UniWeakWProp{m}{n}}}{ \UniAxiomSetA}$ & FEP\cite{Cardona2015} \UniHere{PS}(\ref{def-derive-sets-wc})   & \PSPACE\cite{horcik2011} & \UniHere{$\UniFGHProbOneAppLevel{\omega}$}(\ref{cor:ub-flwewmn-logic}) & FEP\cite{Cardona2015} \UniHere{PS}(\ref{def-derive-sets-wc})   & \UniHere{$\UniFGHProbOneAppLevel{\omega}$}(\ref{c: knotweak+wc=Ack Logics})\tnote{c} & \UniHere{$\UniFGHProbOneAppLevel{\omega}$}(\ref{cor:ub-flwewmn-logic})\\
               & $\UniAxiomExt{\UniFLExtLogic{\UniIProp}}{\UniAxiomSetA}$ & FEP\cite{GalJip13} PS\cite{GreatiRamanayake2024}   & \PSPACE\cite{horcik2011} & $\UniFGHProbOneAppLevel{\omega^\omega}$\cite{GreatiRamanayake2024} & FEP\cite{GalJip13} PS\cite{GreatiRamanayake2024}   & \PSPACE\cite{horcik2011} & $\UniFGHProbOneAppLevel{\omega^\omega}$\cite{GreatiRamanayake2024}\\
             \midrule
             \parbox[t]{2mm}{\multirow{2}{*}[-18pt]{
             \rotatebox[origin=c]{90}{\makecell{$\mathcal{A} \subseteq \mathcal{P}_3^\flat$}}
             }}
             & $\UniAxiomExt{\UniFLeExtLogic{\UniCProp}}{ \UniAxiomSetA}$ & \UniHere{FEP}(\ref{c: algebraicFEP}) PS\cite{BalLanRam21LICS}   &  open & $\UniFGHProbOneAppLevel{\omega^\omega}$\cite{BalLanRam21LICS} & \UniHere{FEP}(\ref{c: algebraicFEP}) PS\cite{BalLanRam21LICS}\tnote{a}  & open &$\UniFGHProbOneAppLevel{\omega^\omega}$\cite{BalLanRam21LICS}\tnote{a}\\
              & $\UniAxiomExt{\UniFLeExtLogic{\UniWeakCProp{m}{n}}}{ \UniAxiomSetA}$ & \UniHere{FEP}(\ref{c: algebraicFEP}) \UniHere{PS}(\ref{fact:flecmnr-ackermannian})  & open & \UniHere{$\UniFGHProbOneAppLevel{\omega^\omega}$}(\ref{cor-ded-Fw}) & \UniHere{FEP}(\ref{c: algebraicFEP}) \UniHere{PS}(\ref{fact:flecmnr-ackermannian})  & open & \UniHere{$\UniFGHProbOneAppLevel{\omega^\omega}$}(\ref{cor-ded-Fw})\\
               & $\UniAxiomExt{\UniFLweExtLogic{\vec a}{\UniWeakCProp{m}{n}}}{ \UniAxiomSetA}$ & \UniHere{FEP}(\ref{c: algebraicFEP}) \UniHere{PS}(\ref{fact:flewmnr-weak-com-ackermannian})  & open & \UniHere{$\UniFGHProbOneAppLevel{\omega^\omega}$}(\ref{c: FLknotcontr+wc<Ack}) & \UniHere{FEP}(\ref{c: algebraicFEP}) \UniHere{PS}(\ref{fact:flewmnr-weak-com-ackermannian})  & open & \UniHere{$\UniFGHProbOneAppLevel{\omega^\omega}$}(\ref{c: FLknotcontr+wc<Ack})\\
              & $\UniAxiomExt{\UniFLeExtLogic{\UniWProp}}{ \UniAxiomSetA}$ & \UniHere{FEP}(\ref{c: algebraicFEP}) PS\cite{BalLanRam21LICS}  &  open & $\UniFGHProbOneAppLevel{\omega^\omega}$\cite{BalLanRam21LICS} & \UniHere{FEP}(\ref{c: algebraicFEP}) \UniHere{PS}(\ref{def-derive-sets})   & open &$\UniFGHProbOneAppLevel{\omega^\omega}$\cite{BalLanRam21LICS}\tnote{a}\\
              & $\UniAxiomExt{\UniFLeExtLogic{\UniWeakWProp{m}{n}}}{ \UniAxiomSetA}$ & \UniHere{FEP}(\ref{c: algebraicFEP}) \UniHere{PS}(\ref{def-derive-sets})  & open & \UniHere{$\UniFGHProbOneAppLevel{\omega^\omega}$}(\ref{c: FLew(m,n) complexity}) & \UniHere{FEP}(\ref{c: algebraicFEP}) \UniHere{PS}(\ref{def-derive-sets})  & open & \UniHere{$\UniFGHProbOneAppLevel{\omega^\omega}$}(\ref{c: FLew(m,n) complexity})\\
              & $\UniAxiomExt{\UniFLweExtLogic{\vec a}{\UniWeakWProp{m}{n}}}{ \UniAxiomSetA}$ & \UniHere{FEP}(\ref{c: algebraicFEP}) \UniHere{PS}(\ref{def-derive-sets-wc})  & open & \UniHere{$\UniFGHProbOneAppLevel{\omega^\omega}$}(\ref{cor:ub-flwewmn-logic}) & \UniHere{FEP}(\ref{c: algebraicFEP}) \UniHere{PS}(\ref{def-derive-sets-wc})  & open & \UniHere{$\UniFGHProbOneAppLevel{\omega^\omega}$}(\ref{cor:ub-flwewmn-logic})\\
              & $\UniAxiomExt{\UniFLExtLogic{\UniIProp}}{\UniAxiomSetA}$ & \UniHere{FEP}(\ref{c: algebraicFEP}) \UniHere{PS}(\ref{def-derive-sets-hfli})  & open  & \UniHere{$\UniFGHProbOneAppLevel{\omega^{\omega^{\omega^\omega}}}$}(\ref{the:calc-compl-results}) & \UniHere{FEP}(\ref{c: algebraicFEP}) \UniHere{PS}(\ref{def-derive-sets-hfli})  & open &\UniHere{$\UniFGHProbOneAppLevel{\omega^{\omega^{\omega^\omega}}}$}(\ref{the:calc-compl-results})\\
             \bottomrule
        \end{tabular}

        \begin{tablenotes}
            \item[a] By a deduction theorem for the corresponding logic.
            %Even though \cite{urquhart1999,BalLanRam21LICS}
            % do not mention deducibility,
            % their results for provability extends
            % easily to deducibility in view of
            % the deduction theorem  (which also holds for
            %  axiomatic extensions).
             \item[b] \cite{tanaka2022} 
             provides
            a proof-search procedure that demands
             a translation to another
             logic, while ours does not demand translations.
             \item[c]
             {Under strict conditions on $\UniAxiomSetA$, described in the linked references/results. For the other cases, the
PSPACE~\cite{horcik2011} lower bound applies {for non-trivial logics (see Lemma~\ref{l: HorcikTerui})}.
}
        \end{tablenotes}
    \end{threeparttable}
%}%
\end{table}

%% file: arxiv-NG-VG-RR-SJ.bbl
\providecommand{\bysame}{\leavevmode\hbox to3em{\hrulefill}\thinspace}
\providecommand{\MR}{\relax\ifhmode\unskip\space\fi MR }
\providecommand{\MRhref}[2]{%
  \href{http://www.ams.org/mathscinet-getitem?mr=#1}{#2}
}
\providecommand{\href}[2]{#2}
\begin{thebibliography}{CDOY11}

\bibitem[AB75]{AB1}
Alan~Ross Anderson and Nuel Belnap, Jr., \emph{Entailment}, Princeton University Press, Princeton, N.J.-London, 1975, Volume I: The logic of relevance and necessity, With contributions by J. Michael Dunn and Robert K. Meyer, and further contributions by John R. Chidgey, J. Alberto Coffa, Dorothy L. Grover, Bas van Fraassen, Hugues LeBlanc, Storrs McCall, Zane Parks, Garrel Pottinger, Richard Routley, Alasdair Urquhart and Robert G. Wolf. \MR{406756}

\bibitem[ABD92]{AB2}
Alan~Ross Anderson, Nuel~D. Belnap, Jr., and J.~Michael Dunn, \emph{Entailment. {T}he logic of relevance and necessity. {V}ol. {II}}, Princeton University Press, Princeton, NJ, 1992, With contributions by Kit Fine, Alasdair Urquhart et al., Includes a bibliography of entailment by Robert G. Wolf. \MR{1223997}

\bibitem[AF88]{AF}
Marlow Anderson and Todd Feil, \emph{Lattice-ordered groups, an introduction}, Reidel Texts in the Mathematical Sciences, D. Reidel Publishing Co., Dordrecht, 1988. \MR{937703}

\bibitem[AF14]{abriola2014}
Sergio Abriola and Santiago Figueira, \emph{A note on the order type of minoring orderings and some algebraic properties of $\omega^2$-well quasi-orderings}, 2014 XL Latin American Computing Conference (CLEI), 2014, pp.~1--9.

\bibitem[AFS15]{abriola2015}
Sergio Abriola, Santiago Figueira, and Gabriel Senno, \emph{Linearizing well quasi-orders and bounding the length of bad sequences}, Theoretical Computer Science \textbf{603} (2015), 3--22.

\bibitem[AG25]{AG}
Kempton Albee and Nikolaos Galatos, \emph{A recipe for canonical formulas in substructural logics.}, submitted.

\bibitem[AGN97]{Andrka1997}
Hajnal Andréka, Steven Givant, and István Németi, \emph{Decision problems for equational theories of relation algebras}, Memoirs of the American Mathematical Society \textbf{126} (1997), no.~604, 0–0.

\bibitem[Avr87]{Avr87}
Arnon Avron, \emph{A constructive analysis of {RM}}, J. of Symbolic Logic \textbf{52} (1987), no.~4, 939--951.

\bibitem[Bal20]{bala2020}
A.~R. Balasubramanian, \emph{Complexity of controlled bad sequences over finite sets of $\mathbb{N}^d$}, ACM International Conference Proceeding Series, Association for Computing Machinery, 7 2020, pp.~130--140.

\bibitem[Bel82]{Be}
Nuel Belnap, Jr., \emph{Display logic}, J. Philos. Logic \textbf{11} (1982), no.~4, 375--417. \MR{691450}

\bibitem[BG11]{Bochman2011}
Alexander Bochman and Dov~M. Gabbay, \emph{Sequential dynamic logic}, Journal of Logic, Language and Information \textbf{21} (2011), no.~3, 279–298.

\bibitem[BLR21]{BalLanRam21LICS}
A.~R. Balasubramanian, Timo Lang, and Revantha Ramanayake, \emph{Decidability and complexity in weakening and contraction hypersequent substructural logics}, 36th Annual {ACM/IEEE} Symposium on Logic in Computer Science, {LICS} 2021, Rome, Italy, June 29 - July 2, 2021, {IEEE}, 2021, pp.~1--13.

\bibitem[BP89]{BlokPig1989}
W.~J. Blok and Don Pigozzi, \emph{Algebraizable logics}, Memoirs of the American Mathematical Society \textbf{77} (1989), no.~396, 1--78.

\bibitem[Bra97]{brauner1997}
Torben Braüner, \emph{A general adequacy result for a linear functional language}, Theoretical Computer Science \textbf{177} (1997), no.~1, 27--58.

\bibitem[BS11]{BuSa00}
S.~Burris and H.~P. Sankappanavar, \emph{A course in universal algebra}, Springer, 2011.

\bibitem[Bus91]{Bu2}
Wojciech Buszkowski, \emph{On generative capacity of the {L}ambek calculus}, Logics in {AI} ({A}msterdam, 1990), Lecture Notes in Comput. Sci., vol. 478, Springer, Berlin, 1991, pp.~139--152. \MR{1099626}

\bibitem[Bus18]{Bu1}
\bysame, \emph{Categorical grammars and their logics}, The {L}vov-{W}arsaw school. {P}ast and present, Stud. Univers. Log., Birkh\"auser/Springer, Cham, 2018, pp.~91--115. \MR{3838711}

\bibitem[BvA02]{BvA02}
W.~J. Blok and Clint van {A}lten, \emph{The finite embeddability property for residuated lattices, pocrims and bck-algebras}, Algebra Universalis \textbf{48} (2002), 253--271.

\bibitem[CDM00]{Mu}
Roberto L.~O. Cignoli, Itala M.~L. D'Ottaviano, and Daniele Mundici, \emph{Algebraic foundations of many-valued reasoning}, Trends in Logic---Studia Logica Library, vol.~7, Kluwer Academic Publishers, Dordrecht, 2000. \MR{1786097}

\bibitem[CDOY11]{CalDisOHeYan2011}
Cristiano Calcagno, Dino Distefano, Peter~W. O’Hearn, and Hongseok Yang, \emph{Compositional shape analysis by means of bi-abduction}, J. ACM \textbf{58} (2011), no.~6.

\bibitem[CG15]{Cardona2015}
R.~Cardona and N.~Galatos, \emph{The finite embeddability property for noncommutative knotted extensions of {RL}}, International Journal of Algebra and Computation \textbf{25} (2015), no.~03, 349--379.

\bibitem[CGT08]{CiaGalTer08}
Agata Ciabattoni, Nikolaos Galatos, and Kazushige Terui, \emph{From axioms to analytic rules in nonclassical logics}, {LICS} 2008, 2008, pp.~229--240.

\bibitem[CGT12]{CiaGalTer12}
Agata Ciabattoni, Nikolaos Galatos, and Kazushige Terui, \emph{Algebraic proof theory for substructural logics: cut-elimination and completions}, Ann. Pure Appl. Logic \textbf{163} (2012), no.~3, 266--290.

\bibitem[CGT17]{CiaGalTer17}
Agata Ciabattoni, Nikolaos Galatos, and Kazushige Terui, \emph{Algebraic proof theory: Hypersequents and hypercompletions}, Ann. Pure Appl. Logic \textbf{168} (2017), no.~3, 693--737.

\bibitem[CH16]{horcik2016}
Karel Chvalovsk\'y and Rostislav Hor{\v c}{\'\i}k, \emph{Full lambek calculus with contraction is undecidable}, The Journal of Symbolic Logic \textbf{81} (2016), no.~2, 524--540.

\bibitem[CM17]{ColMet2017}
Almudena Colacito and George Metcalfe, \emph{Proof theory and ordered groups}, Logic, Language, Information, and Computation (Berlin, Heidelberg) (Juliette Kennedy and Ruy~J.G.B. de~Queiroz, eds.), Springer Berlin Heidelberg, 2017, pp.~80--91.

\bibitem[CMM10]{CiaMetMon10}
Agata Ciabattoni, George Metcalfe, and Franco Montagna, \emph{Algebraic and proof-theoretic characterizations of truth stressers for {MTL} and its extensions}, Fuzzy Sets Syst. \textbf{161} (2010), no.~3, 369--389.

\bibitem[CS14]{CourtSchmitz2014}
Jean-Baptiste Courtois and Sylvain Schmitz, \emph{Alternating vector addition systems with states}, Mathematical Foundations of Computer Science 2014 (Berlin, Heidelberg) (Erzs{\'e}bet Csuhaj-Varj{\'u}, Martin Dietzfelbinger, and Zolt{\'a}n {\'E}sik, eds.), Springer Berlin Heidelberg, 2014, pp.~220--231.

\bibitem[Dar95]{Da}
Michael~R. Darnel, \emph{Theory of lattice-ordered groups}, Monographs and Textbooks in Pure and Applied Mathematics, vol. 187, Marcel Dekker, Inc., New York, 1995. \MR{1304052}

\bibitem[DGK93]{darlington1993}
John Darlington, Yi-ke Guo, and Martin K{\"o}hler, \emph{Functional programming languages with logical variables: A linear logic view}, Progamming Language Implementation and Logic Programming (Berlin, Heidelberg) (Maurice Bruynooghe and Jaan Penjam, eds.), Springer Berlin Heidelberg, 1993, pp.~201--219.

\bibitem[Dic13]{dickson1913}
Leonard~Eugene Dickson, \emph{Finiteness of the odd perfect and primitive abundant numbers with $n$ distinct prime factors}, American Journal of Mathematics \textbf{35} (1913), no.~4, 413--422.

\bibitem[Dil39]{Di}
R.~P. Dilworth, \emph{Non-commutative residuated lattices}, Trans. Amer. Math. Soc. \textbf{46} (1939), 426--444. \MR{230}

\bibitem[DSH93]{dosen1993}
Kosta Dosen and Peter Schroder-Heister (eds.), \emph{Substructural logics}, Clarendon Press, 1993.

\bibitem[EG01]{EstGod01}
Francesc Esteva and Llu{\i}s Godo, \emph{Monoidal t-norm based logic: towards a logic for left-continuous t-norms}, Fuzzy Sets and Systems \textbf{124} (2001), no.~3, 271 -- 288, Fuzzy Logic.

\bibitem[EGN10]{EstGodNog10}
Francesc Esteva, Llu{\i}s Godo, and Carles Noguera, \emph{On expansions of wnm t-norm based logics with truth-constants}, Fuzzy Sets and Systems \textbf{161} (2010), no.~3, 347 -- 368, Fuzzy Logics and Related Structures.

\bibitem[EGRS04]{llcs2004}
Thomas Ehrhard, Jean-Yves Girard, Paul Ruet, and Philip Scott (eds.), \emph{Linear logic in computer science}, Cambridge University Press, November 2004.

\bibitem[FFSS11]{figueira2011}
Diego Figueira, Santiago Figueira, Sylvain Schmitz, and Philippe Schnoebelen, \emph{{A}ckermannian and primitive-recursive bounds with {D}ickson's {L}emma}, Proceedings - Symposium on Logic in Computer Science, 2011, pp.~269--278.

\bibitem[Fis66]{FISCHER66}
Patrick~C. Fischer, \emph{Turing machines with restricted memory access}, Information and Control \textbf{9} (1966), no.~4, 364--379.

\bibitem[FMR68]{FMR68}
Patrick~C. Fischer, Albert~R. Meyer, and Arnold~L. Rosenberg, \emph{Counter machines and counter languages}, Mathematical Systems Theory \textbf{2} (1968), 265--283.

\bibitem[FS01]{finkel2001}
A.~Finkel and Philippe Schnoebelen, \emph{Well-structured transition systems everywhere!}, Theoretical Computer Science \textbf{256} (2001), no.~1, 63--92, ISS.

\bibitem[Gal02]{Ga02b}
Nikolaos Galatos, \emph{The undecidability of the word problem for distributive residuated lattices}, Ordered algebraic structures, Dev. Math., vol.~7, Kluwer Acad. Publ., Dordrecht, 2002, pp.~231--243. \MR{2083042}

\bibitem[Gal04]{Ga04a}
\bysame, \emph{Equational bases for joins of residuated-lattice varieties}, Studia Logica \textbf{76} (2004), no.~2, 227--240. \MR{2072984}

\bibitem[Gen35]{Gentzen1935}
Gerhard Gentzen, \emph{Untersuchungen {\"u}ber das logische schlie{\ss}en. {I}}, Mathematische Zeitschrift \textbf{39} (1935), no.~1, 176--210.

\bibitem[Gen69]{Gen69}
\bysame, \emph{The collected papers of {G}erhard {G}entzen}, Edited by M. E. Szabo. Studies in Logic and the Foundations of Mathematics, North-Holland Publishing Co., Amsterdam, 1969.

\bibitem[Gir87]{Gi}
Jean-Yves Girard, \emph{Linear logic}, Theoret. Comput. Sci. \textbf{50} (1987), no.~1, 101. \MR{899269}

\bibitem[GJ13]{GalJip13}
Nikolaos Galatos and Peter Jipsen, \emph{Residuated frames with applications to decidability}, Trans. Amer. Math. Soc. \textbf{365} (2013), no.~3, 1219--1249. \MR{3003263}

\bibitem[GJKO07]{GalJipKowOno07}
Nikolaos Galatos, Peter Jipsen, Tomasz Kowalski, and Hiroakira Ono, \emph{Residuated lattices: an algebraic glimpse at substructural logics}, Studies in Logic and the Foundations of Mathematics, vol. 151, Elsevier, 2007.

\bibitem[GO10]{GalOno10}
Nikolaos Galatos and Hiroakira Ono, \emph{Cut elimination and strong separation for substructural logics: An algebraic approach}, Annals of Pure and Applied Logic \textbf{161} (2010), no.~9, 1097--1133.

\bibitem[GOR08]{GOR2008}
Nikolaos Galatos, Jeffrey~S. Olson, and James~G. Raftery, \emph{Irreducible residuated semilattices and finitely based varieties}, Rep. Math. Logic (2008), no.~43, 85--108. \MR{2417724}

\bibitem[GR04]{GR04}
Nikolaos Galatos and James~G. Raftery, \emph{Adding involution to residuated structures}, Studia Logica \textbf{77} (2004), no.~2, 181--207. \MR{2080238}

\bibitem[GR24]{GreatiRamanayake2024}
Vitor Greati and Revantha Ramanayake, \emph{Deducibility in the {F}ull {L}ambek {C}alculus with weakening is {HAck}-complete}, Advances in Modal Logic, AiML 2024, Prague, Czech Republic, August 19-23, 2024 (Agata Ciabattoni, David Gabelaia, and Igor Sedl{\'{a}}r, eds.), College Publications, 2024, pp.~401--422.

\bibitem[GSJ22]{galatos2022}
Nikolaos Galatos and Gavin St.~John, \emph{Most simple extensions of $\textbf{FL}_{\textbf{e}}$ are undecidable}, The Journal of Symbolic Logic \textbf{87} (2022), no.~3, 1156–1200.

\bibitem[H\'98]{Ha}
Petr H\'{a}jek, \emph{Metamathematics of fuzzy logic}, Trends in Logic---Studia Logica Library, vol.~4, Kluwer Academic Publishers, Dordrecht, 1998. \MR{1900263}

\bibitem[Han17]{Han17}
Zuzana Hanikov\'{a}, \emph{Complexity of some language fragments of fuzzy logics}, Soft Computing \textbf{21} (2017).

\bibitem[Hei95]{Morphology}
Henk J. A.~M. Heijmans, \emph{Mathematical morphology: a modern approach in image processing based on algebra and geometry}, SIAM Rev. \textbf{37} (1995), no.~1, 1--36. \MR{1327714}

\bibitem[Hig52]{Higman1952}
Graham Higman, \emph{Ordering by divisibility in abstract algebras}, Proc. Lond. Math. Soc. (3) \textbf{s3-2} (1952), no.~1, 326--336 (en).

\bibitem[HMT71]{HMT}
Leon Henkin, J.~Donald Monk, and Alfred Tarski, \emph{Cylindric algebras. {P}art {I}. {W}ith an introductory chapter: {G}eneral theory of algebras}, Studies in Logic and the Foundations of Mathematics, vol. Vol. 64, North-Holland Publishing Co., Amsterdam-London, 1971. \MR{314620}

\bibitem[HNP07]{HorNogPet07}
Rostislav Horc{\'{\i}}k, Carles Noguera, and Milan Petr{\i}k, \emph{On n-contractive fuzzy logics}, Mathematical Logic Quarterly \textbf{53} (2007), 268 -- 288.

\bibitem[Hor15]{horcik2015}
Rostislav Hor\v{c}\'{i}k, \emph{Word problem for knotted residuated lattices}, Journal of Pure and Applied Algebra \textbf{219} (2015), no.~5, 1548--1563.

\bibitem[Hor17]{horcik2017}
\bysame, \emph{On square-increasing ordered monoids and idempotent semirings}, Semigroup Forum (2017), no.~94, 297–313.

\bibitem[HOS94]{hori1994}
Ryuichi Hori, Hiroakira Ono, and Harold Schellinx, \emph{{Extending intuitionistic linear logic with knotted structural rules.}}, Notre Dame Journal of Formal Logic \textbf{35} (1994), no.~2, 219 -- 242.

\bibitem[HSS13]{hasse2013}
Christoph Haase, Sylvain Schmitz, and Philippe Schnoebelen, \emph{The power of priority channel systems}, CONCUR 2013 -- Concurrency Theory (Berlin, Heidelberg) (Pedro~R. D'Argenio and Hern{\'a}n Melgratti, eds.), Springer Berlin Heidelberg, 2013, pp.~319--333.

\bibitem[HT61]{HT}
Leon Henkin and Alfred Tarski, \emph{Cylindric algebras}, Proc. {S}ympos. {P}ure {M}ath., {V}ol. {II}, Amer. Math. Soc., Providence, RI, 1961, pp.~83--113. \MR{124250}

\bibitem[HT11]{horcik2011}
Rostislav Horčík and Kazushige Terui, \emph{Disjunction property and complexity of substructural logics}, Theoretical Computer Science \textbf{412} (2011), no.~31, 3992--4006.

\bibitem[Joh82]{Jo}
Peter~T. Johnstone, \emph{Stone spaces}, Cambridge Studies in Advanced Mathematics, vol.~3, Cambridge University Press, Cambridge, 1982. \MR{698074}

\bibitem[JT02]{JT02}
Peter Jipsen and Constantine Tsinakis, \emph{A survey of residuated lattices}, Ordered algebraic structures, Dev. Math., vol.~7, Kluwer Acad. Publ., Dordrecht, 2002, pp.~19--56. \MR{2083033}

\bibitem[KM94]{KM}
V.~M. Kopytov and N.~Ya. Medvedev, \emph{The theory of lattice-ordered groups}, Mathematics and its Applications, vol. 307, Kluwer Academic Publishers Group, Dordrecht, 1994. \MR{1369091}

\bibitem[Kom86]{Komori1986}
Yuichi Komori, \emph{Predicate logics without the structure rules}, Studia Logica \textbf{45} (1986), no.~4, 393--404.

\bibitem[Koz94]{KoAA}
Dexter Kozen, \emph{On action algebras}, Logic and information flow, Found. Comput. Ser., MIT Press, Cambridge, MA, 1994, pp.~78--88. \MR{1295061}

\bibitem[Kri59]{Kri59}
S.~Kripke, \emph{The problem of entailment (abstract)}, J. of Symbolic Logic \textbf{24} (1959), 324.

\bibitem[KS97]{KoKA}
Dexter Kozen and Frederick Smith, \emph{Kleene algebra with tests: completeness and decidability}, Computer science logic ({U}trecht, 1996), Lecture Notes in Comput. Sci., vol. 1258, Springer, Berlin, 1997, pp.~244--259. \MR{1611502}

\bibitem[KT03]{dexter2003}
Dexter Kozen and Jerzy Tiuryn, \emph{Substructural logic and partial correctness}, ACM Trans. Comput. Logic \textbf{4} (2003), no.~3, 355–378.

\bibitem[Lam58]{lambek1958}
Joachim Lambek, \emph{The mathematics of sentence structure}, The American Mathematical Monthly \textbf{65} (1958), no.~3, 154--170.

\bibitem[Lam93]{La1}
Joachim Lambek, \emph{From categorial grammar to bilinear logic}, Substructural logics ({T}\"ubingen, 1990), Stud. Logic Comput., vol.~2, Oxford Univ. Press, New York, 1993, pp.~207--237. \MR{1283198}

\bibitem[Lam08]{La2}
Joachim Lambek, \emph{Pregroup grammars and {C}homsky's earliest examples}, J. Log. Lang. Inf. \textbf{17} (2008), no.~2, 141--160. \MR{2373257}

\bibitem[LMSS92]{lincoln1992}
Patrick Lincoln, John Mitchell, Andre Scedrov, and Natarajan Shankar, \emph{Decision problems for propositional linear logic}, Annals of Pure and Applied Logic \textbf{56} (1992), no.~1, 239--311.

\bibitem[LS15]{LazSchmitz2015}
Ranko Lazi\'{c} and Sylvain Schmitz, \emph{Nonelementary complexities for branching {VASS}, {MELL}, and extensions}, ACM Trans. Comput. Logic \textbf{16} (2015), no.~3.

\bibitem[Mac94]{Mackie1994}
Ian Mackie, \emph{Lilac: a functional programming language based on linear logic}, Journal of Functional Programming \textbf{4} (1994), no.~4, 395–433.

\bibitem[Mar01]{marcone2001}
Alberto Marcone, \emph{Fine analysis of the quasi-orderings on the power set}, Order \textbf{18} (2001), no.~4, 339--347.

\bibitem[May03]{mayr2003}
Richard Mayr, \emph{Undecidable problems in unreliable computations}, Theoretical Computer Science \textbf{297} (2003), no.~1, 337--354, Latin American Theoretical Informatics.

\bibitem[Min68]{Min68}
G.~E. Minc, \emph{Some calculi of modal logic}, Trudy Mat. Inst. Steklov \textbf{98} (1968), 88--111.

\bibitem[MOG09]{metcalfe2009}
George Metcalfe, Nicola Olivetti, and Dov Gabbay, \emph{Proof theory for fuzzy logics}, Springer Netherlands, 2009.

\bibitem[MP01]{Mu1}
Christopher~J. Mulvey and Joan~Wick Pelletier, \emph{On the quantisation of points}, J. Pure Appl. Algebra \textbf{159} (2001), no.~2-3, 231--295. \MR{1828940}

\bibitem[MP02]{Mu2}
\bysame, \emph{On the quantisation of spaces}, vol. 175, 2002, Special volume celebrating the 70th birthday of Professor Max Kelly, pp.~289--325. \MR{1935983}

\bibitem[MPT23]{MPT}
George Metcalfe, Francesco Paoli, and Constantine Tsinakis, \emph{Residuated structures in algebra and logic}, Mathematical Surveys and Monographs, vol. 277, American Mathematical Society, Providence, RI, 2023. \MR{4684245}

\bibitem[Mul86]{Mu3}
Christopher~J. Mulvey, \emph{\&}, no.~12, 1986, Second topology conference (Taormina, 1984), pp.~99--104. \MR{853151}

\bibitem[NEG08]{NogEstGis08}
Carles Noguera, Francesc Esteva, and Joan Gispert, \emph{On triangular norm based axiomatic extensions of the weak nilpotent minimum logic}, Math. Log. Q. \textbf{54} (2008), 387--409.

\bibitem[NvPR01]{negri2001}
Sara Negri, Jan von Plato, and Aarne Ranta, \emph{Structural proof theory}, Cambridge University Press, 2001.

\bibitem[O'H18]{OHe2018}
Peter~W. O'Hearn, \emph{Continuous reasoning: Scaling the impact of formal methods}, Proceedings of the 33rd Annual ACM/IEEE Symposium on Logic in Computer Science (New York, NY, USA), LICS '18, Association for Computing Machinery, 2018, p.~13–25.

\bibitem[Ono19]{Ono2019}
Hiroakira Ono, \emph{Proof theory and algebra in logic}, 01 2019.

\bibitem[OP99]{OPym}
Peter~W. O'Hearn and David~J. Pym, \emph{The logic of bunched implications}, Bull. Symbolic Logic \textbf{5} (1999), no.~2, 215--244. \MR{1791304}

\bibitem[OT99]{OT99}
Mitsuhiro Okada and Kazushige Terui, \emph{The finite model property for various fragments of intuitionistic linear logic}, The Journal of Symbolic Logic \textbf{64} (1999), no.~2, 790--802.

\bibitem[Pao02]{Pa}
Francesco Paoli, \emph{Substructural logics: a primer}, Trends in Logic---Studia Logica Library, vol.~13, Kluwer Academic Publishers, Dordrecht, 2002. \MR{2039844}

\bibitem[Pot83]{Pot83}
G.~Pottinger, \emph{Uniform, cut-free formulations of {T}, {S4} and {S5} (abstract)}, J. of Symbolic Logic \textbf{48} (1983), no.~3, 900.

\bibitem[PP12]{PP}
Jorge Picado and Ale\v~s Pultr, \emph{Frames and locales, topology without points}, Frontiers in Mathematics, Birkh\"auser/Springer Basel AG, Basel, 2012. \MR{2868166}

\bibitem[Pra91]{Pratt}
Vaughan Pratt, \emph{Action logic and pure induction}, Logics in {AI} ({A}msterdam, 1990), Lecture Notes in Comput. Sci., vol. 478, Springer, Berlin, 1991, pp.~97--120. \MR{1099624}

\bibitem[Pym02]{Pym}
David~J. Pym, \emph{The semantics and proof theory of the logic of bunched implications}, Applied Logic Series, vol.~26, Kluwer Academic Publishers, Dordrecht, 2002, With a foreword by Dov M. Gabbay. \MR{2332826}

\bibitem[Pé67]{roszapeter1967}
Rózsa Péter, \emph{Recursive functions}, Academic Press, 1967.

\bibitem[Rad54]{rado1954}
R.~Rado, \emph{Partial well-ordering of sets of vectors}, Mathematika \textbf{1} (1954), no.~2, 89–95.

\bibitem[Ram20]{revantha2020}
Revantha Ramanayake, \emph{Extended {K}ripke lemma and decidability for hypersequent substructural logics}, {LICS} '20: 35th Annual {ACM/IEEE} Symposium on Logic in Computer Science, Saarbr{\"{u}}cken, Germany, July 8-11, 2020 (Holger Hermanns, Lijun Zhang, Naoki Kobayashi, and Dale Miller, eds.), {ACM}, 2020, pp.~795--806.

\bibitem[Res00]{Restall}
Greg Restall, \emph{An introduction to substructural logics}, Routledge, 2000.

\bibitem[Rey02]{reynolds2002}
J.C. Reynolds, \emph{Separation logic: a logic for shared mutable data structures}, Proceedings 17th Annual IEEE Symposium on Logic in Computer Science, 2002, pp.~55--74.

\bibitem[RR97]{DellaRocca1997}
Simona Ronchi~Della Rocca and Luca Roversi, \emph{Lambda calculus and intuitionistic linear logic}, Studia Logica \textbf{59} (1997), no.~3, 417--448.

\bibitem[Sch02]{schnoebelen2002}
Philippe Schnoebelen, \emph{Verifying lossy channel systems has nonprimitive recursive complexity}, Information Processing Letters \textbf{83} (2002), no.~5, 251--261.

\bibitem[Sch10]{schnoebelen2010}
\bysame, \emph{Revisiting {A}ckermann-hardness for lossy counter machines and reset {P}etri nets}, Mathematical Foundations of Computer Science 2010 (Berlin, Heidelberg) (Petr Hlin{\v{e}}n{\'y} and Anton{\'i}n Ku{\v{c}}era, eds.), Springer Berlin Heidelberg, 2010, pp.~616--628.

\bibitem[Sch16]{schmitz2016hierar}
Sylvain Schmitz, \emph{Complexity hierarchies beyond elementary}, ACM Trans. Comput. Theory \textbf{8} (2016), no.~1.

\bibitem[Sch17]{schmitzthesis}
\bysame, \emph{{Algorithmic Complexity of Well-Quasi-Orders}}, Habilitation {\`a} diriger des recherches, {{\'E}cole normale sup{\'e}rieure Paris-Saclay}, November 2017.

\bibitem[Sed24]{Sedlr2024}
Igor Sedlár, \emph{Implicational {K}leene algebra with domain and the substructural logic of partial correctness}, Mathematical Structures in Computer Science \textbf{34} (2024), no.~7, 645–660.

\bibitem[SJ19]{gavin2019}
Gavin St.~John, \emph{Decidability for residuated lattices and substructural logics}, Ph.D. thesis, University of Denver, 2019.

\bibitem[SS11]{schmitzschnoebelen2011}
Sylvain Schmitz and Philippe Schnoebelen, \emph{Multiply-recursive upper bounds with {H}igman's lemma}, Automata, Languages and Programming (Berlin, Heidelberg) (Luca Aceto, Monika Henzinger, and Ji{\v{r}}{\'i} Sgall, eds.), Springer Berlin Heidelberg, 2011, pp.~441--452.

\bibitem[SS12]{schmitz2012notes}
\bysame, \emph{{Algorithmic Aspects of WQO Theory}}, Lecture, August 2012.

\bibitem[SSJ25]{SpadaStJohn2024}
Luca Spada and Gavin St.~John, \emph{Weil 2-rigs}, Theory and Applications of Categories \textbf{43} (2025), no.~12, 382--402.

\bibitem[SSW20]{schuster2020}
Peter~M. Schuster, Monika Seisenberger, and Andreas Weiermann, \emph{Well-quasi orders in computation, logic, language and reasoning}, vol.~53, Springer Cham, 2020.

\bibitem[Sta79]{statman1979}
Richard Statman, \emph{Intuitionistic propositional logic is polynomial-space complete}, Theoretical Computer Science \textbf{9} (1979), no.~1, 67--72.

\bibitem[Ste12]{Stell}
John~G. Stell, \emph{Relations on hypergraphs}, Relational and algebraic methods in computer science, Lecture Notes in Comput. Sci., vol. 7560, Springer, Heidelberg, 2012, pp.~326--341. \MR{3040175}

\bibitem[Tan23]{tanaka2022}
Hiromi Tanaka, \emph{Tower-complete problems in contraction-free substructural logics}, 31st {EACSL} Annual Conference on Computer Science Logic, {CSL} 2023, February 13-16, 2023, Warsaw, Poland (Bartek Klin and Elaine Pimentel, eds.), LIPIcs, vol. 252, Schloss Dagstuhl - Leibniz-Zentrum f{\"{u}}r Informatik, 2023, pp.~34:1--34:19.

\bibitem[Tar41]{Ta}
Alfred Tarski, \emph{On the calculus of relations}, J. Symbolic Logic \textbf{6} (1941), 73--89. \MR{5280}

\bibitem[Tro92]{Troelstra1992}
Anne~Sjerp Troelstra, \emph{Lectures on linear logic}, Center for the Study of Language and Information Publications, 1992.

\bibitem[Urq84]{Urq1984}
Alasdair Urquhart, \emph{The undecidability of entailment and relevant implication}, The Journal of Symbolic Logic \textbf{49} (1984), no.~4, 1059--1073.

\bibitem[Urq99]{urquhart1999}
\bysame, \emph{The complexity of decision procedures in relevance logic {II}}, The Journal of Symbolic Logic \textbf{64} (1999), no.~4, 1774--1802.

\bibitem[vA05]{vanalten2005}
Clint van Alten, \emph{The finite model property for knotted extensions of propositional linear logic}, Journal of Symbolic Logic \textbf{70} (2005), 84--98.

\bibitem[vAR04]{AltenClintRaft2004}
Clint van Alten and James~G. Raftery, \emph{Rule separation and embedding theorems for logics without weakening}, Studia Logica \textbf{76} (2004), no.~2, 241--274.

\bibitem[vB91]{vBe}
Johan van Benthem, \emph{Language in action}, Studies in Logic and the Foundations of Mathematics, vol. 130, North-Holland Publishing Co., Amsterdam, 1991, Categories, lambdas and dynamic logic. \MR{1102016}

\bibitem[War38]{Wa}
Morgan Ward, \emph{Structure residuation}, Ann. of Math. (2) \textbf{39} (1938), no.~3, 558--568. \MR{1503424}

\bibitem[WD39]{WD}
Morgan Ward and R.~P. Dilworth, \emph{Residuated lattices}, Trans. Amer. Math. Soc. \textbf{45} (1939), no.~3, 335--354. \MR{1501995}

\end{thebibliography}
